# Reconfigurable Structures for Direct Equalisation in Mobile Receivers

Variable Length Linear and Decision Feedback Equalisers for Power-Constrained Receivers Operating in Time-Varying Environments

Felip RIERA  PALOU

Submitted for the degree of Doctor of Philosophy

2002

Department of Electronics and Telecommunications

University of Bradford

# Reconfigurable Structures for Direct Equalisation in Mobile Receivers.

## Variable Length Linear and Decision Feedback Equalisers for Power-Constrained Receivers Operating in Time-Varying Environments.

Any communication channel will usually distort the transmitted signal. This is especially true in the case of mobile systems, where multipath propagation causes the received signal to be seriously degraded.

Over the years, many techniques have been proposed to combat channel effects. Two of the most popular are linear equalisation (LE) and decision feedback equalisation (DFE). These methods offer a good compromise between performance and computational complexity. LE and DFE are implemented using finite impulse response (FIR) filters whose frequency spectrum approximates the inverse of the channel spectrum plus noise. In mobile systems, the equaliser is made adaptable in order to be able to respond to the channel variations. Adaptability is achieved using adaptive FIR filters whose coefficients are iteratively updated.

In principle, an infinite number of filter coefficients would be needed to achieve perfect channel inversion. In practice, the number of taps must be finite. Simulations show that, in realistic scenarios, making the equaliser longer than a certain (undetermined) number of taps will not yield any benefit. Moreover, computation and power will be wasted.

In battery powered devices, like mobile terminals, it would be desirable to have the equaliser properly dimensioned. The equaliser's optimum length strongly depends on the particular scenario, and as channel conditions vary, this optimum is likely to vary.

This thesis presents novel techniques to perform equaliser length adjustment. Methods for the LE and the DFE have been developed. Simulations in many different scenarios show that the proposed schemes optimise the number of taps to be used. Moreover, these techniques are able to detect changes in the channel and re-adjust the equaliser length appropriately.

# Acknowledgments

I want to most gratefully thank my supervisor Dr James M. Noras for his guidance, help, support and encouragement throughout this research project. I also want to thank my lab colleague Kostas Chaikalis for useful discussions and for being such an easy-going lab mate. I hope, and I am sure, that the friendship with both of them will last beyond this project. I also extend my sincere gratitude to all academic and secretarial staff of the Department of Electronics and Telecommunications at the University of Bradford for their help during my time at Bradford. The financial support from MobileVCE during the first two years of this project is gratefully acknowledged.

It is a pleasure to acknowledge the support, during the whole project, of various members from the Department of Electronics and Electrical Engineering at the University of Edinburgh. Very specially, I want to thank Dr. David G. M. Cruickshank for the many fruitful discussions we have had over the last three years. Additionally, I am also grateful to Dr. Ian W. Band who provided crucial advice on different aspects of the simulation environment implementation.

Although not directly related to the PhD work, I want to express my appreciation to my undergraduate project supervisor Dr. Guillem Bernat, currently at the University of York, previously at the University of the Balearic Islands (UIB), for his support and encouragement during my last year at UIB to come to UK for further study.

I thank the many good friends I have made in Bradford for making my time here most enjoyable and to my great friends in Mallorca, who constantly make me wish go home and also, for their visits to Bradford. I consider myself extremely fortunate to have a wonderful family who have always encouraged me to study, especially during my unmotivated years (age 3 to 18), and it is to them that this thesis is dedicated. Finally, I want to thank Isabel for her full-time love, support and understanding.

Felip Riera Palou, Bradford, April 2002.

*A la meva família, i molt especialment, als meus pares*

## List of Acronyms

| | |
|---|---|
| 2G | $2^{nd}$ Generation (of mobile systems) |
| 3G | $3^{rd}$ Generation (of mobile systems) |
| 3GPP | $3^{rd}$ Generation Partnership Project |
| ADC | Analog to Digital Converter |
| ADPCM | Adaptive Differential Pulse Coded Modulation |
| AM | Amplitude Modulation |
| ASIC | Application Specific Integrated Circuit |
| ASK | Amplitude Shift Keying |
| AWGN | Additive White Gaussian Noise |
| BER | Bit Error Rate |
| BPSK | Binary Phase Shift Keying |
| CDMA | Code Division Multiple Access |
| CIR | Channel Impulse Response |
| DFE | Decision Feedback Equaliser |
| DQPSK | Differential Quaternary Phase Shift Keying |
| DSP | Digital Signal Processor/Processing |
| EMSE | Excess Mean Square Error |
| FAEST | Fast A-priori Error Sequential Technique |
| FBF | FeedBack Filter |
| FDD | Frequency Division Duplex |
| FDMA | Frequency Division Multiple Access |
| FFF | FeedForward Filter |
| FM | Frequency Modulation |
| FPGA | Floating Programmable Gate Array |
| FSE | Fractionally Spaces Equaliser |
| FSK | Frequency Shift Keying |
| GMSK | Gaussian Minimum Shift Keying |
| GSM | Global System for Mobile communications |
| LE | Linear Equaliser |
| LMS | Least Mean Square |
| LNA | Low Noise Amplifier |
| LPC | Linear Predictive Coding |
| MASK | M-ary Amplitude Shift Keying |
| MFSK | M-ary Frequency Shift Keying |
| MMSE | Minimum Mean Square Error |
| MPSK | M-ary Phase Shift Keying |
| MSE | Mean Square Error |
| NLMS | Normalised Least Mean Square |
| OQPSK | Offset Quaternary Phase Shift Keying |
| PCM | Pulse Coded Modulation |
| PLL | Phase Locked Loop |
| PM | Phase Modulation |
| PROM | Programmable Read Only Memory |
| PSK | Phase Shift Keying |
| QAM | Quadrature Amplitude Modulation |
| QPSK | Quaternary Phase Keying |
| RF | Radio Frequency |
| RLS | Recursive Least Squares |
| SDMA | Spatial Division Multiple Access |
| SFAEST | Stabilised Fast A-priori Error Sequential Technique |
| SNR | Signal to Noise Ratio |
| SRAM | Static Random Access Memory |
| SRRC | Square Root Raised Cosine |
| SS-MSE | Steady State Mean Square Error |
| TCM | Trellis Coded Modulation |
| TDD | Time Division Duplex |
| TDMA | Time Division Multiple Access |
| UMTS | Universal Mobile Telecommunications System |
| VHDL | VLSI Hardware Description Language |
| VL LE | Variable Length Linear Equaliser |
| VL FBF DFE | Variable Length FeedBack Filter Decision Feedback Equaliser |
| VSLMS | Variable Step size Least Mean Square |

## TABLE OF CONTENTS

# 1 INTRODUCTION

This chapter serves two main purposes. First, to provide an introduction to modern mobile communication systems including concise descriptions of the main components and also the characteristics of mobile channels. This section also includes a short overview of the most common mobile standards in use nowadays. The second main objective of this chapter is to provide some general information about the thesis. In particular, the specific problem tackled in this project, the organisation of the thesis and the novel contributions appearing in this work are presented.

## 1.1 Introduction to mobile digital systems

The objective of this section is to provide the general background to mobile digital systems. First, the main elements of the transmission/reception chain are briefly covered. It is also important to describe in some detail the characteristics of the mobile channel and how this influences the design of transceivers. Finally, a short account of some current mobile standards is presented. It is obviously not possible to cover all these topics here in depth. A thorough treatment of the different areas can be found in most digital communications textbooks. In particular, the information included in the next sections has mainly been compiled from [Proakis96], [Glover98], [Sklar01], [Haykin01] and [Gitlin92].

### 1.1.1 Block diagram of a generic mobile transceiver

Figure 1.1 shows the main components that can be found in any digital mobile communication system. Indeed, the elements shown appear in most modern communication systems, wireless or wired. It is important to anticipate that the divisions shown in this figure might not be so rigid and some subsystems may merge. It should also be mentioned that in the case of wireless communications with carrier frequencies in the UHF band, both transmitter and receiver would have a radio frequency (RF) front end. RF components are implicitly included in the modulator and demodulator. Each subsystem is now briefly treated.

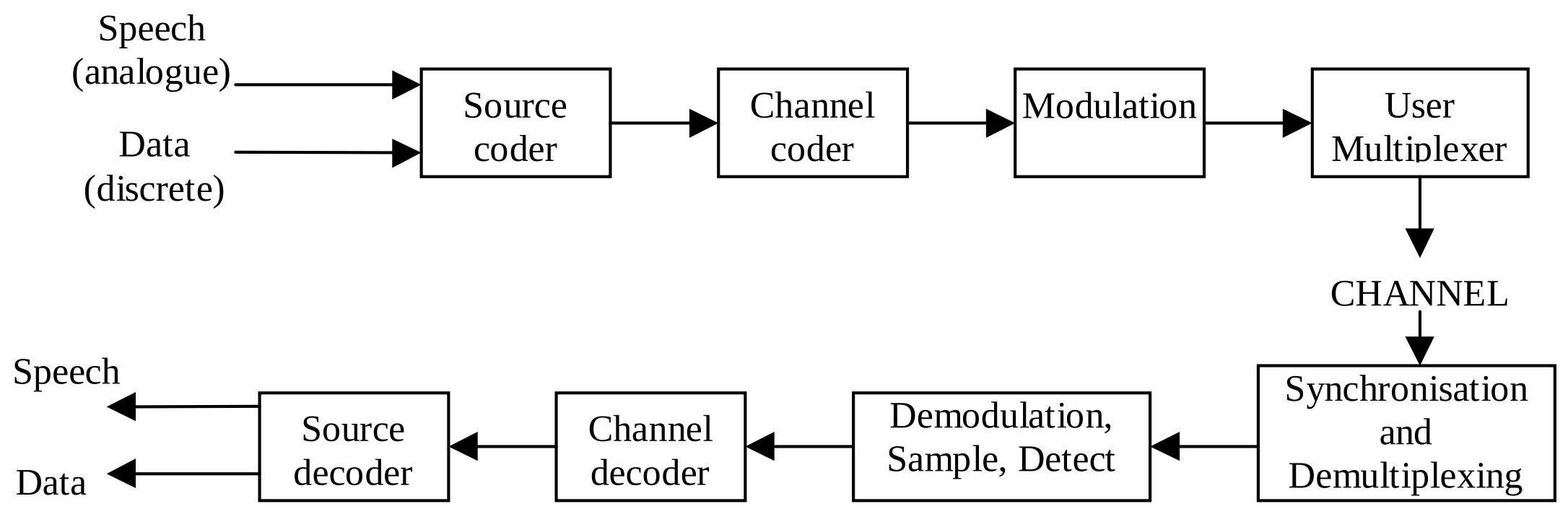

**Figure 1.1**: High level block diagram of a typical digital communication system.

SOURCE CODING / DECODING

In general, bandwidth is a limited resource whose usage can be optimised by properly designing some of the subsystems in Figure 1.1. This is especially true for the case of wireless systems where the same spectrum must be shared among many users and many different services.

One method to maximise bandwidth usage is to compact the information to be transmitted. Source coding, i.e. compression, is one of the principal applications of information theory whose fundamentals results were stated by Shannon in his landmark paper [Shannon48]. Source compression can be achieved if some properties of the original message are exploited. A distinction must be made whether the source of information is a finite discrete source, such as the English alphabet, or is a natural source (typically continuous), such as speech. This distinction is fundamental as, depending on the type of source, different properties are exploited.

In the case of discrete sources, it is very often the case that the different symbols generated by the source do not have the same probability of occurrence. This fact can be effectively utilised by assigning short codes to the most frequent symbols and longer codes to less frequent ones, rather than assigning equal length codes to all symbols. Assuming that the source statistics are known (i.e. the probability of occurrence of each symbol), there are well defined procedures to assign different length codewords to each symbol in a way that reduces the average number of transmitted bits. These coding schemes, generally known as prefix coding, need to be uniquely decodable at the receiver, as otherwise ambiguity would appear with respect to the transmitted message. Huffman coding, a particular class of prefix coding, is the most popular algorithm for assigning codewords to symbols. The algorithm is optimum in the sense that the average number of binary digits required to represent the source symbols is minimum ([Proakis95]), and is described in any of the communications textbooks mentioned at the beginning of this section. If the statistics of the source are not known a-priori, the Huffman algorithm cannot be used and other schemes must be employed. The most widely used technique in this situation is Lempel-Ziv coding which has the ability to adapt to the particular characteristics of the source. The algorithm works by parsing the source data stream into segments that are the shortest subsequences not encountered previously. Again, details of the algorithm can be found in any of the general references given.

If the input to the communication systems is a continuous waveform, typically speech, the signal must first be digitised. When a bandlimited continuous waveform is sampled at a sufficiently high rate (Nyquist sampling), the resulting samples perfectly represent the original signal and this can be reconstructed using a lowpass filter. Fundamental in the context of source coding is the fact that adjacent samples of a signal coming from a natural source will typically exhibit a high degree of correlation. Correlation is a sign of redundancy and in the case of continuous signals, the objective of the source coder is to remove this redundancy, transmitting only the essential information. Broadly speaking two different approaches are at hand to perform the process of redundancy removal in speech signals, namely, waveform coding and channel vocoding.

Waveform codes try to mimic as closely as possible the original speech signal. Examples of waveform coding are pulse code modulation (PCM) and adaptive differential pulse code modulation (ADPCM). PCM does not exploit signal redundancy and can only be regarded as a method of coding speech into a 64 kbit/s data stream. ADPCM does exploit the signal redundancy to compress the speech signal. ADPCM is based on the use of linear predictors. A linear predictor can be thought of a filter whose output predicts the current input to the filter as a function of the past samples of the incoming signal. Logically, the prediction will not be perfect and by comparing the prediction with the current sample, an error signal can be formed. This error signal, whose variance is much smaller than that of the original samples, can be quantised and PCM encoded with fewer bits. At the receiver, the incoming error signal is used to correct the output of a predictor identical to that used in the transmitter, achieving in this way perfect reconstruction of the original signal. Given that the statistics of speech are non-stationary, the predictor coefficients vary with time, and need to be periodically transmitted so that reconstruction can take place at the receiver. Using ADPCM, a speech signal can be coded into a 32 kbit/s bit stream without any quality degradation with respect to the conventional PCM encoding.

Vocoding systems, on the other hand, assume that the signal (speech) being coded conforms to a certain model. The vocoder works by analysing a segment of speech in order to extract the parameters of the model corresponding to that particular segment. These parameters are then transmitted and the receiver uses them to synthesise the original speech segment. One of the most common vocoders in use is the linear predictive coder (LPC). LPC assumes that the speech generation mechanism can be modelled by an all-pole filter (vocal tract) excited by either white noise (unvoiced sounds) or frequency tones (voiced sounds). In this case, the LPC encoder estimates the filter coefficients for a short speech segment using linear prediction techniques. It also decides whether the sound is voiced or unvoiced. This information is then sent and at the other end, the receiver uses it to reconstruct the original speech. Rates as low as 4.8 Kbits/s can be achieved using LPC with an acceptable voice quality.

Nearly all current mobile standards such as GSM, IS-54 or IS-95 rely on some variant of LPC to compress speech. This is due to the necessity, in a mobile environment, of reducing as much as possible the transmitted bit rate. Detailed treatments of speech compression schemes can be found in Chapter 3 of [Steele99] and Chapter 7 of [Rappaport96]. A thorough treatment of adaptive linear prediction can be found in chapter 6 of [Haykin96].

CHANNEL CODING / DECODING

Channel coding, also known as error control coding, is the part of information theory that deals with the design of coding schemes which allow the receiver (i.e. the decoder) to detect and/or recover errors introduced by the transmission medium. The coding techniques briefly presented in the following paragraphs all have the common feature of adding some form of redundancy, i.e., extra bits, to the original message. The ratio between the number of input bits to the number of output (coded) bits is called the coding rate. In a way, channel coding acts in the opposite direction to source coding, where the purpose is to eliminate redundancy; however now the redundancy added has a very well defined structure in order to help in the control of errors.

It is important to recognise that adding bits to the original information sequence with the purpose of lowering the bit error rate has the effect of increasing the bandwidth needed to transmit the message. Alternatively, just the original bits could be sent, but with more signal power transmitted to overcome the channel impairments. This illustrates the classical trade-off between signal power and transmission bandwidth when designing a system to achieve a given BER. In the context of mobile systems, both resources must be used carefully as spectrum is scarce and strictly limited and power should be kept as low as possible to reduce interference among different users and extend battery times.

Coding theory is a highly mathematical subject whose presentation falls out of the scope of this introductory chapter. Therefore the most important coding techniques will be just briefly described. The coding schemes covered are: linear block codes, cyclic codes, convolutional codes and compound codes.

An (n, k)-block code accepts k information bits and adds to them (n-k) redundant parity check bits producing a stream of n bits (coding rate = k/n). Depending on how many bits are included for parity, the decoder will be able to perform some checking procedures and to detect and/or correct one or more erroneous bits. A code is said to be linear if the modulo-2 addition of any two code words produces another code word belonging to the code. Code linearity simplifies the decoding process making it very efficient.

Cyclic codes are a particular subclass of linear block codes, which satisfy, apart from the linearity property, the cyclic property. The cyclic property states that any cyclic shift of a code word in the code is also a code word. When both properties are satisfied, the resulting code has a very well defined mathematical structure, whose properties simplifiy the encoding process and also allow the use of very efficient decoding schemes.

Next on the list are convolutional codes. These codes, unlike block codes, have as input a continuous (serial) stream of information bits and produce another continuous stream, typically at a higher rate, of coded bits. The input bits are fed into a shift register and the output bits are obtained from linear combinations (XOR) of the bits in the shift register. The span in bits of the shift register is termed the constraint length. From the encoding process just described, it is obvious that the generated code words are not independent of each other as is the case in block codes. The decoding of convolutional codes is normally performed using the Viterbi algorithm [Forney73]. This is a very efficient method of implementing a nearest neighbour decoding strategy (also known as maximum likelihood sequence estimation). A complete example illustrating the functioning of the Viterbi algorithm can be found in [Sklar01].

Compound, or concatenated, codes combine two or more coding schemes (component codes) such as the ones introduced in previous paragraphs. The resulting codes have large coding gains and error-correction capabilities of longer single codes. Moreover, their inner structure allows a relatively simple decoding. Among the different compound codes, Turbo-Codes ([Berrou93]) have recently gained wide acceptance for achieving a performance close to the limit predicted by Shannon ([Shannon48]). A Turbo encoder is typically made of two convolutional encoders operating on two different interleaved versions of the same information sequence. At the receiver, the incoming bits are subject to a double decoding process by the decoders of the two component codes. These decoders rather than performing hard decisions as is the case of single codes, generate soft decisions and information is exchanged between them. Typically, this exchange of information between decoders is iterated several times, with each iteration producing more reliable solutions.

In a GSM speech frame (260 bits), bits are classified according to their importance in the speech quality. A 3-bit cyclic code is calculated for 50 of the 182 most significant bits. These 182 bits are then encoded using a ½-convolutional encoder with a constraint length of

5. The remaining bits in the frame are left unprotected as they do not affect speech quality significantly. In data frames, all bits are coded using the same convolutional encoder as in the case of speech.

In UMTS, various coding schemes are supported. Depending on the type of service assigned to a particular link, convolutional encoding with coding rates of ½ or 1/3 and constraint length of 8 can be used. Support for Turbo-Codes with 1/3 coding rate and constraint length of 4 is also provided in the UMTS standard.

MODULATION

Modulation is the process of modifying one signal in sympathy with another ([Glover98]). In the context of digital communications, an information signal is used to modify the characteristics of a carrier. In baseband systems, the carrier is a rectangular pulse train whereas in radio communications, where our interest is, the carrier is a sinusoid.

The three basic methods in RF modulation correspond to variations of one of the three parameters of a sinusoid, namely, amplitude, frequency and phase, giving rise to amplitude modulation (AM), frequency modulation (FM) and phase modulation (PM) respectively. In the case of digital modulation schemes, the number of levels of each of these parameters is discrete and finite and they are renamed as amplitude shift keying (ASK), frequency shift keying (FSK) and phase shift keying (PSK). FSK and PSK signals can be demodulated in a coherent or non-coherent way depending on whether the receiver has perfect knowledge of the phase of the transmitted signal. Coherent receivers achieve lower bit error rates but at the cost of increased hardware complexity, as they need to use some form of carrier synchronisation. The advantages and inconveniences of each type of modulation can be found in the communication textbooks mentioned above.

In most radio communication systems, spectrum is a rather scarce resource whose use must be optimised. One method to increase the spectral efficiency of the system is to use high order modulation schemes, where the carrier parameter being varied can take more than 2 values (usually a power of 2). The resulting modulation schemes, called multilevel ASK (MASK), FSK (MFSK) and PSK (MPSK), offer the possibility of transmitting more bits of information using the same channel bandwidth. The penalty paid is an increased bit error rate or a higher power transmission. Some important multilevel schemes are quaternary PSK (QPSK) and quadrature amplitude modulation (QAM). QAM is a multilevel scheme that combines amplitude and phase modulation providing as many levels as 1024 or 2048, achieving in this way a high spectral efficiency.

Gaussian minimum shift keying (GMSK) is a particularly important modulation format based on a form of QPSK (OQPSK) in which the information bits are first Gaussian-shaped before modulation. This shaping has the effect of avoiding the abrupt envelope variation which is present at the output of practical (i.e. filtered) OQPSK modulators. Changes in the envelope increase interference when the modulated signal is then amplified by a low noise amplifier (LNA) operated in the non-linear region, such as is the case in power restricted environments (satellites transponders, mobile handsets). As in some of the other schemes, GMSK modulated signals can be coherently or non-coherently detected.

Concluding this brief review of modulation methods it is important to mention a scheme that combines modulation and channel coding as a single process. Trellis coded modulation (TCM) is a technique that achieves coding gains of 3-6 dB without any bandwidth expansion. Details of TCM can be found in the general references given previously and in [Ungerboeck87].

GSM uses GMSK as its modulation scheme. This decision was made in light of the very low adjacent channel interference generated by this type of modulation which allow the use of amplifiers in the non-linear region (i.e. power efficient). In UMTS the selected modulation formats are OQPSK (offset quaternary phase shift keying) for the uplink and QPSK (quaternary phase shift keying) for the downlink. Notice that in the case of the downlink, envelope transitions of the output signals are of no importance because the base station can operate the LNAs in the linear region as there are no power constraints.

MULTIPLE ACCESS

Multiuser communication systems, such as mobile systems, allow a transmission medium to be shared among different users simultaneously. Clearly, there has to be some means of assigning transmission resources to users in a structured way, as otherwise mutual user interference would greatly impair communication. The technique employed to do this strongly affects the design of the rest of the network and many of its characteristics. The three most common methods to allow simultaneous use of the communication channel by different users are: frequency division multiple access (FDMA), time division multiple access (TDMA) and code division multiple access (CDMA). Each of them is now introduced.

In FDMA, each user is assigned a subband of the total frequency band. Guard bands are left between users in order to reduce mutual interference, as the filters separating users are non-ideal. FDMA was the multiple access technique used in early (analogue) mobile systems.

TDMA allows a user to employ the entire available spectrum but only for a fraction (slot) of the total time. In analogy with FDMA, guard times are inserted between users to allow for the imperfections of the synchronisers. The standards GSM and IS-54 use a combination of TDMA with FDMA in which users are divided among carriers, and the users on each carrier are time multiplexed.

CDMA allows the utilisation of the whole spectrum all the time by any single user. Of the different types of CDMA, the one covered here is the direct sequence CDMA (DS-CDMA) which is the one selected by all forthcoming 3G systems. In DS-CDMA, each user is assigned a code. This code is used to multiply (spread) the user information stream. Separation of users is achieved by the special properties of the codes. The most important is that all codes are orthogonal to each other. In this way, the transmission from a single user can be recovered by multiplying the incoming signal, which contains the signals from all users, by an in-phase replica of the code used at the transmitter. This will pull out the information from the desired user and cancel out the rest of signals.

It is not a coincidence that all 3G cellular standards currently being deployed, in Japan, US and Europe, have chosen CDMA as the multiple access technique. CDMA has proved to have certain advantages over FDMA and TDMA ([Gilhousen91], [Lee91], [Pickholtz91] and [Jung93]) the most important being the increased system capacity it offers. This improvement is due to the total frequency re-use of CDMA systems, that is, the same frequencies can be used in all cells. It is important to mention that for this frequency re-use to be feasible, an accurate power-control mechanism must be employed, to ensure that transmitters far away from the base station are not shadowed by users close to the base station (near-far problem).

Recent advances in spatial signal processing and antenna technology ([Winters98], [Kohno98]) have paved the way for a new method of multiple access called space division multiple access (SDMA). In this scheme, users are separated by virtue of their location. Separation of users is achieved with the use of highly directive antennas which are able to track the position of receivers (smart antennas) and direct the beams in their directions. The UMTS standard gives the possibility of using this type of antenna at the base station with the goal of increasing the system capacity.

SYNCHRONISATION

Two sequences of events are said to be synchronised when the events in one sequence and the corresponding events in the other occur simultaneously. The process of making a situation synchronous and maintaining it in this condition is called synchronisation ([Haykin01]).

In general, three levels of synchronisation are required:

- Phase synchronisation
- Symbol synchronisation
- Frame synchronisation

Phase synchronisation only applies to receivers that make use of coherent demodulation. The basic device to perform this synchronisation is the phase locked loop (PLL), a device that generates a carrier whose phase difference in minimised with that of the incoming signal. Detailed descriptions of PLL and its variants can be found in [Stremler90]. Increasingly, classical PLL circuits are being replaced by DSP systems that perform the synchronisation in the digital domain. This algorithmic approach formulates the carrier recovery process as a problem of maximum likelihood estimation of a parameter (i.e. the phase). Descriptions of these techniques can be found in Chapter 10 of [Sklar01] and Chapter 6 of [Haykin01].

Symbol synchronisation is the process of determining when a symbol starts and when it finishes or, in other words, at which point the modulation can change state. This knowledge is essential in order to determine the correct interval over which the received power must be integrated prior to making a symbol decision. There are two basic methods to achieve symbol synchronisation: open loop and closed loop. In open loop synchronisation, the receiver recovers a replica of the transmitter data clock using only the incoming signal. On the other hand, in closed loop systems the clock recovery is performed by comparative measurements of the incoming signal and the locally generated clock. Both methods are covered in detail in [Sklar01].

The last level of synchronisation, frame synchronisation, is required in any system in which information is grouped in blocks. As an example, in a system where users are time-multiplexed like GSM, boundaries between users must be known in order to route the information appropriately. In this case, notice that synchronisation is mainly a requirement for the transmitter who must send the symbols at very precise instants so that symbols reach the central node (where multiplexing takes place) at the appropriate time. Again, more information on this type of synchronisation can be found in Chapter 10 of [Sklar01].

DEMODULATION

The demodulation process encompasses several subsystems in charge of converting the incoming RF signal into a sequence of detected bits or symbols. The different blocks of a demodulator are shown in figure 1.2.

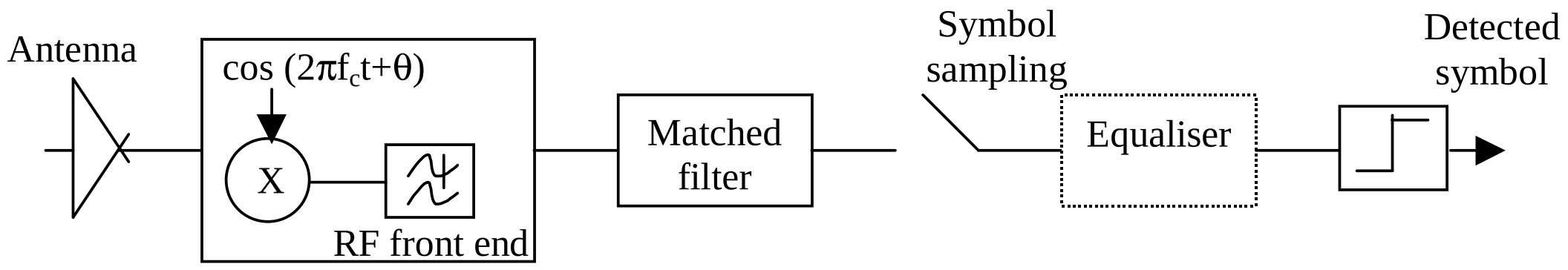

**Figure 1.2**: Block diagram of a typical demodulator.

The RF energy is picked up by the antenna. Handset antennas are invariably omni-directional radiating elements while the characteristics of base station antennas depend very much on their location although commonly they provide a sectorised coverage with 120° sectors.

The RF front end is in charge, among other things, of the conversion from RF frequency to an intermediate frequency or baseband. This down conversion process is illustrated in figure 1.2 by the multiplication of the incoming signal with a sinusoid whose frequency is the carrier frequency (mixing). There are other elements forming part of the RF front end whose purpose is to amplify the signal power while keeping as low as possible noise and interference. Extensive information on RF architectures and techniques can be found in [Smith86] and [Stremler90].

The matched filter is defined as follows ([Glover98]): *A filter which immediately precedes the decision circuit is said to be matched to a particular symbol pulse if it maximises the output SNR at the sampling instant when that pulse is present at the filter input.* Typically the filter is matched to the transmitted symbol shape, which in turn has been generated by shaping an impulse with a pulse shaping filter. The function of the pulse shaping filter is to limit the required bandwidth to transmit the information. The most usual choice for matched filter and pulse shaping filter is the square root raised cosine filter (SRRC) whose combination forms a raised cosine filter. Matched filters are implemented as integrators (with an appropriate shaping) whose output is sampled at symbol rate.

In this previous definition of a matched filter, it is assumed that the channel is ideal in the sense that although it may introduce additive white Gaussian noise (AWGN), it does not induce any other type of interference. Under these conditions, the matched filter is the optimum detection structure. If the channel does introduce distortion, such as due to multipath propagation in wireless environments, the matched filter is no longer adequate. In this situation, the optimum receiver structure can be shown to consist of a matched filter followed by an infinite tap-delay line which weights the outputs of the matched filter ([Gitlin92]). This infinite tap-delay line is what is normally called an equaliser. The reason why the equaliser appears in a dashed box in figure 1.2 is to denote that this subsystem is optional. As has been mentioned before, if there is no distortion, no equaliser is needed. Obviously, when implementing the equaliser only a finite tap-delay line can be used. For the time being nothing else is said about the equaliser as equalisation methods are explained in detail in the next chapter.

As a last step in the demodulation phase, the symbol detector will map the soft output of the equaliser or matched filter to the nearest possible symbol.

### 1.1.2 Mobile channels

Most communication channels degrade the transmitted signal in one form or another. The most obvious type of degradation is AWGN, which is caused by the thermal motion of electrons present in all dissipative components such as resistors.

A simplified model for a wireless channel could be constructed assuming there are no interposing objects between transmitter and receiver. In this case, the only two effects to be taken into account would be the AWGN and the free space path loss. Such a model could be appropriate for a satellite link where a direct path, free of echoes, exists between transmitter and receiver.

In the case of mobile radio communications, the simple model described previously is not adequate, as transmission takes place near the ground. This implies that the propagation of the signal is subject to reflection, diffraction and scattering from objects in the environment where communication take place ([Rappaport96]). Reflection results when the propagated wave hits objects whose dimensions are large when compared with its wavelength. This is the case when waves hit buildings, walls or the earth surface. Diffraction occurs when the transmitted signal encounters objects with sharp edges, giving rise to a bending of waves around the obstacle. The last effect, scattering, is due to the propagating wave encountering objects which are small in comparison with its wavelength. Trees, lamp posts and other small objects cause the transmitted wave to be scattered.

All these propagating mechanisms make the transmitted signal arrive at the receiver via multiple paths. This phenomenon is called multipath propagation. Moreover, as a mobile user or some of the objects in his surroundings move, the multipath profile changes, causing fluctuations in the received signal's amplitude or phase, giving rise to what is called multipath fading. The characterisation of both phenomena and their relation were first described in [Bello63].

Multipath propagation and fading pose important limitations on the transmission of information over the mobile channel. Both mechanisms are now described in some more detail. This discussion follows very closely the one presented in [Sklar97a] and [Sklar97b].

The most obvious way to characterise a communication channel is by its channel impulse response (CIR). In the power domain, the channel can be seen as a power delay profile showing the maximum excess delay, denoted by $T_m$, between the first and last received components. This is shown on the left-hand side of figure 1.3.

The relation of $T_m$ to the symbol period, $T_s$, is of fundamental importance as it largely determines the type of degradation introduced by the channel. When $T_m<T_s$, all the received multipath components arrive within the symbol period. In this case, the channel is said to be frequency non-selective. This condition means that there is no interference between successively transmitted symbols. Still, some degradation occurs as the different components may add up destructively and cause a drop in the received SNR. In order to compensate this effect, some form of diversity or channel coding should be used.

If $T_m > T_s$, neighbouring symbols interfere with each other causing intersymbol interference (ISI). This type of channel is frequency selective. Frequency selectivity can be reversed by the action of an equaliser whose operation is described in detail in the next chapter.

Taking the Fourier transform of the power delay profile, the channel can be described in the frequency domain. The resulting function, shown on the right-hand side of figure 1.3, represents the correlation between the channel's response to two sinusoids as a function of their frequency difference. The range of frequencies over which the channel passes all spectral components with the same gain and linear phase is termed the coherence bandwidth, denoted as $f_0$.

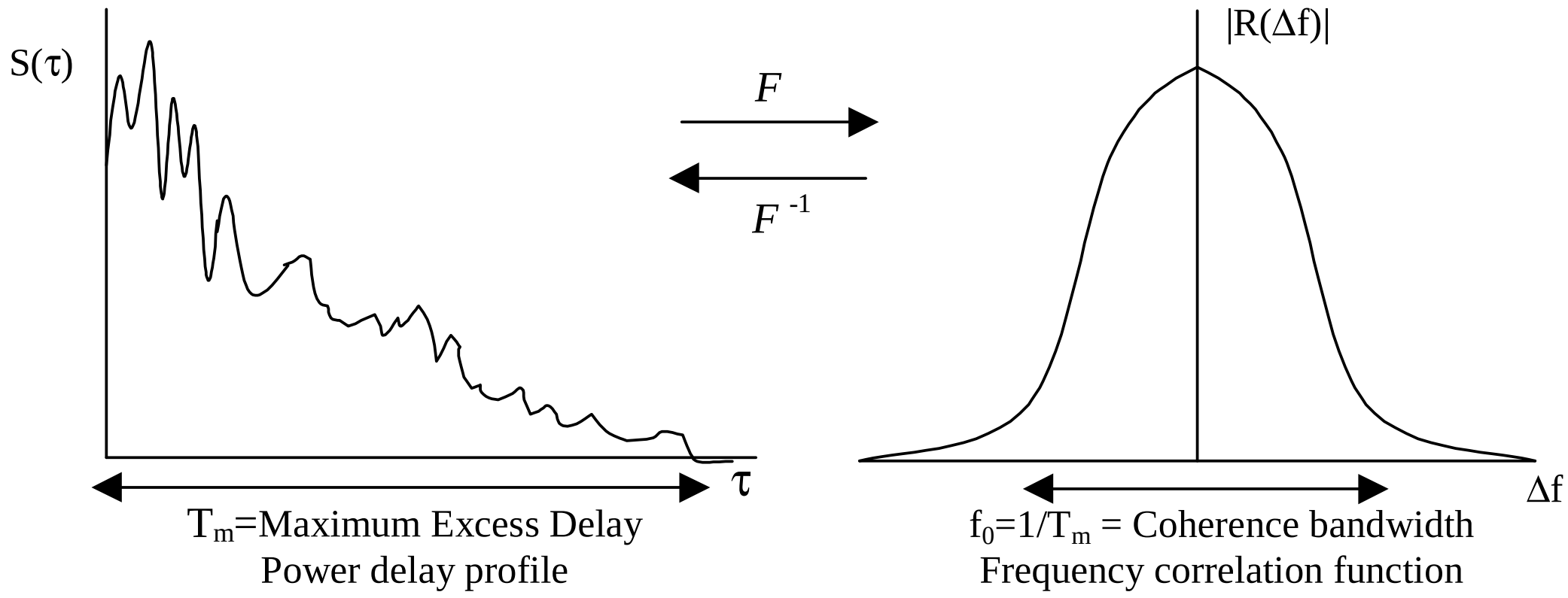

**Figure 1.3**: Time and frequency domain effects of multipath propagation.

Defining the signal bandwidth, W, as the inverse of the symbol period, $W=1/T_s$, the frequency selectivity of the channel can now be expressed in frequency domain terms. When $f_0 > W$, all the spectral components of the transmitted signal go undistorted through the channel and the channel is frequency non-selective. If on the other hand, $f_0 < W$, the channel will distort some of the frequency components of the signal making the channel frequency selective.

So far, the effects of multipath propagation have been described. These would apply to any wireless channel, mobile or fixed. In the specific case of a mobile channel, another effect, time variance, needs also to be considered.

The effects of time variance can be characterised in a similar manner to those caused by multipath. The time-domain effect of time variations can be seen by means of the space-time correlation function depicted on the left-hand side of figure 1.4. This function shows the correlation between identical sinusoids sent over the channel with a time difference $\Delta t$. The coherence time, $T_0$, is defined as the time span over which the channel can be considered to be invariant. The faster the environment changes, the shorter will be the time coherence. . Environment variations will cause changes in the number and phases of paths reaching the receiver causing fluctuations in the received signal power.

Taking the Fourier transform of the time-spaced correlation function, a frequency domain characterisation of the time variance phenomenon is found. The graph on the right-hand side of figure 1.4 shows what is typically known as the Doppler spectrum. This function plots the spectrum broadening suffered by a carrier with frequency $f_c$, as a function of the Doppler shift, $f_d$. The Doppler shift is given by $f_d=v/\lambda$ where v is the relative speed between transmitter and receiver and $\lambda$ is the wavelength of the carrier.

It is well known that in the absence of line of sight (LOS) between transmitter and receiver, the received signal amplitude of a flat fading signal or the amplitude of an individual component in a multipath profile follows Rayleigh statistics. This explains why mobile radio channels are often named multipath Rayleigh fading channels. When there is a dominant non-fading component, such as when LOS is present, the received signal amplitude follows Ricean statistics.

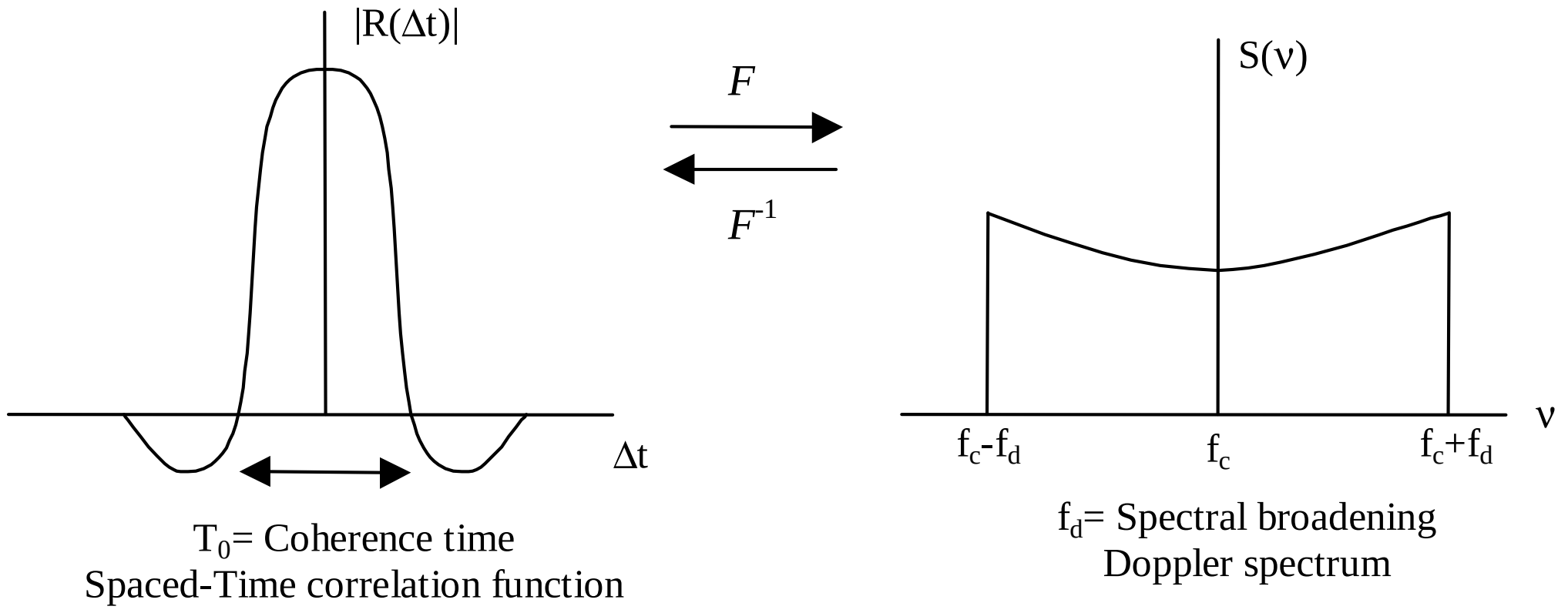

**Figure 1.4**: Time and frequency domain effects of channel time variance.

Rayleigh and Ricean fading effects receive the name of fast fading (or small-scale fading) as they can be observed even when only small positional changes take place between transmitter and receiver. In the context of mobile systems, apart from the fast fading effects, slow fading (or large-scale fading) needs also to be taken into account. This results when the distance between transmitter and receiver changes significantly. Measurements show that slow fading follows a Log-normal distribution ([Rapport96], [Sklar97a]).

Numerous measurements campaigns have been conducted in order to measure the characteristics of mobile radio channels and contrast the results with the theoretical predictions. One of the most important studies was performed during the definition phase of the GSM system under the COST action 207. The output of this study was the COST-207 channel models. One of them, the 6-path Typical Urban channel model, is used in Chapter 7 to test the performance of the novel equalisation techniques proposed in this thesis in a realistic scenario.

Extended coverage of the characterisation of mobile channels can be found in Chapters 3 and 4 of [Rappaport96], Chapter 2 of [Steele99] and [Braun91].

### 1.1.3 Review of mobile Standards

Concluding this short introduction to mobile digital systems, the most important features of some common mobile standards currently in use or being deployed are given in table 1.1. The standards included are GSM, IS-54, IS-95 and UMTS.

Some comments about the information provided about table 1.1 are relevant here. First, no information is provided about 1G standards as most of these systems are, nowadays, in the process of being switched off. The Duplexing parameter refers to the way in which the uplink (mobile to base station) and downlink (base station to mobile) channels are multiplexed. All standards use frequency division duplexing (FDD) which assigns a different carrier to each of the links. UMTS provides also the possibility of using time division duplexing (TDD), dividing the transmission time in slots which are used alternately for the uplink and downlink channels.

| | GSM | IS-54 | IS-95 | UMTS |
|---|---|---|---|---|
| Deployment Area | Europe, South America, Asia, Australia | USA | USA | Europe, Japan |
| Carrier Frequency | 900 MHz | 850 MHz | 900 MHz | 2 GHz |
| Multiple Access | TDMA/FDMA | TDMA/FDMA | CDMA | CDMA |
| Duplexing | FDD | FDD | FDD | FDD/TDD |
| Modulation | 0.3 GMSK | π/4-DQPSK | 4-QAM | OQPSK/QPSK |
| Channel Coding | CRC+ 1/2 Conv. (K=5) | 7-bit CRC ½-Conv. (K=6) | ½-Conv. (K=9) | ½,1/3-Conv. (K=9), Turbocode |
| Speech Coding | RPE-LPC | VSELP | QCELP | AMR |
| Multipath mitigation | Viterbi Eq. | DFE | RAKE | RAKE+Advanced Detectors |

**Table 1.1**: Main characteristics of some important cellular standards.

RPE-LPC, VSELP and QCELP are all variants of the LPC technique covered in the previous section. Adaptive multirate (AMR) is the speech coding system specified in the UMTS standard. Different speech coders can be integrated into the AMR system, allowing the coding of speech at different rates. In this way, different qualities can be produced and compatibility with codecs from other standards (in particular GSM) is eased.

Another important point to mention is the difference in the CDMA implementation between IS-95 and UMTS. In the IS-95 standard, each of the two links occupies a bandwidth of 1.25 MHz whereas in the case of UMTS the bandwidth is 5 MHz (including guard bands). This is the reason why UMTS is sometimes referred as wideband CDMA (WCDMA). This difference in bandwidth is the result of the different chip rates in the systems, 1.2288 Mchip/s in IS-95 compared to the 3.840 Mchips/s in UMTS.

The different multipath mitigation techniques are covered in detail in the next chapter, but the particular technique to counteract multipath propagation is not specified by the standard. The mechanisms shown are just the options that are most commonly implemented in commercial devices.

Detailed discussions of the different standards can be found in [Rappaport96] (IS-54, GSM, IS-95 and others) and [Steele00] (GSM, UMTS). [Baier94] is a generic article presenting the design of a CDMA-based cellular network without focusing on any particular standard. [Holma01] and [Kaaranen01] are entirely devoted to UMTS. [Dahlman98], [Adachi98] present short overviews of WCDMA in the context of UMTS although they are now somewhat outdated as the standard has evolved considerably since 1998. For this reason, the best source for accurate and timely UMTS information is the specifications generated by the 3GPP consortium ([3GPP00]) which is in charge of the UMTS standardisation.

# 1.2 Reconfigurable communication systems

The general area in which this research project started was reconfigurable signal processing for mobile terminals. The term “reconfiguration” involves two different concepts. The first one would refer to reconfiguration among different standards. The second one would deal with performance reconfiguration, that is, the adjustment of the parameters of a radio transceiver according to the specific conditions in which it is operating.

The diversity of mobile standards around the world makes the idea of inter-standard reconfiguration for mobile handsets attractive. Terminals able to function in different standards have already appeared in the market, an example being the dual band GSM phones which are able to work in the 900 MHz and 1.8 GHz (PCS system) bands. However, this can be regarded as a very limited form of standard reconfiguration, as in both systems the baseband processing is identical. Moreover, the band reconfigurability is achieved with the crude approach of duplicating the RF front end, one for each of the frequency bands. Reconfiguration among completely different standards such as GSM, IS-95 or UMTS is far more problematic given the differences among the protocol stacks of these systems and the baseband functions performed on each of them. The approach of integrating a corresponding transceiver for each of the supported standards would invariably lead to bulky and heavy handsets, therefore a more elegant approach is needed.

A high degree of reconfiguration could be achieved with the concept of software radio ([Mitola95], [Tutlebee98]). The ultimate goal of software radio is to be able to digitise the received signal directly from the RF band and implement all the radio processing functions and protocol stacks in software. This software would execute on DSPs, microcontrollers and/or general microprocessors. Using this architecture, reconfiguration of the transceiver to adjust it to a particular standard would be just a matter of downloading the appropriate software into the radio. Architectures to handle the download of reconfiguration software in the context of mobile terminals have already been proposed ([Noblet98]). At the moment, however, there are still important technological limitations for the implementation of a pure software radio. Two of the most important are the need for much faster analogue to digital (ADC) converters ([Wepman95]), and faster processors able to handle the huge amount of data that would result from RF digitisation ([Baines95]). Nevertheless, advances in these two technologies make the idea of software radio feasible, at least partially, in the near future.

In the area of performance reconfiguration, some techniques have recently been developed in order to exploit the variability of the wireless channel. The objective is to increase the rate of transmission when the channel properties are favourable and reduce it when the channel degrades. To this end, adaptive modulation schemes ([Alouini99], [Webb95]) and adaptive channel coding ([Vucetic91]) have been proposed. These methods rely on the collection of channel state information by the receiver, which is then sent, via a feedback channel, to the transmitter to allow the adjustment of the modulation order and coding level. In a recent paper ([Duel-Hallen00]), a technique has been presented in which the transmitter performs a long-range prediction of up to 2 ms in advance of the state of the channel. This information is then used by the transmitter to adjust the modulation order without having to wait for the information from the receiver.

The work on this thesis, focuses in the latter type of reconfiguration, performance reconfiguration, albeit in our case, the reconfiguration takes place at the receiver. A detailed explanation of the objectives of this work follows next.

## 1.3 Objectives of the thesis

The subject of this thesis is equalisation for mobile terminals. As has been mentioned in section 1.1.1 (demodulation), implementation of equalisers uses finite length tap-delay lines rather than ideal infinite ones. This transition from an infinite to a finite number of taps introduces problems which are particularly important in the case of power restricted devices such as mobile handsets.

Recently, an article has appeared ([Yates00][1]) which states the problem to whose solution this thesis is contributing. In this paper, there is one paragraph which is especially relevant for our work and is, therefore, reproduced here ([Yates00], pp. 97):

---

[1] This publication is from May 2000, well after this PhD project was started in November 1998.

*Variable Structure Equalizers: On the downlink, high data rates will require equalization at the mobile. At different distances from the infostation, there will be different demands on the equalizer. It is anticipated that at 20 meters there will almost certainly be a line-of-sight path. How this can aid the equalizer is an interesting problem. In addition, as the mobile moves away from the infostation, but is still within range for high-data rate transmission, the equalizer will have to perform better. In analogy to the proposed transition in modulation schemes, it may be useful to design equalizers whose structure grows or changes with distance from the infostation.*

The above paragraph was written in the context of a proposed new type of mobile network based around the idea of infostations ([Goodman97]). However, the problem it poses is equally challenging in the context of conventional cellular systems. The question can be generically stated as: How can the equaliser in a mobile handset be reconfigured in order to exploit the specific environmental conditions in which it is operating at one specific moment in time? The term "exploit" in this question can mean improvement in the performance of the equaliser (reduction of the BER) or reduction of the computational complexity which in turn implies a decrease in the power consumption. In our work, we focus on the latter objective, as power consumption is a scarce resource in any mobile handset. The specific type of reconfiguration studied is the variation of the length of the filters making up the equaliser.

To further reinforce the motivation and importance of this work, another very significant fragment from another paper, [Treichler96] pp. 73-75, is also reproduced:

*…it is not unusual for the adaptive equalizer in a voiceband modem to consume more than 80 percent of the multiply/add cycles needed to demodulate and decode a 256-QAM trellis-coded signal. Thus, it is important to limit the length of the equalizer to that required to handle adequately the range of propagation channels expected. […] This brings into high profile the question of how long must the equalizer's filter be to satisfactorily compensate for the channel's*

*dispersion. In fact, no clear cut answers are available. It depends on the type of channel to be equalized and the sample rate relative to the transmitted signal bandwidth, but little more can be said.*

Although again in a different context (voiceband modems) from the one of our interest, this paragraph illustrates the importance and difficulty of properly adjusting the equaliser length. A trivial approach would be to make the equaliser long enough to be able to compensate the worst scenario the mobile can encounter. However, in the context of mobile radio communications, where there is a great variety of possible channel profiles, this strategy is too pessimistic and, most of the time, would imply a great waste of resources. Moreover, in the thesis it will be shown that under certain conditions, short equalisers perform better than longer ones.

The objective of this thesis is to present dynamic and efficient techniques to adjust the equaliser length of a mobile handset according to the specific environmental conditions. The aim of these methods is to reduce the number of computations used in the equalisation process without introducing any performance degradation. The focus is on the two most common structures used for adaptive equalisation: linear equalisers (LE) and decision feedback equalisers (DFE).

## 1.4 Organisation of the thesis

Chapter 1 has introduced different general aspects of mobile systems such as the main elements of the transceivers, characteristics of mobile channels and a brief overview of the most popular cellular standards.

In chapter 2, a detailed description of equalisation is provided. The most common equalising structures are presented alongside with the adaptive algorithms used to adjust their coefficients. The chapter concludes by showing how equalisers can be used in different types of systems (TDMA and CDMA).

Chapter 3 describes the simulation environment implemented. Simulations results are presented for the algorithms and structures introduced in chapter 2.

In chapter 4, mean squared error (MSE) expressions are derived for two popular adaptive algorithms used to drive a linear equaliser. The derivations follow closely [Widrow76] and [Eleftheriou86] for the system identification case, but adapt the results to the system inversion problem.

The derived MSE expressions in chapter 4 provide the basis for the development of several algorithms controlling the length of a linear equaliser (LE). These algorithms are presented in chapter 5, including simulation results that verify their performance.

Chapter 6 extends the MSE expressions to the decision feedback equaliser (DFE) and discusses the importance of the length of the feedforward and feedback filters. Following this discussion, an algorithm is derived to control the length of the feedback filter.

In chapter 7, the performance of the length update algorithms for the LE and DFE is verified in a much more realistic environment using accurate mobile channel models.

Chapter 8 introduces some important aspects that need to be considered when implementing signal processing algorithms on real hardware. It then describes how the length update algorithms have been implemented on a commercial digital signal processor (DSP). Simulation results obtained using the algorithms on the DSP are also shown.

In chapter 9, the main conclusions obtained from this project are stated alongside with some suggestions for further work.

The thesis finishes with a complete list of the bibliography used for this work.

## 1.5 Thesis notation

Given the large number of equations and mathematical formulas over the thesis, some notational conventions are now in place. Following the common notation found in the literature, scalars are denoted by lower case non-bold letters, vectors use lower-case bold letters and matrices are denoted by upper-case bold letters. If the dimensions of vectors and matrices are not specified it is assumed that they are arbitrary.

## 1.6 Original contributions

This thesis presents novel results on equalisation that are now summarised:

- Chapter 4, Sections 4.4 and 4.6. Expressions for the steady state MSE when using the LMS (4.4) and RLS (4.6) algorithms are derived for the case of system inversion using a linear equaliser. Known expressions were only valid in the context of system identification. We do not claim that the derived expressions are exact, and it will be seen that numerous approximations have been applied. Still, simulation results show that in most cases the derived equations are fairly accurate. Moreover, the importance of these expressions lies in the fact that they show, in a qualitative way, the dependence of the MSE level with the equaliser length.

- Chapter 5, Section 5.2. A novel filter structure and algorithm controlling its length are introduced. Simulations show that the resulting variable length linear equaliser (VL LE) is able to adjust its length according to the channel conditions.
- Chapter 6, Sections 6.3 and 6.6. The steady state MSE expressions for the LMS and RLS algorithms are expanded to the case of the decision feedback equaliser (DFE). Again, the resulting expressions show the dependence of the MSE with the lengths of the feedforward and feedback filters.
- Chapter 6, Section 6.5. A relation between the choice of decision delay and convergence properties of gradient DFEs is presented. It is shown how the decision delay can be set in order to reduce the convergence time of the equaliser when using the LMS algorithm.
- Chapter 6, Section 6.10. An algorithm is presented to adjust the length of the feedback filter of a DFE. Simulation results are shown verifying the advantages of using a variable length feedback filter DFE (VL FBF DFE).

## 1.7 Author's publications

- F. Riera-Palou, J. M. Noras, D. G. M. Cruickshank, "A DFE with a variable length feedback filter", Submitted to Electronics Letters.

- F.Riera-Palou, J. M. Noras, D. G. M. Cruickshank, "Analysis of the Decision Delay Effect on the Convergence of Gradient Recursive Decision Feedback Equalizers", IEEE International Conference on Acoustics, Speech and Signal Processing (ICASSP), May 2002, Orlando (FL, USA).

- F.Riera-Palou, J. M. Noras and D. G. M. Cruickshank, "Linear equalisers with dynamic and automatic length selection", Electronic Letters, Vol. 37, No. 25 , 6th December 2001.

- F. Riera-Palou, J. M. Noras and D. G. M. Cruickshank, "Segmented Equalizers with Efficient Length Selection", 35th IEEE Asilomar Conference on Signals, Systems and Computers, Pacific Grove (CA, USA), November 2001.

- F. Riera-Palou, J. M. Noras and D. G. M. Cruickshank, "Variable Length Equalizers for Broadband Mobile Systems", 42th IEEE Vehicular Technology Conference (Fall), Boston (MA, USA), September 2000.

- F. Riera-Palou, J. M. Noras and D. G. M. Cruickshank, "HF Equalizers based on the SFAEST Algorithm", 8th IEE International Conference on HF Radio Systems and Techniques, Guilford (England, UK), July 2000.

- F. Riera-Palou, C. Chaikalis and J. M. Noras, "Analysis of Reconfigurable Mobile Terminals Requirements for 3rd Generation Applications", IEE Colloquium on Software Radio and UMTS Terminals, Glasgow (Scotland, UK), April 1999.

- PATENT APPLICATION: F.Riera-Palou and J. M. Noras, “Elastic Linear Equaliser”, Initial Filing April 28th 2001. Pending Decision.

# 2 ADAPTIVE EQUALISERS: STRUCTURES, ALGORITHMS AND APPLICATIONS

This chapter presents a thorough coverage of equalisation techniques. In section 2.1, the concept of optimum filtering, also known as Wiener filtering, is introduced. This important topic forms the theoretical basis upon which the design of equalisers, and many other devices, is based. Section 2.2 covers the main equaliser structures used in digital communications. In the specific case of wireless applications, where the channel is unknown and/or time varying, the equaliser needs to be able to adjust and track the specific channel conditions, therefore its parameters need to be made adaptive. The most common algorithms to perform this filter adaptation are introduced in section 2.3. Finally, section 2.4 presents the role equalisers play in different types of receivers, in particular, TDMA and CDMA systems.

## 2.1 Wiener filter theory

The design of conventional digital filters is usually performed in the frequency domain. Typically, filter specifications are given specifying which frequency bands should be present at the output of the filter. Then the designer uses one of the many techniques available to design the filter ([Mitra01]) which ends up producing a set of time-domain coefficients giving the desired frequency response. An example of this type of procedure could be the design of a filter to eliminate the 50 Hz interference coming from the mains.

On the other hand, there are situations where the filter is required to minimise a certain error measure. In this case, the filter is normally designed using time-domain methods. The design of optimum filters falls into this latter category. A filter is said to be optimum in a given criterion, when the error between its output and a required output is minimised with respect to that particular criterion. The most generic configuration of the problem of optimum filter design is depicted in figure 2.1.

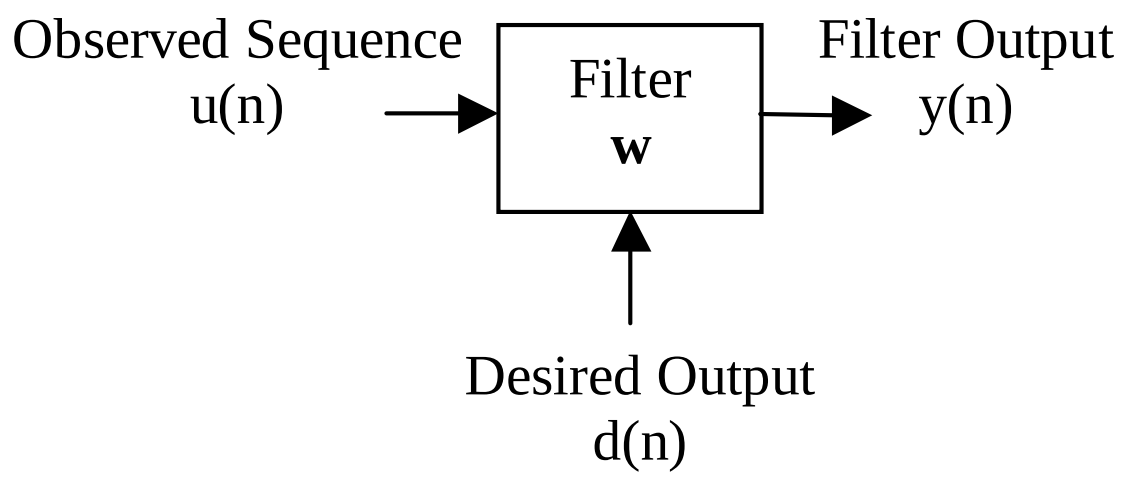

**Figure 2.1**: Generic configuration for optimal filtering.

The filter accepts as inputs an observed sequence, u(n), and a desired or required sequence, d(n). By desired sequence is meant the sequence that ideally should be produced at the filter output given the observed sequence. Instead the filter output is y(n). The difference between the desired output and the filter output is used to form an error signal, e(n)=d(n)-y(n), which gives an idea of how well the filter is performing. The task of the designer in this case is to design a filter which a minimises a particular function of the error, f(e).

A common choice for the error function is the mean squared error (MSE), defined as:

$$MSE=E[|y(n)-d(n-D)|^2=E[e(n)^2] \qquad (2.1)$$

where E denotes the statistical expectation operator and D is an integer accounting for the delay (in samples) of the filter. For simplicity and without loss of generality we assume from now that the delay is zero. Filters which minimise the MSE between its output and the required output are called Wiener filters.

The MSE can be expressed as a function of the filter weights:

$$MSE=E[|\mathbf{u}(n)\mathbf{w}(n)-d(n)|^2] \qquad (2.2)$$

It is now possible to minimise the error function by deriving the MSE with respect to the filter weights and equating the resulting expression to zero:

$$\frac{\partial \, \mathrm{MSE}}{\partial \, \mathbf{w}} = 0 \qquad (2.3)$$

Solving this partial derivative equation and after some manipulations (see [Haykin96] or [Mulgrew98]), the following expression is found:

$$\mathbf{Rw}=\mathbf{p} \qquad (2.4)$$

where **R** is the autocorrelation matrix of the input vector, $\mathbf{R}$=E[$\mathbf{u}$(n)$\mathbf{u}$(n)$^T$] and **p** is the crosscorrelation between the desired output and the input vector, $\mathbf{p}$=E[$\mathbf{u}$(n)d(n)]. This equation is called the Wiener-Hopf equation and the values of the optimum set of weights minimising the MSE, $\mathbf{w}_{opt}$, can be found by solving it:

$$\mathbf{w}_{opt}=\mathbf{R}^{-1}\mathbf{p} \qquad (2.5)$$

The MSE level achieved when using this set of filter weights is called the minimum MSE (MMSE) and is given by:

$$\mathrm{MMSE}=E[u(n)^2]-\mathbf{p}^T\mathbf{w}_{opt} \qquad (2.6)$$

The MSE as a function of the filter weights can be represented as a multidimensional bowl-shaped surface. The Wiener-Hopf equation computes the filter coefficients whose output MSE is the bottom (minimum) of this surface.

The Wiener-Hopf equation is a very general method for solving linear estimation problems and can be applied to many apparently different problems such as system identification, system inversion, linear prediction and interference cancellation (see introductory chapter in [Haykin96]).

Channel equalisation is a particular case of a more general problem known as system inversion, which is also found in other contexts, such as acoustic processing or control theory.

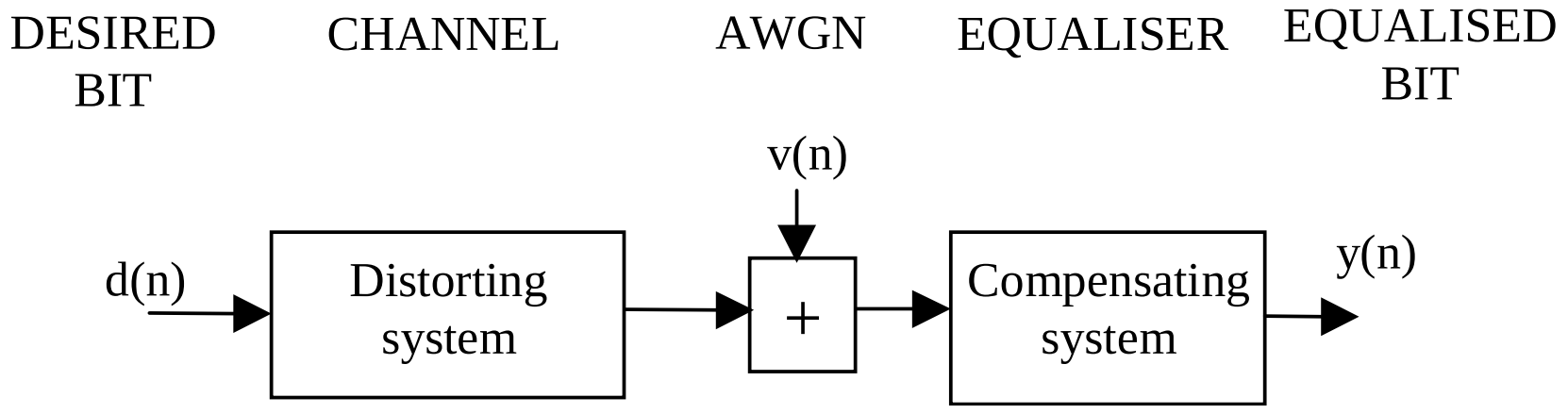

**Figure 2.2**: Generic system inversion problem with upper words indicating how it applies to channel equalisation.

Figure 2.2 shows a diagram of the system inversion set-up and how it relates to the specific case of channel equalisation (upper case words). It is important to relate figure 2.2 to the most general estimation problem shown in figure 2.1. The equaliser is the filter being designed. The observed sequence corresponds to the output of the channel onto which AWGN samples have been added. The desired output would be the transmitted bit and finally, the filter output corresponds to the sample coming out of the equaliser. With this information, the Wiener-Hopf equation could be used to design an equaliser that minimises the MSE between its output and the transmitted (i.e. desired) symbol.

Concluding this overview of Wiener filtering, three important remarks must be made. First, in order to make use of the Wiener-Hopf equation, knowledge of the **R** and **p** is needed. In the case of equalisation, it can be shown that this corresponds to the fact that the channel must be known.

Secondly, it must be pointed out that, in principle, Wiener filters can be implemented using finite impulse response (FIR) or infinite impulse response (IIR) filters. Adaptive and Wiener IIR filters is a current topic of research. [Shynk89] and Chapter 9 in [Farhang-Boroujeny98] provide reviews of the work done on this area. At present, adaptive IIR filters have two serious inconveniences. First, they suffer from the typical stability problems of IIR filters ([Mitra01]) and second, the IIR MSE surface presents local minima. Both problems make FIR filters the almost exclusive choice to implement Wiener and adaptive filters. Consequently, in this work only FIR structures are utilised.

Finally, it is also important to mention that, although the most common, the MSE is not the only criterion used in the design of optimum filters. In this chapter, some other criteria will also be applied and their pros and cons with respect to the MSE will be discussed.

## 2.2 Equaliser structures

The ultimate performance criterion in a digital communication system is the bit error rate (BER). However, due to its non-linearity, it is difficult to analyse a system on the basis of the BER, therefore alternative measures are needed. The most commonly used is the MSE just introduced. Over the years, different relations have been established between MSE level and BER, one of them being the Saltzberg bound ([Saltzberg68], [Gitlin92]). The Saltzberg bound indicates that a decrease in the MSE level will usually, although not always, lead to a reduction in the BER. This allows us to work with the MSE as a performance metric with the reassurance that benefits observed from an MSE point of view will, most likely, imply reductions in the BER.

In this section, different equaliser structures are introduced whose purpose is the minimisation of the MSE although another criterion is also briefly covered (zero-forcing equalizers).

Most of the structures shown next are suitable for adaptive and non-adaptive implementation. In the non-adaptive case, the designer, with perfect knowledge of the channel, can calculate the optimum[2] equaliser a-priori as has been shown in the previous section by solving the Wiener-Hopf equation. When the equaliser is adaptive, algorithms like those presented in the next section are used to adjust the filter coefficients.

The information of this section has been extracted from [Proakis96], [Gitlin92], [Haykin01], [Benedetto99] and [Qureshi85]. This latter reference is particularly useful as it provides a comprehensive survey of all the classic equalisation methods. Chapters 7 and 8 in [Gitlin92] are also especially helpful on the topic of equalisation as they compile many of the ideas that appeared in the Bell Systems Technical Journal during the 70s and early 80s. These papers form the original bulk of information about equalisation. For historical completeness, most of the original references, where the different techniques first appeared, are also included.

### 2.2.1 Linear equaliser

Linear equalisers (LE), also known as transversal equalisers, are the simplest equalising structure and the most commonly used. The introduction of its adaptive version by Lucky in 1965 ([Lucky65]) paved the way for the development of a high speed voice band modem. It just consists of a finite impulse response (FIR) filter ([Lyons97], [Mitra01]) which is then followed by the threshold detector. This structure is shown in figure 2.3. To put it into context, the figure 2.3 corresponds to the dashed box (equaliser) in the demodulator block diagram of figure 1.2 (section 1.1.1). In particular, the signal being is sampled is the output of the matched filter at the receiver.

---

[2] Optimum with respect to one defined performance criterion such as the MSE.

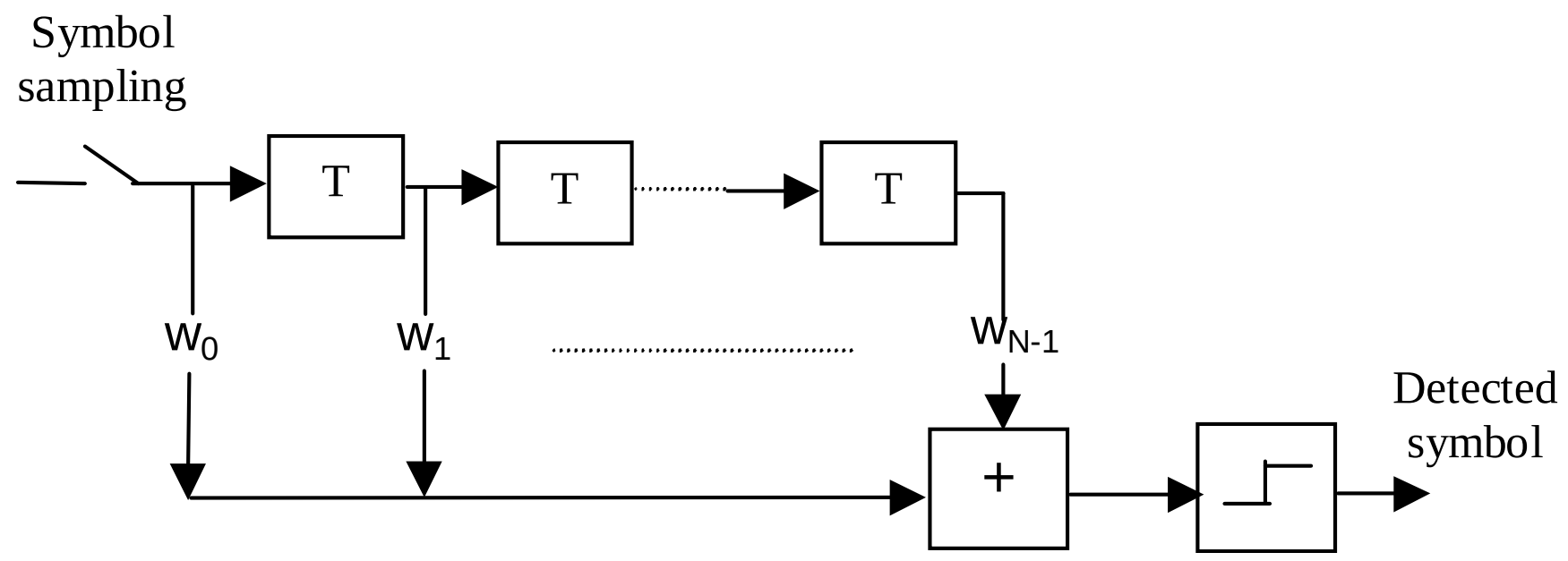

**Figure 2.3**: Linear equaliser.

For analytical purposes, it is useful to consider an equaliser with an infinite number of taps. This allows an easy translation of the Wiener-Hopf equation to the frequency domain. The whole procedure for solving the Wiener-Hopf equation in the case of an infinite length equaliser can be found in [Proakis96], [Benedetto99] and [Qureshi85]. Here, only the final result is presented, which in the frequency domain is given by:

$$W(f) = \frac{1}{S(f) + N_0} \tag{2.7}$$

where W(f) represents the equaliser response in the frequency domain, that is, the discrete-time Fourier transform (DTFT) of the values obtained from solving the Wiener-Hopf equation. Note the important fact that, because the equaliser has an infinite number of taps, the frequency response is continuous. S(f) is the frequency response of the convolution of the pulse shaper, channel and matched filter, or equivalently, the DTFT of the samples entering the equaliser. $N_0$ is the power spectral density of the AWGN.

The DTFT of the equalised signal, Y(f), is then given by:

$$Y(f) = \frac{S(f)}{S(f) + N_0} \tag{2.8}$$

Equation (2.8) shows that, in the event of very low noise level ($N_0$ close to zero), Y(f)=1, which means that the system does not introduce any distortion. Notice that Y(f) represents the frequency transform of all the elements between the symbols source and the detector at the receiver. Y(f)=1 corresponds, in the time domain, to an impulse response function which is the unit operator for the convolution, indicating that what is sent, gets to the receiver undistorted.

The problem arises when $N_0$ is significant. In this case, $Y(f)\neq 1$, which in the time domain implies the presence of ISI terms in the detection process. This problem is especially important when the channel presents spectral nulls in the passband, a situation common in mobile radio channels. As it will be seen in later chapters, this strongly limits the action of the equaliser.

When the equaliser is limited to a finite number of taps, a performance degradation (increased MSE) occurs with respect to the infinite case. The amount of this degradation is difficult to quantify a-priori as it depends very much on the particular channel, equaliser length and noise level. In chapter 5, a detailed coverage of the effect the equaliser length has on the MSE performance will be presented.

For the sake of completeness, it will be mentioned that another popular criterion to design linear equalisers is known as the peak criterion ([Proakis96]). In this case, the equaliser is forced to produce zero-ISI samples at its output. For this reason, they are normally called zero-forcing equalisers. Their frequency response is given by:

$$W_{ZF}(f) = \frac{1}{S(f)} \quad (2.9)$$

where S(f) is again the frequency domain expression for the convolution of the pulse shaper, channel and matched filter. When there is no noise, the MSE-LE and ZF-LE have the same expression as can be inferred from equations (2.7) and (2.9). However, when $N_0$ is significant, the resulting equaliser expressions are quite different as the ZF-LE does not take into account the noise level. Consequently, the ZF-LE may induce large noise enhancements while trying to cancel out all the ISI. In general, the MSE-LE offers better performance than the ZF-LE and is usually the preferred method to design equalisers.

The main advantages of linear equalisers are their simplicity, their tolerance to changing phase channels (going from minimum to maximum phase) and their immunity to decision errors. This last point will be better understood when the LE is compared with decision feedback equaliser (DFE). The main drawback of LE is its sensitivity to noise level which in some situations can make the equaliser completely useless.

## 2.2.2 Decision feedback equaliser

The idea of decision feedback equalisation (DFE) was introduced by Austin in 1967 ([Austin67]). The MSE design of DFEs was presented in [Salz73]. Some more properties, along with comparisons with maximum sequence likelihood receivers (see section 2.2.4), were presented in [Belfiore79]. The review provided by [Qureshi85] also covers the basics of DFE.

In order to understand the principle of DFE, it is first important to have an idea of a typical channel impulse response. Figure 2.4 presents a hypothetical impulse response of a communications channel when sampled at the symbol rate. Generally, the main pulse of energy, the one onto which detection is performed, gets to the receiver with a certain delay. Due to the multipath effect, additional energy arrives before and after the main pulse causing precursor and postcursor ISI respectively.

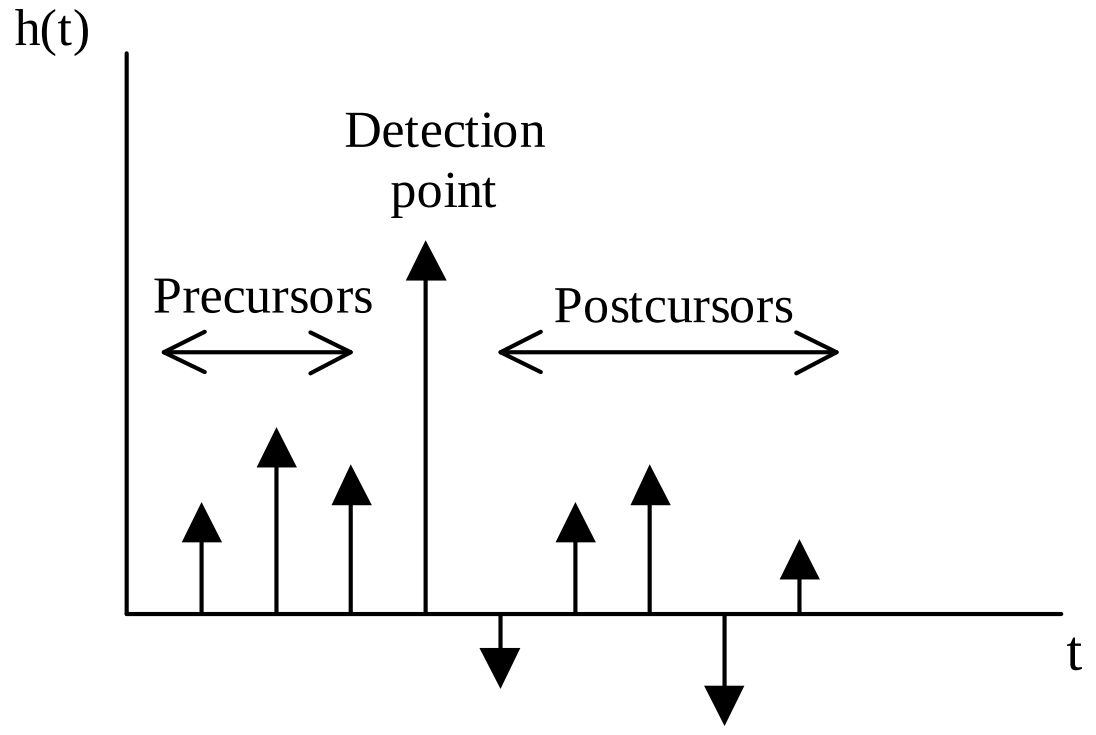

**Figure 2.4**: Example of discrete channel impulse response.

The idea behind DFE is that once a symbol has been detected, its postcursor interference can be perfectly cancelled, as the symbol is already known. The precursor ISI is handled with a conventional linear equaliser. Figure 2.5 shows the structure of a DFE. It has two FIR filters; one a feedforward filter (FFF) which acts as a LE, and the other a feedback filter (FBF) which performs postcursor cancellation. The inputs to the FFF are the samples coming out of the matched filter whereas the inputs to the FBF are the previously detected symbols.

It is important to recognise that the DFE is a non-linear structure because of the symbol detection used to form the inputs to the FBF.

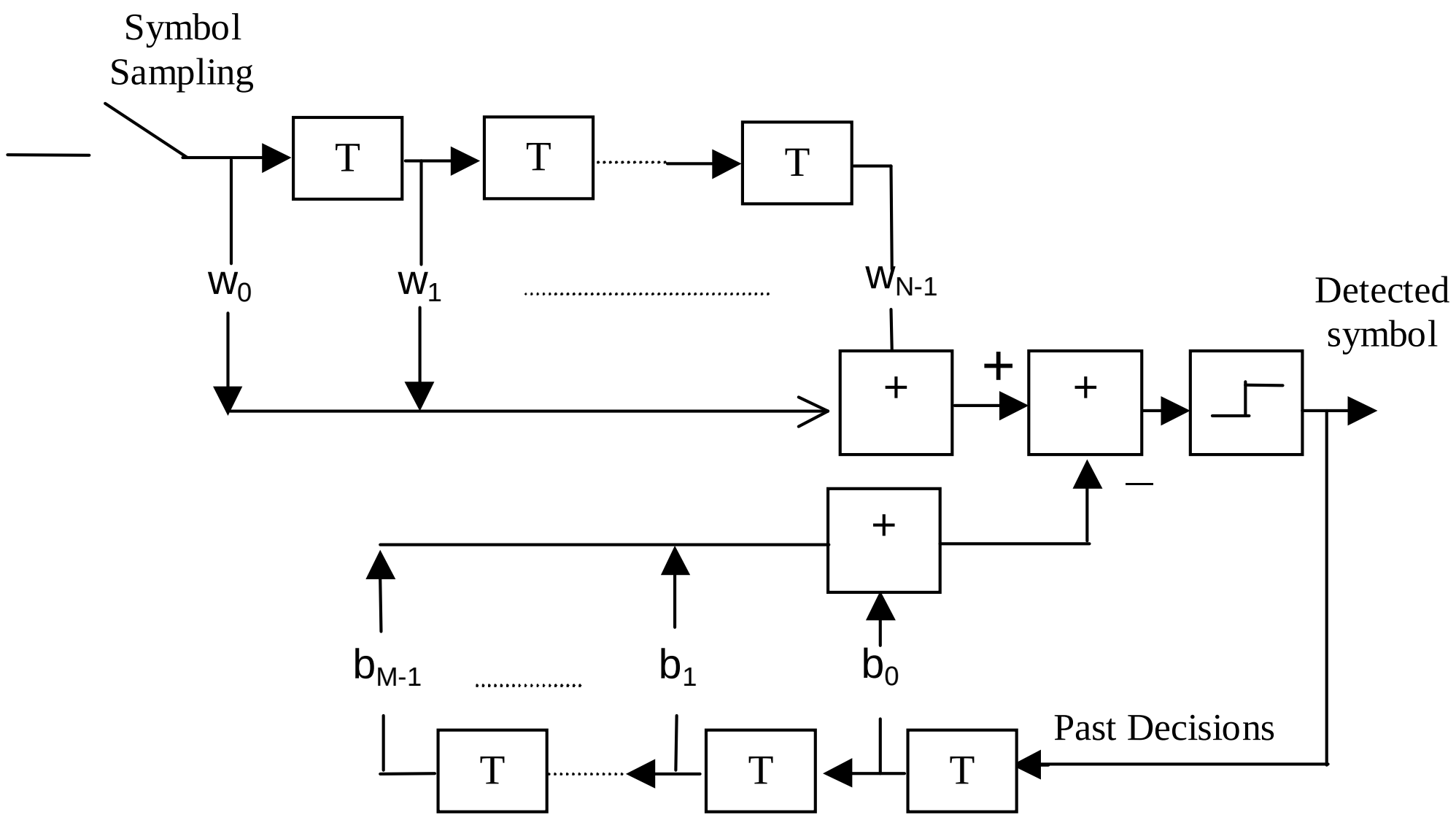

**Figure 2.5**: Decision Feedback Equaliser.

The MSE analysis of the DFE is difficult because of its non-linearity. However, the structure shown in figure 2.5 can be linearised by assuming that the decisions being made by the detector are correct[3]. This assumption allows the Wiener-Hopf equation to be applied in a similar way as for the LE. This derivation is rather long and not included here, but can be found in [Salz73] or [Gitlin92].

The greatest advantage of DFE over LE is the avoidance, in great measure, of any noise enhancement. This is explained by the observation that the FBF works on noise-free samples. As a consequence, the MMSE-DFE achieves a much lower value than the corresponding MMSE-LE when the channel is heavily distorted, i.e. has a spectral null, as it will not raise the noise floor when compensating spectral gaps.

In a real scenario not all symbols will be correctly detected. In this situation, the DFE inherits a new problem, namely, error propagation. When a symbol is wrongly detected it will also be fed back to the FBF producing a wrong cancellation tail. This, in turn, will reduce the noise margin for the detection of subsequent samples. Fortunately error-propagation is not catastrophic ([Benedetto99]) and although its effects tend to make the real DFE performance worse than predicted by the Wiener-Hopf equation, in most cases, its performance is still better than that of the LE. Still, it should be noted that there is no conclusive proof that the DFE will outperform the LE under any conditions. More will be said about error propagation in section 6.9.

[3] This issue is treated in more detail in section 6.1.

The application of DFEs to multipath fading radio channels is treated in detail in [Foschini82]. More recently, due to increasing bit rates, there has been a renewed interest in DFE as a method to compensate effectively the channel effects with a moderate computational complexity. [Al-Dhahair95], [Al-Dhahair96] and [Voois96] have presented new results in the area of DFE specifically related to practical implementation issues such as finite-length effects and choice of decision delay. The relation between DFE equalisation and coding has been studied in [Cioffi95].

### 2.2.3 Fractionally spaced equaliser

Fractionally spaced equalisation (FSE) is a modification applicable to any of the two structures introduced previously, LE and DFE, rather than a structure by itself. However, given its importance it is treated separately. It was first proposed in 1969 in an unpublished work by A. Gersho of Bell Labs ([Benedetto99]) and later popularised by [Ungerboeck76] and [Gitlin81]. A more recent tutorial on FSE appeared in [Treichler96].

In figures 2.3 and 2.5, where the structures of the LE and DFE were shown, it can be appreciated that the equaliser taps are spaced at symbol rate, a direct consequence of the symbol sampling at the input of the equaliser. Some important benefits are achieved if the input to the equaliser is sampled at a faster rate and the equaliser taps are correspondingly spaced closer. An example of a LE with a tap-spacing of T/2, where T is the symbol period, is shown in figure 2.6.

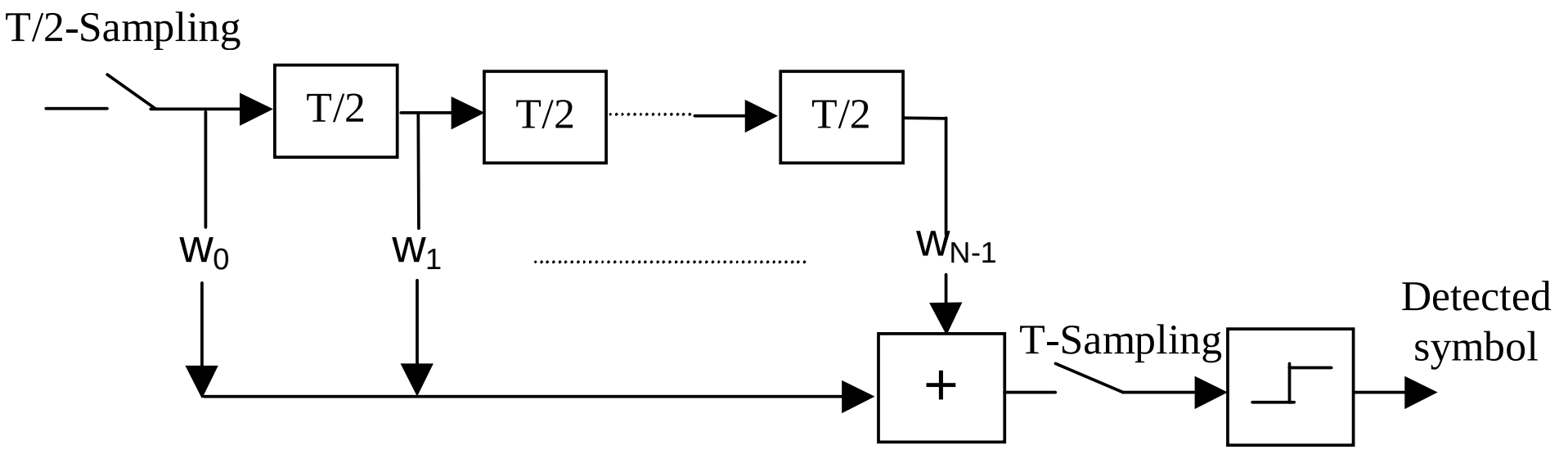

**Figure 2.6**: T/2-Fractionally spaced linear equaliser.

Before describing its benefits, there are two important points to note about the structure shown in figure 2.6. First, notice that although the sampling rate at the input of the equaliser is double the symbol rate, the equaliser output is sampled at symbol rate. In figures 2.3 and 2.5, the sampler before the detector was not shown as it was clear that the whole system was working at symbol rate. Now the sampler is made explicit in order to emphasise that the equaliser inputs and outputs are sampled at different rates.

The other important point to mention is that FSEs need more taps than symbol spaced equalisers to cover a certain time span. For example, if N taps are used in a conventional equaliser, 2N taps will be needed in a T/2-FSE covering the same time span. This has important consequences from a computational point of view, as FSEs tend to be much longer than symbol spaced equalisers.

FSEs offer two very important advantages, so important, that nearly all equalisers implemented in real communications receivers are fractionally spaced.

The first of these advantages is related to the bandwidth of the transmitted signal. In principle, given a particular bit rate, $f_0=1/T_0$, a signal with minimum bandwidth occupancy could be realised by transmitting sinc pulses (chapter 8 in [Glover98]). In this case, the signal bandwidth would be limited to $1/2T_0$. In practice, this is not possible, as it would require the use of ideal filters, which are not realisable, and an extremely precise sampler at the receiver. The alternative is to use filters (pulse shapers) that produce a bandwidth expansion (excess bandwidth) with respect to the minimum case. If these pulses are transmitted and they are sampled at the receiver symbol rate, aliasing will occur as the Nyquist sampling criterion is not satisfied because of the bandwidth expansion. This is the situation that takes place in a symbol-spaced equaliser. On the other hand, if samples are taken fast enough so as to satisfy the Nyquist sampling theorem, no aliasing results at the input of the equaliser, allowing the equaliser to modify independently any frequency band of the received signal. Fractionally spaced equalisers are designed so that the Nyquist sampling theorem is satisfied at their input. In this way, MSE levels lower than for equivalent symbol-spaced equalisers are achieved.

The second important property of FSEs lies in their capability to correct timing phase without raising the noise floor. This compensates any delay (phase) distortion without noise enhancement and relaxes the synchronisation mechanisms at the receiver.

Both facts, avoidance of aliasing and compensation of delay distortion, make FSEs the preferred choice, rather than symbol spaced filters, to implement LEs or the FFF of DFEs. Notice in this latter case that the FFF is the only filter in the DFE whose taps can be fractionally spaced. Given that the outputs of the DFE (inputs to the FBF) are generated at symbol rate, the FBF must also be symbol spaced.
Mathematical derivations proving the advantages of FSEs can be found [Ungerboeck76], [Gitlin81] and [Gitlin92]. A recent treatment on the optimum design of MSE-FSE-LE appeared in [Casas97].

### 2.2.4 Other multipath mitigation structures

Over the years, many different techniques to counteract the effects of the channel have appeared. Some of them are variations of the LE or DFE, while some others rely on completely different principles. Our work in later chapters is solely based on LE and DFE. However, for the sake of completeness, a brief account of some of these other techniques is presented. The first methods described below, pre-equalisers and cancellers, are modifications of the DFE. The application of maximum likelihood sequence estimation as a method to compensate ISI is presented next. The last structure, the RAKE receiver, is mentioned briefly, as it will be covered in more detail in section 2.4.2. This section concludes with a quick list of and brief comments on some other methods that have recently appeared in the literature.

PRE-EQUALIZATION

Pre-equalisation, also known as Tomlinson-Harashima precoding, was simultaneously proposed by [Tomlinson71] and [Harashima72]. The main idea of this scheme is to move the feedback filter of a DFE to the transmitter side so that the symbols sent through the channel have their post cursor ISI already suppressed (i.e. pre-equalised). Given that the symbols are known at the transmitter, error propagation is avoided.
For this technique to work, the transmitter should have perfect knowledge of the channel impulse response, which limits its usage to systems whose CIR is fixed and known or to systems in which there is a feedback channel between transmitter and receiver. In this latter case, the receiver has the additional task of estimating the CIR.

The combination of Trellis coded modulation with pre-equalisation has been studied in [Forney91].

CANCELLERS

Cancellers are another extension to the DFE original structure proposed in [Gersho81]. The idea is to cancel the precursor ISI in addition to the postcursor ISI. To this end, the canceller is preceded by a conventional LE that provides tentative decisions about future samples whose ISI (precursor) is then subtracted from the symbol being detected. Cancellation of the postcursor ISI is achieved in a similar fashion to the DFE by feeding back the symbols already detected.

MLSE

Maximum likelihood sequence estimation is a rather general block technique which, given an observed sequence, selects the one which most likely has been transmitted. It has already been introduced in Chapter 1 as a method for decoding convolutionally encoded sequences. The application of MLSE to the detection of symbols corrupted by a linear distortive channel was proposed in [Forney72]. Using the Viterbi algorithm ([Forney73]) the complexity of this nearest neighbour detection can be shown to be $L^M$ where L is size of the alphabet and M the channel dispersion or channel memory.

The main drawback of this approach is that its computational complexity grows exponentially with the channel dispersion, therefore its use is only feasible in channels with a short impulse response.

The application of MLSE for ISI removal is often known as Viterbi equalisation and is the preferred equaliser structure in the GSM system. It is important to mention that the symbol rate in GSM is relatively low and the channel dispersion is limited to 4-6 symbol intervals. This means that the Viterbi equaliser only has to perform the search and detection among $2^4$-$2^6$ possible sequences. Full explanation of the GSM Viterbi equaliser can be found in [Sklar97b]. The exponential dependence of the computational complexity on channel dispersion precludes this method from being used in very high bit rate systems where channel dispersion extends over many symbol periods.

RAKE

RAKE[4] receivers were introduced in [Price58] to exploit the multipath components as a form of diversity in spread spectrum systems. The details of RAKE receivers will be presented in section 2.4.2 in the context of multipath mitigation techniques applied in mobile radio systems.

OTHER STRUCTURES

Many other equalisation techniques have appeared and continue to appear in the technical literature. Some of them are now mentioned briefly. In [Duel-Halen89] it was proposed to substitute the FFF in a DFE for Viterbi sequence estimator. [Ariyavisitakul92] introduced the concept of DFE with time-reversal structure. The idea is to store blocks of received samples and time-reverse them prior to equalisation; as a result, the channel seen by the equaliser is a time-reversed version of the real channel. Selective time-reversal allows a DFE with a very short FFF (potentially just 1 tap) to work equally well in minimum and maximum phase channels. In [Raghavan93] the design of linear equalisers with non-uniform tap spacing was presented, the resulting techniques offering computational reductions with respect to conventional LE. In [Ariyavisitakul97], the DFE with a non-uniformly spaced FFF was introduced. The proposed equaliser was able to deal satisfactorily with channels of very different delay spreads.

Concluding this quick review of different equalising structures, it must be mentioned that the combination of equalisation and diversity reception has also been covered by different authors, the main references being the thorough treatments given in [Balaban92a] and [Balaban92b]. In these papers, it is shown that this combination can lead to reductions of up to two orders of magnitude in the BER with respect to single branch receivers. In [Lo91] a combination of DFE and diversity was applied to the particular scenario of a TDMA mobile system (IS-54). Additional results have been recently published in [Ariyavisitakul97]. Unfortunately, in the context of mobile systems, use of diversity in the downlink is rather limited due to the reduced physical dimensions of mobile handsets.

---

[4] RAKE is not short for anything, however in the communications literature has historically been written in capital letters.

# 2.3 Adaptive algorithms

Most of the structures described in the previous section can be used to mitigate the effects of known static channels. In that case, their parameters (i.e. filter taps) can be computed a-priori. However, in most practical situations, certainly in the case of mobile radio systems, the channel is unknown and time varying. Under these conditions, the equaliser parameters must be adaptable and able to respond to the channel changes. This is the reason why equalising systems are typically implemented using adaptive filters, which can be defined as digital filters whose coefficients vary with time.

In order to drive the equaliser coefficients, an adaptive algorithm must be used. This section describes the most common algorithms to perform this task. These algorithms can be broadly divided into gradient methods and least squares methods. They are covered in sections 2.3.1 and 2.3.2 respectively. Section 2.3.3 introduces a particular low computational complexity least squares method.

Before each technique is presented in some detail, a few general comments must be made. All the methods presented, have as an objective the computation of the Wiener-Hopf equation in a recursive way. This procedure has two important benefits. First, no knowledge of the channel is needed a-priori as the algorithms are able to learn from the observed data. Second, a huge reduction in the computational complexity of the solution to the Wiener-Hopf equation is achieved when compared with the direct solution[5] of equation (2.5). These benefits, however, do not come for free. The penalty to pay is that the filter will have to undergo a convergence process before it achieves its near optimum values. Moreover, the steady-state filter values will not exactly coincide with those of the Wiener-Hopf solution as, in general, adaptive algorithms add some form of error (excess error). This will be covered in detail in chapter 4.

The presentation of the algorithms in the following subsections is done in the context of FIR filtering. As has already been mentioned, IIR filtering present some inconveniences that, at the moment, greatly limit their application in real systems. For tutorial review of adaptive IIR filtering see [Shynk89].

---

[5] Matrix inversion procedures have a computational complexity of $O(M^3)$ where M is the matrix dimension (see [Press92]).

Finally, as a last general comment about the algorithms to be covered, it must be pointed out that they all work in the time domain. Different schemes have appeared that perform the filter adaptation in the frequency domain. These methods offer some computational reductions over time domain techniques when the filters involved are very long. Frequency domain algorithms are typically used in applications such as acoustic echo cancellation, where adaptive filters with orders of a few thousands taps are not uncommon. In [Shynk92] a review of frequency domain techniques can be found.

The main reference for the information presented in next sections is the comprehensive textbook by Haykin ([Haykin96]), and also [Farhang-Boroujeny98] and [Mulgrew99].

## 2.3.1 Least mean squares algorithm and variants

The least mean squares (LMS) algorithm is the most widely used adaptive scheme due to its simplicity and robustness. It was originally proposed in [Widrow60] and since then a huge number of variants and applications have appeared in the technical literature.

The LMS algorithm is derived from a general technique known as steepest descent (SD), which is often used in optimisation problems. In section 2.1 it has already been stated that the MSE, as a function of the filter weights, has the shape of a multidimensional bowl-shaped paraboloid with a single minimum. The SD algorithm finds the minimum MSE (MMSE) point by computing the gradient vector of the MSE, that is, the direction of maximum MSE increase, and then updating the coefficients one step in the opposite direction (maximum MSE decrease). By iterating this basic procedure, the filter weights will converge to the values producing the MMSE. The algorithm is given by the equations:

$$\nabla\, \mathbf{MSE}(n) = -2\,\mathbf{p} + 2\,\mathbf{R}\,\mathbf{w}(n) \qquad \text{Gradient computation} \qquad (2.10)$$

$$\mathbf{w}(n+1) = \mathbf{w}(n) - \mu \nabla \mathbf{MSE}(n) \qquad \text{Filter update} \qquad (2.11)$$

In the above equations, $\nabla\mathbf{MSE}(n)$ denotes the gradient vector of the MSE, **R** is the autocorrelation matrix of input data, **p** is the cross-correlation vector between the desired and observed data and **w** is the filter weights vector. All these vectors have length equal to the filter order (N). The parameter $\mu$ is an arbitrary step size that defines the amount of correction applied to the filter weights. The derivation of equation (2.10) can be found in pp. 342 of [Haykin96].

From a practical point of view, the SD algorithm has one important problem: typically, **R** and **p** are unknown in real situations, and therefore the exact gradient of equation (2.10) cannot be computed.

Fortunately, an estimation of the gradient vector can be computed from the available data. This estimated gradient, $\hat{\nabla}\mathbf{MSE}(n)$, is given by ([Haykin96]):

$$\hat{\nabla}\mathbf{MSE}(n) = -2\mathbf{u}(n)d(n) + 2\mathbf{u}(n)\mathbf{u}^T(n)\hat{\mathbf{w}}(n) \tag{2.12}$$

where d(n) is the desired output value. We use a different notation for the filter weights used, $\hat{\mathbf{w}}(n)$, to emphasise that they are computed using an estimated (noisy) gradient.

Using this estimated gradient vector, rather than the exact one, in equation (2.11), gives rise to the LMS algorithm:

$$e(n) = d(n) - \hat{\mathbf{w}}^T(n)\mathbf{u}(n) \tag{2.13}$$

$$\hat{\mathbf{w}}(n+1) = \hat{\mathbf{w}}(n) + \mu\,\mathbf{u}(n)e(n) \tag{2.14}$$

Notice that the complexity of the algorithm given by equations (2.13) and (2.14) is only O(2N) where N is the order of the filter.

The general analysis of the LMS algorithm in stationary and non-stationary scenarios was presented in [Widrow76]. Other papers have presented additional analytic results for particular applications. In the case of equalisation, [Ungerboeck72] is particularly important as it analysed for the first time the transient behaviour of adaptive linear equalisers employing the LMS algorithm. [Gitlin79] analysed the behaviour of the LMS linear equaliser when the filter is implemented in finite-precision arithmetic.

Complete derivation for most of the analytical results for the LMS can be found in a unified way in chapter 9 of [Haykin96]. Additionally, in chapters 4 and 6 of this thesis, some of these results are re-derived for the specific case of equalisation. For the time being, only the main conclusions of the different analyses are enounced:

- *Stability*. The LMS algorithm can be shown to converge as long as the following condition on the step size holds ([Farhang-Borujeny98]):

$$0 < \mu < \frac{2}{3\|\mathbf{u}(n)\|^2}$$

- *Steady-state MSE error*. The algorithm, once converged, will always exhibit an MSE level superior to the MMSE predicted by the Wiener-Hopf equation. This excess mean squared error (EMSE) is caused by the use of a step size greater than 0. In fact, the EMSE level is proportional to the value of $\mu$. The larger $\mu$ is, the larger is the EMSE component.

- *MSE Convergence*. The MSE presents a convergence that follows a decaying exponential pattern. The time constant in the exponential (i.e. rate of convergence) is directly influenced by two factors. First, the step size: the larger $\mu$ is, the faster the algorithm converges. Second, a direct relation exists between the eigenvalue spread of the autocorrelation matrix of the input data, $\chi(\mathbf{R})$, and speed of convergence. The larger $\chi(\mathbf{R})$ is, the more time it takes the algorithm to achieve the steady state.

This last result about convergence has a special relevance for the case of equalisation. It can be proved ([Gray72], [Ungerboeck72]) that, in the case of equalisation, $\chi(\mathbf{R})$ is bounded by the maximum and minimum values of the channel power spectral density. This in turn implies that in channels presenting frequency selectivity, the equaliser may exhibit a slow convergence. This effect is very noticeable if the channel spectrum contains a null in the passband.

Also, notice the conflicting requirements for the selection of $\mu$. If the step-size is large, convergence might be accelerated at the cost of a higher steady-state MSE. On the contrary, if $\mu$ is chosen to be small, the steady-state MSE will approach that of the Wiener solution but convergence time will be much longer. Often a compromise $\mu$ is chosen in the middle of the stability region, i.e.:

$$\mu = \frac{1}{3\|\mathbf{u}(n)\|^2}$$

A step size is typically chosen with the help of computer simulations. Its value is also dependent on the non-stationarity of the channel as the step size, $\mu$, needs to be large enough to allow an adequate channel tracking.

In order to improve the performance of the LMS algorithm, a number of variants have appeared, improving some of its deficiencies. Two of them are explained below as they have been extensively used throughout this project; they are the normalised LMS (NLMS) and the variable step size LMS (VSLMS).

The NLMS algorithm differs from the conventional LMS in that the correction applied to the filter weights is normalised with respect to the power of the input vector. In this way, gradient noise amplification due to large input samples values is avoided. The algorithm is given by:

$$e(n) = d(n) - \hat{\mathbf{w}}^T(n)\mathbf{u}(n) \tag{2.15}$$

$$\hat{\mathbf{w}}(n+1) = \hat{\mathbf{w}}(n) + \frac{\tilde{\mu}}{a + \|\mathbf{u}(n)\|^2}\mathbf{u}(n)e(n) \tag{2.16}$$

Notice the normalisation factor $\|\mathbf{u}(n)\|^2$ dividing the step size. The parameter a is a small positive number used to avoid numerical problems if $\|\mathbf{u}(n)\|^2$ is close to zero. For the NLMS to be stable (i.e. MSE convergent), the step size must satisfy the following relation:

$$0 < \tilde{\mu} < 2$$

The main advantage of the NLMS is that by avoiding the gradient noise amplification, the algorithm has the potential to convergence faster. Simulation results in the next chapter confirm this fact. The major drawback is an increase in computational complexity to O(3N), and the need to perform a division. This latter point might be an issue in custom hardware structures.

The VSLMS algorithm, as its name suggests, dynamically varies the step-size in order to overcome the conflicts of a static step-size. Initially, $\mu$ has a large value (close to the upper stability limit), so that convergence is accelerated. As the filter converges toward its steady state, the value of $\mu$ is decreased in order to reduce the steady state MSE. Algorithms implementing this idea have appeared in [Kwong92], [Mathews93], [Aboulnasr97] and [Woo98]. The one implemented in our system is the algorithm in [Kwong92], which adds an extra equation to those given by the conventional LMS (equations 2.13 and 2.14). This additional equation is the step size update and is given by:

$$\mu'(n+1) = a\mu'(n) + \rho e(n)^2 \tag{2.17}$$

The parameter a is an exponential forgetting factor used to give more importance to recent data. The variable $\rho$ is typically a very small positive value determined by the steady-state level required. In our work, their values have been determined by simulation.

Over the years, many other LMS variations have appeared in the literature. The sign-error-LMS algorithm introduced by ([Gersho84]) only uses the sign of the error to update the filter weights, that is, multiplication by either 1 or –1. Performance degradation can be observed in the sign-LMS algorithm when compared to the plain LMS. However it has the advantage of being implemented in a very efficient way on hardware. The self-orthogonalising-LMS algorithm proposed in [Gitlin77] pre-multiplies the weight correction term of the LMS algorithm (equation 2.14) by the inverse of **R**. This has the effect of whitening the input to the equaliser which in turn reduces its eigenvalue spread and accelerates its convergence. The main drawback is its computational complexity, which grows quadratically with the filter length. In [Sari82], a variation of the self-orthogonalising-LMS is presented with reduced computational burden. Finally, several variants of the LMS algorithm ([Aboulnar99] and [Douglas97]) have appeared where only a subset of all the coefficients in the filter is updated.

### 2.3.2 RLS algorithm

The recursive least squares (RLS) algorithm overcomes most of the performance inconveniences of the LMS algorithm. In particular, it has a very rapid convergence rate as in approximately 2N iterations (N = filter length) finds the set of filter weights approaching the Wiener solution. Moreover, in the case of static channels, it achieves the Wiener solution without any excess error. Additionally, the convergence rate is independent of the eigenvalue spread of the input signal.

All these nice features are achieved by solving the Wiener-Hopf equation in a rather different form than the LMS algorithm. In order to explain the RLS algorithm, first the least squares principle will be introduced and then it will be explained how the algorithm can be made recursive. A thorough review of least squares techniques has recently appeared in [Glentis99].

Recall that LMS finds the MMSE solution by applying a gradient descent along the MSE surface using estimates of the required autocorrelation and cross-correlation functions. The need to estimate certain correlation functions is because, in an unknown channel, such functions will not be available. Least squares techniques face the same problem of lack of knowledge of **R** and **p** in equation (2.4). They tackle it in a different way by substituting ensemble averages for time averages.

The time averaged squared error is defined as:

$$\varepsilon(n) = \sum_{i=0}^{N} |e(i)|^2 \qquad (2.18)$$

In the above equation, N is an arbitrary number of iterations and e(i)=y(i)-d(i). Notice that $\varepsilon(n)$ is really a scaled version of the time average as it is not divided by the total number of iterations, N. Nevertheless this factor is typically ignored in the literature.

Similar to the derivation of the Wiener filter, it is possible to minimise $\varepsilon(n)$ as a function of the filter weights and get a temporal version of equation (2.4). The resulting equation has a very similar form to the Wiener-Hopf equation:

$$\mathbf{\Phi}\,\hat{\mathbf{w}} = \mathbf{z} \qquad (2.19)$$

where $\mathbf{\Phi} = \sum_{i=0}^{N} \mathbf{u}(i)\mathbf{u}^T(i)$ is the temporal correlation matrix of the input data and **z** is the temporal cross-correlation between the input vector and the desired output. The optimum filter weights are given by:

$$\hat{\mathbf{w}} = \mathbf{\Phi}^{-1}\mathbf{z} \qquad (2.20)$$

This set of filter coefficients is called the least squares solution in allusion to the error function that minimises (equation (2.18)). An important point to notice is that if the underlying process is ergodic, equations (2.20) and (2.4) are equivalent, and consequently, the least squares solution is the same as the Wiener solution.

Equation (2.20) is practical to compute, as it is based on time-averages of observed samples. However, notice that for every new sample of data that becomes available, the filter weights need to be recomputed to exploit the new information. This calculation implies the inversion of a NxN matrix, a process which has a computational complexity of $O(N^3)$.

Fortunately, a result known as the matrix inversion lemma[6] allows the recursive computation of the inverse of the correlation matrix and equation (2.20) to be put in the form of a recursive algorithm known as recursive least squares (RLS). The transformation from (2.20) to the RLS algorithm is rather long and can be found in [Haykin96], [Mulgrew98] or [Farhang-Boroujeny98]. Here only the final form of the algorithm is shown.

---

[6] The origins of this result are unknown, see pp. 566 in [Haykin96] for some information.

$$\mathbf{k}(n) = \frac{\lambda^{-1}\mathbf{P}(n-1)\mathbf{u}(n)}{1+\lambda^{-1}\mathbf{u}^T(n)\mathbf{P}(n-1)\mathbf{u}(n)} \quad \text{Kalman gain calculation} \quad (2.21)$$

$$\zeta(n) = d(n) - \hat{\mathbf{w}}^T\mathbf{u}(n) \quad \text{Error calculation} \quad (2.22)$$

$$\hat{\mathbf{w}}(n) = \hat{\mathbf{w}}(n-1) + \mathbf{k}(n)\zeta(n) \quad \text{Filter update} \quad (2.23)$$

$$\mathbf{P}(n) = \lambda^{-1}\mathbf{P}(n-1) - \lambda^{-1}\mathbf{k}(n)\mathbf{u}^T(n)\mathbf{P}(n-1) \quad \text{Inversion of the input correlation matrix} \quad (2.24)$$

The vector **k**(n), historically called the Kalman gain, represents the gain applied to the error signal, $\zeta(n)$, to update the filter weights. **P**(n) is the inverse of the input correlation matrix, i.e. $P(n) = \mathbf{\Phi}^{-1}(n)$. Its recursive computation appears in equation (2.24). The parameter $\lambda$ is a forgetting factor that weights more importantly new data over older data[7]. If the channel is static then $\lambda$ is set to 1. If the channel is time-varying, $\lambda$ will have a value close to, but less than 1. It can be shown (pp. 563-564 in [Haykin96]) that when $\lambda<1$, the RLS algorithm no longer converges exactly to the Wiener solution since, as with the LMS algorithm, an excess error is introduced.

There are several methods to initialise the RLS algorithm. The procedure used in this thesis is called soft-constrained initialisation. This technique initialises the algorithm by setting $\hat{\mathbf{w}}(0) = \mathbf{0}$ and $\mathbf{P}(0) = \delta^{-1}\mathbf{I}$, where **I** is the NxN identity matrix and $\delta$ is a number which should be small compared to the power of the input samples. The specific value of $\delta$ does not affect the steady-state performance of the algorithm but it influences its convergence. A thorough treatment of initialisation methods for the RLS algorithm can be found in [Hubing91].

The RLS algorithm represents a major reduction in complexity with respect to the original formulation of the least squares problem given by equation (2.19). However, from equations (2.21)-(2.24) it is found that its complexity[8] is still $O(2N^2)$, far larger than that of the LMS, which is linear. In applications where the filter has a moderate to large number of taps, such a complexity order may become an issue, especially when the device has power consumption limitations (like a mobile handset). In the next section, a particular class of least squares algorithms is presented that reduces the computational complexity of the RLS algorithm.

---

[7] This particular form of the RLS algorithm is usually known as the exponentially weighted RLS.

[8] Exploiting some symmetry properties that **P**(n) must satisfy.

Another general problem of the least square approach is that the recursive algorithms tend to present numerical stability problems when implemented on finite-precision arithmetic (especially fixed-point). Sources and solutions to RLS instability have been presented in [Bottomley89] and [Bottomley91].

### 2.3.3 Fast Kalman algorithms

The previous section showed that the RLS algorithm offers some performance advantages over the LMS algorithm at the expense of a significant increase in computational complexity. This last issue has motivated the search for variations of the least squares scheme with a reduced, preferably linear, computational complexity. The resulting algorithms are generally known as Fast Kalman algorithms. The first such technique found, [Morf76], dates from 1976. Since then many variations have appeared offering different advantages and complexities. It is worth mentioning [Falconer77], where a Fast Kalman algorithm with a complexity of O(10N) was applied for the particular case of channel equalisation. In [Carayannis83], another variation called fast a-posteriori error sequential technique (FAEST) with a complexity of O(7N) was introduced.

Much of the complexity of the RLS algorithm resides in the computation of the Kalman gain (equation (2.21)). All the Fast Kalman algorithms mentioned have in common the exploitation of the shift invariant property (SIP) of the input data in order to reduce the number of computations in calculating the Kalman gain. The SIP states that the input vector to the adaptive filter at instant n, is the same as the one at instant n-1, except that there is a new sample, the oldest sample has been dropped out and the rest of elements in the tap-delay line have shifted along one position.

When the SIP is satisfied, computation of the Kalman gain is greatly simplified with the use of linear forward and backward prediction techniques. A detailed explanation of the FAEST algorithm is beyond the scope of this introduction to adaptive filters. The reader is referred to the original reference ([Carayannis83]) for its original formulation and simulation results.

At this point, it must be reinforced the fact that Fast Kalman and RLS schemes are all the same algorithm but using different formulations.

Fast Kalman algorithms inherit and exacerbate the stability problems of the RLS algorithm when implemented on finite precision arithmetic. The instability may appear in two forms [Moustakides89]. In the first one, explosive divergence, the Kalman gain suddenly becomes extremely large and so do the filter weights. In the second one, lock-up divergence, the Kalman gain tends to zero and the filter coefficients stop being updated. The simplest way to tackle instability is by periodically restarting the filter and algorithm. However, this approach will degrade the performance of the system due to the resets.

More elaborate approaches have appeared, and continue to appear, that algorithmically stabilise the Fast Kalman algorithm. Examples are [Cioffi84], [Botto89], [Moustakides89] and [Slock91]. The common approach in these techniques is to compute some of the variables in the algorithm in two different ways. In infinite precision arithmetic, both methods would produce the same result. In finite precision arithmetic, some discrepancy appears between the two results. This difference is then utilised to apply a correction term to particular variables.

In all these references, the adaptive filter is configured as a channel estimator and the reported results indicate that these approaches successfully prevent the algorithm instability.

In this project, the stabilised FAEST algorithm (SFAEST, [Moustakides89]) has been implemented and extensively tested for the case of channel equalisation. We have found that it works most of the times but occasionally goes unstable. Additionally, the algorithm parameters were difficult to tune to obtain satisfactory performance. Some simulation results with the SFAEST scheme will be shown in the next chapter.

### 2.3.4 Comparison of gradient and least squares techniques

Concluding this short introduction to adaptive algorithms, it is worth paying some attention at how gradient (LMS) and least squares (RLS) algorithms compare.

During the transient phase, the RLS algorithm clearly offers some advantages over the LMS algorithm, especially if the eigenvalue spread of the autocorrelation matrix is large. For the particular scenario of equalisation, this condition is likely to happen when the E/No is large and the channel is heavily frequency selective. If E/No is low, the spectral gaps of the channel will be less pronounced because of the large noise level, reducing in this way the eigenvalue spread.

Once the filter has converged, the filter will start to track the channel fluctuations. It is important to remark now that tracking is a steady-state phenomenon. Comparisons of the tracking capabilities of LMS and RLS are rather more involved than in the case of convergence and previous studies ([Eweda94] and chapter 16 in [Haykin96]) have shown that the superiority of one or the other depends on the particular application.

In [Cioffi86] the tracking ability of both algorithms is analysed. It is concluded that in scenarios with large time variations (i.e. abrupt changes), the RLS algorithm has better tracking properties than the LMS. However, it mentions that for this to be true, the variations need to be much larger than those encountered in practical systems.

Also interesting is the work in [Rupp97b] where the LMS-DFE and RLS-DFE are compared in a mobile radio environment. The conclusion is that the LMS-DFE can track as well as the RLS-DFE, provided that the RLS algorithm was used during the training. This suggests a switching of algorithm between training and decision directed mode.

Finally, [Eleftheriou86] observed that if the adaptive filter is very long, the LMS step size will have to be very small to satisfy the stability criterion. This will limit the tracking ability of the LMS filter. Under these conditions, the RLS algorithm has better tracking capabilities.

## 2.4 Applications of equalisers to mobile radio systems

Finishing this introduction to adaptive equalisers structures and algorithms it is important to look at how equalisation is used in modern mobile systems. Mobile systems are mainly characterised by the technique used to multiplex users. As mentioned in chapter 1, TDMA and CDMA are the two mostly in use nowadays. Different forms of equalisation have traditionally been used in TDMA systems as described in section 2.4.1. In contrast, the application of equalisers to CDMA systems has just been proposed. In section 2.4.2 it will be shown how adaptive equalisers fit into a spread spectrum system. The section concludes with some comments about the training procedures of equalisers.

## 2.4.1 TDMA systems

TDMA systems can be considered as the classical scenario where all the techniques introduced in the previous sections of this chapter are applied. Equalisation has typically been performed in TDMA systems in two different ways: direct equalisation and Viterbi (MLSE) equalisation.

The direct equalisation approach utilises the adaptive equaliser to reverse the channel distortion. An example of system where this technique is usually applied is the American TDMA IS-54 system ([Rappaport96]). Most receivers operating in this standard incorporate an adaptive DFE with 3 feedforward taps (T/2-spaced) and 1 feedback tap. Given that the equaliser is very short, utilisation of the RLS algorithm is affordable.

The other technique, Viterbi or MLSE equalisation, uses an adaptive filter to get an estimate of the channel impulse response. It then passes a sequence of known bits (training sequence) through the estimated channel. The difference between the original training sequence and that obtained after being filtered by the channel estimate provides the receiver with information on how the channel is distorting the transmitted bits. This information is then used to perform a process of maximum likelihood sequence estimation (MLSE) on the received bits by means of the Viterbi algorithm. The GSM receivers usually resort to Viterbi equalisation to eliminate the ISI ([Sklar01]).

Surveys of equalisation methods and algorithms for TDMA systems can be found in [Proakis91] and [Tijdhof97].

Not long ago, a new approach combining the two techniques just described appeared in the literature ([Lo91], [Shukla91]). This technique, called channel-estimation-based equalisation, first estimates (and tracks) the channel and then, periodically, uses the channel estimate to compute the Wiener-Hopf equation in order to obtain the optimum equaliser settings. These optimum coefficients are then plugged into the equaliser that processes the incoming signal. The advantage this method offers is that, usually, the channel is easier to track than its inverse (direct equaliser).

Over the years many publications have appeared reporting results on equalisation for TDMA or similar systems. A brief account is provided next. In [Lo95] it is shown that the DFE can be used to reduce not only the multipath ISI, but also the effects of co-channel interference. Similar results were also reported in [Niger91]. In [Lo99] the same topic, reduction of ISI and co-channel interference, is addressed but in this case using a blind (no training sequence) method. In [Fukawa91], [Rupp97], [Alberi98] and [Rupp99] results are given on the performance of the different algorithms and structures described in this chapter in various mobile scenarios. Finally, in [Husson99] a method is presented to reduce power consumption by equalising the incoming signal only when it is significantly distorted.

### 2.4.2 CDMA systems

Spread spectrum systems have typically used the RAKE receiver to exploit diversity reception using the received signal through different paths ([Turin80]). The effectiveness of the RAKE receiver depends on the properties of the sequence used to spread the original information signal. If the sequence is well designed, its autocorrelation function will be just an impulse located at zero delay. This means that the original signal, when correlated with any delayed version of itself will be zero.

This autocorrelation correlation property is exploited in the architecture of the RAKE receiver shown in figure 2.7. The RAKE works as follows: first, the path delays are estimated and the P strongest paths are selected (in the figure P=3). A correlator is then assigned to each of these paths. The function of each correlator is to multiply the incoming signal with the same spreading sequence used at the transmitter. The outputs of the correlators are delayed appropriately to synchronise them and then added up together. Finally, symbol detection is performed. In order for the RAKE to be effective in a mobile scenario, the path delays must be tracked continuously to follow the variations in the channel profile. Using this method, the signal energy coming from the P strongest paths is independently received through different branches, called fingers, and coherently summed. Simulation results ([Chan94]) suggest that 3 fingers are enough to capture most of the energy reaching the receiver.

The RAKE receiver is also able to separate the users in a CDMA system thanks to the orthogonality (zero cross-correlations) among the different spreading codes used by the different users. Most IS-95 equipment uses RAKE receivers ([Rappaport96], [Sklar97b]) and it is also the recommended basic reception technique in UMTS ([3GPP00]).

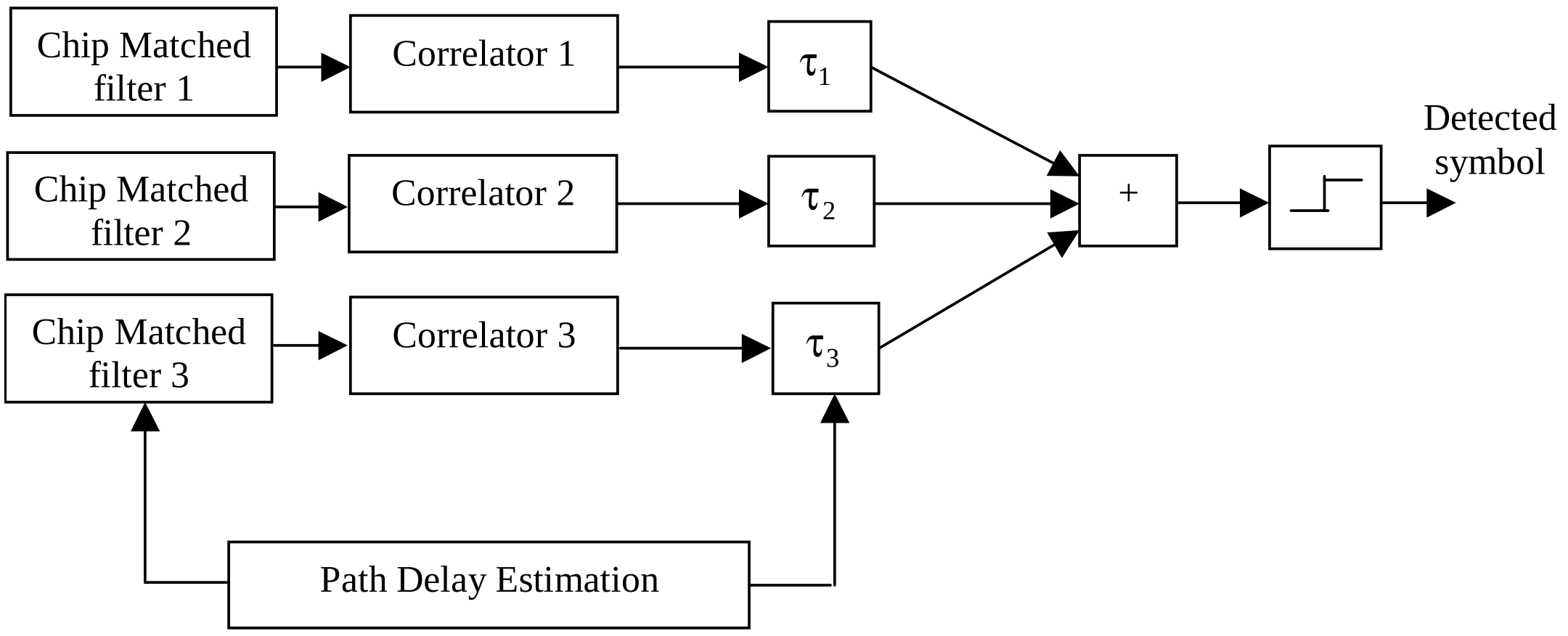

**Figure 2.7**: RAKE receiver architecture.

The RAKE structure just described ignores two effect caused by the multipath propagation. First, multipath may induce interchip interference (ICI[9]) degrading the codes' autocorrelation properties. Secondly, and also due to multipath, the orthogonality among different user codes is destroyed. This second phenomenon causes what is called multiple access interference (MAI). MAI is the responsible for the near-far effect (commented in chapter 1) making necessary the use of accurate power control schemes.

In an effort to relax the power control requirements and improve the system capacity, multiuser detectors have been introduced. Multiuser detectors exploit the structure of the inter-user interference in order to cancel it. [Lupas89] is regarded as the seminal publication proposing a receiver architecture to eliminate or reduce the effects of MAI. Since then, many different techniques have appeared on this topic. Concise introductions to this area generically called "multiuser detection" can be found in [Duel-Hallen95] and [Moshavi96]. A more comprehensive coverage appears in [Woodward98].

Many of the multiuser detection techniques published rely on the fact that the receiver has knowledge of all the user codes. This situation is reasonable on the base station side but it is certainly not realistic on a mobile terminal. For this reason, alternative structures have recently been researched for the downlink multiuser detection.

In [Madhow94] and [Abdulrahman94] it is shown how classical adaptive equalisation structures such as the LE or DFE, operating at chip level (rather than symbol), can be used to reduce significantly the amount of MAI and also ICI. Moreover, they do so just having knowledge of the code of the user being detected. The main idea is to substitute the correlator in a typical receiver by an adaptive structure (LE or DFE) which, after training, will perform the correlation and multiuser interference removal in a single step. As in conventional equalisation this structure is designed to minimise the MSE between a received and a desired sequence.

[9] ICI can be regarded as ISI at chip level

This concept of chip level equalisation has been recently treated in [Grant99], [Petre00], [Hooli00] and [Wilson93]. Simulation results show that the performance of these structures is clearly superior to that offered by a conventional RAKE receiver.

### 2.4.3 Training aspects

Concluding this chapter, some attention is devoted to training aspects. Most commercial communication systems operating in unknown environments use some form of training to gain knowledge about the state of the channel. This is especially true in mobile and wireless environment where periodic training is needed in order for the receiver to be aware of the channel variations. Some specific training methods have been proposed such as in [Qureshi77] and references therein. However these methods were proposed in the context of wireline communications where a partial knowledge of the channel spectrum can be assumed. No such a-priori knowledge can be presupposed in wireless scenarios.

Obviously the transmission of a training sequence implies a reduction in the transmission bandwidth of information bits. This fact has spurred the investigation of blind equalisation methods whose origins can be traced back to the publication of [Godard80] where the most widely used blind algorithm, the constant modulus algorithm (CMA), was proposed. A short introduction to blind equalisation principles can be found in [Litwin99]. For a more detailed overview see chapter 18 in [Haykin96].

Recently, two references ([Boss98], [Labat98]) have appeared studying the use of blind methods in wireless systems. In [Boss98] the feasibility of performing the channel estimation in the GSM system without using the training sequence is considered. In there, it is shown that the proposed blind channel estimation performs only 1.2-1.3 dB worse than the conventional Viterbi equaliser. Clearly, the gain in transmission bandwidth (22% more) would make up for the slight degradation in performance. However, it must be said that the amount of computation for the blind method was not addressed.

In [Labat98] a reconfigurable DFE structure is proposed that is able to adjust its coefficients without any form of training. Simulation results are presented in the context of wireless over-water channels showing that the equaliser does not suffer any performance degradation with respect to the trained counterpart and computational costs remains unaffected. In principle, it seems viable to apply the same structure to mobile scenarios.

Despite these promising results obtained using blind methods, all modern mobile standards, 2G and 3G, rely on some form of training to adapt the receiver to the channel in which they are operating.

# 3 SIMULATION ENVIRONMENT AND BACKGROUND RESULTS

Nowadays, nearly all communications research relies on computer simulations in order to verify the performance of new systems in an efficient and cost effective way ([Jeruchim92]). The first objective of this chapter is to present an overview of the simulation environment developed to test the techniques described in later chapters. This rather generic and simple environment is not focused on a particular standard, but the techniques developed in chapters 5 and 6 are applicable to a wide variety of systems. The different channel models used in chapters 4, 5 and 6 are also described in detail here. Results presented in chapter 7, where the environment is configured according to UMTS-like specifications, use more complex system and channel models and will be described there.

The second objective of this chapter is to present some simulation results obtained using the algorithms and structures previously introduced, using the channel models introduced in this chapter. These simulations serve to contrast the theory previously presented with experimental evidence. The results also help us to take some decisions regarding which structures and algorithms to use as a foundation for further study.

## 3.1 Simulation tools

During the first stages of this project a dilemma was whether to use one of the commercial simulation tools (Cossap, SPW, Simulink, SystemView) or to implement a new simulation environment in a high level language such as C. Each option had its pros and cons. Simulation packages are easy to use and offer a wide range of basic blocks making it easy to put together a whole system for simulation. On the other hand, simulations typically take longer to execute and eventually, C code needs to be written to implement "non-typical" signal processing operations. Implementing the whole environment in C from scratch is a fairly complex task but it gives complete freedom on how to assemble the system. In addition, simulations run considerably faster.

It was finally decided to implement a whole new environment in C, mainly because it was anticipated that many of the new techniques developed would need to be hand coded anyway and later integrated into the simulation tool. The C code was generated using the GNU C/C++ compiler for Solaris Sun workstations. The simulations were mainly run on a Pentium II-350MHz running Linux. Analysis of the results was done using Matlab 5.3.1. Additionally, Matlab was used for some particular computational tasks such as the calculation of Wiener filters.

## 3.2 System description

The basic model used in the next three chapters is shown in figure 3.1. In the context of this thesis, the words bit and symbol will be used interchangeably as binary phase keying (BPSK) is used throughout the work. We recognise that many of today's mobile/wireless systems allow the use of higher order modulation schemes such as QPSK or QAM which provide superior spectral efficiency in terms of b/s/Hz when compared to plain BPSK. However, the use of BPSK is justified on the basis that the novel techniques proposed in this work are independent of the modulation order utilised. Use of BPSK eased in great measure the implementation of the simulation environment and adaptive algorithms.

The generated symbols are drawn from a discrete uniform source where both symbols have equal probability of occurrence. The different channel models used are explained in detail in the next section. The noise components are samples of additive white Gaussian noise (AWGN). At the receiver, the equaliser is in charge of compensating for the channel effects in order to provide the threshold detector with better estimates than those coming out of the channel. This subsystem, the equaliser, is where the focus will be in all subsequent chapters in this thesis.

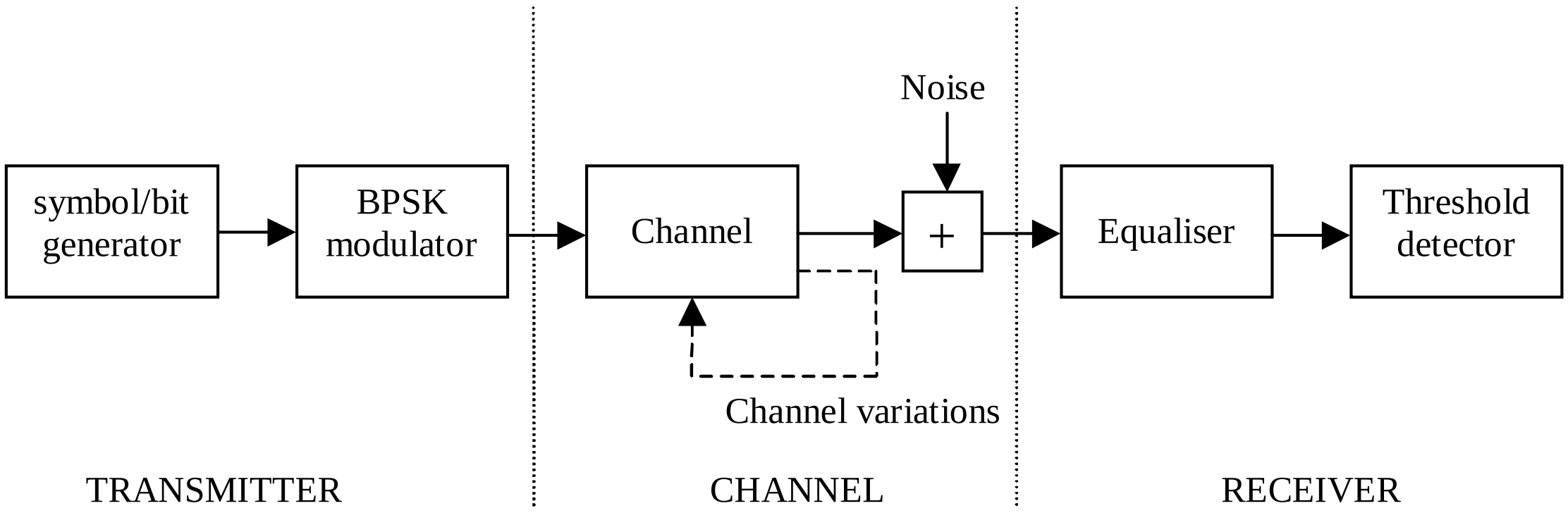

**Figure 3.1:** Generic system model.

Note that in the basic model of figure 3.1, several elements that would be essential in a practical system are missing. For example, at the transmitter, a pulse shaping filter, typically a square root raised cosine (SRRC) filter, would be required for limiting the bandwidth of the transmitted pulse. At the receiver, a matched filter, for example another SRRC, would also be required to maximise the SNR at the decision instant. Additionally the receiver would need a subsystem, like a phase locked loop (PLL), to achieve carrier, or at least symbol, synchronisation. During this work, it is assumed that the receiver is appropriately synchronised, which by no means is a trivial task (see [Gitlin92] or [Sklar01]]). Another important subsystem not shown in figure 5.1 is a channel coding system. Nearly all communication systems incorporate nowadays some form of channel coding, ranging from simple parity tests to sophisticated turbo-codes. Channel coding is very effective in combating errors in AWGN channels, acting in a complementary way to equalisation whose primary goal is to remove the correlation introduced by the channel. Given that the equaliser acts as front end to the channel decoder, it is possible to study its performance independently of the decoding technique used. However, it must be emphasised that the BER curves shown in this and future chapters would be considerably improved by the action of the decoder.

# 3.3 Channel models

One of the key elements of the system model described in the previous section is the channel profile used, especially as the objective of the whole model is to study the performance offered by various types of equaliser. As will be shown later in this chapter, the equaliser performance is greatly conditioned by the particular channel being compensated. Therefore in order to get a complete picture of the behaviour of a particular equaliser, it is necessary to check its performance with different classes of channels. In this and following chapters (except in chapter 7), four channel profiles have been thoroughly used. Each of these channels has different characteristics and introduces different levels of distortion. The first three channels are static while the last one is time varying and serves to study the tracking properties of the different algorithms and structures. These profiles have been selected to simulate the different scenarios, which at a given instant a mobile wireless receiver may have to confront.

As a last general comment on the channels used in this and forthcoming chapters, it is important to mention that although any physical channel has a continuous nature, the channel models used here are sampled versions of this continuous response. This is motivated by the fact that any modern wireless communications system is nowadays digital, and therefore all the continuous signals involved in the system can be modelled in a discrete manner ([Bello63] and [Treichler96]). Obviously, the bit/symbol rate influences the number of samples required to model a given continuous response. For the work in chapters 4, 5 and 6, the bit/symbol rate is assumed to be normalised to one. Additionally, all the discrete channel impulse responses presented in the following subsections are normalised so that their total power is unity.

### 3.3.1 Channel model 1

Channel model 1 is shown in figures 3.2 and 3.3 in the time and frequency domain respectively. This rather short length type of channel arises when transmitter and receiver are fairly close to each other and/or there are not too many obstacles between them. This could be the scenario when a mobile terminal is in the proximity of a base station and there is line of sight (LOS) propagation. The time domain profile reveals the presence of a non-delayed strong component followed by only one echo. These characteristics intuitively tell us that this channel does not heavily distort the channels and a short equaliser (few taps) should be enough to compensate its distortion.

Even more revealing than the time profile is the channel frequency response shown in figure 3.3. Looking at magnitude response, the low-pass characteristics of this channel are clear, letting the low frequencies pass and mildly attenuating the high frequency components.

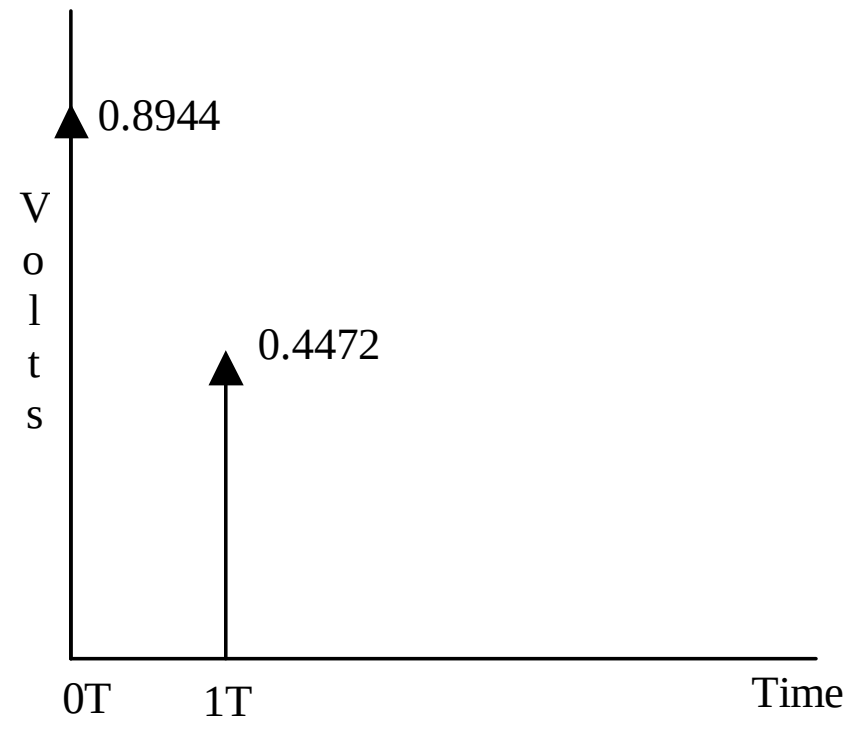

**Figure 3.2**: Multipath profile (time domain) corresponding to Channel model 1.

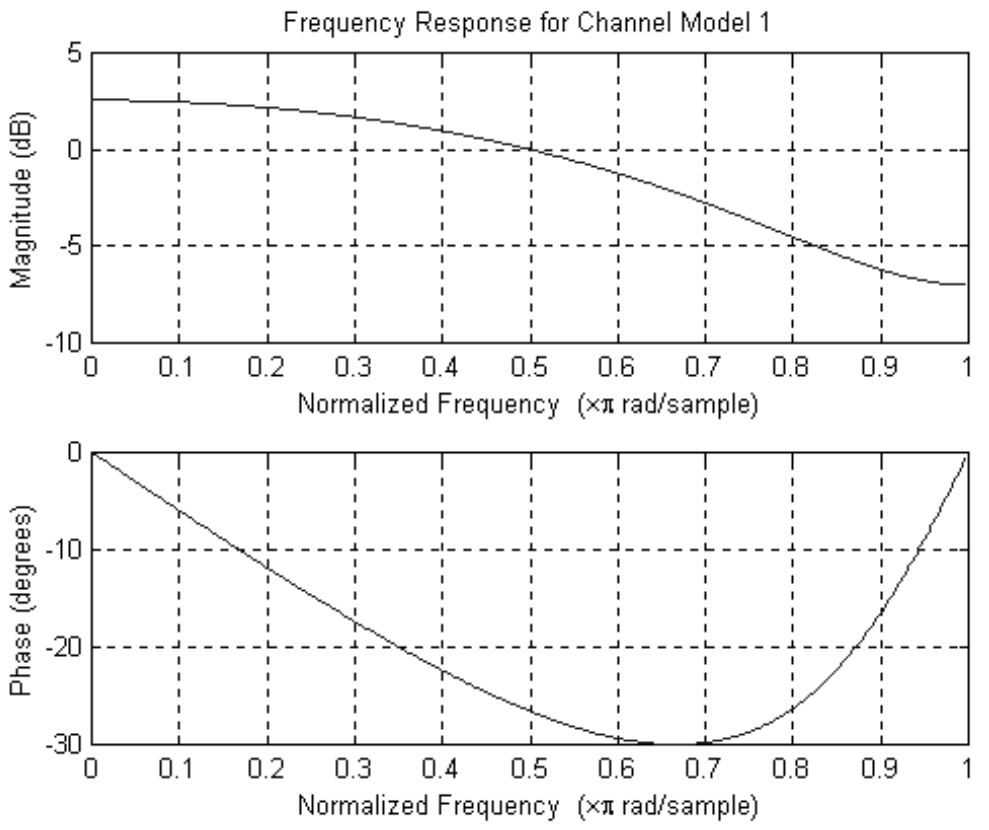

**Figure 3.3**: Frequency response for Channel model 1. Upper graph: Magnitude response. Lower graph: Phase response.

## 3.3.2 Channel model 2

Figures 3.4 and 3.5 show the impulse and frequency responses of what is defined as Channel model 2. This profile has been obtained from sampling the impulse response of a typical urban environment (as described in the COST207 models). The channel impulse response reveals a large delay spread, which implies that the received symbol, if left unequalised, will suffer interference from many neighbouring symbols when being detected.

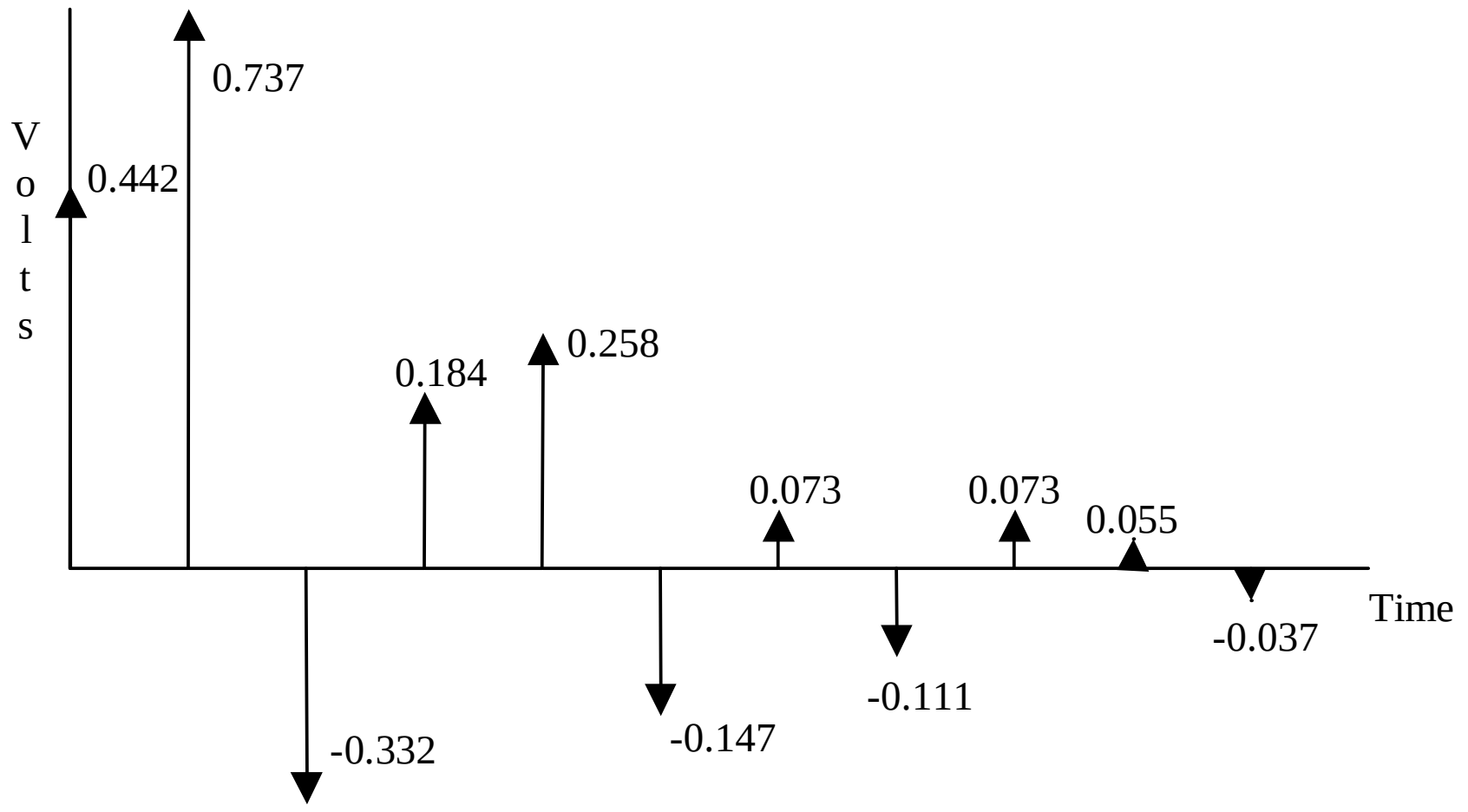

**Figure 3.4:** Multipath profile corresponding to Channel model 2.

The magnitude frequency response reveals some moderate attenuation (-5 dB) in the middle of the passband whereas the phase non-linearity implies that not all frequencies will be delayed uniformly. Intuitively it can be inferred that a longer equaliser than in the previous channel model will be required to equalise this channel impulse response.

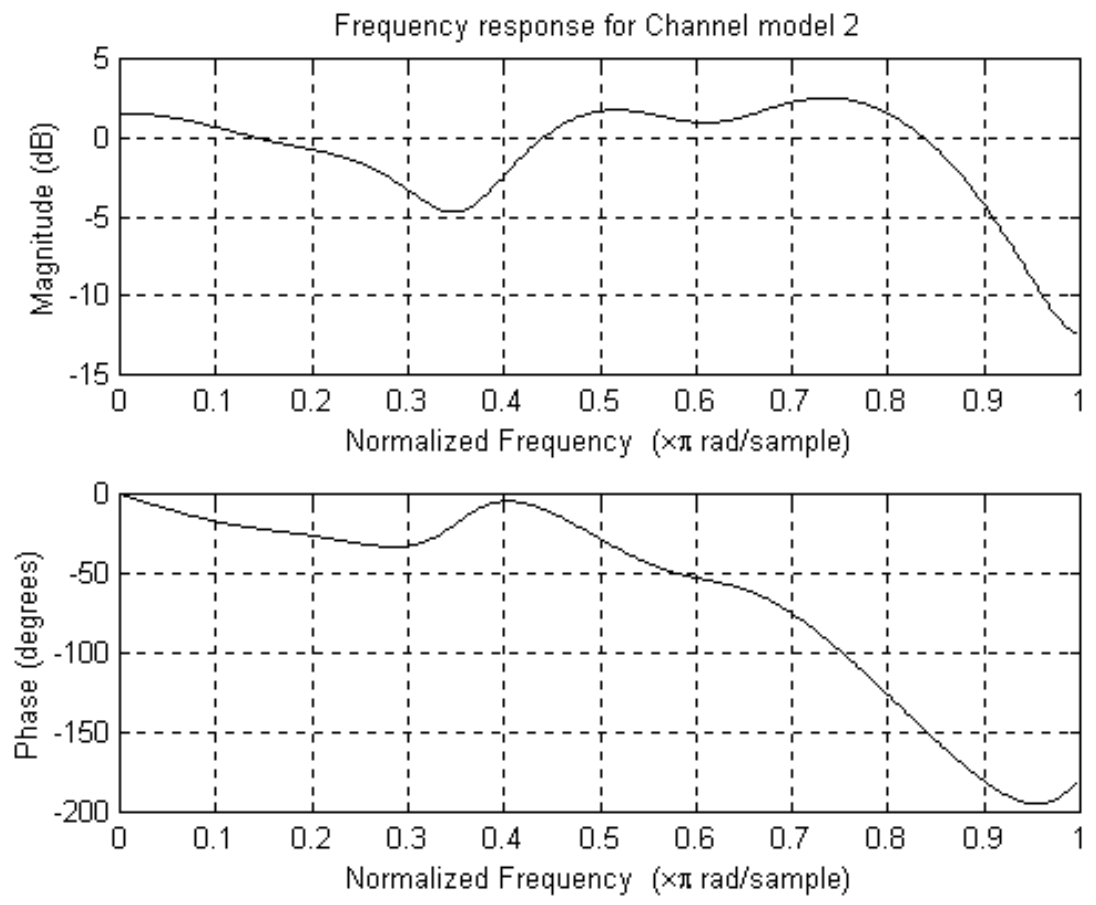

**Figure 3.5:** Frequency response for Channel model 2. Upper graph: Magnitude response. Lower graph: Phase response.

## 3.3.3 Channel model 3

The last static channel used regularly in this work, referred to as Channel model 3, is depicted in figures 3.6 and 3.7 in the time and frequency domains respectively. This channel is used in various important books ([Proakis95], [Haykin96]) as an example of a bad quality channel. The profile is defined by a moderately short delay spread where the main component is delayed by two samples. From the large value of the interfering paths, it can be conjectured that this channel will limit very much the transmission of information. The frequency response shows the poor spectral characteristics of the channel. In particular note the large spectral null of –80 dB in the passband and the non-linear phase response. Channels with such spectral nulls do occur in practice in radio communication systems ([Proakis91]).

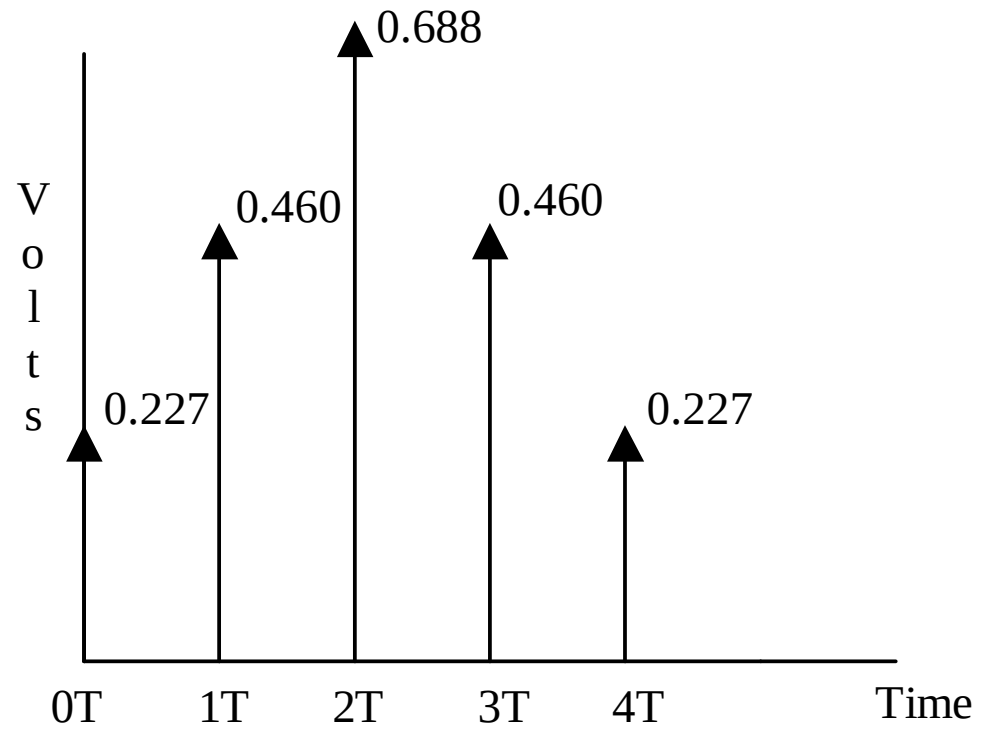

**Figure 3.6**: Multipath profile (time domain) corresponding to Channel model 3.

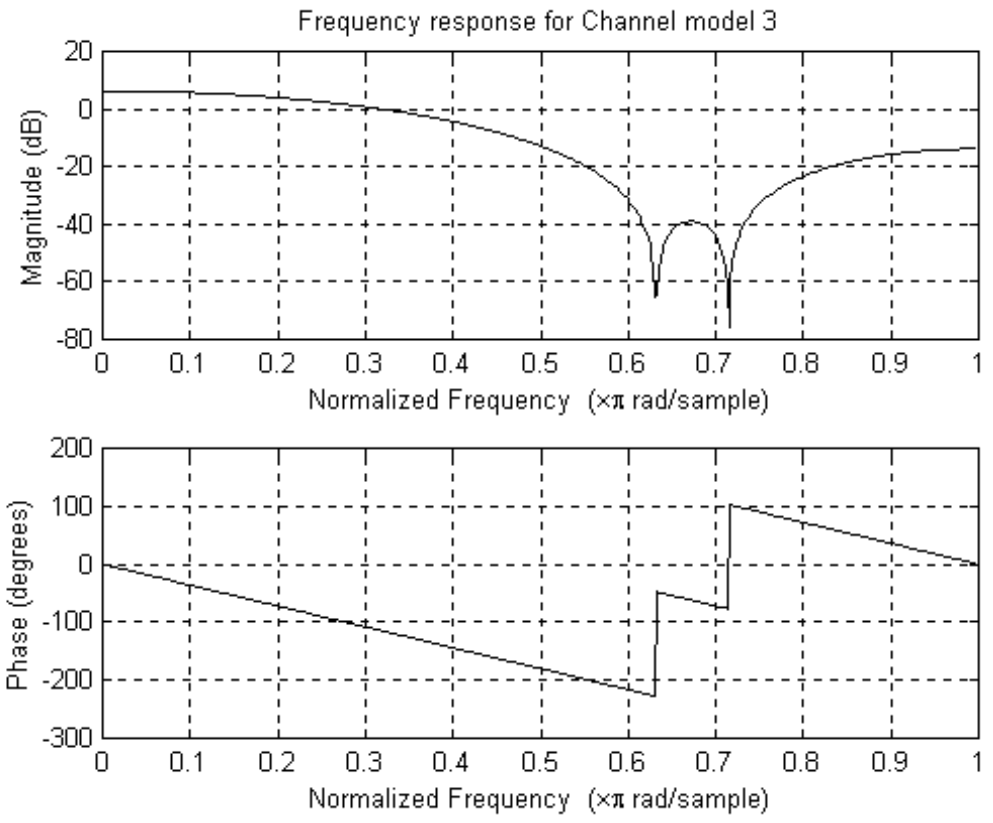

**Figure 3.7**:Frequency response for Channel model 3. Upper graph: Magnitude response. Lower graph: Phase response.

### 3.3.4 Time varying channel

The time varying channel model used is largely based on the COST 207 – Typical Urban profile, which will be described in detail in chapter 7 where a more realistic environment has been implemented. For the time being, we just say that it is a multipath Rayleigh fading channel with 11 taps obtained from the sampling of a reduced[10] version of the COST207-TU profile at 3.48 Msamples/s. The resulting channel has a length of 11 taps. The carrier frequency was set to 2 GHz.

## 3.4 Background simulation results for LE

The next subsections present the performance of different adaptive algorithms (LMS, NLMS, VSLMS, RLS, SFAEST) when using a linear equaliser (LE) to compensate the channel models introduced in the previous section. The performance measures used are the bit error rate (BER) and the mean squared error (MSE). Although BER is the ultimate performance measure for a communication system, more emphasis is put on the second criterion, MSE, as this is analytically more tractable ([Gitlin92]) and will be the basis for most of the work in later chapters.

[10] The last path of the COST207-TU profile has been discarded as it enlarged the impulse response very significantly (and consequently the number of computations) while the results showed only small variations.

The results shown in the following subsection assume that the system operates permanently in training mode. This means that the adaptive algorithm can always make use of the correct symbol to perform the equaliser adaptation. In a practical system, it is obvious that at some point the receiver would be switched to a decision directed mode of operation. However, operating the system in a training mode is more appropriate when studying the theoretical properties of the different algorithms and structures. The number of samples required by the adaptive algorithm to make the equaliser converge gives an idea of the length of the training sequence to be used.

Another important parameter to be chosen is the equaliser length. This is the fundamental topic of this research project and is treated extensively in chapters 5 and 6. For the simulations presented in this chapter a rather common rule of thumb ([Treichler96]) has been used which, assuming the channel has N taps, makes the equaliser 2N+1 taps long in the case of linear equalisation.

This section concludes with a discussion of the effect that the decision delay has on the attained MSE level and the criteria used to select it.

## 3.4.1 Channel model 1

The results obtained for this channel are plotted in figures 3.8 to 3.16. In this case and given the short delay spread of the channel, only a 5-tap equaliser was used with decision delay being set to 2 samples. The simulation length was set to 100,000 samples, although only the first 50,000 are shown in order to show the transient stage clearly. The averaged (30 independent runs) MSE curves when the equaliser is being updated with the LMS algorithm are shown in figures 3.8 (E/No=5 dB) and 3.9 (E/No=25 dB). Three different curves are shown in each figure corresponding to different values of the step size ($\mu$). There are a few important points to notice in these graphs. First, as theory predicts in the case of static channels, the smaller the step size, the smaller is the steady state MSE (SS-MSE) but the longer it takes for the equaliser to converge. In general, this trade off between convergence time and SS-MSE (see Chapter 2) is more evident in large E/No scenarios as in this case convergence time is much longer. A compromise $\mu$ needs to be chosen in order to obtain satisfactory performance in both environments.

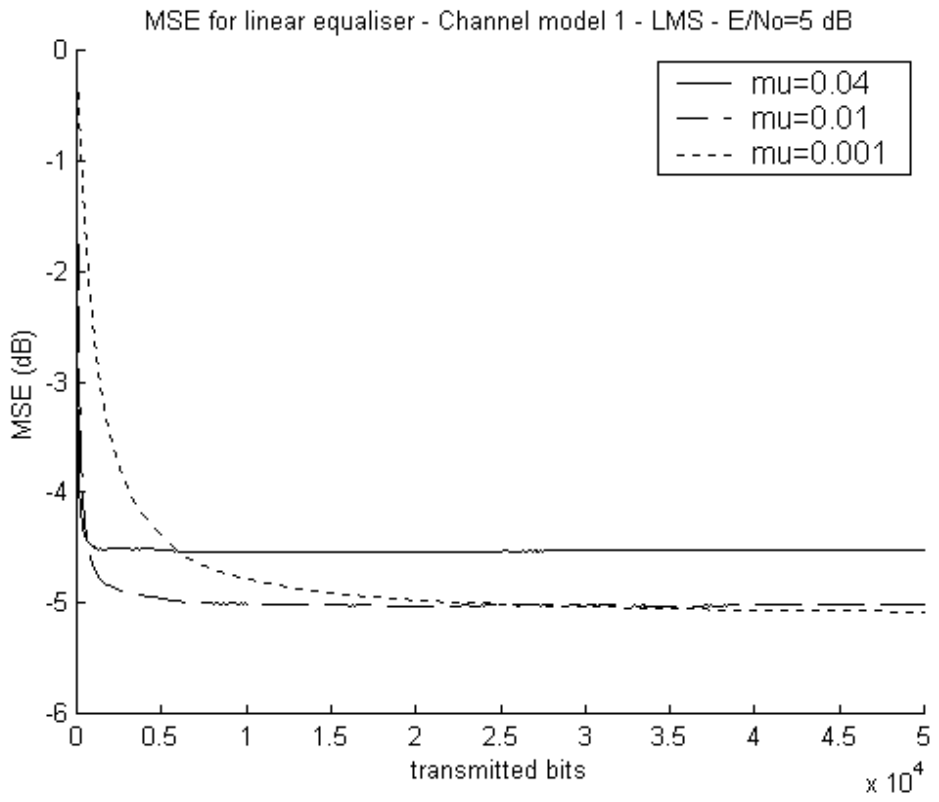

**Figure 3.8**: MSE for LE using LMS with different μ in Channel model 1. E/No=5 dB.

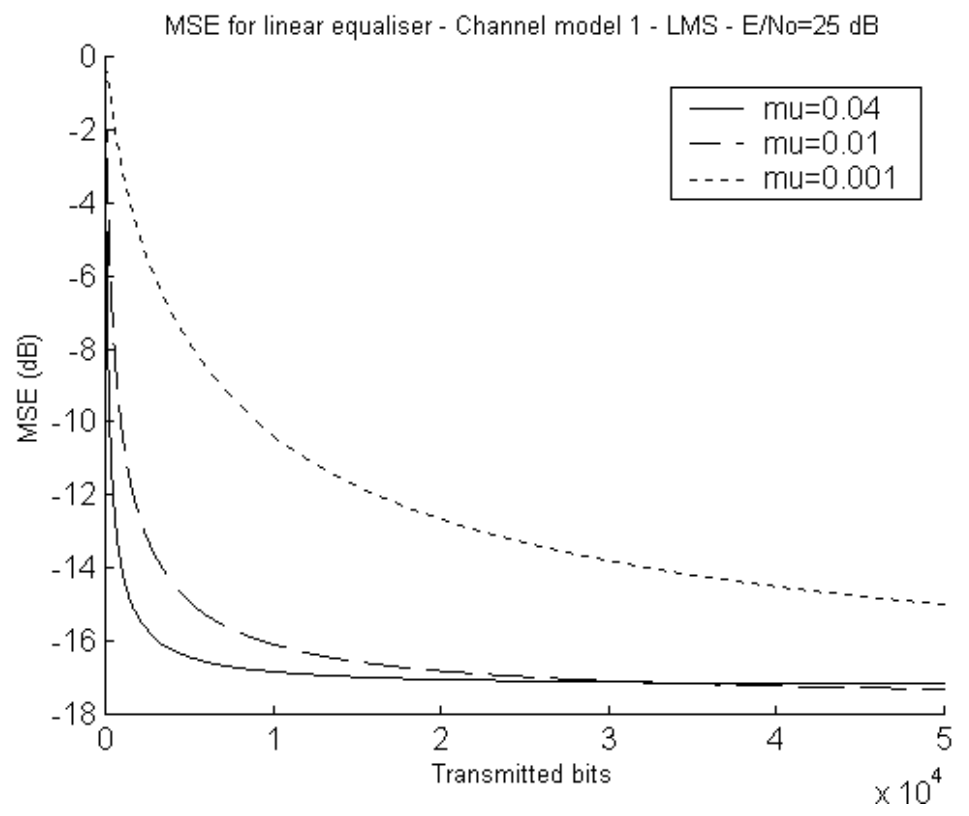

**Figure 3.9**: MSE for LE using LMS with different μ in Channel model 1. E/No=25 dB.

In the next two figures, 3.10 and 3.11, the MSE curves are again shown, but in this case the equaliser is being updated with the NLMS algorithm. Similar conclusions to those drawn for the LMS step size apply also to the NLMS step size ($\tilde{\mu}$). There is some improvement in the rate of convergence with respect to the LMS. This improvement is more evident when comparing small values of $\mu$ and $\tilde{\mu}$.

Figure 3.12 shows the MSE curves obtained when using the VSLMS in three different E/No levels. By judiciously choosing the VSLMS parameters (ρ and a), the algorithm is able to operate very efficiently under any E/No conditions, achieving fast convergence and low SS-MSE. For the results shown here, the settings chosen were ρ=0.0001 and a=0.99.

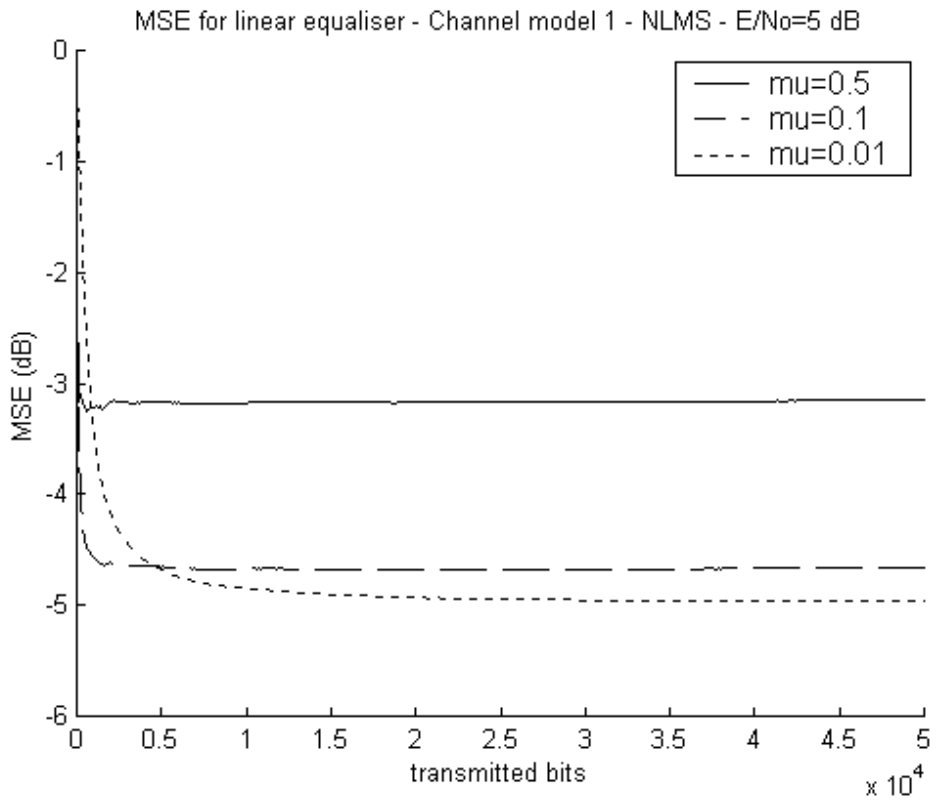

**Figure 3.10**: MSE for LE using NLMS with different $\tilde{\mu}$ in Channel model 1. E/No=5 dB.

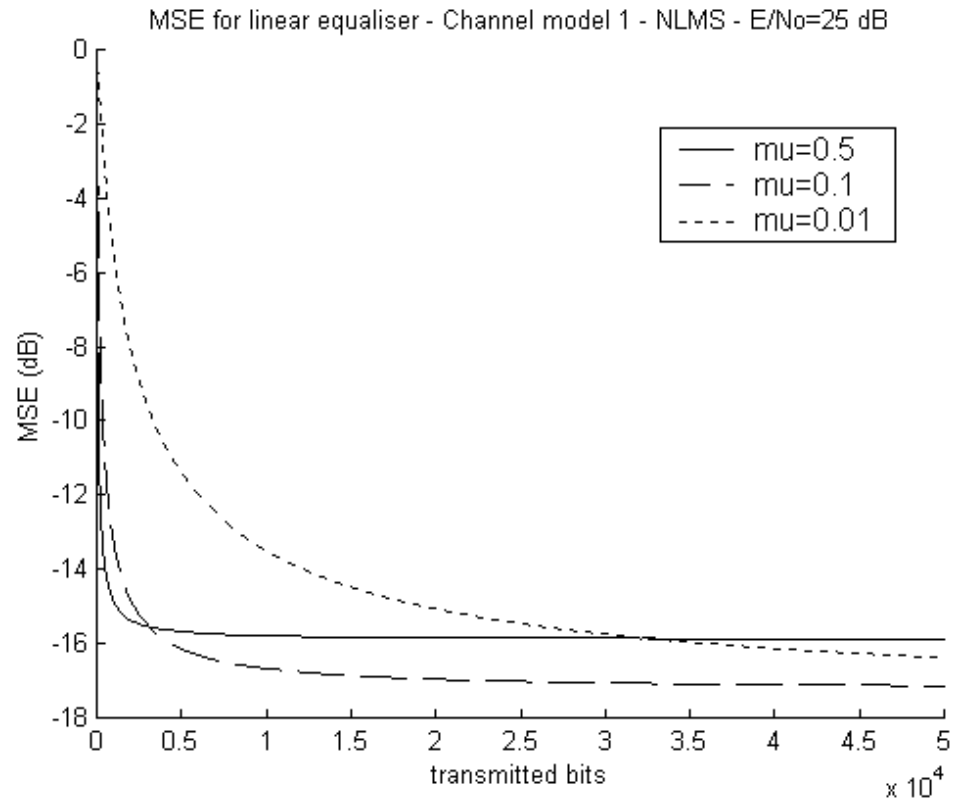

**Figure 3.11**: MSE for LE using NLMS with different $\tilde{\mu}$ in Channel model 1. E/No=25 dB.

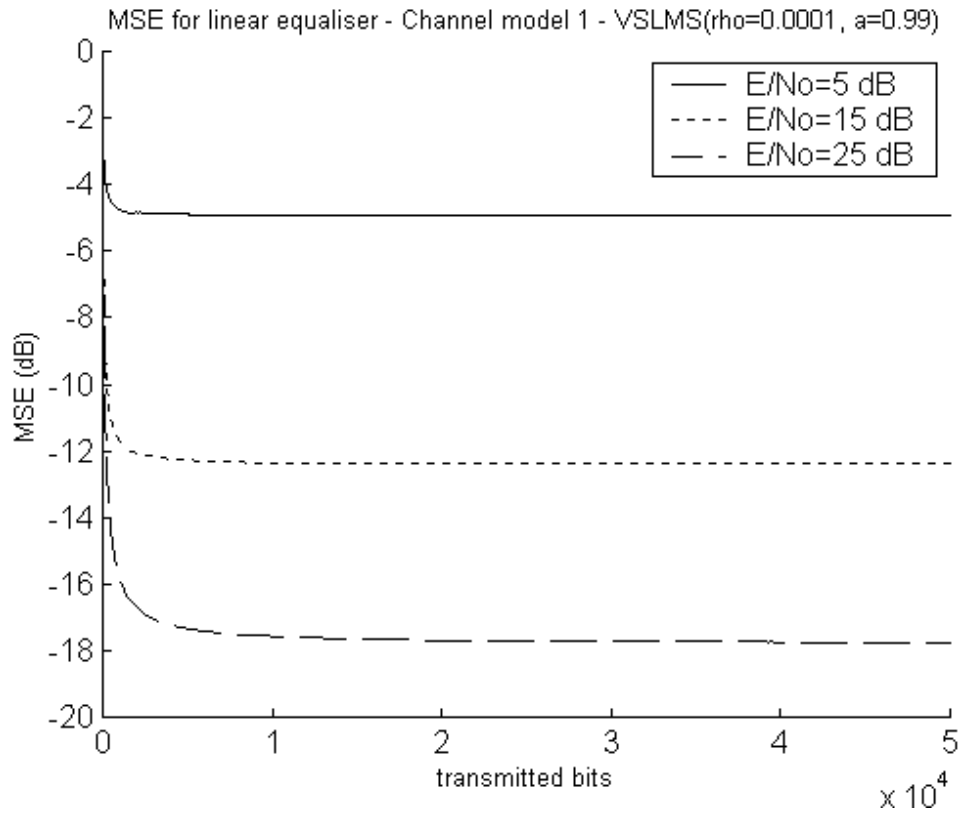

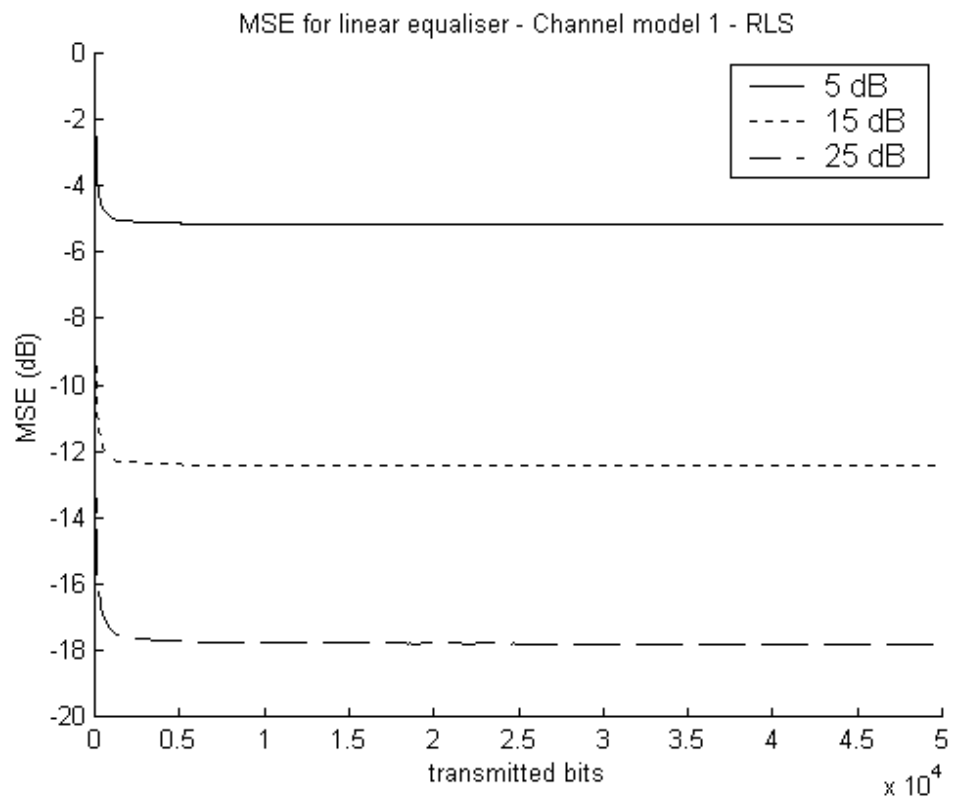

**Figure 3.12**: MSE for LE using VSLMS (ρ=0.0001, a=0.99) in Channel model 1.

**Figure 3.13**: MSE for LE using RLS ($\lambda$=1.0) in Channel model 1.

The RLS results are shown in figure 3.13. In this particular case the forgetting factor ($\lambda$) was set to 1 as the channel is static. In this case (static environment), the SS-MSE attained by the RLS algorithm would correspond to the Wiener-Hopf solution, that is, the lowest minimum mean squared error (MMSE). Particularly important to notice is the extremely fast convergence rate when E/No is large.

In figure 3.14, the learning rate is shown for the SFAEST algorithm for the three different E/No values. As has already been discussed in section 2.3.3, the settings of the SFAEST parameters strongly affect its performance and stability. In particular we have found that decreasing the value of $\mu$ improves its convergence but increases the chances of the algorithm going unstable. For these particular simulations the chosen settings were: $\lambda$=1 (static environment), ρ=1 and $\mu$=10. This value of $\mu$ tends to slow down slightly the MSE convergence but it has proved satisfactory to keep the algorithm indefinitely stable without the need of reset mechanisms.

In order to get an idea of the complexity associated with each algorithm, figure 3.15 presents the number of products performed during the whole simulation by the different adaptive algorithms. Only the most significant term of the complexity order was taken into account in the results shown in this figure. The difference in number of computations between the LMS-type algorithms (LMS, NLMS and VSLMS) and the least-squares schemes (RLS, SFAEST) albeit significant, is not enormous. This is due to the short length of the equaliser used in this particular scenario.

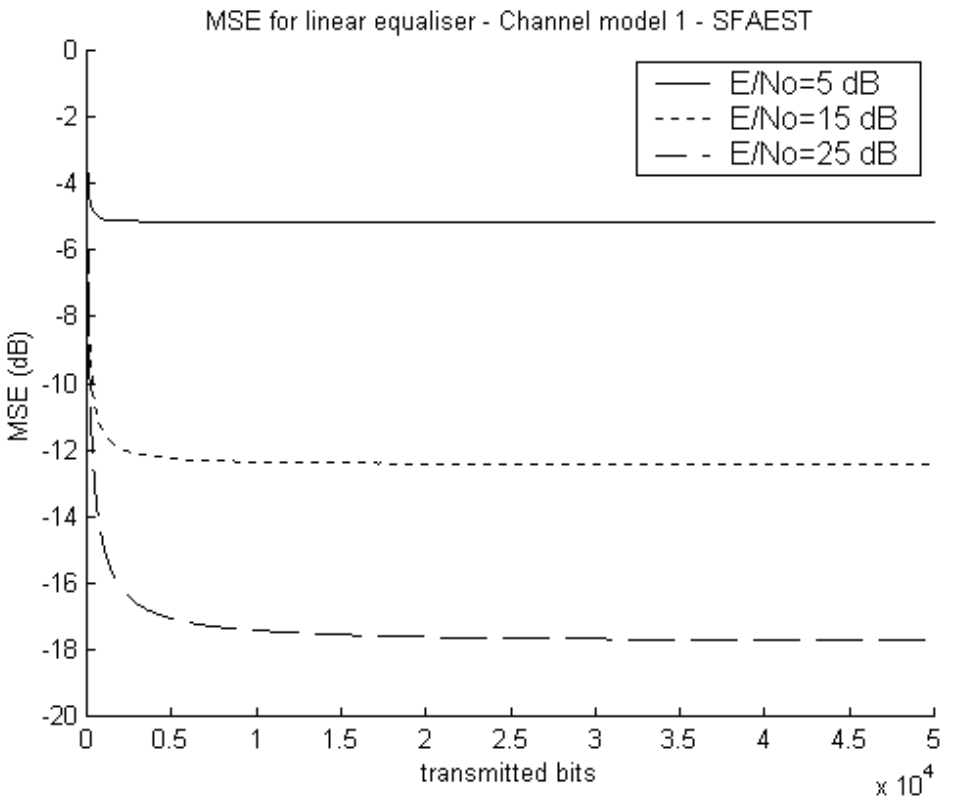

**Figure 3.14**: MSE for LE using SFAEST ($\lambda$=1, $\rho$=1, $\mu$=10) in Channel model 1.

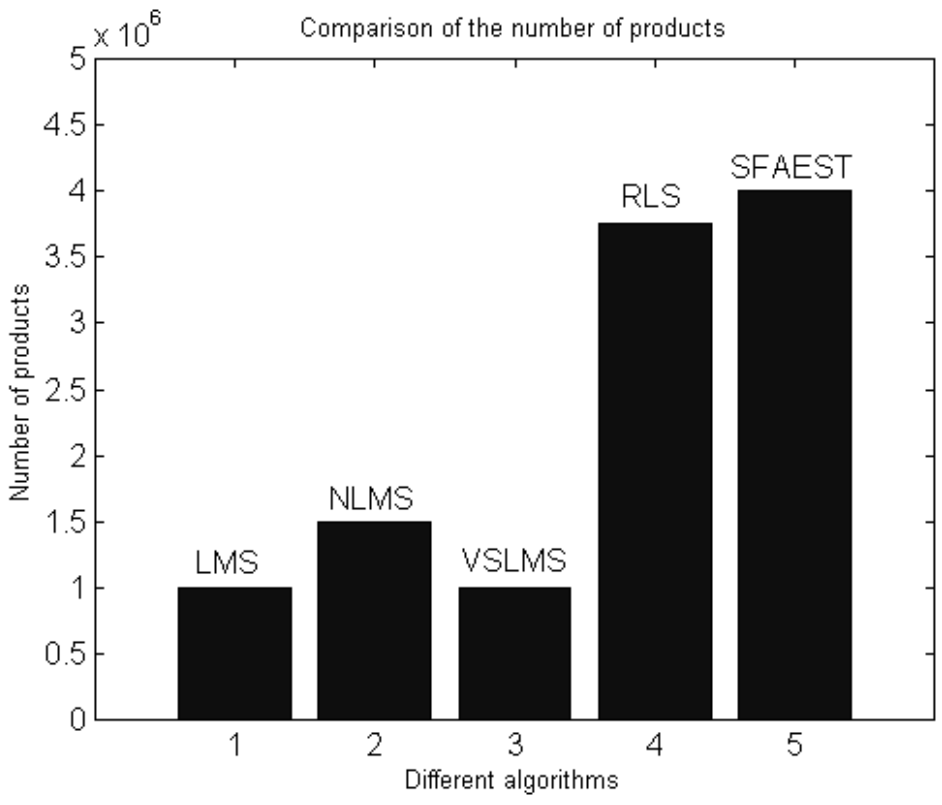

**Figure 3.15**: Number of products performed during the simulation (100,000 samples). Channel model 1.

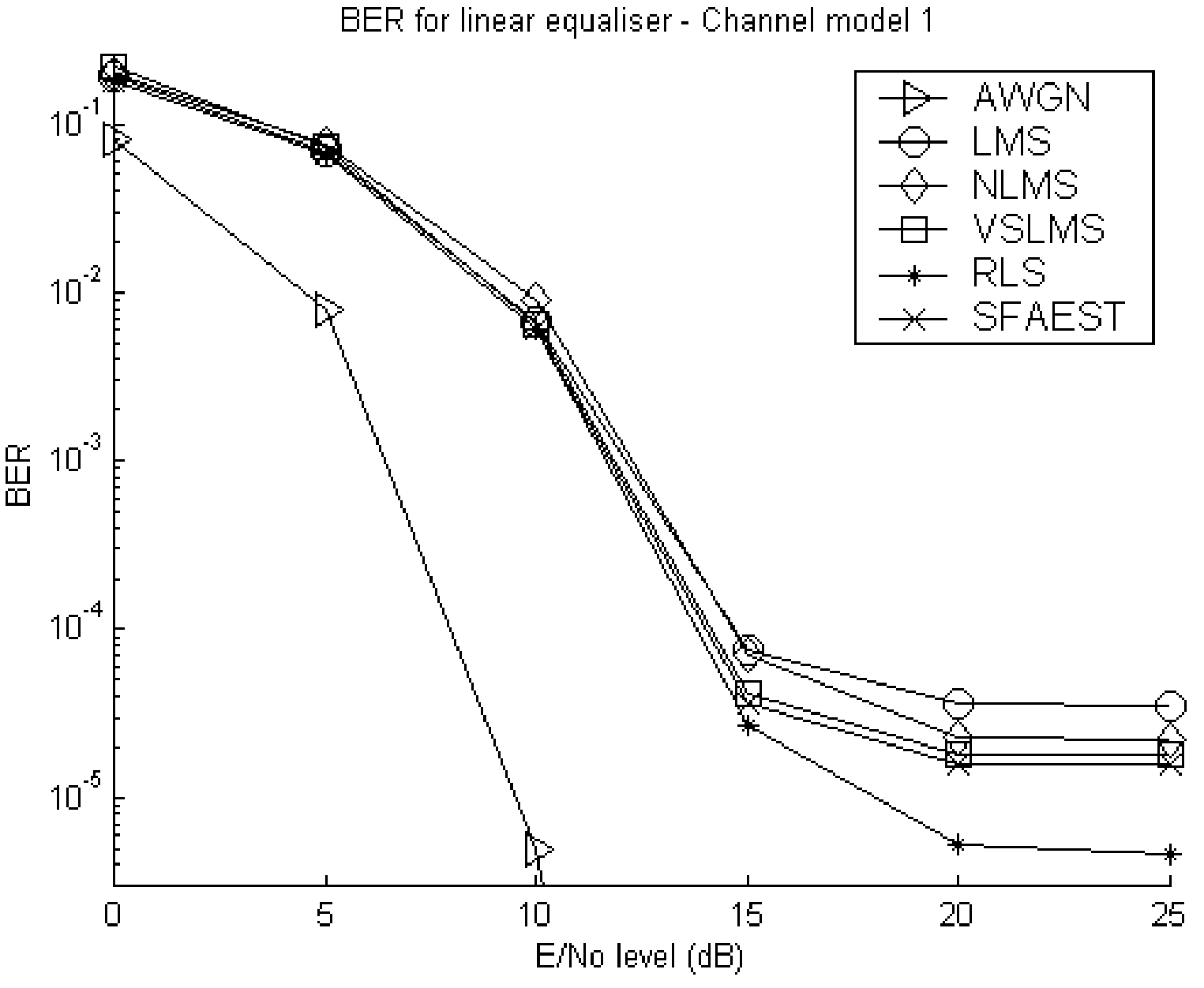

**Figure 3.16:** BER achieved with different algorithms equalising Channel model 1. Simulation length=100,000 samples.

Finally, a comparison of the different BER obtained with each of these algorithms is given in figure 3.16. For purposes of comparison, the BER obtained with an AWGN channel is also included. For the LMS and NLMS algorithms, the BER curves shown have been obtained when using $\mu$=0.01 and $\tilde{\mu}$=0.1 respectively. These step sizes have proved adequate over the whole range of E/No covered (see figures 3.8-3.11). Given that the channel is static, the BER is directly related to the convergence time. Notice that this particular channel allows the LMS-type algorithms to use small step sizes without enlarging significantly the convergence time. Consequently, they are also able to get very close to the optimum solution. Looking at the BER curves in figure 3.16 and focusing on the very low E/No levels (0 to 10 dB), no differences can be appreciated among the different algorithms. This is because, as theory predicts, in low SNRs, gradient and least-squares algorithms present similar convergence rates ([Haykin96]). As the E/No increases, RLS tends to achieve a lower BER for the same SNR. In theory SFAEST should achieve the same BER levels as RLS but due to the choice of its parameters, its convergence is somewhat compromised in favour of stability. NLMS and VSLMS slightly outperform the conventional LMS due to their faster convergence.

### 3.4.2 Channel model 2

Figures 3.17 to 3.25 show the results obtained when equalising Channel model 2. This channel model, despite its large delay spread, can be considered a good channel in view of its spectral properties. The equaliser in this case had 23 taps with the decision delay set to 12 samples. Results when using the LMS algorithm are very similar to those obtained in Channel model 1 although in this case it takes more time to converge due to the larger eigenvalue spread of the channel.

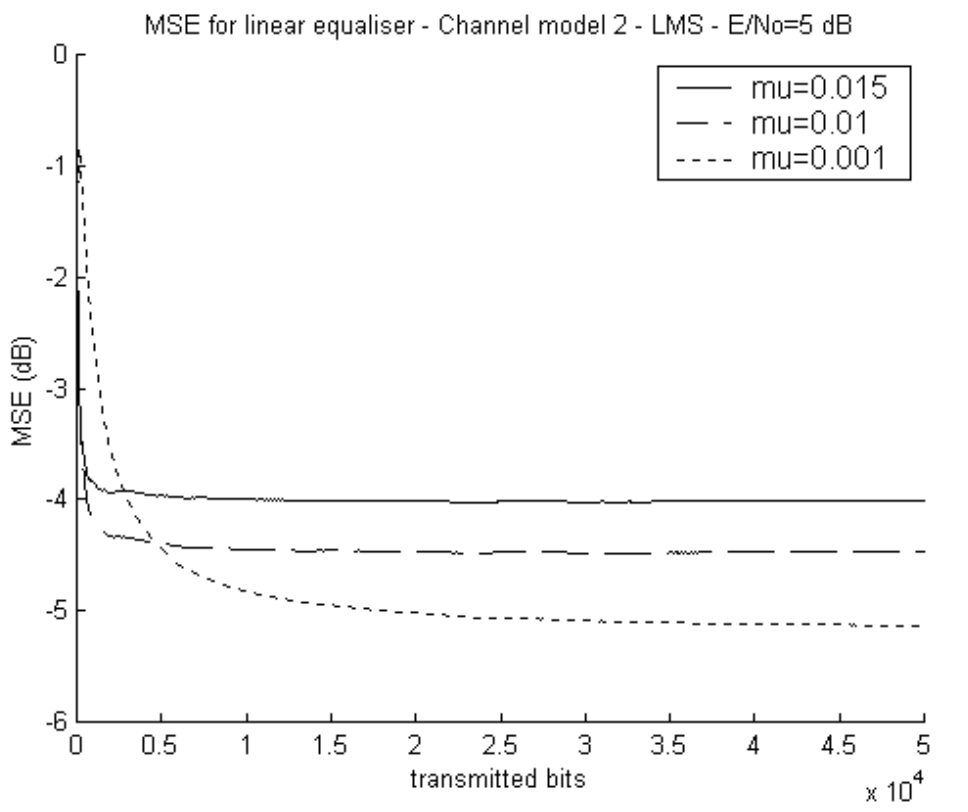

**Figure 3.17**: MSE for LE using LMS with different μ in Channel model 2. E/No=5 dB.

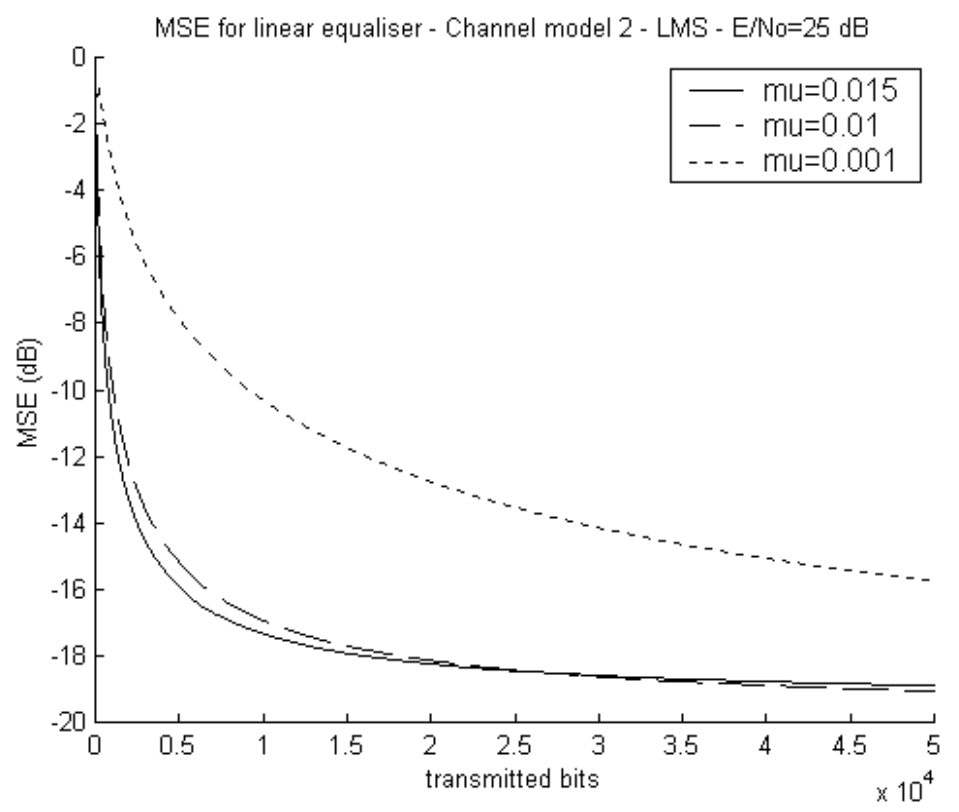

**Figure 3.18**: MSE for LE using LMS with different μ in Channel model 2. E/No=25 dB.

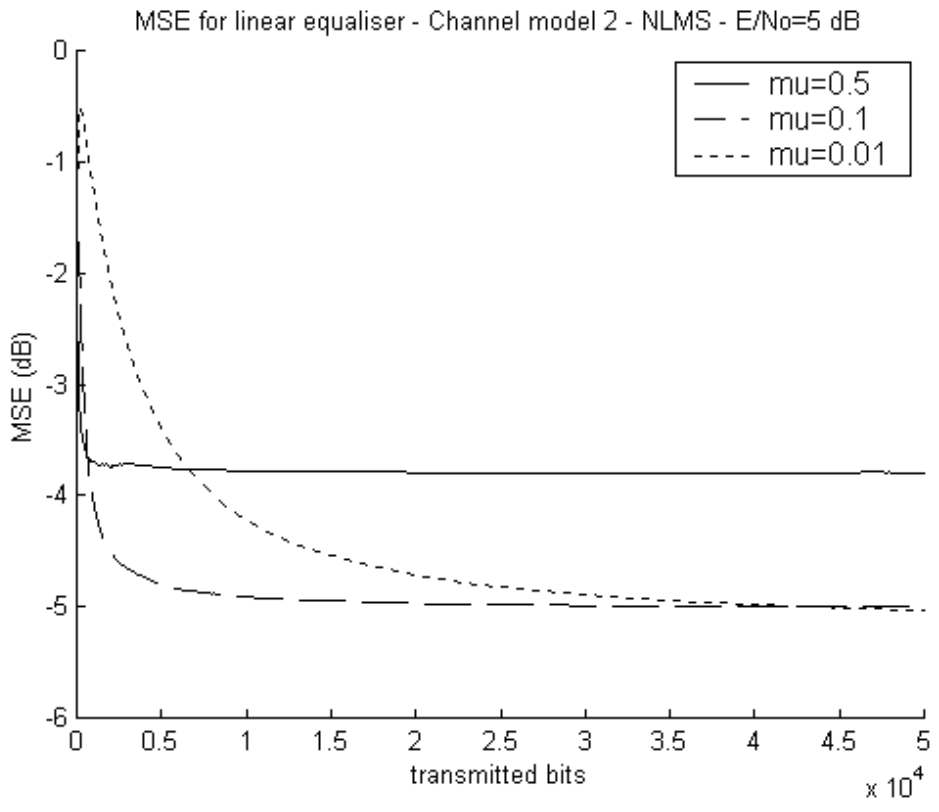

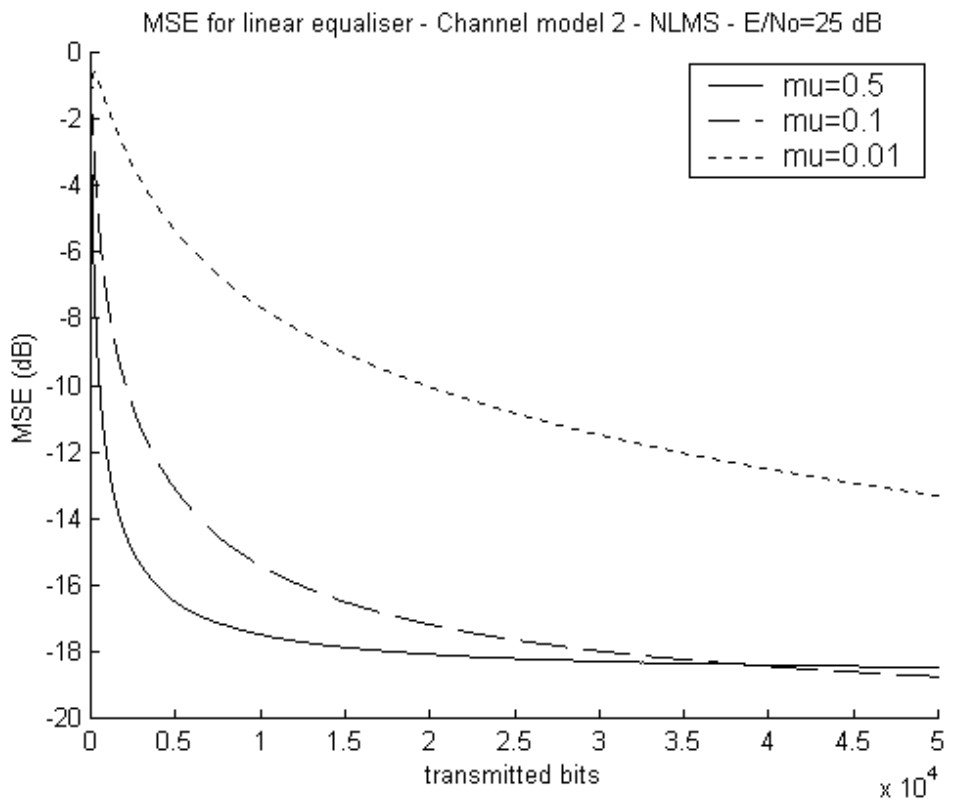

**Figure 3.19**: MSE for LE using NLMS with different $\tilde{\mu}$ in Channel model 2. E/No=5 dB.

**Figure 3.20**: MSE for LE using NLMS with different $\tilde{\mu}$ in Channel model 2. E/No=25 dB.

Again, notice the difficulty in choosing a step size that works well under any E/No level. Also recall that the upper limit for $\mu$ is inversely proportional to the equaliser length if stability is to be guaranteed. This greatly limits the choice of $\mu$ when medium/long equalisers are used. The NLMS algorithm presents the same problem regarding $\tilde{\mu}$ and its dependence with E/No level, although a slight improvement in convergence time with respect to the LMS can be appreciated.

Figure 3.21 presents the MSE curves obtained when using the VSLMS algorithm. Looking at the VSLMS MSE curve for E/No=25 dB it can be noticed that it is significantly slower than the corresponding NLMS MSE obtained when $\tilde{\mu}$=0.5 (figure 3.20). The VSLMS algorithm tends to reduce very much the step size when the error gets very small. This does not seem to be a problem as the algorithm is able to converge very quickly to a point where performance is often acceptable. Notice that convergence up to an MSE level of –12 dB takes place very fast. Moreover, unlike the LMS and NLMS, the same algorithm settings ($\rho$=0.0001, a=0.99) work well in any E/No level.

The MSE performance of the RLS algorithm shown in figure 3.22 reveals its superiority in terms of both convergence speed and SS-MSE level, when E/No is large. In a low E/No condition, as has already been noted for the Channel model 1, it behaves similarly to the other algorithms. The same applies to the SFAEST algorithm whose convergence, shown in figure 3.23, is nearly as good as the RLS and clearly much better than that of the gradient type algorithms in large E/No environments.

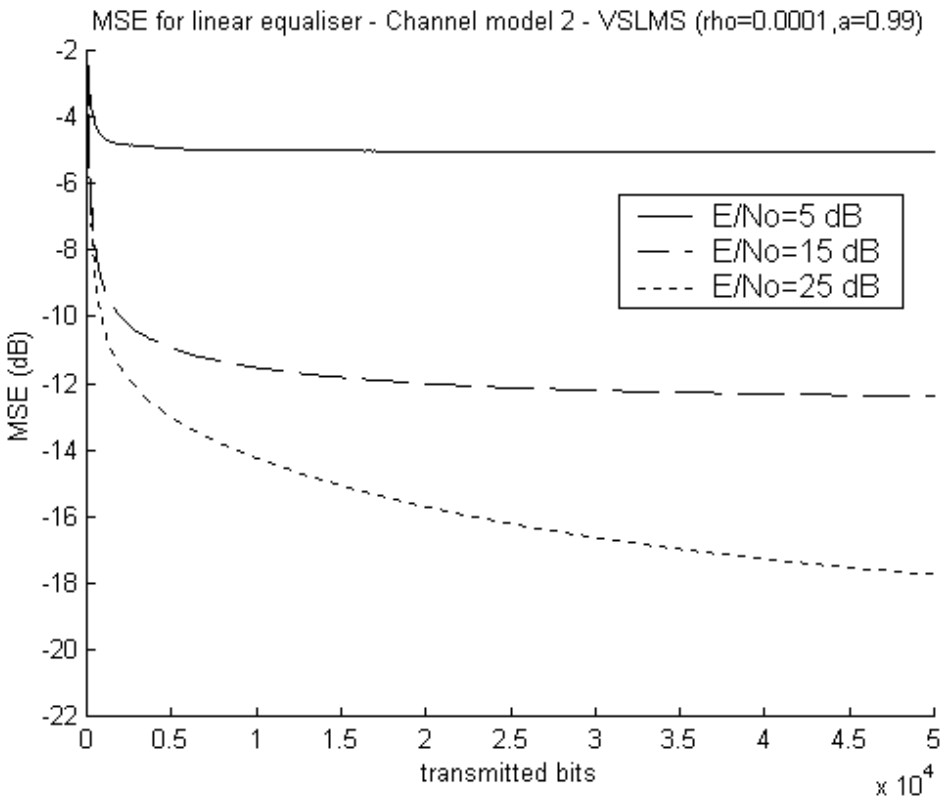

**Figure 3.21**: MSE for LE using VSLMS (ρ=0.0001, a=0.99) in Channel model 2.

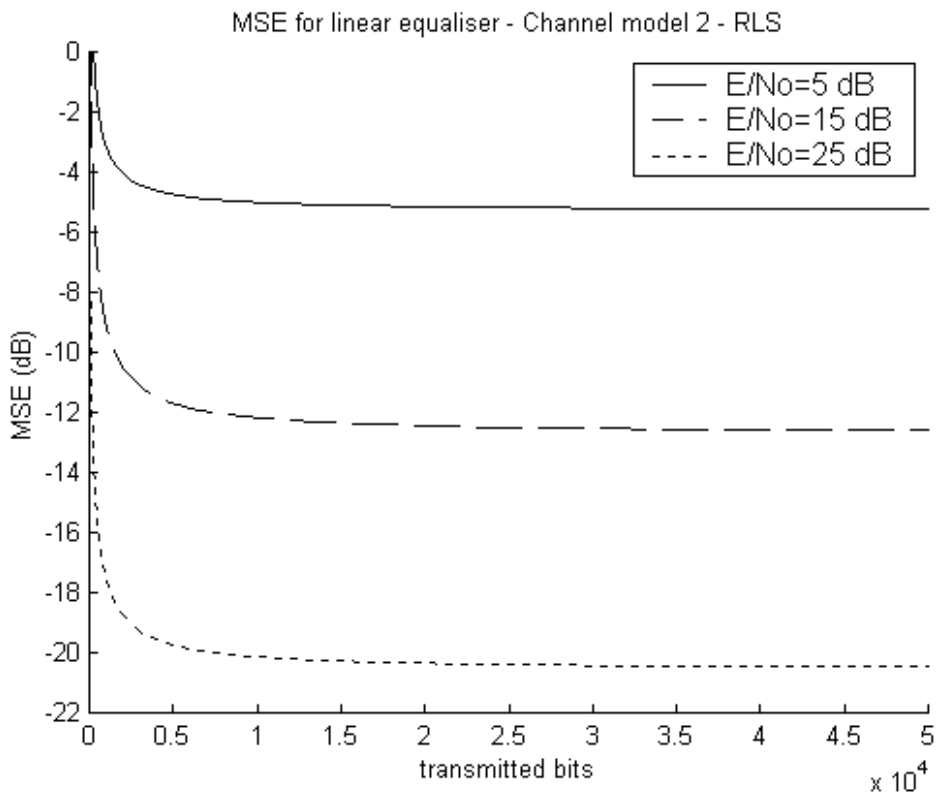

**Figure 3.22**: MSE for LE using RLS (λ=1.0) in Channel model 2.

Figure 3.24 is especially important in the context of this channel as it shows the computational complexity difference between the RLS (squared complexity) and the rest of algorithms (linear complexity) when the equaliser has many taps. The BER curves shown in figure 3.25, as in the previous section, only exhibit significant differences when E/No is moderate to large. In this case, the RLS algorithm offers an improvement of up to one order of magnitude in BER reduction with respect to gradient algorithms. The advantages of the least square over gradient schemes are also noticeable in the case of the SFAEST algorithm although not as remarkable as with the RLS algorithm.

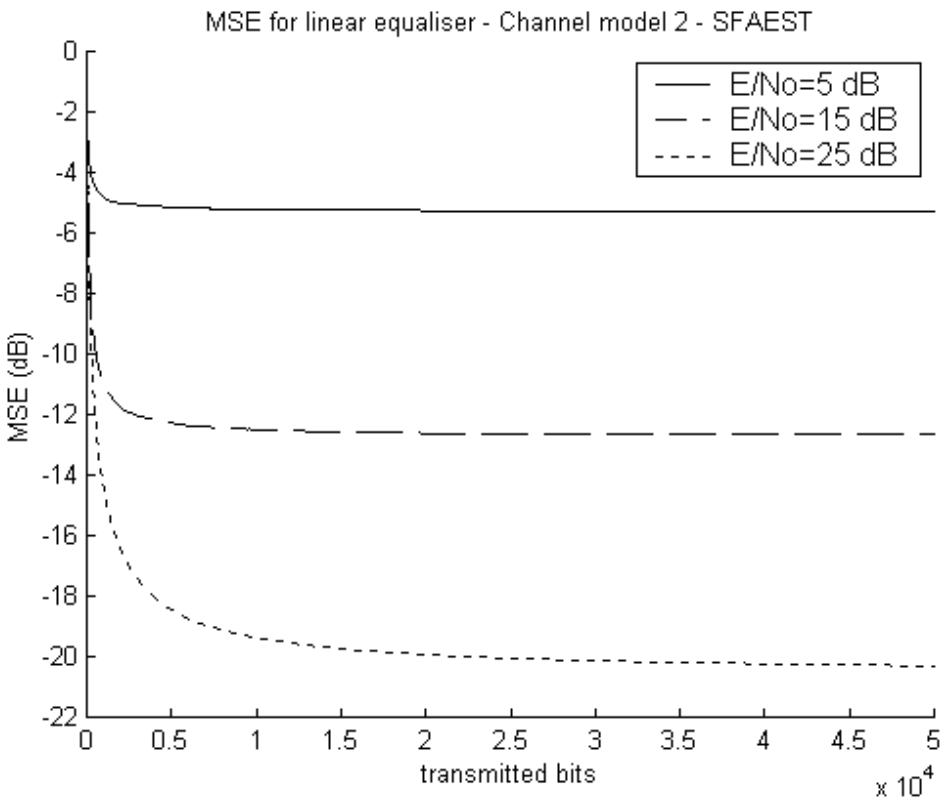

**Figure 3.23**: MSE for LE using SFAEST (λ=1, ρ=1, μ=10) in Channel model 2.

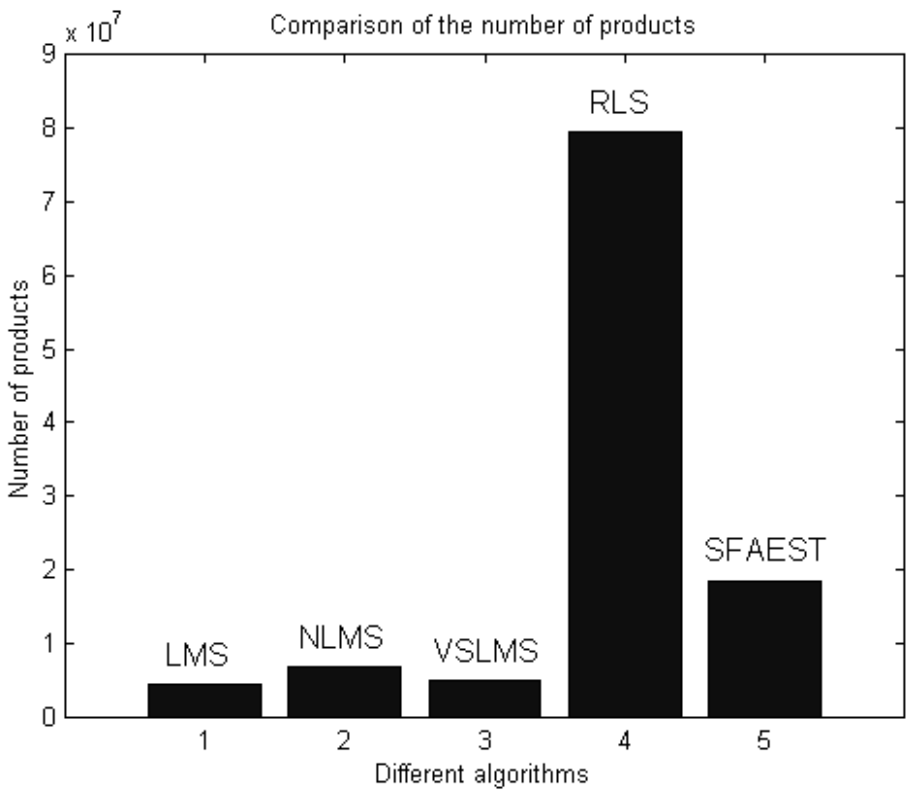

**Figure 3.24**: Number of products performed during the simulation (100,000 samples). Channel model 2.

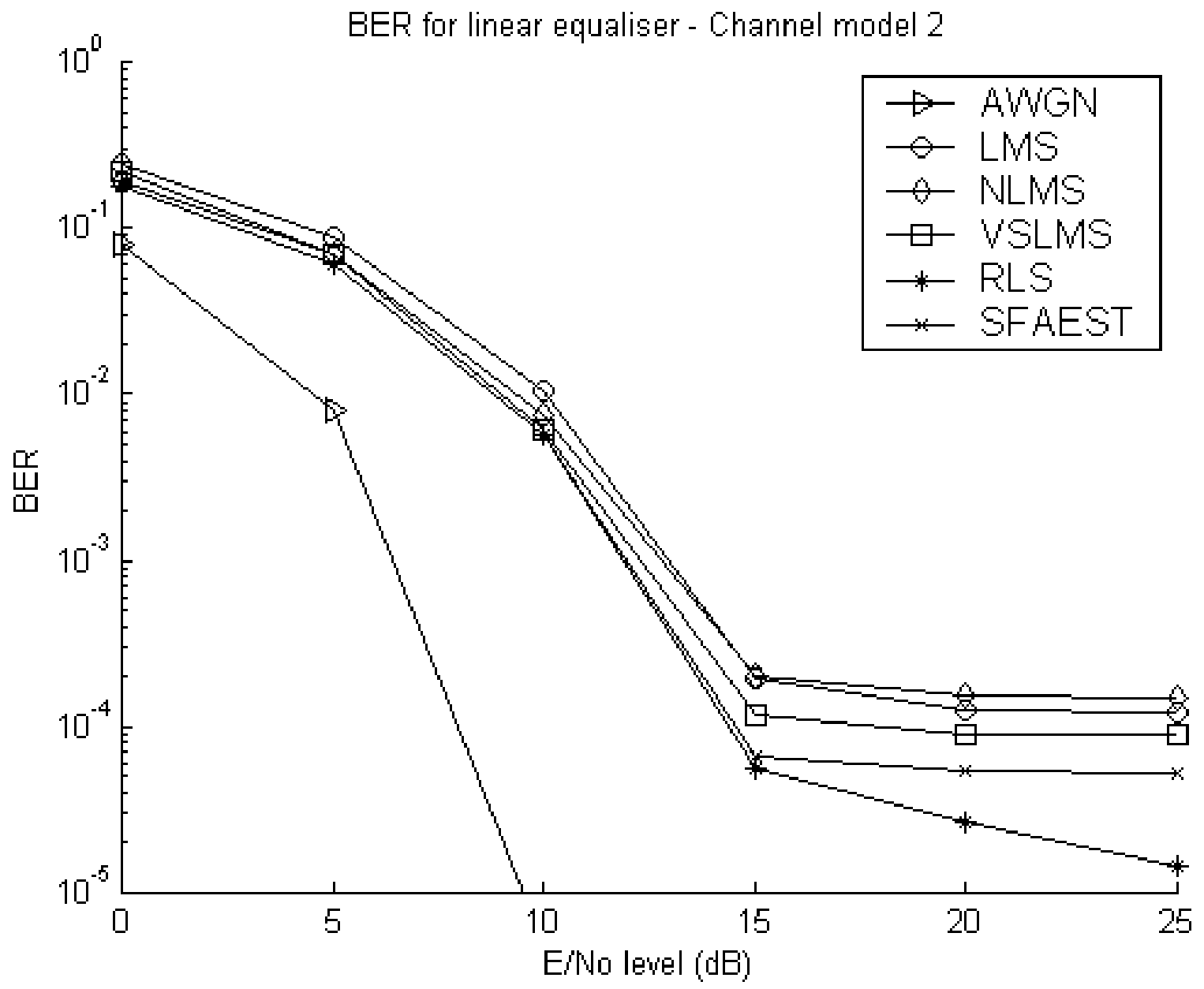

**Figure 3.25:** BER achieved with different algorithms equalising Channel model 2. Simulation length=100,000 samples.

It is an engineering decision to evaluate whether the reduction in BER offered by the RLS makes up for the huge increase in computational complexity of this algorithm. Obviously, in situations where power/computations is not a concern, the RLS would typically be the preferred option. In power restricted systems, the SFAEST could very well be a good alternative if the application requires a very low probability of error.

## 3.4.3 Channel model 3

Simulation results with the last of the static channel models, Channel model 3, are presented in figures 3.26-3.34. Recall that this channel presented a spectral null (-80dB) in the passband. This already suggests (see section 2.3.1) that the gradient type algorithms will require many iterations to converge. For this reason, in this channel model the MSE evolution is shown over the entire simulation (100,000 samples). Figures 3.26 and 3.27 display the results for the LMS algorithm for E/No=5 dB and E/No=25 dB respectively. The first point to notice is the much larger value of the SS-MSE when compared with that obtained from the other channels. Convergence time has increased considerably when E/No=25 dB. Comparing figure 3.27 with 3.18 (corresponding to Channel model 2) shows that they have a very similar appearance, but in figure 3.27 the x-axis spans twice as many samples as in figure 3.18. On the other hand, when E/No is low convergence time is hardly affected at all (figure 3.26). In this case, the large value of the noise power spectral density (No) fills the spectral null of the channel, reducing the eigenvalue spread of the data input to the equaliser and accelerating the convergence of the gradient type algorithms.

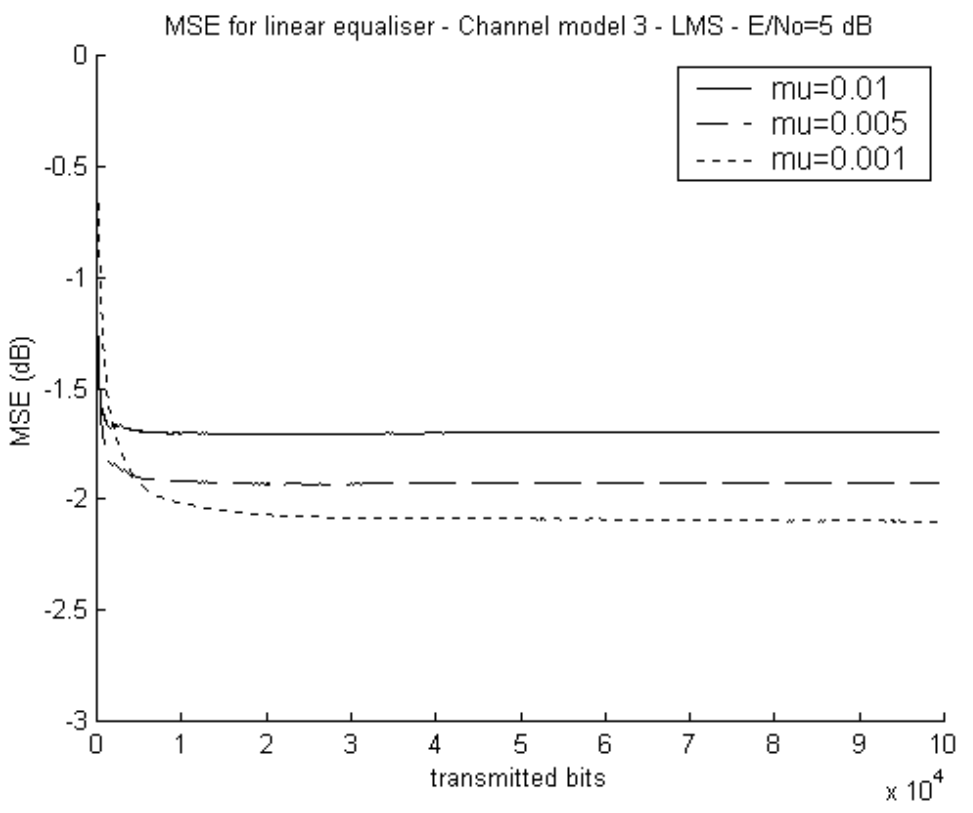

**Figure 3.26**: MSE for LE using LMS with different $\mu$ in Channel model 3. E/No=5 dB.

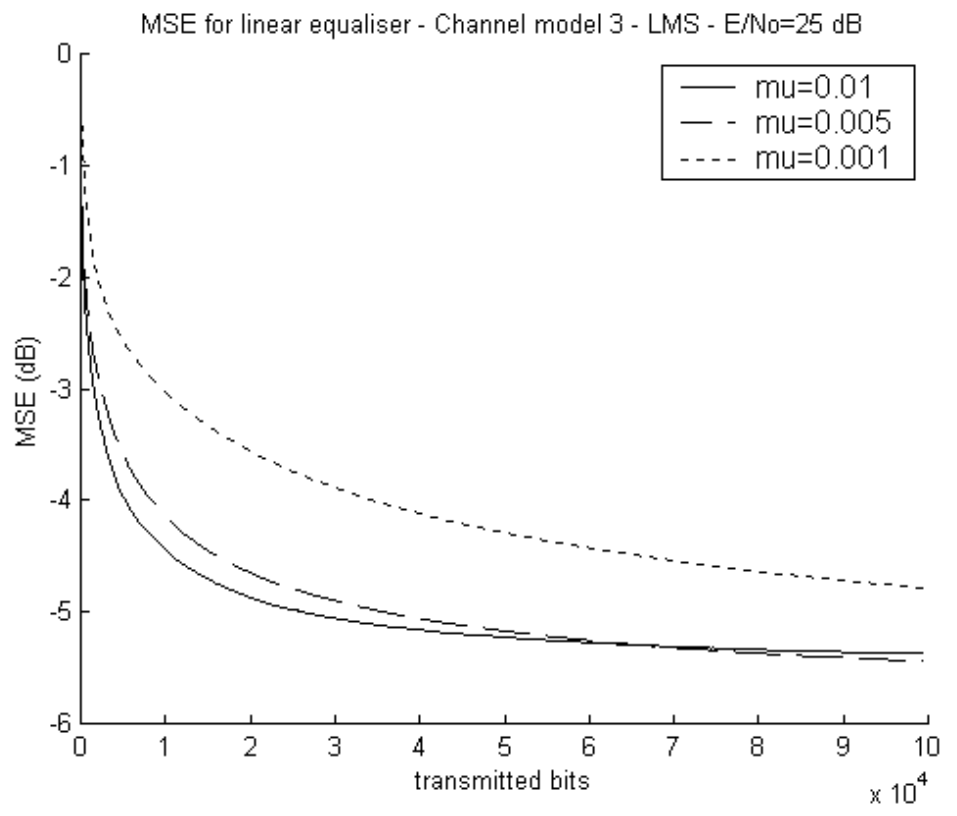

**Figure 3.27**: MSE for LE using LMS with different $\mu$ in Channel model 3. E/No=25 dB.

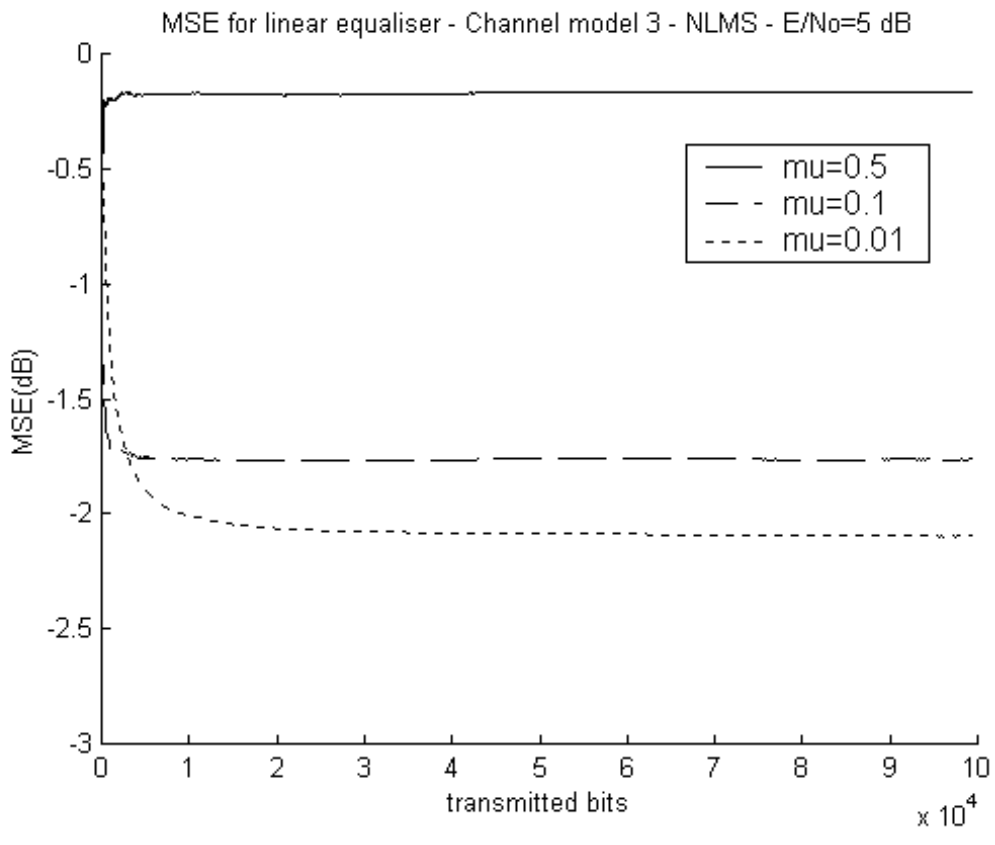

**Figure 3.28**: MSE for LE using NLMS with different $\tilde{\mu}$ in Channel model 3. E/No=5 dB.

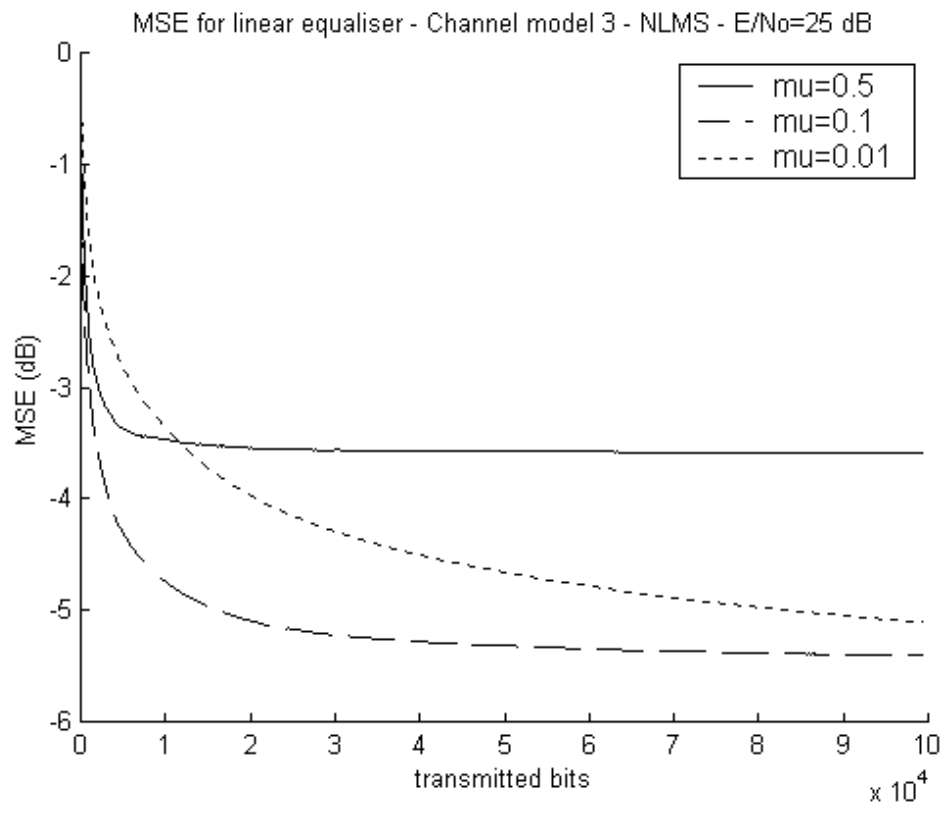

**Figure 3.29**: MSE for LE using NLMS with different $\tilde{\mu}$ in Channel model 3. E/No=25 dB.

In figure 3.29, a moderate improvement in convergence and SS-MSE of the NLMS ($\tilde{\mu}$=0.1) with respect to the LMS ($\mu$=0.01) can be observed. However, the problem is that the appropriateness of $\tilde{\mu}$ clearly depends on the E/No level. Notice in figure 3.28 that when E/No=5 dB, the step size offering the best performance is $\tilde{\mu}$=0.01 where as for E/No=25 dB is $\tilde{\mu}$=0.1. Again, if this algorithm is to be used in a variety of E/No scenarios, a compromise $\tilde{\mu}$ needs to be chosen.

In figure 3.30, the main advantage offered by the VSLMS is again observed. That is, the same set of parameters is able to operate very satisfactorily with regard to convergence time and attained SS-MSE level irrespective of the E/No level. The superiority of the least squares algorithms over the gradient ones, when operating in channels with large spectral variations, is clearly seen in figures 3.31 and 3.32. Notice that both RLS and SFAEST converge much faster and achieve an SS-MSE nearly one dB lower that that of the LMS-type algorithms.

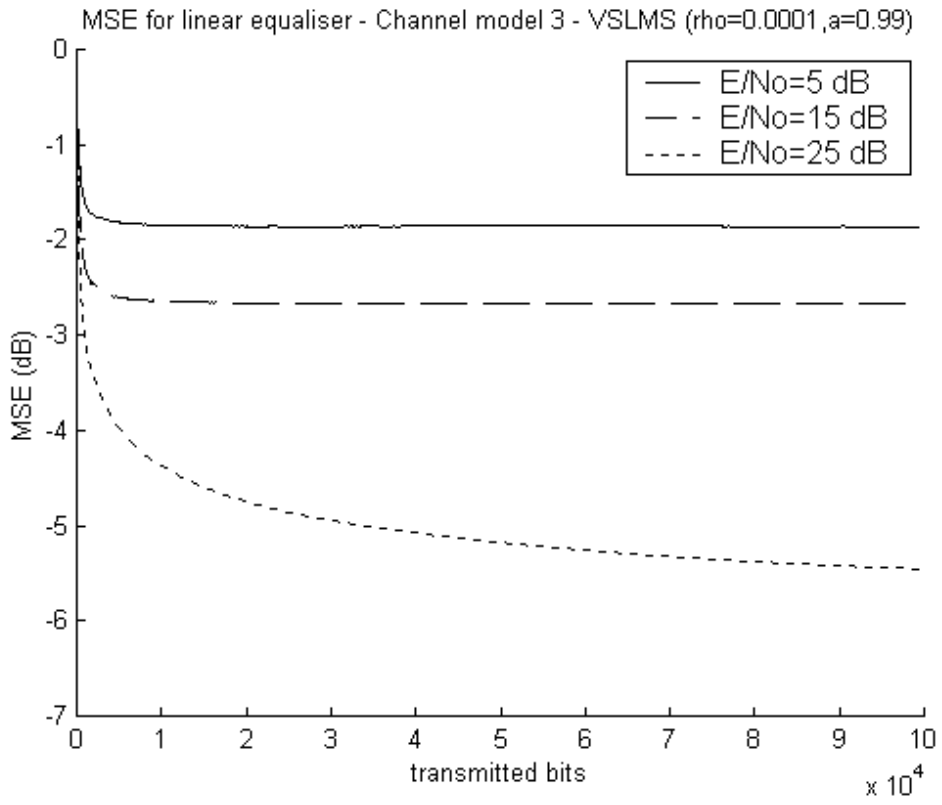

**Figure 3.30**: MSE for LE using VSLMS (ρ=0.0001, a=0.99) in Channel model 3.

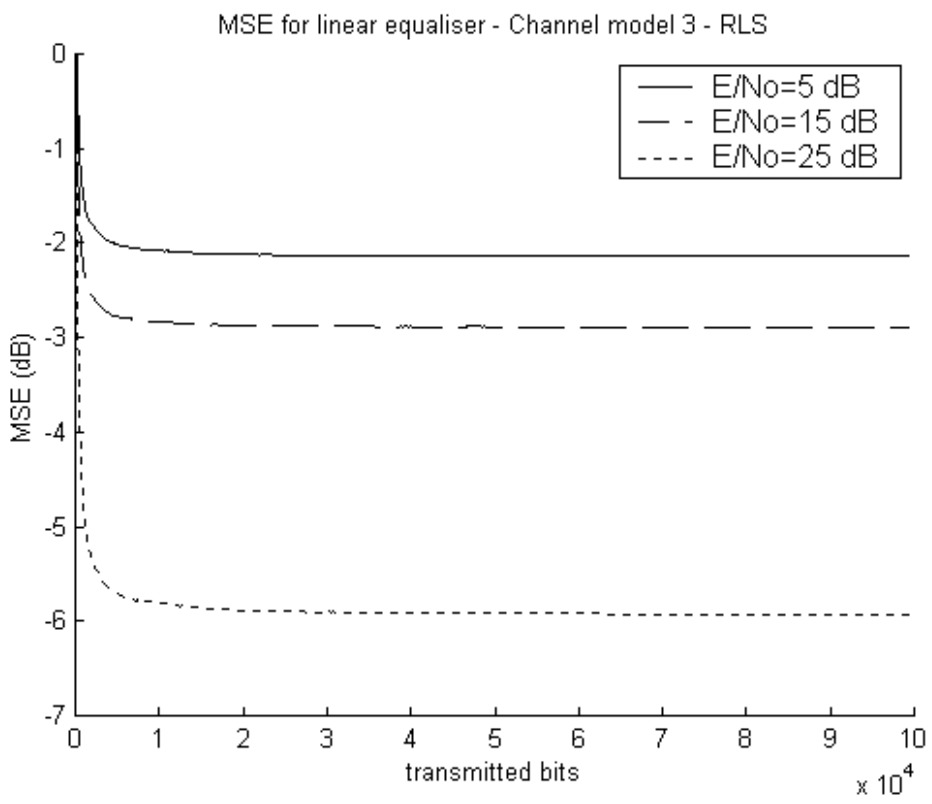

**Figure 3.31**: MSE for LE using RLS ($\lambda$=1.0) in Channel model 3.

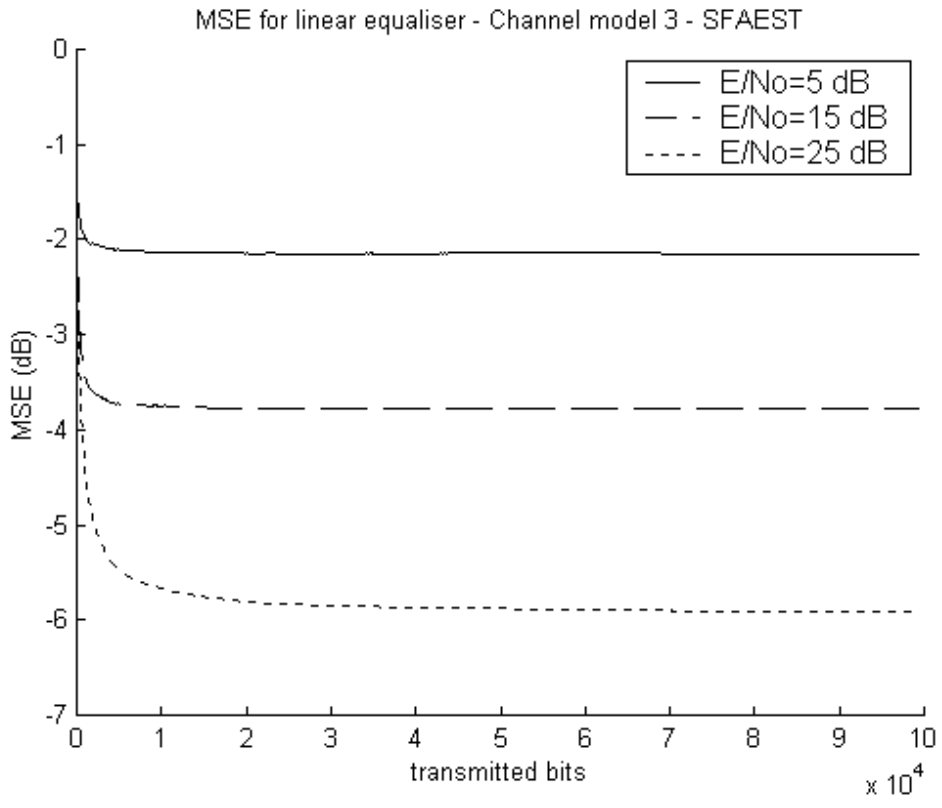

**Figure 3.32**: MSE for LE using SFAEST ($\lambda$=1, ρ=1, $\mu$=10) in Channel model 3.

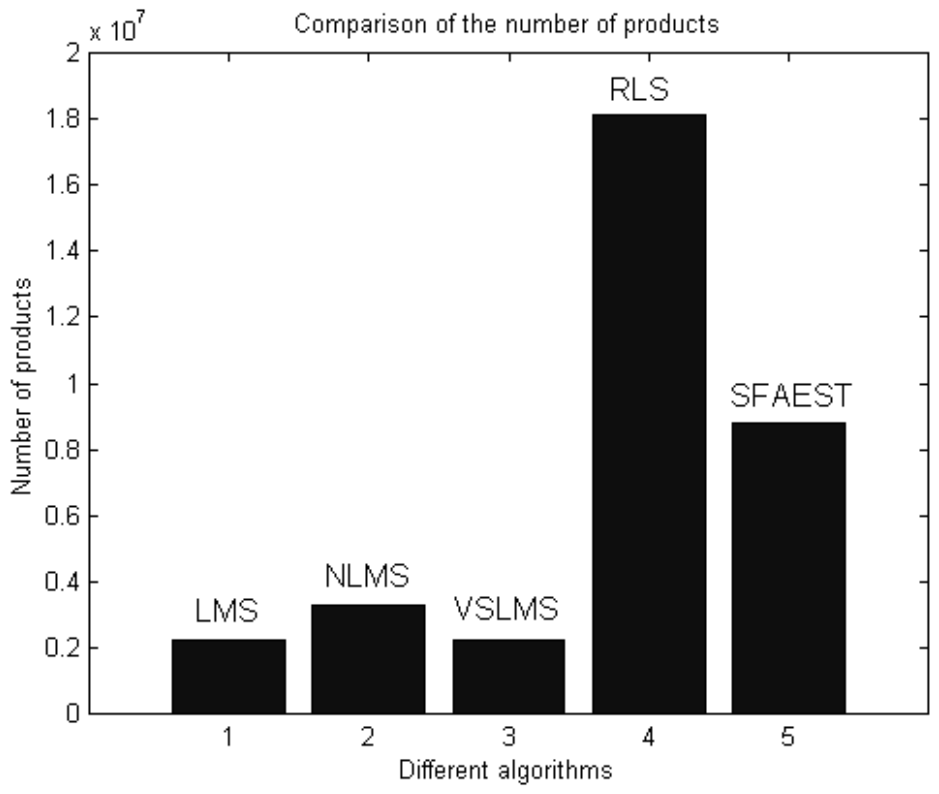

**Figure 3.33**: Number of products performed during the simulation (100,000 samples). Channel model 3.

From figure 3.33 the large differences in the number of computations of the RLS algorithm with respect to all the others can again be observed. Also notice that although the SFAEST algorithm has linear complexity, the large factor of the linear term (O(8N)) makes the number of products considerably greater than for the rest of linear algorithms

.

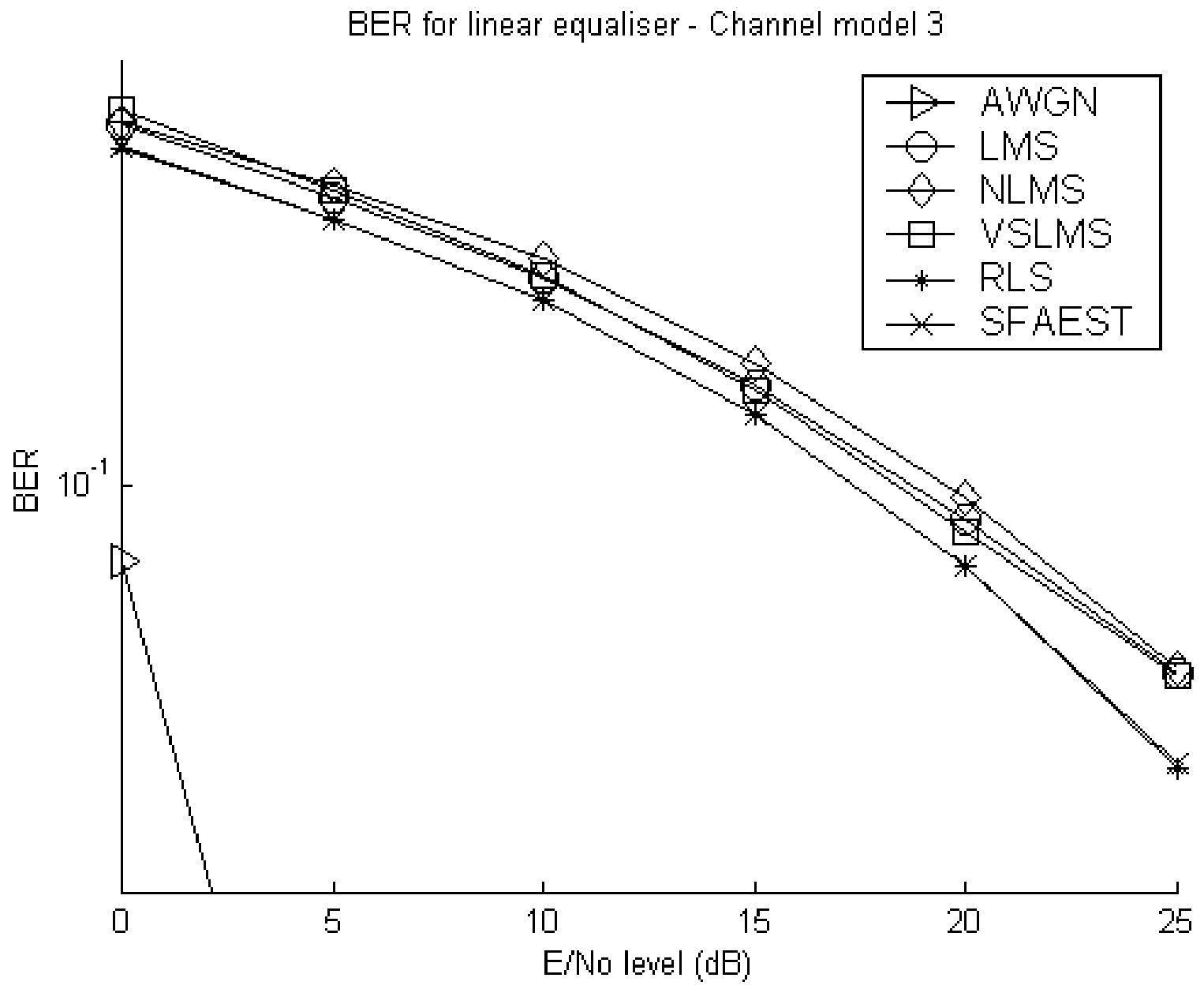

**Figure 3.34:** BER achieved with different algorithms equalising Channel model 3. Simulation length=100,000 samples.

The measured BER curves shown in figure 3.34 reveal the extremely poor characteristics of this channel; even at an E/No of 25 dB, the BER does not even get to $10^{-2}$. Obviously, for a system to operate in this channel it would require the use of some other countermeasures to enhance the BER such as error coding or some form of diversity.

Notice in this figure that the curves corresponding to the RLS and SFAEST overlap and therefore cannot be distinguished. At a large E/No, these two algorithms offer a reduction of nearly 2.5 dB in the required E/No to achieve a given probability of error with respect to the gradient algorithms.

### 3.4.4 Time-varying channel

Concluding with the background simulations using a LE, some results are now presented which show the performance of the different algorithms in the time varying model briefly introduced in section 3.3.4. The motivation to use a time varying channel is to check the tracking abilities of the different algorithms. The results shown next were obtained when the Doppler spread was set to 300 Hz (which correspond to a mobile speed of 45 m/s). The results presented here can be roughly compared with those obtained from Channel model 2. The static Channel model 2 was obtained by time-averaging several snapshots of this time-varying channel. Therefore, they have the same number of taps and relatively similar profiles. The simulation length was set to 150,000 samples. This increase in simulation length is motivated with the desire to check the tracking performance of the different algorithms for a longer time. Additionally, and given that in this case the stress is on the tracking capabilities of the algorithms, the BER are calculated using only the last 100,000 samples. In this way, the transient effects can be separated from the tracking performance.

Figures 3.35 to 3.38 present the results when using the LMS algorithm (figs. 3.35, 3.36) and the NLMS algorithm (figs. 3.37, 3.38). The first important point to mention, when comparing these results with those for the similar profile Channel model 2, is the significant degradation in performance. The SS-MSE levels in the time-varying channel are considerably higher than those presented in section 3.4.2. It is also remarkable in the case of large E/No level that larger step sizes achieve the same SS-MSE as smaller ones. For example in the case of the LMS, the curves for $\mu$=0.01 and $\mu$=0.015 overlap because they achieve the same SS-MSE. This is due to the fact that although a larger step size implies a loss in accuracy, it improves very much the tracking capabilities of the filter. Notice that in case of an equal SS-MSE level for two step sizes, it would be advisable to utilise the larger one, as this would reduce the MSE convergence phase allowing the use of shorter training sequences. The same effect can be observed when using the NLMS algorithm (E/No=25 dB) for $\tilde{\mu}$=0.5 and $\tilde{\mu}$=0.1. Both algorithms, LMS and NLMS, achieve the same SS-MSE when using an appropriate step size and their convergence time is nearly identical.

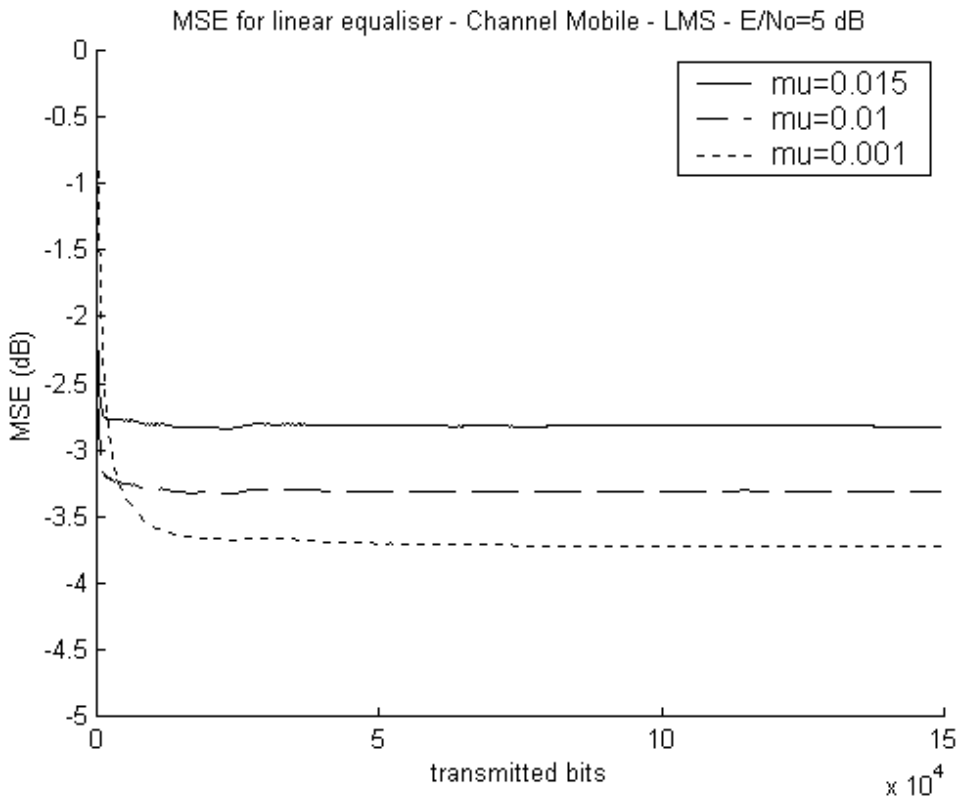

**Figure 3.35**: MSE for LE using LMS with different μ in Channel Time varying. E/No=5 dB.

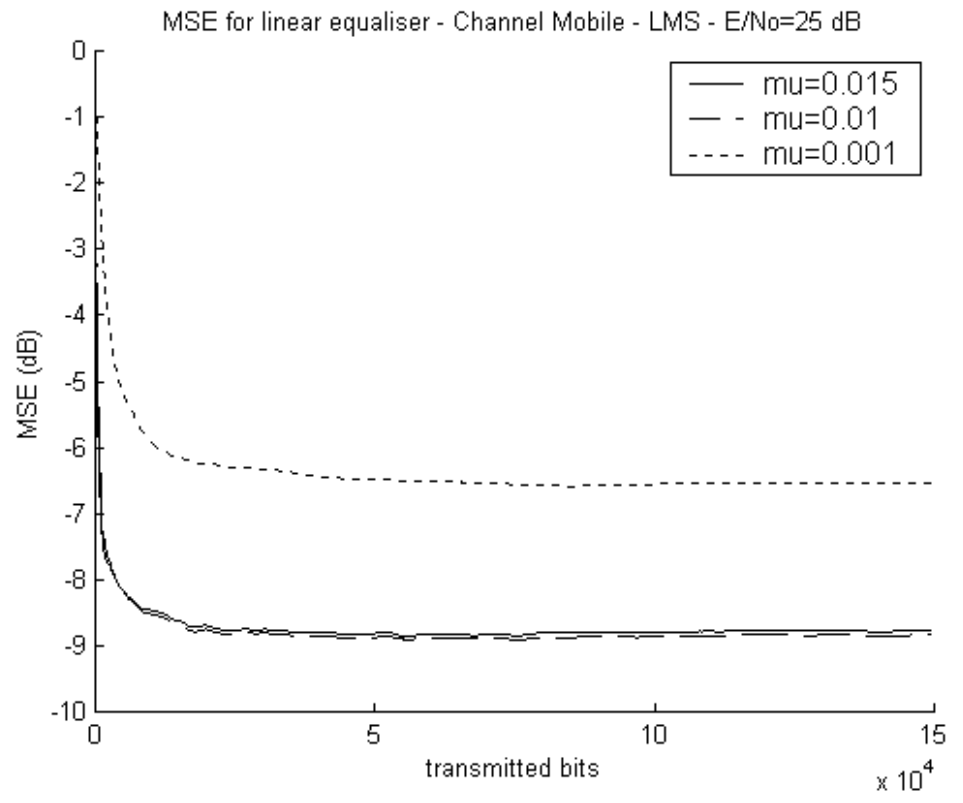

**Figure 3.36**: MSE for LE using LMS with different μ in Channel Time varying. E/No=25 dB.

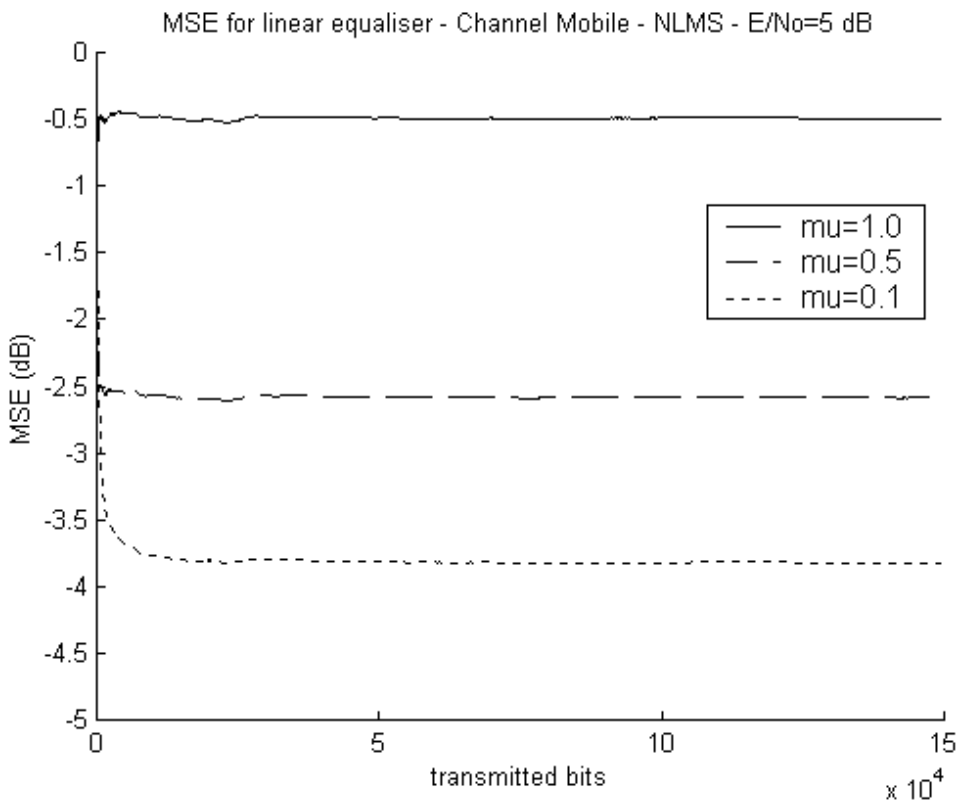

**Figure 3.37**: MSE for LE using NLMS with different $\tilde{\mu}$ in Channel Time varying. E/No=5 dB.

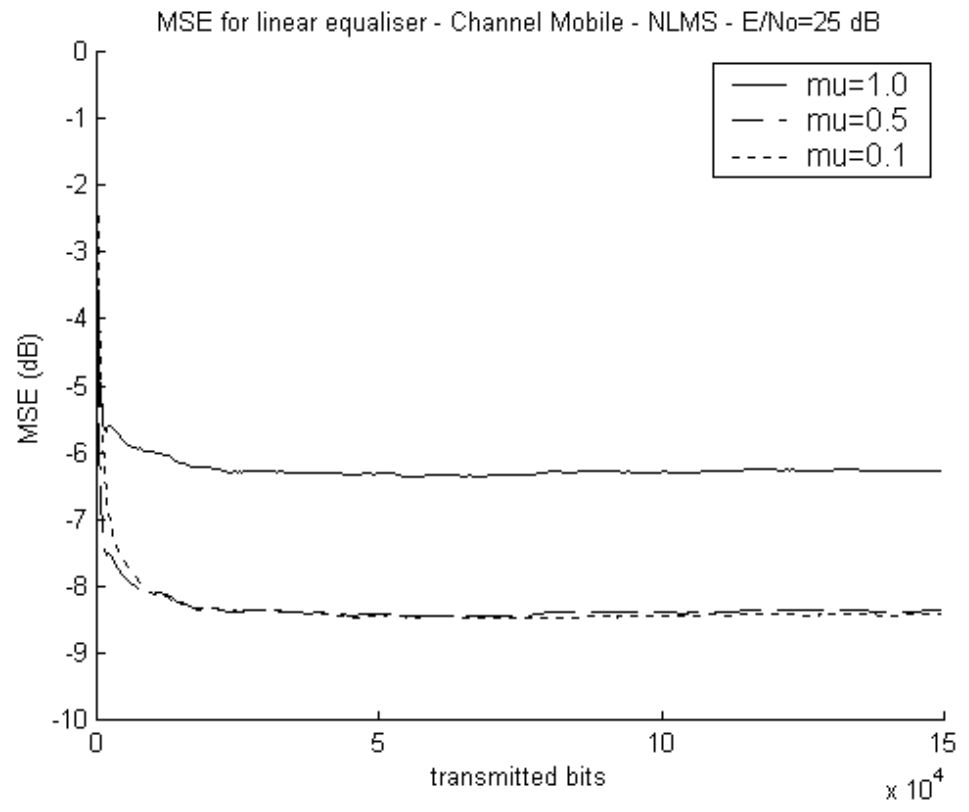

**Figure 3.38**: MSE for LE using NLMS with different $\tilde{\mu}$ in Channel Time varying. E/No=25 dB.

Figure 3.39 shows the MSE performance obtained when using the VSLMS algorithm. Its performance is comparable to that of the NLMS and VSLMS. The VSLMS parameters have been adjusted to provide better tracking and are now set to ρ=0.0005 and a=0.97. In figure 3.40, the evolution of the step size of the VSLMS is compared for the Time-varying model and for Channel model 2. It can be clearly appreciated that in the case of the dynamic channel, the step size remains nearly an order of magnitude larger than when operating in the static scenario. Also, and focusing in the steady state, while in the static environment the step size suffers only tiny (not visible) changes in its value, in the dynamic environment the changes in the step size are quite apparent.

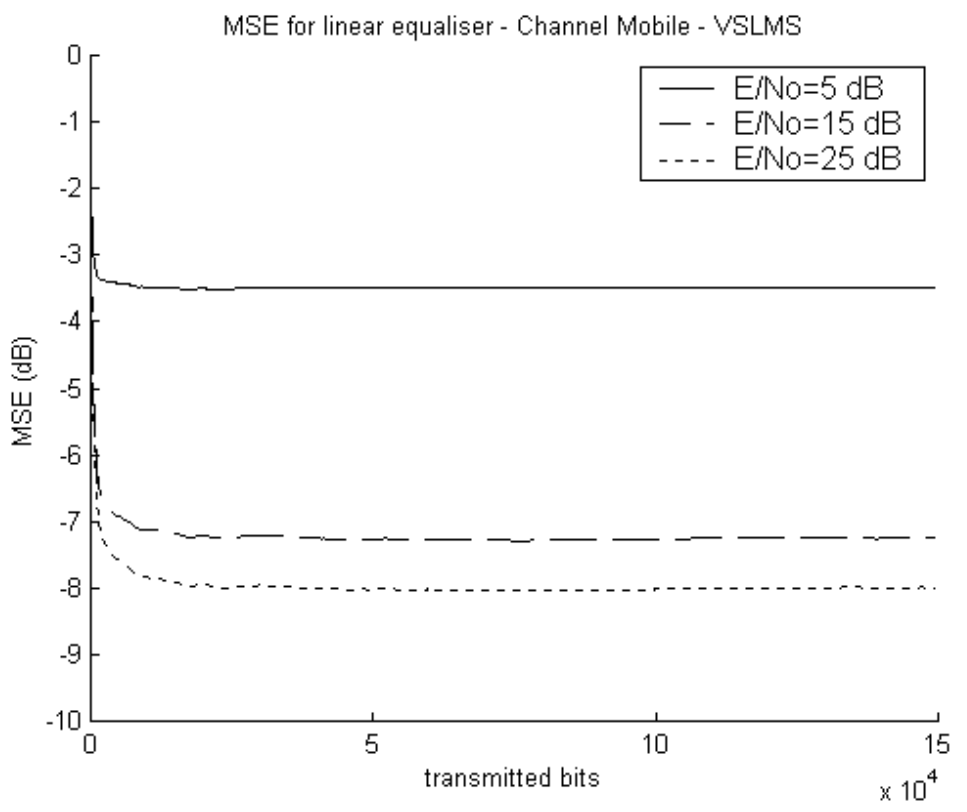

**Figure 3.39**: MSE for LE using VSLMS (ρ=0.0005, a=0.97) in Channel Time-varying.

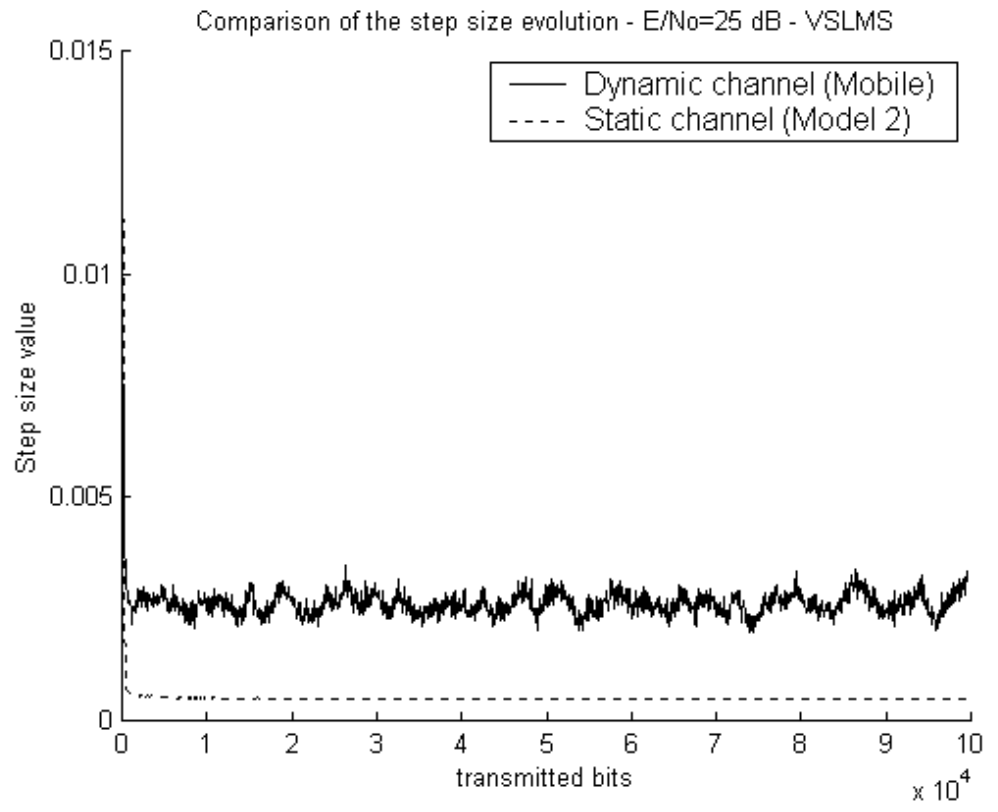

**Figure 3.40**: Comparison of the evolution of the VSLMS step size: Channel Time-varying vs Channel model 2. E/No=25 dB.

Figures 3.41 to 3.44 presents the results obtained when using least squares algorithms. Both algorithms, RLS and SFAEST, perform nearly identically, reinforcing the fact that they are indeed different formulations of the same algorithm. Reductions in the MSE level of 0.5 and 1 dB can be observed with respect to the gradient schemes for E/No levels of 5 and 25 dB respectively. It is interesting to point out an effect visible in all these four graphs but particularly in figures 3.42 and 3.44 (E/No=25 dB). The MSE curves for $\lambda$=0.9999 increase significantly after convergence. This increase is due to the inability of the RLS to track the channel variations adequately when $\lambda$ is so close to one ($\lambda$=0.9999). The best performance is obtained for $\lambda$=0.995 which allows the algorithm to follow the channel fluctuations. Lower values of the forgetting factor also tended to increase the MSE level.

Finalising the treatment of linear equalisation for the time-varying channel, the BER curves obtained for the different algorithms are shown in figure 3.45. Comparing these curves with those of figure 3.25 (BER for Channel model 2), a degradation of nearly two orders of magnitude in BER can be appreciated for moderate to large E/No levels. In these channel conditions, other mechanisms (FEC, diversity) would also be required to reduce the BER to an acceptable level.

In this figure, it can be seen that all gradient type algorithms perform very similarly in terms of BER. Least squares techniques provide a significant improvement with respect to LMS algorithms. Notice that the BER curves of the RLS and SFAEST cannot be distinguished in the graph as they have nearly identical values.

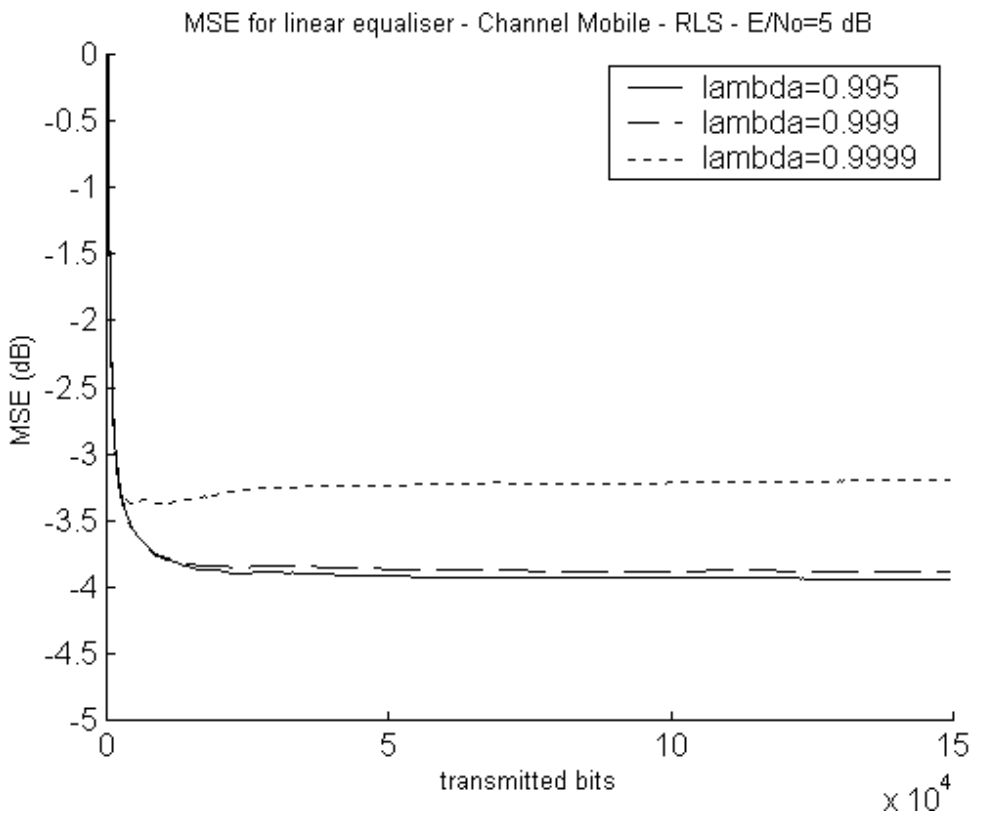

**Figure 3.41**: MSE for LE using RLS with different λ in Channel Time varying. E/No=5 dB.

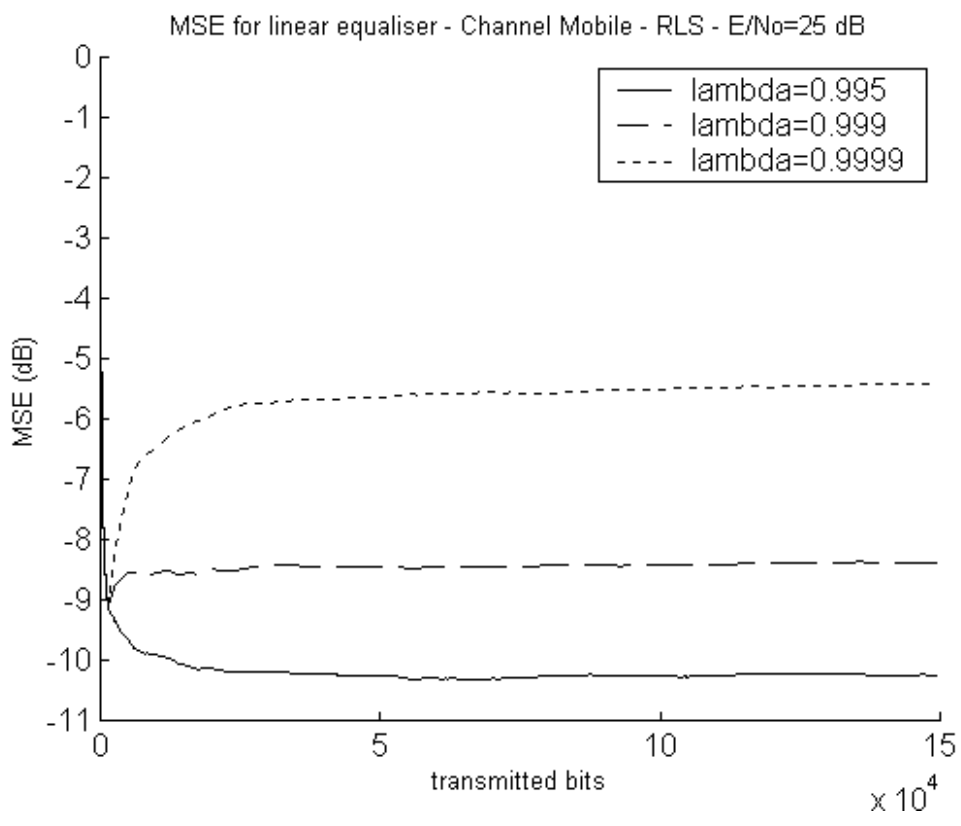

**Figure 3.42**: MSE for LE using RLS with different λ in Channel Time varying. E/No=25 dB.

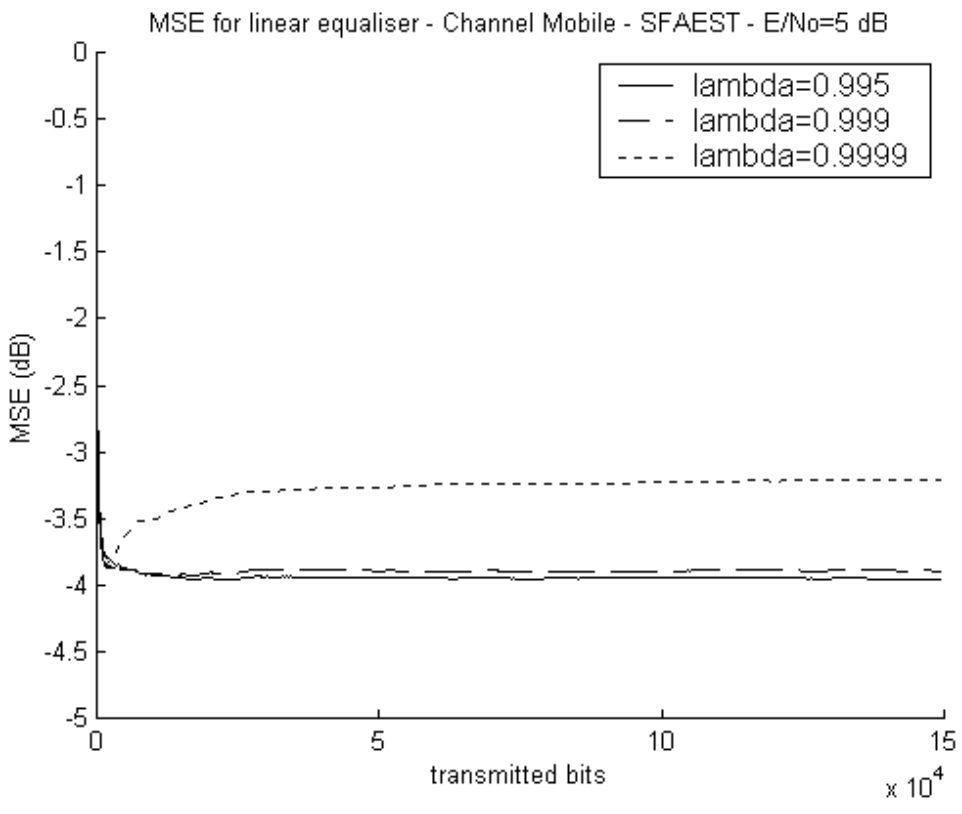

**Figure 3.43**: MSE for LE using SFAEST with different λ in Channel Time varying. E/No=5 dB.

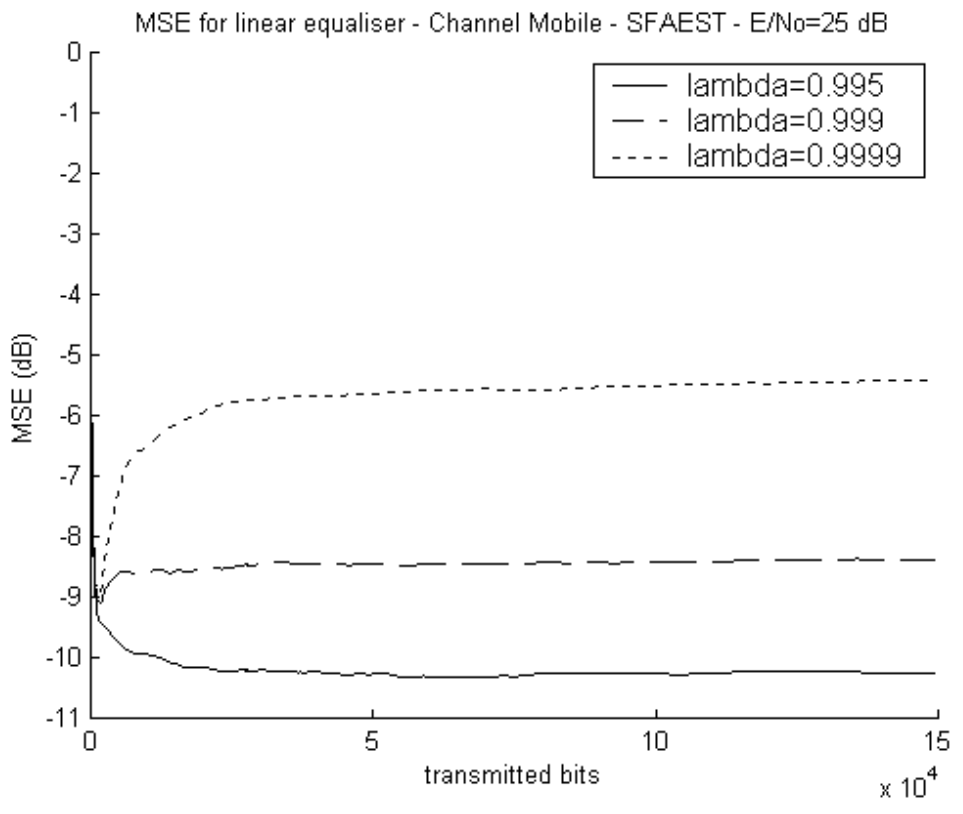

**Figure 3.44**: MSE for LE using SFAEST with different λ in Channel Time varying. E/No=25 dB.

As has already been explained in section 2.3.4, general comparison of the tracking capabilities of the gradient and least squares algorithms is difficult as each of them proves to be superior in certain scenarios. In this scenario, least squares outperform gradient type algorithms. One reason for this superiority can be sought in the long length of the equaliser, which greatly limits the choice of the step size for the LMS algorithm ([Cioffi86]).

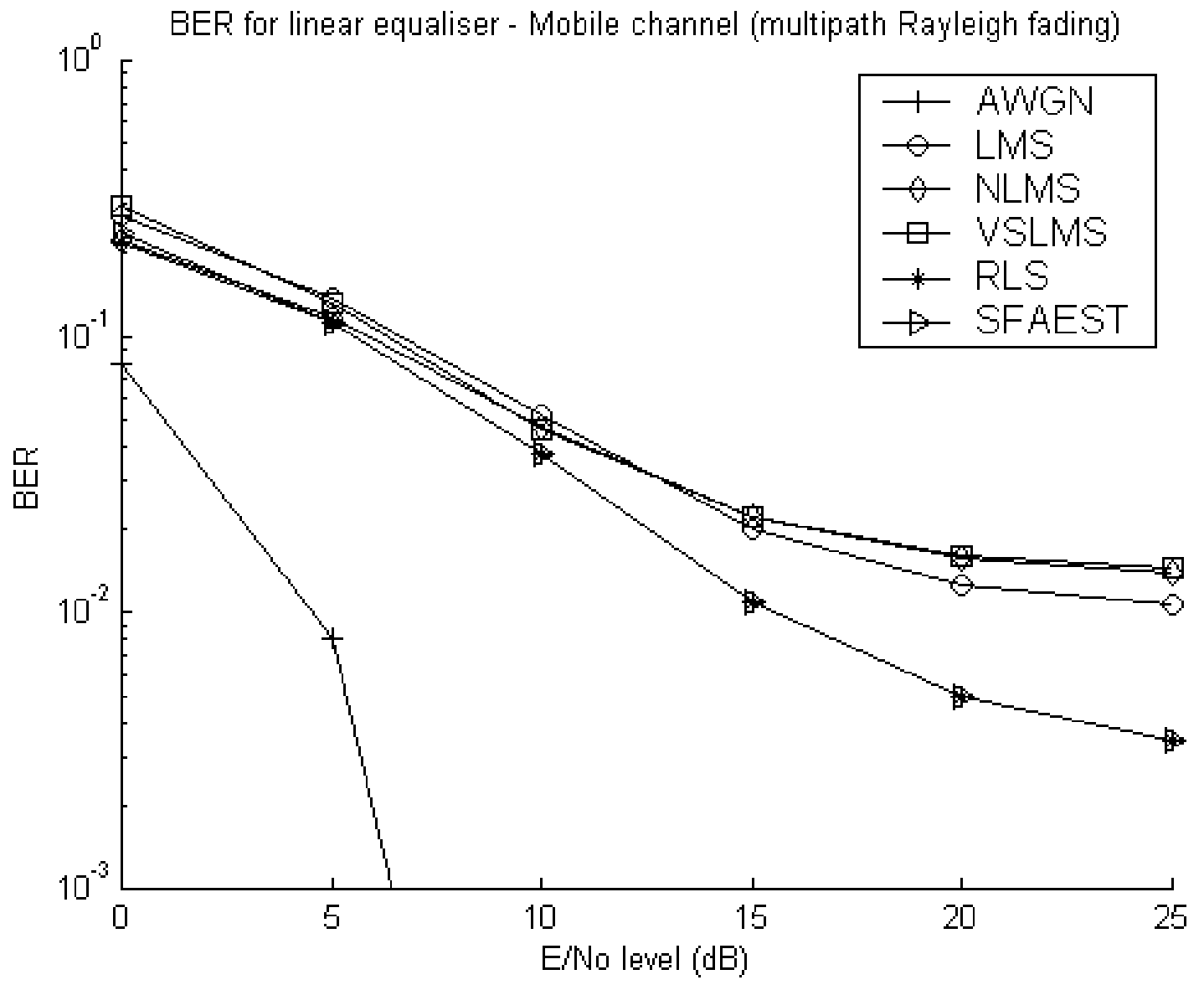

**Figure 3.45:** BER achieved with different algorithms equalising the Time-varying model (multipath Rayleigh fading). Simulation length=100,000 samples (excluding the 50,000 samples from initial convergence).

## 3.4.5 Effect of the decision delay in the LE

One of the parameters to select when implementing an equalisation system is the delay between the sample being detected and the sample being transmitted. There are not many references on how to select this parameter because, even when the channel is known, there are no definite rules to set the delay optimally[11]. [Qureshi73] proposed performing an exhaustive search to determine the optimum delay in the context of gradient type algorithms. A similar scheme proposed by [Mueller75], termed cyclic equalisation, implements in an efficient way this exhaustive testing of all delays by using rotations of a cyclic training sequence whose length equals the filter length. In [Manolakis83] an algorithm is presented to determine the optimum delay. This algorithm however, given its complexity, is not suitable for real-time implementation. In [RUPP], some simulation results using different choices for the delay are presented in the particular context of mobile systems but no conclusions are drawn on how to select it. Notice that in a time-varying channel, the problem becomes more complex because the optimum delay may vary as the channel changes.

[11] Optimally in an MSE sense.

In the context of this work, we have resorted to simulation to set the delay for the different channels. Figures 3.46 and 3.47 show the MMSE level computed via the Wiener-Hopf equation for channel models 2 and 3 when using various delays under different E/No conditions. For the Channel model 2 the equaliser length was fixed at 23 taps. Note that as the E/No level is increased, the equaliser becomes more sensitive to the choice of the delay. What this graph shows is that a wrong choice of the delay may cause a significant loss of performance. In figure 3.47 the same information is shown but now for Channel model 3. In this case, the equaliser length was set to 11 taps. This channel is less sensitive to the delay than Channel model 2 but notice again that the equaliser becomes more sensitive to the delay at large E/No, although now in a much less pronounced way. What is also clear from both graphs is that the delay should be set so that the main paths of the channel are captured, i.e. the decision on a sample is made after most of the energy corresponding to that sample has arrived.

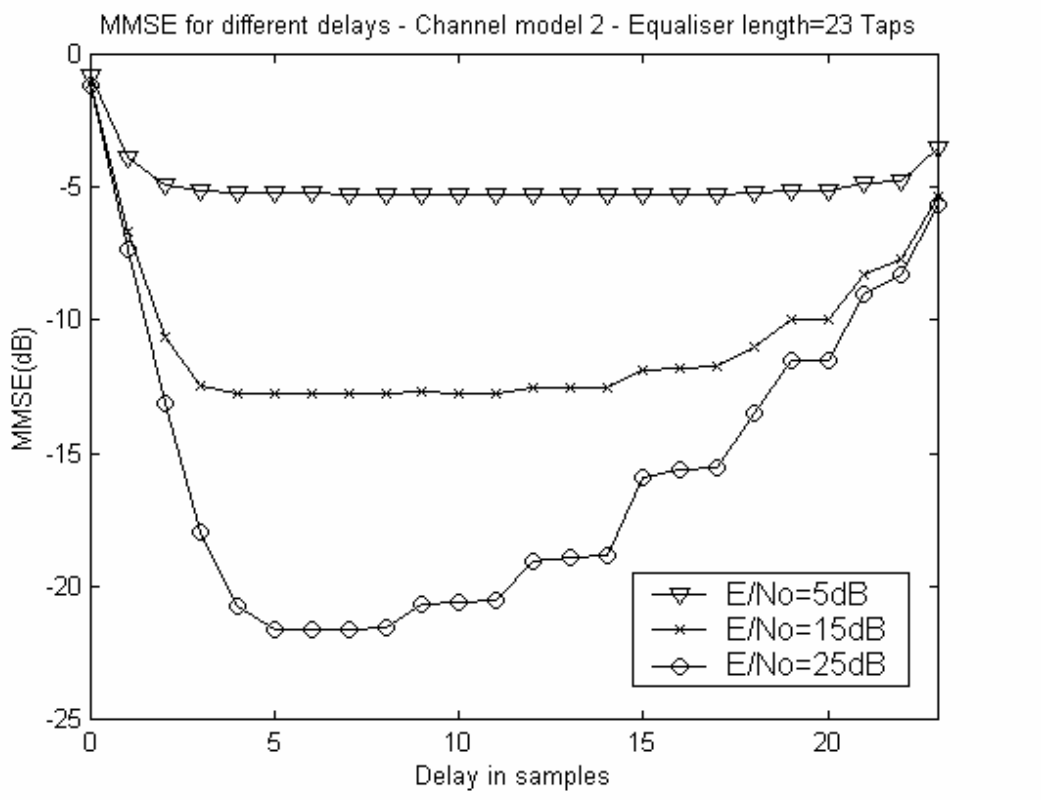

Figure 3.46: Effect of the decision delay on the MMSE level. Channel model 2. 23-tap LE.

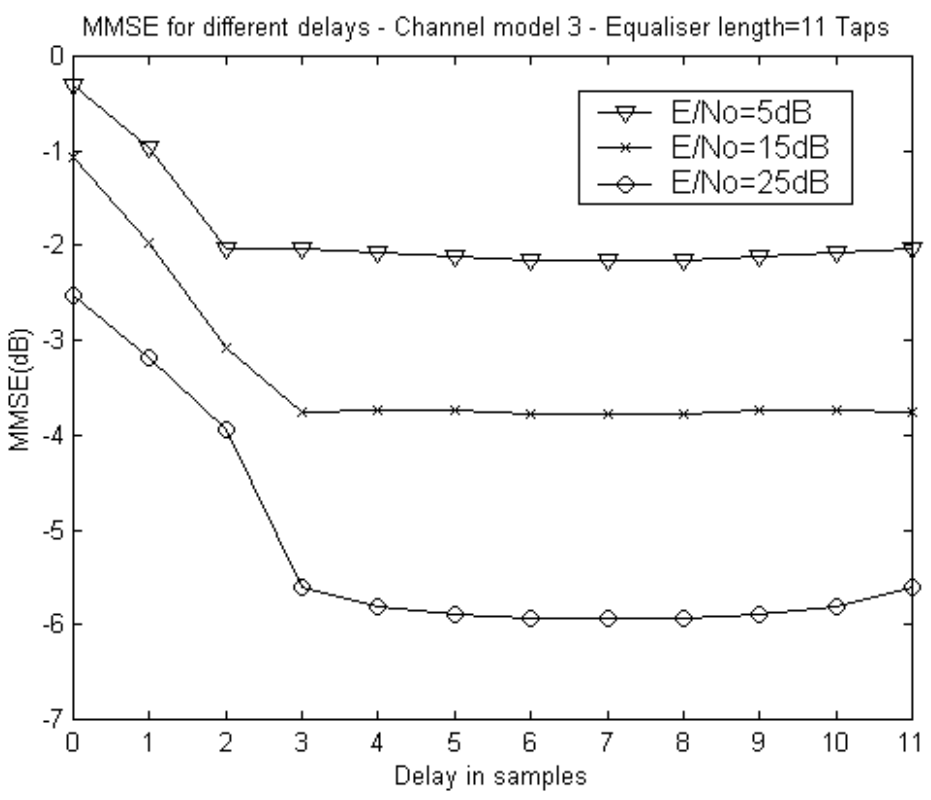

Figure 3.47: Effect of the decision delay on the MMSE level. Channel model 3. 11-tap LE.

In the simulations in this and later chapters, the delay has been set so that the dominant path of the equaliser lies roughly in the middle of the tap delay line (TDL). This has proved to be a convenient strategy when the channel is time varying and the dominant paths of the channel may change. When this happens, the dominant equaliser tap will also vary its position but will still be far away from the extremes of the TDL where, in some cases, performance is severely degraded as has been shown in the previous graphs. Varying the decision delay during operation has not been considered, as this would require some form of resynchronisation mechanism, which would further complicate the design of the receiver.

In real channels, it is not possible to know what will be the position of the main channel taps. However, there are channel profiles available such as the COST207 models, which can serve as a guideline to help in deciding an appropriate value for the decision delay.

# 3.5 Background simulation results for DFE

In this section the performance of the DFE is analysed using the same channel models as for the LE although now only Channel models 2 and 3 and the Time varying model are tested. The feedforward filter (FFF) has arbitrarily[12] been set to 6 taps whereas the feedback filter (FBF) is chosen to have as many taps as the channel impulse response. Regarding the algorithms, it is important to point out that the SFAEST algorithm cannot be used on the DFE. Recall that the SFAEST algorithm (and other Fast Kalman variants) is based on the exploitation of the shift invariant property of the input vector to the adaptive filter to re-formulate the least squares equations (see section 2.3.3). In the case of a DFE, the input vector is a combination of the channel samples and previously detected symbols. This input vector structure does not satisfy the shift invariant property, thus precluding thus the use of the SFAEST algorithm.

As a last general comment on these simulations it must be said that, as in the LE simulations, the DFE is assumed to operate all the time in training mode. In the case of DFE, this must be born in mind when drawing conclusions from the results, as in a practical system (decision directed), performance will be worse due to error propagation.

## 3.5.1 Channel model 2

Figures 3.48 and 3.49 show the DFE MSE when using the LMS algorithm. The steady state MSE is now about 1-2 dB lower than the equivalent LE simulation. The fact that the DFE has fewer taps than the LE used previously allows the algorithm to use larger step sizes like

---

[12] This choice is arbitrary in the context of this chapter. In chapter 6 where DFE length is treated in detail, an explanation is given of why such a length is chosen.

$\mu$ =0.03. As before, the ideal step size to use is very much influenced by the E/No level at which the receiver is operating.

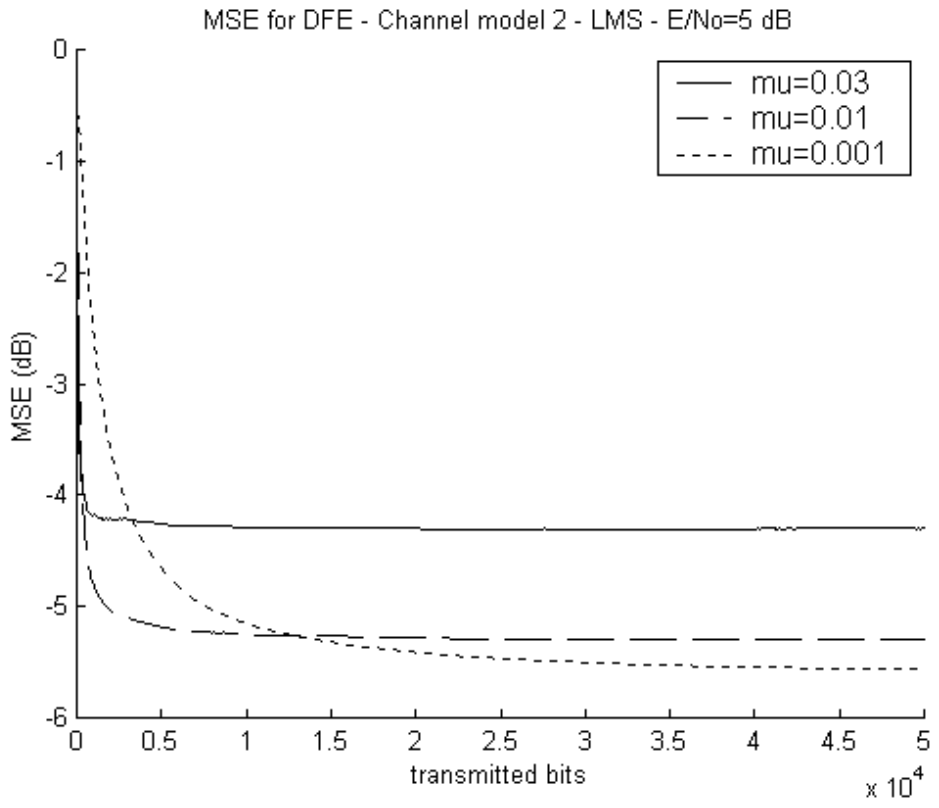

Figure 3.48: MSE for DFE using LMS with different $\mu$ in Channel model 2. E/No=5 dB.

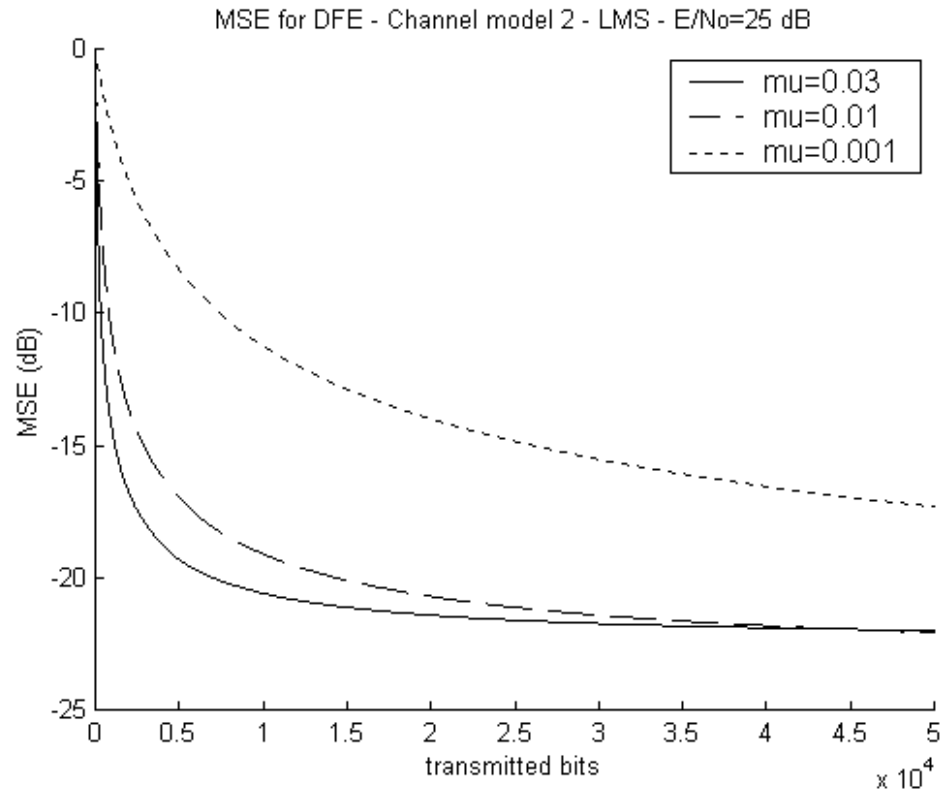

**Figure 3.49**: MSE for DFE using LMS with different $\mu$ in Channel model 2. E/No=25 dB.

These comments for the LMS algorithm apply also to the simulation results obtained when using the NLMS, VSLMS and RLS algorithms, whose results are presented in figures 3.50 to 3.53. A 1-2 dB MSE reduction is observed with respect to the LE. The convergence time seems to be independent of the equalisation structure used.

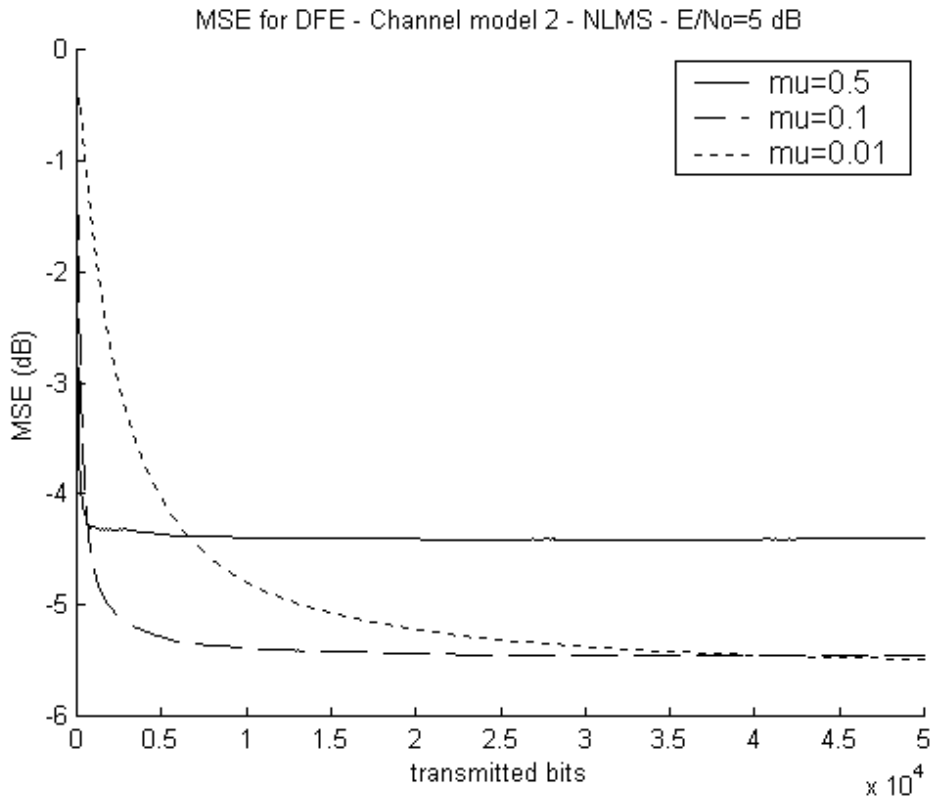

**Figure 3.50**: MSE for DFE using NLMS with different $\tilde{\mu}$ in Channel model 2. E/No=5 dB.

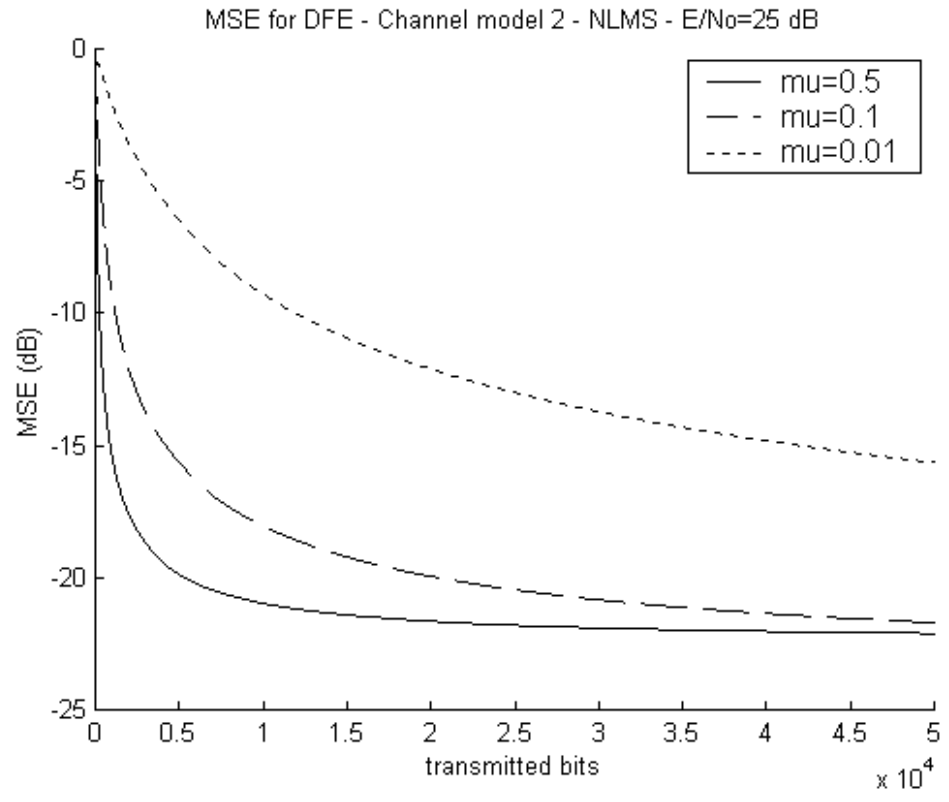

**Figure 3.51**: MSE for DFE using NLMS with different $\tilde{\mu}$ in Channel model 2. E/No=25 dB.

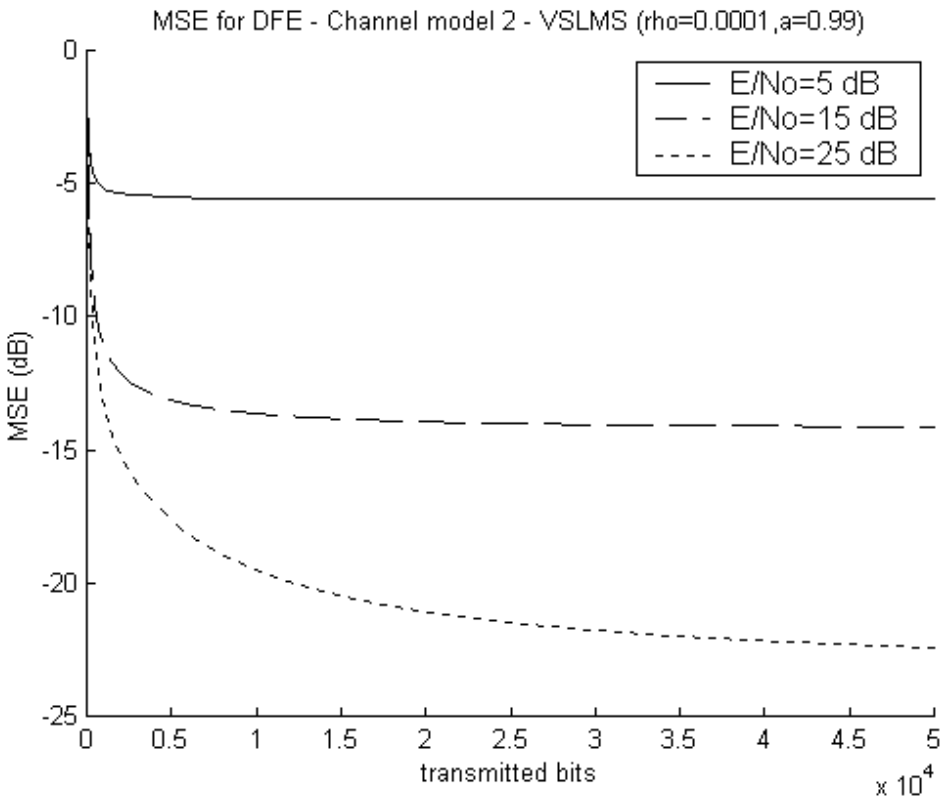

**Figure 3.52**: MSE for DFE using VSLMS (ρ=0.0001, a=0.99) in Channel model 2.

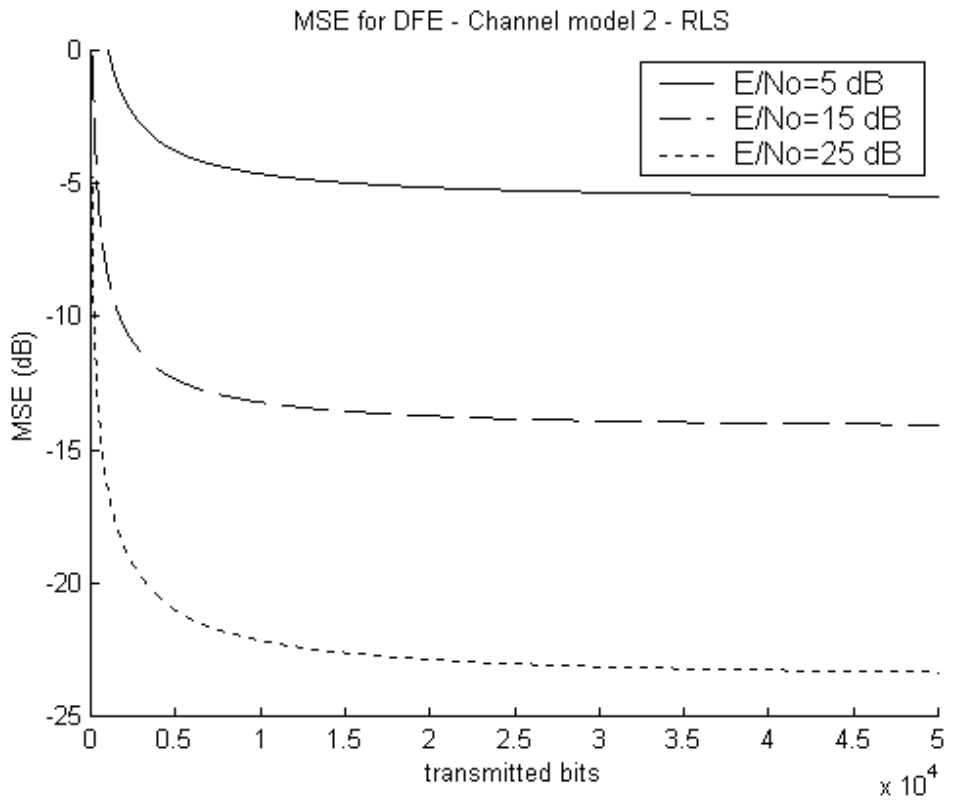

**Figure 3.53**: MSE for DFE using RLS ($\lambda$=1.0) in Channel model 2.

The BER curves corresponding to all the simulated algorithms are presented in figure 3.54. The main conclusion that can be drawn is that in this particular channel the use of a DFE does not significantly improve the BER with respect to the LE. Moreover, error propagation may significantly decrease the BER with respect to the curves shown in 3.54. In order to get a quantitative idea of the degradation caused by error propagation, see pp. 624-625 in [Proakis95].

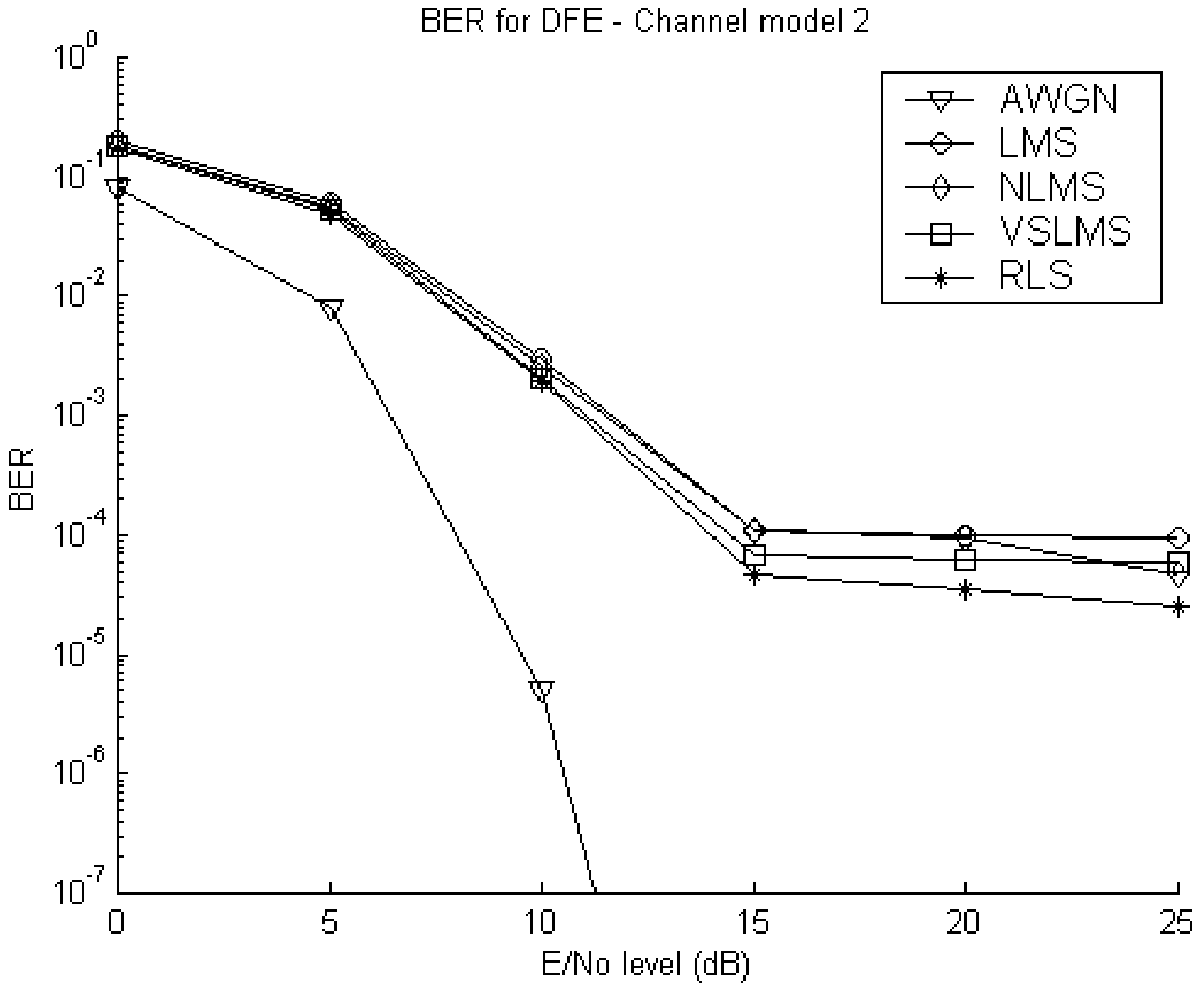

**Figure 3.54:** BER achieved with different algorithms equalising Channel model 2. Simulation length=100,000 samples.

### 3.5.2 Channel model 3

Remember from section 3.3.3 where Channel model 3 was introduced, that this channel contains a spectral null in the passband making its equalisation difficult. It is in this channel model that the DFE shows its potential benefits.

Figures 3.55 to 3.58 present the MSE curves for the LMS and NLMS algorithms. When E/No=5 dB (figures 3.55 and 3.57) reductions of about 1.5 dB with respect to the LE in the steady state MSE are obtained for both algorithms when an appropriate step size is used. When E/No=25 dB, the DFE completely outperforms the LE by decreasing the steady state MSE by as much as 10 dB independently of the algorithm (compare the figures 3.56 and 3.58 with 3.26 to 3.29).

Notice that because of the large eigenvalue spread of the channel, the algorithm takes a long time to achieve full convergence, but now the SS-MSE is much lower than in the LE case. Again, it can be observed when using the DFE structure that the NLMS algorithm converges much faster than the plain LMS.

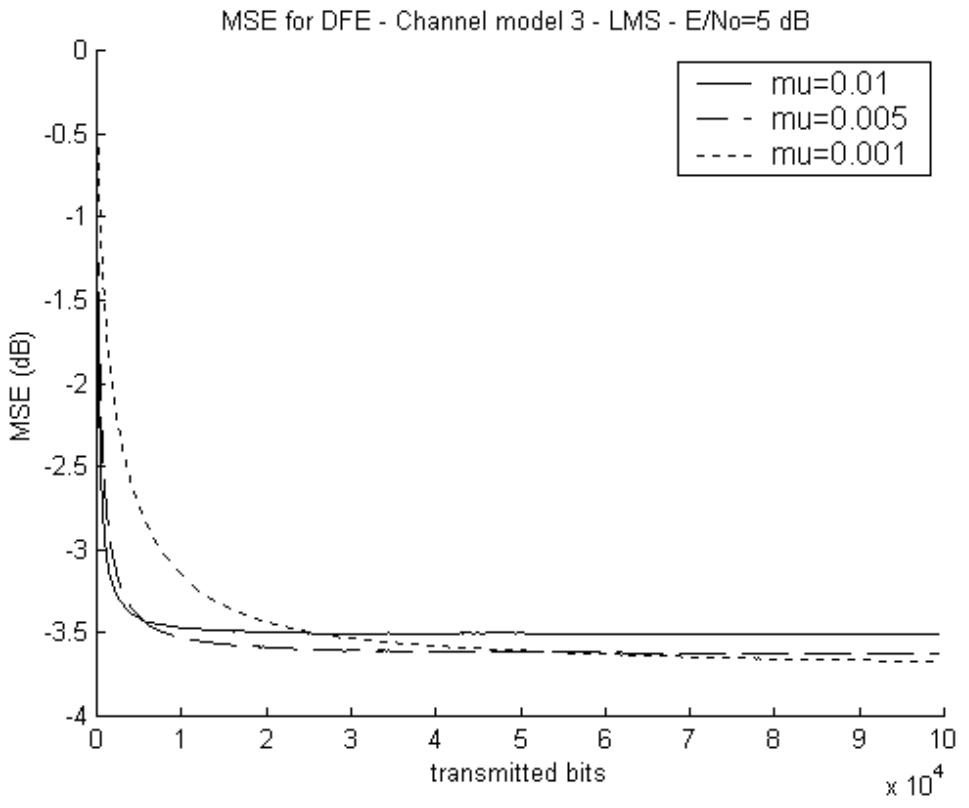

**Figure 3.55**: MSE for DFE using LMS with different μ in Channel model 3. E/No=5 dB.

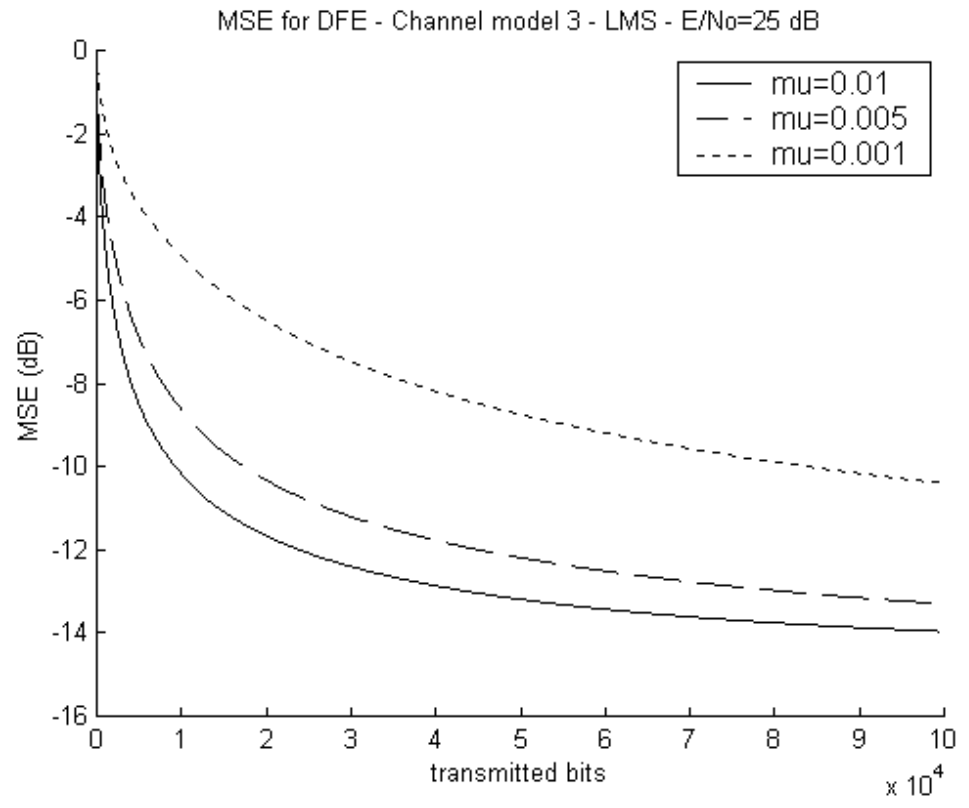

**Figure 3.56**: MSE for DFE using LMS with different μ in Channel model 3. E/No=25 dB.

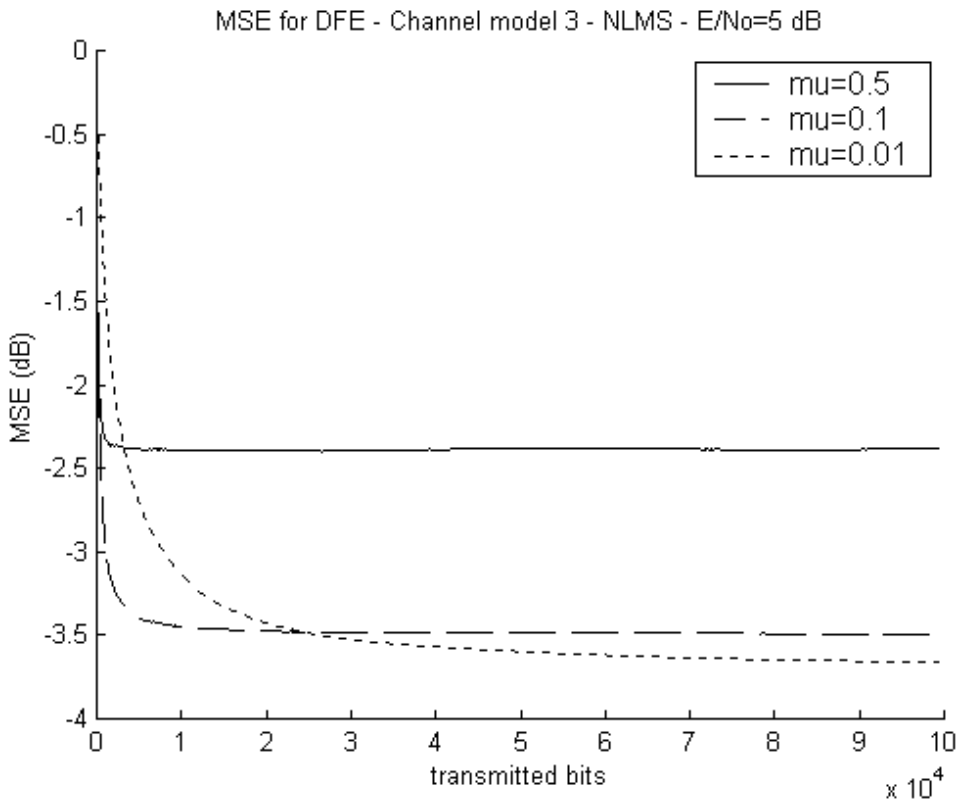

**Figure 3.57**: MSE for DFE using NLMS with different $\tilde{\mu}$ in Channel model 3. E/No=5 dB.

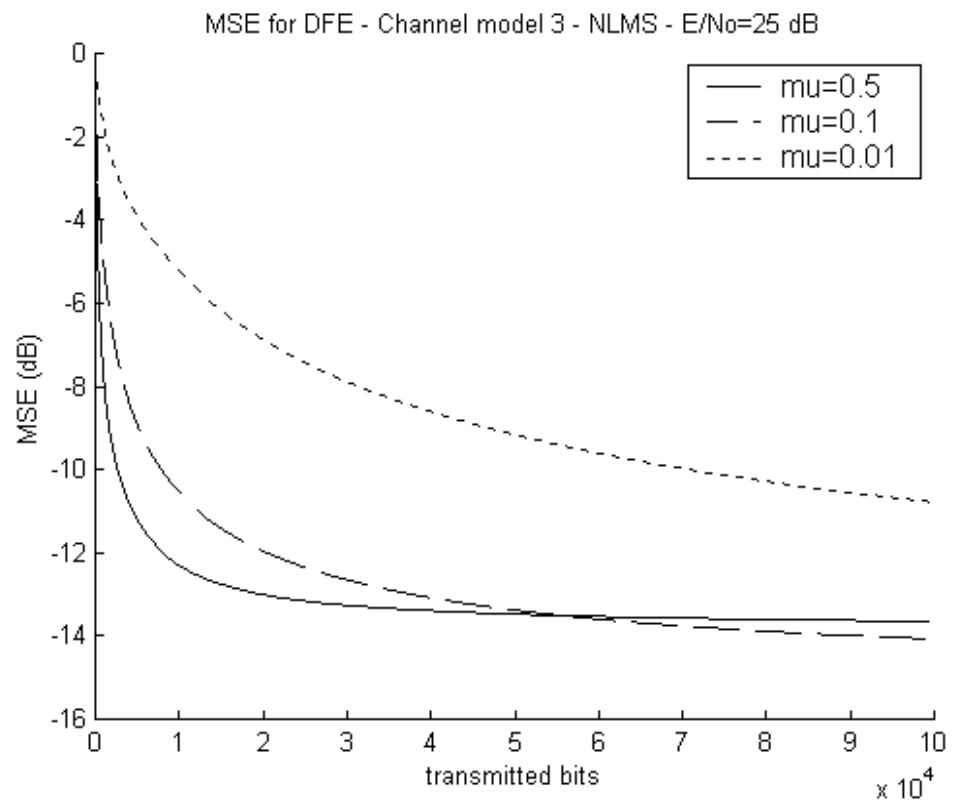

**Figure 3.58**: MSE for DFE using NLMS with different $\tilde{\mu}$ in Channel model 3. E/No=25 dB.

Figure 3.59 shows the learning curve of the VSLMS algorithm. It can be seen that after an initial very fast convergence (up to –6 dB in the MSE level), it then slows down significantly. Still the very fast initial reduction of the MSE, which is due to the large step size in the first iterations of the algorithm, helps very much in reducing the number of errors during the transient phase. Notice again that the same VSLMS parameters are used irrespective of the E/No level.

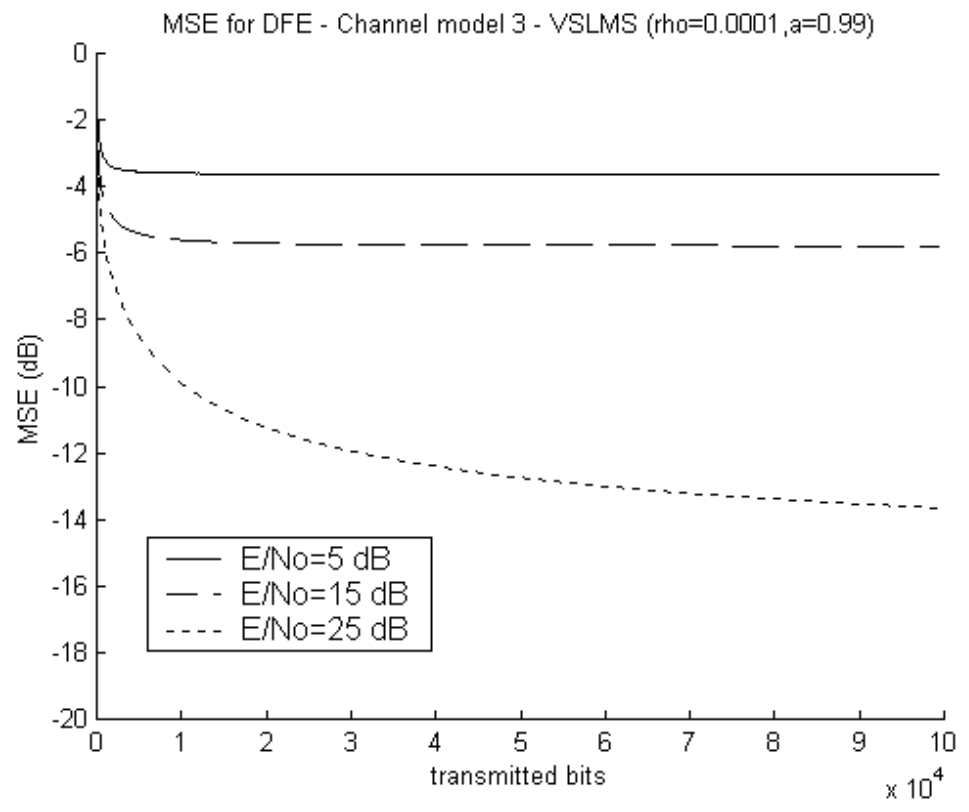

**Figure 3.59**: MSE for DFE using VSLMS ($\rho$=0.0001, a=0.99) in Channel model 2.

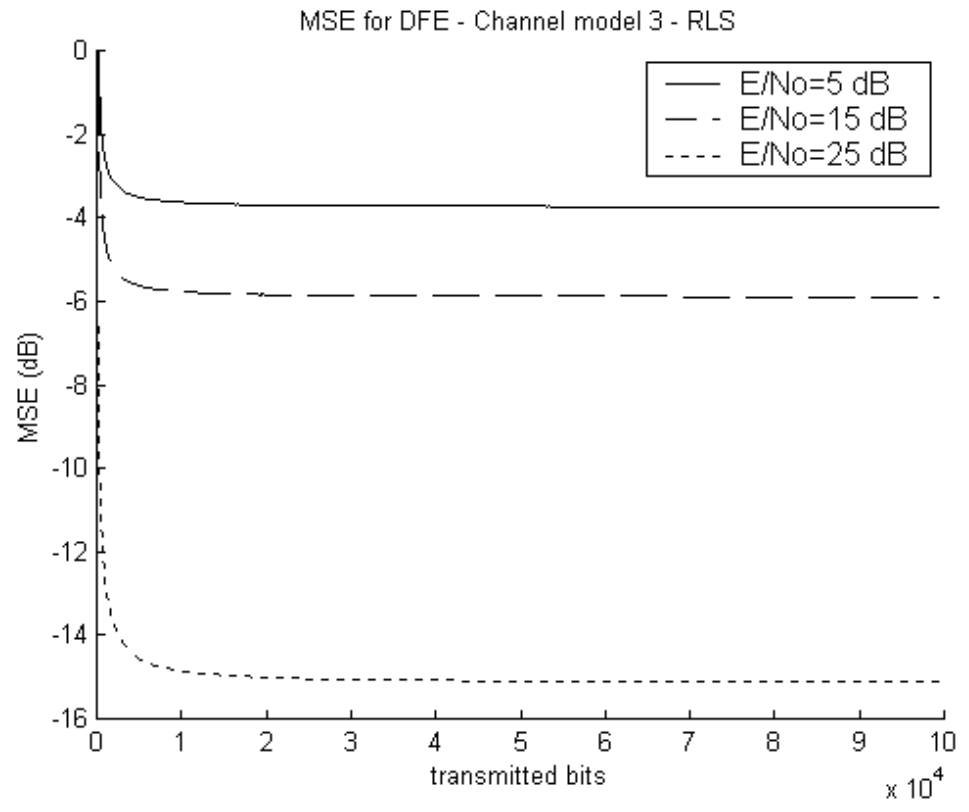

**Figure 3.60**: MSE for DFE using RLS ($\lambda$=1.0) in Channel model 2.

In figure 3.60 the MSE curves corresponding to the RLS algorithm are presented. Convergence time is clearly superior to any of the gradient schemes previously shown. Since the channel is static, the forgetting factor can be set to one and therefore the equaliser weights converge to the Wiener-Hopf solution as there is no excess mean squared error.

Although the DFE MSE curves show an improvement, sometimes very significant, with respect to the equivalent LE MSE ones, the graph below (figure 3.61) showing the BER for the DFE gives a clearer idea of its potential. It is important to compare the curves shown here with those shown in figure 3.34 (LE). The DFE produces a reduction in the probability of error for a given E/No of up to two orders of magnitude for the gradient type schemes and even three for the RLS algorithm. The maximum improvements can be observed in large E/No situations (20-25 dB). Over the range from 0 to 10 dB, the DFE improvement over the LE is rather modest and error propagation would probably cause the DFE to perform even worse than the LE. This situation is due to the spectral null in the channel, which limits the action of the equaliser, LE or DFE, when the noise level is large.

Once more it is worth noting that the RLS algorithm only provides an advantage over the other algorithms when E/No is large. This phenomenon could already be inferred from the MSE curves shown before as when E/No=5 dB, all algorithms exhibit similar convergence times and SS-MSE. Notice also that compared to the AWGN channel, about 10-15 additional dB are now necessary to achieve the same probability of bit error.

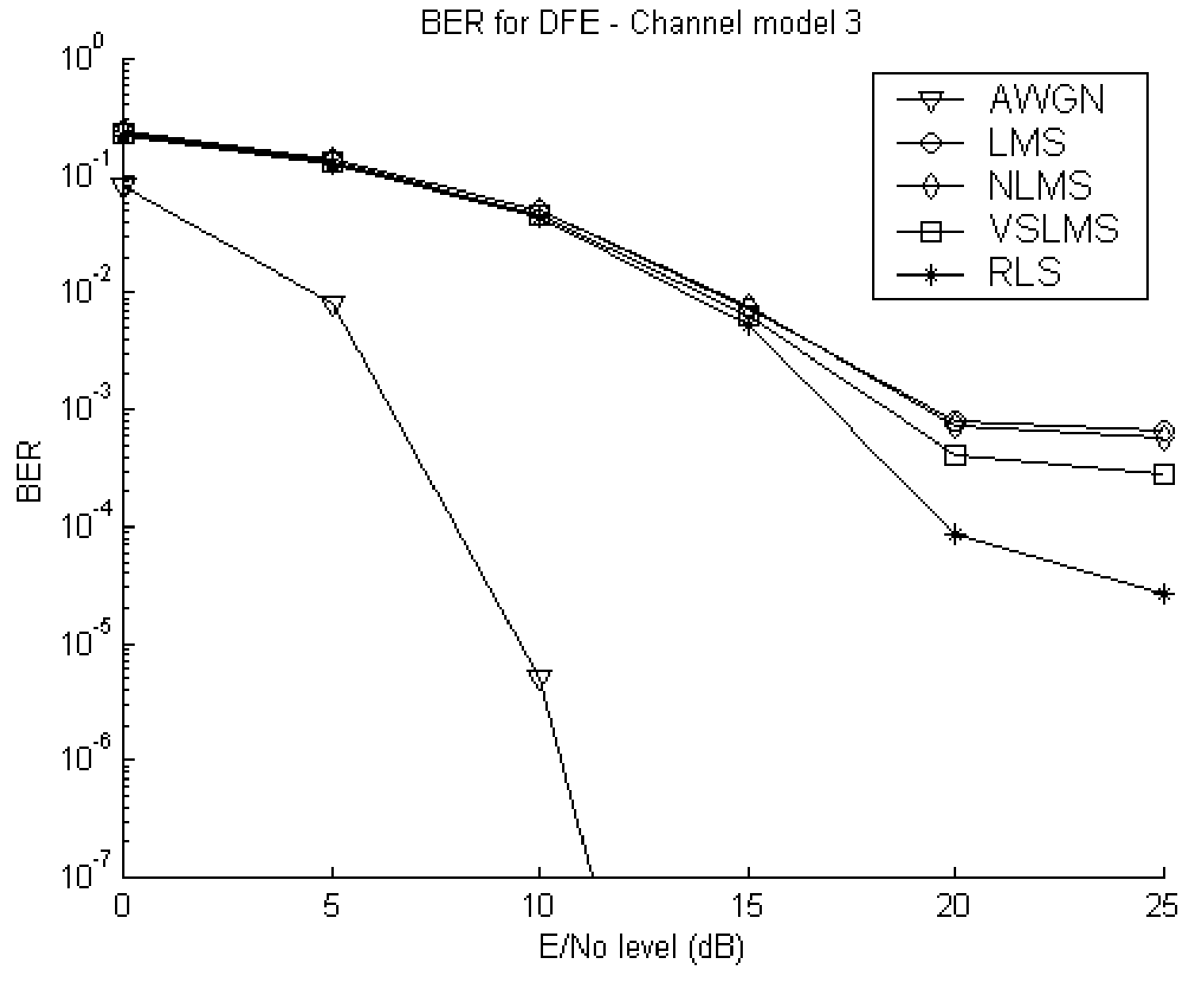

**Figure 3.61:** BER achieved with different algorithms equalising Channel model 3. Simulation length=100,000 samples.

## 3.5.3 Channel Time-varying

Concluding the background DFE simulations, results are now presented for the time-varying model briefly commented in section 3.4.4. The MSE curves for any of the algorithms shown below (figures 3.62 to 3.68) indicate that the DFE is able to withstand the fluctuations of the channel much more robustly than the LE. The MSE curves presented here are 9-10 dB lower than those shown in section 3.4.4 for the LE.

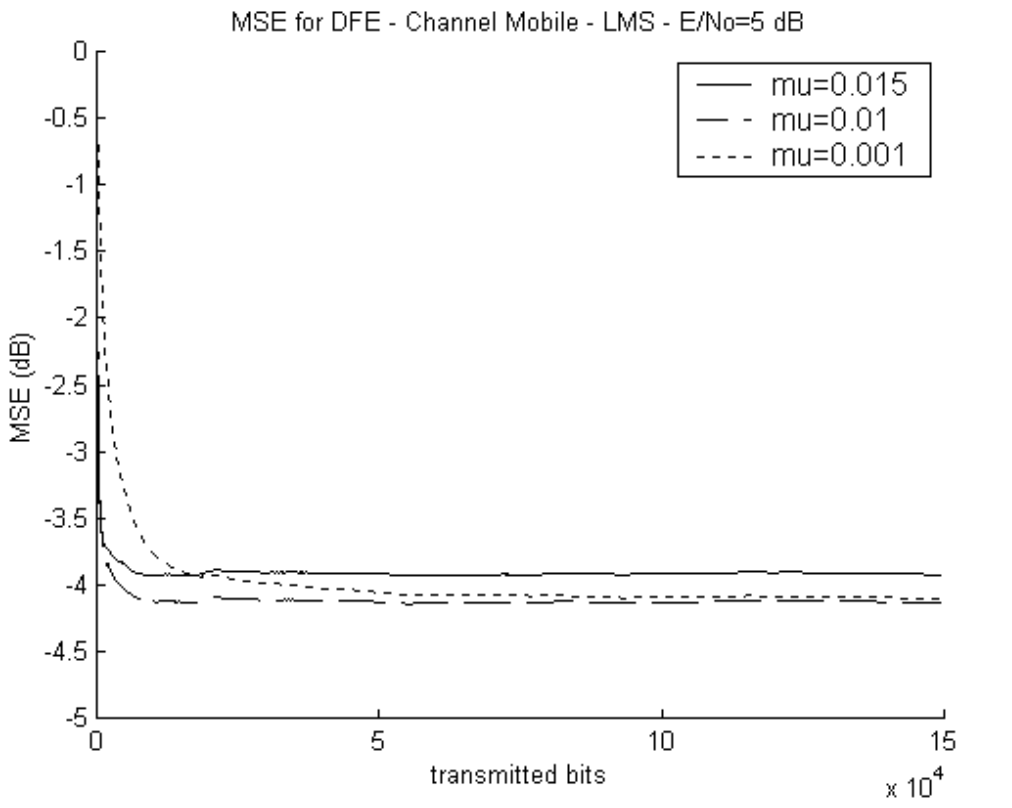

**Figure 3.62**: MSE for LE using LMS with different $\mu$ in Channel time-varying. E/No=5 dB.

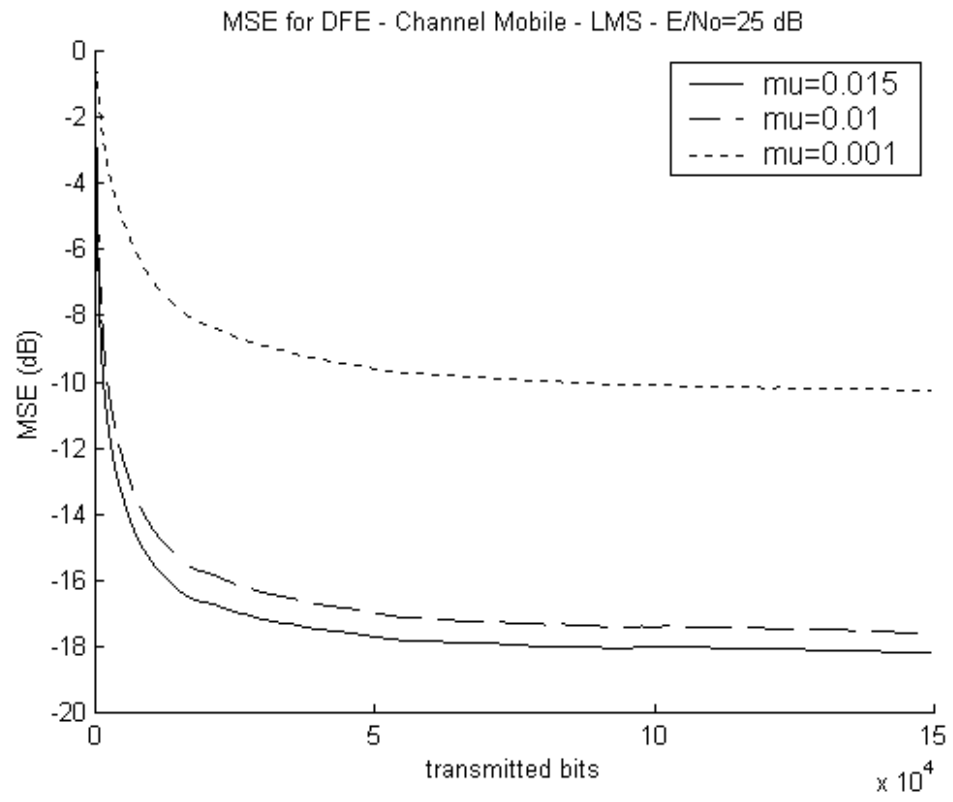

**Figure 3.63**: MSE for LE using LMS with different $\mu$ in Channel time-varying. E/No=25 dB.

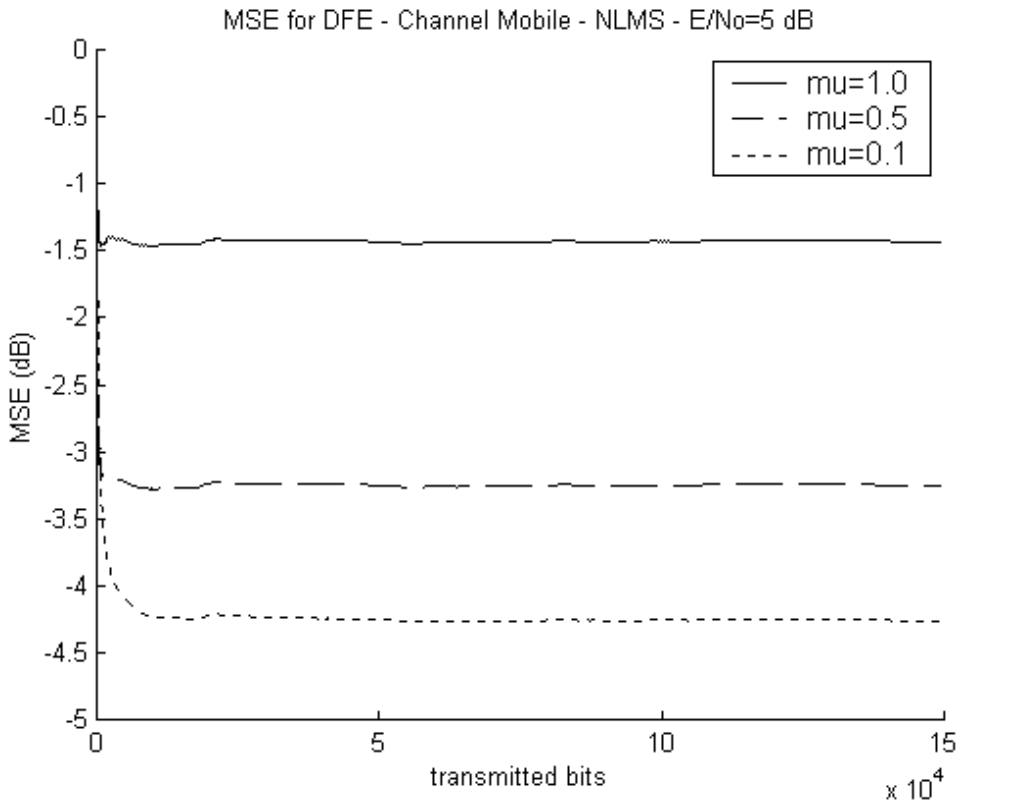

**Figure 3.64**: MSE for DFE using NLMS with different $\tilde{\mu}$ in Channel time-varying. E/No=5 dB.

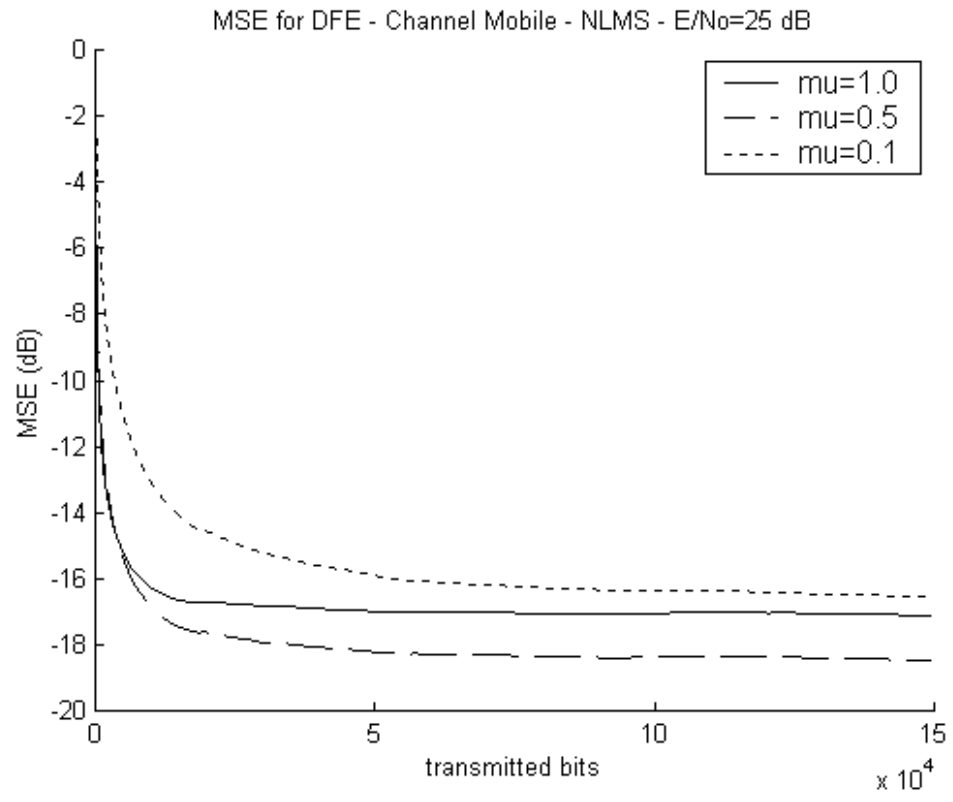

**Figure 3.65**: MSE for DFE using NLMS with different $\tilde{\mu}$ in Channel time-varying. E/No=25 dB.

In figure 3.66 it can be seen that the steady state MSE achieved by the VSLMS algorithm when E/No=25 dB is somewhat larger (around 2 dB) than when using the LMS or NLMS algorithm. Despite this slight, albeit significant, increase in the SS-MSE, the VSLMS offers the advantage of being able to work, unlike the other LMS-like algorithms, with the same parameters in any E/No level.

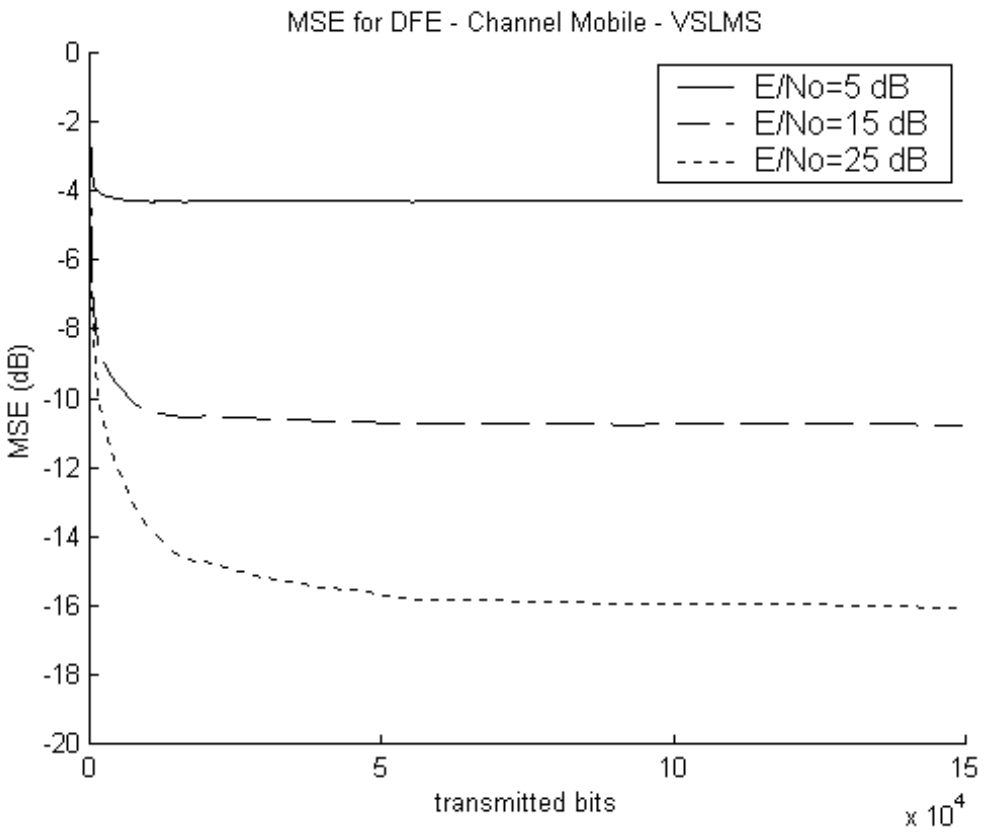

**Figure 3.66**: MSE for LE using VSLMS (ρ=0.0005,a=0.97) in Channel time-varying.

The performance of the RLS algorithm for three different forgetting factors is shown in figures 3.67 and 3.68. When E/No=25 dB, the same phenomenon that was found using the LE is again obvious here: if the forgetting factor is chosen too close to 1, the MSE increases because of its inability to track the channel variations.

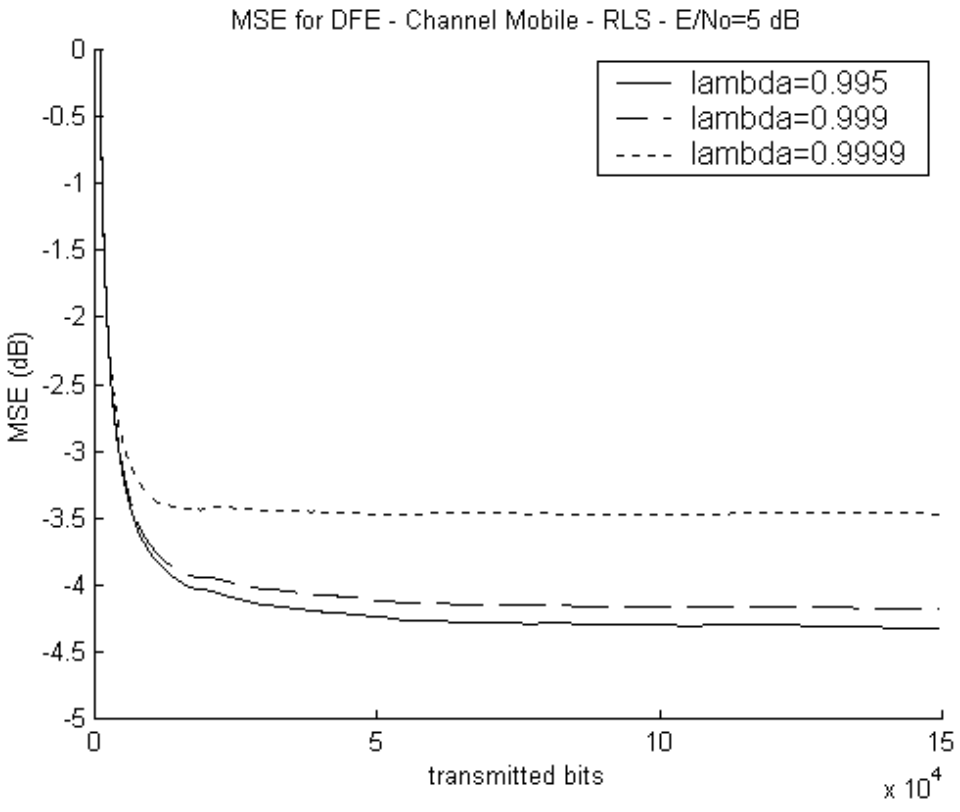

**Figure 3.67**: MSE for DFE using RLS with different λ in Channel time-varying. E/No=5 dB.

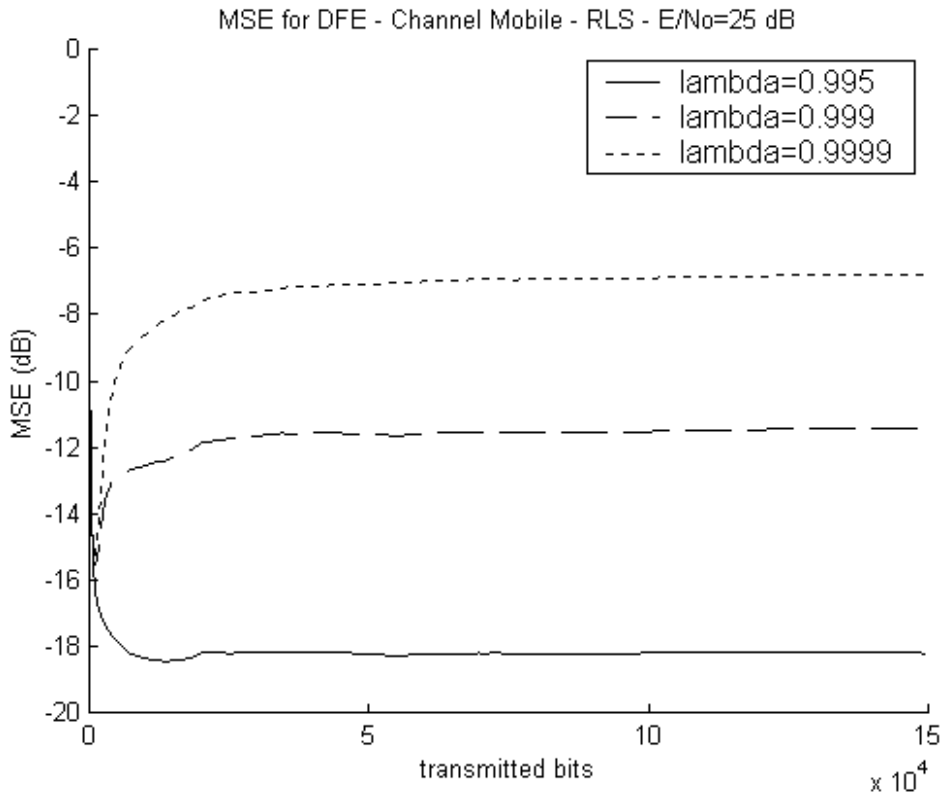

**Figure 3.68**: MSE for DFE using RLS with different λ in Channel time-varying. E/No=25 dB.

The BER confirms the superiority of the DFE over the LE by reducing the probability of BER up to two orders of magnitude with respect to a LE operating at the same E/No level.

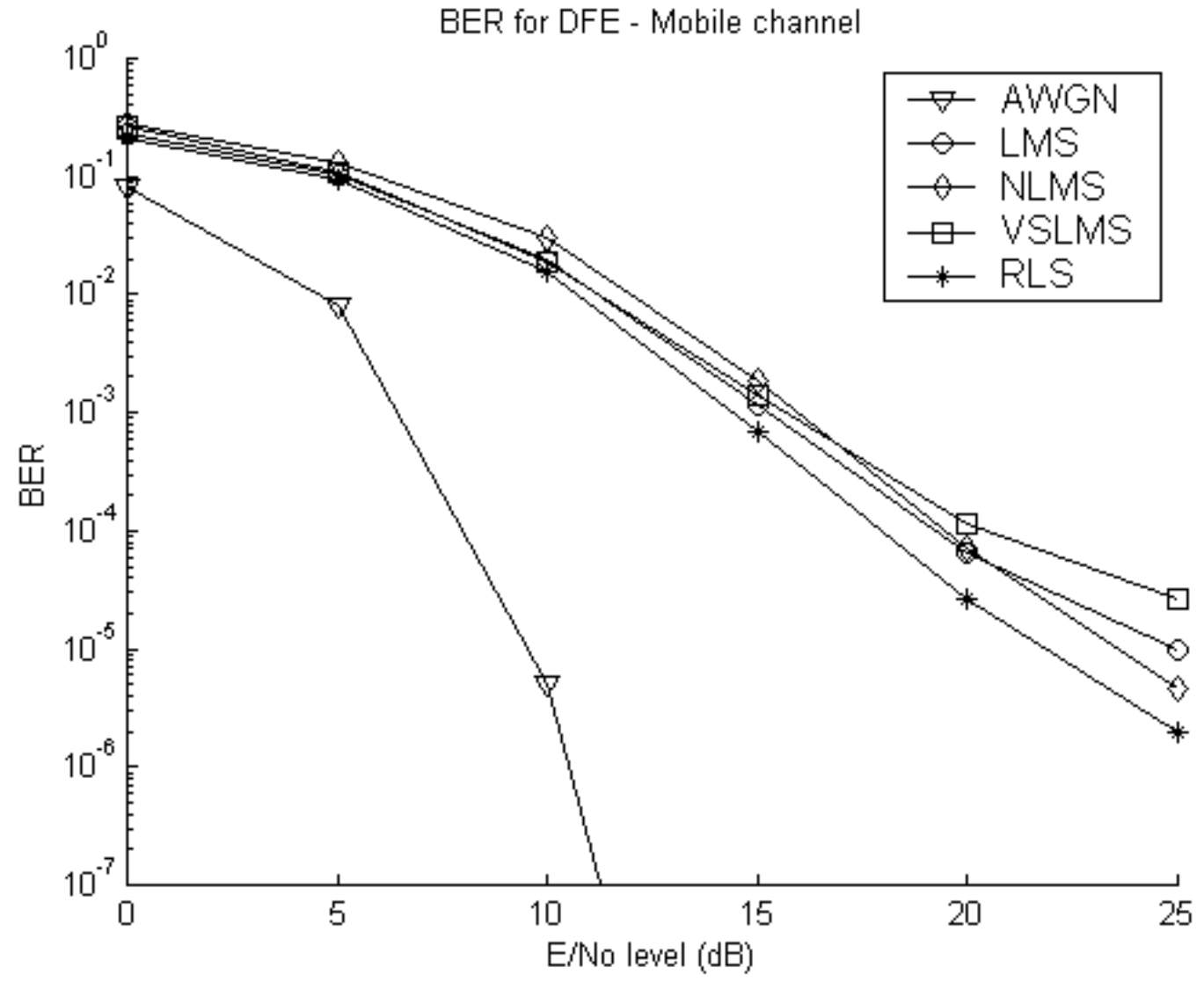

**Figure 3.69:** BER achieved with different algorithms equalising the time-varying model (multipath Rayleigh fading). Simulation length=100,000 samples (excluding the 50,000 samples from initial convergence).

There is one point to clarify regarding figure 3.69 if its results are compared with the ones obtained for the static Channel model 2. It may seem that the BER for the time-varying channel is lower than that of the static model whose profile is very similar. It should be recalled that the BER computed for the time-varying profile excludes the initial 50,000 iterations, as the objective is to measure the tracking capabilities of the different algorithms. In the static profiles, the BER is computed using also the initial 50,000 iterations where most of the errors take place, raising in this way the BER.

## 3.5.4 Effect of the decision delay in the DFE

As in linear equalisation, a given decision delay must be chosen between the sample being detected and the currently transmitted symbol. Unlike in the LE situation, for the DFE there are some precise rules on how to select the delay in order to attain the lowest MSE level. In [Al-Dhahair96] and [Voois96] it is shown how in minimum phase channels, the optimum delay is given by $N_f$-1, where $N_f$ is the length of the feedforward filter. When the channel is not minimum phase, as is often the case, an exhaustive search for the optimum delay must be performed. [Al-Dhahair96] shows a computationally efficient way of performing this exhaustive search of the optimum delay. This method is feasible in static or quasi-static environments, but in the case of time-varying channels, where the optimum delay is likely to vary very often, the complexity of this process will become prohibitive.

In the simulations shown so far and the ones to be presented in future chapters, the decision delay has been set to $N_f$-1. This choice is based on the fact that although the channel may not be minimum phase, the main paths will still be located at the beginning of the channel impulse response (see COST207 models in pp. 728 [Steele99]). This implies that the channel will be close to minimum phase and the chosen decision delay will also be close to its optimum value.

# 3.6 Conclusions from the background results

The objective of the simulations presented in this chapter was to give a general idea of the performance of the different algorithms and structures in different scenarios. For detailed comparisons and analysis, the reader is referred to the references given in this and the previous chapter.

Some conclusions can be drawn from these results. Regarding the algorithms, the RLS has proven to be superior to the gradient algorithms when operating in high E/No conditions. However, this performance benefit comes at the cost of a much larger computational complexity, especially when the equaliser has many taps. Within the gradient schemes, the enhanced LMS versions, namely, NLMS and VSLMS, have been shown to offer some significant advantages with respect to the plain LMS algorithm with a modest increase in complexity. The VSLMS algorithm, in certain scenarios, has been found not to converge as quickly as the NLMS algorithm with an appropriate step size. Still it provides a very important advantage over the NLMS and LMS algorithms, which is the ability to work very well independently of the E/No level. This is a very important property if the receiver must operate under different levels of noise or interference.

In terms of complexity, between the RLS and gradient algorithms lies the SFAEST algorithm, which in the simulations presented, has been shown to offer a performance similar to the RLS algorithm but with a much lower computational complexity. However, it must be mentioned, that in some other simulations not shown here, the SFAEST has proved to be an algorithm difficult to tune. If its parameters are not chosen properly, it is likely that the algorithm will become unstable during execution. Additionally, it is important to realise that the simulations were executed using double precision floating point arithmetic. If fixed-point arithmetic is used, as is typically done in mobile terminals, instability problems will become more severe.

For the reasons mentioned above, the algorithms used in the rest of this work are:

- The LMS algorithm, as it is the most common adaptive scheme.
- The VSLMS algorithm, because it has proved to be a robust and enhanced version of the LMS algorithm with a minimal increase in complexity.
- The RLS algorithm, given its excellent convergence and tracking properties independently of the channel characteristics.

Concerning the equalising structures, both LE and DFE are widely used in commercial systems and therefore it is worth examining the potential reconfigurability of the two structures.

# 4 MSE ANALYSIS OF LINEAR EQUALISERS

The most common measure of the performance of a communication system is typically the BER. However, designing or analysing a system is often difficult to do by trying to minimise the BER, basically because it is a non-linear parameter. Alternative measures need to be found. In the case of adaptive filters, as mentioned in chapter 2, the mean squared error (MSE) is the usual cost function to be optimised.

MSE analysis of many different adaptive filter algorithms has been addressed by several authors in the past, two of the most important studies being [Widrow76] and [Eleftheriou86]. [Widrow76] presents a steady state MSE expression when using the LMS to drive an FIR adaptive filter in a time varying scenario. In [Eleftheriou86] the counterpart RLS MSE equation is derived.

Both studies (and many others) configure the adaptive filter as an estimator of an unknown and time varying channel. In this chapter, derivations of expressions for the MSE are presented when the FIR adaptive filter is configured as a channel inverter (i.e. linear equaliser). Channel equalisation is a harder problem to analyse than channel estimation due to the correlation of the input data at the input of the adaptive filter. By no means do we claim that the analysis presented herein is completely accurate. As will be seen, numerous approximations and assumptions need to be made. Nonetheless, the resulting expressions are valuable because they show the different parameters influencing the steady state MSE and, in particular, the effect of the equaliser length. The derivations presented here follow closely those in [Widrow76] and [Eleftheriou86], but some key observations allow the results to be applied to the equalisation scenario. The first section of this chapter introduces the system model used for the analysis. Section 4.2 establishes some assumptions about the model, which simplify the mathematical development. Sections 4.3 and 4.4 present the derivations of the MSE expressions for the LMS and RLS respectively. Some simulation results validating the final equations are also shown. The chapter concludes showing which are the limitations, in terms of MSE accuracy, of the resulting equations and discusses their relevance. It should be emphasised that in this chapter the focus lies on linear equalisation. Non-linear structures such as decision feedback equalisers are covered in a later chapter.

# 4.1 Analytical system model for LE

Figure 4.1 shows the system model used in this analysis. The use of this very basic model eases in great measure the derivation of the MSE expressions derived in later sections. The system is considered to be in a fully digitised form: x(n) corresponds to the value of the continuous time signal x(t) at instant t=nT, where T is the sampling period and perfect sampling synchronisation is assumed. This notation allows the equations to be written in a more compact form.

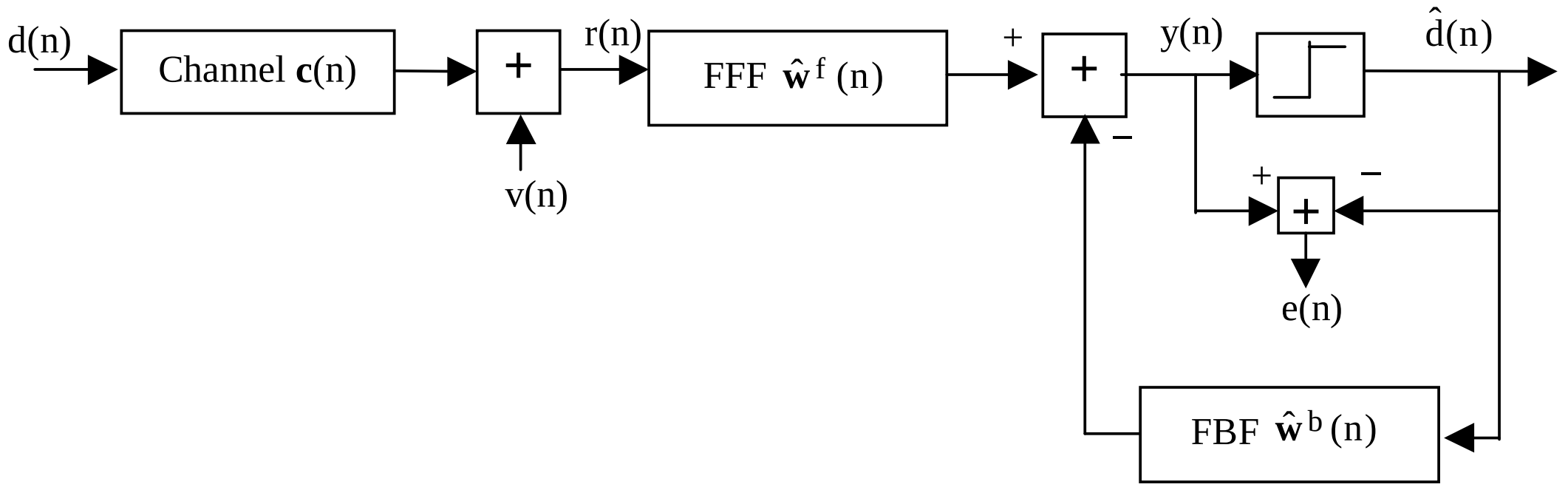

**Figure 4.1**: System model.

The variables have the following meaning:

d(n) = Transmitted data symbol

$\mathbf{c}(n) = [c_o(n)\ \ c_1(n)\ \ c_2(n)\ \ \ldots.\ \ c_{N-1}(n)]$ = N-tap Channel impulse response

$\hat{\mathbf{w}}(n) = [w_0(n)\ \ w_1(n)\ \ w_2(n)\ \ \ldots.\ \ w_{M-1}(n)]$ = M-tap Equaliser impulse response

$v(n)$ = Noise sample

$u(n)$ = Input signal to the equaliser

$y(n)$ = Output of the equaliser

$\hat{d}(n)$ = Estimated data symbol

$e(n)$= Error signal between equalised and detected symbol.

The data symbols, d(n), are drawn from an independent and identically distributed[13] (uniform) process with zero mean and variance $\sigma_d^2$. The noise samples, v(n), correspond to additive white Gaussian noise (AWGN) with zero mean and variance $\sigma_v^2$.

The communication channel is modelled as an N-tap FIR filter, $\mathbf{c}(n)$, whose coefficients represent the different multipath components arriving at the receiver. In realistic scenarios the channel coefficients vary with time. In the particular case of wireless mobile channels, the variations of the coefficient amplitudes follow a Rayleigh or Rice distribution [Rappaport96]. The linear equaliser consists of an M-tap FIR filter, $\hat{\mathbf{w}}(n)$, followed by a threshold detector that produces an estimate, $\hat{d}(n)$, of the original transmitted symbol.

In a real implementation of a digital transceiver, a pulse shaping filter would be included in the transmitter in order to limit the bandwidth of the signal. At the receiver, a filter matched to the transmit filter would maximise the signal to noise ratio at the sampling instant. In order to simplify the analysis, both filters are assumed to be included in the channel ([Proakis96]), that is, the channel represents the convolution of the transmitter filter, the physical channel and the matched filter.

## 4.2 Equalisation vs channel estimation

As has been mentioned in the introduction, the analysis of adaptive algorithms is usually done for the case of channel estimators. It is important to point out what are the differences that make channel estimation a simpler situation. Fig. 4.2 shows the structure of a channel estimator.

---

[13] This implies that the statistical parameters of the source are stationary.

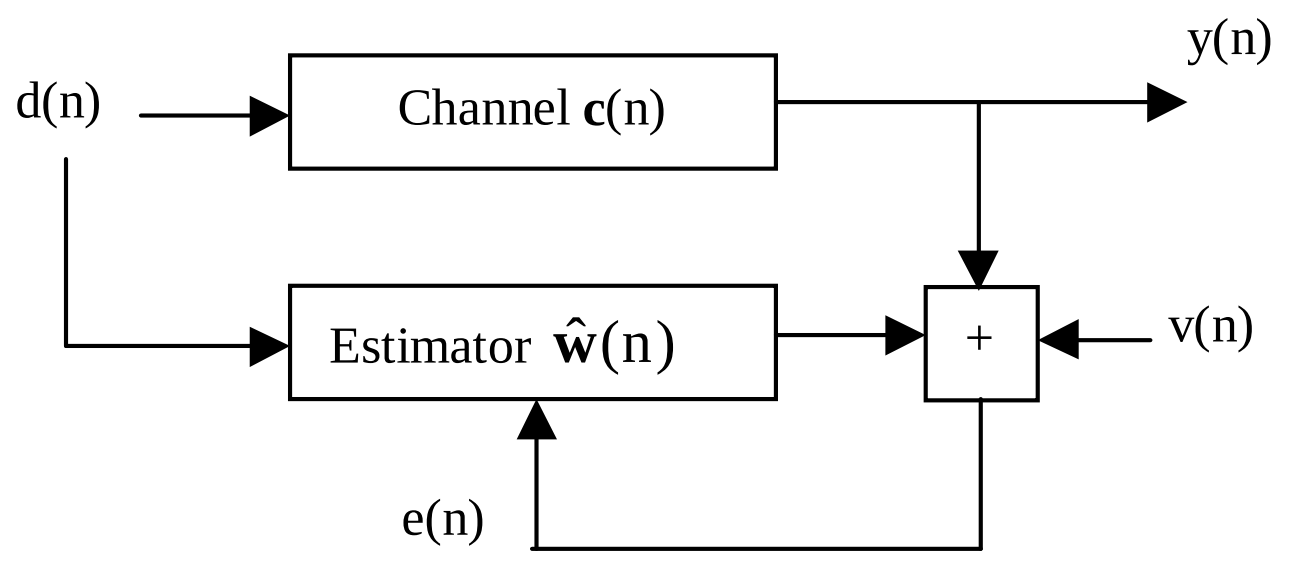

**Figure 4.2**: Channel estimator.

The main difference between the system in Fig. 4.1 and the one in Fig. 4.2 lies in the input to the adaptive filter. Assuming that in both cases data, d(n), are drawn from statistically equivalent sources, in the case of the channel estimator d(n) enter the adaptive filter without any distortion. In the equaliser configuration, d(n) are linearly distorted by the channel before entering the filter, therefore the input samples to the equaliser are no longer uncorrelated. Later it will be shown in detail why the correlation of the filter inputs poses some additional problems to the analysis of the equaliser performance.

It is also important to notice that the problem of channel estimation consists of modelling an FIR filter with another FIR filter, typically both having the same length. The lowest achievable MSE level of the channel estimator is only bounded by the noise level $\sigma_v^2$. Linear channel equalisation, on the other hand, consists of modelling the inverse of an FIR (i.e. the channel) filter with another finite length FIR filter. Notice that a more general solution to the equalisation problem would be to use an IIR filter as an equaliser. Unfortunately, adaptive IIR filters have some inconveniences that make their use rather limited (see [Shynk89]). Perfect equalisation is generally not achievable with an FIR finite length equaliser and therefore the minimum MSE is affected, not only by the noise level, but also by the residual ISI left in the samples.

# 4.3 LE model assumptions

The simple, albeit realistic, model described in section 4.1 still poses some difficulties to the derivation of expressions describing the MSE performance achieved by the equaliser. Some hypotheses need to be established to facilitate this analysis; in particular, four assumptions are stated concerning:

- Channel model
- Channel power response
- Optimum equaliser behaviour
- Independence assumption

The following subsections describe and justify each of these important assumptions.

## 4.3.1 Channel model

Realistic channel models such as the ones based on Rayleigh or Rice processes complicate the analysis to the point of making it impossible, therefore a more tractable model is needed. A commonly used channel model ([Eweda94]) is the one based on a first order Markov model described recursively by the equation:

$$\mathbf{c}(n) = \mathbf{c}(n-1) + \mathbf{q}(n) \tag{4.1}$$

The vector **c**(n) corresponds to the channel coefficients. The vector **q**(n), called the process noise vector (PNV), is an N-tap vector where every $q_i(n)$ is an independent uniform random variable with zero mean and power $\sigma_q^2$. This array models the channel variations with time. Large values for $\sigma_q^2$ would model a fast Doppler channel while a low $\sigma_q^2$ would correspond to a near static channel.

The validity of this model has been verified by comparing the autocorrelation function of a tap evolution using the Markov model for different values of $\sigma_q^2$ with the one using a Rayleigh fading generator.

Figure 4.3 shows the autocorrelation function of a channel tap generated using the Markov model for four different $\sigma_q^2$ values (in the graphs $\sigma_q^2$ is denoted by PNV). Ten thousand samples of the channel tap were generated using equation (4.1) iteratively and then the autocorrelation function of the resulting sequence of tap values was computed. Increasing $\sigma_q^2$ makes the channel coefficients less correlated in time, in other words, the channel varies more rapidly.

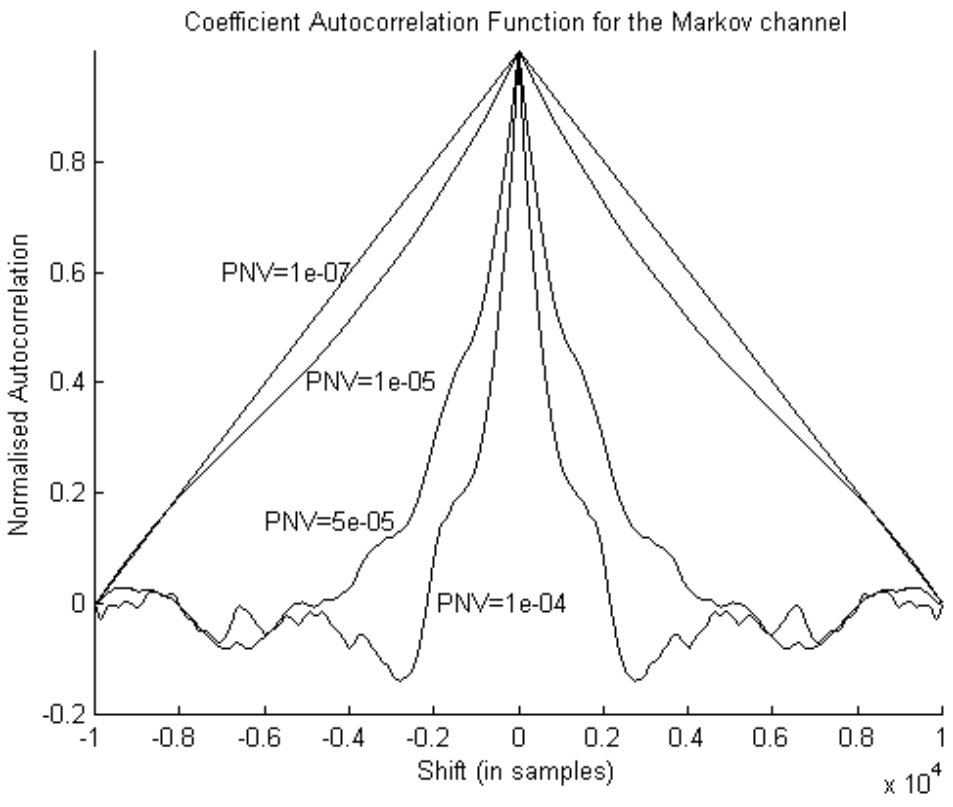

**Figure 4.3**: Tap autocorrelation using the Markov model.

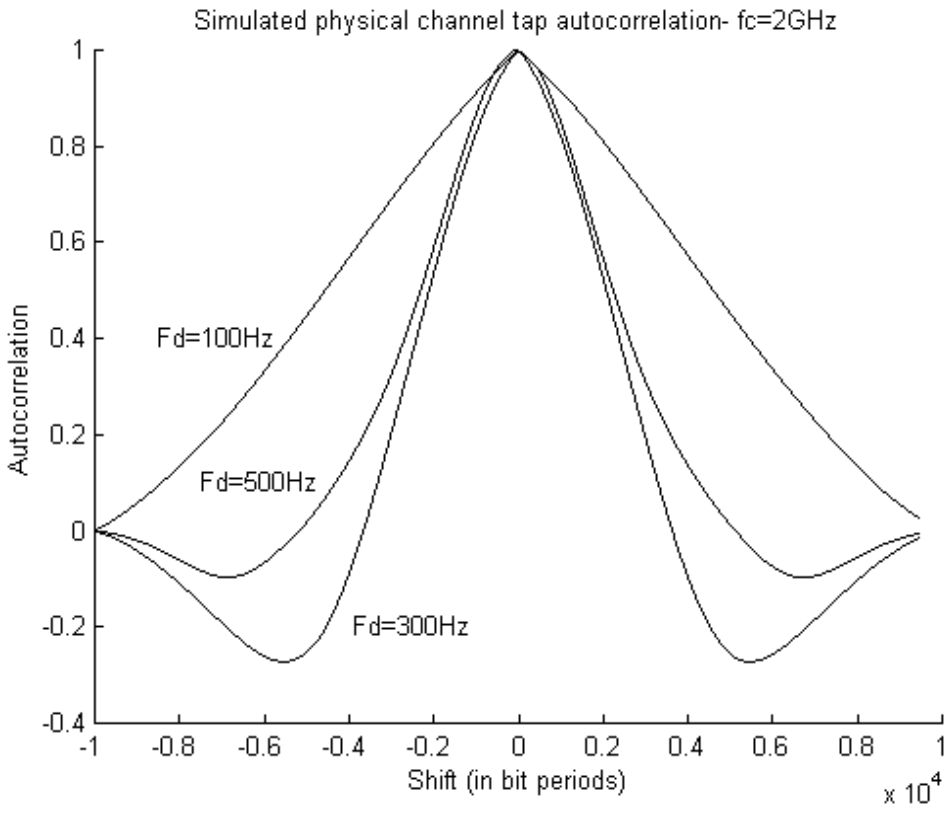

**Figure 4.4**: Tap autocorrelation using the Sum of Sines model.

In order to contrast the Markov model with a realistic wireless mobile channel, a multipath fading channel with a carrier frequency of 2 GHz and a sampling rate of 3.84 Msamples/s (UMTS parameters) has been simulated using the Sum of Sines method ([Jakes74]). Figure 4.4 shows the autocorrelation function, for three different Doppler spreads ($F_d$), of one of the channel taps. As with the Markov model, ten thousand tap channel samples were used to compute the autocorrelation function. It can be observed now that as the Doppler spread increases, the channel fades more rapidly. Comparing figures 4.3 and 4.4 it can be concluded that the Markov model, using reasonable values for the process noise vector ($\sigma_q^2 < 10^{-4}$), serves as a rough approximation of the fading phenomena found in realistic scenarios. Direct relations between values of $\sigma_q^2$ and $F_d$ are impossible to be established. However, it can be inferred from figures 4.3 and 4.4 that PNV values like $\sigma_q^2 = 10^{-4}$ would serve to model very large Doppler spreads (300-500 Hz) and $\sigma_q^2 = 10^{-6}$ would model small Doppler spreads (<100Hz).

## 4.3.2 Channel power response

It is assumed that the channel has unit power response therefore the channel coefficients, $c_i(n)$, always satisfy:

$$\sum_{i=0}^{N-1} c_i^2(n) = 1 \qquad (4.2)$$

The normalised power response of the channel implies that the received signal power, $\sigma_u^2$, is constant and equal to $\sigma_d^2 + \sigma_v^2$. Although (4.2) may seem a very restrictive constraint, it is usually satisfied in real systems by means of the power control mechanisms established between the transmitter and the receiver. Power control mechanisms are normally implemented in cellular systems to control interference among nearby users. In the case of CDMA systems, this problem is normally known as the near-far effect.

### 4.3.3 Optimum equaliser behaviour

Recall from section 2.2.1 that given a static channel impulse response, **c**, it is possible to compute the optimum MSE linear equaliser **w** by solving the Wiener-Hopf equation. Alternatively, the optimum equaliser can also be approximated using one of the recursive algorithms available that minimise the MSE. In many realistic scenarios like those found in nearly all wireless systems (cellular, fixed wireless or WLANs), the channel impulse response will be time varying, and in the particular context of this chapter, its time evolution follows (4.1). As a consequence, the optimum linear equaliser, **w**(n), will also vary with time. This optimum equaliser produces at each instant the minimum error, $e_o(n)$:

$$e_o(n)=d(n)-\mathbf{w}^T(n)\mathbf{u}(n) \qquad (4.3)$$

It is assumed that the fluctuations of the optimum equaliser follow:

$$\mathbf{w}(n)=\mathbf{w}(n-1)+\mathbf{p}(n) \qquad (4.4)$$

where **p**(n) is an M-tap vector with every $p_i(n)$ having zero mean and power $\sigma_p^2$. This assumption is used in [Rupp97b] for the analysis of equalisers in flat Rayleigh channels.

In our context, it is further assumed that **p**(n) and **q**(n) are related by:

$$\|\mathbf{q}(n)\|_2 \approx \|\mathbf{p}(n)\|_2 \qquad (4.5)$$

where $\|\mathbf{x}\|_2$ denotes the 2-norm of **x** and the symbol $\approx$ is used to denote similar magnitude. In fact during the derivation the symbol $\approx$ will be further approximated as an equality symbol assuming both quantities have the same magnitude. Equation (4.5) is equivalent to saying that large changes in the channel will produce large changes in the optimum equaliser and small channel changes will barely perturb the optimum equaliser settings. It is important to note that **q**(n) and **p**(n), albeit having equal power, have different lengths (N and M respectively). This allows the equaliser in the model to have a different number of taps (typically more) from the channel filter. Some simulation results are now presented that justify the use of (4.5) but also show the limitations of this model.

Fifty consecutive samples of different channels, whose initial static profiles were described in section 3.3, were generated using (4.1). For each channel sample, the optimum equaliser was computed and the norm of the two update vectors (**q**(n) and **p**(n)) was calculated. Figures 4.5 to 4.8 show the evolution of the $\|\mathbf{q}(n)\|_2$ and $\|\mathbf{p}(n)\|_2$ over the 50 samples.

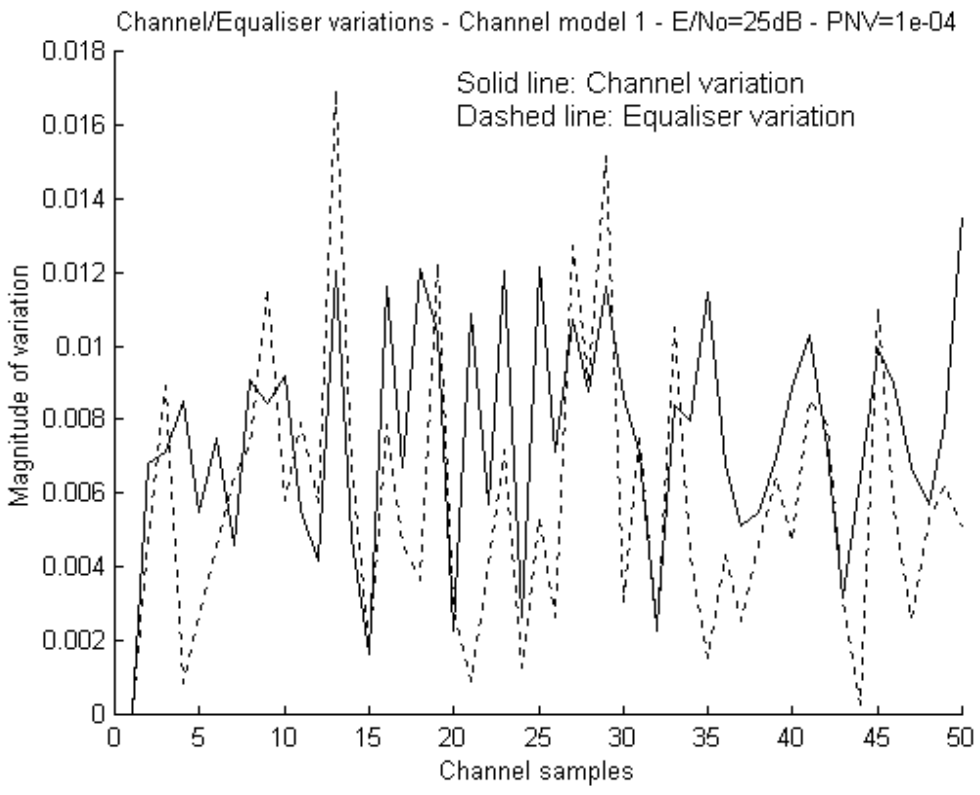

**Figure 4.5:** Comparison of the norm of the channel and equaliser variations. Channel model 1. E/No=25dB. $\sigma_q^2 = 10^{-4}$

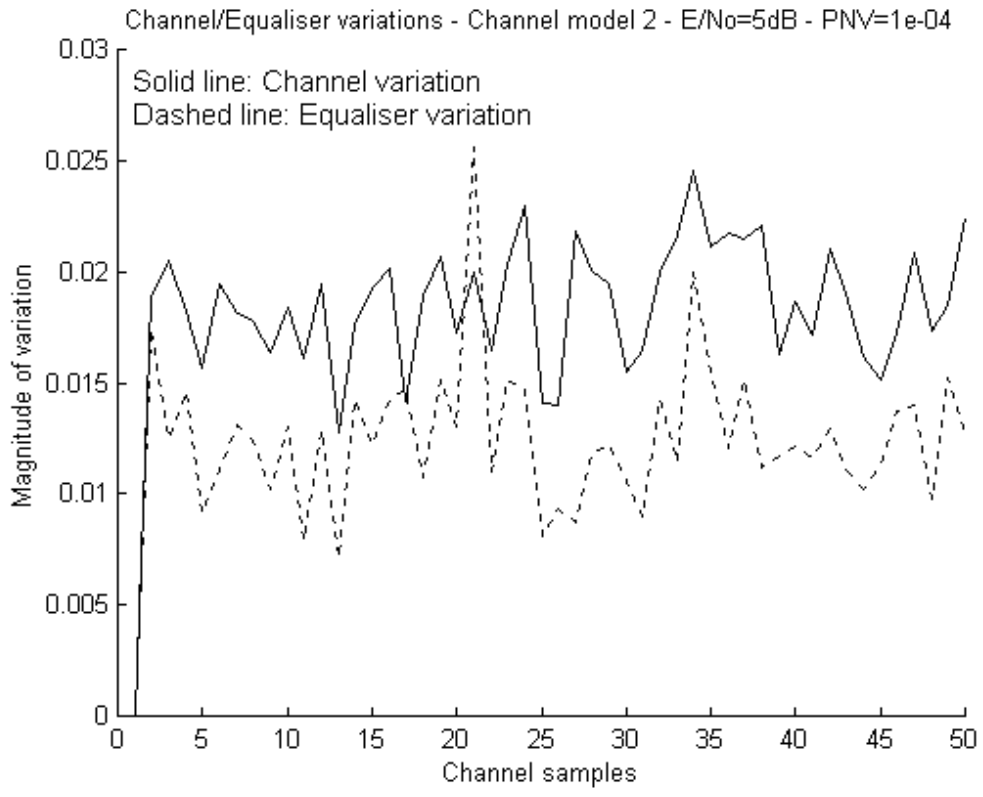

**Figure 4.6:** Comparison of the norm of the channel and equaliser variations. Channel model 2. E/No=5dB. $\sigma_q^2 = 10^{-4}$

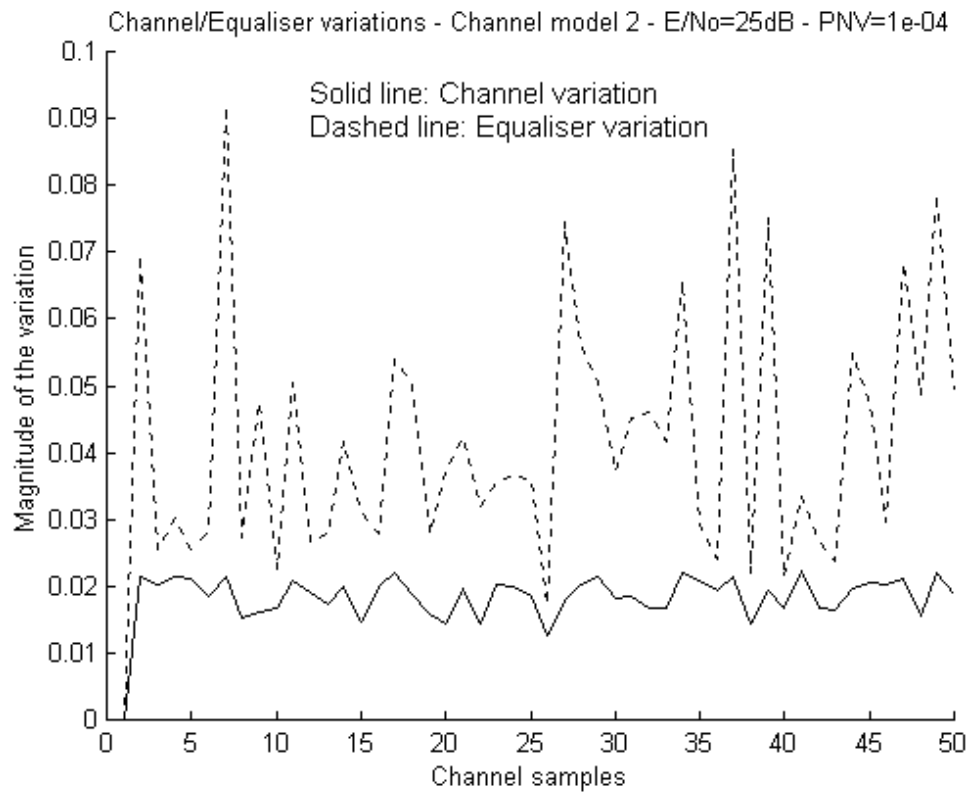

**Figure 4.7:** Comparison of the norm of the channel and equaliser variations. Channel model 2. E/No=25dB. $\sigma_q^2 = 10^{-4}$.

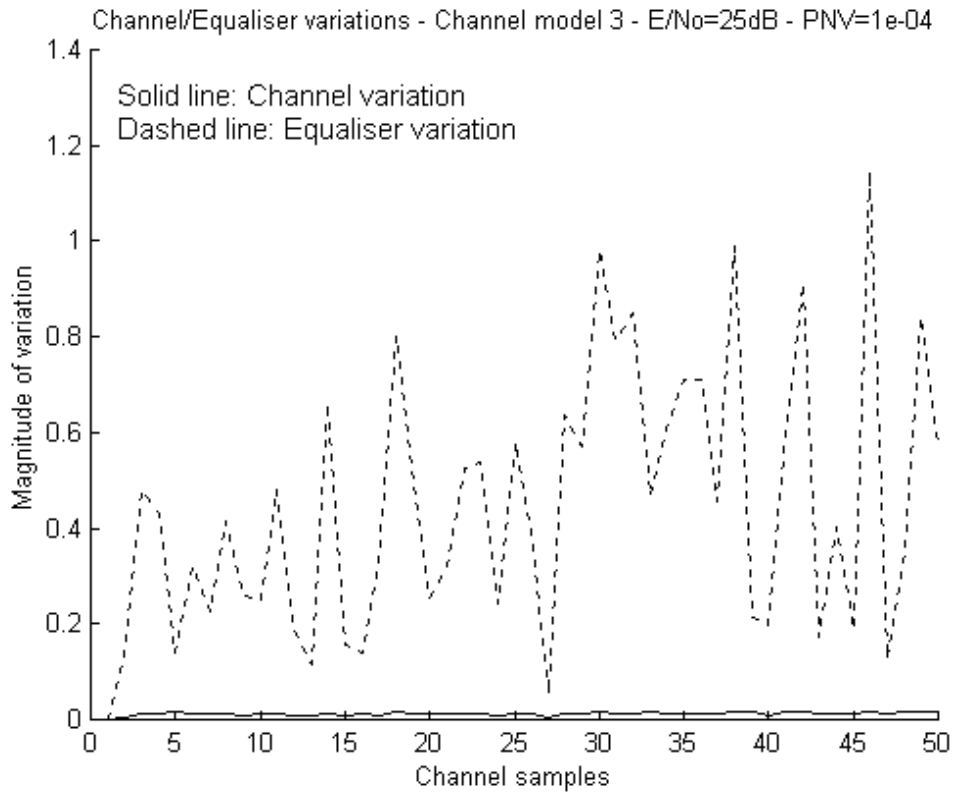

**Figure 4.8:** Comparison of the norm of the channel and equaliser variations. Channel model 3. E/No=25dB. $\sigma_q^2 = 10^{-4}$.

Figures 4.5 and 4.6 show how equation (4.5) describes quite accurately the variations of the optimum equaliser ($\|\mathbf{p}(n)\|_2$, dashed line) with respect to the channel variations ($\|\mathbf{q}(n)\|_2$, solid line). Figure 4.7 illustrates a situation where equation (4.5) holds only in a loose way. Note how changes in the channel cause larger variations in the optimum equaliser, however they are still within an order of magnitude of the channel variations. Figure 4.8 depicts an extreme situation that completely invalidates equation (4.5) as small changes in the channel provoke radical variations in the equaliser coefficients.

The channel spectrum is the factor responsible for the different degrees of accuracy of hypothesis. As pointed out in chapter 2, the eigenvalue spread (ES) of the autocorrelation matrix of the input data (**R**(n)) is controlled by the channel spectrum. Large spectral variations will cause a proportionally large ES. An important result of Perturbation Theory ([Noble88]) states that a matrix will be ill-conditioned if the ES of the matrix is large. Ill-conditioning of a matrix describing the coefficients to a set of linear equations (as is the case of **R**(n) in the Wiener-Hopf equation) means that small perturbations to any of the matrix coefficients will to lead to a completely different solution with respect to the solution before the perturbation.

Figures 4.5 and 4.6 correspond to fairly flat channel spectrums (see figures 3.3 and 3.5), therefore small variations in the channel coefficients do not affect significantly the value of the corresponding optimum equaliser, making equation (4.5) a reasonable assumption. On the other hand, figure 4.7 corresponds to a channel with a spectral null in the passband (see figure 3.7) which causes a large ES in the input autocorrelation matrix. As a consequence, even the slightest variation in the channel coefficients provokes a radical change of the values of the optimum equaliser coefficients, thus invalidating the assumption.

For most of the time[14], the channel behaviour in a realistic wireless scenario will be closer to profiles 1 and 2 (figures 3.2 and 3.4) than to profile 3 making equation (4.5) a sensible approximation. Note also that if a system is known to operate permanently in channels like Channel model 3, linear equalisation is not an adequate technique to mitigate distortion as LE cannot compensate for spectral nulls.

[14] As can be inferred from the channel profiles in the COST207 models from the ITU-T.

### 4.3.4 Independence assumption

The independence assumption is an hypothesis often used in adaptive signal processing to facilitate mathematical derivations that otherwise would be extremely complicated or impossible. It states that the current inputs to the adaptive algorithm, **u**(n) and d(n), are independent of the past sets of inputs: **u**(n-1), **u**(n-2), **u**(n-3), … and d(n-1), d(n-2), d(n-3),… .The degree of accuracy of this assumption depends very much on the application of the adaptive filter. For example, in the case of spatial beamformers, the assumption models reality very well as successive input vectors correspond to distinct snapshots of data acquired at different instants. In the context of temporal filters such as equalisers or channel estimators, it is easy to see that the inputs to the filter (samples in the tap delay line) at successive instants are dependent on each other. The input vector at time n is a shifted version of the one at time n-1 so **u**(n) and **u**(n-1) have M-1 terms in common.

In spite of this dependence, previous analysis of channel estimators and other systems based on adaptive filters using the independence assumption matches accurately with the experimental results. The reason for this apparent contradiction is that the independence assumption is only used at certain points in the development where there is no other way to progress in the analysis. However, if the analysis can continue without making use of the independence assumption, the data dependence is preserved. It turns out that this strategy retains most of the influence of the dependence of the successive input vectors, making the analysis fairly accurate when confronted with the measured results. A detailed explanation of the independence assumption can be found in [Widrow76] and [Mazo79].

## 4.4 Steady state LMS-MSE analysis for the LE

In this section a detailed analysis of a recursive linear equaliser using the least mean square (LMS) algorithm is presented. An analytical expression for the steady state MSE is derived. The steady state is far more significant than the transient, as under normal circumstances the transient is just a small fraction of the total time of the operation of the equaliser. The transient stage will be briefly considered in section 4.9. Some simulation results show the agreement between the predicted values and experiments.

The derivation of an MSE expression starts by stating the LMS equations:

$$e(n) = d(n) - \mathbf{u}^T(n)\hat{\mathbf{w}}(n) \qquad (4.6)$$

$$\hat{\mathbf{w}}(n+1) = \hat{\mathbf{w}}(n) + \mu\mathbf{u}(n)e(n) \qquad (4.7)$$

$\hat{\mathbf{w}}(n)$ is defined as the estimated equaliser coefficients while $\mathbf{w}(n)$ is used to denote the optimum equaliser value at instant *n*. From these two definitions it is possible now to define the error vector as:

$$\mathbf{t}(n) = \hat{\mathbf{w}}(n) - \mathbf{w}(n) \qquad (4.8)$$

Using the definition of the error vector, (4.6) can be rewritten as

$$e(n) = \mathbf{u}^T(n)[\mathbf{w}(n) + \mathbf{t}(n)] = e_o(n) - \mathbf{u}^T(n)\mathbf{t}(n) \qquad (4.9)$$

where $e_o(n)$ is the minimum error signal, produced when the equaliser uses the optimum coefficients. Recalling the definition of the mean squared error introduced in previous chapters, MSE(n)=E[|e(n)|$^2$], it is possible now to reformulate this definition using (4.9):

$$\begin{aligned} MSE(n) &= E[|e(n)|^2] = E[(e_o(n) - \mathbf{t}^T(n)\mathbf{u}(n))(e_o(n) - \mathbf{u}^T(n)\mathbf{t}(n))] \\ &= E[|e_o(n)|^2] + E[\mathbf{t}^T(n)\mathbf{u}(n)\mathbf{u}^T(n)\mathbf{t}(n)] - 2E[e_o(n)\mathbf{u}^T(n)\mathbf{t}(n)] \end{aligned} \qquad (4.10)$$

The first term in (4.10) corresponds to the minimum mean squared error, MMSE(n), which, in the case of equalisation, varies with time as the channel varies. The last term in (4.10) will cancel out as a result of the orthogonality between the input vector and the minimum error signal ([Haykin96]). Therefore the MSE evolution can be expressed as the sum of two components. One is due to the MMSE and the other due to what is usually called the excess mean squared error (EMSE):

$$MSE(n) = MMSE(n) + E[\mathbf{t}^T(n)\mathbf{u}(n)\mathbf{u}^T(n)\mathbf{t}(n)] = MMSE(n) + EMSE(n) \quad (4.11)$$

The MMSE(n) solely depends on the channel conditions and on the number of taps of the equaliser. The second term, the EMSE(n), is directly related to the algorithm being used, in this case the LMS. As the equaliser coefficients converge, the amount of EMSE reduces. However in the case of the LMS algorithm it will not disappear completely even when the equaliser has fully converged.

Making use of the independence assumption it is possible to refine the expression for the EMSE:

$$\begin{aligned} EMSE(n) &= E[\mathbf{t}^T(n)\mathbf{u}(n)\mathbf{u}^T(n)\mathbf{t}(n)] = E[Tr[\mathbf{t}^T(n)\mathbf{u}(n)\mathbf{u}^T(n)\mathbf{t}(n)]] \\ &= E[Tr[\mathbf{u}(n)\mathbf{u}^T(n)\mathbf{t}(n)\mathbf{t}^T(n)]] = Tr[E[\mathbf{u}(n)\mathbf{u}^T(n)]E[\mathbf{t}(n)\mathbf{t}^T(n)] \\ &= Tr[\mathbf{R}(n)\mathbf{K}(n)] \end{aligned} \qquad (4.12)$$

where $\mathbf{R}$(n) and $\mathbf{K}$(n) are the autocorrelation matrices of the input ($\mathbf{u}$(n)) and error ($\mathbf{t}$(n)) vectors respectively, and Tr[$\mathbf{X}$] denotes the trace of matrix $\mathbf{X}$. Equation (4.12) will be useful at a later point in this derivation.

Using (4.9) it is possible to write the equaliser coefficients update (4.7) in another form:

$$\begin{aligned}\hat{\mathbf{w}}(n+1) &= \hat{\mathbf{w}}(n)+\mu\,\mathbf{u}(n)[e_o(n)-\mathbf{u}^T(n)\,\mathbf{t}(n)] \\ &= \hat{\mathbf{w}}(n)+\mu\,\mathbf{u}(n)e_o(n)-\mu\,\mathbf{u}(n)\hat{\mathbf{w}}^T(n)\mathbf{u}(n)+\mu\,\mathbf{u}(n)\mathbf{w}^T(n)\mathbf{u}(n)\end{aligned} \qquad (4.13)$$

Subtracting $\mathbf{w}$(n+1) from both sides of (4.13) and using the definitions of (4.4) and (4.8) the following expression is obtained:

$$\begin{aligned}\mathbf{t}(n+1) = \hat{\mathbf{w}}(n) + \mu\,\mathbf{u}(n)e_o(n) - \mu\,\mathbf{u}(n)\hat{\mathbf{w}}^T(n)\mathbf{u}(n) + \\ \mu\,\mathbf{u}(n)\mathbf{w}^T(n)\mathbf{u}(n) - \mathbf{w}(n) - \mathbf{p}(n+1)\end{aligned} \qquad (4.14)$$

All the common factors are now grouped together, resulting in:

$$\mathbf{t}(n+1) = [\mathbf{I}_M - \mu\,\mathbf{u}(n)\mathbf{u}^T(n)]\mathbf{t}(n) + \mu\,\mathbf{u}(n)e_o(n) - \mathbf{p}(n+1) \qquad (4.15)$$

where $\mathbf{I}_M$ denotes the MxM identity matrix. The assumption made in channel estimation ([Haykin96]) is also assumed here: it is supposed that under the realistic assumption of $\mu$ (step size) being small, $\mu\,\mathbf{u}(n)\mathbf{u}^T(n) \cong \mu\,E[\mathbf{u}(n)\mathbf{u}^T(n)] = \mu\mathbf{R}(n)$. This assumption is called the Direct Averaging approximation. This approximation allows us to write (4.15) as:

$$\mathbf{t}(n+1) = [\mathbf{I}_M - \mu\,\mathbf{R}(n)]\mathbf{t}(n) + \mu\,\mathbf{u}(n)e_o(n) - \mathbf{p}(n+1) \qquad (4.16)$$

In contrast with channel estimation, now the autocorrelation matrix of the input data is time varying, hence its time dependence.

Defining $\mathbf{K}$(n) as the autocorrelation matrix of the error vector $\mathbf{t}(n)$, it can be shown from (4.16):

$$\mathbf{K}(n+1) = [\mathbf{I}_M - \mu\,\mathbf{R}(n)]\mathbf{K}(n)[\mathbf{I}_M - \mu\,\mathbf{R}(n)] + \mu^2 E[|e_o(n)|^2]\mathbf{R}(n) + \mathbf{P} \qquad (4.17)$$

where $\mathbf{P}$ is defined as $\mathbf{P}$=E[$\mathbf{p}$(n+1)$\mathbf{p}^T$(n+1)]. Due to the fact that the statistics of the process noise vector are stationary, $\mathbf{P}$ does not depend on time. Realising that the factor $E[|e_o(n)|^2]$ corresponds to the time varying MMSE and resolving the first term in (4.17), the following expression is obtained:

$$\begin{aligned}\mathbf{K}(n+1) = \mathbf{K}(n) - \mu\,\mathbf{K}(n)\mathbf{R}(n) - \mu\,\mathbf{R}(n)\mathbf{K}(n) + \\ \mu^2\mathbf{R}(n)K(n)\mathbf{R}(n) + \mu^2 MMSE(n)\mathbf{R}(n) + \mathbf{P}\end{aligned} \qquad (4.18)$$

To simplify the mathematical treatment, the 4$^{th}$ term in (4.18) is discarded. This simplification is justified because the $\mu^2$ factor makes this term small compared with all the other terms containing **K**(n). In the steady state it is reasonable to assume that $\mathbf{K}(n) \cong \mathbf{K}(n+1)$. Taking all of this into account and rearranging, (4.18) becomes:

$$\mathbf{K}(n)\mathbf{R}(n) + \mathbf{R}(n)\mathbf{K}(n) = \mu \text{MMSE}(n)\mathbf{R}(n) + \frac{\mathbf{P}}{\mu} \tag{4.19}$$

Taking now the trace of both sides of (4.19):

$$\text{Tr}[\mathbf{K}(n)\mathbf{R}(n) + \mathbf{R}(n)\mathbf{K}(n)] = \text{Tr}[\mu\ \text{MMSE}(n)\mathbf{R}(n) + \frac{\mathbf{P}}{\mu}] \tag{4.20}$$

Now use is made of two results from matrix theory [Noble88]:

- Tr[**AB**]=Tr[**BA**]
- Tr[**A**+**B**]=Tr[**A**]+Tr[**B**]

where **A** and **B** are product compatible matrices. Recognising that according to (4.12) Tr[**R**(n)**K**(n)] corresponds to the EMSE(n) and making use of the above two algebraic results, equation (4.20) becomes:

$$\text{EMSE}(n) = \frac{\mu}{2}\text{MMSE}(n)\text{Tr}[\mathbf{R}(n)] + \frac{1}{2\mu}\text{Tr}[\mathbf{P}] \tag{4.21}$$

In order to move forward in the derivation, the matrices **R**(n) and **P** must be carefully examined in order to compute their traces. The starting point is the definition of **R**(n):

$$\mathbf{R}(n) = E[\mathbf{u}(n)\mathbf{u}^T(n)] =$$

$$= E\begin{bmatrix} u(n)u(n) & u(n)u(n-1) & \cdots & u(n)u(n-M+1 \\ u(n-1)u(n) & u(n-1)u(n-1) & \cdots & u(n-1)u(n-M+1) \\ \vdots & \vdots & \ddots & \vdots \\ u(n-M+1)u(n) & u(n-M+1)u(n-1) & \cdots & u(n-M+1)u(n-M+1) \end{bmatrix} \tag{4.22}$$

The form of the matrix **R**(n) in (4.22) is fairly different from the one that would arise in the channel estimation scenario. There **R**(n) would have a diagonal form as the inputs to the adaptive filter would be uncorrelated. It is also important to point out that if the channel was static **R**(n) would have Toeplitz structure[15] and the analysis would be much simpler. In general, this is not true if the channel is time varying.

Given that solely the trace of **R**(n) is needed, only the diagonal elements need to be calculated:

[15] A matrix is said to be Toeplitz when all its Northwest to Southeast diagonals have a constant value.

$$E[u(n-i)u(n-i)] = E\begin{bmatrix}[d(n-i)c_0(n-i) + d(n-i-1)c_1(n-i) \\ + \cdots + d(n-i-N+1)c_{N-1}(n-i) + v(n-i)]^2\end{bmatrix}$$
$$= E\begin{bmatrix}[d(n-i)c_0(n-i)]^2 + [d(n-i-1)c_1(n-i)]^2 + \\ \cdots + [d(n-i-N+1)c_{N-1}(n-i)]^2 + v^2(n-i)\end{bmatrix}$$
$$= \sigma_d^2 E[c_0^2(n-i) + c_1^2(n-i) + \cdots + c_{N-1}^2(n-i)] + \sigma_v^2 \quad 0 \le i < M \qquad (4.23)$$

In (4.23) advantage is taken of using the independence and zero mean of the data symbols to set to zero the expectation of the product of different symbols and also with the noise samples. Now the assumption of a unitary normalised channel power response is used. This assumption corresponds to the following expression:

$$c_0^2(n) + c_1^2(n) + \cdots + c_{N-1}^2(n) = 1 \qquad \forall n \qquad (4.24)$$

so that (4.23) becomes:

$$E[u_i(n)u_i(n)] = \sigma_d^2 + \sigma_v^2 \qquad 0 \le i \le M-1 \qquad (4.25)$$

Using (4.25) it is easy to observe now that

$$\mathrm{Tr}[\mathbf{R}(n)] = M[\sigma_d^2 + \sigma_v^2] \qquad (4.26)$$

It is important to stress that (4.25) and, indeed, the whole derivation from this point forward are possible due to the assumption of the normalised channel power response, otherwise it would not be possible to find a closed expression for the trace of **R**(n).

A similar procedure is followed for the matrix **P**:

$$\mathbf{P} = E[\mathbf{p}(n)\mathbf{p}^T(n)] = E\left[\begin{bmatrix} p_0(n) \\ p_1(n) \\ \vdots \\ p_{M-1}(n)\end{bmatrix}\begin{bmatrix} p_0(n) & p_1(n) & \cdots & p_{M-1}(n)\end{bmatrix}\right] =$$
$$= E\begin{bmatrix} p_0(n)p_0(n) & p_0(n)p_1(n) & \cdots & p_0(n)p_{M-1}(n) \\ p_1(n)p_0(n) & p_1(n)p_1(n) & \cdots & p_1(n)p_{M-1}(n) \\ \vdots & \vdots & \ddots & \vdots \\ p_{M-1}(n)p_0(n) & p_{M-1}(n)p_1(n) & \cdots & p_{M-1}(n)p_{M-1}(n)\end{bmatrix} \qquad (4.27)$$

Given that all $p_i(n)$ have zero mean and variance $\sigma_p^2$, computing the trace of **P** is straightforward:

$$\mathrm{Tr}[\mathbf{P}] = E[p_0^2(n) + p_1^2(n) + \cdots + p_{M-1}^2(n)] = M\sigma_p^2 \qquad (4.28)$$

It is known from equation (4.5) (assumption 1) that $p_0^2(n)+p_1^2(n)+\cdots+p_{M-1}^2(n)=q_0^2(n)+q_1^2(n)+\cdots+q_{N-1}^2(n)$ therefore (4.28) can also be written as:

$$\mathrm{Tr}[\mathbf{P}]=N\sigma_q^2 \tag{4.29}$$

Using (4.26) and (4.29) it is now possible to rewrite (4.21) as:

$$\mathrm{EMSE(n)}= M[\sigma_d^2+\sigma_v^2]\frac{\mu}{2}\mathrm{MMSE}(n)+\frac{N\sigma_q^2}{2\mu} \tag{4.30}$$

Finally, recognising that MSE(n)=MMSE(n)+EMSE(n), the final MSE expression for the LMS equaliser operating in the steady state can be found:

$$\mathrm{MSE(n)=MMSE(n)}+ M[\sigma_d^2+\sigma_v^2]\frac{\mu}{2}\mathrm{MMSE}(n)+\frac{N\sigma_q^2}{2\mu} \tag{4.31}$$

Equation (4.31) is the equaliser counterpart to the channel estimation equation in [Widrow76]. It can be observed that it has three terms. The first one corresponds to the MMSE, the middle term is due to the noise level and equaliser imperfection and the final term corresponds to the error due to tracking. From the form of (4.31) it can be easily seen that the predicted MSE will be a replica of the MMSE multiplied by a factor greater than one $(1+M[\sigma_d^2+\sigma_v^2]\frac{\mu}{2})$, with the added constant $\frac{N\sigma_q^2}{2\mu}$. Logically, the MSE(n) will always be larger than the MMSE(n).

Although (4.31) looks very similar to the LMS channel estimation equation, there are some significant differences. First, the equation depends on the instantaneous MMSE, which is related to the particular channel characteristics. Secondly, (4.31) depends on both the channel length, N, and the equaliser length M. Given a set of channel characteristics, this latter dependence on equaliser length provides the designer with an extra parameter, apart from the step size, to adjust the equaliser performance.

From equation (4.31) it is possible to compute the optimum step size that minimises the MSE:

$$\frac{\partial\,\mathrm{MSE(n)}}{\partial\mu}=0 \qquad \rightarrow \qquad \mu_{opt}(n)=\sqrt{\frac{N\sigma_q^2}{M[\sigma_d^2+\sigma_v^2]\mathrm{MMSE}(n)}} \tag{4.32}$$

The dynamic nature of the step size is apparent from (4.32). In practical situations, the instantaneous MMSE is not known but it would be possible to use one of the variable step size LMS algorithms available, such as [Kwong92], to perform the step adaptation.

# 4.5 Validation of the LMS equations

The objective of this section is to present simulation results that validate equations (4.31) and (4.32). Results are presented for different channel profiles with different degrees of non-stationarity (from $\sigma_q^2=10^{-4}$ to $\sigma_q^2=10^{-7}$) and different E/No conditions (5dB, 15dB and 25dB). The channel models and simulation conditions are those introduced in chapter 3.

## 4.5.1 Channel model 1

This channel corresponds to the delay profile shown in figure 3.2. Its channel spectrum (figure 3.3) reveals a fairly mild distortion in nearly all the in-band spectrum, becoming slightly more severe near the band edges. Given the channel's low eigenvalue spread, the convergence of the LMS algorithm is not very much hindered. The equaliser used in these simulations had 5 taps and the delay was set to 2 samples.

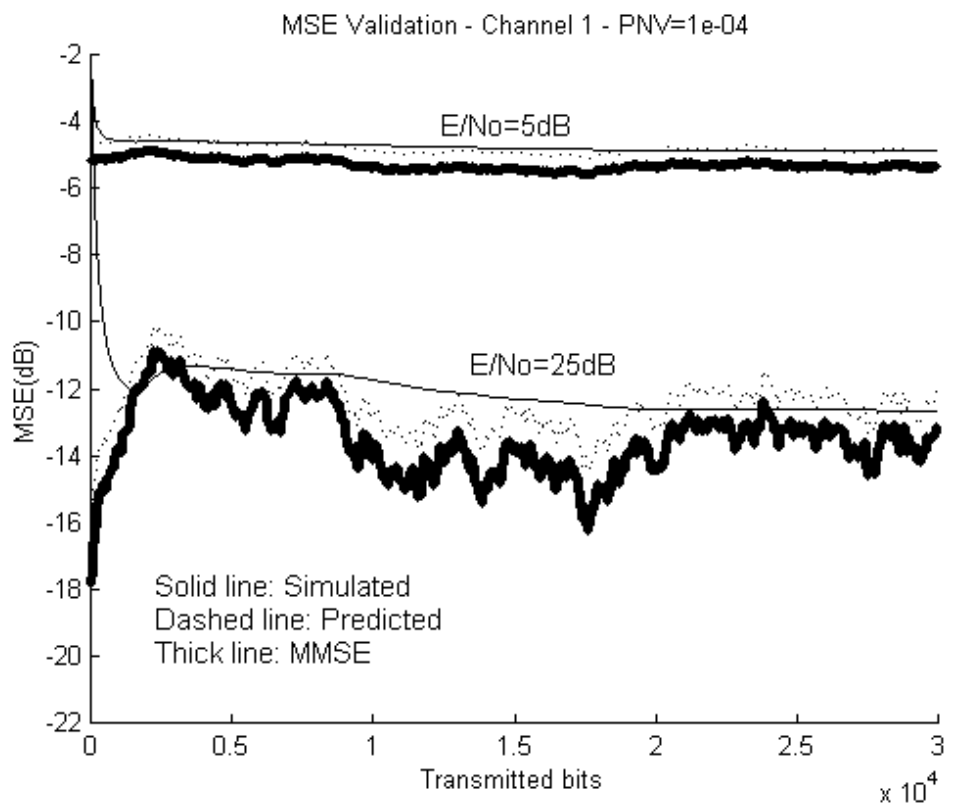

**Figure 4.9**: LMS MSE Validation, Channel model 1, $\sigma_q^2=10^{-4}$.

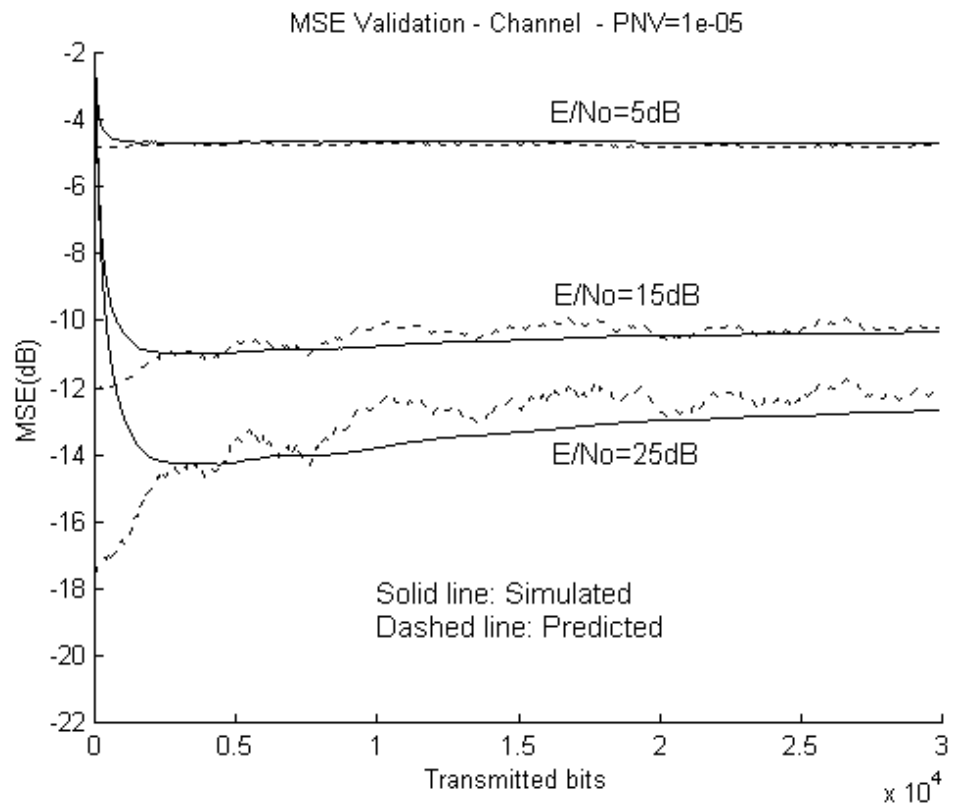

**Figure 4.10**: LMS MSE Validation, Channel model 1, $\sigma_q^2=10^{-5}$.

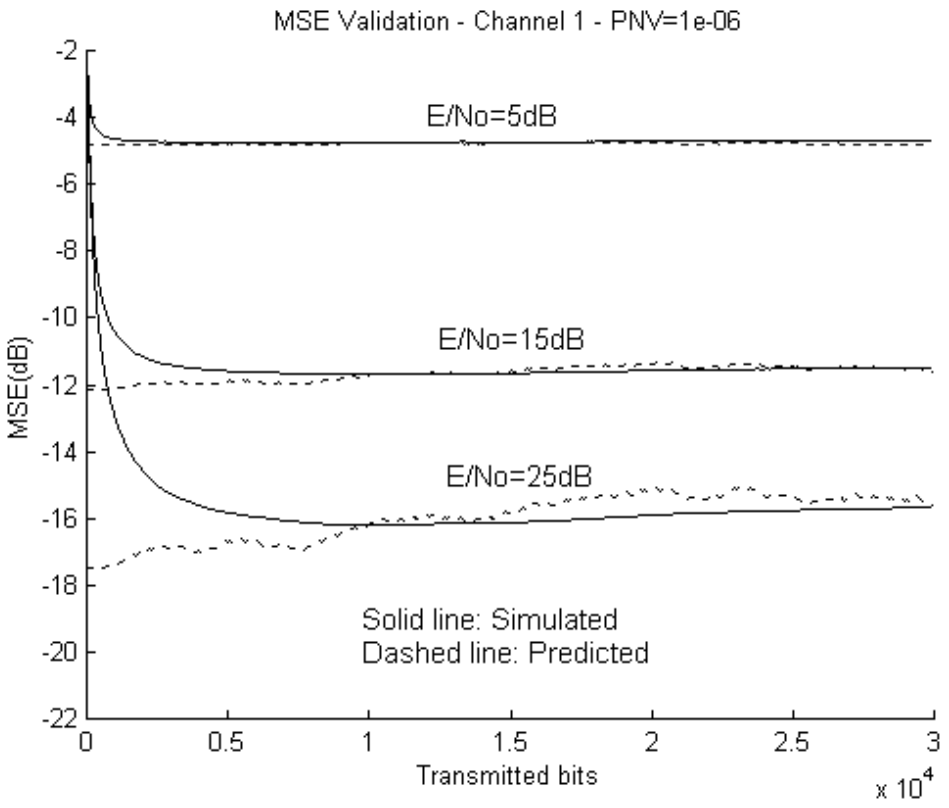

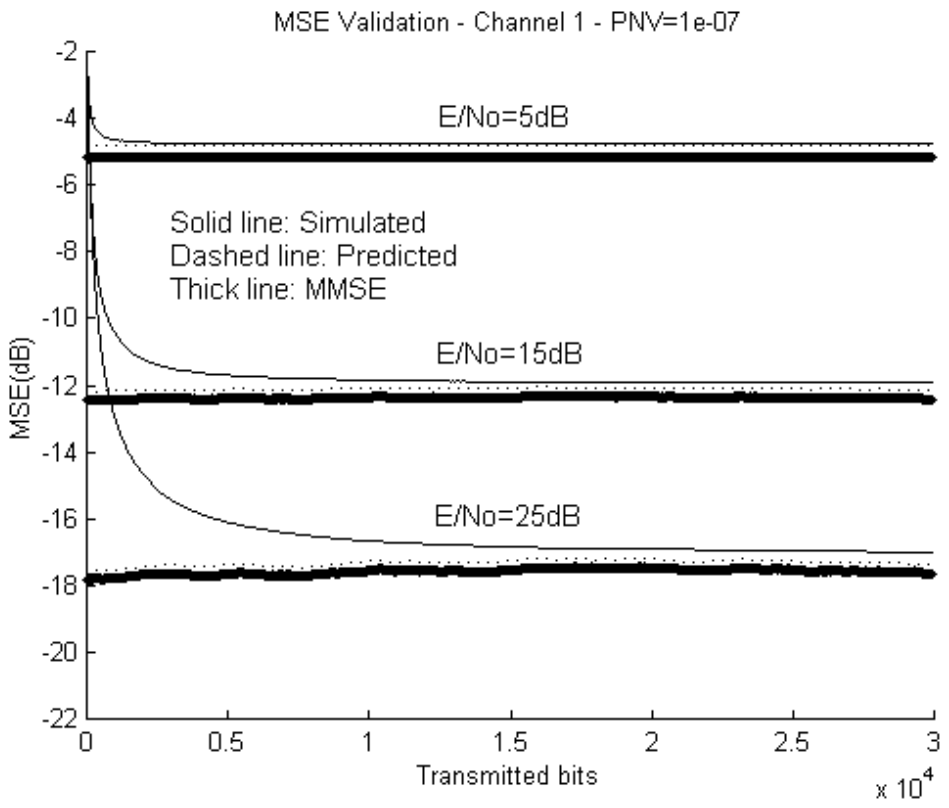

**Figure 4.11**: LMS MSE Validation, Channel model 1, $\sigma_q^2=10^{-6}$.

**Figure 4.12**: LMS MSE Validation, Channel model 1, $\sigma_q^2=10^{-7}$.

Figures 4.9, 4.10, 4.11 and 4.12 show the predicted and simulated MSE evolution. In figures 4.9 and 4.12 the MMSE is also shown. The most important fact to be appreciated in these figures is the agreement between the predicted MSEs using (4.31) and the simulation results. Notice the differences in MSE level oscillations as $\sigma_q^2$ takes very large values ($10^{-4}$), corresponding to very rapidly fading channels, and very small ones ($10^{-7}$) which would correspond to a nearly stationary channel.

There is another peculiar phenomenon worth mentioning: observe for example in Fig. 4.9, the E/No=25dB curve, how at some instants (at 2,500 or 24,000 samples), the simulated MSE using the LMS algorithm seems to operate below the MMSE level. This paradox is a consequence of the inability of the adaptive algorithm to track such rapid changes in the environment, so the simulated curve is a "low pass" version of the predicted and MMSE curves.

It can also be deduced from the different figures that at low E/No levels, the dominant term is the noise estimation error. Observe that the MSE level reached for an E/No=5dB is independent of $\sigma_q^2$. On the contrary, at high E/No, the tracking error becomes much more significant, as shown by the difference in MSE level reached for the different $\sigma_q^2$.

## 4.5.2 Channel model 2

The channel model 2 is illustrated in the time and frequency domains in figures 3.4 and 3.5. As mentioned before, the initial profile of Channel model 2 corresponds to a snapshot of an urban mobile channel sampled at 3.84 Msamples/s. Its spectral characteristics makes it a relatively good channel because although there is significant fading in the middle of the passband, it is not a complete spectral null. The equaliser used in this case had 23 taps and the delay was set to 8 samples. Different delays were tested. As long as the delay was greater than 4 samples it did not influence significantly the MSE results.

Notice in this case that the MSE curves (figures 4.13 to 4.16) converge much more slowly than with the previous channel model, especially at high E/No ratios. This is due to the greater distortion introduced now by the channel, which provokes a moderately high eigenvalue spread of the autocorrelation matrix of the input data.

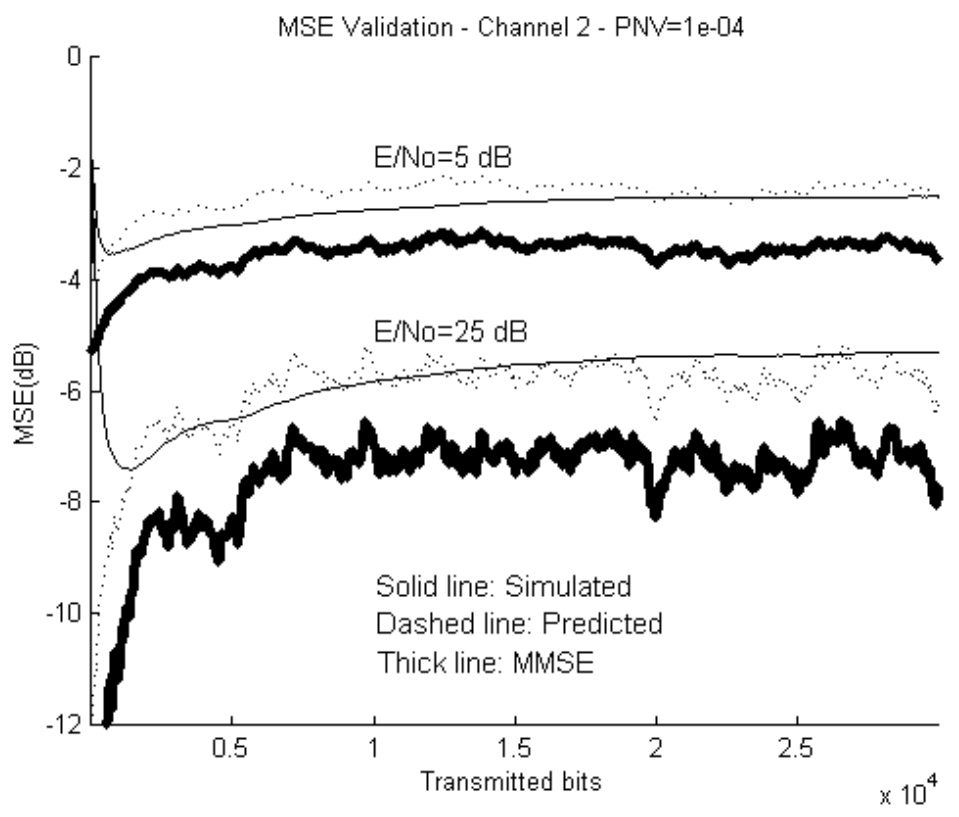

**Figure 4.13**: LMS MSE Validation, Channel model 2, $\sigma_q^2 = 10^{-4}$.

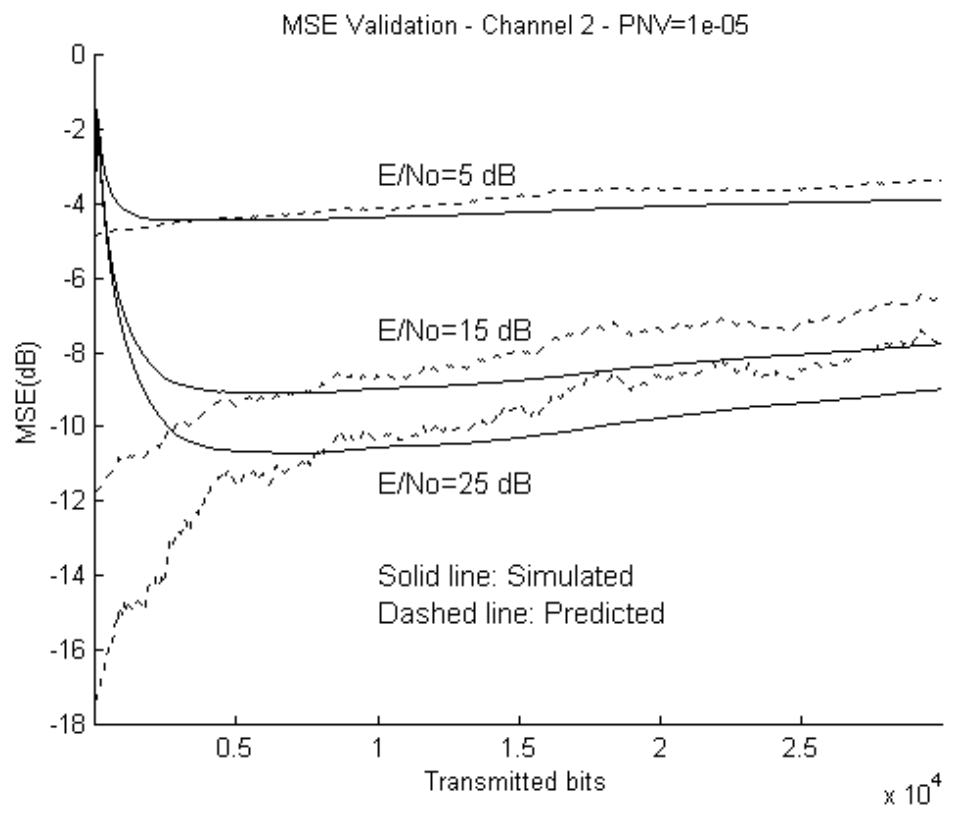

**Figure 4.14**: LMS MSE Validation, Channel model 2, $\sigma_q^2 = 10^{-5}$.

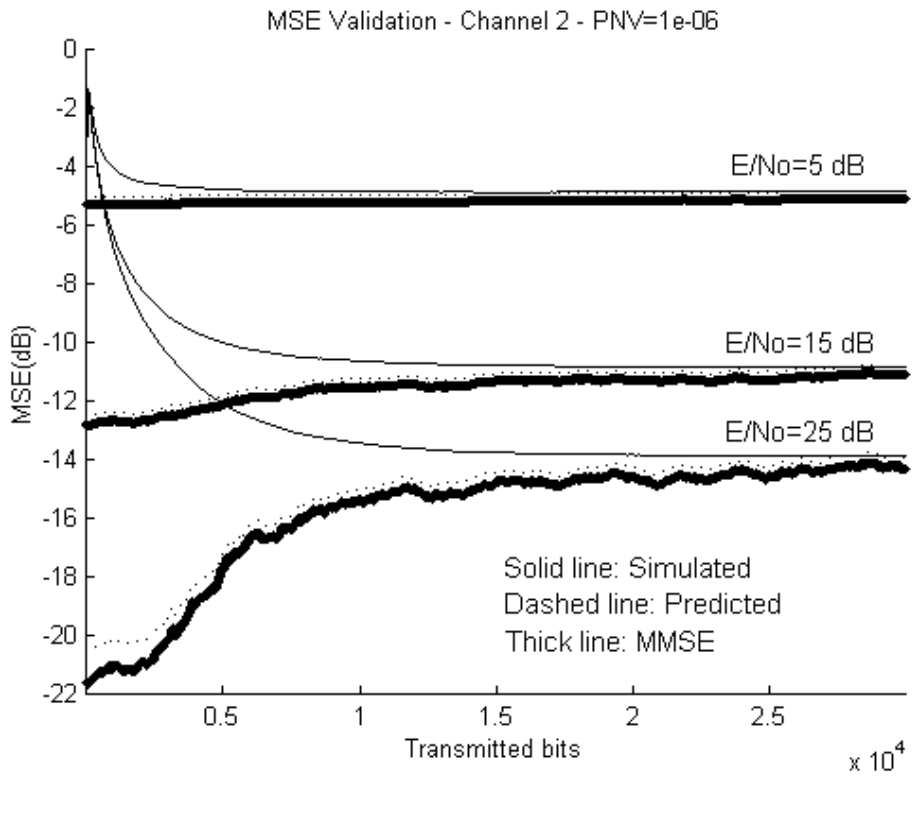

**Figure 4.15**: LMS MSE Validation, Channel model 2, $\sigma_q^2 = 10^{-6}$.

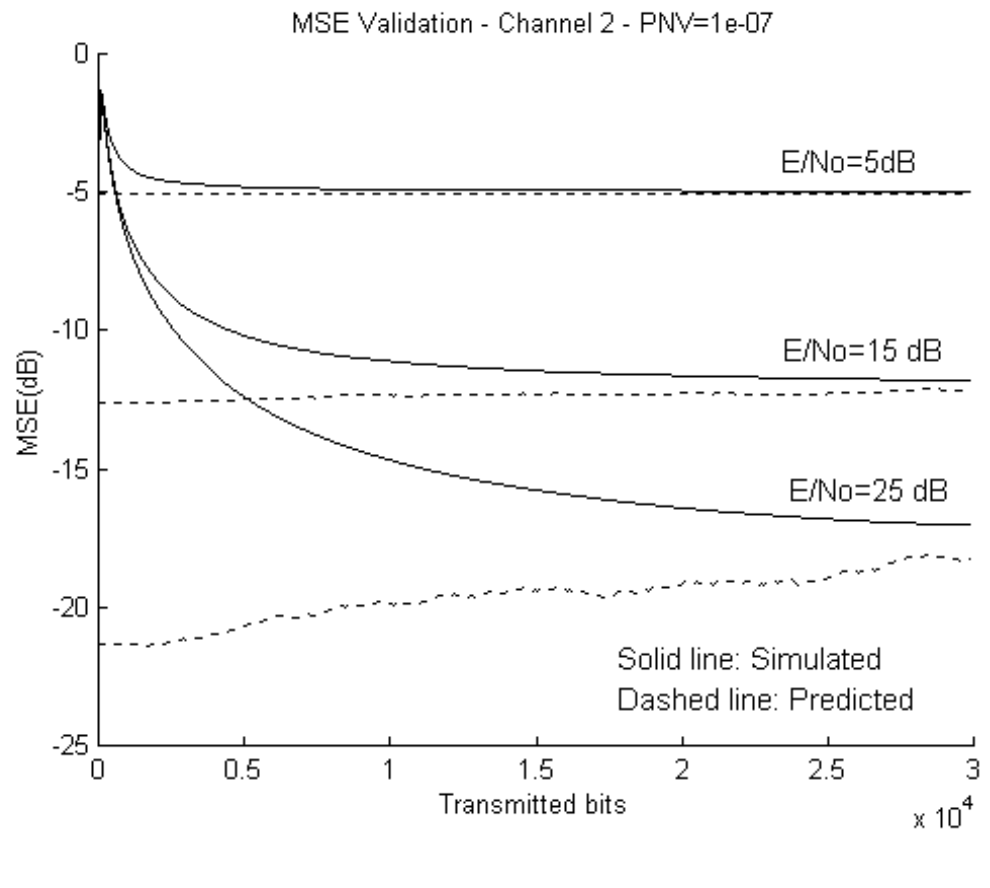

**Figure 4.16**: LMS MSE Validation, Channel model 2, $\sigma_q^2 = 10^{-7}$.

Notice that in the four pictures the different MSE curves tend to increase. The process noise vector (**p**(n)) added to the channel coefficients at every iteration worsens the fairly good initial channel profile. This fact, although it is observable at any E/No level, is much more noticeable at high E/No, where the residual error is mainly due to the equaliser imperfection.

In order to evaluate the agreement between the predicted and simulated MSE, the curves need to be examined in the steady state, that is, in the region covering the last iterations (from 25,000 iterations onwards). Overall the MSE prediction lies always within a fraction of 1 dB of the true LMS MSE obtained through simulation once the algorithm has fully converged. This holds true for all E/No and $\sigma_q^2$ levels checked.

One could argue that the simple computation of the MMSE could serve as an estimate of the performance the LMS equaliser will achieve. It can be observed in figure 4.15 that in situations where the channel does not vary very rapidly, the LMS nearly achieves the MMSE level. However, when the channel suffers very rapid fading, as in figure 4.13, the prediction is much more accurate than the MMSE alone. Moreover, if the equaliser has a large number of taps, the prediction will account for the estimation noise and therefore produces a more accurate result, which will help in deciding the optimum number of taps the equaliser should have. The effect of the equaliser length has also been investigated using the same channel model. The next two figures show the MSE evolution using different length equalisers under different levels of channel variation.

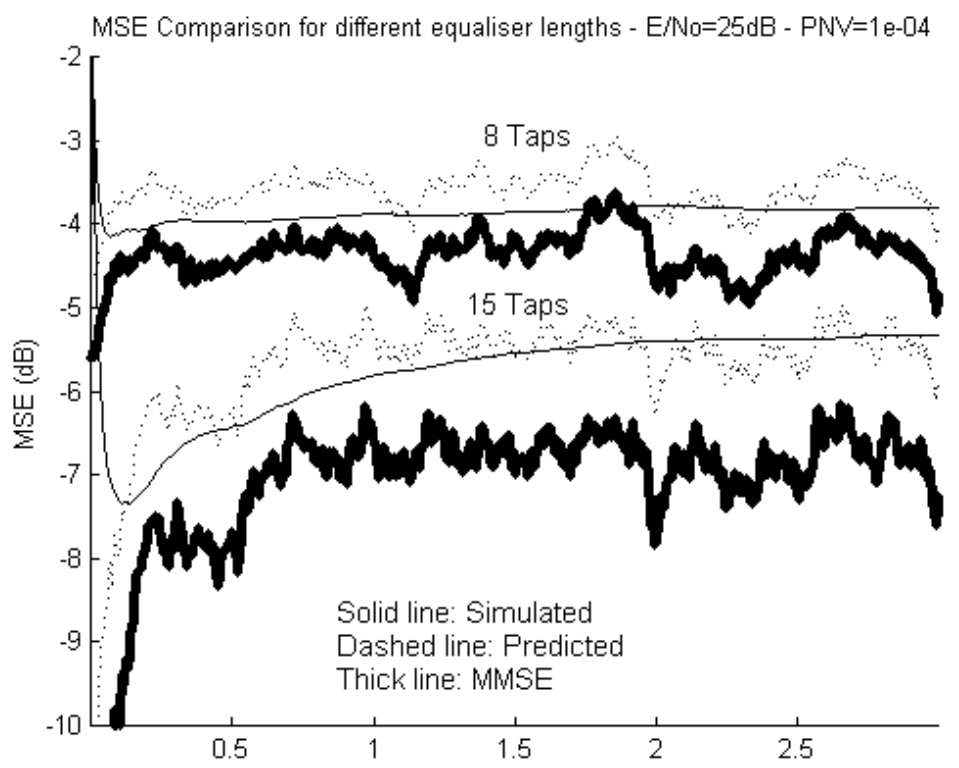

**Figure 4.17**: LMS MSE comparison for different length equalisers. Channel model 2. E/No=25dB. $\sigma_q^2=10^{-4}$.

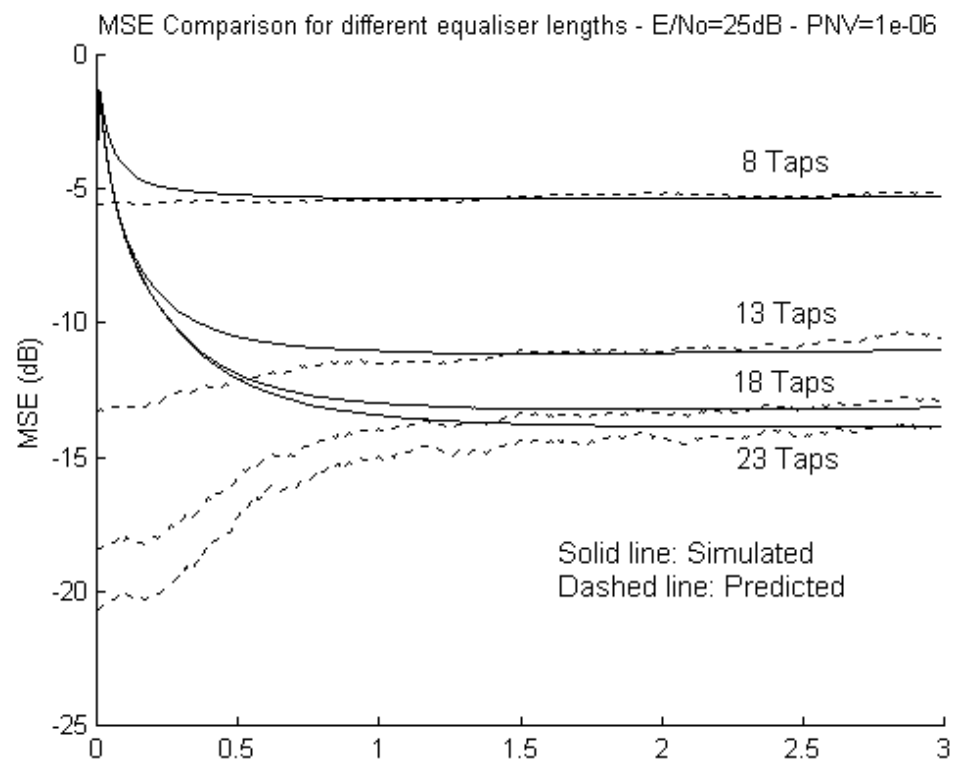

**Figure 4.18**: LMS MSE comparison for different length equalisers. Channel model 2. E/No=25dB. $\sigma_q^2=10^{-6}$.

There are no definite rules to select the equaliser length. A quite common rule of thumb consists of making the equaliser 2N+1 taps long where N is the channel length. Often designers resort to computer simulations to evaluate the equaliser performance for different lengths. Next chapter covers length selection in detail and a novel technique is presented to choose dynamically the most appropriate number of taps without any previous knowledge of the channel.

Using the same channel model, the accuracy of the step size prediction equation has also been investigated. Figures 4.19 and 4.20 compare the MSE curves obtained when using arbitrary step size values with a 0.005 difference and the mean predicted optimum step size value obtained from equation (4.32). Figure 4.19 shows that the performance of the equaliser using the mean predicted optimum step size ($\mu$=0.0139) is very close to the one found by simulation ($\mu$=0.01). Notice how the MSE worsens if the step size is made smaller ($\mu$=0.005). In figure 4.20, it can be observed that during the transient stage, the curve corresponding to the mean predicted optimum step size takes significantly more time to converge than with larger step sizes. It should be recalled at this point that in the derivation of equation (4.32), steady state operation is assumed. In the steady state, the predicted step size achieves the lowest MSE.

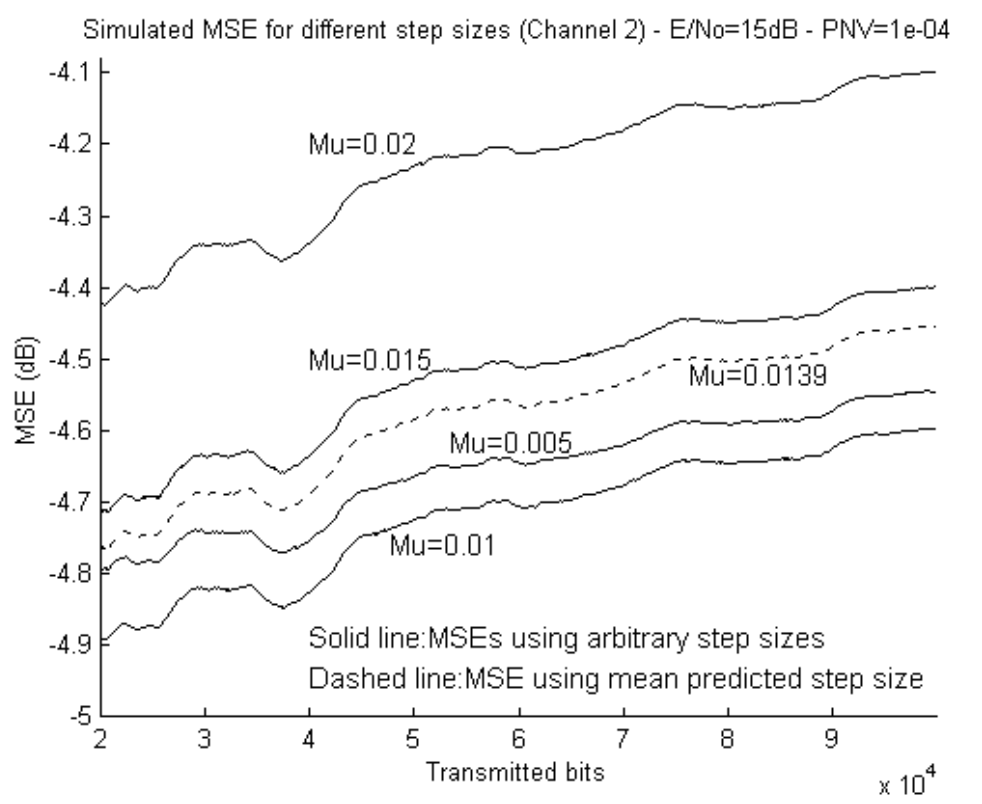

**Figure 4.19**: LMS MSE comparison using arbitrary and predicted optimum step sizes. Channel model 2. E/No=15dB. $\sigma_q^2=10^{-4}$.

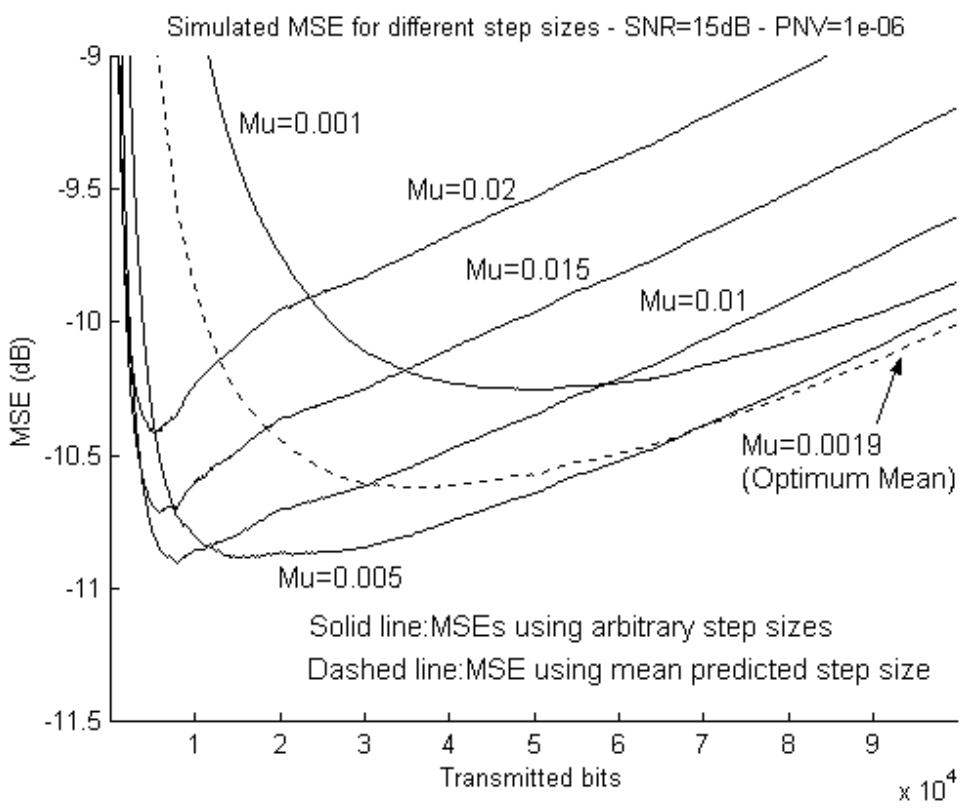

**Figure 4.20**: LMS MSE comparison using arbitrary and predicted optimum step sizes. Channel model 2. E/No=15dB. $\sigma_q^2=10^{-6}$.

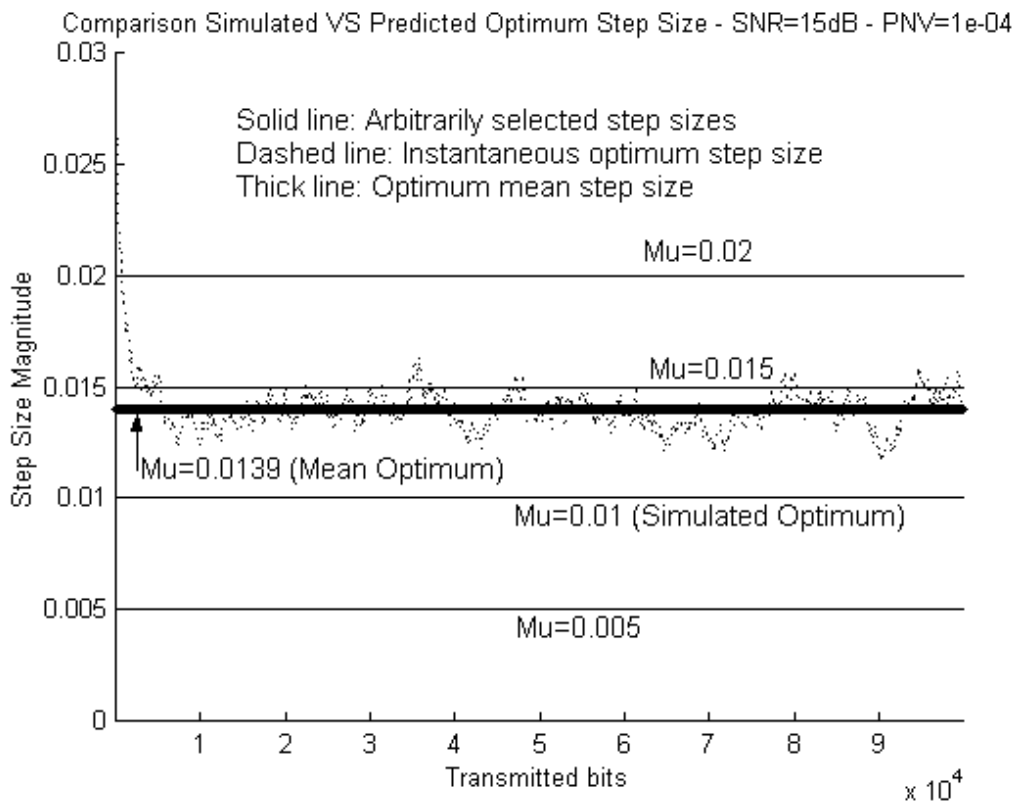

**Figure 4.21**: Step size magnitude comparison between predicted step size and arbitrarily selected ones. Channel model 2. E/No=15dB. $\sigma_q^2=10^{-4}$.

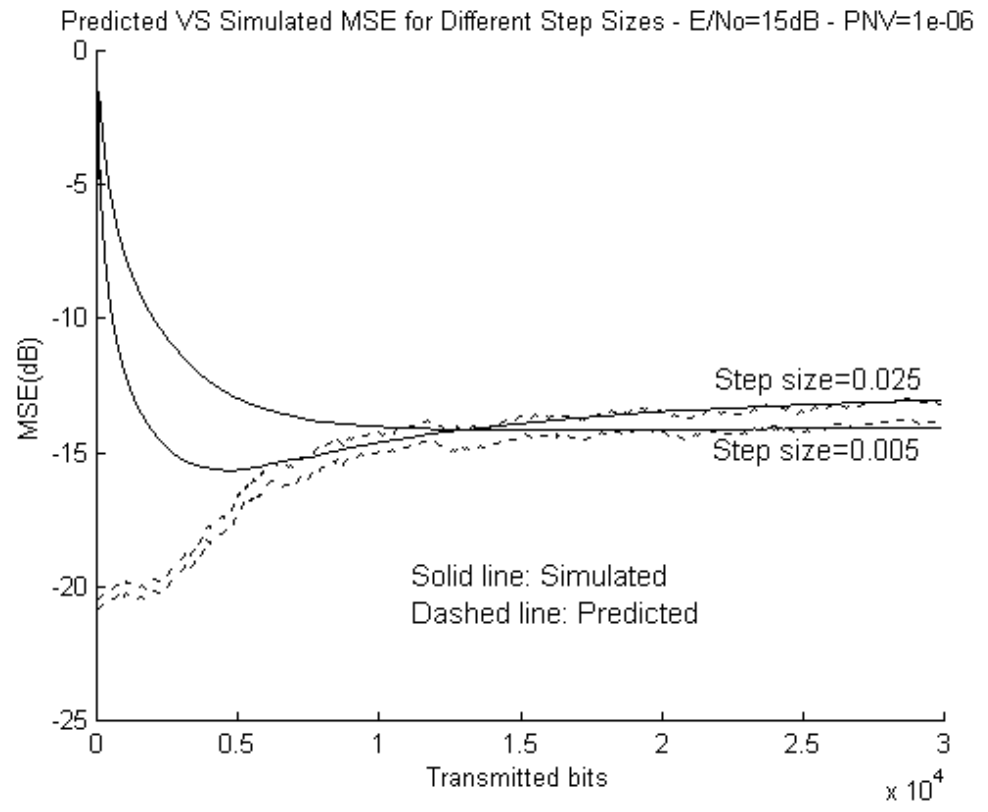

**Figure 4.22**: LMS MSE comparison for two different step sizes. Channel model 2. E/No=15dB. $\sigma_q^2=10^{-6}$.

In slowly varying channels, like that in figure 4.19, it is arguable how small the step size should be. Obviously some compromise needs to be found between the residual error achieved and the speed of convergence. This compromise can be avoided if a variable step size algorithm like the one presented in chapter 2 is used. The important point to be recognized from these figures is that equation (4.32) predicts fairly accurately which step size should be used.

Figure 4.21 is included to give an idea of the closeness of the prediction to the simulated optimum for the scenario of Figure 4.19. In the graph is also included the instantaneous predicted optimum step size. Note how the instantaneous prediction of the step size evolves in a fashion similar to the step size in the VSLMS algorithm. The step size has initially a large value to achieve rapid convergence and then it is progressively reduced in order to attain a small steady state error. Once the steady state has been reached the variations of the optimum step size are not very big, in spite of the large channel channels.

Comparing the optimum step sizes, both predicted and simulated, of figures 4.19 and 4.20, it can be observed that, logically, the faster the channel variations the larger is the optimum step size in order to improve the tracking ability of the equaliser.

Figure 4.22 shows the difference in MSE performance for two distinct step sizes and how equation (4.31) is able to predict them. Note the difference in convergence time between the equaliser with $\mu$=0.025 and the one with $\mu$=0.005.

## 4.5.3 Channel model 3

The last channel covered in this section corresponds to Channel model 3, described by figures 3.6 (time domain) and 3.7 (frequency domain). This channel constitutes one of the worse possible scenarios in which a linear equaliser can operate, due to the spectral null in the passband. Recall from section 4.4.3 (optimum equaliser assumption), that for this particular channel, the assumption relating the channel and equaliser dynamics (equation (4.5)) does not hold. Therefore small channel variations may produce very significant changes in the equaliser coefficients. Consequently, the derived MSE prediction equation will hold only in an approximate sense.

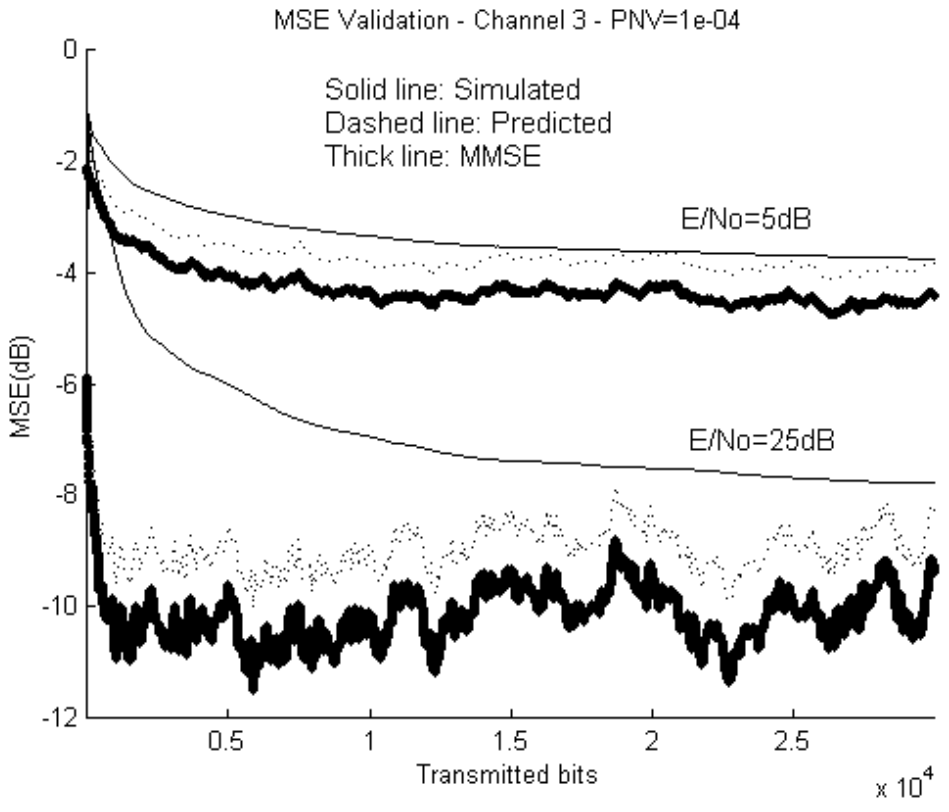

**Figure 4.23:** LMS MSE Validation, Channel model 3, $\sigma_q^2$=10$^{-4}$.

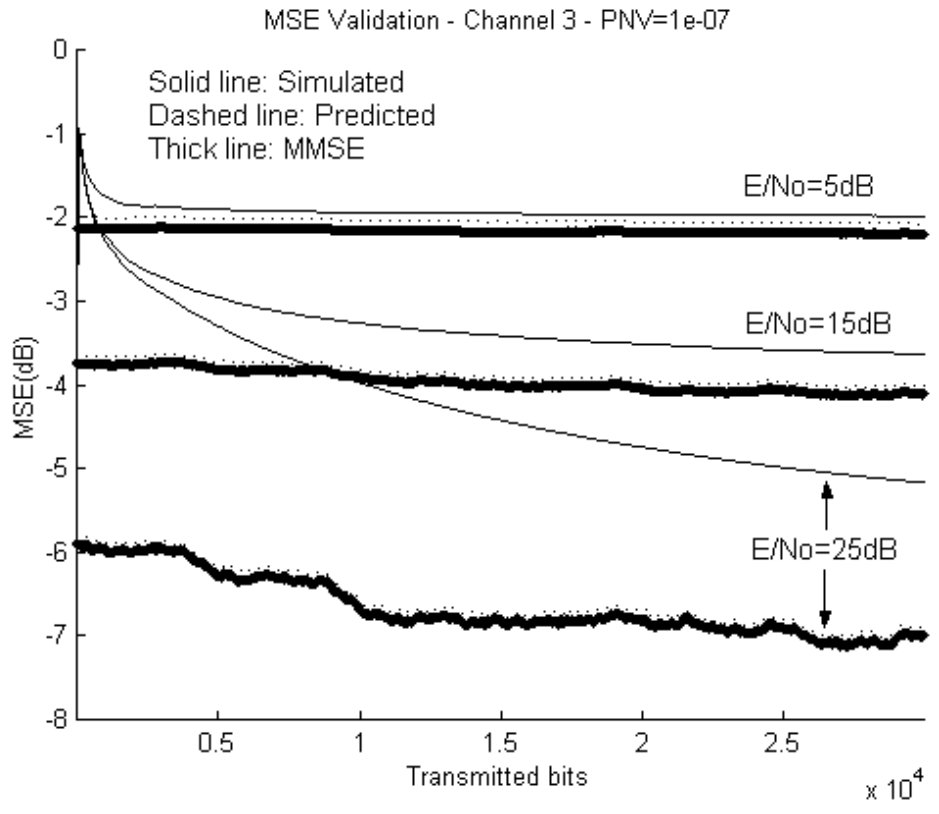

**Figure 4.24:** LMS MSE Validation, Channel model 3, $\sigma_q^2$=10$^{-7}$.

Figures 4.23 and 4.24 show a phenomenon that may look paradoxical at first sight. Comparing the curves corresponding to, for example, E/No=25dB from both graphs, the one corresponding to the faster varying channel achieves a lower MSE than the slower channel. The reason for this result is the initial channel profile. Given the extremely distortive characteristics the channel has at the beginning, large magnitude variations make the channel quickly drift away from the poor initial conditions. Notice the steep decrease in the MMSE level during the first iterations in figure 4.23 showing how the channel is improving (i.e. the spectral null disappears). When the variations are small the channel conditions remain constantly bad, hence the larger MSE.

Another point worth mentioning is the superior accuracy of the prediction equation in low E/No's which is due to the smoothing effect of the noise; the more noise there is in the system, the more flat the spectrum is. The worst possible scenario for the MSE prediction equation derived in this chapter is shown in figure 4.24 for the E/No=25 dB curve, where the combination of a poor channel response and a high E/No makes equation (4.5) completely invalid.

There is another factor contributing to the mismatch between the predicted and simulated results. As has been stated before, the prediction equation assumes steady state operation. In heavily distorting channels such as this one, the LMS algorithm takes a long time to converge (see section 2.3.1). The next two figures show the same simulation scenario but with different simulation lengths.

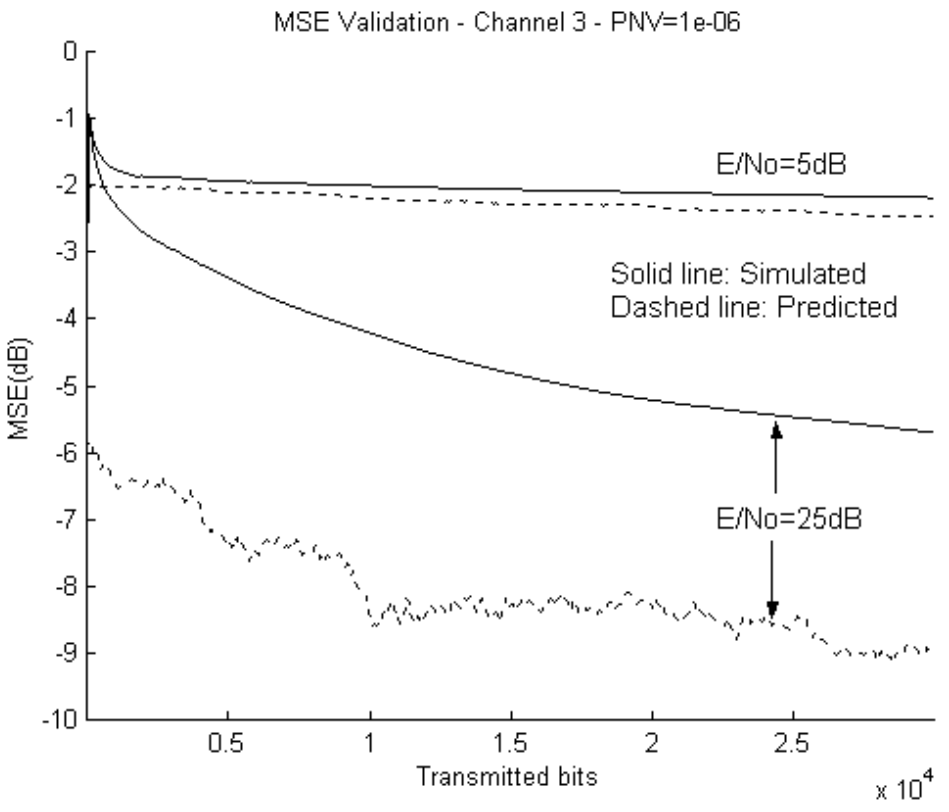

**Figure 4.25:** LMS MSE Validation, channel model 3, $\sigma_q^2=10^{-6}$.

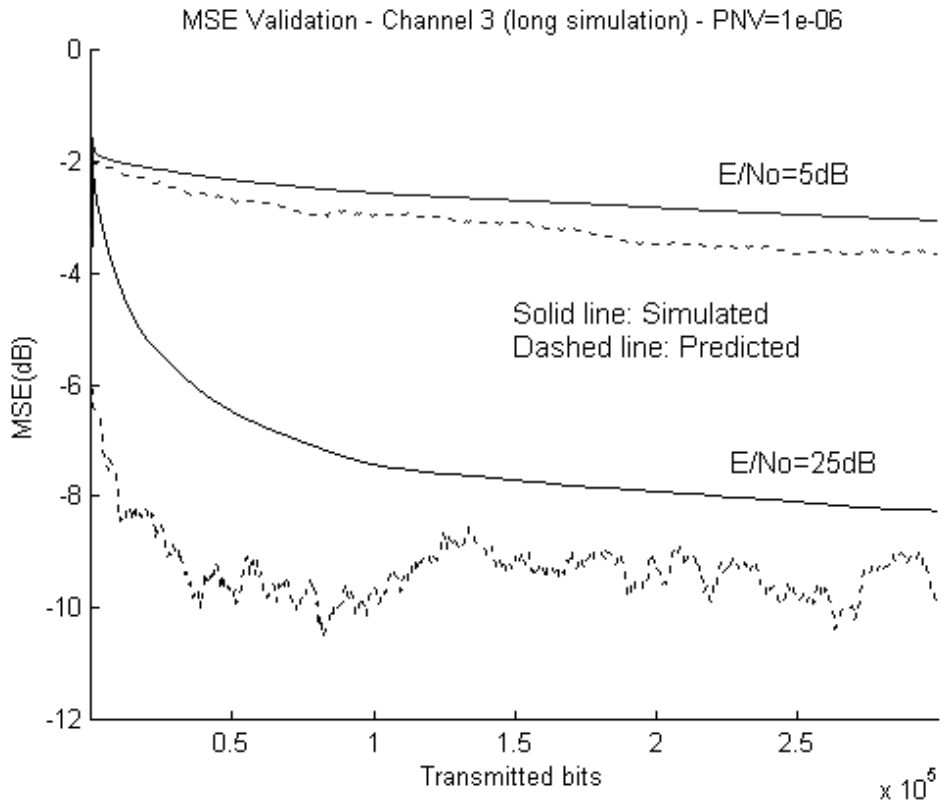

**Figure 4.26:** LMS MSE Validation, Channel model 3, $\sigma_q^2=10^{-6}$. Long simulation.

Figure 4.25 shows a simulation with the standard length of 30,000 iterations. On E/No=25 dB curve, a very significant gap can be noticed between prediction and simulation (around 4 dB). However, it can also be appreciated that the MSE has not fully converged. In figure 4.26 the simulation time was increased by a factor of 10 (300,000 iterations). This large number of simulation samples allows the LMS to converge fully. The prediction has significantly improved its accuracy. Although it is still far from the simulated value, it now lies within 2 dB of the true curve.

## 4.6 Steady state RLS-MSE analysis for the LE

In this section the equivalent analysis is now presented for the RLS algorithm. First an equation for the MSE expression is derived. Notice that although the RLS algorithm minimises a different cost function (least squares), it is not a problem to measure how well it performs with respect to another criterion, in this case the MSE. In this way, direct comparisons with the LMS algorithm are possible. An equation is also presented for the optimum forgetting factor. As in the LMS case, this analysis applies only when the adaptive filter has achieved its steady state. In the subsequent section, the derived equations are compared with simulation results.

The RLS algorithm is defined by the following equations:

$$\mathbf{k}(n) = \frac{\lambda^{-1}\mathbf{P}(n-1)\mathbf{u}(n)}{1+\lambda^{-1}\mathbf{u}^T(n)\mathbf{P}(n-1)\mathbf{u}(n)} \tag{4.33}$$

$$\xi(n) = d(n) - \hat{\mathbf{w}}^T(n-1)\mathbf{u}(n) \tag{4.34}$$

$$\hat{\mathbf{w}}(n) = \hat{\mathbf{w}}(n-1) + \mathbf{k}(n)\xi(n) \tag{4.35}$$

$$\mathbf{P}(n) = \lambda^{-1}\mathbf{P}(n-1) - \lambda^{-1}\mathbf{k}(n)\mathbf{u}^T(n)\mathbf{P}(n-1) \tag{4.36}$$

In these equations $\lambda$ is the forgetting factor of the algorithm, $\mathbf{k}(n)$ is the so-called Kalman gain and $\xi(n)$ is the error signal[16]. The matrix $\mathbf{P}(n)$ is the inverse of the correlation matrix of the input data $\mathbf{\Phi}(n)$ which is defined as:

$$\mathbf{\Phi}(n) = \sum_{i=0}^{n}\lambda^{n-i}\mathbf{u}(i)\mathbf{u}^T(i) \tag{4.37}$$

It can easily be shown that

$$\mathbf{k}(n) = \mathbf{P}(n)\mathbf{u}(n) = \mathbf{\Phi}^{-1}(n)\mathbf{u}(n) \tag{4.38}$$

Using (4.34) and (4.38) the weights' update can be rewritten as:

$$\hat{\mathbf{w}}(n) = \hat{\mathbf{w}}(n-1) + \mathbf{\Phi}^{-1}(n)\mathbf{u}(n)\left[d(n) - \hat{\mathbf{w}}^{\mathbf{T}}(n-1)\mathbf{u}(n)\right] \tag{4.39}$$

---

[16] Often $\xi(n)$ is called the a-priori error as it uses the filter weights from the previous iteration, $\hat{\mathbf{w}}(n-1)$. In the RLS algorithm, the mean squared error is defined as function of $\xi(n)$, MSE(n)=E[| $\xi(n)$ |[2]].

The minimum instantaneous error, denoted by $e_o(n)$, is defined as the error produced when using the optimum equaliser settings:

$$e_o(n) = d(n) - \mathbf{w}^T(n-1)\mathbf{u}(n) \tag{4.40}$$

Substituting the above definition into (4.39) the following expression is obtained:

$$\hat{\mathbf{w}}(n) = \hat{\mathbf{w}}(n-1) + \mathbf{\Phi}^{-1}(n)\mathbf{u}(n)\left[e_o(n) + \mathbf{w}^T(n-1)\mathbf{u}(n) - \hat{\mathbf{w}}^T(n-1)\mathbf{u}(n)\right] \tag{4.41}$$

Now **w**(n) is subtracted from both sides of (4.41) and use is made of equation (4.5) from the optimum equaliser assumption:

$$\begin{aligned}\hat{\mathbf{w}}(n) - \mathbf{w}(n) = {} & \hat{\mathbf{w}}(n-1) + \mathbf{\Phi}^{-1}(n)\mathbf{u}(n)\left[e_o(n) + \mathbf{w}^T(n-1)\mathbf{u}(n) - \hat{\mathbf{w}}^T(n-1)\mathbf{u}(n)\right] \\ & - \mathbf{w}(n-1) - \mathbf{p}(n-1)\end{aligned} \tag{4.42}$$

As in the LMS derivation, the error vector $\mathbf{t}(n)$ is defined as:

$$\mathbf{t}(n) = \hat{\mathbf{w}}(n) - \mathbf{w}(n) \tag{4.43}$$

Using (4.43) and rearranging terms (4.42) becomes

$$\mathbf{t}(n) = \left[\mathbf{I} - \mathbf{\Phi}^{-1}(n)\mathbf{u}(n)\mathbf{u}^T(n)\right]\mathbf{t}(n-1) + \mathbf{\Phi}^{-1}(n)\mathbf{u}(n)e_o(n) - \mathbf{p}(n) \tag{4.44}$$

The analysis done so far is independent of whether the adaptive filter is a channel estimator or a channel equaliser. In order to further develop (4.44), $\mathbf{\Phi}^{-1}(n)$ must be computed, and this step is where our analysis differs from that in [Eleftheriou86] and [Haykin96]. Assuming that $\mathbf{\Phi}(n) \cong E[\mathbf{\Phi}(n)]$ it can be written:

$$E[\mathbf{\Phi}(n)] = E\left[\sum_{1}^{n} \lambda^{n-i}\mathbf{u}(i)\mathbf{u}^T(i)\right] = \sum_{i=1}^{n} \lambda^{n-i} E\left[\mathbf{u}(i)\mathbf{u}^T(i)\right] = \sum_{i=1}^{n} \lambda^{n-i}\mathbf{R}(i) \tag{4.45}$$

Note that now, in the equalisation case, the autocorrelation matrix is time dependent, making the analysis more complex. As in the LMS analysis, the structure of the autocorrelation matrix needs to be examined in detail; in fact in deeper detail as now all its elements are needed, as opposed to the LMS case where only the diagonal elements were required. The objective is to find a recursion for the evolution of the autocorrelation matrix in order to compute the summation appearing in (4.45).

The definition of the autocorrelation matrix of the input data, **R**(n), is given by (4.22). From the assumption of the normalised power response and using the same analysis as in the LMS case (equation (4.22) ), the diagonal elements of **R**(n) are given by:

$$E[u(n-i)u(n-i)] = \sigma_d^2 + \sigma_v^2 \qquad 0 \le i \le M-1 \tag{4.46}$$

**R**(n) is symmetric as $E[u_i(n)u_j(n)] = E[u_j(n)u_i(n)]$ holds for all i, j, so only half of the elements need to be computed. The elements of the upper triangular submatrix are given by:

$$E[u(n-i)u(n-j)] = \sigma_d^2 \begin{bmatrix} c_0(n-j)c_{j-i}(n-i) + c_1(n-j)c_{j-i+1}(n-i) \\ + \cdots + c_{N-(j+i+1)}(n-j)c_{N-1}(n-i) \end{bmatrix} + \sigma_v^2 \quad \text{with } j > i \quad (4.47)$$

Substituting the sampling index n by n+1 in (4.46) and (4.47), it is possible to find the elements of **R**(n+1), and then to use the channel model equation (equation (4.1) ) to expand the channel coefficients at time n+1 as a function of the coefficients at instant n plus samples of the process noise vector. The corresponding equations to (4.46) and (4.47) at time instant n+1 are given by:

$$\begin{aligned} E[u(n+1-i)u(n+1-i)] &= \sigma_d^2 E\begin{bmatrix} [c_0^2(n) + c_0^2(n) + \cdots + c_{N-1}^2(n)] + \\ [q_0^2(n) + q_0^2(n) + \cdots + q_{N-1}^2(n)] \end{bmatrix} + \sigma_v^2 \\ &= \sigma_d^2 + N\sigma_d^2\sigma_q^2 + \sigma_v^2 \qquad 0 \le i < M \end{aligned} \quad (4.48)$$

$$E[u(n+1-i)u(n+1-j)] = E[u(n-i)u(n-j)] \qquad j > i \quad (4.49)$$

In order to derive (4.48) and (4.49), use is made of the independence between the channel coefficients, the independence of sample to sample process noise vector coefficients, and also the independence between the channel and the process noise vector. Noting the relation between (4.46) and (4.48) and between (4.47) and (4.49) it is possible to define the autocorrelation matrix in a recursive manner:

$$\mathbf{R}(n+1) = \mathbf{R}(n) + N\sigma_d^2\sigma_q^2\mathbf{I}_M \quad (4.50)$$

This recursion allows the summation in (4.45) to be expressed in a different way:

$$\mathbf{\Phi}(n) \cong E[\mathbf{\Phi}(n)] = \sum_{i=0}^{n} \lambda^{n-i}\mathbf{R}(i) = \lambda^n\mathbf{R}(0) + \lambda^{n-1}\mathbf{R}(1) + \cdots + \lambda\mathbf{R}(n-1) + \mathbf{R}(n) \quad (4.51)$$

Now the different **R**(i) terms are expanded using (4.50):

$$\mathbf{\Phi}(n) \cong \lambda^n\mathbf{R}(0) + \lambda^{n-1}(\mathbf{R}(0) + N\sigma_d^2\sigma_q^2\mathbf{I}_M) + \cdots + (\mathbf{R}(0) + (n-1)N\sigma_d^2\sigma_q^2\mathbf{I}_M) \quad (4.52)$$

Notice that all the summing terms in (4.52) are expressed as a function of **R**(0). In order to obtain a close expression for this summation it is more convenient to express the terms in the "opposite" direction, that is as a function of **R**(n). This can be done by expressing **R**(n) as a function of **R**(n+1) in (4.50). The resulting expression is:

$$\Phi(n) \cong \lambda^n(\mathbf{R}(n) - nN\sigma_d^2\sigma_q^2\mathbf{I}_M) + \cdots + \lambda(\mathbf{R}(n) - N\sigma_d^2\sigma_q^2\mathbf{I}_M) + \mathbf{R}(n) \quad (4.53)$$

The above expression corresponds to the sum of two arithmetic progressions, one corresponding to $\lambda^i \mathbf{R}(n)$ and the other one to $N\sigma_d^2\sigma_q^2\mathbf{I}_M i\lambda^i$. Using the formula for the summation of an arithmetic progression for both progressions, $\mathbf{\Phi}(n)$ becomes:

$$\mathbf{\Phi}(n) \cong \frac{1}{1-\lambda}\mathbf{R}(n) - \frac{\lambda N\sigma_d^2\sigma_q^2}{(1-\lambda)^2}\mathbf{I}_M \qquad (4.54)$$

The second term matrix has non-zero elements in the main diagonal and they are given by $\frac{\lambda N\sigma_d^2\sigma_q^2}{(1-\lambda)^2}$ whereas the diagonal elements of the first matrix are given by $\frac{\sigma_d^2+\sigma_v^2}{1-\lambda}$.

Under plausible conditions, the elements of the second term matrix in (4.54) will be at least an order of magnitude smaller than those in the first term. By "plausible conditions" is meant that the forgetting factor is chosen correspondingly (roughly within an order of magnitude of its optimum value) given a determined value of channel change, $\sigma_q^2$.

An example of non-plausible choice of forgetting factor would be, for example, $\lambda$=0.9999 (very long memory) for a channel change rate $\sigma_q^2=10^{-4}$ (very fast changing). In this situation the magnitude of the second term of (4.54) is comparable to the first term.

The inverse of $\mathbf{\Phi}(n)$ is easily calculated if the second term in (4.54) is discarded:

$$\mathbf{\Phi}^{-1}(n) = (1-\lambda)\mathbf{R}^{-1}(n) \qquad (4.55)$$

Notice that now, unlike the channel estimation scenario, (4.55) depends on a time varying autocorrelation matrix. Substituting (4.55) into (4.44) and assuming that $\mathbf{u}(n)\mathbf{u}^T(n) \cong E[\mathbf{u}(n)\mathbf{u}^T(n)] = \mathbf{R}(n)$ the following expression is obtained:

$$\mathbf{t}(n) = \left[\mathbf{I}_M - (1-\lambda)\mathbf{R}^{-1}(n)\mathbf{R}(n)\right]\mathbf{t}(n-1) + (1-\lambda)\mathbf{R}^{-1}(n)\mathbf{u}(n)e_o(n) - \mathbf{p}(n) \qquad (4.56)$$

Rearranging (4.56):

$$\mathbf{t}(n) = \lambda\mathbf{t}(n-1) + (1-\lambda)\mathbf{R}^{-1}(n)\mathbf{u}(n)e_o(n) - \mathbf{p}(n) \qquad (4.57)$$

The derivation now follows a similar course to the LMS one in section 4.4.1. The autocorrelation of the error vector is computed:

$$\mathbf{K}(n) = E[\mathbf{t}(n)\mathbf{t}^T(n)] = \lambda^2\mathbf{K}(n-1) + (1-\lambda)^2\mathbf{R}^{-1}(n)\mathrm{MMSE}(n) + \mathbf{P} \qquad (4.58)$$

In the steady state it is reasonable to assume $\mathbf{K}(n) \cong \mathbf{K}(n-1)$, so (4.58) becomes:

$$\mathbf{K}(n) = \frac{(1-\lambda)^2}{1-\lambda^2}\mathbf{R}^{-1}(n)\mathrm{MMSE}(n) + \frac{1}{1-\lambda^2}\mathbf{P} \qquad (4.59)$$

Now using the common approximation $1-\lambda^2 \cong 2(1-\lambda)$ for $\lambda \cong 1$, and multiplying both sides of (4.59) by $\mathbf{R}(n)$:

$$\mathbf{R}(n)\mathbf{K}(n) = \frac{(1-\lambda)\text{MMSE}(n)}{2}\mathbf{I}_M + \frac{1}{2(1-\lambda)}\mathbf{R}(n)\mathbf{P} \tag{4.60}$$

Finally the trace is taken of both sides of (4.60) in order to obtain the EMSE(n) (see equation (4.12) ):

$$\text{EMSE}(n) = \frac{(1-\lambda)\text{MMSE}(n)}{2}\text{Tr}[\mathbf{I}_M] + \frac{1}{2(1-\lambda)}\text{Tr}[\mathbf{R}(n)\mathbf{P}] \tag{4.61}$$

The computation of the two traces is straightforward, $\text{Tr}[\mathbf{I}_M] = M$ and $\text{Tr}[\mathbf{R}(n)\mathbf{P}] = M\sigma_p^2(\sigma_d^2 + \sigma_v^2)$, so (4.61) becomes:

$$\text{EMSE}(n) = \frac{M(1-\lambda)}{2}\text{MMSE}(n) + \frac{M\sigma_p^2(\sigma_d^2 + \sigma_v^2)}{2(1-\lambda)} \tag{4.62}$$

Finally, using (4.5), it is possible to write $M\sigma_p^2 = N\sigma_q^2$, so the final expression for the RLS MSE is given by:

$$\text{MSE}(n) = \text{MMSE}(n) + \frac{M(1-\lambda)}{2}\text{MMSE}(n) + \frac{N\sigma_q^2(\sigma_d^2 + \sigma_v^2)}{2(1-\lambda)} \tag{4.63}$$

The resulting MSE expression has three terms. The first one corresponds to the instantaneous MMSE level. The second and third terms constitute the excess mean squared error of the RLS algorithm and they appear as a result of using a forgetting factor smaller than one. Notice that when $\lambda = 1$ and $\sigma_q^2 = 0$ (i.e. the channel is static), and using l'Hopital rule to solve the resulting 0/0 condition, the EMSE terms vanish and the steady state MSE turns out to be the MMSE as least squares theory predicts ([Haykin96]).

Equation (4.63) is very similar to equation 4.14 in [Eleftheriou86], but again two significant differences arise. First the time varying MMSE makes the MSE equation channel specific and time varying. Secondly, as in the LMS case, it is clear that the MSE performance depends on the channel length and also on the equaliser length.

The optimum forgetting factor can be obtained by differentiating (4.63) with respect to $\lambda$ and equating the result to zero:

$$\frac{\partial MSE(n)}{\partial \lambda} = 0 \qquad \rightarrow \qquad \lambda_{opt}(n) = 1 - \sqrt{\frac{N\sigma_q^2(\sigma_d^2 + \sigma_v^2)}{M\ MMSE(n)}} \qquad (4.64)$$

Note by looking at the magnitude of the variables in (4.64) that $\lambda_{opt}$ will always be less than, but very close to, 1. In particular notice that the larger $\sigma_q^2$ is, the smaller $\lambda_{opt}$ becomes. As in the LMS case, the optimum parameter of the algorithm, in this case the forgetting factor, is time varying. Although some variable forgetting factor algorithms have appeared in the literature, their complexity is significantly higher than that of RLS (of the order of $O(M^3)$ where M is the filter length), therefore these algorithms are not widely used.

# 4.7 Validation of the RLS equations

Some simulation results are now presented to show the validity of equations (4.63) and (4.64). As with the LMS analysis, the equations have been tested using the models described in chapter 3.

## 4.7.1 Channel model 1

This channel model corresponds to the one described by figures 3.2 (time domain) and 3.3 (frequency domain) in chapter 3. The results are shown in figures 4.27 to 4.30. It should be recalled that if the channel is stationary, the RLS algorithm does not introduce any excess mean squared error (EMSE) and therefore is capable of achieving the minimum mean squared error. When the channel is nearly stationary ($\sigma_q^2$=$10^{-6}$ or $\sigma_q^2$=$10^{-7}$) the forgetting factor will be extremely close to 1 and the RLS is able to achieve near MMSE performance. On the other hand, in rapidly varying channels, the forgetting factor will take a smaller value (still close to 1) which introduces EMSE. The forgetting factors used in these simulations have been set to $\lambda$=0.99 for $\sigma_q^2$=$10^{-4}$, $\lambda$=0.995 for $\sigma_q^2$=$10^{-5}$, $\lambda$=0.999 for $\sigma_q^2$=$10^{-6}$ and $\lambda$=0.9995 for $\sigma_q^2$=$10^{-7}$. These near optimum values were determined by simulation.

The MMSE curves are shown in figure 4.27-4.30 along with the predicted and simulated MSEs. On average the prediction curve is closer to the simulated value than the MMSE curve. This is because the prediction equation takes into account the EMSE introduced as a result of using a forgetting factor smaller than 1.

As in the LMS case, notice how at some points the instantaneous MSE is larger than the simulated MSE. Again this is due to the inability of the adaptive algorithm to track instantaneously the large sudden changes of the channel. It is clear from figures 4.27-4.30 that the prediction equation matches the simulation very accurately.

It is worth comparing this set of figures with those corresponding to the same channel model but using the LMS algorithm (figures 4.9-4.12). The steady-state MSE error level is approximately the same for both algorithms, but the convergence time of the adaptive filter is very much reduced when using the RLS algorithm, particularly in high E/No scenarios.

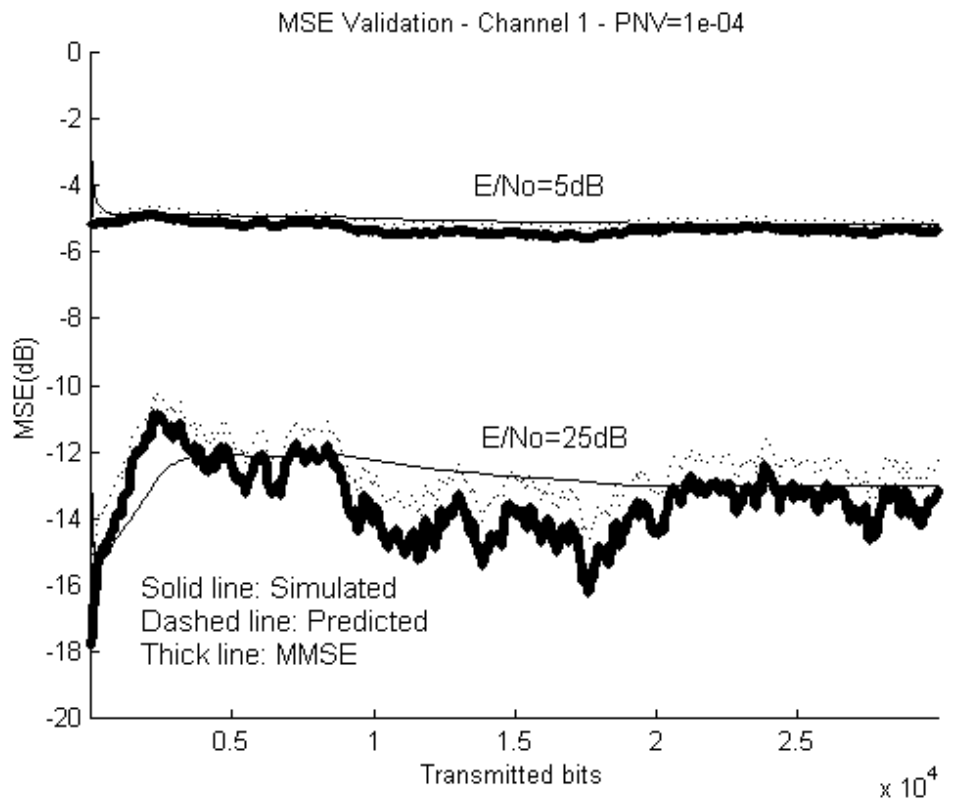

**Figure 4.27**: RLS MSE Validation, Channel model 1, $\sigma_q^2=10^{-4}$.

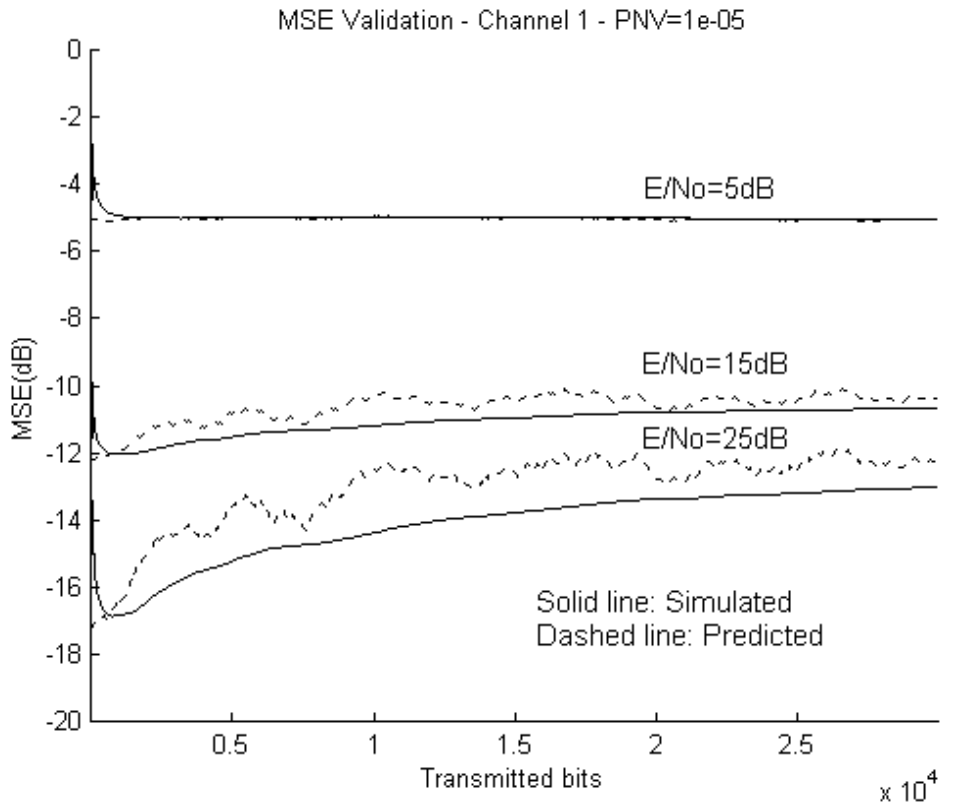

**Figure 4.28**: RLS MSE Validation, Channel model 1, $\sigma_q^2=10^{-5}$.

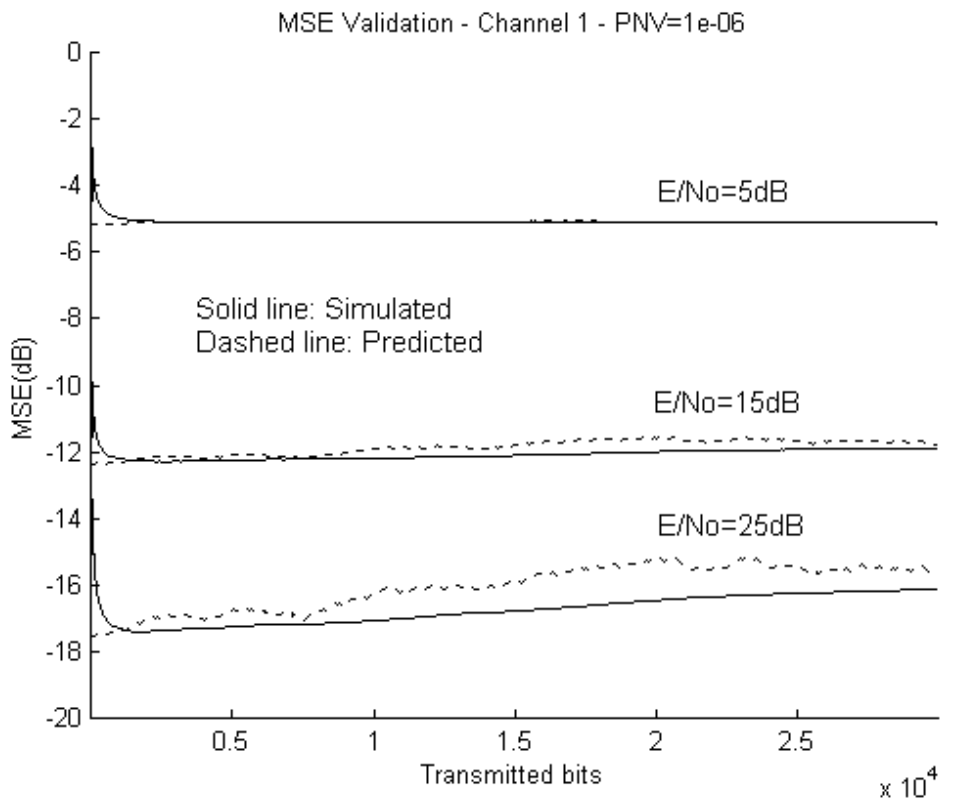

**Figure 4.29**: RLS MSE Validation, Channel model 1, $\sigma_q^2=10^{-6}$.

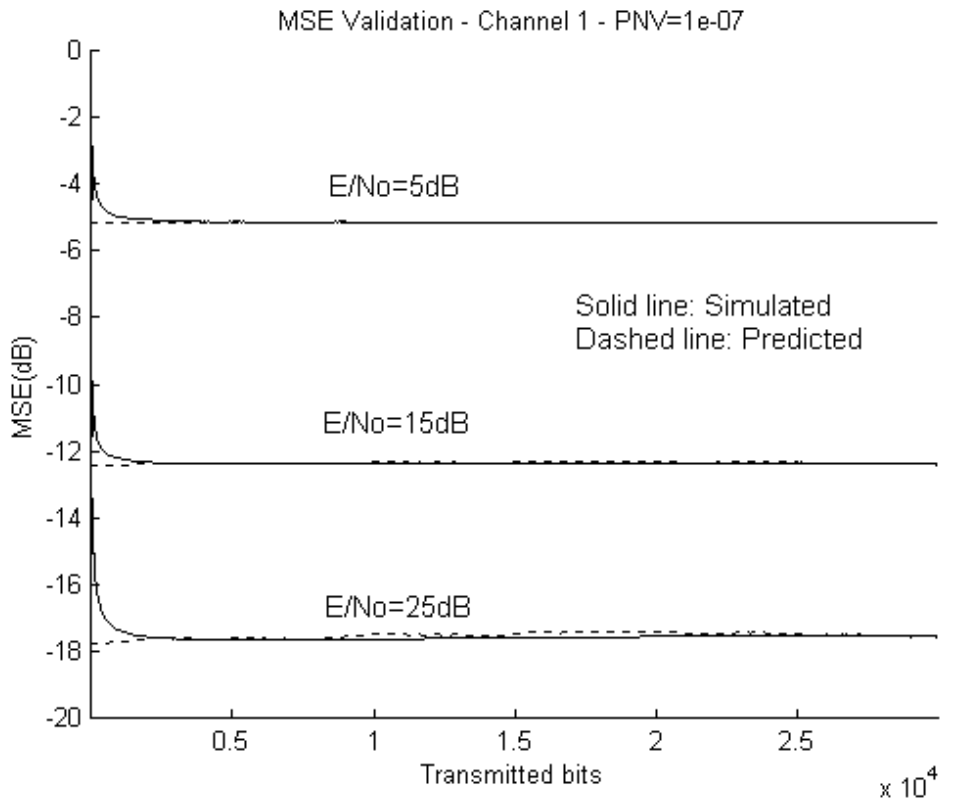

**Figure 4.30**: RLS MSE Validation, Channel model 1, $\sigma_q^2=10^{-7}$.

## 4.7.2 Channel model 2

This typical wireless channel profile is shown in figure 3.4. The equaliser was 23 tap long and the delay was set to 8 samples. Figures 4.30 to 4.33 compare the predicted MSE performance with the simulated one for different values of $\sigma_q^2$ and E/No. In general, there is good agreement between the predicted and measured MSE, the difference between the two being within 1.5 dB (in most cases much less than that). However, it can be seen also that at higher E/No the prediction becomes less accurate. There is an explanation for this inaccuracy. As has been stated before, in stationary environments the RLS algorithm is able to attain the MMSE level.

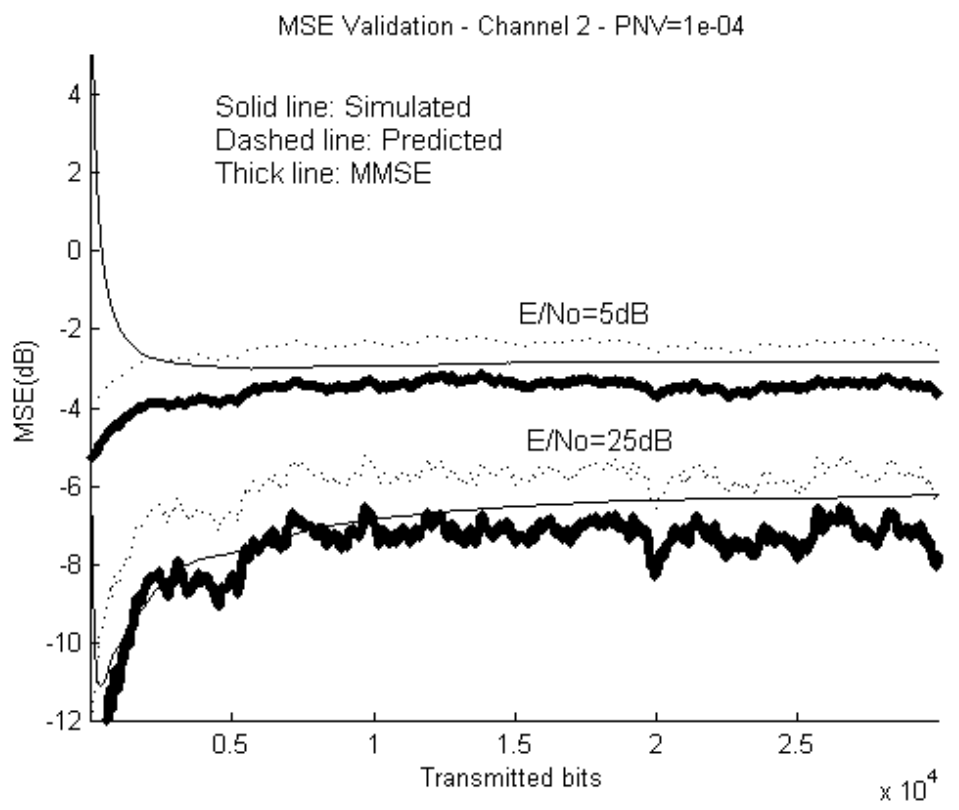

**Figure 4.31**: RLS MSE Validation, Channel model 2, $\sigma_q^2=10^{-4}$.

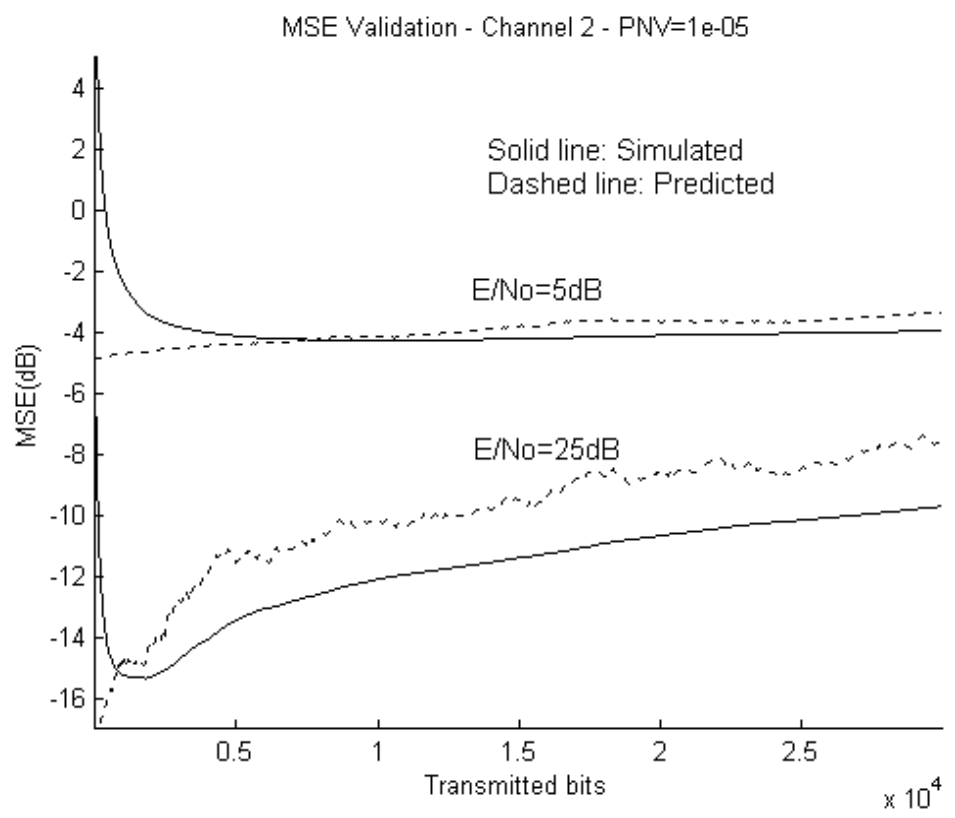

**Figure 4.32**: RLS MSE Validation, Channel model 2, $\sigma_q^2=10^{-5}$.

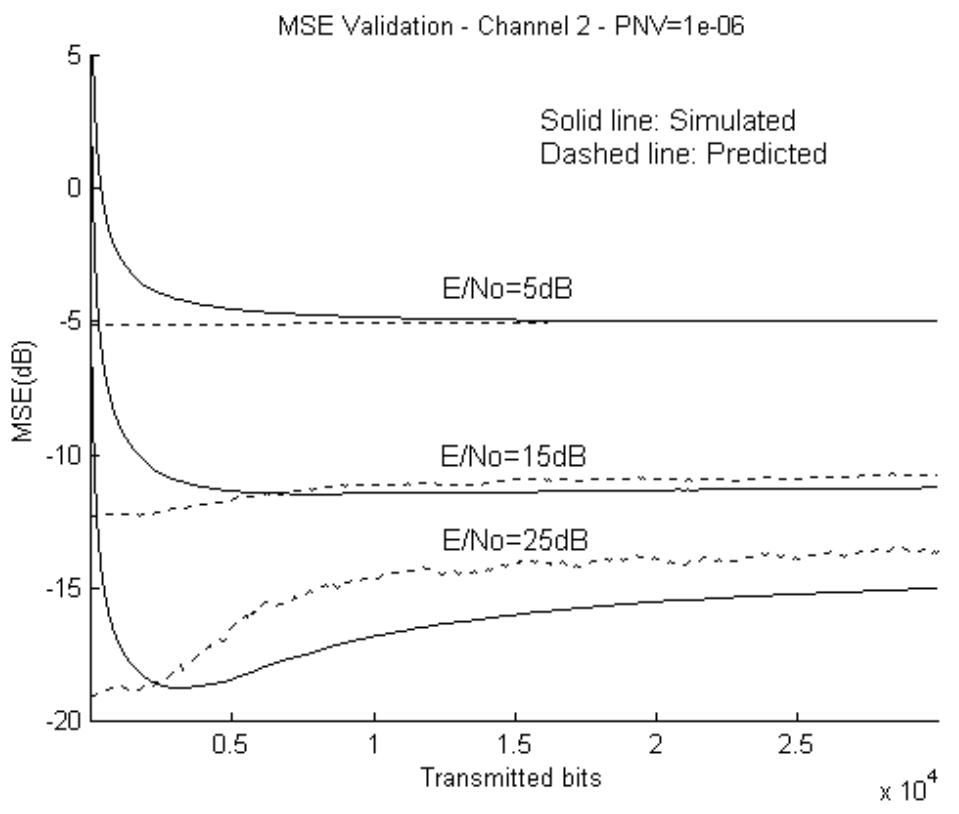

**Figure 4.33**: RLS MSE Validation, Channel model 2, $\sigma_q^2=10^{-6}$.

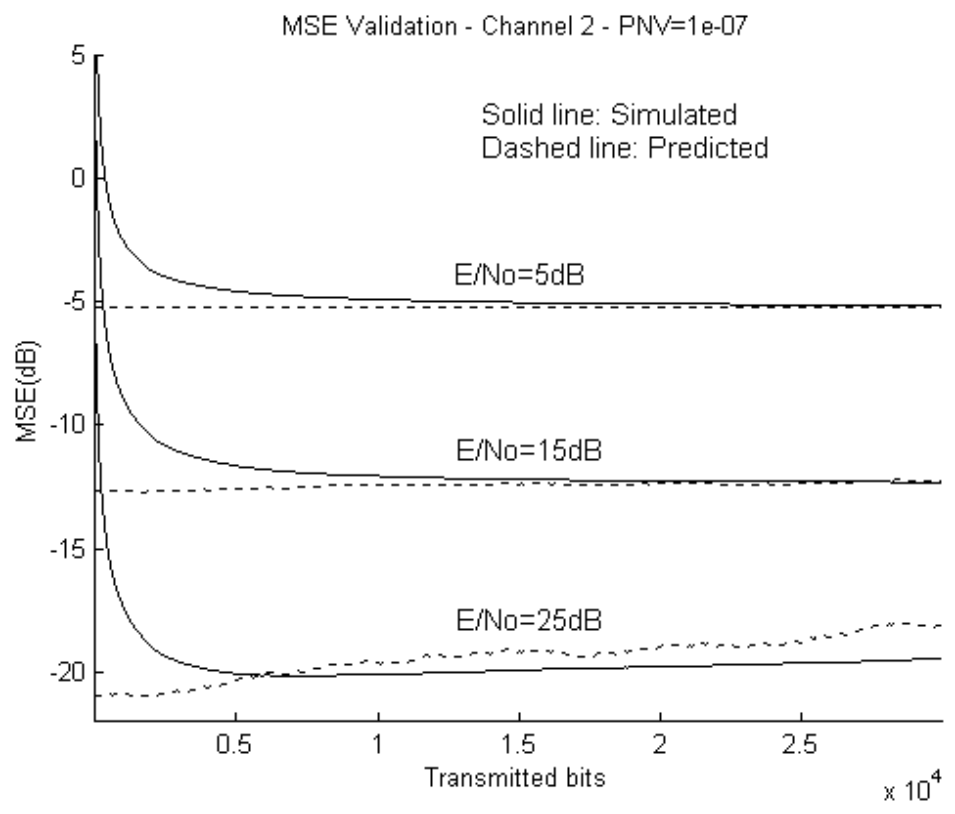

**Figure 4.34**: RLS MSE Validation, Channel model 2, $\sigma_q^2=10^{-7}$.

In the derivation of equation (4.63) some simplifications have been made, therefore the final expressions cannot be 100% accurate. In particular, the inaccuracies affect the EMSE terms of (4.63). In near static scenarios the adaptive filter is able to operate at the MMSE level or very close to it, thus under these conditions the MMSE level serves as a good indicator of the performance the equaliser will achieve. In particular, notice that if $\sigma_q^2$ is very small (near static channel), then the forgetting factor should be adequately set very close to one otherwise the EMSE terms in (4.63) will become significant and will distort the prediction because the filter is able to operate at the MMSE level. Thus the mismatch between predicted and measured MSE can be attributed to an inexact value of $\lambda$ for a particular $\sigma_q^2$ level. This phenomenon affects the adaptive filter at any E/No, but is much more significant at high E/No because the inaccuracies become of the order of the MMSE. Notice how at a high $\sigma_q^2$ level, as shown in figure 4.31, the prediction equation is more accurate than the plain MMSE level. Under these conditions, the longer the equaliser, the more precise the prediction equation becomes. This can be observed in figure 4.35 where the equaliser length is extended to 28 taps. The graph clearly shows that equation (4.63) models the real MSE performance much better than the MMSE level alone.

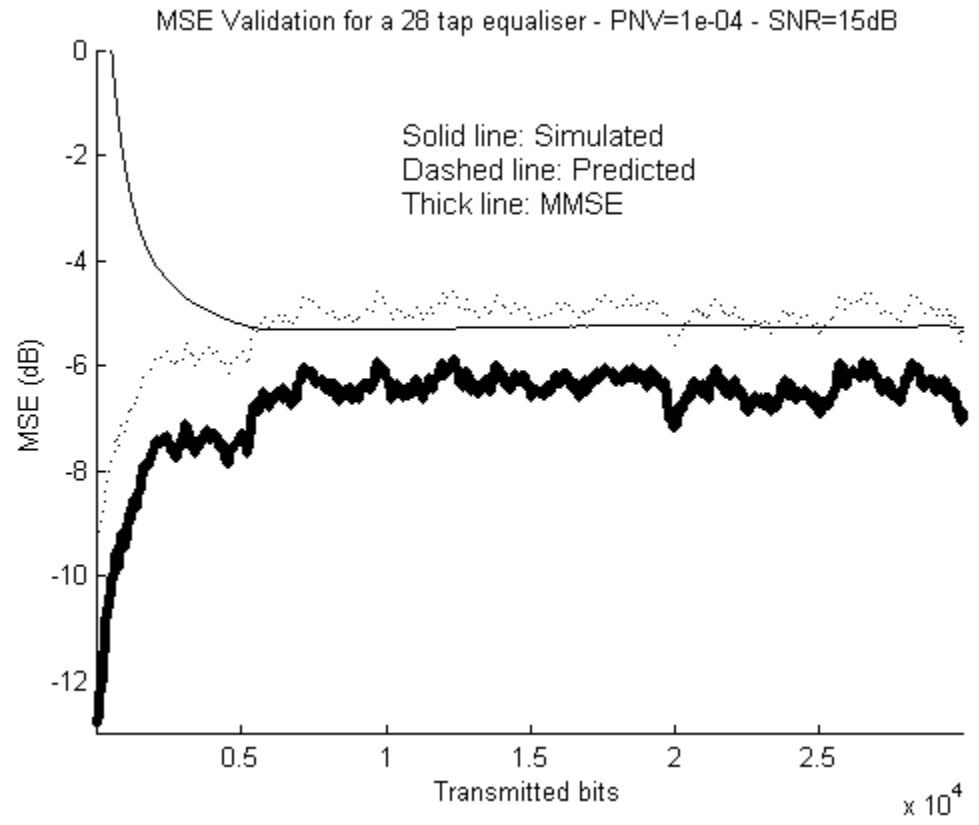

**Figure 4.35**: RLS MSE Validation, Channel Model 2, $\sigma_q^2 = 10^{-4}$, E/No=15dB, 28 Taps equaliser.

The effect of varying the equaliser length has also been considered to check that equation (4.63) is also able to predict such changes in the equaliser. Figures 4.36 and 4.37 show the predicted and measured MSE curves for 3 different length equalisers under two $\sigma_q^2$ levels. It can be seen in both graphs that the prediction lies in all cases within 1 dB of the true MSE value. Moreover, despite small deviations from the true MSE curves, the relative improvements of expanding the equaliser are accurately predicted. Notice the non-linearity of the reduction in the MSE level when enlarging the equaliser. Expanding the equaliser from 8 to 13 taps is far more significant than the expansion from 13 to 18 taps.

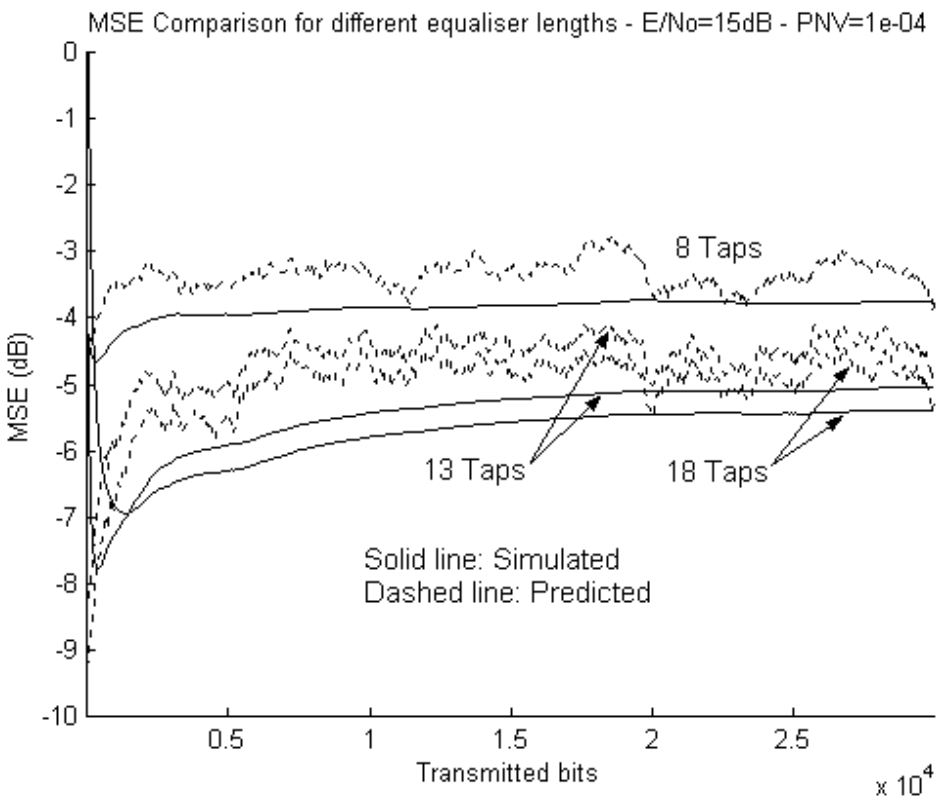

**Figure 4.36**: RLS MSE comparison for different length equalisers. Channel model 2. E/No=15dB. $\sigma_q^2=10^{-4}$.

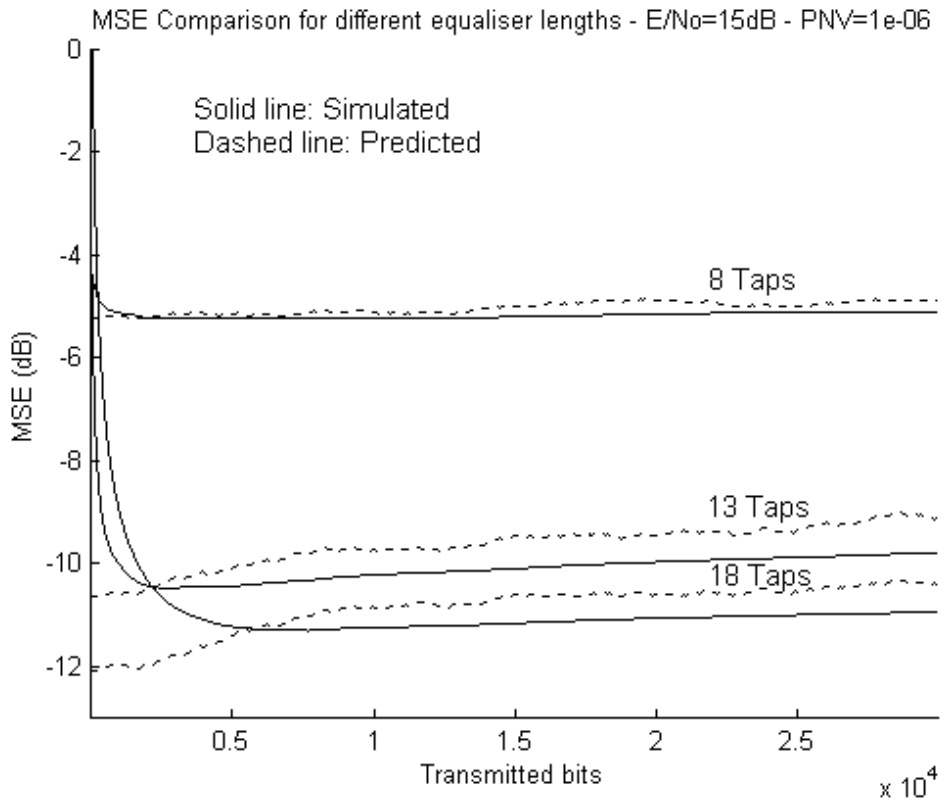

**Figure 4.37**: RLS MSE comparison for different length equalisers. Channel model 2. E/No=15dB. $\sigma_q^2=10^{-6}$.

Finally, this channel model has also been used to validate the equation for the optimum forgetting factor. Figure 4.38 shows the MSE performance in simulations using different arbitrary values of $\lambda$ and the one corresponding to the mean value of the prediction using equation (4.64). Clearly the one using average predicted optimum $\lambda$ performs very close to the optimum determined by simulation. If it was possible to plot a curve having as Y-axes the MSE level and as X-axes a continuum of $\lambda$ values, this curve would have a bowl shaped waveform with the values above and below the optimum attaining higher MSE levels. This graph can be found in [Eleftheriou86] for static channels. For dynamic channels, this graph would need to be a 3-D plot having axes MSE level, $\lambda$ and time, due to the time varying nature of the optimum $\lambda$. Unfortunately such a graph becomes extremely confusing and therefore is not shown here. However, observing the selected arbitrary values in figure 4.38, it is important to notice that slightly smaller and bigger values than the optimum perform worse than the optimum, thus this 3-D bowl shape can be inferred. The values $\lambda$=0.98 and $\lambda$=0.999 were also simulated, however their MSE curves were much worse than the ones shown and displaying them made it impossible to distinguish the rest of curves which are closer to the optimum value.

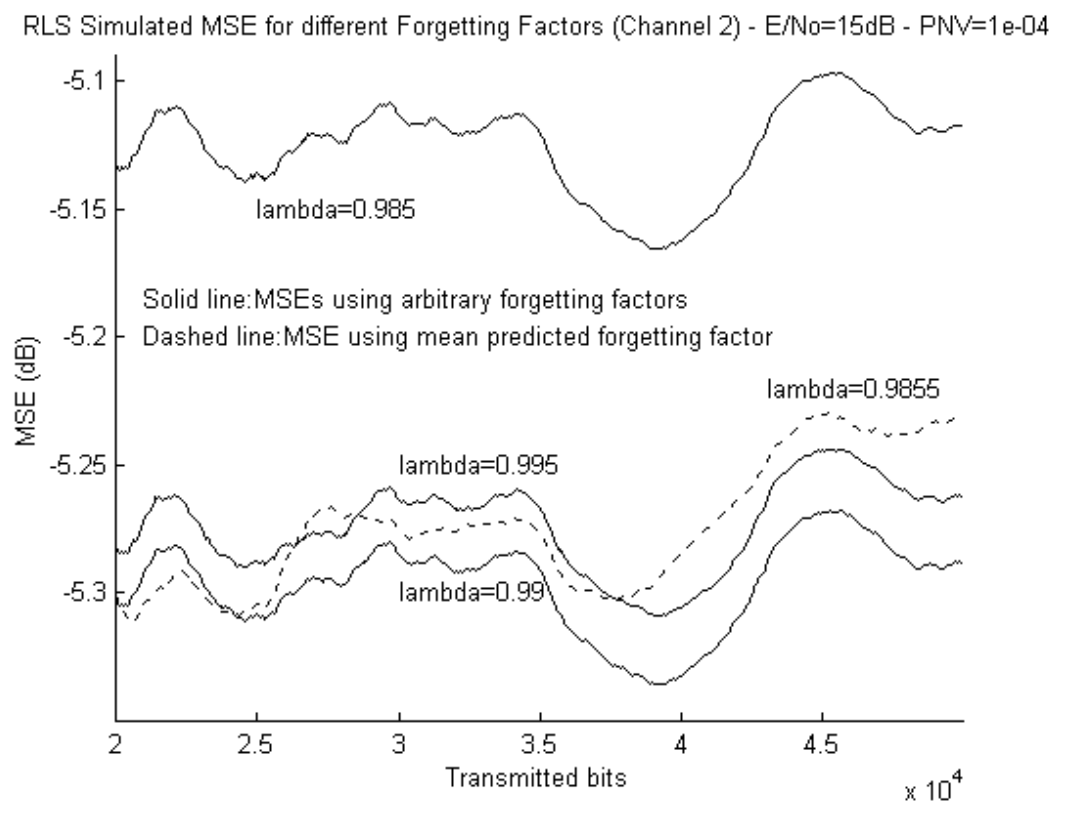

**Figure 4.38**: LMS MSE comparison using arbitrary and predicted optimum step sizes. Channel model 2. E/No=15dB. $\sigma_q^2=10^{-4}$.

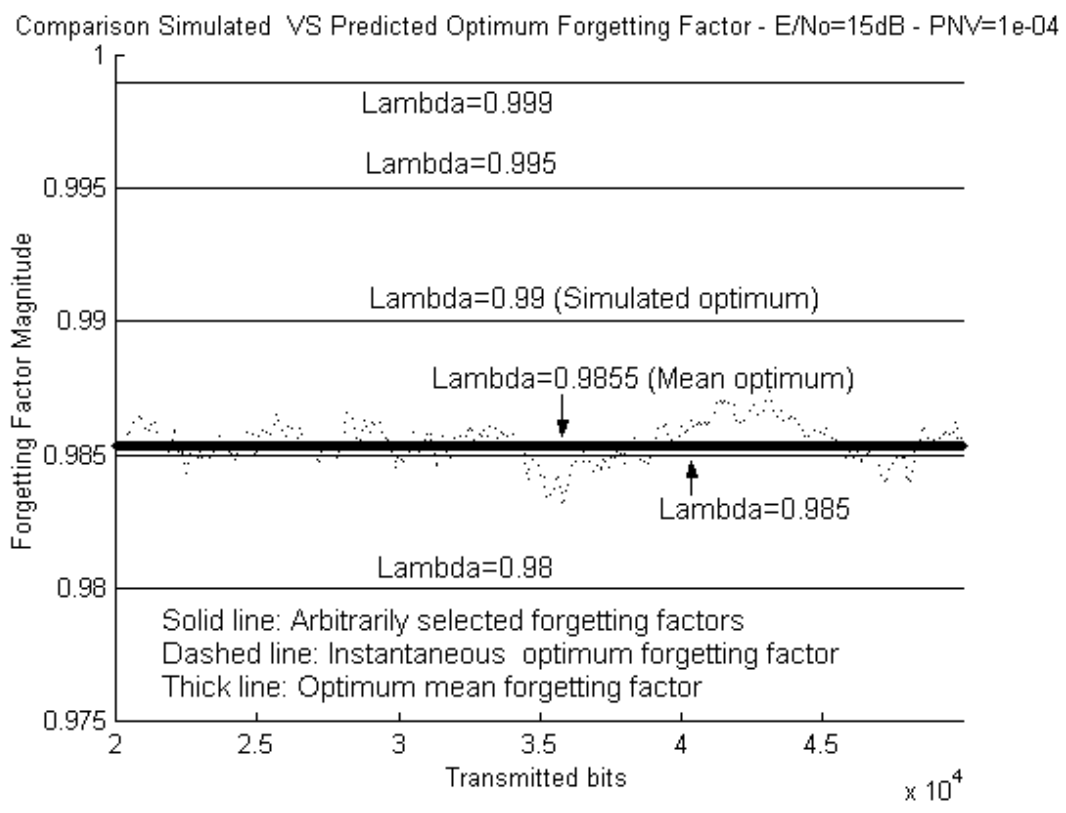

**Figure 4.39**: Forgetting factor magnitude comparison between predicted $\lambda$ and arbitrarily selected values. Channel model 2. E/No=15dB. $\sigma_q^2=10^{-4}$.

Other channel conditions ($\sigma_q^2=10^{-6}$ and different E/No levels) have been used to check equation (4.64) obtaining in all of them similar results, in terms of prediction accuracy, to those shown in the figures below.

Figure 4.39 gives an idea of the closeness of the predicted optimum to the simulated optimum. The picture also shows the instantaneous optimum predicted value. Notice how despite the channel variations being very large ($\sigma_q^2=10^{-4}$), the optimum forgetting factor does not experience very big changes. This supports the idea of using a fixed forgetting factor given the increase in complexity that supposes the use of variable forgetting factor algorithms.

## 4.7.3 Channel model 3

The last channel used for the verification of equations (4.62) and (4.63) is the channel model 3 described by figures 3.6 and 3.7. As has already been mentioned the very poor characteristics (the passband contains a spectral null) of this channel makes it extremely difficult to equalise using a linear structure.

The following figures (4.40-4.43) show the level of agreement between the predictions and the simulations. What has been said in the LMS section about the Channel model 3 is also applicable with the RLS algorithm, that is, the fact that the initial channel profile does not meet assumption 4.3.3 (see figure 4.8) compromises the accuracy of the predictions. When $\sigma_q^2 = 10^{-4}$ (figure 4.40), the channel tends to improve quickly and the prediction become very accurate. On the other hand when the channel variations are small, the spectral null remains in the passband and the predicted MSE is not very precise. Despite the inaccuracy, the prediction stays in all cases within 2 dB of the true value.

Comparing the set of figures above with those corresponding to the same channel model using the LMS algorithm (figures 4.23 to 4.25). The superiority in convergence speed of the RLS algorithm when the channel is heavily distorted is clearly seen, especially when E/No is large.

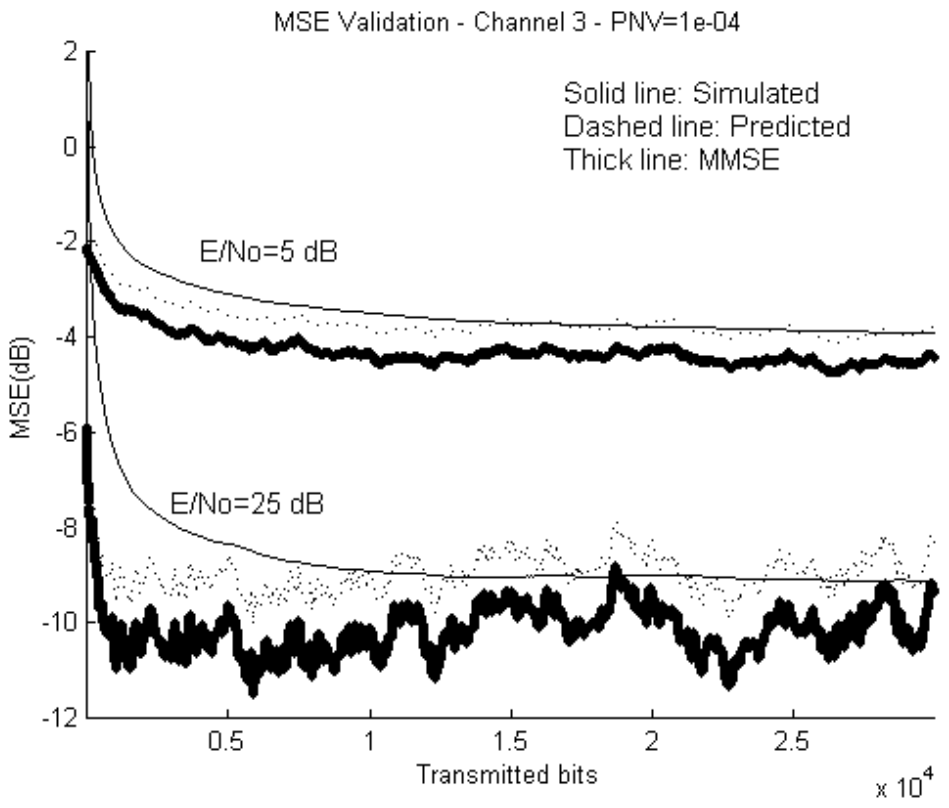

**Figure 4.40**: RLS MSE validation for Channel model 3. $\sigma_q^2 = 10^{-4}$.

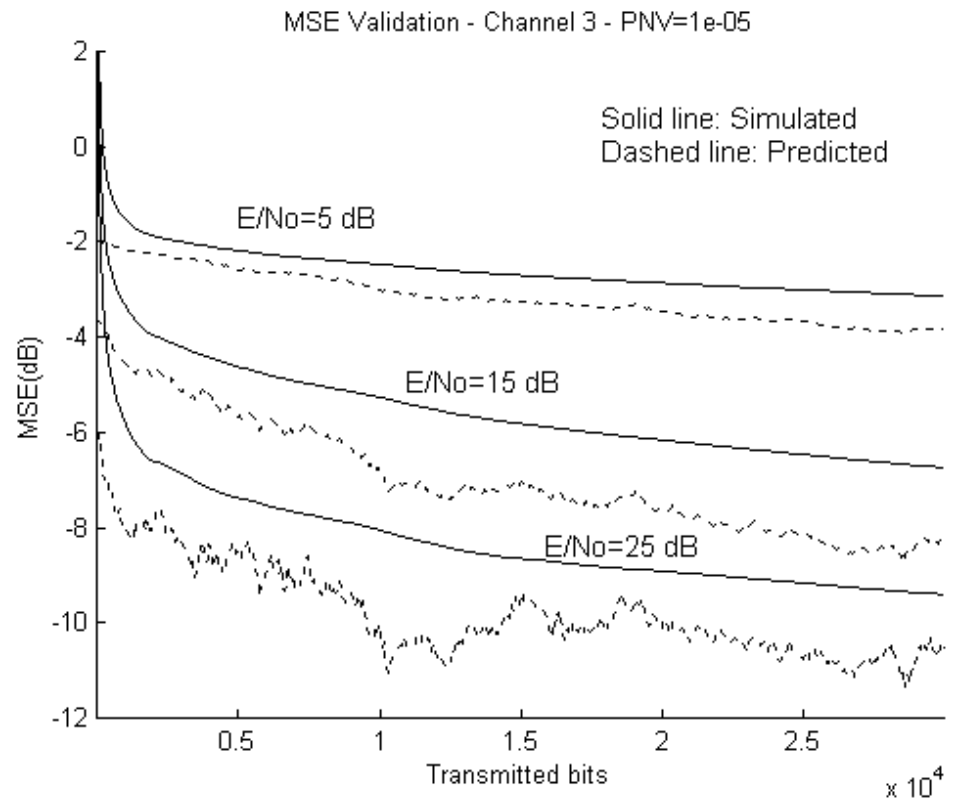

**Figure 4.41**: RLS MSE validation for Channel model 3. $\sigma_q^2 = 10^{-5}$.

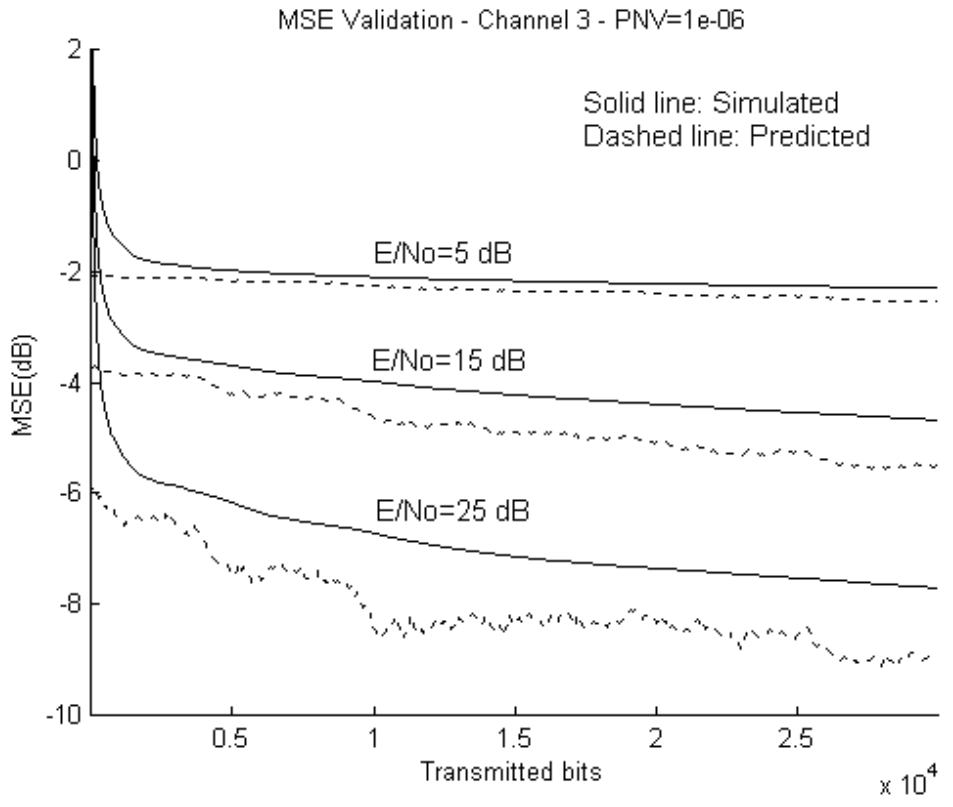

**Figure 4.42**: RLS MSE validation for Channel model 3. $\sigma_q^2 = 10^{-6}$.

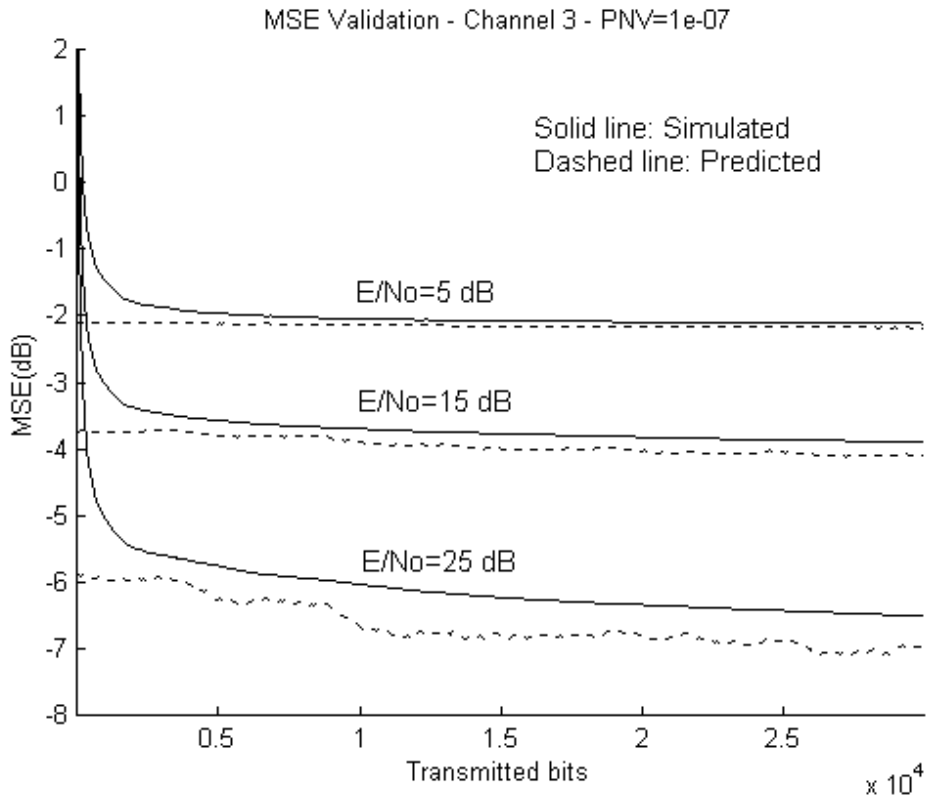

**Figure 4.43**: RLS MSE validation for Channel model 3. $\sigma_q^2 = 10^{-7}$.

# 4.8 Limitations of the LMS/RLS MSE equations

Over the previous sections of this chapter simulation results have been presented showing that, in general, the derived MSE expressions model quite accurately the real/simulated MSE behaviour of the LMS and RLS linear equalisers. It is worth now pointing out the accuracy limitations of the derived equations, and when the real MSE deviates significantly from the prediction.

The developments presented in sections 4.4 and 4.6 are based in the set of assumptions stated in 4.3. Obviously if the assumptions do not hold then equations (4.31) and (4.63) will not precisely model the true MSE of the equaliser. The assumptions concerning the channel model and its power response can be assumed to be true, as explained in detail in sections 4.3.1 and 4.3.2.

The hypotheses concerning the optimum equaliser variations and the independence of the input data vector to the equaliser may compromise the matching of equations (4.31) and (4.63) with the measured results. Figure 4.8 is an important one as it shows when the assumption 4.3.3 (optimum equaliser variation) will not hold at all. The mismatch is caused because the channel spectrum contains a deep spectral null in the passband (see section 4.3.3 for a detailed explanation). Spectral nulls are caused when the channel introduces a high degree of correlation in the originally uncorrelated signal, introducing large ISI components in the sample to be detected. At the same time, the introduction of very large ISI causes the successive data vectors at the input of the equaliser to become more strongly dependent, compromising in this way the independence assumption. Hence it is clear that the satisfaction of these two assumptions is jeopardised by the same phenomenon, the existence of spectral nulls in the channel frequency response. Therefore the validity of (4.31) and (4.63) is constrained to channels whose spectrum do not contain deep nulls. As an example of this effect it has already been shown in previous sections that the MSE prediction equations are not very accurate for channel model 3 (this channel contains a –80dB null in the passband).

Fortunately, most channels are relatively well behaved and for most of the time they do not contain very deep spectral nulls. Otherwise alternative structures (DFE, MLSE) should be used, as linear equalisation will not be able to improve very much the performance of the receiver.

Additionally, during the mathematical development of (4.31) and (4.61) some approximations have been made in order to make the derivations feasible. In general these simplifications are justified and they do not compromise greatly the accuracy of the MSE expressions. The only exception is the approximation in the RLS derivation, discarding the second term in (4.54). Ideally, this term should have been taken into account, but no way has been found to compute a closed form expression for the inverse of $\mathbf{\Phi}(n)$ as it appears in (4.54). Nonetheless simulation results have shown that this simplification, albeit crude, does not affect drastically the accuracy of the final equation.

# 4.9 Transient considerations

The focus of this chapter has been so far on the steady state behaviour of the MSE. This is justified because in typical links, the transient phase is just an extremely small part of the total transmission time.

Some attention is now given to the MSE convergence stage of a linear equaliser for the sake of completeness, and also because some of the techniques presented in following chapters are also applicable to the transient stage. The convergence of the LMS and RLS algorithms differ radically from each other as described briefly next.

## 4.9.1 LMS transient

The transient MSE for the LMS algorithm has been extensively studied since the first published works in equalisation ([Ungerboeck72], [Gitlin79]). These papers present the convergence analysis for static channels, and a recursion describing the transient MSE evolution is obtained.

The basic approach to obtain this expression consists of diagonalizing the autocorrelation matrix of the input data, **R**, using the Unitary Similarity Transformation ([Noble88]). The mean squared error can thus be expressed in terms of the eigenvalues and eigenvectors of **R** and, from the connection between eigenvalue structure and channel spectrum, some bounds to the convergence time can be established for particular channel characteristics. The final recursive expression for the MSE evolution during the transient stage is given by:

$$\begin{aligned} MSE(n+1) \le & [1-2\mu\lambda_{av}+\lambda_M M(\sigma_d^2+\sigma_v^2)]MSE(n) \\ & +\lambda_M M(\sigma_d^2+\sigma_v^2)\mu^2 MMSE + MMSE \end{aligned} \qquad (4.65)$$

where M is the equaliser length, $\mu$ is the step size, MMSE is the minimum mean squared error, $\sigma_d^2$ is data signal power, $\sigma_v^2$ is the noise power and $\lambda_{av}$, $\lambda_M$ denote the average and largest eigenvalue of **R**. The first term on the right hand side of (4.65) constitutes the transient component, the second term corresponds to the steady state EMSE and the third to the MMSE.

Unfortunately when the channel is time varying, the autocorrelation matrix of the input data is also time dependent (see 4.22 and 4.23) and the same approach cannot be followed. However simulation results reveal that during the initial period of convergence the MSE is not very sensitive to the magnitude of the channel variations. This can be observed in figures 4.44 and 4.45 below.

In the case of E/No=5dB (figure 4.43) it can be seen that during the first 500 iterations all four curves converge at nearly the same rate. When E/No=25dB and focusing over the first 700/800 bits, the differences become slightly more notable, specially between the $\sigma_q^2=10^{-4}$ curve and the rest of $\sigma_q^2$ values, but are still not very significant.

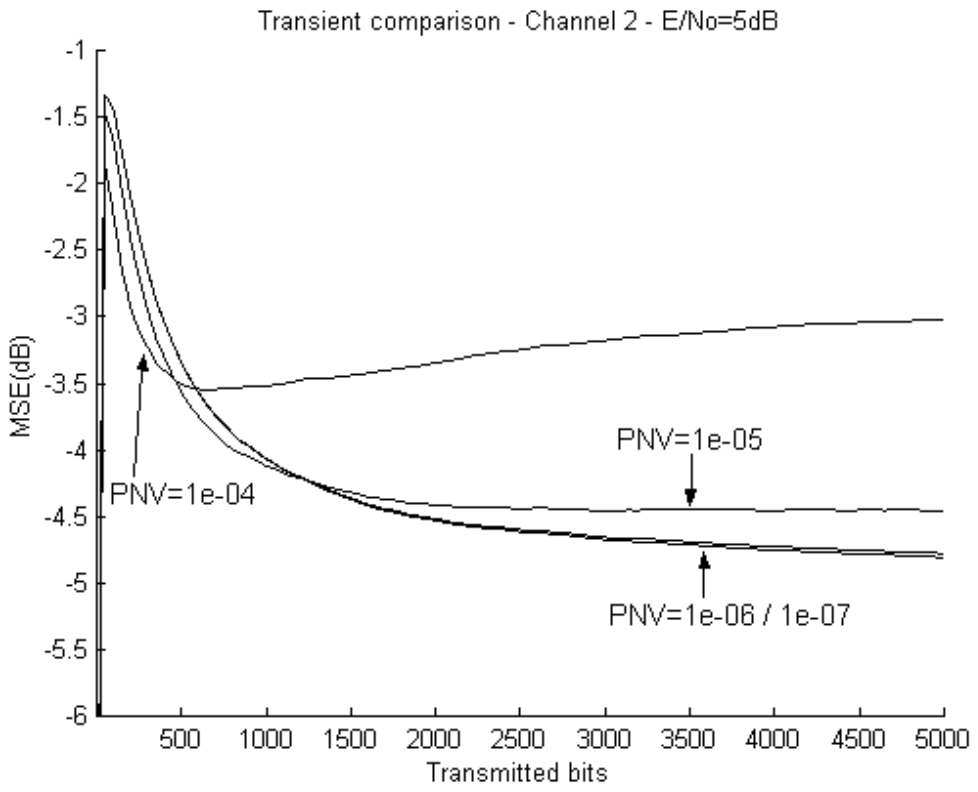

**Figure 4.44**: LMS Transient comparison for different $\sigma_q^2$ levels. Channel model 2. E/No=5dB.

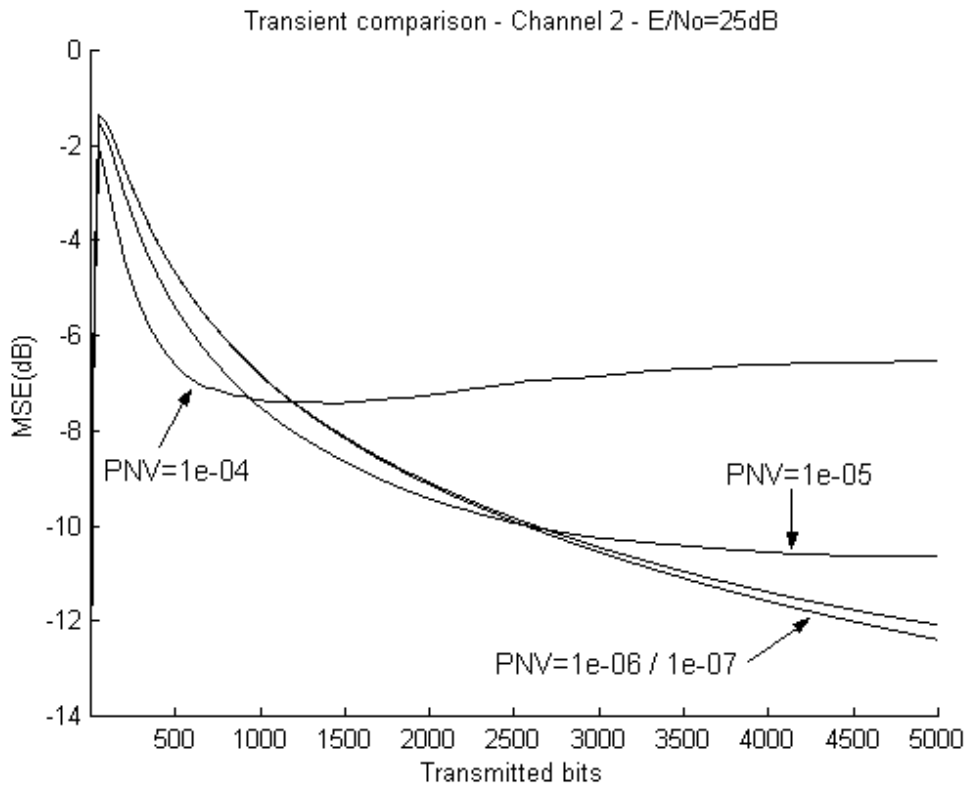

**Figure 4.45**: LMS Transient comparison for different $\sigma_q^2$ levels. Channel model 2. E/No=25dB.

This supports the thesis that, during the initial part of the transient, MSE convegence is only weakly dependent on the channel dynamics. This is equivalent to saying that during the first $\delta$ iterations, the channel coefficients can be assumed to be nearly static, which in turn implies that $\mathbf{R}(0) \cong \mathbf{R}(\delta)$, where $\mathbf{R}(0)$ is the autocorrelation corresponding to the initial channel profile and $\mathbf{R}(\delta)$ the autocorrelation corresponding to the channel after $\delta$ iterations have ellapsed. Obviously, the larger $\sigma_q^2$ is, the smaller the value for $\delta$ for which this assumption holds true. With the assumption of $\mathbf{R}(n)$ being nearly stationary over the first $\delta$ transmitted bits, (4.65) can be now used to model the MSE evolution over this period of time (from 0 to $\delta$).

## 4.9.2 RLS transient

The RLS transient performance is easier to analyse given the extremely good convergence properties of least squares algorithms. In stationary environments, the MSE convergence of the RLS is given by [Haykin96]:

$$\mathrm{MSE}(n) = \mathrm{MMSE} + \frac{M}{n - M - 1}\mathrm{MMSE} \qquad (4.66)$$

Convergence is achieved in about 2M iterations, where M is the order of the equaliser.

As in the LMS case, the channel is considered to be near stationary during the first steps of the convergence process. Clearly for any reasonable value of $\sigma_q^2$ the channel does not vary significantly in 2M iterations, so equation (4.66) can be assumed to be true also for dynamic channels.

# 4.10 Conclusions on the derived MSE equations

This fairly long chapter has presented the derivations of steady state MSE expressions for a linear equaliser using the LMS and RLS algorithms. Simulation results show that in normal channel conditions, they model very accurately the true MSE performance. The resulting equations have the same structure as their channel estimation counterparts but with two significant differences, namely, the variation of the MMSE level with time and the dependence on the equaliser length (which in general will be different from the channel length). Given their importance both equations are shown again here:

$$MSE(n)=MMSE(n)+M[\sigma_d^2+\sigma_v^2]\frac{\mu}{2}MMSE(n)+\frac{N\sigma_q^2}{2\mu} \qquad \text{LMS MSE}$$

$$MSE(n)=MMSE(n)+\frac{M(1-\lambda)}{2}MMSE(n)+\frac{N\sigma_q^2(\sigma_d^2+\sigma_v^2)}{2(1-\lambda)} \qquad \text{RLS MSE}$$

Observing both expressions it should be clear that the instantaneous MSE level clearly depends on the instantaneous MMSE level. One may ask what is the utility of these expressions when the instantaneous MMSE will not typically be known. Their usefulness can be justified from two distinct points of view.

First, although the instantaneous MMSE might not be known, in many situations an approximate average value can be extracted from the initial channel profile. Notice that the Markov channel model used in this chapter is indeed a pessimistic approximation to real radio channels, as in the Markov model the channel variations are completely random. In realistic mobile wireless systems, channels, although being time varying, tend to conform to a determined channel profile (such as the COST207 channel models). From these profiles, an estimate of the mean MMSE can be computed. If an average MMSE is used in place of MMSE(n), both equations become time independent and they are reduced to a single numerical value. Notice however that the time varying nature of the channel is still taken into account with the 3rd term of each equation.

The second, and from the point of view of this project, more important reason why the derived expressions are valuable is due to the qualitative information they reveal, in particular, the parameters upon which the MSE level depends. Looking at the equations and assuming a given signal power level, only two parameters can be adjusted; the step size/forgetting factor and the equaliser length. The rest of the parameters are either given by the environment itself (N, $\sigma_v^2$ and $\sigma_q^2$) or they are not directly tractable (instantaneous MMSE). Of the two available variables, one, the algorithm parameter ($\mu$ or $\lambda$) has already been considered extensively in the literature in the form of the variable step size/forgetting factor algorithms ([Kwong92], [Mathews93]). The other parameter, the equaliser length (M), has received little attention and therefore is thoroughly studied in this project.

The equaliser length has an explicit relation to the steady state MSE level as it appears in the second term of the two derived MSE expressions. Of equal or even greater importance is the implicit relation of the equaliser length with the MMSE, as the MMSE is determined by the specific channel impulse response, noise level ($\sigma_v^2$) and length of the adaptive filter (M). This last variable is the only one that can be manipulated. Notice that the MMSE is independent of any algorithm parameters, and indeed, is independent of the algorithm being used. This implies that equaliser length adjustment can be as crucial as the choice of step size/forgetting factor to achieve satisfactory performance. Also, in the case of the LMS algorithm, the range of values of the step size is directly determined from the number of taps of the adaptive filter ([Widrow76]).

The key point to be appreciated with respect to the equaliser length is that increasing the equaliser length will not always result in a reduction in the overall MSE level. Increasing the number of taps of the equaliser will result in a reduction of the MMSE level but will also imply an increase in the EMSE level that may be greater than the MMSE reduction. This should be taken into account when choosing the equaliser length. Additionally, the filter order directly influences the complexity of the equalisation process, so the equaliser length can serve as a "tuner" to trade performance with complexity (i.e. power consumption).

In the next chapter, a thorough study of the equaliser length is presented along with techniques to adjust it in an efficient manner.

# 5 VARIABLE LENGTH LINEAR EQUALISERS

The steady state MSE equations for the LMS and RLS algorithms derived in the previous chapter reveal a strong dependence on the equaliser length, therefore, care should be taken to choose an adequate number of taps of the adaptive filter. In this chapter, first the problem of length selection is covered in detail for both static and dynamic channels. Then, a novel technique is presented for adjusting the equaliser length automatically in static and time varying scenarios. Extensive simulations in a wide variety of channel conditions are presented showing the robustness of the proposed technique. In this chapter, the performance measure used is the MSE. The advantages of this technique from the BER point of view are deferred until chapter 7 where the technique is used in a more realistic scenario. As in the previous chapter, the focus of the work is on the steady state phase of operation although the advantages of using the proposed technique during the transient stage are briefly treated towards the end of the chapter.

## 5.1 Equaliser length selection problem

Given a known channel impulse response with N taps and a given E/No level, there are no rules predicting how many taps will be needed to achieve an irreducible MMSE level. One possible solution to this problem is to set the equaliser length, M, to a very large number of taps, typically several times N. This approach presents two problems. First, in many cases the equaliser will be overdimensioned as a much shorter equaliser would achieve an almost identical MMSE level. The computational waste provoked by system overdimensioning might be a concern in systems with restricted power consumption such as mobile terminals. Second, although increasing the equaliser length will never result in an increase in the MMSE level, it may very well cause an increase in the overall MSE level due to the EMSE introduced if an adaptive algorithm such as the LMS is used.

Another possibility would consist of successively computing the Wiener-Hopf equation for different equaliser lengths and determining from the results what the adequate length should be. The next two figures show the computed MMSE for a wide range of equaliser lengths for the Channel models 2 and 3 introduced in chapter 3 under different E/No conditions. The decision delays have been set to 8 samples for Channel model 2 and 5 samples for Channel model 3.

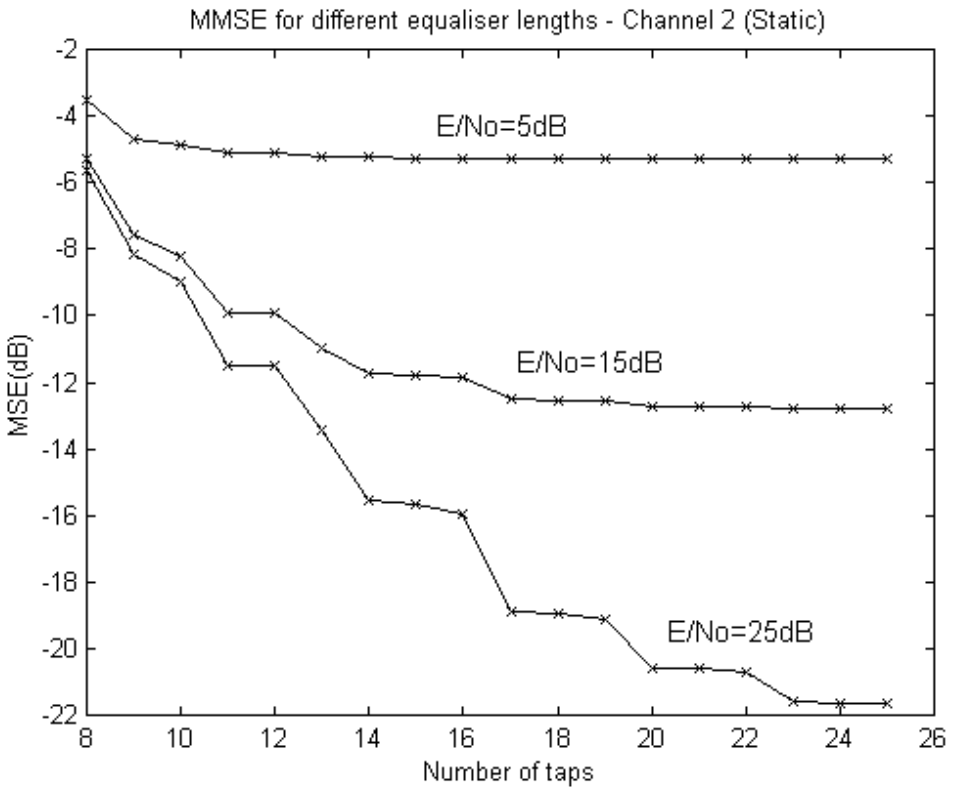

**Figure 5.1**: MMSE level for different length equalisers. Channel model 2 (Static).

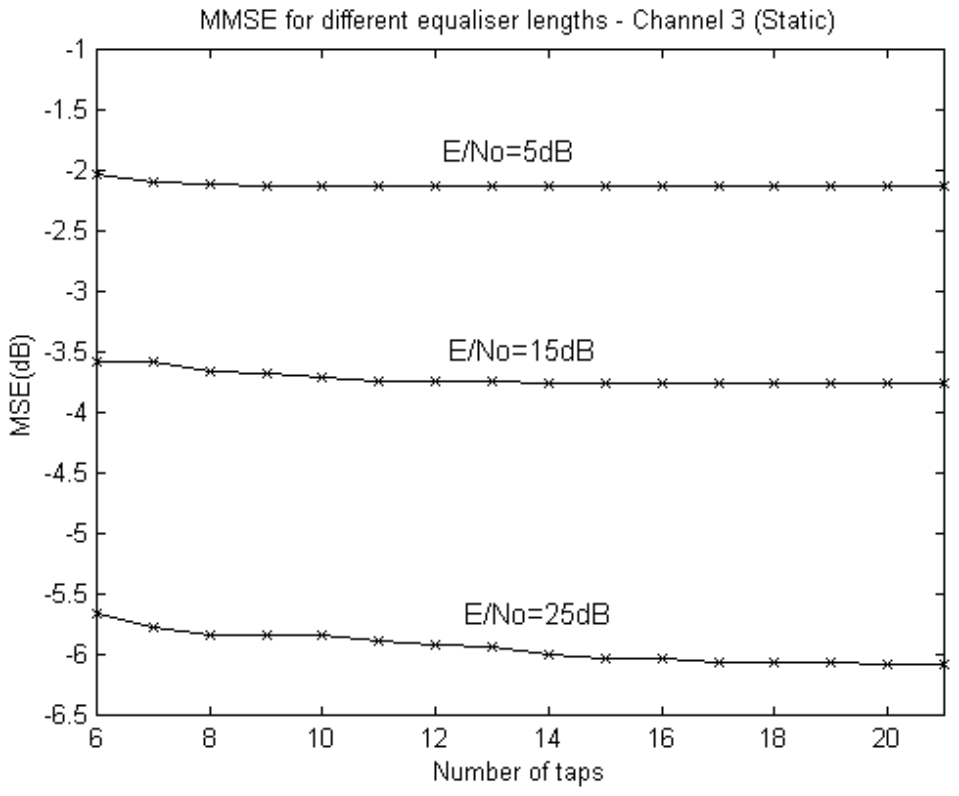

**Figure 5.2**: MMSE level for different length equalisers. Channel model 3 (Static).

From the curves in figures 5.1 and 5.2 can be deduced the optimum[17] lengths for different channels and E/No levels. Clearly this optimum varies from curve to curve. This method would avoid the first of the problems mentioned above (overdimensioning) as the length is chosen very accurately but the algorithm imperfection is still not taken into account. In the next set of figures, the overall MSE level obtained using the LMS and RLS algorithms is compared with the theoretical achievable MMSE levels for different equaliser lengths using the same channel models as in the previous figures. The LMS-MSE and RLS-MSE values have been obtained by simulation, letting the equaliser converge fully before recording the MSE level attained.

[17] By optimum is meant the minimum length that achieves the minimum MMSE level in each curve.

When the RLS algorithm is used (figures 5.4 and 5.6) the steady state MSE achieves the MMSE level for the different lengths. The tiny divergences between the MMSE and the RLS-MSE shown in the curves when the equaliser has many taps are basically due to numerical inaccuracies in the computation of the MMSE. Although the RLS-MSE level does not increase when the equaliser is made very long, notice that it does not decrease either, therefore useless calculations are being made. The RLS computational complexity is $2M^2$ products and $1.5M^2$ additions per iteration (see chapter 2).

Given this high computational load, it is advisable to keep the equaliser as short as possible. Take for example the curve for E/No=15 dB in figure 5.4; the MSE level for a 17-tap equaliser is basically the same as with a 25-tap equaliser. The operations corresponding to these 8 extra taps, 128 products and 96 additions per iteration, do not yield any significant benefit and are therefore wasted. This waste of resources is even more evident for lower E/No and for all curves in Channel model 3 (figure 5.6).

When the LMS algorithm is used, making the equaliser too long will end up increasing the MSE level. This effect can be observed in the curve for E/No=5 dB in figure 5.3 (channel model 2) as equalisers longer than 15 taps increase the overall MSE level. The phenomenon is much more evident in figure 5.5 (Channel model 3) for any E/No level, especially for E/No=5 dB, where the MSE is about 1dB worse for the 21 taps than it is for the 6 tap equaliser. Therefore, in the LMS case there are two deleterious effects of excessively long equalisers: increase in the overall MSE level and waste of computations.

So far it has been assumed that channels were static and known. In practical scenarios, such as mobile or wireless systems, the channel will be unknown and probably time varying, therefore the number of taps the equaliser cannot be determined a-priori and can also vary with time as conditions in the environment change.

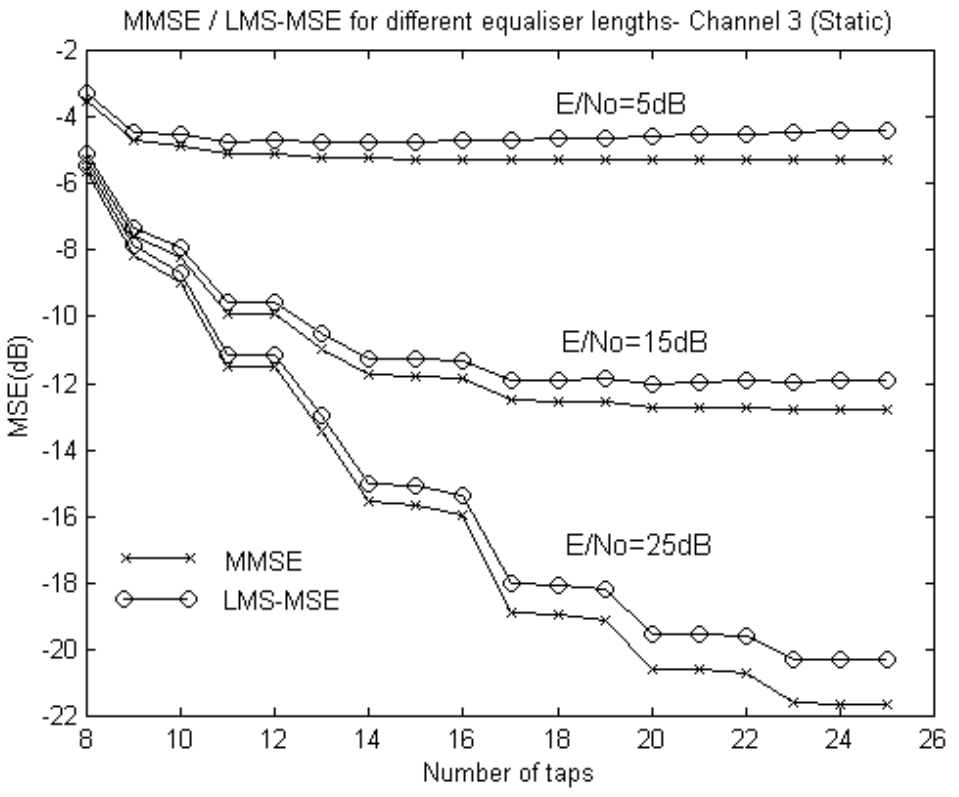

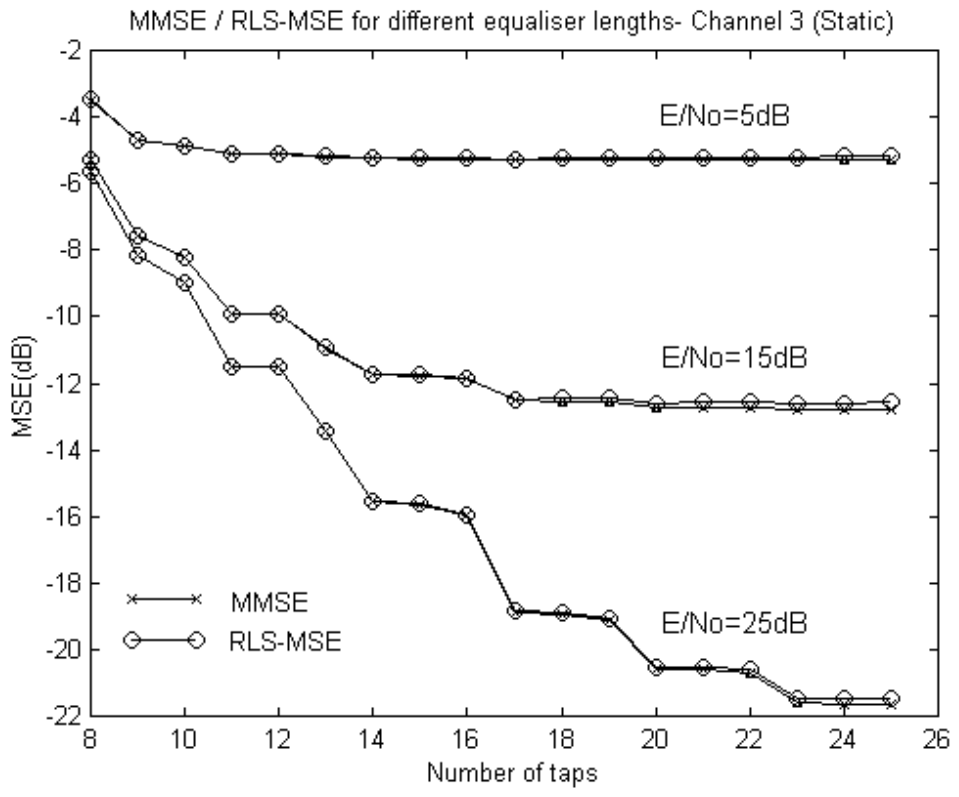

**Figure 5.3**: MMSE / LMS Steady State MSE comparison for different length equalisers. Channel model 2 (Static).

**Figure 5.4**: MMSE / RLS Steady State MSE comparison for different length equalisers. Channel model 2 (Static).

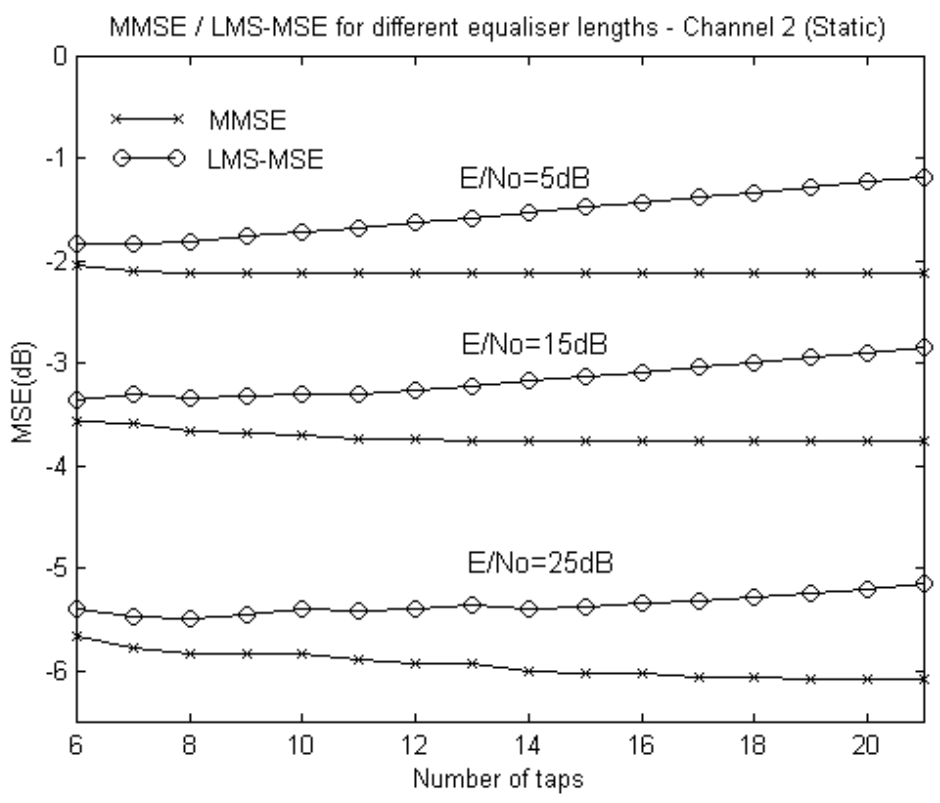

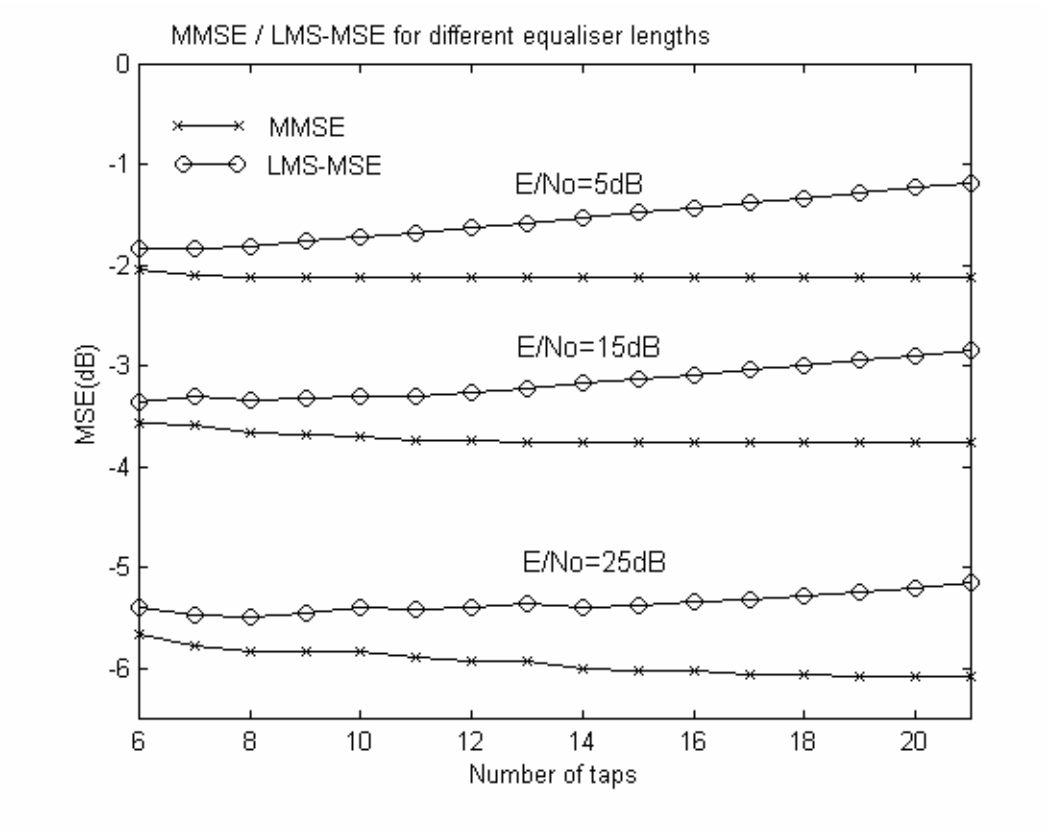

**Figure 5.5**: MMSE / LMS Steady State MSE comparison for different length equalisers. Channel model 3 (Static).

**Figure 5.6**: MMSE / RLS Steady State MSE comparison for different length equalisers. Channel model 3 (Static).

This fact has been the main motivation of the study of reconfigurable equalisation structures; to help in better tracking of channel conditions and to avoid unnecessary waste of computational power.

## 5.2 Variable length linear equalisation principles

All adaptive equalisation subsystems are composed of a filtering structure and an adaptive algorithm. Typical filtering structures and adaptive algorithms have been covered in sections 2.2 and 2.3 in chapter 2 respectively. So far, most of the research effort on adaptive equalisation has concentrated on the algorithmic side whereas the work presented in this thesis focuses more on the structures. In this section a novel reconfigurable structure based on a linear equaliser is presented along with an algorithm to update it. This new type of equalisation correctly guesses the appropriate number of taps the equaliser should have at every instant depending on the channel conditions. This flexibility can potentially lead to improved performance (from an MSE point of view) with respect to fixed equalisers and a potential reduction in the number of computations of the equalisation process. First the structure and algorithm are presented and then some practical rules in order to choose appropriately its parameters are stated.

## 5.2.1 Segmented FIR filters for linear equalisation

A typical linear equaliser consists of an FIR filtering structure followed by a threshold detector as depicted in figure 2.3. The concept of segmented filtering introduced in this work is based on the idea of partitioning the original FIR equalising filter into K smaller and concatenated P-tap subfilters where M=KP. Figure 5.7 illustrates this idea:

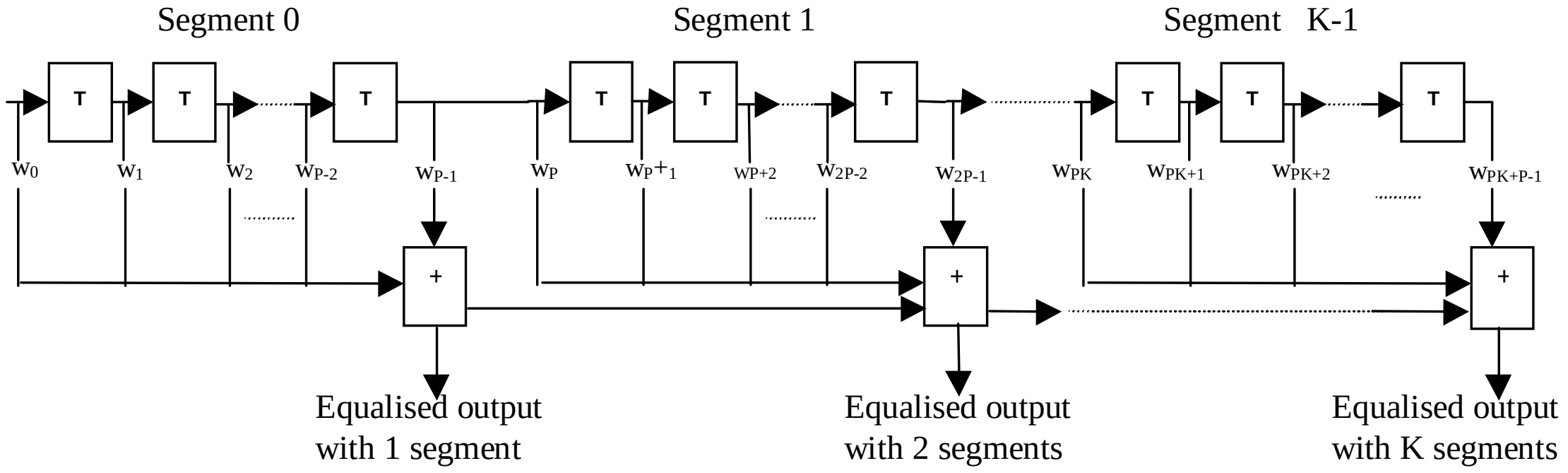

**Figure 5.7:** Structure of a segmented FIR filter.

Having decomposed the original filter into K segments, it is now possible to compute the output of each subfilter separately at any given instant. Connecting serially the adders of each segment to the following one, the output of one segment corresponds to the accumulated filtering operation up to that segment.

Notice that provided KP=M, the output of the last segment of the segmented equaliser gives the same results as the output of a conventional M-tap FIR filter. Comparing the structure of figure 5.7 with the one from an ordinary filter, the segmented filter needs K-1 extra adders. Later on it will be shown that combining this structure with the algorithm presented in the next subsection only requires one extra adder.

Any other properties the original equaliser may have, such as linearity, are preserved when it is converted into a segmented structure. Basically what this new structure allows is to "monitor" how the filtering operation is progressing.

Figure 5.8 shows how segmented filters can be used in a linear equaliser. From this graph it is clear that each segment produces a partially equalised output. Ideally, each successive segment should improve the equalisation process making its output closer to the detected bit than the previous segment's output, in this case $e_o(n) \geq e_1(n) \geq \cdots \geq e_{K-1}(n)$. In practice it may happen, as has been shown through figures 5.3 to 5.6, that the last segments/taps may increase the error level. However with the partitioned architecture it is possible to detect when this happens and do something about it. The next section introduces an algorithm that controls the number of segments in the equaliser to avoid this situation.

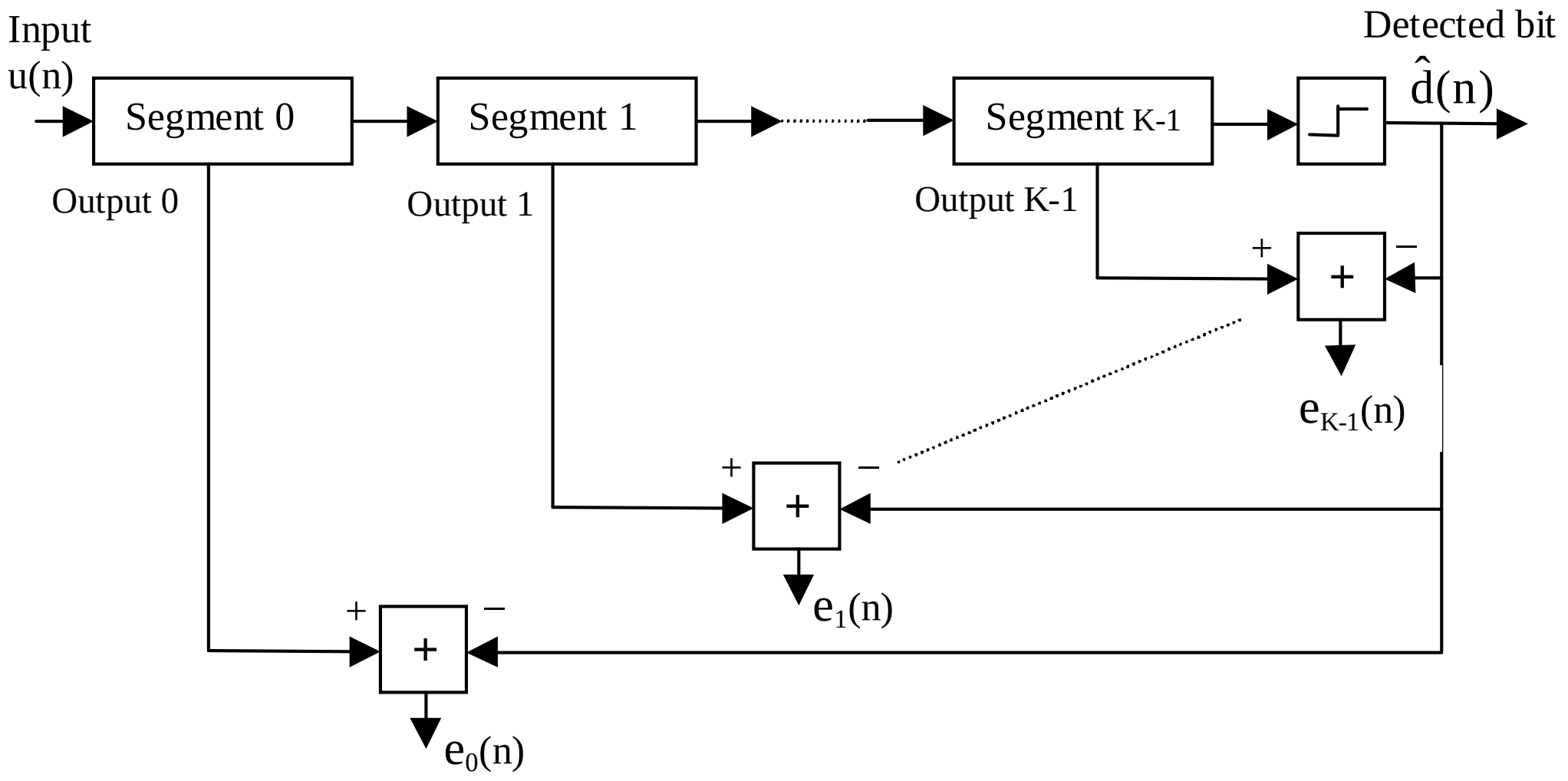

**Figure 5.8:** Segmented linear equaliser.

## 5.2.2 Structure update algorithm

The segmented equaliser structure introduced in the previous section does not offer any significant benefit over the conventional linear equaliser unless there is some means to modify the equaliser length dynamically. The structure depicted in figure 5.8 has the potential to increase or decrease the number of taps used in the equalisation process, however there is still the question of what criterion to follow in order to reconfigure the filter in an efficient manner. This algorithm should have two qualities:

- It should only use the information available to the equaliser subsystem, that is, no external information should be required (such as noise level, Doppler spread).

- It should be easy to compute. There is no point in using criteria that require lot of computations, as then the benefits of reducing the number of taps might be lost.

The criterion proposed here, termed accumulated squared error (ASE), is given by:

$$ASE(n)=\sum_{i=0}^{n}\left|d(n)-y(n)\right|^2=\sum_{i=0}^{n}e(n)^2 \tag{5.1}$$

Notice that the ASE(n) has exactly the same definition as the least squares criterion from which the RLS algorithm is derived. It has been decided to refer to it with a different name because its use is rather different from that in the least squares algorithms. Note that the ASE(n) and MSE(n), assuming ergodicity and reasonably large n, are related by:

$$MSE(n)=\frac{ASE(n)}{n} \tag{5.2}$$

Looking again at figure 5.8, it is clear that a distinctive $ASE_i(n)$ value can be computed for every segment, using the different error signal $e_i(n)$ available at every segment's output. In this way the measures $ASE_0(n)$, $ASE_1(n)$,…$ASE_{K-1}(n)$ are obtained.

It is now possible to state an algorithm that dynamically updates the length of the equaliser. Suppose that at any given moment an equaliser made of K segments of P taps/segment is only using the first L segments with $L \le K$, the rest of the segments (K-L) have their taps set to zero and they are not being updated. At every iteration, the $ASE_L(n)$ and $ASE_{L-1}(n)$ are being computed and the following decision algorithm is executed:

$$ASE_{L-1}(n)=\sum_{i=0}^{n}\left|d(n)-y_{L-1}(n)\right|^2 \tag{5.3}$$

$$ASE_{L}(n)=\sum_{i=0}^{n}\left|d(n)-y_{L}(n)\right|^2 \tag{5.4}$$

If $ASE_L(n) \le \alpha_{up} ASE_{L-1}(n) \quad \rightarrow \quad$ Add one segment (P extra taps) (5.5)

If $ASE_L(n) \ge \alpha_{dw} ASE_{L-1}(n) \quad \rightarrow \quad$ Remove one segment (P less taps) (5.6)

with $0 < \alpha_{up}, \alpha_{dw} \le 1$ and $\alpha_{up} \le \alpha_{dw}$.

**Algorithm 5.1**: Basic equaliser length control.

The function of the variables $\alpha_{up}$ and $\alpha_{dw}$ is to determine the amount of improvement or worsening necessary to force the equaliser to expand or contract.

There are three possible outcomes when algorithm 5.1 is executed. If the comparison in (5.5) is true, the equaliser will be expanded by P taps. If the comparison in (5.6) holds true then the equaliser will be reduced by P taps. If none of the comparisons hold true, the equaliser will be left as it is. Provided that $\alpha_{up} < \alpha_{dw}$, both comparisons cannot be true at the same time (expansion and contraction).

This algorithm can be translated into plain words as: If an equaliser with L segments performs significantly better than one with L-1 segments then add another segment. If an equaliser with L segments performs similarly to one with L-1 segments remove the last segment. This simple heuristic limits the number of segments to those that really make a significant contribution in the equalisation process. Notice that algorithm 1 satisfies the qualities sought: it only relies on information available to the equalising subsystem (only uses the different $e_i(n)$), and its computational complexity is low, requiring only four products and four additions per iteration. Also important is the fact that with the given algorithm only the performance of the last two segments is monitored, therefore only one extra adder is required.

### 5.2.3 Parameter selection

Combining the segmented FIR structure and the basic length control algorithm requires four parameters to be determined, namely, K (maximum number of segments), P (taps per segment), $\alpha_{up}$ and $\alpha_{dw}$. The maximum number of taps of the equaliser, M, is given by KP. At this point attention is only given to the qualities and relations of each of the parameters. Simulation results with variations of the parameters will be presented later on.

The parameter K determines, jointly with P, the maximum number of taps of the equaliser. For a given value of P, K defines the amount of equalisation the system can potentially perform. K´s choice will be mainly conditioned by the computational/power resources available. The parameter P is more interesting from a system point of view as it provides the degree of "equalisation granularity" of a segment. By equalisation granularity is meant the amount of reduction in MSE level a segment can achieve. This concept is better explained with an example. Suppose the total number of taps of the equaliser (KP) is 12, then the equaliser can be partitioned into segments of length 1, 2, 3, 4, 6 or 12 taps. For simplicity, only two of the partitions are considered: 2 segments of 6 taps/segment (referred as 2_6 equaliser) or 4 segments of 3 taps/segment (4_3 equaliser). The 2_6 equaliser generates two error signals, one for each segment, denoted by $e_0^{2-6}(n)$ and $e_1^{2-6}(n)$. Similarly the 4_3 equaliser generates four error signals: $e_0^{4-3}(n)$, $e_1^{4-3}(n)$, $e_2^{4-3}(n)$ and $e_3^{4-3}(n)$. Assuming both equalisers are fully in use then $e_0^{2-6}(n) = e_1^{4-3}(n)$ and $e_1^{2-6}(n) = e_3^{4-3}(n)$ but notice that the 4_3 equaliser offers extra information as the signal $e_0^{4-3}(n)$ provides how the equalisation is progressing during the first 3 taps. The same applies for the signal $e_2^{4-3}(n)$. This extra information aids in deciding if the equaliser length should be altered.

Large values of P imply an equaliser with very few partitions and big differences (coarse granularity) among the distinct error signals due to the large number of taps in each segment. On the other hand small P values mean that the contribution of each segment to the MSE performance is not very big and therefore the equaliser can be adjusted more precisely (fine granularity). When P=KP $\rightarrow$ K=1, this corresponds to an equaliser with just one segment, i.e. a conventional linear equaliser. If P=1 $\rightarrow$ K=KP, each tap contribution is assessed individually and there are as many segments as taps.

The two other parameters to be specified are $\alpha_{up}$ and $\alpha_{dw}$ and their values should be set in accordance with the chosen value of P. The effects of modifying $\alpha_{up}$ and $\alpha_{dw}$ are best understood by looking at equations (5.5) and (5.6) in algorithm 1. If $\alpha_{up}$ is set to 1 it can be seen from (5.5) that even the slightest reduction in ASE level will provoke the equaliser to expand by one segment. Typically $\alpha_{up}$ will be set to values less than 1 so that the equaliser only expands when the ASE improvement is significant. As an example, $\alpha_{up}$ =0.9 means that the equaliser will expand only if the difference in the MSE (ASE) reduction between the last two segments is greater than 10%.

The parameter $\alpha_{dw}$ controls the sensitivity of the equaliser to reduce its length. This is useful to prevent the equaliser from being kept unnecessarily long if channel conditions change. Usually $\alpha_{dw}$ will have a value very close to one. In this way it is detected when the last segment has started increasing the MSE level and the equaliser is shrunk. Lower values of $\alpha_{dw}$ will increase the tendency of the equaliser to contract.

When the values of $\alpha_{up}$ and $\alpha_{dw}$ are very close to each other the equaliser length will tend to oscillate. The further apart $\alpha_{up}$ and $\alpha_{dw}$ are, the less likely is the equaliser will change its length. Notice though, that length oscillation does not cause any significant trouble. An exception to this occurs when the RLS algorithm is being used, an issue treated in the next section.

Finally it is worth noting that $\alpha_{up}$ and $\alpha_{dw}$ have to be set jointly with P (taps per segment). As has been mentioned, when the segment length is short, the effect of adding or removing a segment will be far less significant than when a long segment length is used, therefore $\alpha_{up}$ and $\alpha_{dw}$ have to be chosen appropriately. Short segments will tend to use larger values for $\alpha_{up}$ and long segments smaller values for $\alpha_{up}$. The values for $\alpha_{dw}$ are not as sensitive as for $\alpha_{up}$ and values close to 1 work well for any segment length.

From the explanations of all the parameters to be chosen it may be inferred that parameter selection is a rather difficult process, however, practical experimentation with this system have shown us this is not the case. There are many combinations of the settings offering similar levels of performance and therefore tuning the equaliser is fairly simple. In section 5.4 simulation results are presented for a very wide range of values for each of the parameters. However some typical values are given now in table 5.1.

| P | K | $\alpha_{up}$ | $\alpha_{dw}$ |
|---|---|---|---|
| 3 | 10 | 0.8 | 0.99 |

**Table 5.1**: Typical parameter values.

The values in table 5.1 corresponds to a 10_3 equaliser, consequently this equaliser has the potential to use up to 30 taps in the equalising process which will be enough to compensate a wide range of channel profiles.

## 5.3 Variable length LE enhancements

The basic structure update algorithm described in section 5.2.2 works well in channels whose conditions are fairly stable during operation. By stable is meant that in spite of the time varying nature of the channel, the variations are approximately of constant magnitude. However this structure may have trouble reconfiguring when there are sudden changes after prolonged intervals of stationarity. The algorithm needs a minor modification in order to be able to cope with these situations. This modification is presented in section 1.3.1. Additionally the variable length LE may have some stability problems if the RLS algorithm is used to drive the filter coefficients, so a method of stabilisation when using the RLS algorithm is introduced in 1.3.2. Finally some attention is paid to the choice of the delay in the detection process in the segmented equaliser.

### 5.3.1 Robust variable length linear equalisation

Simulations using the basic algorithm reveal that algorithm 1 loses efficiency in detecting channel changes after long periods of operation. This problem arises because with algorithm 5.1, the measured ASE weights equally all the information from the start of operation.

When dealing with time-varying channels it is better to stress the information obtained from recent data as the scenario may have changed and old data may not be very relevant to the current situation. Similarly to the exponentially weighted RLS algorithm, a forgetting factor, denoted as $\beta$, is introduced in the ASE measurement. This makes the equaliser much more robust to large and irregular channel changes.

Algorithm 5.1 can be modified with the inclusion of the forgetting factor to obtain the algorithm shown below.

$$ASE_{L-1}(n)=\sum_{i=0}^{n}\beta\left|d(n)-y_{L-1}(n)\right|^2 \quad (5.7)$$

$$ASE_{L}(n)=\sum_{i=0}^{n}\beta\left|d(n)-y_{L}(n)\right|^2 \quad (5.8)$$

$$\text{If} \quad ASE_L(n) \le \alpha_{up} ASE_{L-1}(n) \quad \rightarrow \quad \text{Add one segment (P extra taps)} \quad (5.9)$$

$$\text{If} \quad ASE_L(n) \ge \alpha_{dw} ASE_{L-1}(n) \quad \rightarrow \quad \text{Remove one segment (P less taps)} \quad (5.10)$$

with $0 < \alpha_{up}, \alpha_{dw} \le 1$ and $\alpha_{up} \le \alpha_{dw}$, $\beta \le 1$.

**Algorithm 5.2**: Robust equaliser length control.

This modification represents a modest increase in computational complexity with respect to the basic algorithm as it only requires 2 additional products per iteration. The total complexity of algorithm 2 consists of 6 products and 4 additions per iteration.

## 5.3.2 Stabilisation of the RLS variable length LE

As has been described in chapter 2, least squares algorithms are much more prone to suffer stability problems than gradient type algorithms (like LMS). RLS is no exception and it was found that its use in conjunction with the variable length structure occasionally led to severe stability problems. In order to understand where the instability comes from some insight into the algorithm is needed. For convenience the RLS algorithm is repeated here:

$$\mathbf{k}(n) = \frac{\lambda^{-1}\mathbf{P}(n-1)\mathbf{u}(n)}{1+\lambda^{-1}\mathbf{u}^{T}(n)\mathbf{P}(n-1)\mathbf{u}(n)} \tag{5.11}$$

$$\zeta(n) = d(n) - \hat{\mathbf{w}}^{T}\mathbf{u}(n) \tag{5.12}$$

$$\hat{\mathbf{w}}(n) = \hat{\mathbf{w}}(n-1) + \mathbf{k}(n)\zeta(n) \tag{5.13}$$

$$\mathbf{P}(n) = \lambda^{-1}\mathbf{P}(n-1) - \lambda^{-1}\mathbf{k}(n)\mathbf{u}^{T}(n)\mathbf{P}(n-1) \tag{5.14}$$

The recursive nature of the RLS algorithm imposes the need for consistency between successive iterations. When the equaliser length changes all the vectors and matrices also change their dimensions and this inter-iteration consistency is seriously compromised. The critical equation in the context of this work is equation (5.14), the update of the **P**(n), which corresponds to the inverse of the correlation matrix $\mathbf{\Phi}(n)$. Two different situations may occur, contraction and expansion of the equaliser.

Suppose that at instant n, the equaliser has $M_1$ taps, and at that particular instant it is decided to contract the equaliser to $M_2$ taps ($M_2<M_1$). When this happens, the $M_1xM_1$ autocorrelation matrix, denoted by $\mathbf{\Phi}_{M1xM1}(n)$ is reduced to dimension $M_2xM_2$ ($\mathbf{\Phi}_{M2xM2}(n)$) by simply discarding the components outside the range [0..$M_2$-1, 0..$M_2$-1]. At this moment, equation 5.14 becomes inconsistent as assuming that $\mathbf{P}_{M1xM1}(n) = \mathbf{\Phi}_{M1xM1}(n)^{-1}$, it does not hold that $\mathbf{P}_{M2xM2}(n) = \left[\mathbf{\Phi}_{M1xM1}(n)^{-1}\right]_{M2xM2}$. In plain words, the inverse of a truncated matrix is not simply the truncation of the inverse. This fact completely invalidates (5.14) making the filter go unstable whenever the equaliser contracts.

There seems to be only one solution in order to solve this problem: restart the filter every time it is contracted. This may seem a very drastic solution, however, given the extremely good convergence properties of the RLS algorithm, the restart does not affect the MSE performance greatly, provided the number of resets is not too large.

The other possibility occurs when the equaliser is expanded. In this case and using the same notation as in the paragraph above, $M_2 > M_1$. When this happens, the $M_1 x M_1$ correlation matrix $\mathbf{\Phi}_{M1xM1}(n)$ is expanded in the following way into the expanded correlation matrix $\mathbf{\Phi}_{M2xM2}(n)$:

$$\mathbf{\Phi}_{M1xM1}(n) = \begin{bmatrix} \Phi^{0,0}_{M1xM1}(n) & \Phi^{0,1}_{M1xM1}(n) & \cdots & \Phi^{0,M_1-1}_{M1xM1}(n) \\ \Phi^{1,0}_{M1xM1}(n) & \ddots & \ddots & \vdots \\ \vdots & \ddots & \ddots & \vdots \\ \Phi^{M_1-1,0}_{M1xM1}(n) & \cdots & \cdots & \Phi^{M_1-1,M_1-1}_{M1xM1}(n) \end{bmatrix}$$

$$\Downarrow$$

$$\mathbf{\Phi}_{M2xM2}(n) = \begin{bmatrix} \Phi^{0,0}_{M1xM1}(n) & \Phi^{0,1}_{M1xM1}(n) & \cdots & \Phi^{0,M_1-1}_{M1xM1}(n) & 0 & \cdots & 0 \\ \Phi^{1,0}_{M1xM1}(n) & \ddots & \ddots & \vdots & \vdots & \vdots & \vdots \\ \vdots & \ddots & \ddots & \vdots & \vdots & \vdots & \vdots \\ \Phi^{M_1-1,0}_{M1xM1}(n) & \cdots & \cdots & \Phi^{M_1-1,M_1-1}_{M1xM1}(n) & 0 & \vdots & \vdots \\ 0 & \cdots & \cdots & 0 & \delta & 0 & \vdots \\ \vdots & \cdots & \cdots & \cdots & 0 & \ddots & 0 \\ 0 & \cdots & \cdots & \cdots & \cdots & 0 & \delta \end{bmatrix}$$

where δ is the RLS initialisation constant (see section 2.3.2). The important point now is to realise that the inverse of the expanded matrix, $\mathbf{P}_{M2xM2}(n) = \mathbf{\Phi}_{M2xM2}(n)^{-1}$, can be obtained from $\mathbf{P}_{M1xM1}(n)$ by just adding zeros to all positions in the range $[M_1..M_1+M_2-1, M_1..M_1+M_2-1]$, so that:

$$\mathbf{P}_{M2xM2}(n) = \begin{bmatrix} \mathbf{P}_{M1xM1} & 0 & \cdots & 0 \\ 0 & \delta^{-1} & \ddots & \vdots \\ \vdots & \ddots & \ddots & 0 \\ 0 & \cdots & 0 & \delta^{-1} \end{bmatrix} \quad (5.15)$$

It has not been possible to find this result in the matrix theory literature (which may exist in a different form), however its validity has been verified using the Symbolic Computation Toolbox of Matlab. Using this result, when the equaliser is expanded, the new **P**(n) (after expansion) can be directly and easily calculated from the old **P**(n) (before expansion), therefore equation (5.14) can be kept consistent by simply "zero padding" **P**(n) with the appropriate number of zeros whenever the filter length is increased.

The RLS stabilisation procedure can be summarised as: "Restart the adaptive filter when the number of taps is reduced. Do nothing when it is increased". Of course it is advisable to minimise the number of contractions. Simulation results shown later prove the efficacy of this method.

### 5.3.3 Decision delay in segmented equalisers

The last theoretical topic related to segmented equalisation requiring some attention is the choice of the decision delay. Decision delay in conventional equalisers has been covered in section 3.4.5 and the main conclusion extracted is the lack of techniques to optimise the choice of the decision instant other than performing an exhaustive search. Simulations with typical wireless channels show that the MSE performance is quite insensitive to the delay if this is bigger than a small number of taps.

There are two important points concerning the delay to be aware of when using the segmented structure. First, the initial number of taps of the equaliser has to be greater than the delay (also applies for conventional equalisers). This means that the number of active segments at any given instant has to be large enough to include the delay. This is obvious as otherwise the equaliser would not be able to capture the peak of signal energy of the reference tap.

The second important issue is concerned with the way taps/segments are added and subtracted from the equaliser. According to the techniques presented in 5.2, taps are always added/subtracted at the end (tail) of the equaliser. In principle it would be possible to arbitrarily add/subtract taps at either end (head or tail), however this poses the problem that when taps are added/subtracted at the head of the equaliser, the decision delay should also be modified by an equal number of taps. Changes in the decision delay implies the need for some form of resynchronisation of the bit stream making the system more complex. An alternative solution would be to add some zeroed segments before the first active segment. These zeroed coefficients would initially be used only to introduce an "artificial" delay in the received bit stream, acting as simple buffers. If the equaliser needs to be expanded then some of these segments might be switched on and start contributing to the equalising process. Similarly, if the equaliser has to be shortened, the first segment can be switched off (zeroed) again. With this method the delay does not need to be changed if taps are added or removed from the head of the equaliser as it is already taken into account by the zeroed coefficients. This technique is useful to keep the reference tap in the centre of the equaliser even when the number of taps changes.

In the context of wireless and mobile systems, the reference tap does not need to be centred as channels tend to be minimum or near minimum phase, i.e. the strongest components arrive first. Consequently, segments/taps are only added at the tail of the equaliser, making unnecessary the technique described in the previous paragraph.

# 5.4 Simulation results of the variable length LE

In this section extensive simulation results are presented using the segmented filtering structure with the algorithms introduced in sections 5.2.2 (basic) and 5.3.1 (robust). Both techniques are tested using the LMS and RLS algorithms. The channel profiles used in the validation are those presented in chapter 3. Additionally, time varying E/No profiles and sudden changes in the channel profile have also been tested. The general simulation conditions are also those described in chapter 3.

## 5.4.1 Channel model 2 (Static)

This channel, corresponding to a “frozen” typical urban mobile profile, is shown in the time and frequency domains in figures 3.4 and 3.5. This profile has been used to test extensively the variable length equaliser with different combinations of parameter values. Due to the channel and environment being static, the equaliser length tends to change only during the first stages and then remain constant for the rest of the simulation. Different fixed length equalisers have also been simulated in order to compare them with the variable length structure. Notice that in the figures $\alpha_{up}$ and $\alpha_{dw}$ are denoted by a+ and a- respectively. Figures 5.9, 5.11 and 5.13 compare the MSE performance of the fixed length and variable length equalisers using the LMS algorithm. In these simulations, the segments were three taps long, $\alpha_{dw}$ =0.99 and $\beta$=1 (static channel). The equaliser had initially 9 taps (3 segments) and the delay was set to 8 samples. The LMS step size was fixed to 0.01.

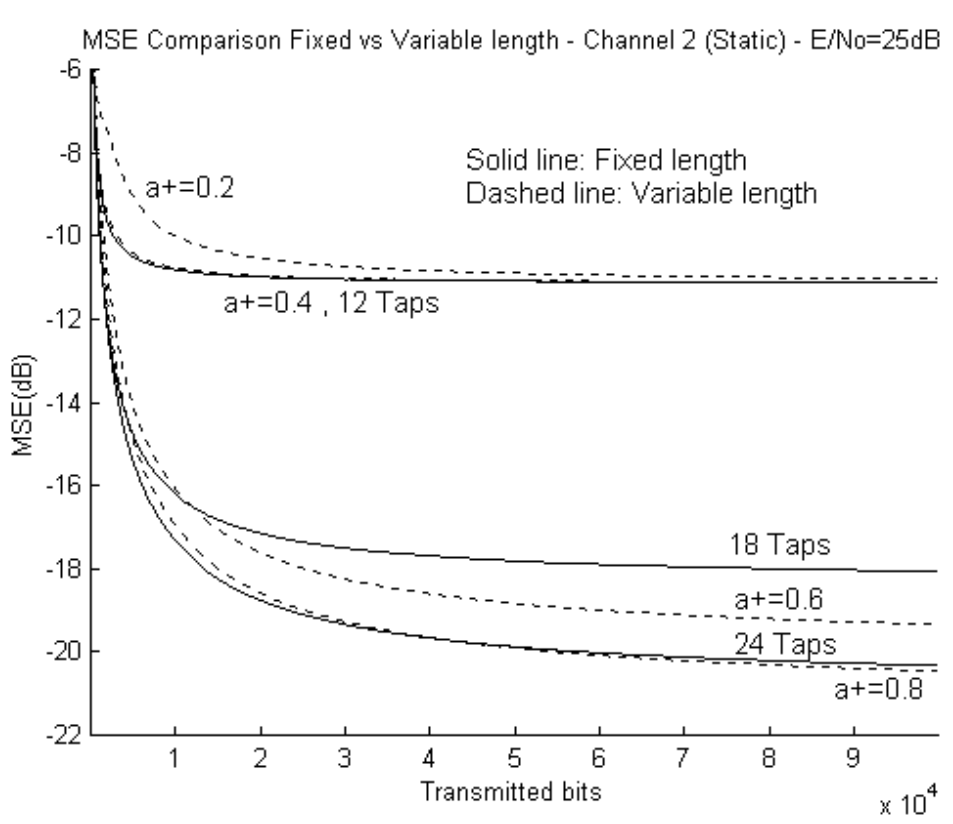

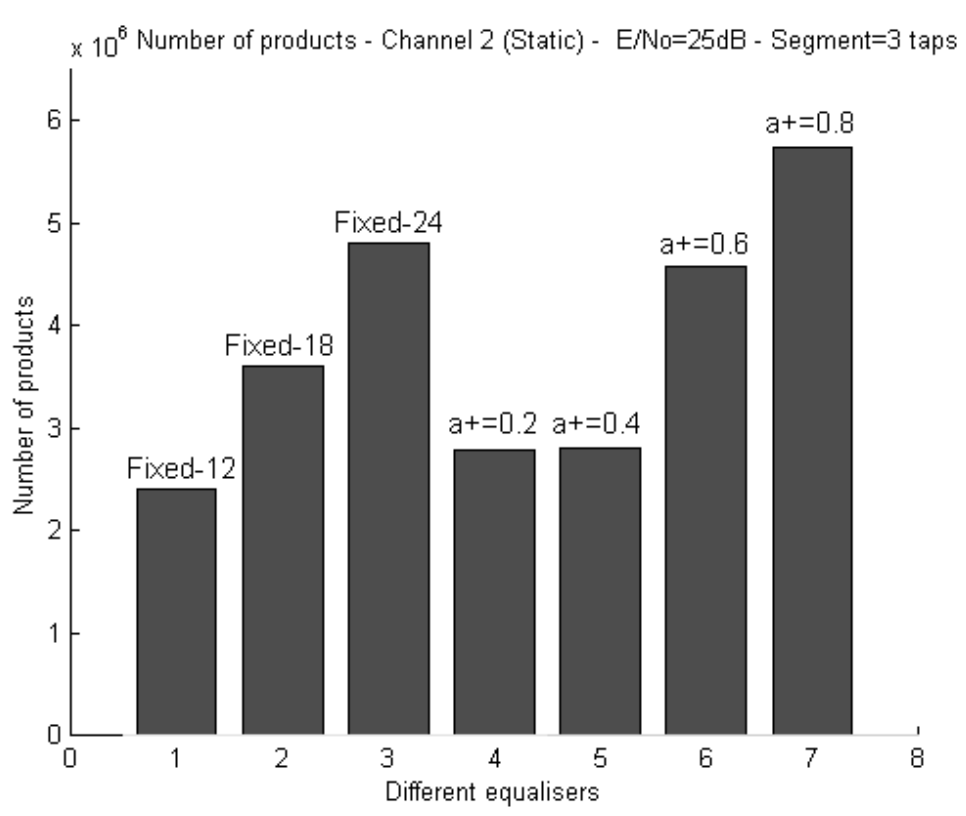

**Figure 5.9**: LMS MSE comparison of variable and fixed length equalisers. Channel model 2 (Static). E/No=25dB.

**Figure 5.10**: Comparison of the number of products computed over 100,000 iterations. Channel model 2 (Static). E/No=25dB.

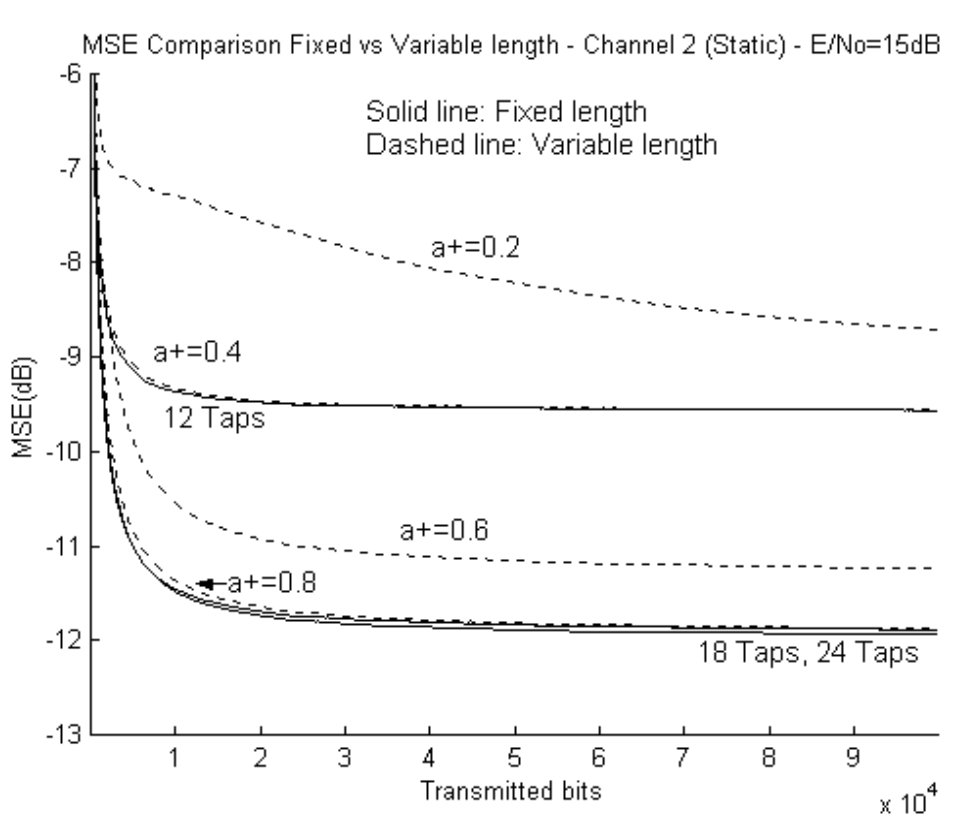

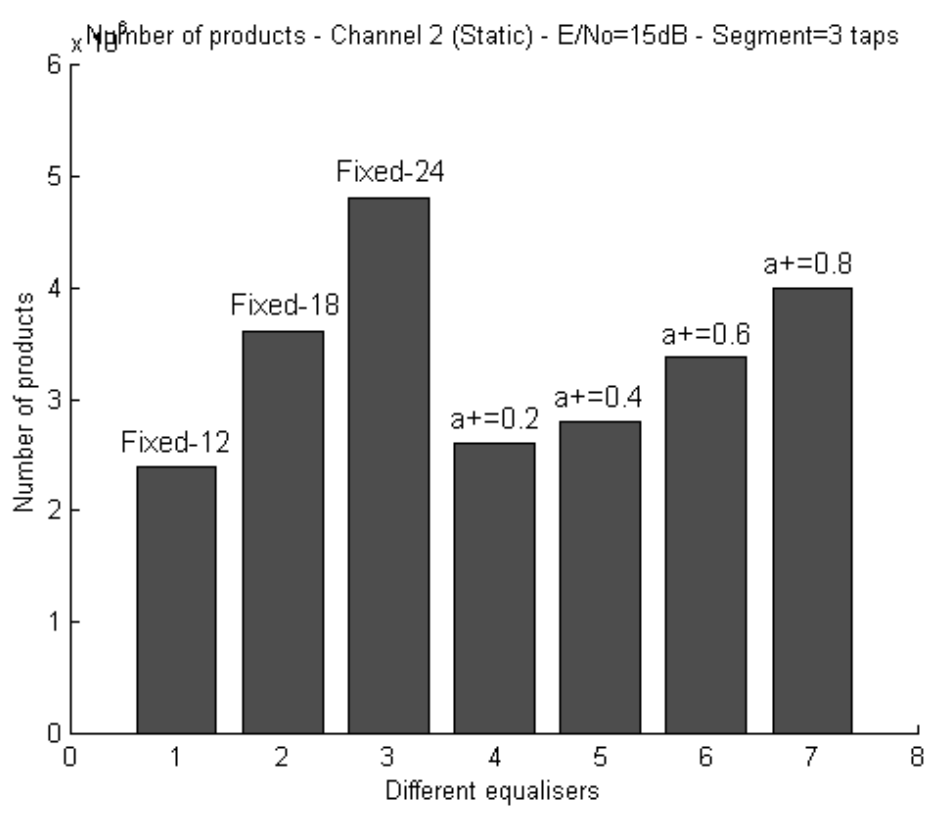

**Figure 5.11**: LMS MSE comparison of variable and fixed length equalisers. Channel model 2 (Static). E/No=15dB.

**Figure 5.12**: Comparison of the number of products computed over 100,000 iterations. Channel model 2 (Static). E/No=15dB.

For each of the MSE graphs there is a corresponding figure comparing the number of products calculated during the simulation whose length was 100,000 bits. It would also be possible to compare the number of additions, however given that the algorithm controlling the equaliser length computes more products than additions per iteration, it is fairer to compare the number of products.

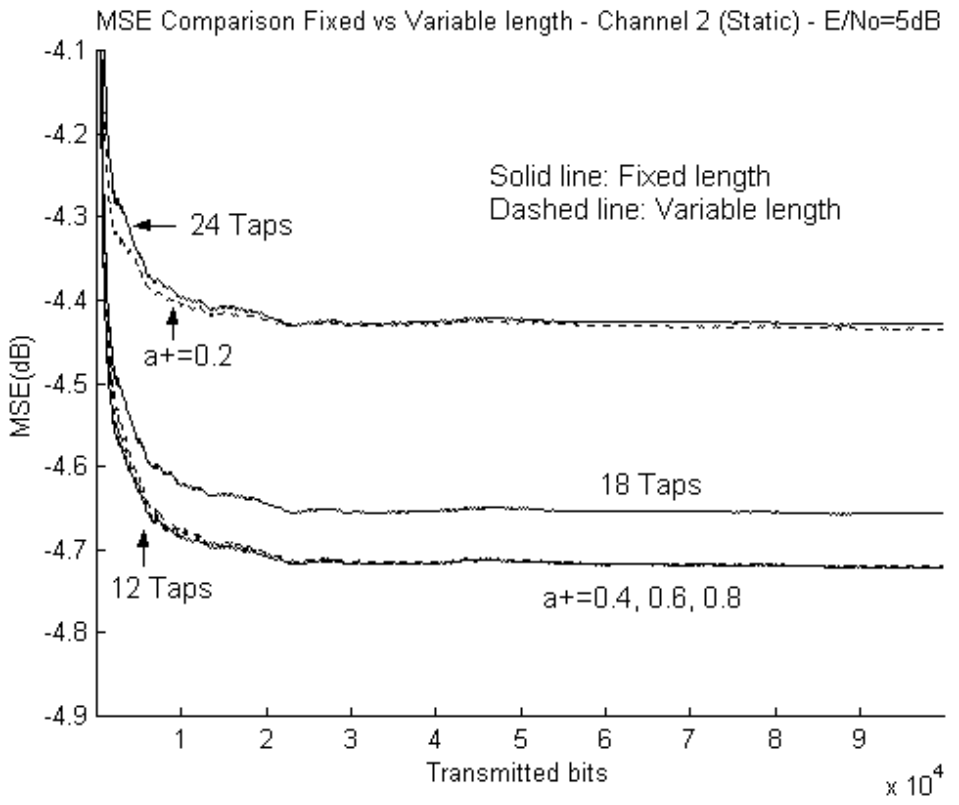

**Figure 5.13**: LMS MSE comparison of variable and fixed length equalisers. Channel model 2 (Static). E/No=5dB.

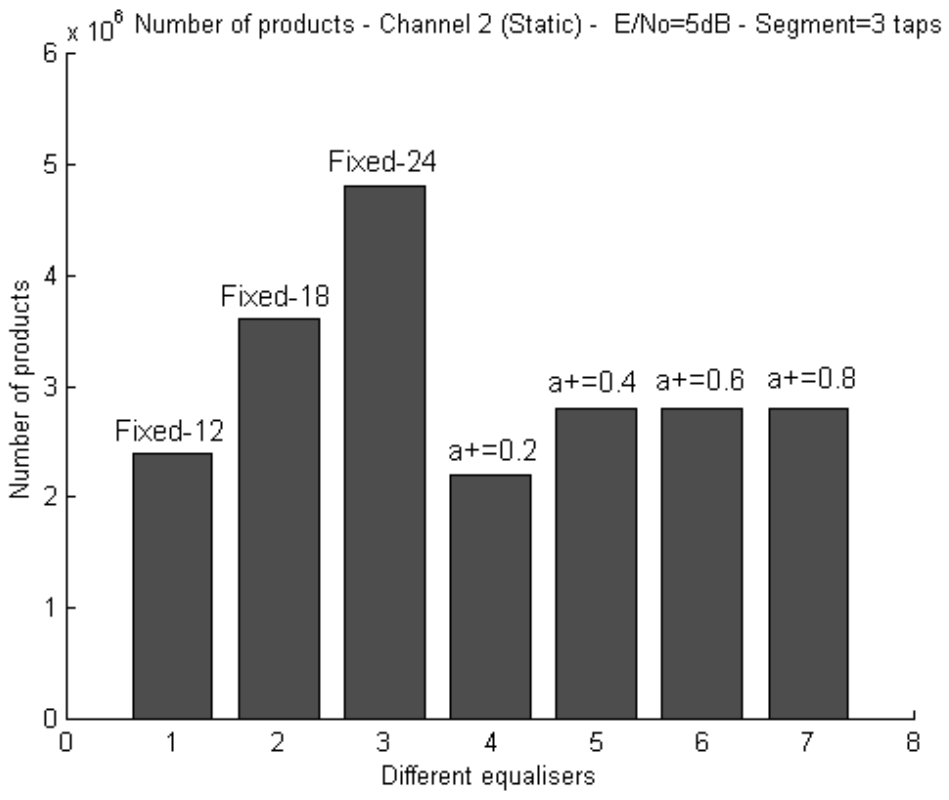

**Figure 5.14**: Comparison of the number of products computed over 100,000 iterations. Channel model 2 (Static). E/No=5dB.

Starting with figure 5.9 and focusing on the fixed length equalisers (solid lines) it can be seen how for this noise level (E/No=25dB), increasing the equaliser length produces significant reductions in the MSE level. The performance of the variable length equaliser changes according to the value of $\alpha_{up}$. Low values of $\alpha_{up}$ (0.2 and 0.4) tend to keep the equaliser short. Notice that $\alpha_{up}$ =0.2 implies that another segment will be added if and only if the difference in MSE level between the last two active segments is of at least 80%. This explains why these equalisers (low $\alpha_{up}$) are likely to remain short, achieving similar MSE level as the 12 tap fixed equaliser.

In realistic systems $\alpha_{up}$ will have a value closer to 1 such as 0.6, or more likely, in the range [0.8-0.9]. The curves for $\alpha_{up}$ =0.6 and $\alpha_{up}$ =0.8 show how the equaliser expands enough to achieve or even surpass the performance of the fixed long equalisers (18 and 24 taps). Figure 5.10 shows additional information concerning the MSE curves in figure 5.9. In particular, from the number of products computed over the simulation can be inferred the average length attained by the different variable length equalisers. Knowing that the associated overhead with the variable length structure is 4 products per iteration and the LMS computational load is 2M products per iteration, the average length of the filter ($M_{av}$) can be computed. Take for example, in figure 5.10, the bar for $\alpha_{up}$ =0.8. The number of products over 100,000 iterations is 5,737,190, subtracting 400,000 (100,000 iterations x 4 products/iteration) corresponding to the overhead results in 5,337,190 products. It should be clear now from the LMS complexity that $2M_{av}(100{,}000)=5{,}337{,}190$ therefore $M_{av}=26.6$ taps.

Most of the comments for figures 5.9 and 5.10 also apply to figures 5.11 and 5.12 corresponding to an E/No level of 15 dB. Nonetheless, notice how in this scenario figure 5.11 shows that the 24 tap fixed equaliser achieves the same performance level as the 18 tap. Notice also that the variable length equaliser with $\alpha_{up}=0.8$ achieves the same MSE level as the longest fixed equaliser. As $\alpha_{up}$ is reduced, performance degrades because the equaliser is not sufficiently expanded. As in the E/No=25 dB case, the average length can be computed from the total number of products. Repeating the procedure described previously it is found that $M_{av}=17.9$ taps.

More interesting is the situation shown in figures 5.13 and 5.14. The very low E/No provokes an increase in the MSE whenever the equaliser length is increased beyond a small number of taps. The fixed equaliser achieving the lowest MSE is the one with 12 taps, whereas the 24 taps equaliser produces the highest level of MSE. On the other hand all the variable length equalisers for any value of $\alpha_{up}$, except for $\alpha_{up}=0.2$, achieve the same MSE level as the 12 tap equaliser indicating that only one extra segment is added to the original 3 segments (9 taps). For $\alpha_{up}=0.2$ the equaliser does not grow at all. Computing the average equaliser length as with the previous E/No shows that $M_{av}=11.9$ taps.

Notice how the value of $M_{av}$ reduces as the E/No level decreases, for E/No of 5, 15 and 25dB. The resulting average equaliser lengths are 11.9, 17.9 and 26.6 taps respectively. It is of fundamental importance to compare these $M_{av}$ values for $\alpha_{up}=0.8$ with the LMS curves (circles) in figure 5.3. Essentially, for any E/No level, the corresponding $M_{av}$ value matches very accurately the optimum equaliser length that can be extracted from figure 5.3. Recall the idea that by optimum is meant the smallest equaliser length achieving the minimum MSE level. The main conclusion to be drawn from these simulations is that the variable length equaliser using a large $\alpha_{up}$ (for example 0.8) successfully guesses the number of taps worth having in the equaliser to compensate an unknown static channel under any noise conditions. Adding further taps will not help in reducing the MSE level in a significant way and will increase the computational burden of the equalisation process.

This ability to predict the useful number of taps of the equaliser does not come for free, as the price to be paid is an increase of 4 products and 2 additions per iteration. However the potential savings obtained from using an equaliser with the optimum number of taps clearly outweighs this modest increase in the computational complexity. Moreover the problem of a rise in the MSE level due to the EMSE generated by the LMS algorithm is also avoided.

The same results are now presented but using the RLS algorithm. The system parameters are the same as in the LMS simulations. The RLS forgetting factor, given that the channel is static, was fixed to 1.

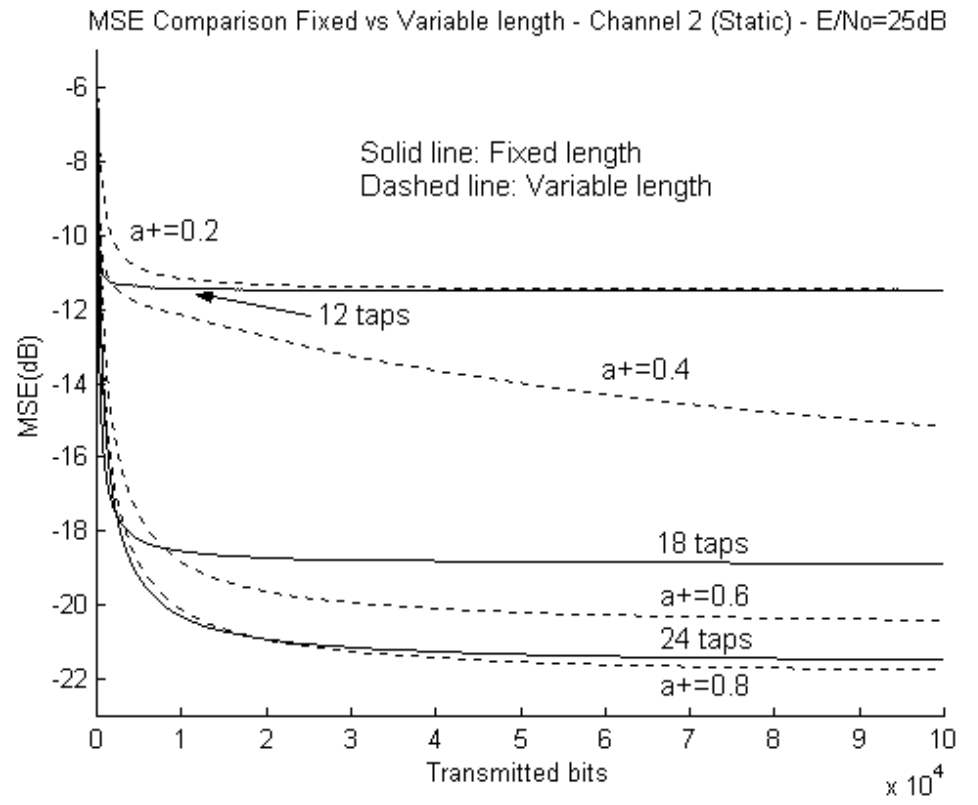

**Figure 5.15**: RLS MSE comparison of variable and fixed length equalisers. Channel model 2 (Static). E/No=25dB.

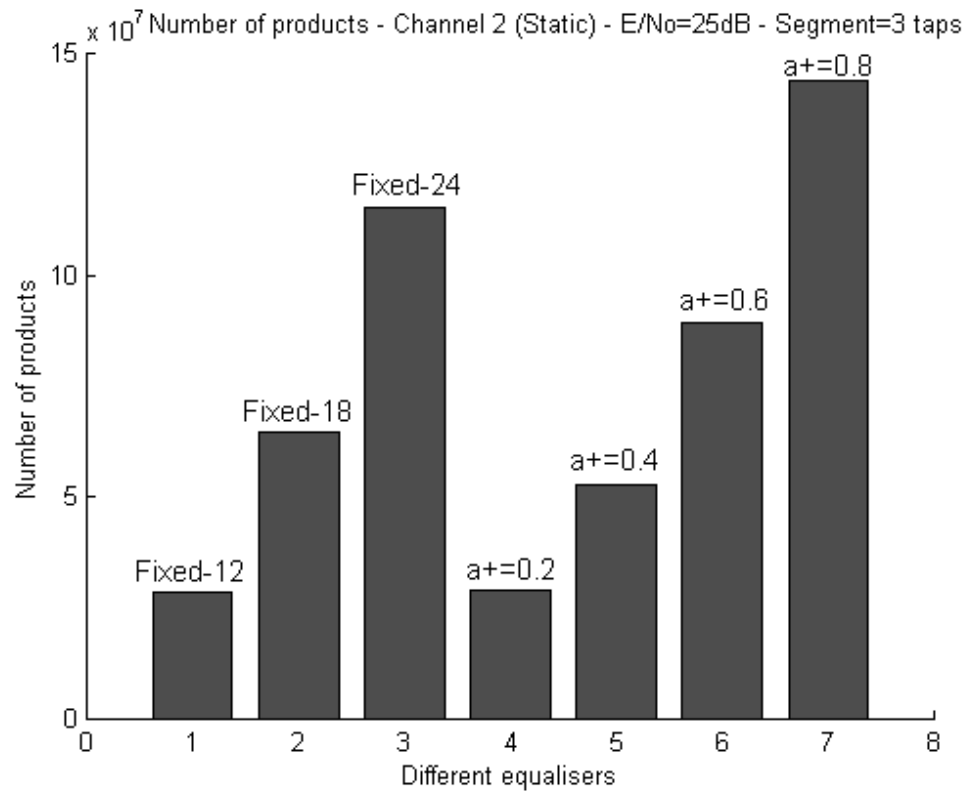

**Figure 5.16**: Comparison of the number of products computed over 100,000 iterations. Channel model 2 (Static). E/No=25dB.

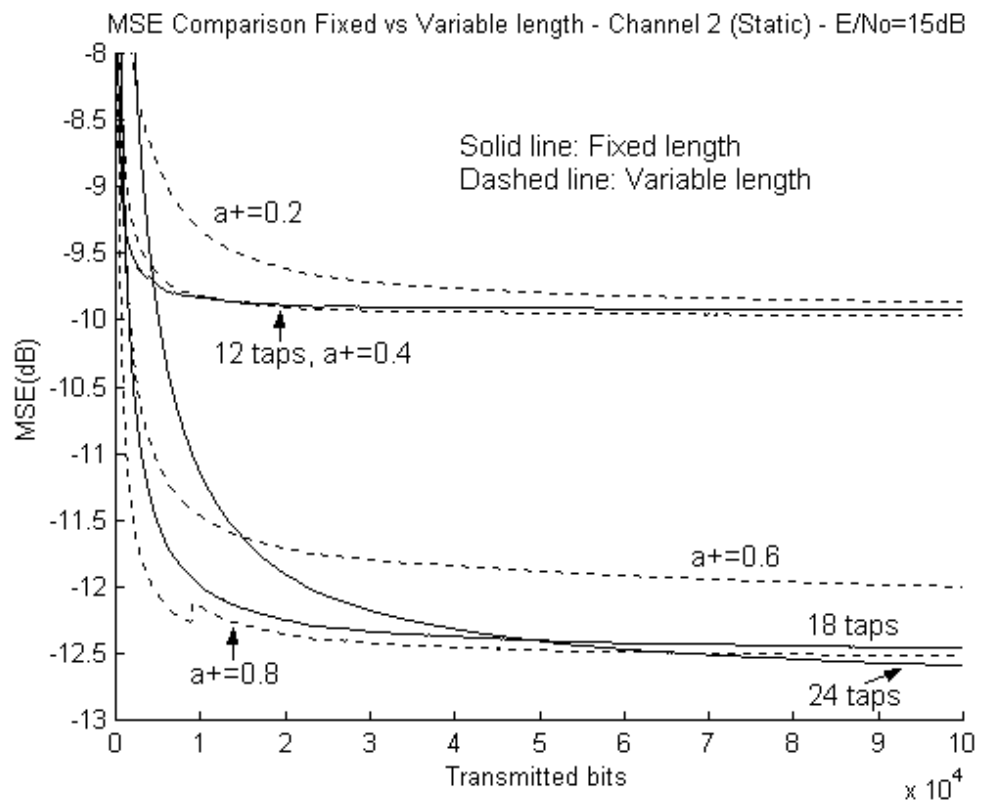

**Figure 5.17**: RLS MSE comparison of variable and fixed length equalisers. Channel model 2 (Static). E/No=25dB.

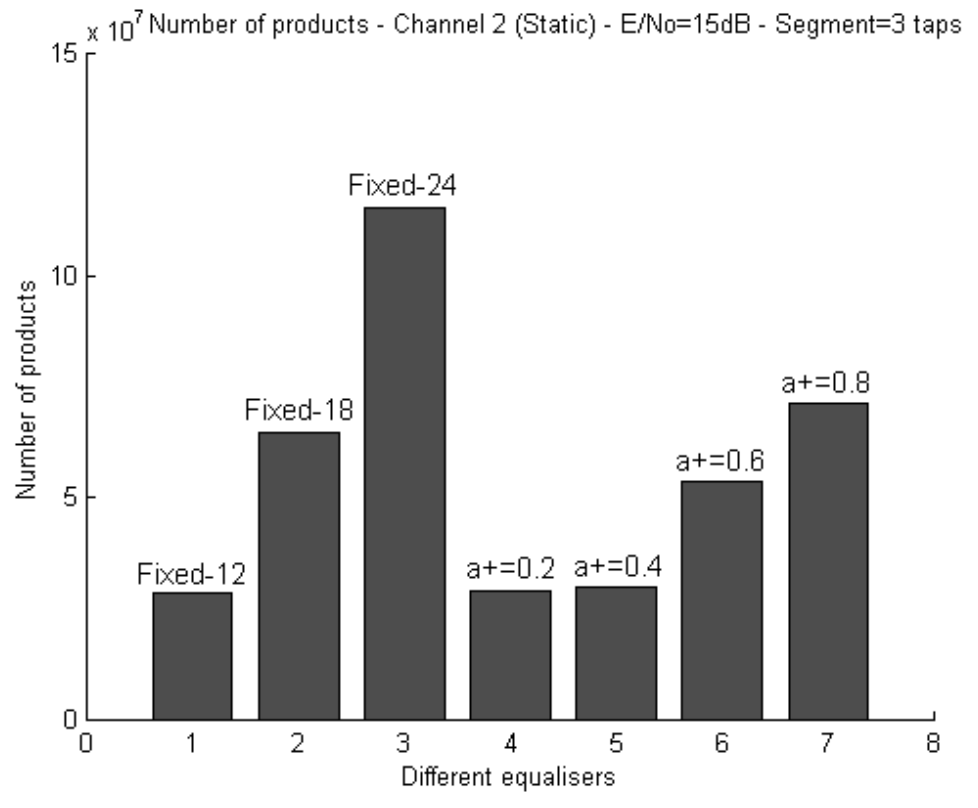

**Figure 5.18**: Comparison of the number of products computed over 100,000 iterations. Channel model 2 (Static). E/No=25dB.

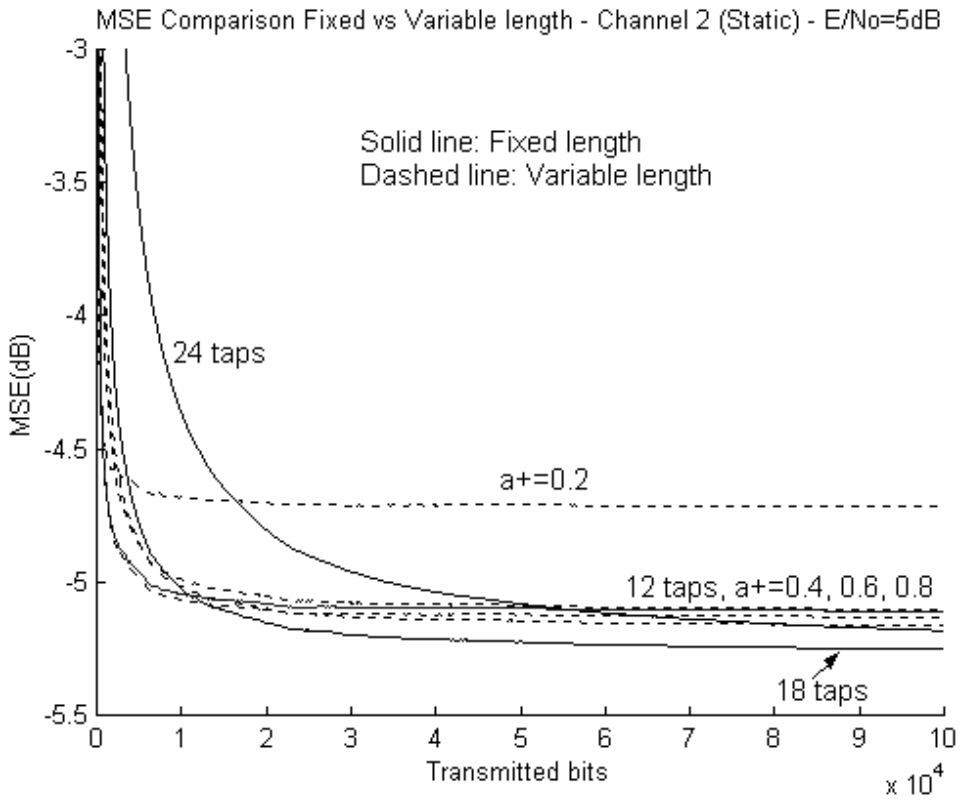

**Figure 5.19**: RLS MSE comparison of variable and fixed length equalisers. Channel model 2 (Static). E/No=25dB.

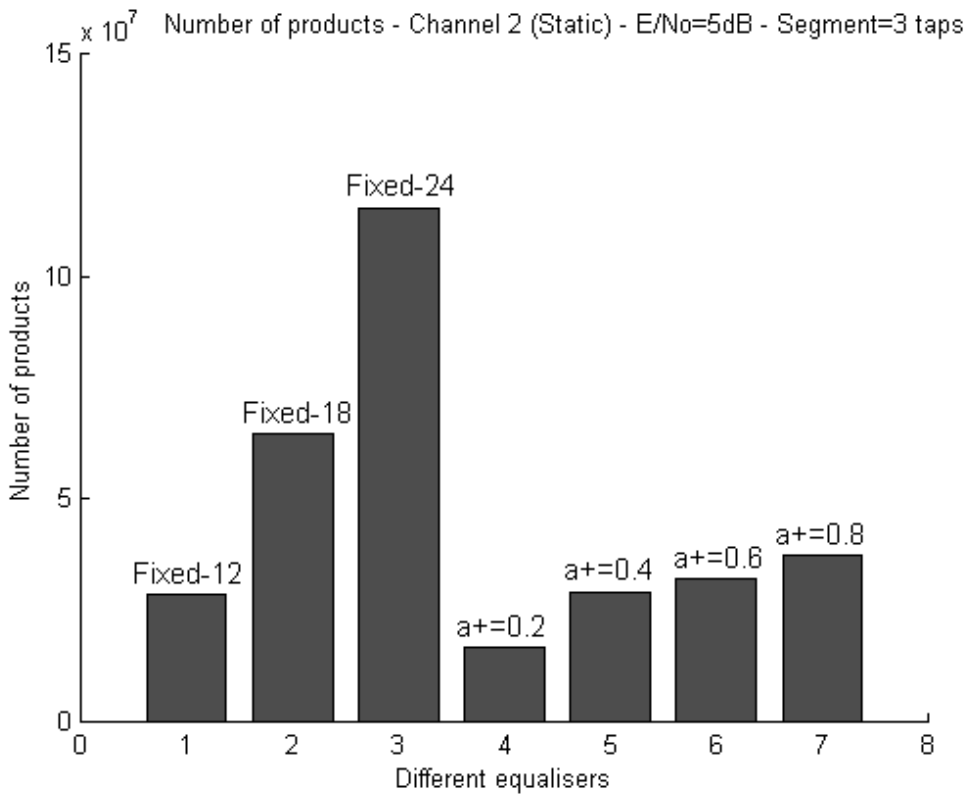

**Figure 5.20**: Comparison of the number of products computed over 100,000 iterations. Channel model 2 (Static). E/No=25dB.

Nearly everything said for the LMS simulation also applies to the RLS case. Notice how under any noise conditions, the RLS algorithm is able to attain a slightly lower MSE than the LMS. This is due to the fact that the RLS does not produce any EMSE when operating in static environments. As in the LMS algorithm, and using a similar procedure, it is possible to compute average equaliser lengths from figures 5.16, 5.18 and 5.20. The resulting average lengths, $M_{av}$, are 13.6, 18.8 and 26.7 taps for 5, 15 and 25 dB respectively. These $M_{av}$ values are slightly greater than the corresponding LMS values (11.9, 17.9 and 26.6 taps). This is also attributed to the non-existence of any EMSE component when using the RLS algorithm, allowing the equaliser to grow without any performance penalty (remember from the previous chapter that the LMS EMSE is proportional to the number of taps). Again these average length values match precisely the optimum length values that can be observed in figure 5.4.

Given the large computational complexity of the RLS algorithm, the computational savings arising from estimating the right number of taps to be used in the equalisation process are more noticeable than in the LMS case. Take for example the results shown in figure 5.20. The variable equaliser with $\alpha_{up}$ =0.8 uses less than a third of the products required for the 24 tap equaliser and less than a half with respect to the 18 tap filter.

Using the same channel profile the effect of the segment length has also been studied. The objective was to determine which segment lengths offer better performance. Four different segment lengths have been tested: 4, 3, 2 and 1 tap per segment. Simulations have been run using LMS although any other algorithm could have been used. The E/No level has been set to 25dB as it is in low noise scenarios where the effects of the segment length become more evident. The results are shown in figures 5.21 to 5.24.

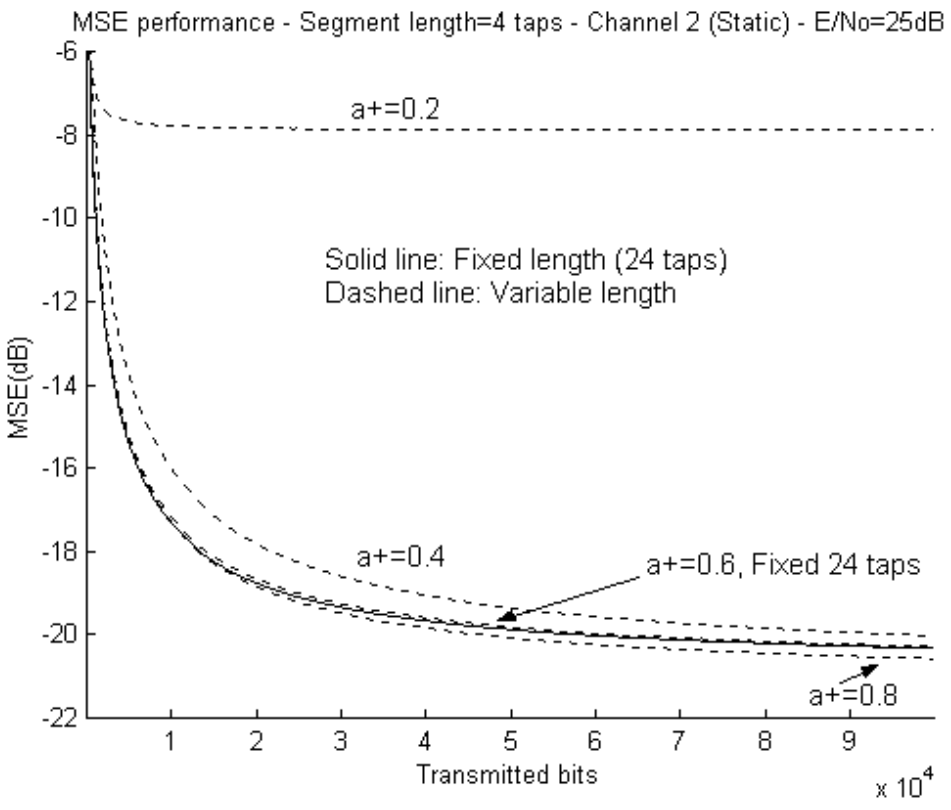

**Figure 5.21**: LMS MSE using 4-tap segments. Channel model 2 (Static). E/No=25dB.

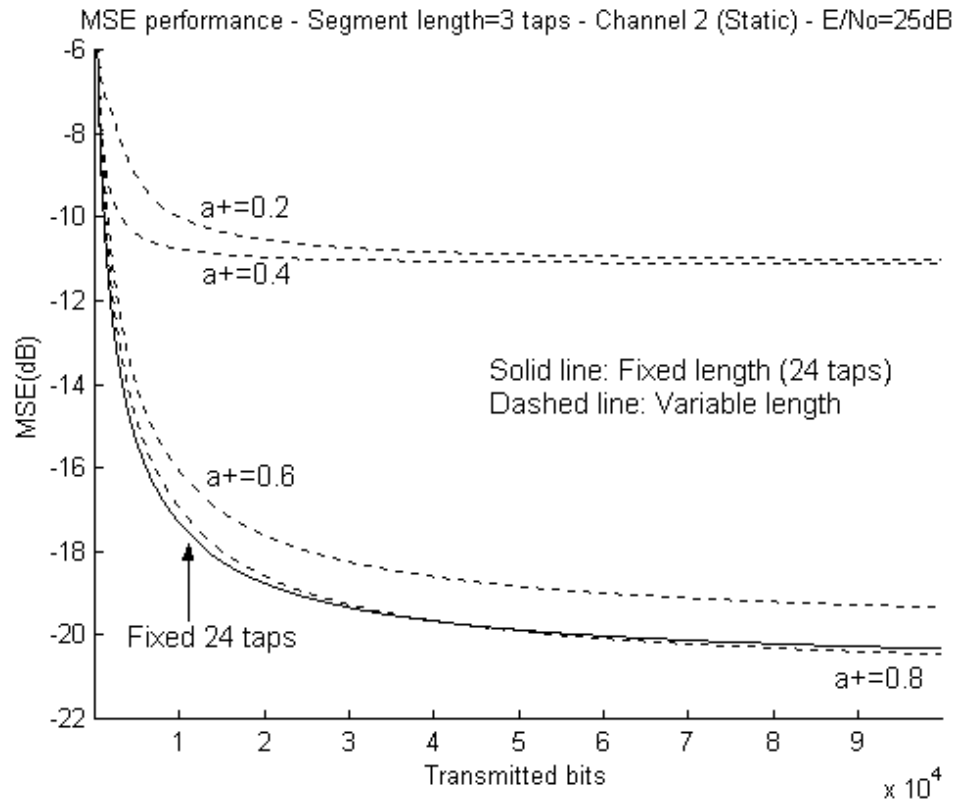

**Figure 5.22**: LMS MSE using 3-tap segments. Channel model 2 (Static). E/No=25dB.

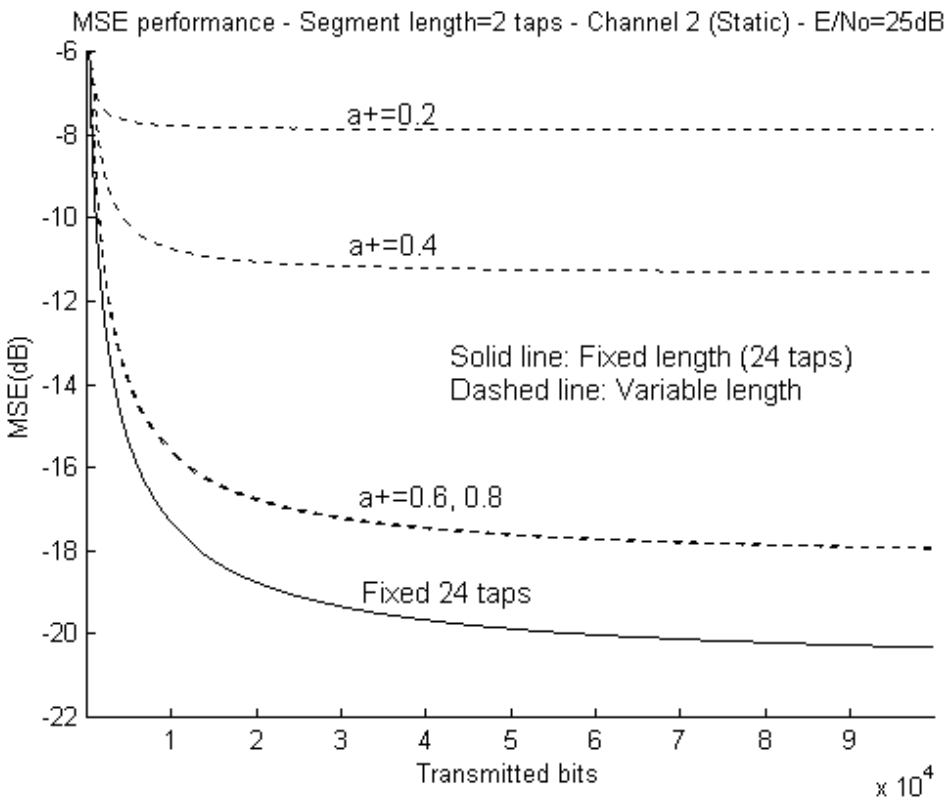

**Figure 5.23**: LMS MSE using 2-tap segments. Channel model 2 (Static). E/No=25dB.

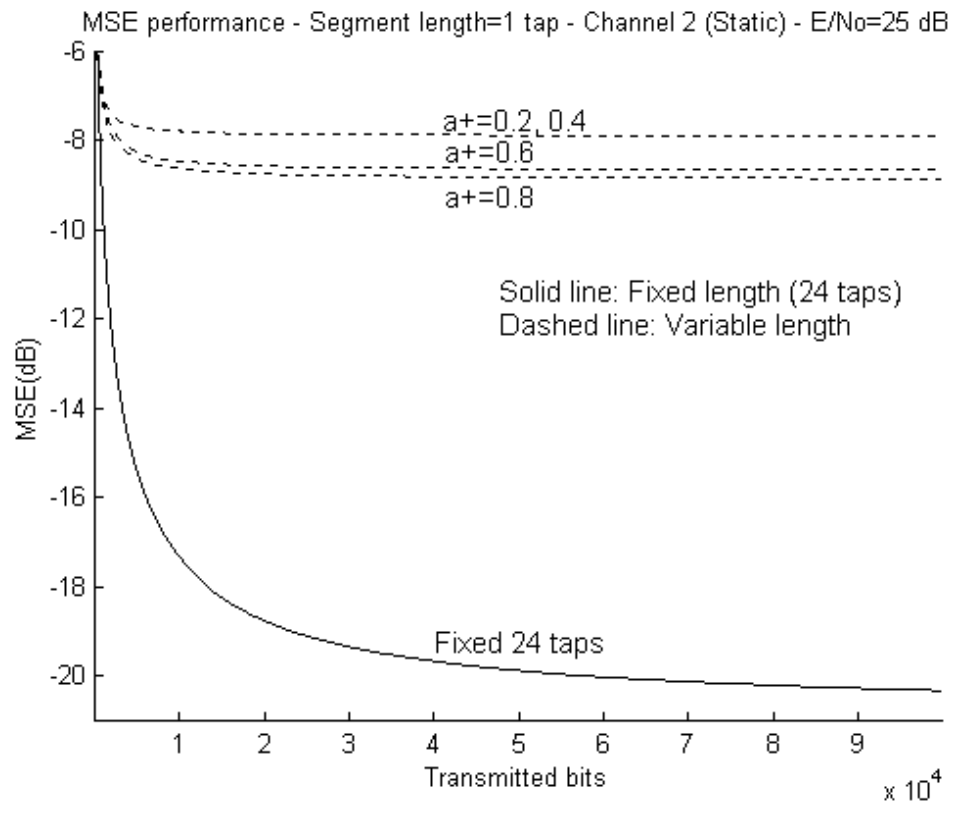

**Figure 5.24**: LMS MSE using 1-tap segments. Channel model 2 (Static). E/No=25dB.

The figures above show the MSE performance for variable length equalisers using different $\alpha_{up}$ values with different segment lengths. For comparison purposes, the MSE curve of the 24-tap fixed equaliser is also plotted. Notice how the longer the length, the smaller is the difference among the curves for different $\alpha_{up}$.

This loss of granularity is most apparent in figures 5.21 and 5.22. A large segment length implies that the difference between the last two segments is more likely to be big and therefore, the equaliser tends to expand. Except for the smallest value of $\alpha_{up}$, notice how in figure 5.21 (4-tap segments) the rest of $\alpha_{up}$ values tend to expand the equaliser up to about 24 taps. For the results in figure 5.23, 2-tap segments were used. In this case different $\alpha_{up}$ values produce a wider range of MSE curves.

In general, using large segment lengths will produce a coarser approximation to the optimum number of taps than short segment lengths. This is simply due to the larger differences between the last two segments. In theory this would push for the use of short segment lengths, however short segment lengths have also a drawback which can be clearly appreciated in figure 5.24. Notice how where 1-tap segments have been used the MSE, remain very high for all values of $\alpha_{up}$, even for $\alpha_{up}$ =0.8. The reason for this phenomenon is that the very small differences in performance between two successive segments (in this case, just 1 tap) constrain the equaliser from expanding as the proposed algorithm (algorithm 5.1) considers that the performance limit has been reached. This can also be inferred from figure 5.3 where it can be seen that many pairs of successive taps achieve the nearly identical MSE level (for example lengths 11 and 12 in the E/No=25dB curve). One solution to this problem could be to use closer values to 1 for $\alpha_{up}$, in the range [0.95..0.99]. This approach works well but some simulation work has shown that the adjustment of the system needs to be far more accurate (and time consuming) than in the case of longer segments. Additionally, an $\alpha_{up}$ value very close to 1 implies that it would also be extremely close to $\alpha_{dw}$ which would inevitably provoke continuous oscillations in the number of taps of the equaliser which in turn could cause stability problems if the RLS algorithm was being used (see section 5.3.2).

Extensive simulations using different segment lengths with a variety of channels have shown that segment lengths of 3 or 4 taps provide good performance and easiness of adjustment of the system parameters. Segment lengths greater than 4 make the equaliser length changes too steep and inherently inaccurate to find the right number of taps. Segment lengths of 1 or 2 taps need very big values for $\alpha_{up}$ and are difficult to adjust. In light of these results, for the rest of the simulations presented in this chapter the segment length has been fixed to 3 taps.

### 5.4.2 Channel model 3 (Static)

The next channel profile tested is channel model 3 whose characteristics are shown in figures 3.6 and 3.7. The extremely bad spectral properties of this channel manifest some of the limitations of linear equalisation.

For this type of channel, as it will be shown in the next chapter, DFE is a far more effective solution. However it is important to show that the proposed variable length equaliser detects the uselessness of adding taps and keeps the equaliser rather short in all cases. A hint of the limitations of linear equalisation is given by the fact that all graphs showing the measured MSE level for this particular channel appear somewhat messy as all curves are very close to each other. In other words, no matter what parameters are changed the performance does not improve.

Figures 5.25 and 5.27 present the LMS MSE curves for different fixed and variable length equalisers under two different noise conditions (5 dB and 25 dB). In this channel the equaliser had initially 6 taps and the delay was set to 5 samples. The LMS step size was 0.01.

As has been done with the previous channel profile, the average length of the variable length equaliser can be computed from figures 5.26 and 5.28. Using the $\alpha_{up}$ =0.8 data, for E/No=25 dB the calculated $M_{av}$ is 10.9 taps. For E/No=5 dB the average length is 8.9 taps. Again is important to compare these average lengths with the results from section 5.1, in particular the circled curves in figure 5.5, and notice that the variable length equaliser does not expand very much because adding taps does not help in a reduction of the MSE level. In fact, figure 5.27 clearly shows that by increasing the length, the performance of the equaliser is degraded.

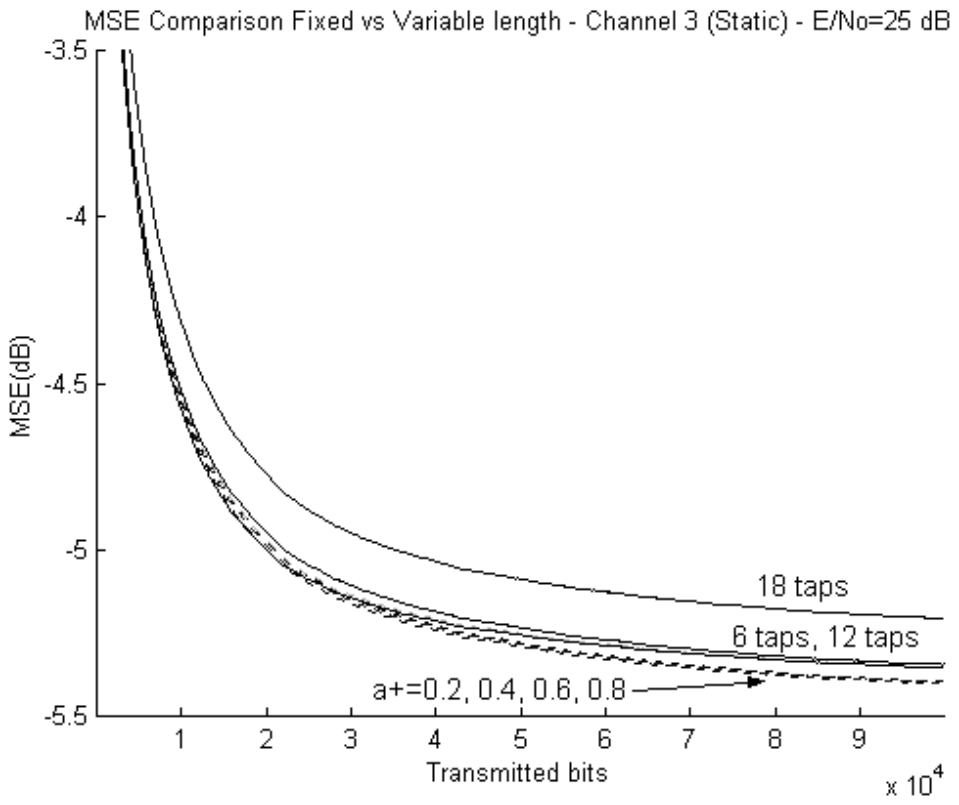

**Figure 5.25**: LMS MSE comparison of variable and fixed length equalisers. Channel model 3 (Static). E/No=25dB.

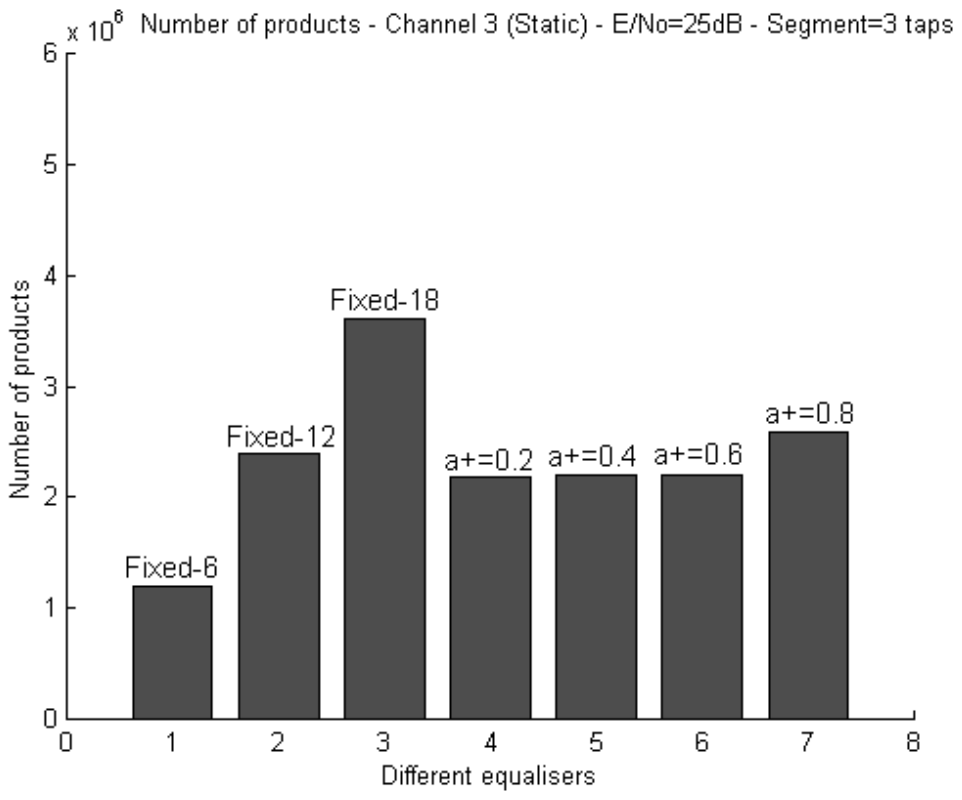

**Figure 5.26**: Comparison of the number of products computed over 100,000 iterations. Channel model 3 (Static). E/No=25dB.

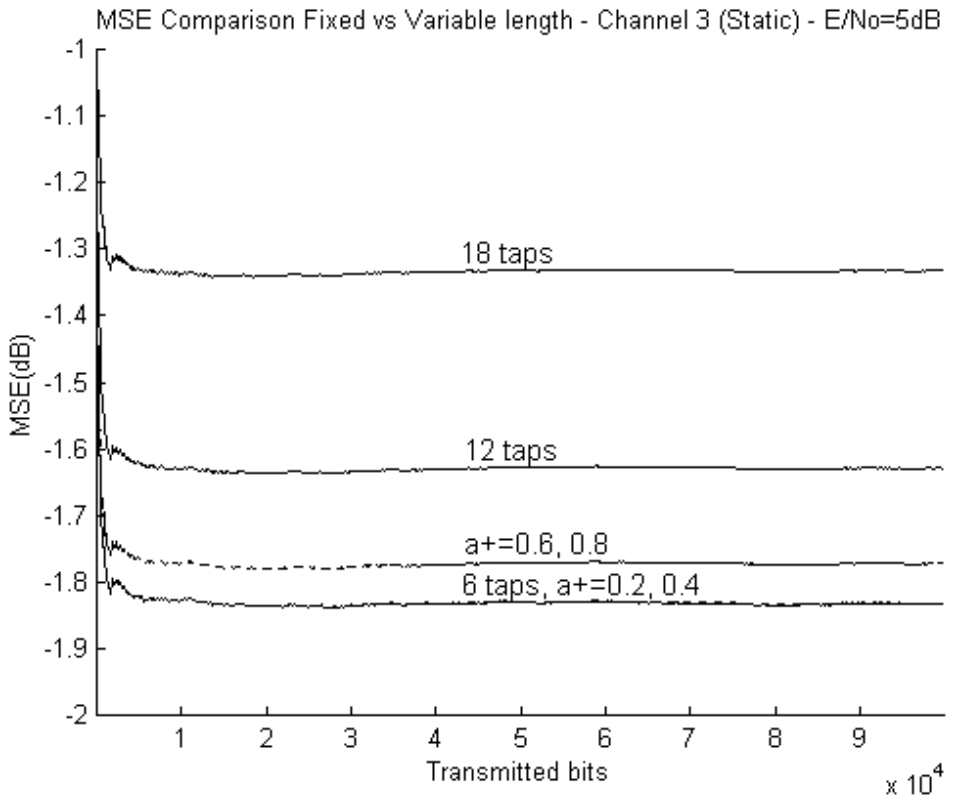

**Figure 5.27**: LMS MSE comparison of variable and fixed length equalisers. Channel model 3 (Static). E/No=5dB.

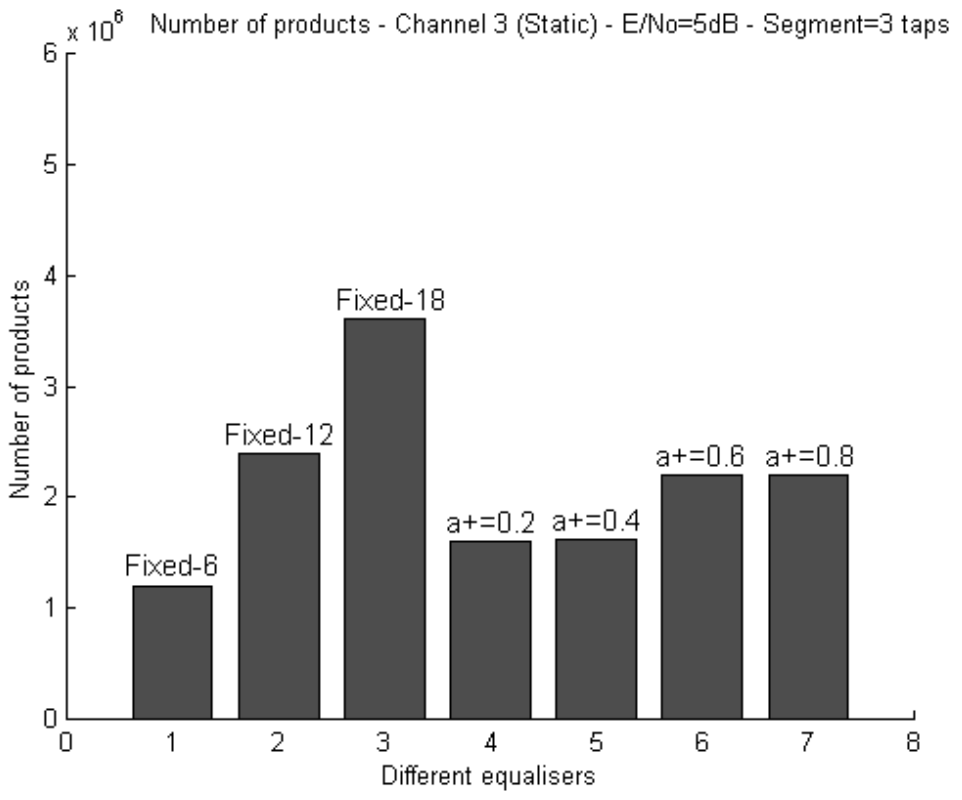

**Figure 5.28**: Comparison of the number of products computed over 100,000 iterations. Channel model 3 (Static). E/No=5dB.

Similar results are obtained when the RLS algorithm is used to drive the equaliser coefficients. The results are shown in figures 5.29 through 5.32. Using the data from figures 5.30 and 5.32 the average length when using the RLS algorithm can be computed, resulting in 11.02 taps for E/No=5 dB and 14.3 taps for E/No=25 dB.

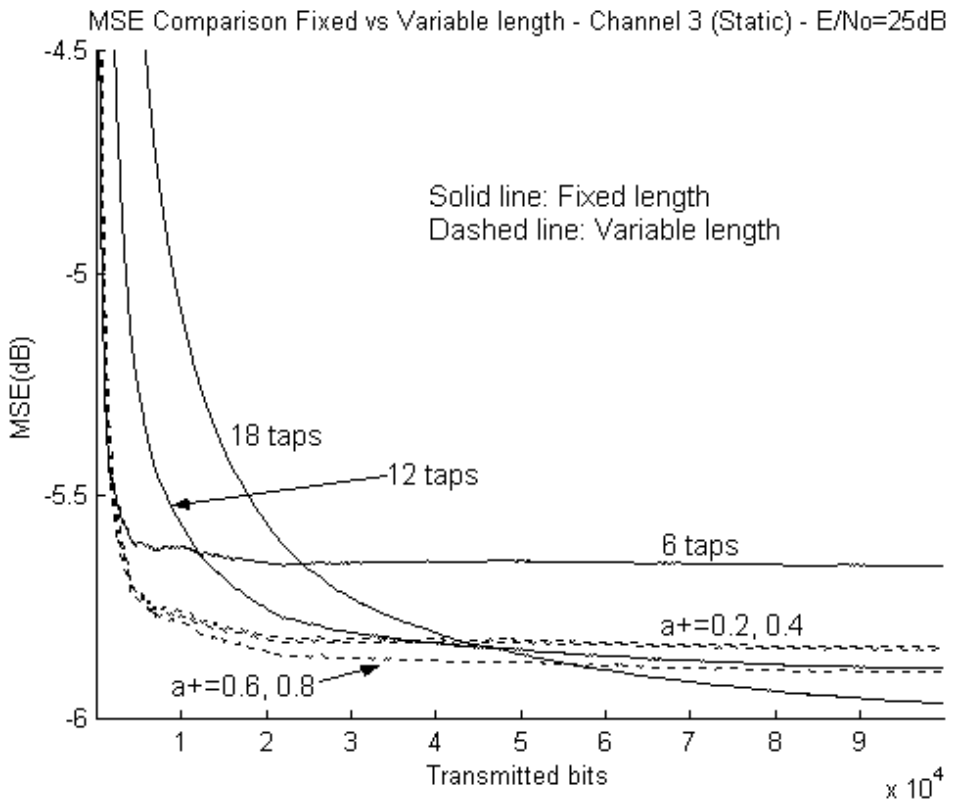

**Figure 5.29**: RLS MSE comparison of variable and fixed length equalisers. Channel model 3 (Static). E/No=25dB.

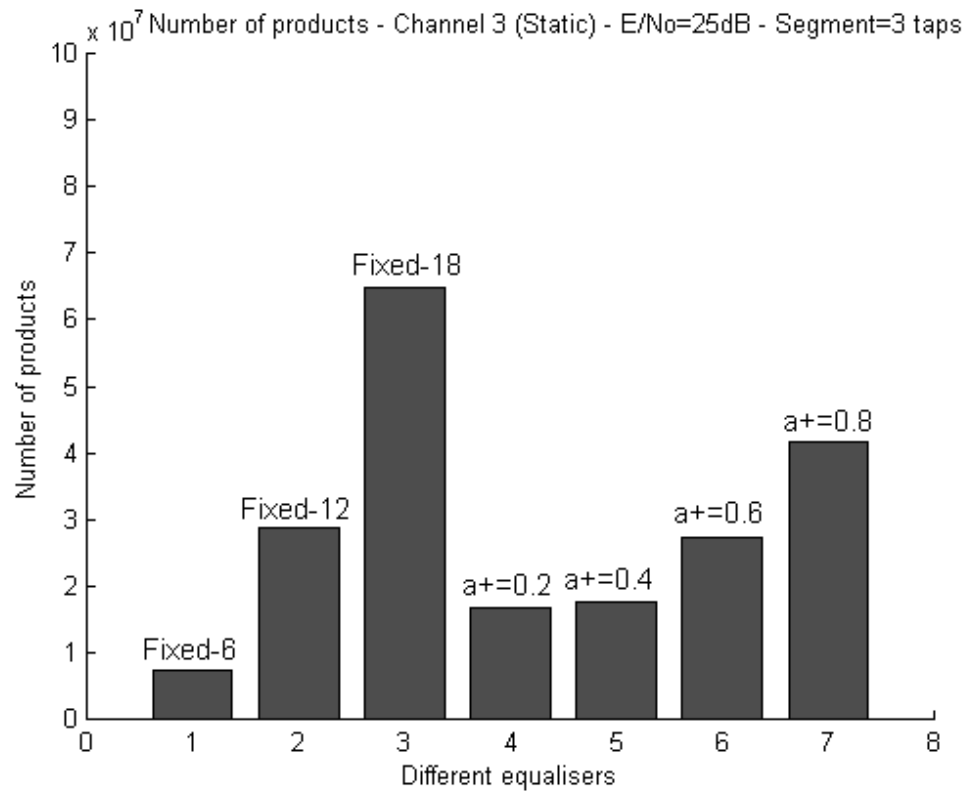

**Figure 5.30**: Comparison no. of products computed over 100,000 iterations. Channel model 3 (Static). E/No=25dB.

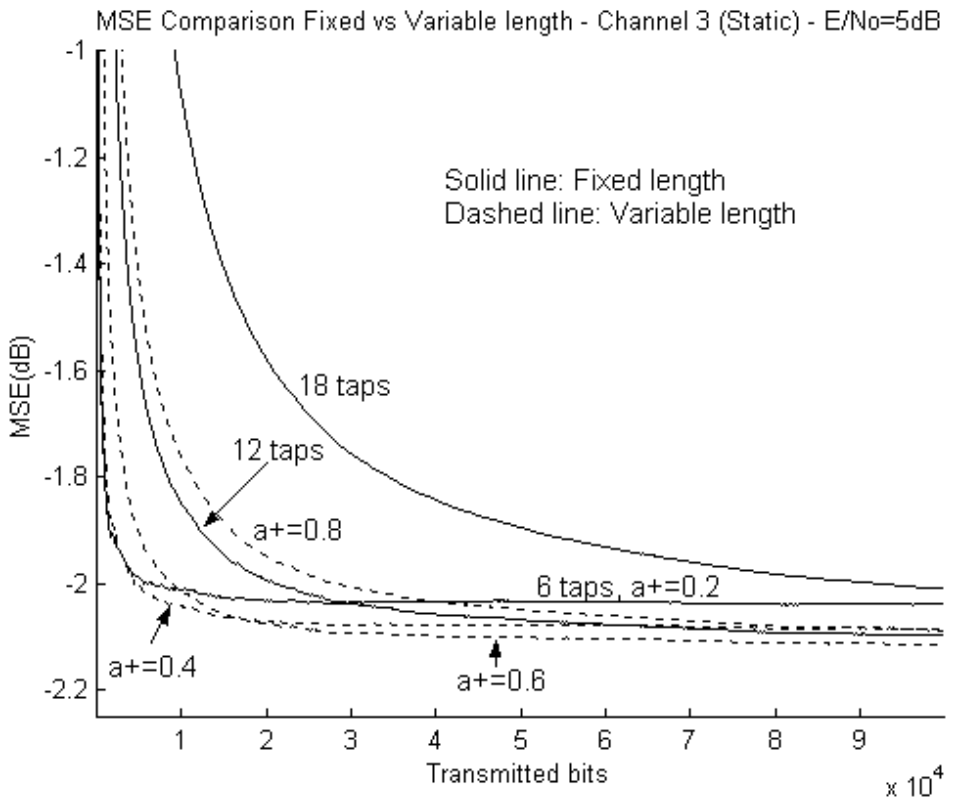

**Figure 5.31**: RLS MSE comparison of variable and fixed length equalisers. Channel model 3 (Static). E/No=5dB.

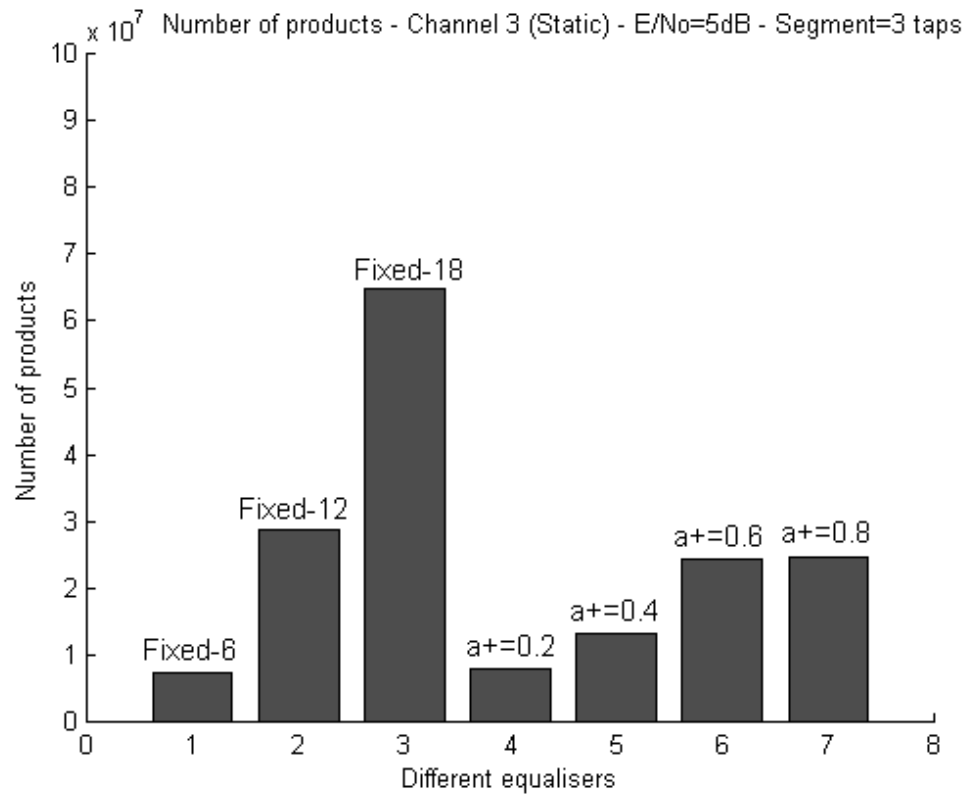

**Figure 5.32**: Comparison of the number of products computed over 100,000 iterations. Channel model 3 (Static). E/No=5dB.

A recurring effect that can be observed when comparing the resulting $M_{av}$ values for the LMS and RLS algorithms is that normally, when using the RLS algorithm, the equaliser tends to grow more than when the LMS is used. Again this is due to the ability of the RLS to avoid any EMSE. In connection with this, and contrary to the LMS case, when using the RLS algorithm, increasing the equaliser length always decreases the MSE level. However after a certain number of taps, the reduction becomes minimal. Therefore it is possible, or even advisable, to use a lower $\alpha_{up}$ such as 0.6 to avoid this slight over-expansion of the equaliser with respect to the optimum number of taps when using the RLS algorithm.

### 5.4.3 Channel model 3 (Dynamic)

The simulation results presented so far were based on static channel models. This section and the next two show results obtained by operating the variable length equaliser in dynamic environments.

In this particular section, the time varying nature of the system comes from fluctuations in the channel profile coefficients. In order to generate the channel variations, the Markov model update equation (equation 4.1) introduced in the previous chapter has been used. Unlike the previous simulations, the algorithm used here is the robust version of the variable length equaliser (algorithm 5.2), a choice motivated by the varying scenario.

The initial profile of the channel is the one defined by Channel model 3 (figures 3.6 and 3.7), as in the previous section. There is an important point worth noting about this channel model. As has been said, the spectrum of the initial channel profile contains a deep spectral null in the passband that severely limits the action of a linear equaliser (this has been shown in 5.4.2). So it can be said that, initially, the channel is a "worst case scenario". The successive addition of random components, following the Markov recursion, to the channel coefficients will tend to improve the spectral properties of the channel making linear equalisation more and more effective. In fact, this is the main point of this simulation, to show that the variable length equaliser is able to detect the channel improvement and increases its length accordingly. The variable length equaliser parameters have been chosen as follows: $\alpha_{up}$ =0.8, $\alpha_{dw}$ =0.99 and $\beta$=0.999. The segment length was 3 taps per segment and initially the equaliser had 6 taps. The delay, as in the static case, was set to 5 samples. The adaptive algorithm used was the LMS with step size 0.01. The noise level was set at E/No=25dB.

Figure 5.33 shows the MSE evolution of a variable length equaliser with the specified parameters and compares it with that of various fixed length equalisers. Comparing this figure with figure 5.25 (same profile but static) it is clear that the channel variations help significantly in improving the channel respect the initial profile. Note that the lowest MSE level achieved in 5.33 is about –8 dB and is still descending. In contrast, in figure 5.25 the lowest MSE level converges to about –5 dB. Another salient feature of this comparison is that in figure 5.25 the fixed equalisers with 6 and 12 taps achieve the best performance (nearly identical) while the 18 tap equaliser performs less well. Now in figure 5.33, the 6 tap equaliser is the one with the worst performance whereas the 12 and 18 taps equalisers attain practically the same MSE level. Remember that the average equaliser length in the static profile under identical conditions and using the same parameters was deemed to be $M_{av}$=10.9 taps. The length progression is shown in figure 5.34 and it can be clearly seen that the variable length equaliser tends to increase its order up to nearly 14 taps. That is, the variable length equaliser is able to infer that the channel conditions are becoming favourable and increase the equaliser length to further reduce the MSE level.

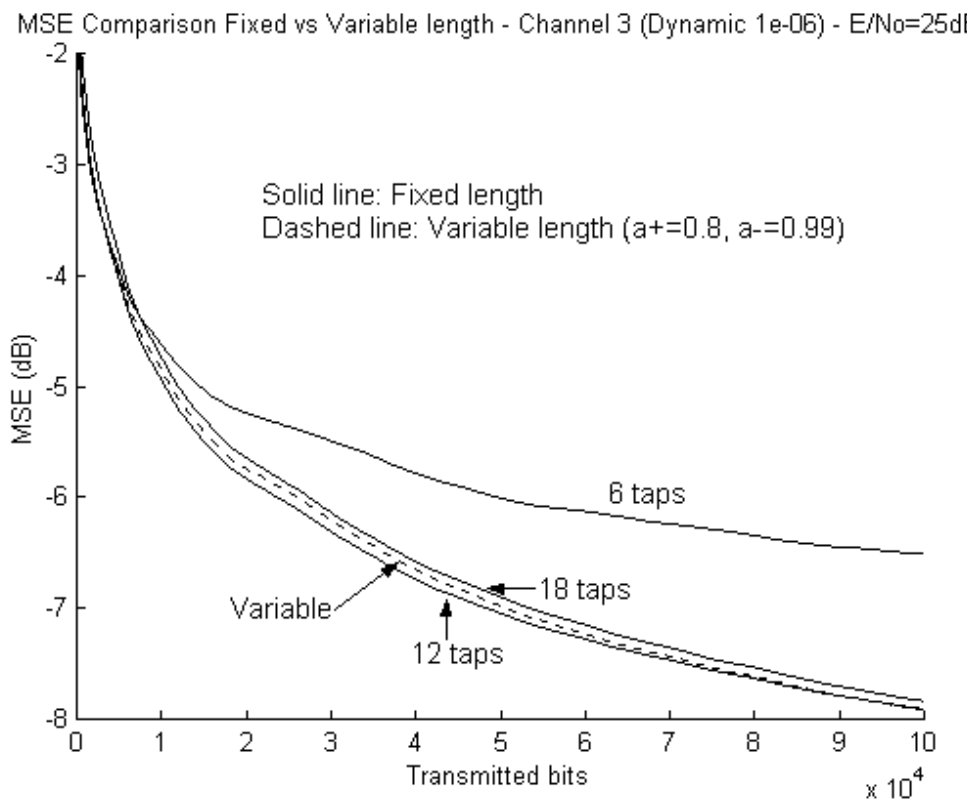

**Figure 5.33**: LMS MSE comparison of variable and fixed length equalisers. Channel model 3 (Dynamic $\sigma_q^2=10^{-6}$). E/No=25dB.

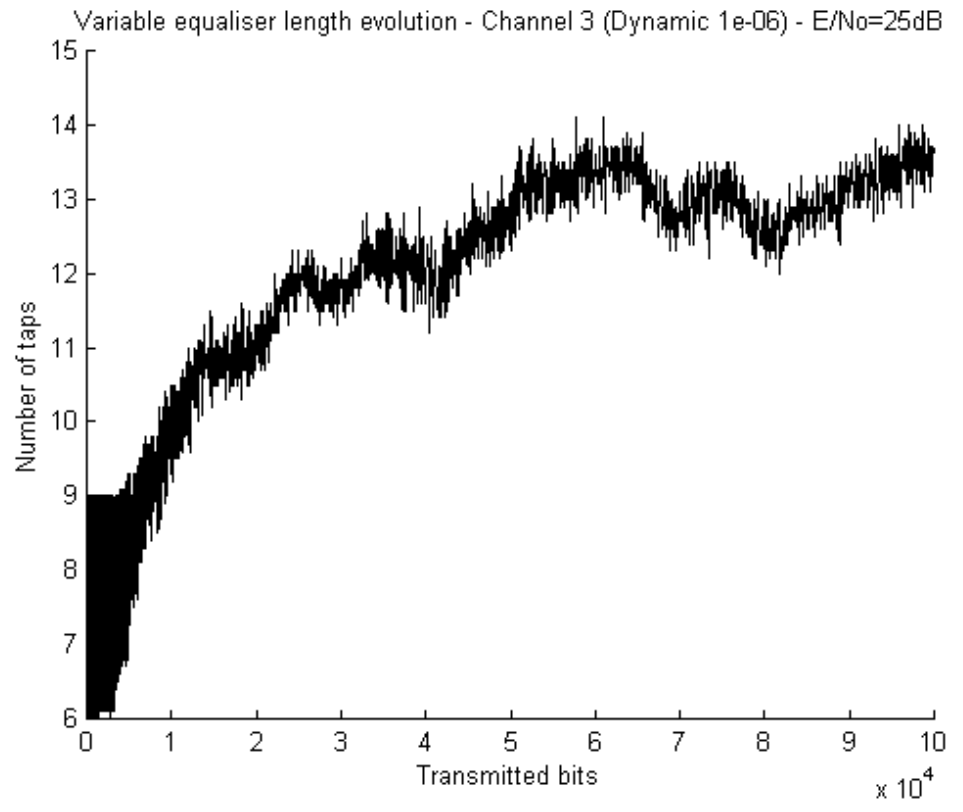

**Figure 5.34**: Evolution of the variable equaliser length in a dynamic environment. Channel model 3 (Dynamic $\sigma_q^2=10^{-6}$). E/No=25dB.

For the RLS algorithm, similar results were found. The only difference is that, as in the case of the static channels, the achieved length with the RLS algorithm was slightly longer than the LMS values presented above.

### 5.4.4 Variable E/No profile

The objective of this next set of simulations is to observe the variable length equaliser behaviour when there are sudden changes in the environment. In this case the variations are in the E/No level. Abrupt changes in the noise level can come from different physical phenomena such as sudden shadowing of the signal by an object, for example, a big building. Also in the context of CDMA systems, the signal from other users can be roughly modelled as AWGN ([Wilson93]). In this case, the switching on and off of users in the system will provoke a variation in the E/No level. This also applies for GSM systems when accounting for the co-channel interference.

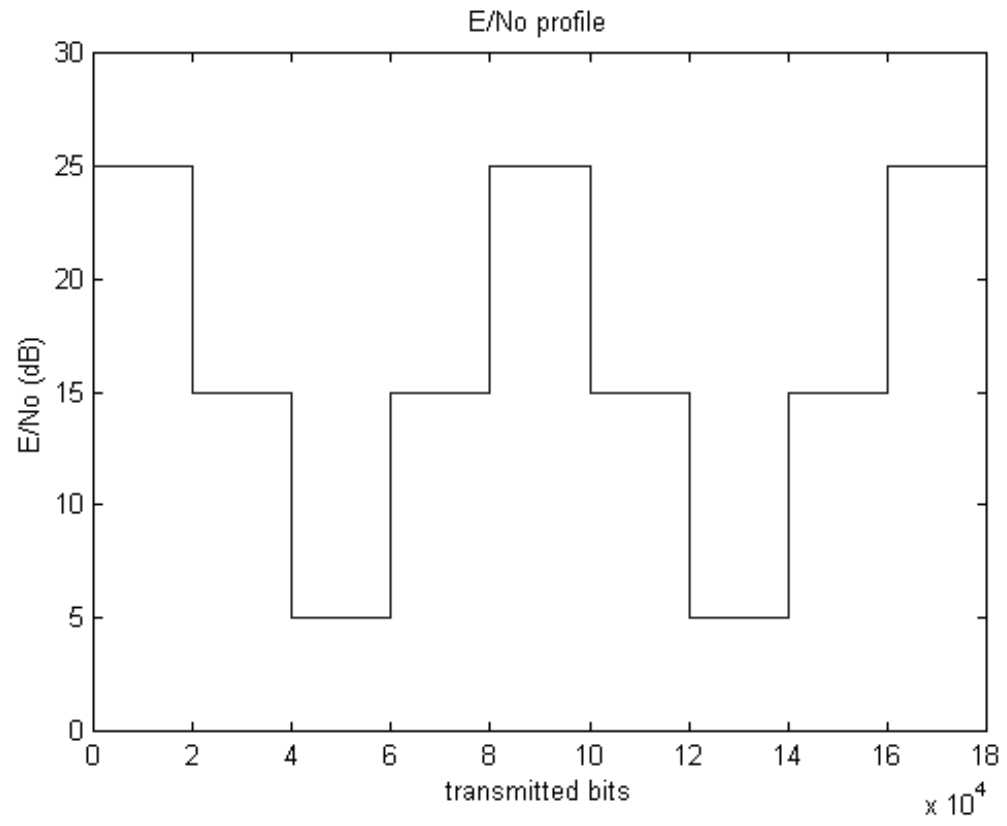

**Figure 5.35:** Variable E/No profile.

The E/No profile used in this set of simulations is shown in figure 5.35. As can be seen in the figure, the E/No level oscillates between 5 dB and 25 dB in steps of 10 dB at regular instants (every 20,000 samples). This profile helps to highlight how the variable length equaliser confronts these variations.

At this point, it is important to clarify a particular aspect of the performance measure used in these simulations. So far the quantity measured has been the MSE. Initially the same quantity was used in this particular scenario however a “visualisation” problem arises due to the different MSE levels involved in the same graph.

In particular, when the E/No level is 5 dB, the MSE curve attains its maximum value, which is much bigger than the minimum value achieved when E/No=25 dB. The problem comes from the fact that the MSE is recorded continually from the start of the simulation. Once the MSE has gone through its maximum value (E/No=5 dB) it is impossible to appreciate the subsequent part of the MSE as the largest values dominate from that point onwards the rest of the curve.

In order to solve this problem the recorded MSE is a windowed version of the original MSE, i.e., only the values inside the window corresponding to the most recent data are taken into account in the MSE computation. This has the effect of “forgetting” old data and the resulting MSE curve reflects more accurately what it is happening in the system. The window length clearly affects the shape of the resulting MSE curve. Large windows will tend to present smoother MSE curves than short windows.

For the simulations presented here, and after several trials, the window length has been chosen to be of 2,000 samples. The following two figures show the windowed MSE obtained when using fixed length equalisers to compensate the static Channel model 2 with the variable E/No profile superimposed.

The simulation length was set to 180,000 samples, so during the simulation, 8 changes in the E/No level occurred as shown in figure 5.35. Figure 5.36 shows the whole simulation whereas in figure 5.37 the MSE evolution corresponding to one of the intervals when E/No=5 dB is zoomed as it is difficult to appreciate it in the general view of the simulation.

As can be seen in figure 5.36, the windowed MSE level closely follows the variations in the E/No level (figure 5.35). Clearly when E/No is high, the longer the equaliser, the lower is the attained MSE level. On the contrary when E/No is low, the shortest equaliser of the three works best as shown in figure 5.37 and also requires fewest computations.

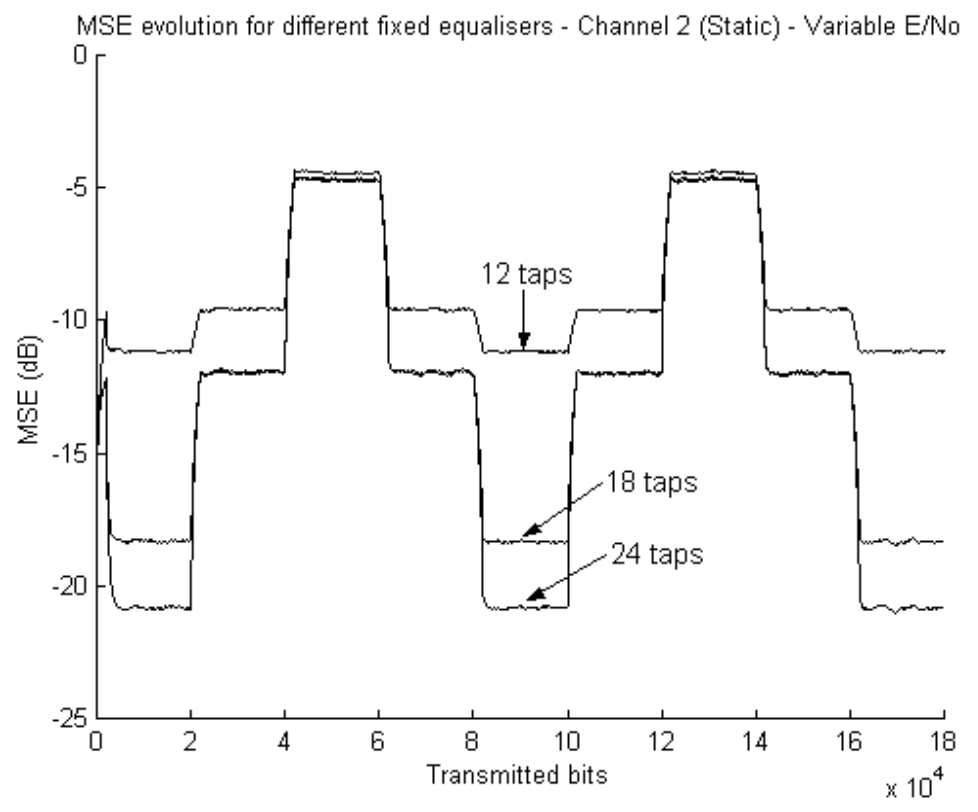

**Figure 5.36**: LMS MSE for different fixed length equalisers. Channel model 2 (Static). Varying E/No profile. Full view.

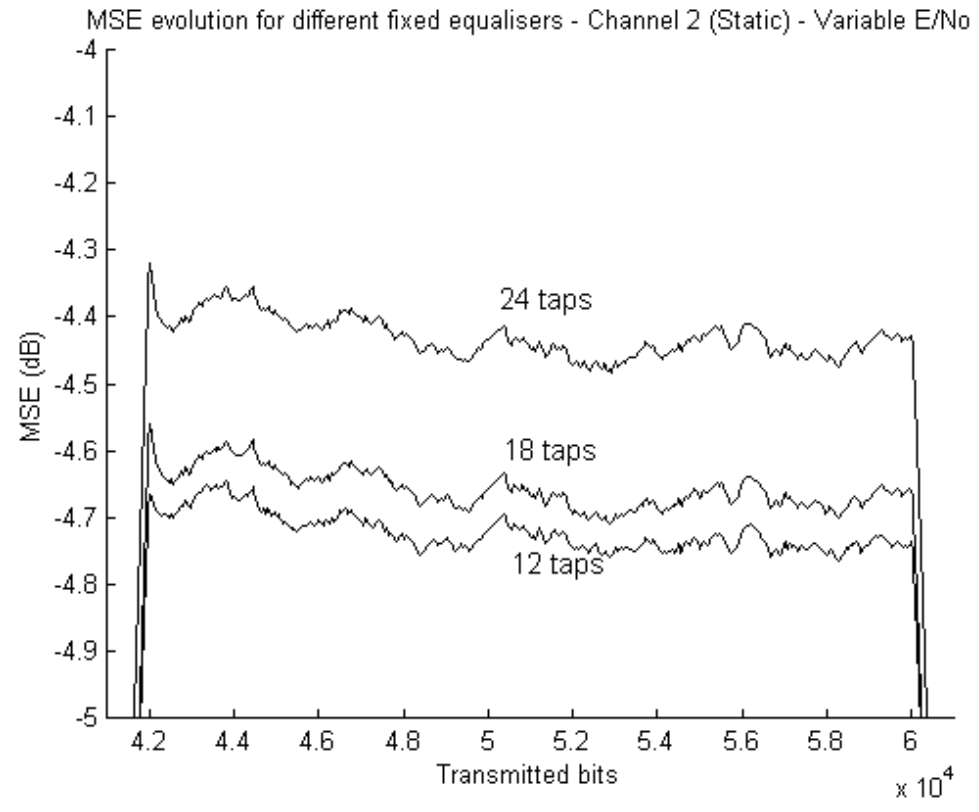

**Figure 5.37**: MSE for different fixed length equalisers. Channel model 2 (Static). Varying E/No profile. Zoomed view [41,000-61,000].

The next figures (5.38 and 5.39) show how the variable length equaliser copes with this environment. Its parameters are chosen as: $\alpha_{up}$ =0.8, 0.6, 0.4, $\alpha_{dw}$ =0.99, $\beta$=0.999 and 3 taps per segment. The LMS step size was kept to 0.01.

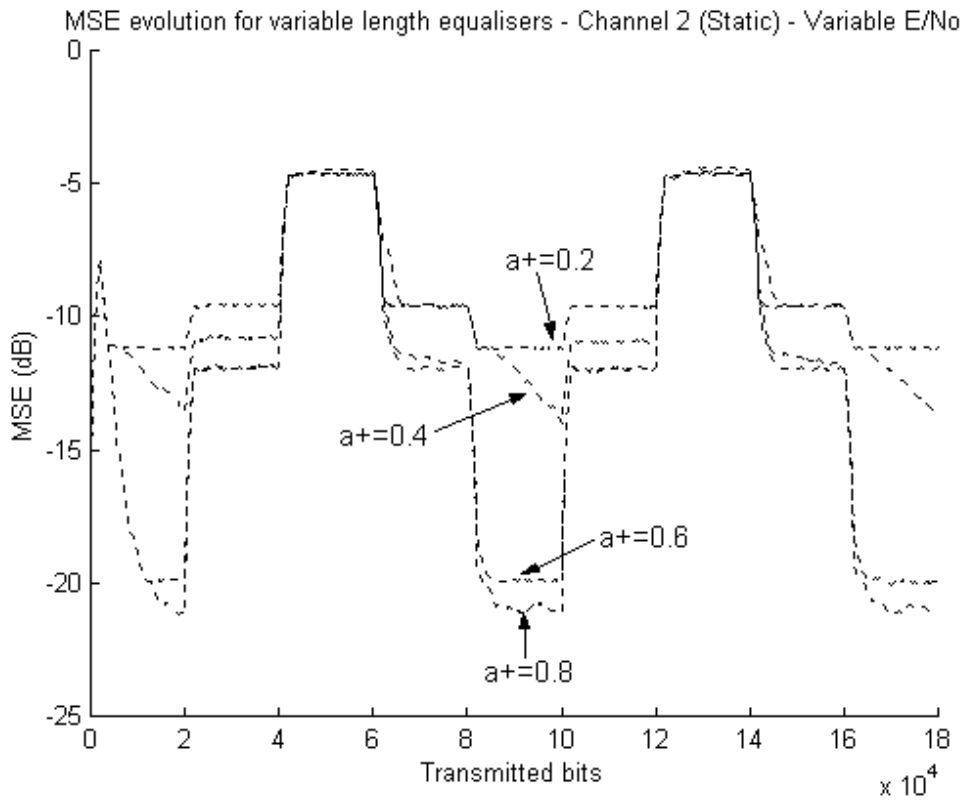

**Figure 5.38**: LMS MSE for different variable length equalisers. Channel model 2 (Static). Varying E/No profile.

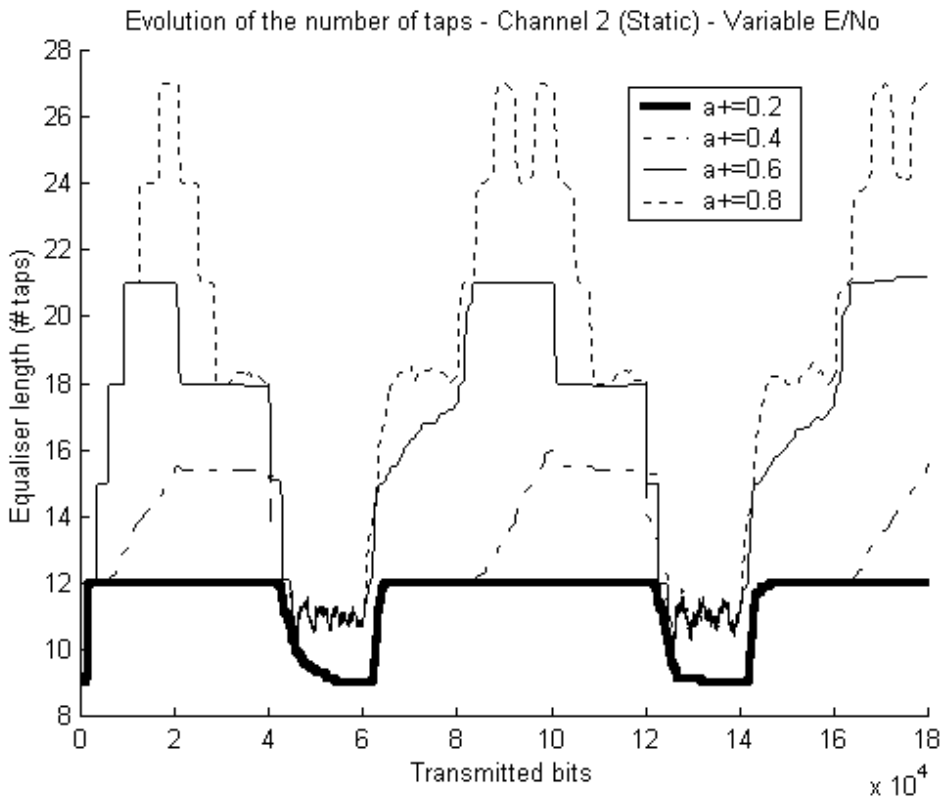

**Figure 5.39**: Equaliser length evolution with different variable length equalisers (LMS). Channel model 2 (Static). Varying E/No profile.

Comparing figure 5.38 with 5.36 it is clear that when the variable length equaliser is used with a large $\alpha_{up}$ (for example 0.8), the MSE when E/No is high resembles that of the long fixed equalisers. Notice that a zoomed version of 5.38 for a low E/No interval is not shown as even zooming cannot be used to distinguish the different curves because they all have nearly identical MSE curves (except for $\alpha_{up}$ =0.2). The reason for such a similar behaviour is that the different equalisers contract when detecting that the conditions are not favourable and hence achieve the same MSE. The length evolution for different $\alpha_{up}$ values is shown in figure 5.39. This can be considered as one of the most important figures of the chapter as it illustrates how the equaliser performs a self reconfiguration as channel conditions varies.

There are several points worth noting in 5.39. First and mainly, the equaliser length tracks the E/No profile, expanding when the noise level is low and contracting when it is high. Second, different values of $\alpha_{up}$ offer different levels of performance. Notice for example the difference between $\alpha_{up}$ =0.8 and $\alpha_{up}$ =0.6 in terms of MSE achieved and computational complexity (which can be deduced from the length evolution). Clearly low $\alpha_{up}$ values will not be used as their performance is very limited, however the designer may choose to use a value in the range [0.6 **..** 0.9] taking into account the computational power available.

The MSE performance for $\alpha_{up}$ =0.6 is only slightly worse than for $\alpha_{up}$ =0.8 and on average uses about 4-5 taps less saving some computations. It is up to the designer to decide whether it is worth increasing the computational complexity of the equalisation process to get this "last drop" of performance. Finally, notice that when $\alpha_{up}$=0.8, looking at the equaliser length achieved for each E/No level, these values again match with the optimum lengths deduced from the curves in figure 5.3.

Graphs comparing the computational complexity of fixed and variable length equalisers (similar to figures 5.26, 5.28) are not shown now as the computational savings the variable structure offers depend very much on the E/No profile. However the simulations presented above give an idea of the potential benefits of this approach.

The same profile has been tested using the RLS algorithm. The results obtained in this scenario show the stability problems arising from the combination of the RLS algorithms and the variable length structure. The RLS forgetting factor was set to 0.999 for all simulations.

RLS MSE performance curves using fixed length equalisers are shown in figure 5.40. Notice that this figure is very similar to figure 5.36 corresponding to the LMS. In particular, notice that in this scenario RLS does not offer any convergence improvement when there is a sudden E/No change. This is because the channel coefficients are not altered and therefore the equaliser coefficients do not change drastically (they are only modified to avoid enhancement of the new noise floor).

The MSE curves when using different variable length equalisers are shown in figure 5.41. Logically, increasing the value for $\alpha_{up}$ lowers the MSE, however notice that the curve corresponding to $\alpha_{up}$=0.6 presents a fairly degraded aspect, clearly much worse than the equivalent curve using the LMS algorithm. In section 5.2.3 it was mentioned that the nearer $\alpha_{up}$ and $\alpha_{dw}$ are, the more prone the equaliser is to length oscillation.

If the LMS algorithm is used, length oscillation does not cause any trouble at all. On the other hand, when using the RLS algorithm, and according to the stabilisation procedure described in section 5.3.2, the RLS algorithm is re-initialised every time the equaliser contracts. Therefore, if the length is oscillating often, the equaliser is continually being restarted, causing an increase in the MSE level. In order to alleviate this problem, $\alpha_{dw}$ was set to 1.0 (instead of the 0.99 used so far) and, as can be observed in figure 5.41, the problem was not completely solved. The MSE curve corresponding to $\alpha_{up}$=0.8 is not shown in figure 5.41 as its performance was far worse than the one corresponding to $\alpha_{up}$ =0.6 due to an even higher restart rate. Figure 5.42 presents the equaliser length evolution corresponding to the different $\alpha_{up}$ values shown in figure 5.41. The higher value used for $\alpha_{dw}$ (compared to the simulations displayed so far), in order to minimise the number of restarts, makes the equaliser less likely to contract, it only does so when conditions are very favourable. This is reflected in figure 5.42 by the fact that when E/No drops from 25 to 15 dB, the length is still kept constant, and only when it drops to 5 dB does the equaliser contract.

It should be point out that results are not as bad as figure 5.41 suggests. The MSE curves shown in 5.41, as almost any other MSE curve shown in this thesis, result from the averaging of 30 independent runs. The filter is restarted whenever the equaliser contracts, therefore after each contraction there is a short period of time where the equaliser is re-converging. In each simulation run, the contractions take place at different instants in time. When the results are averaged, all the MSE peaks pertaining to different runs tend to appear on the averaged curve.

Figure 5.43 show the RLS MSE from a single realisation using the same scenario as in figure 5.41 with $\alpha_{up}$ =0.6. The curve in figure 5.43 resembles closely that for the LMS algorithm shown in figure 5.38 (in fact, for high E/No the RLS achieves a lower MSE), except for the MSE peaks due to the equaliser restarts. Notice that the convergence stage after the restart is fairly short. Certainly, the appearance of figure 5.43 is much more acceptable than the averaged curve for $\alpha_{up}$ =0.6 in figure 5.41.

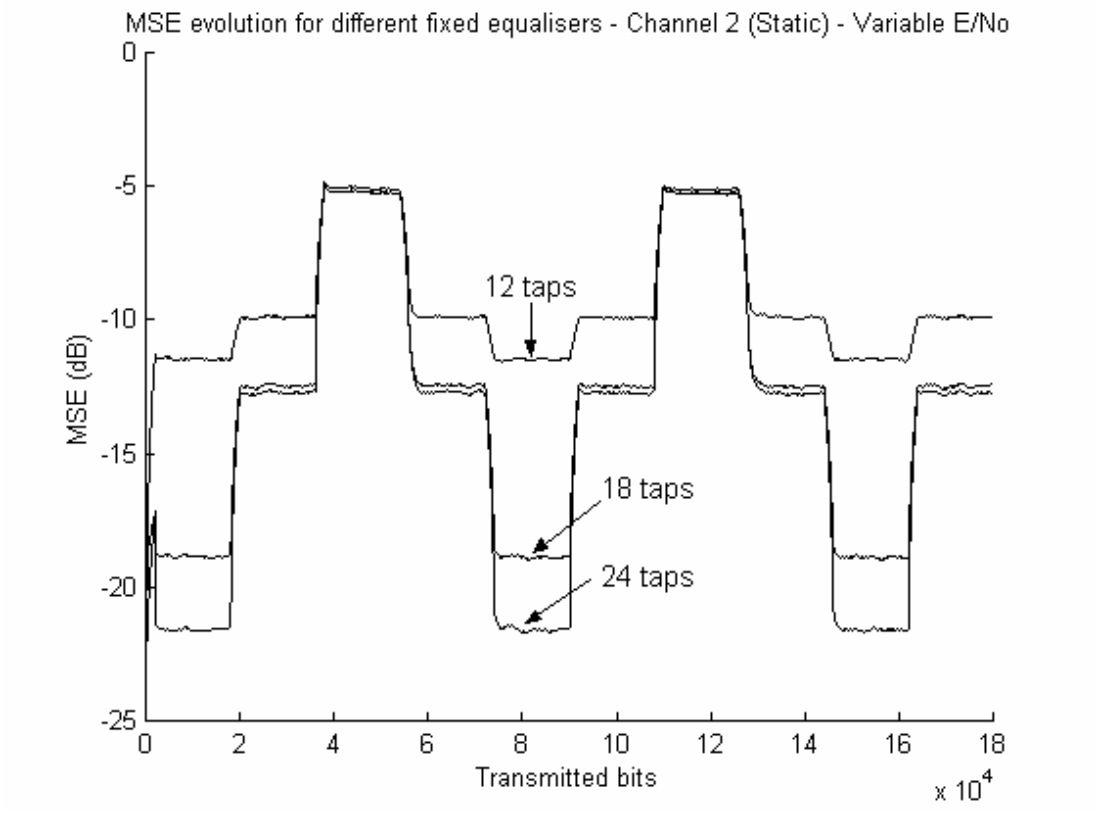

**Figure 5.40**: RLS MSE for different fixed length equalisers. Channel model 2 (Static). Varying E/No profile. Full view.

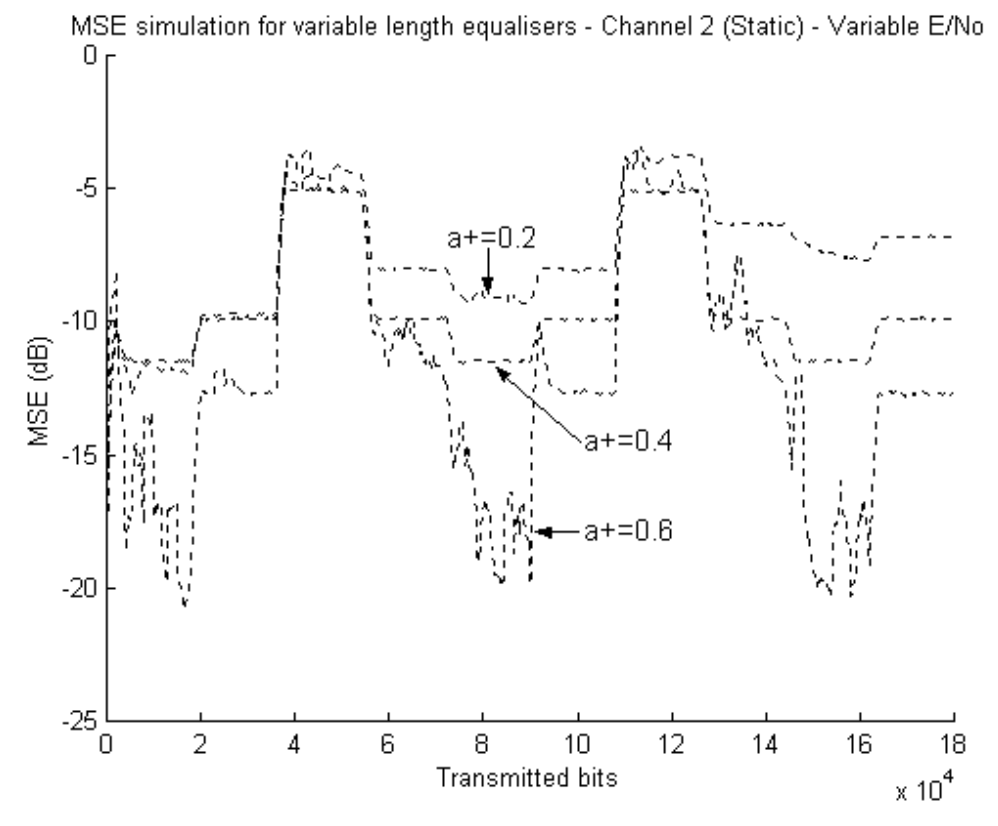

**Figure 5.41**: RLS MSE for different variable length equalisers. Channel model 2 (Static). Varying E/No profile.

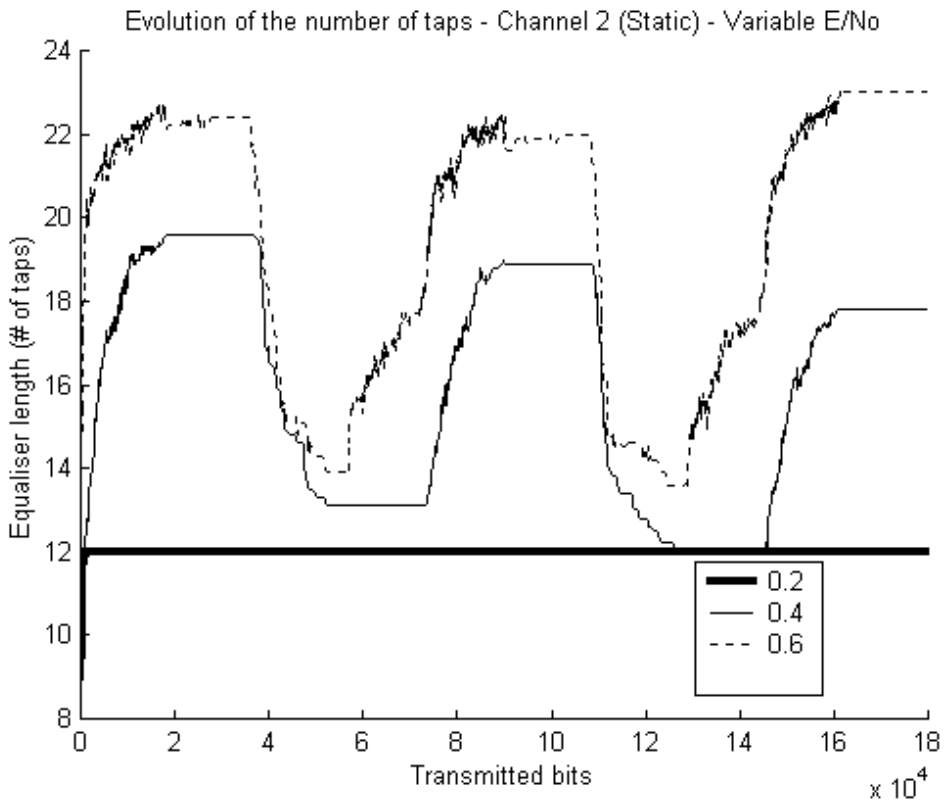

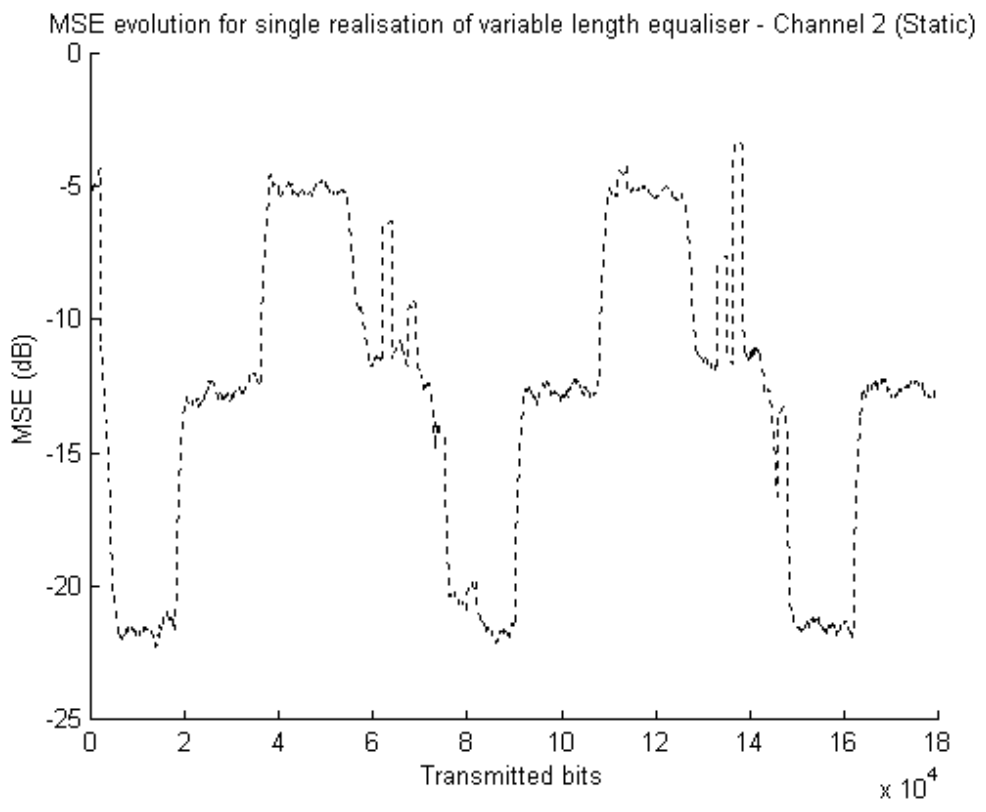

**Figure 5.42**: Equaliser length evolution with different variable length equalisers (RLS). Channel model 2 (Static). Varying E/No profile.

**Figure 5.43**: Single realisation of RLS variable length equaliser using $\alpha_{up}$ =0.6, $\alpha_{dw}$ =1.0. Channel model 2 (Static). Varying E/No profile.

## 5.4.5 Abrupt changes in the channel profile

The last scenario used for highlighting the performance of the variable length equaliser consists of simulations containing abrupt changes in the channel profile. In mobile/cellular systems such changes are very common and might be caused, for example, by the handover from one base station to another or by some sudden change such as going from an outdoor to an indoor environment.

In the simulations presented here two combinations of channels have been used. The first profile starts with the impulse response corresponding to Channel model 2, then switches to Channel model 1 and later on goes back to Channel model 2. This would correspond to an average quality channel (Channel model 2) with a sudden improvement in its characteristics (Channel model 1) which may be due, for example, to the appearance of direct line of sight with the transmitter. Finally the channel goes back to its original characteristics (Channel model 2).

The second profile used also starts using channel model 1 but then switches to Channel model 3, which could simulate a strong sudden shadowing by a nearby object. Towards the end, the channel goes back to Channel model 2.

For all simulations in this section, the length was set to 180,000 bits, with the switching instants fixed at 60,000 bits (1st change) and 120,000 bits (2nd change). The E/No has been set to 35 dB. Note that this value of E/No is higher than in previous simulations. This value has been used in order to “isolate” the behaviour of the equaliser in front of a channel profile change as it has been seen that the E/No level also influences the response of the variable length equaliser. In this set of simulations the MSE has also been windowed as otherwise the whole curve would be dominated by the largest MSE level achieved.

Figures 5.44 and 5.45 show the results obtained with the LMS algorithm using the first abrupt channel profile (Channel 2 – Channel 1 – Channel 2). In 5.44, the MSE curves are shown. The MSE curves clearly reflect the changes in the channel profile as the equaliser needs to re-converge after the switching instant.

More interesting is to observe, in figure 5.45, the length evolution when variable length equalisers are used. Initially the equaliser expands up to 30 or 33 taps (depending on the $\alpha_{up}$ used). When the channel changes to Channel model 1, the equaliser immediately detects that fewer taps are needed and shrinks to 18 taps. Finally when the channel goes back to channel model 2, it expands again up to 30 or 33 taps. It is worth noting that when the number of taps is reduced the MSE is still further reduced.

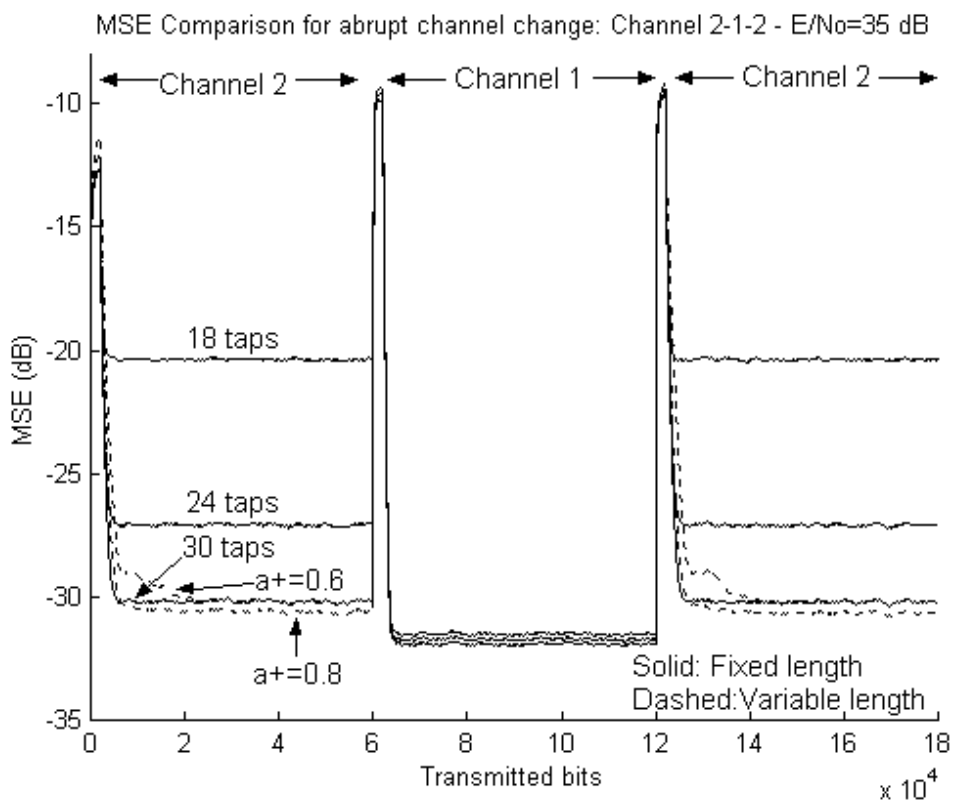

**Figure 5.44**: LMS MSE for different fixed and variable length equalisers. Abrupt channel change: 2-1-2. E/No=35 dB.

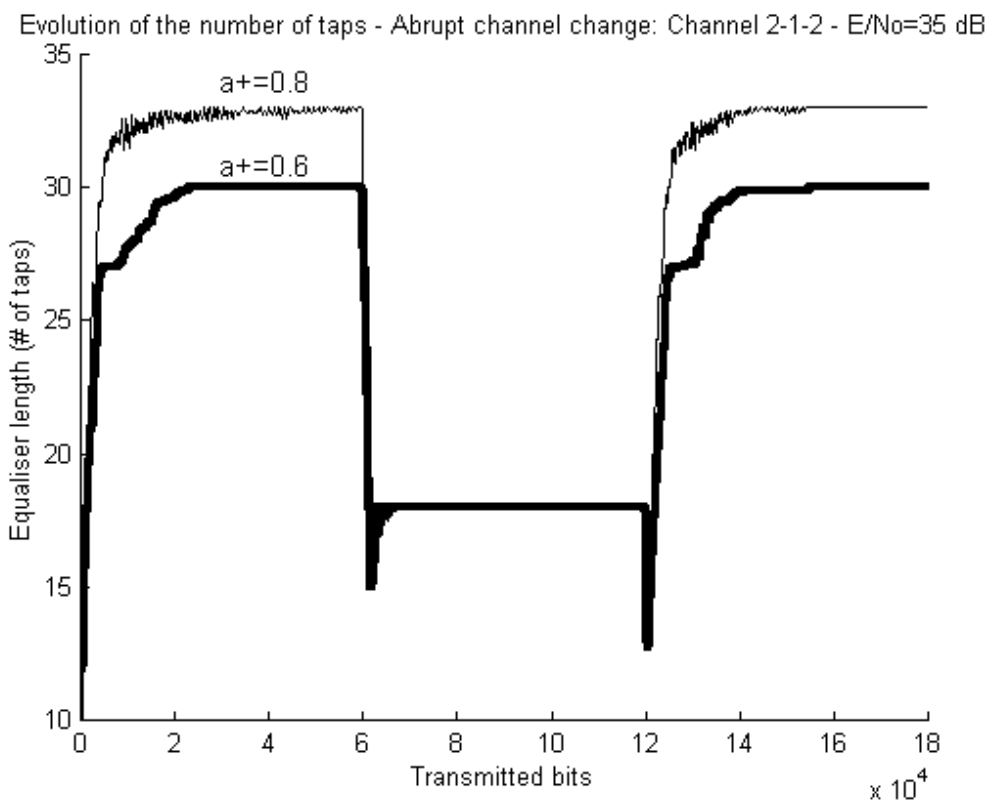

**Figure 5.45**: Equaliser length evolution for different variable length equalisers (LMS). Abrupt channel change: 2-1-2. E/No=35 dB.

The next two figures, 5.46 and 5.47, show respectively the MSE and equaliser length evolution for the second abrupt channel profile (Channel 2 – Channel 3 – Channel 2). The poor quality of Channel model 3 is shown by the huge increase in the MSE level when the channel changes from 2 to 3. Notice also in the length evolution graph (figure 5.47) that the equaliser contracts when channel 2 is switched on. It has been shown in previous sections that linear equalisation is not very effective in combating the profile of Channel model 3, therefore the variable length equaliser adequately reduces the number of taps. When Channel model 2 is restored, the MSE and length return to their respective values as in the first stage of the simulation.

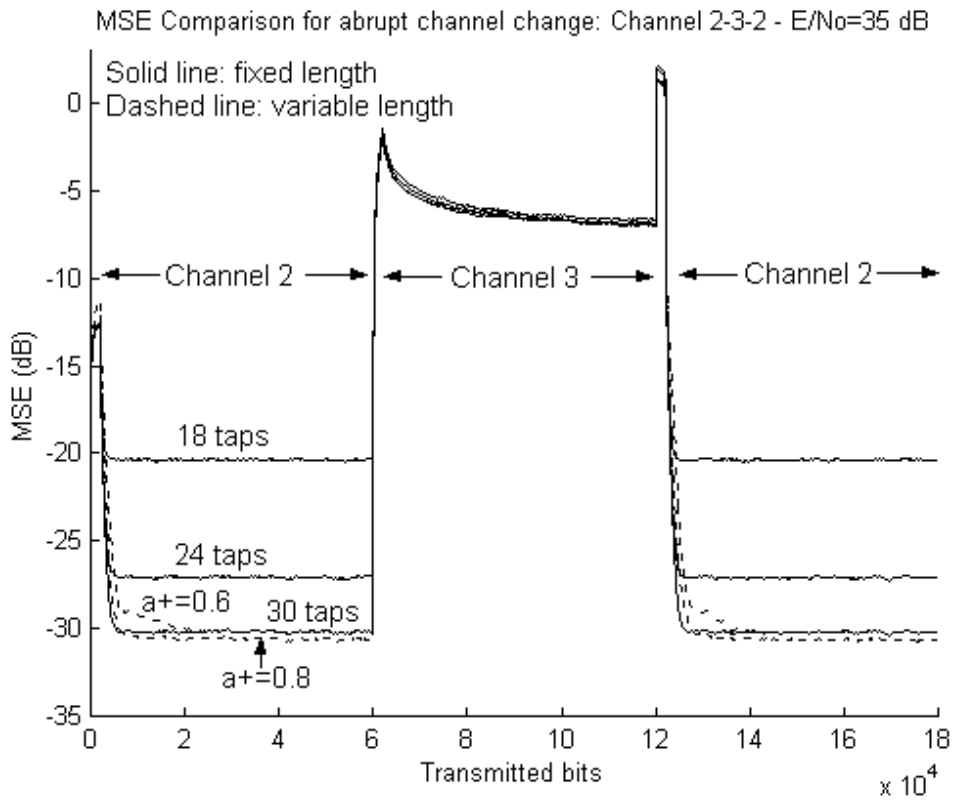

**Figure 5.46**: LMS MSE for different fixed and variable length equalisers. Abrupt channel change: 2-3-2. E/No=35 dB.

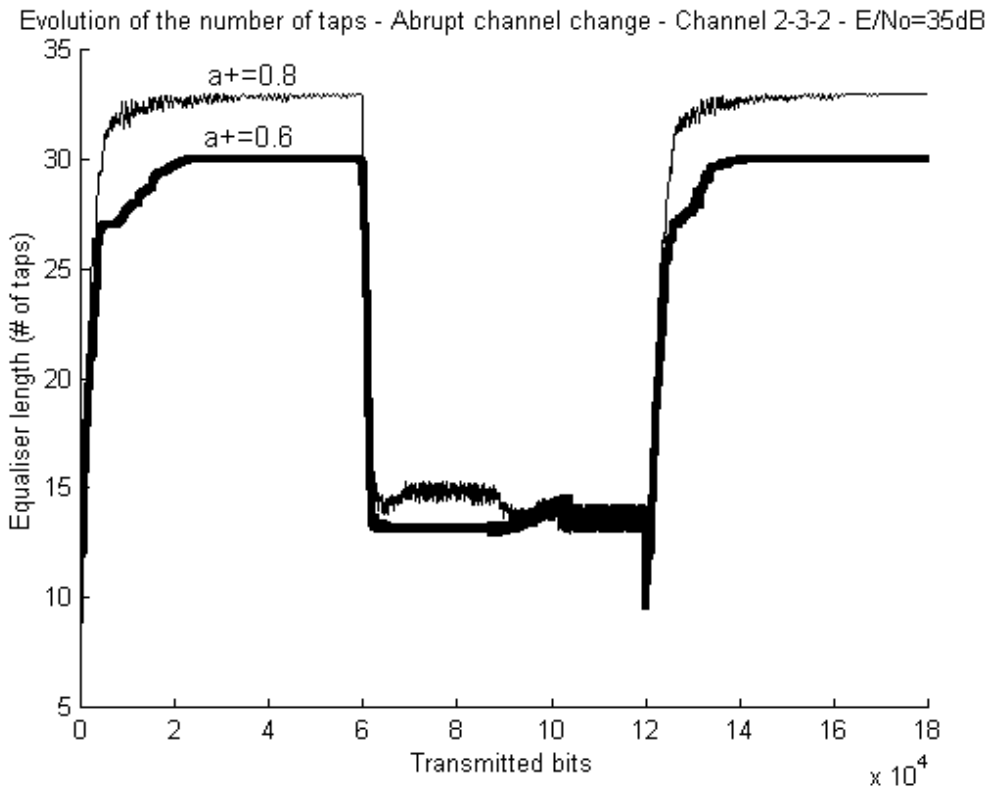

**Figure 5.47**: Equaliser length evolution for different variable length equalisers (LMS). Abrupt channel change: 2-3-2. E/No=35 dB.

One last point with respect to figure 5.46 is the fairly long time it takes to the equaliser to re-converge after the switching to Channel model 3. This is due to the large eigenvalue spread of the input autocorrelation matrix to the equaliser that slows down the operation of the LMS algorithm (see chapter 2). Notice on the other hand, that when Channel model 2 is switched on again, convergence takes place very rapidly as the eigenvalue spread is low.

The same abrupt change profile has been tested using the RLS algorithm with fixed and variable length equalisers. The results are summarised in figures 5.48 and 5.49.

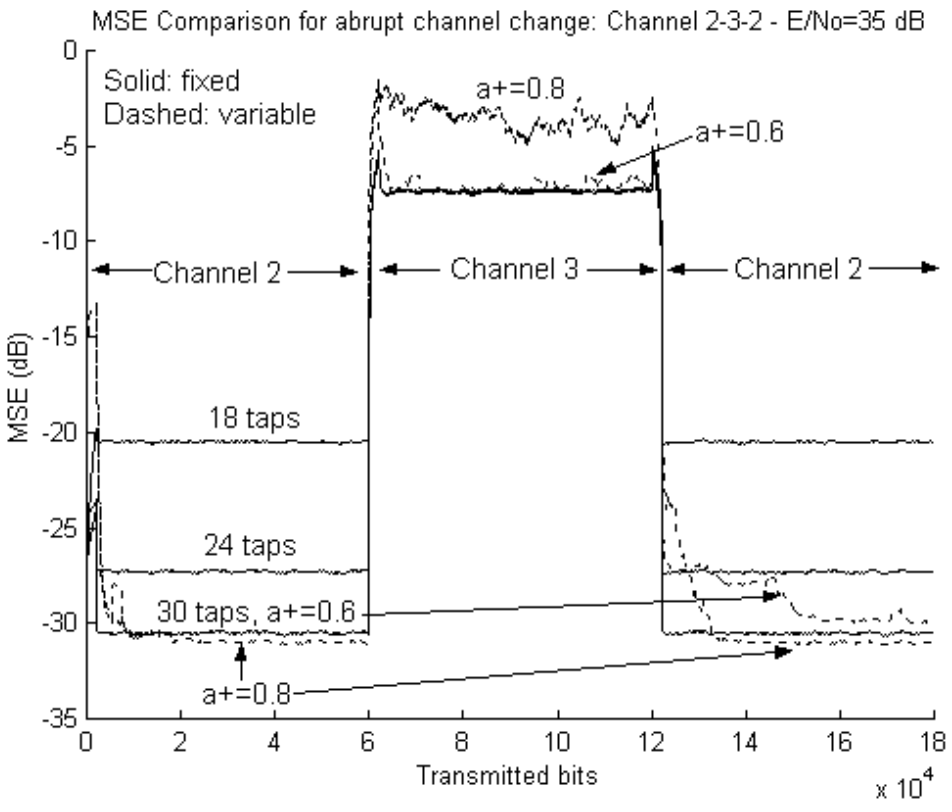

**Figure 5.48**: RLS MSE for different fixed and variable length equalisers. Abrupt channel change: 2-3-2. E/No=35 dB.

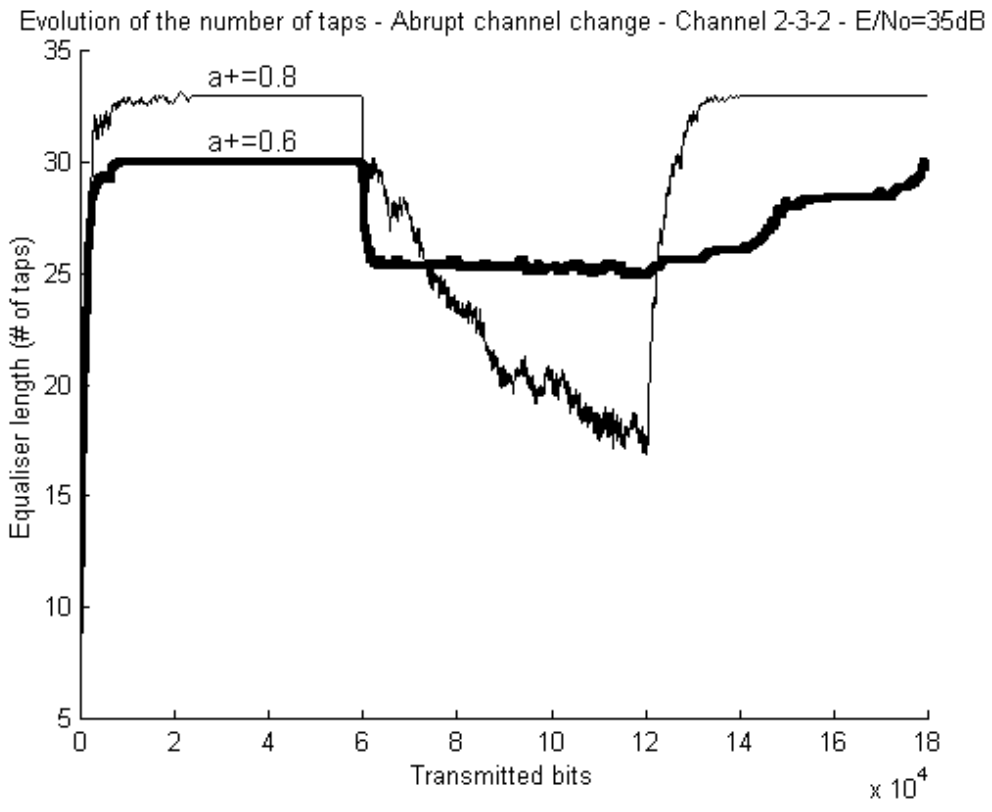

**Figure 5.49**: Equaliser length evolution for different variable length equalisers (RLS). Abrupt channel change: 2-3-2. E/No=35 dB.

As in the variable E/No profile simulations, the results for the variable length equaliser using the RLS are worse than with the LMS due to the restarting of the filter every time the equaliser is contracted. However, as has already explained, the results in figures 5.48 and 5.49 made worse by the averaging of the different runs of the simulation.

In particular notice that for $\alpha_{up}$ =0.8, during the interval the Channel model 3 profile is on, its MSE is far worse than for the rest of equalisers. In a single realisation there were only one or two MSE peaks corresponding to equaliser resets with very rapid convergence to the same MSE level as the rest of equalisers (-7.5 dB). However, when the curves are averaged, all the peaks from all the curves appear in the resulting graph.

The same applies to figure 5.49. A positive aspect of the RLS simulation in comparison with the LMS is the rapid convergence of the different equalisers even when the systems changes to Channel model 3.

# 5.5 Joint VSLMS and variable length equalisation

The simulations presented so far have concentrated on the most common algorithms, LMS and RLS. Now some attention is given to a "flavour" of the LMS algorithm, the variable step size LMS (VSLMS), whose characteristics are well complemented by that of the variable length equaliser.

The properties of the VSLMS algorithm have been described in section 2.3.1. It has been shown in previous studies ([Kwong92], [Mathews93]) and also in some of the simulations in chapter 3, that the possibility of dynamically changing the step size can potentially provide a faster convergence time and a smaller steady state error when compared with the conventional LMS algorithm. It has also been pointed out in the background chapter that all the algorithms of the LMS family (including the VSLMS) have the same constraint regarding the maximum value the step size can take in order to have stability guaranteed. For convenience the condition is rewritten here:

$$0 < \mu_{max} \leq \frac{2}{3M(\sigma_d^2 + \sigma_v^2)}$$

As can be seen, the maximum step size bears a direct relation to the equaliser length. The variable length equaliser, with its ability to reduce its number of taps, allows the step size to achieve higher values which may potentially improve the convergence of the equaliser.

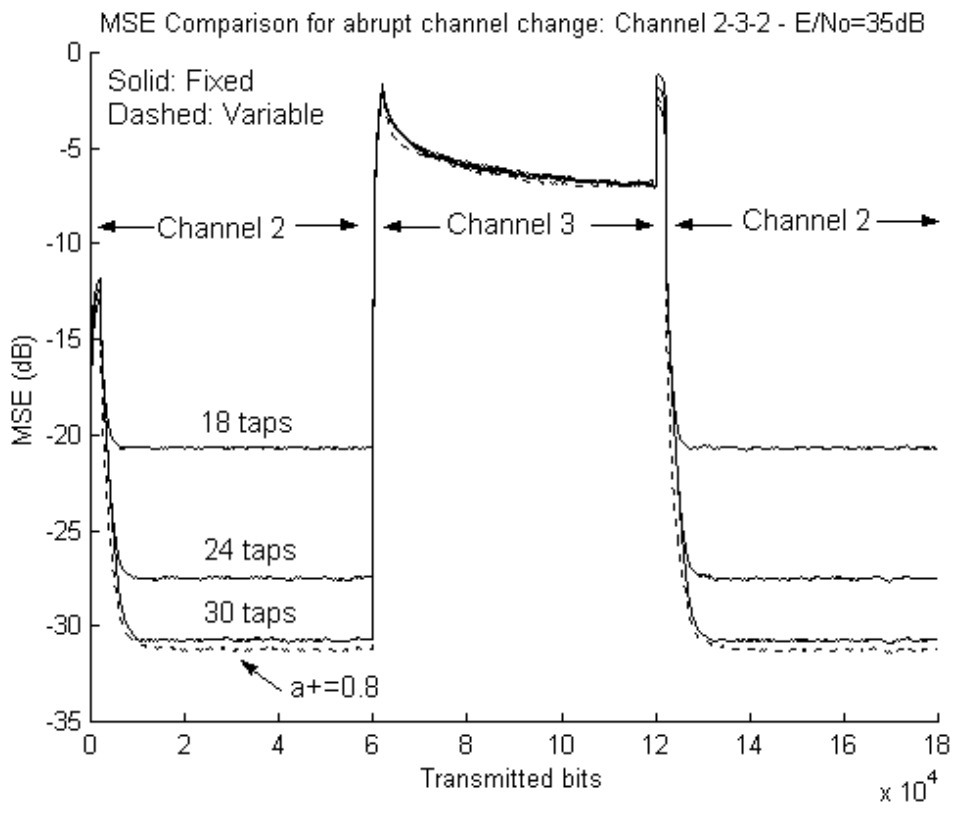

**Figure 5.50**: VSLMS MSE for different fixed and variable length equalisers. Abrupt channel change: 2-3-2. E/No=35 dB.

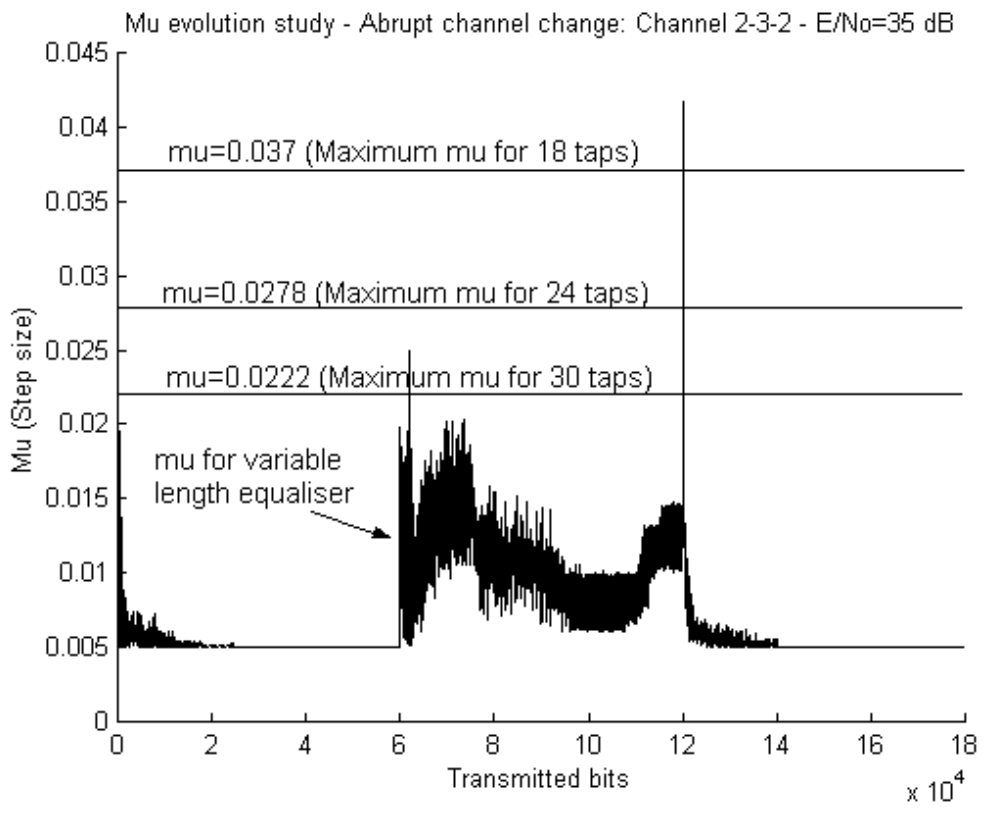

**Figure 5.51**: Comparison of maximum $\mu$ for fixed length equalisers and $\mu$ evolution for variable length equaliser. Abrupt channel change: 2-3-2. E/No=35 dB.

Some simulations have been run using the variable length in the abrupt channel change scenarios. Some of the results are shown in figures 5.50 and 5.51. Although it is more symbolic rather than significant, it can be observed in figure 5.51 that the variable length equalise converges slightly faster than any of the fixed length equalisers.

In figure 5.51 the maximum $\mu$ for the three fixed equalisers have been plotted, additionally the $\mu$ evolution for the variable length equaliser has also been plotted. Notice that when the channel goes from Channel model 3 back to Channel model 2, for a short period of time the step size of the variable length equaliser is able to achieve values unattainable for any of the fixed length equalisers. These higher $\mu$ values are responsible for the slight convergence superiority of the variable length structure.

An extra benefit from the joint use of the VSLMS and the variable length equaliser lies in the fact that they both use the square of the error signal ($e(n)^2$) to perform their respective optimisations (of the filter coefficients and the filter length respectively), as can be seen in the VSLMS equations (section 2.3.1) and in algorithm 5.1 presented in this chapter. Obviously this computation needs to be done only once and may serve for the iteration of both algorithms.

# 5.6 Restriction of the equaliser expansion

This chapter presented techniques to minimise the MSE level while also optimising the number of computations to be performed. In theory the MSE level should be kept as low as possible, but in practice the system may get to an MSE level where any further reduction will not improve the BER of the system. Therefore it might be useful to set a lower bound, $N_\tau$, on the achievable MSE level not allowing any other equaliser expansion when that threshold has been reached.

Algorithm 5.3 includes a new condition for the equaliser to be expanded. This condition should be of the form "expand if MSE(n)> $N_\tau$", using the relation between MSE(n) and ASE(n) (equation 5.2) the new constraint takes the form shown in algorithm 5.3.

$$ASE_{L-1}(n)=\sum_{i=0}^{n}\beta\left|d(n)-y_{L-1}(n)\right|^2 \tag{5.15}$$

$$ASE_{L}(n)=\sum_{i=0}^{n}\beta\left|d(n)-y_{L}(n)\right|^2 \tag{5.16}$$

$$\begin{aligned}&\text{If}\quad ASE_L(n)\le\alpha_{up}ASE_{L-1}(n)\\&\qquad\text{If}\quad ASE_L(n)>nN_\tau\rightarrow\text{Add one segment (P extra taps)}\end{aligned} \tag{5.17}$$

$$\text{If}\quad ASE_L(n)\ge\alpha_{dw}ASE_{L-1}(n)\quad\rightarrow\quad\text{Remove one segment (P fewer taps)} \tag{5.18}$$

with $0 < \alpha_{up}, \alpha_{dw} \leq 1$ and $\alpha_{up} \leq \alpha_{dw}$, $\beta \leq 1$.

**Algorithm 5.3**: Robust equaliser length control with expansion limit.

Notice that the new condition is only evaluated once the equaliser satisfies the other requirements for expansion, therefore, the computational complexity added by the new constraint to the overall algorithm is negligible.

The specific value for $N_\tau$ will mainly depend on other system parameters, that is the modulation scheme used and the channel coding capability of the system. Noting that the steady state MSE can be taken as a rough approximation of AWGN ([Haykin96]), the appropriate value for $N_\tau$ can then be extracted from the modulation and/or coding curves for AWGN channels.

# 5.7 Transient of the variable length equaliser

This chapter has so far focused on the steady-state operation of the equaliser as this would be the normal operating condition of the system for most of the time. Recall that tracking is a steady-state phenomenon. Therefore, results like the ones shown in sections 5.4.4 and 5.4.5 corresponding to sudden changes in the environment cannot be considered, at least in an strict sense, as representative of the transient properties of the segmented architecture.

For the sake of completeness, it is worth commenting on the convergence aspects of the variable length equaliser. In fact, in [Wesolowski92] and [Wesolowski95][18] the idea of variable length adaptive filters was introduced as a way to improve their convergence properties when using the LMS algorithm in static channels. The proposed technique works by using only a fraction of the total number of taps of the filter during the first stages of convergence. After an arbitrary number of iterations, the filter is expanded to its maximum number of taps.

In the context of linear equalisers, it is easy to see why this strategy speeds up the convergence of the equaliser. Recalling the recursive LMS convergence expression given by equation (4.65) in section 4.9.1, it can be observed that the equaliser length (M) appears as a multiplicative factor on its r.h.s. . Therefore, it is reasonable to assume that by shortening the filter, the convergence will take place more rapidly. Simulation results in the two mentioned references confirm this assumption.

[18] These are the only two references found by the author considering filter length changes.

The main drawback of the scheme just described is that the time to switch on the rest of elements of the equaliser must be known a-priori. In the context that work is described, telephone lines, this is a reasonable pre-requisite as, although the channel is unknown, average channel models can be used to calculate this switching instant. In the case of wireless mobile systems, this is certainly not the case as the channel impulse response cannot be accurately predicted beforehand.

In our simulations, the convergence time of the VL LE depended strongly on the parameter $\alpha_{up}$. When $\alpha_{up}$ is small (0.2-0.4), convergence is typically worse than that of a fixed length equaliser with a number of taps equal to the final length achieved by the VL LE. This is not surprising as when using small $\alpha_{up}$ performance is compromised in order to keep the computational complexity low. If $\alpha_{up}$ is made larger (0.6-0.8), the VL LE converges as quickly as a fixed length equaliser.

# 6 DECISION FEEDBACK EQUALISERS: MSE ANALYSIS AND RECONFIGURATION

In chapter 3 it has been shown that linear equalisers do not perform well when the channel to be equalised is heavily distorted. In these situations, DFE has been shown to work significantly better than LE. In this chapter, the steady state MSE derivations presented in chapter 4 are extended to include the effect of the feedback filter present in a DFE. These DFE derivations follow a nearly identical path to the LE ones, however there are also some important differences. The derivations presented in this chapter skip the common steps and highlight the differences.

The derived DFE-MSE expressions are then used to examine the reconfiguration potential of the DFE. Previous studies ([Ariyavisitakul97]) and our own simulations indicate that in mobile/wireless environments the required time span of the feedforward filter is considerably shorter than that of the feedback section. This is due to the fact that, in all practical scenarios (see pp. 728 in [Steele99]), the main peak of energy is always located close to the beginning of the channel impulse response (CIR).

The power and span of postcursor energy depends very much on the specific environment. It is therefore interesting to consider the reconfiguration of the FBF according to the instantaneous amount of postcursor ISI. As we will show, this reconfiguration may lead to important reductions in the number of computations without compromising the MSE performance.

This chapter is structured as follows: in section 6.1 the model used in the DFE analysis is introduced. Section 6.2 revisits the assumptions already established in 4.2. Sections 6.3 and 6.4 derive and validate respectively the steady LMS-MSE equation for the case of a DFE. In section 6.5 the relation between the decision delay and convergence of the LMS-DFE is analysed. Sections 6.6 and 6.7 present the MSE analysis for the DFE using the RLS algorithm. In section 6.8, the main conclusions to be extracted from the derived MSE expressions are stated. Section 6.9 summarises some of the known results regarding the MMSE-DFE. Section 6.10 gives the rationale for the use of an algorithm to control the length of the FBF. The FBF length update algorithm and simulations showing its performance are presented in sections 6.11 and 6.12 respectively. Finally, In section 6.13 isome concluding remarks about the variable length FBF DFE are given.

# 6.1 Analytical system model for DFE

The generic model used for the MSE performance analysis is shown in figure 6.1. As in the case of LE, we assume this system model is in a fully digitised form.

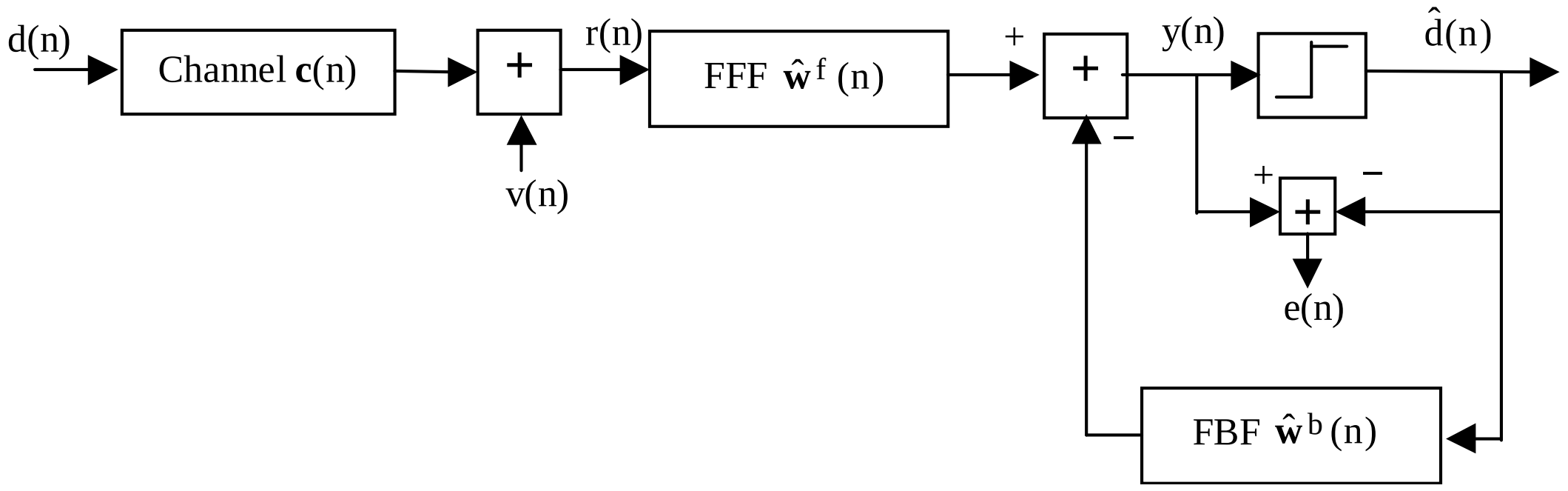

**Fig. 6.1:** DFE system model.

The variables have the following meaning:

$d(n)$ = Transmitted data symbol

$\mathbf{c}(n) = [c_o(n)\ \ c_1(n)\ \ c_2(n) \ldots c_{N-1}(n)]$ = Channel impulse response

$\hat{\mathbf{w}}^f(n)(n) = [\, \hat{w}_0^f(n)\ \hat{w}_1^f(n) \ldots \hat{w}_{Nf-1}^f(n)\,]$ = $N_f$-tap Feedforward filter (FFF)

$\hat{\mathbf{w}}^b(n)(n) = [\,\hat{w}_0^b(n)\;\hat{w}_1^b(n)\ldots\hat{w}_{Nb-1}^b(n)\,]$ = $N_b$-tap Feedback filter (FBF)

v(n) = Noise sample

r(n) = Input signal to the FFF

y(n) = Output of the equaliser

$\hat{d}(n)$ = Estimated data symbol and input signal to the FBF

e(n)= Error signal between equalised and detected symbol.

The noise samples, v(n), as in the case of the LE, are taken from an AWGN process with power $\sigma_v^2$. The source symbols, d(n), are independent and identically distributed samples with power $\sigma_d^2$ .

In chapter 2 the phenomenon of error propagation in DFE has been briefly discussed. From figure 6.1 it is clear that any erroneous decision made by the threshold detector will also affect posterior decisions. This fact complicates very much the treatment of the DFE. Analysis of the DFE taking into account error propagation has been presented in [Dutweiler74], [Altekar93] and [Smee98]. In these references, the performance is analysed from the point of view of BER. On the contrary, in the work presented here, our objective function is the MSE. This measure, as has been shown in the previous chapter, can serve as an indicator to efficiently reconfigure the equaliser. However, in order to make the DFE-MSE analysis feasible it is necessary to assume correct past decisions. This assumption lets us re-draw figure 6.1 as the linearised system shown in figure 6.2.

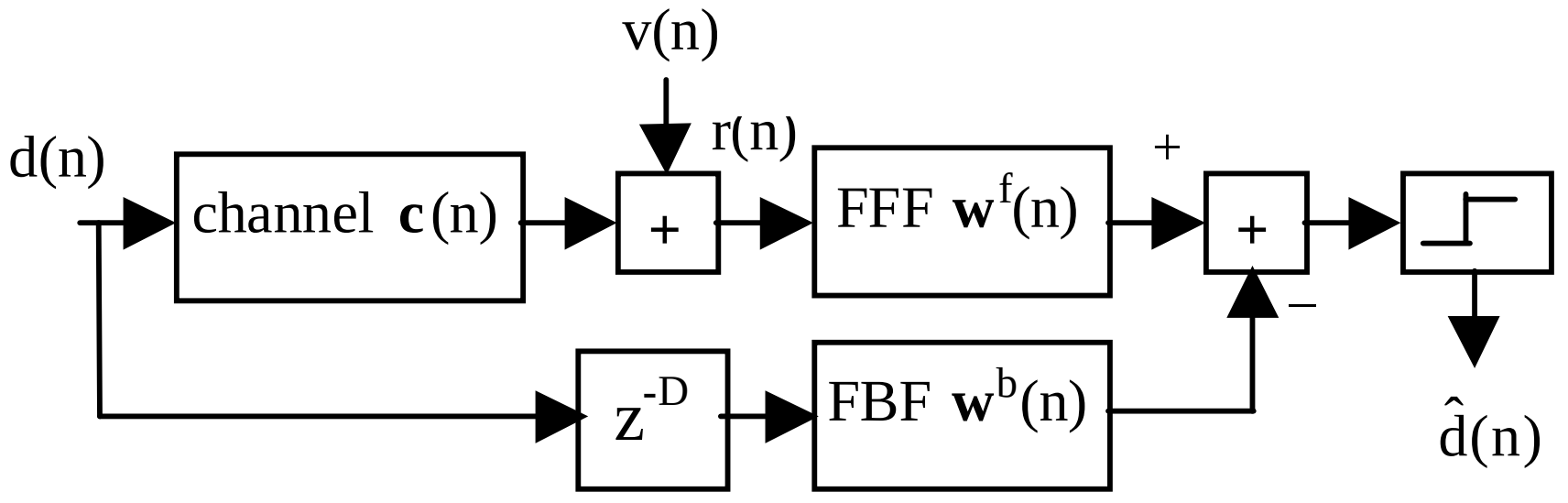

**Fig. 6.2:** DFE linearised system model.

The pure delay element preceding the feedback filter, D, accounts for the decision delay.

# 6.2 DFE model assumptions

All the assumptions stated in 4.3 for the linear equalisation study can also be applied without any change to the case of the DFE model. Notice that assumptions 4.3.1 (channel model) and 4.3.2 (channel power response) refer to the channel characteristics, and therefore they are independent of the structure of the receiver. The independence assumption will also be used at some points of the LMS-MSE derivation.

It is only left to verify that the assumption about the optimum equaliser variations also holds for the case of the DFE. The following figures show the relation between the channel variations and the optimum DFE variations. As in the LE case, the hypothesis has been tested with the three channel profiles introduced in chapter 3. Figures 6.3 and 6.4 show situations where the assumption holds quite accurately and the magnitude of the variations of the equaliser are correlated with the changes in the channel. On the other hand, figure 6.5 depicts a case where the equaliser variations are a magnification of the equaliser changes. Finally, figure 6.6 shows the worst possible scenario where the equaliser changes have little to do with the channel changes.

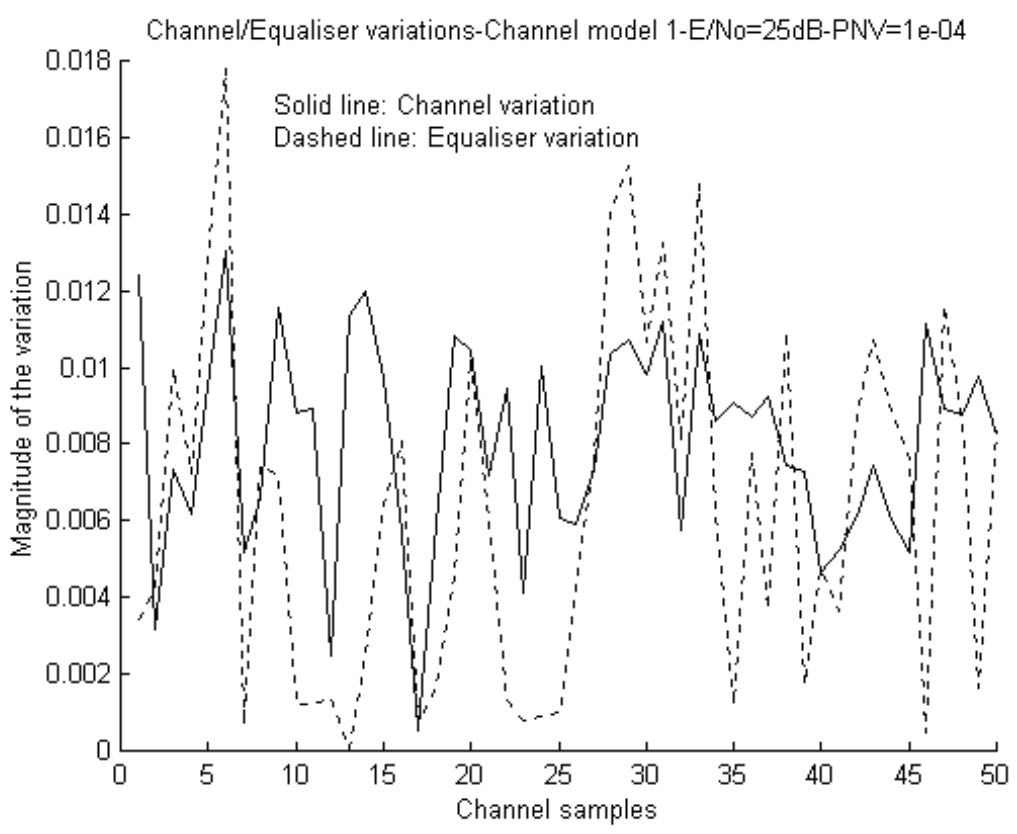

**Figure 6.3:** Comparison of the norm of the channel and DFE variations. Channel model 1. E/No=25dB. $\sigma_q^2=10^{-4}$

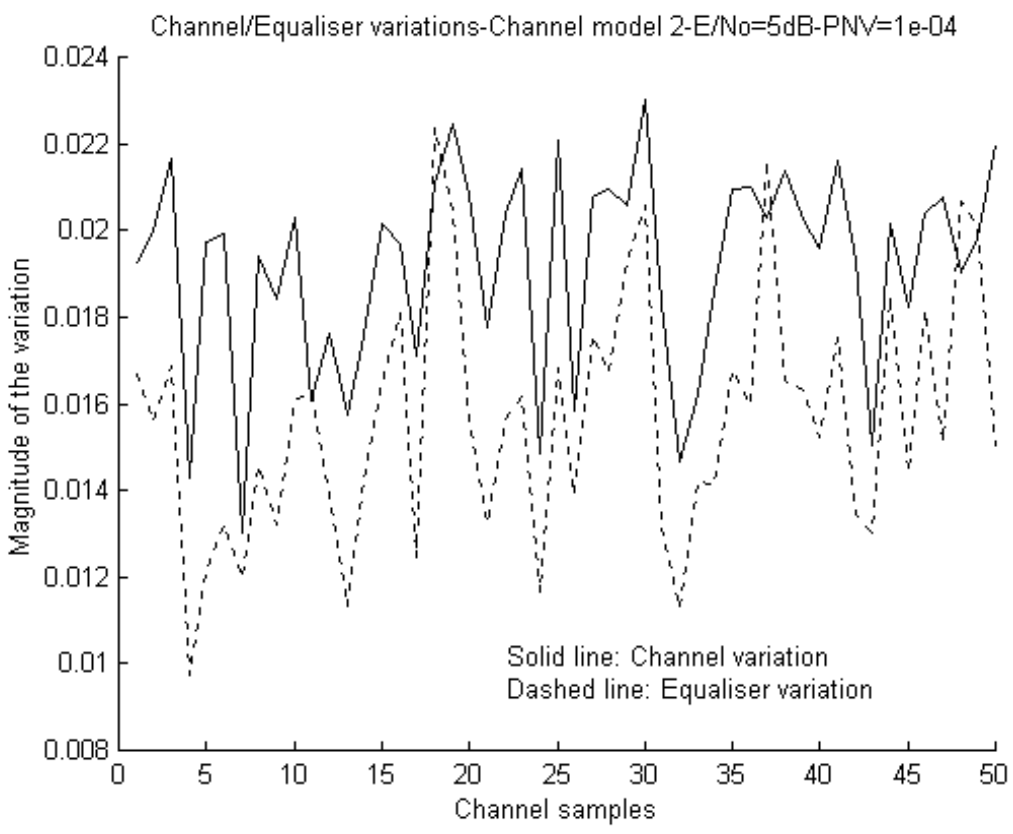

**Figure 6.4:** Comparison of the norm of the channel and DFE variations. Channel model 2. E/No=5dB. $\sigma_q^2=10^{-4}$

As has been already explained in section 4.3.3, the accuracy of this assumption depends on the spectral properties of the input data autocorrelation matrix. If this matrix has a small eigenvalue spread, which happens when the channel spectra is flat or the SNR low, the assumption is very accurate. On the contrary, when the channel spectrum contains nulls or the SNR is very high, the assumption becomes a loose approximation.

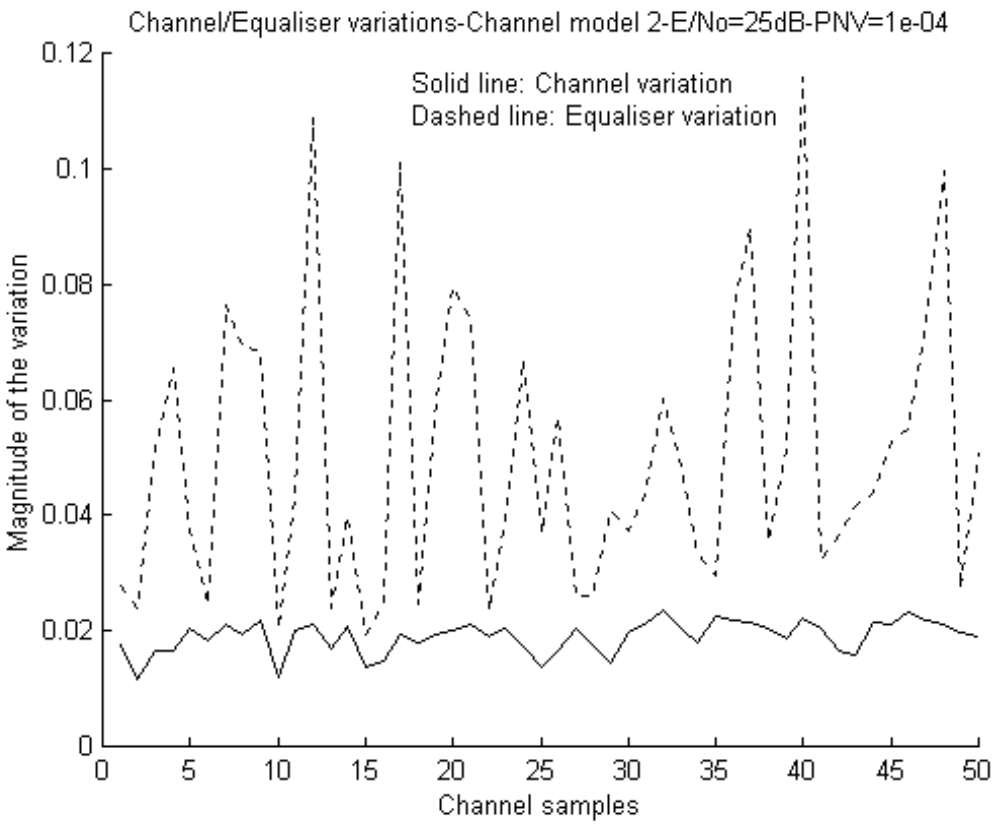

**Figure 6.5:** Comparison of the norm of the channel and DFE variations. Channel model 2. E/No=25dB. $\sigma_q^2 = 10^{-4}$.

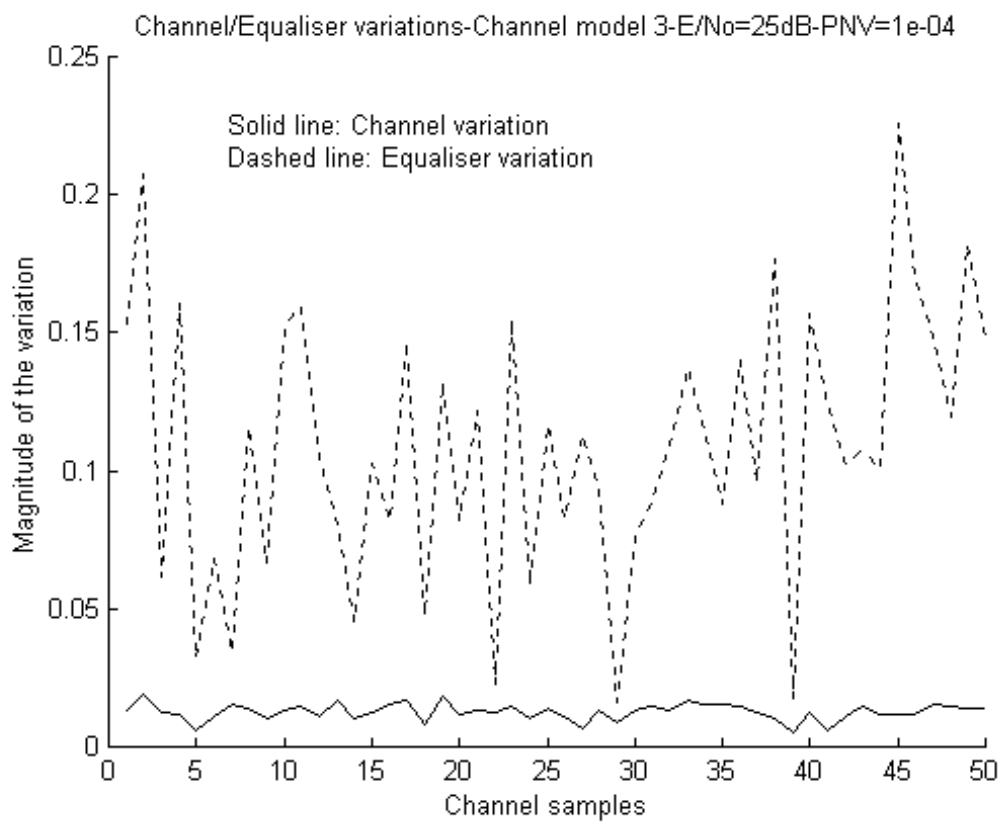

**Figure 6.6:** Comparison of the norm of the channel and DFE variations. Channel model 3. E/No=25dB. $\sigma_q^2 = 10^{-4}$.

As a last comment on this assumption, it is important to compare figure 6.6 with figure 4.8 (corresponding to the LE) and notice that in the case of the DFE, the equaliser variations are more correlated with the channel changes than in the case of the LE. This is due to the fact that the variations in the feedback filter are generally highly correlated with the variations in the channel impulse response.

# 6.3 LMS-DFE analysis

The MSE LMS analysis for the case of the DFE follows exactly the same path as the LE derivation and therefore only the different parts will be covered now in detail.

The LMS algorithm for the case of the DFE is given by:

$$e(n) = d(n) - \mathbf{u}^T(n)\hat{\mathbf{w}}(n) \quad (6.1)$$

$$\hat{\mathbf{w}}(n+1) = \hat{\mathbf{w}}(n) + \mu\mathbf{u}(n)e(n) \quad (6.2)$$

These equations are identical to those given by (4.6) and (4.7) except that now **u**(n) and $\hat{\mathbf{w}}(n)$ have a different structure. The DFE input vector, **u**(n), is given by:

$$\mathbf{u}(n) = [\mathbf{r}(n)\ \mathbf{d}(n)] \quad (6.3)$$

where $\mathbf{r}(n)$ is the vector formed by $N_f$ consecutive channel samples stored in the FFF and $\mathbf{d}(n)$ is made of the last $N_b$ detected symbols. The equaliser coefficient vector, $\hat{\mathbf{w}}(n)$, can also be split as:

$$\hat{\mathbf{w}}(n) = [\,\hat{\mathbf{w}}^f(n) \;\; \hat{\mathbf{w}}^b(n)\,] \tag{6.4}$$

where $\hat{\mathbf{w}}^f(n)$ and $\hat{\mathbf{w}}^b(n)$ represents the $N_f$ feedforward and $N_b$ feedback taps respectively. The MSE derivation now follows exactly the same path as the one for the LE given by equations (4.8) to (4.21) with the resulting expression being:

$$\text{EMSE}(n)= \frac{\mu}{2}\text{MMSE}(n)\text{Tr}[\mathbf{R}(n)] + \frac{1}{2\mu}\text{Tr}[\mathbf{P}] \tag{6.5}$$

Equation (6.5) albeit identical in form to (4.21) presents some important differences as the structure of $\mathbf{R}(n)$ and $\mathbf{P}$ is not the same in the DFE and in the LE. We now examine in detail the characteristics of both matrices in order to obtain the final expression for the DFE-EMSE.

The autocorrelation matrix of the input data is defined as:

$$\mathbf{R}(n)=E[\mathbf{u}(n)\mathbf{u}^T(n)]=E[[\mathbf{r}(n)\ \mathbf{d}(n)][\mathbf{r}(n)\ \mathbf{d}(n)]^T] \tag{6.6}$$

Expanding the vectors $\mathbf{y}(n)$ and $\mathbf{d}(n)$ we obtain:

$$\mathbf{R}(n) = E\left[\begin{bmatrix} r(n-N_f+1) \\ r(n-N_f+2) \\ \vdots \\ r(n) \\ d(n+1) \\ d(n+2) \\ \vdots \\ d(n+N_b) \end{bmatrix} \begin{bmatrix} r(n-N_f+1) & r(n-N_f+2) & \cdots & r(n) & d(n+1) & d(n+2) & \cdots & d(n+N_b) \end{bmatrix}\right]$$

$$= \begin{bmatrix} \mathbf{F}(n) & \mathbf{V}(n) \\ \mathbf{V}(n)^T & \mathbf{D} \end{bmatrix} \tag{6.7}$$

where $\mathbf{F}(n)$ is an $N_f x N_f$ matrix given by $\mathbf{F}(n)=E[\mathbf{r}(n)\mathbf{r}(n)^T]$, $\mathbf{V}(n)$ is an $N_f x N_b$ matrix given by $\mathbf{V}(n)=E[\mathbf{r}(n)\mathbf{d}(n)^T]$ and $\mathbf{D}$ is an $N_b x N_b$ diagonal matrix given by $\mathbf{D}=E[\mathbf{d}(n)\mathbf{d}(n)^T]$. The diagonal form of the matrix $\mathbf{D}$ stems from the fact that source symbols are independent, if we further assume that the source symbol power is normalised ($\sigma_d^2=1$), then $\mathbf{D}$ becomes the $N_b x N_b$ identity matrix. Computing the expectations of the submatrices $\mathbf{F}(n)$, $\mathbf{V}(n)$ and $\mathbf{D}$, the final form of $\mathbf{R}(n)$ is shown in the next page.

$$
\mathbf{R}(n)=\left[\begin{array}{cccc|cccc}
\gamma_0(n-N_f+1)(1+\sigma_v^2) & \gamma_1(n-N_f+1) & \cdots & \gamma_{Nf-1}(n-N_f+1) & c_{Nf}(n-N_f+1) & c_{Nf+1}(n-N_f+1) & \cdots & c_{Nf+Nb}(n-N_f+1) \\
\gamma_1(n-N_f+1) & \gamma_0(n-N_f+2)(1+\sigma_v^2) & \cdots & \gamma_{Nf-2}(n-N_f+2) & c_{Nf-1}(n-N_f+1) & c_{Nf}(n-N_f+2) & \cdots & c_{Nf+Nb-1}(n-N_f+2) \\
\vdots & \vdots & \ddots & \vdots & \vdots & & & \vdots \\
\gamma_{Nf-1}(n-N_f+1) & \gamma_{Nf-2}(n-N_f+2) & \cdots & \gamma_0(n)(1+\sigma_v^2) & c_1(n-N_f+1) & c_2(n-N_f+2) & \cdots & c_{Nb}(n) \\
\hline
c_{Nf}(n-N_f+1) & c_{Nf-1}(n-N_f+1) & \cdots & c_1(n-N_f+1) & 1 & 0 & \cdots & 0 \\
c_{Nf+1}(n-N_f+1) & c_{Nf}(n-N_f+2) & \cdots & c_2(n-N_f+2) & 0 & \ddots & \ddots & \vdots \\
\vdots & & \ddots & \vdots & \vdots & \ddots & \ddots & 0 \\
c_{Nf+Nb}(n-N_f+1) & c_{Nf+Nb-1}(n-N_f+2) & \cdots & c_{Nb}(n) & 0 & \cdots & 0 & 1
\end{array}\right] \tag{6.8}
$$

where the coefficients $\gamma_i(n)$ are given by:

$$\gamma_i(n) = \sum_{j=0}^{N-1} c_j(n)c_{j+i}(n+i) \quad \text{for} \quad 0 \le i \le N_f - 1 \tag{6.9}$$

The dashed lines in (6.8) are used to mark the four submatrices composing **R**(n). It is important to keep in mind that any $c_i$ with $i \notin [0..N-1]$ is zero by virtue of the finite length of the channel impulse response. Consequently, many of the entries in **R**(n), especially the ones on the top-right and bottom-left corners, are potentially zero.

Given that only the trace of **R**(n) is needed, only the diagonal elements in **R**(n) are of concern here. Recalling the assumption of a unitary normalised channel impulse response, it is straightforward to compute the diagonal correlation coefficients given by $\gamma_0(n)$:

$$\gamma_0(n) = \sum_{j=0}^{N-1} c_j(n)c_j(n) = c_0(n)^2 + c_1(n)^2 + \cdots + c_{N-1}(n)^2 = 1 \qquad \forall n \tag{6.10}$$

Using (6.10) the trace of **R**(n) is given by:

$$\mathrm{Tr}[\mathbf{R}(n)] = N_f[1+\sigma_v^2] + N_b \tag{6.11}$$

The structure of the matrix **P** corresponding to the autocorrelation of the process noise vector **p**(n) is far simpler than that of **R**(n). Given the properties of the elements of **p**(n) (see section 4.3.3), **P** is an $(N_f+N_b)$x$(N_f+N_b)$ squared diagonal matrix with entries:

$$\mathbf{P} = \begin{bmatrix} \sigma_p^2 & 0 & \cdots & 0 \\ 0 & \sigma_p^2 & \ddots & \vdots \\ \vdots & \ddots & \ddots & 0 \\ 0 & \cdots & 0 & \sigma_p^2 \end{bmatrix} \tag{6.12}$$

The trace of **P** is:

$$\mathrm{Tr}[\mathbf{P}] = [N_f+N_b]\,\sigma_p^2 \tag{6.13}$$

Using the assumption of similar channel and equaliser variations, we can write:

$$[N_f + N_b]\sigma_p^2 \cong N\sigma_q^2 \tag{6.14}$$

With (6.11), (6.13) and (6.14), a closed-form expression can be written for the EMSE of the LMS-DFE:

$$\mathrm{EMSE}(n) = \frac{\mu}{2}\mathrm{MMSE}(n)\left[N_f\left[1+\sigma_v^2\right] + N_b\right] + \frac{1}{2\mu}N\sigma_q^2 \tag{6.15}$$

The overall MSE, which the sum of the MMSE and EMSE, is given by:

$$MSE(n) = MMSE(n) + \frac{\mu N_f \left[1+\sigma_v^2\right]}{2} MMSE(n) + \frac{\mu N_b}{2} MMSE(n) + \frac{N\sigma_q^2}{2\mu} \qquad (6.16)$$

Equation (6.16) is the expression for the steady-state MSE for a DFE when using the LMS algorithm to update the equaliser settings. The first term on the r.h.s. of (6.15) corresponds to the minimum mean squared error, the second term is due to the noise misadjustment of the FFF, the third term results from the misadjustment of FBF and the last term results from the tracking misadjustment. Notice that the derived expression shows that the FBF does not produce any noise enhancement as the EMSE introduced by the FBF does not depend on $\sigma_v^2$. This is one of the main advantages of the DFE over LE ([Qureshi85]).

It is also intuitively satisfying to check that when $N_b$=0, the resulting expression reduces to equation (4.31) corresponding to the MSE for the LE. Equation (6.16) shows that the EMSE depends on several variables that can be set by the designer, namely, $N_f$, $N_b$ and $\mu$.

The value for $\mu$ needs to be chosen as a compromise between steady state accuracy and tracking capability. The lengths of the FFF and FBF also need careful consideration and will be discussed extensively later on in the chapter. Nonetheless, it can be anticipated from equation (6.16) that making the filters too long may end up increasing the overall MSE due to an increase in the misadjustment. In the next section, simulation results are presented showing the accuracy of equation (6.16).

# 6.4 Validation of the LMS-DFE equation

The simulations in this section use the same channel models and simulation conditions as in the LE case. The Markov variations of the channel impulse response (CIR) imply that any of the coefficients in the CIR may become the dominant path. In order to achieve an acceptable performance, the FFF must be able to gather the main peak of energy, that is, the FFF span must be long enough to include the dominant path. Therefore, and for the case of the Markov channel, the FFF must have at least the same number of taps as the CIR. For the simulations presented next, the FFF length has been set to N+1 (N= number of channel taps) and the FBF to N-1 taps. In practical channels, the energy is always concentrated at the beginning of the CIR, and consequently the FFF can be much shorter than the CIR length. DFE performance in realistic radio environments is covered in chapter 7.

## 6.4.1 Channel model 1

Channel model 1, already described in chapter 3, can be considered as a very well behaved channel introducing a modest amount of distortion. The DFE used in these simulations consisted of 3 forward taps and 1 feedback tap. The step size was set to the rather large value of 0.025 to accelerate convergence.

Figures 6.7 and 6.8 show the MSE curves obtained for $\sigma_q^2=10^{-4}$ and $\sigma_q^2=10^{-6}$ respectively. The results obtained experimentally are in very close agreement with those predicted by equation (6.16). In figure 6.7 ($\sigma_q^2=10^{-4}$) the time varying MMSE is also shown. It can be appreciated that for a low E/No, the MSE is very close to the optimum MSE. On the other hand, when E/No is large, the measured MSE can be fairly different from the MMSE but still equation (6.10) is able to perform an accurate prediction. When $\sigma_q^2=10^{-6}$ (Figure 6.8), the channel varies very slowly, and accordingly there are not the abrupt MSE changes present in figure 6.7. Again, equation (6.10) successfully estimates the steady state MSE value.

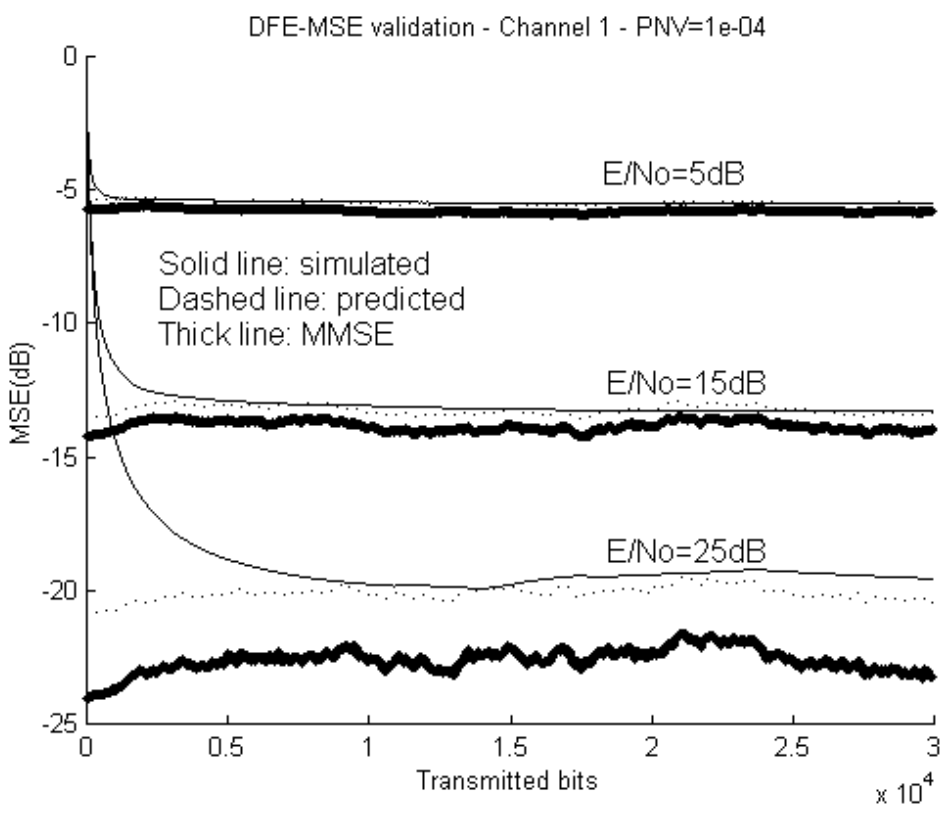

**Figure 6.7**: DFE LMS MSE Validation, Channel model 1, $\sigma_q^2=10^{-4}$.

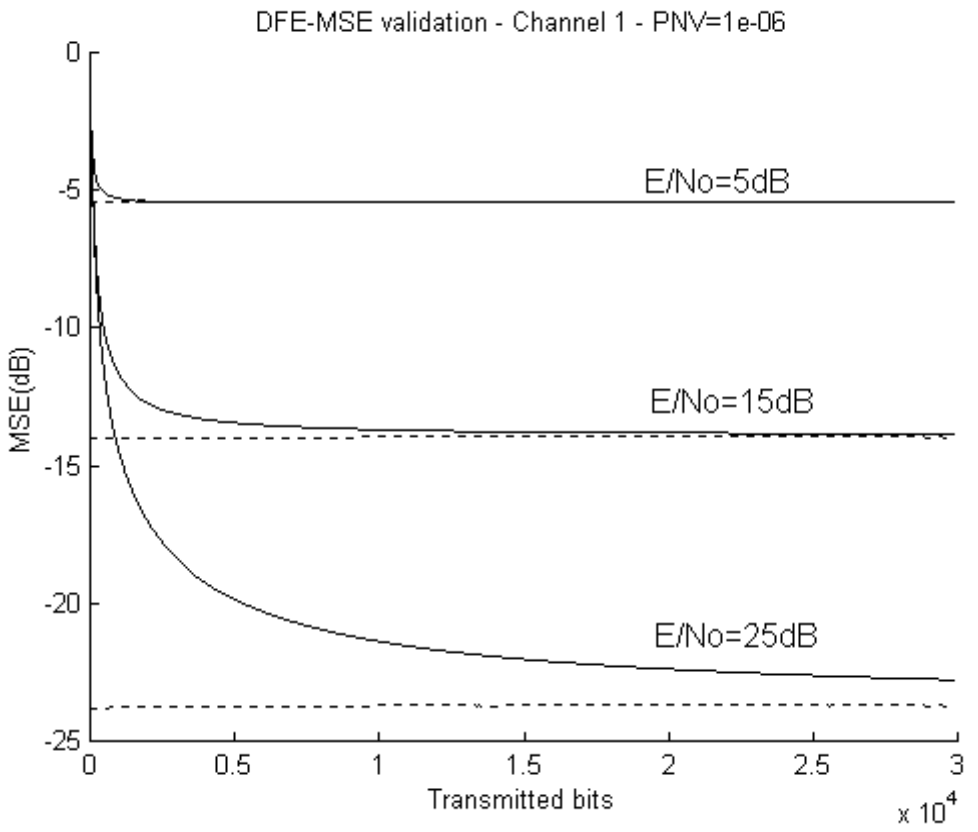

**Figure 6.8**: DFE LMS MSE Validation, Channel model 1, $\sigma_q^2=10^{-6}$.

## 6.4.2 Channel model 2

The Characteristics of channel model 2 are presented in chapter 3. In figures 6.9 and 6.10 the MSE predicted values are compared with simulation results for $\sigma_q^2=10^{-4}$ and $\sigma_q^2=10^{-6}$ at various E/No levels. As in the previous channel model, the predicted MSE curves agree very accurately with the experimental results. As in Channel model 1, notice the big gap between the MMSE curve and the predicted/simulated MSE when E/No=25 dB. This indicates that using the MMSE as an approximation to the real MSE value can be, in certain circumstances such as in high E/No, very inaccurate. The fairly large value of the MSE with respect to the MMSE is due to the use of a large step size in order to track the fast channel variations. This large step size increases very significantly the EMSE component and makes the overall MSE much larger than the theoretical minimum.

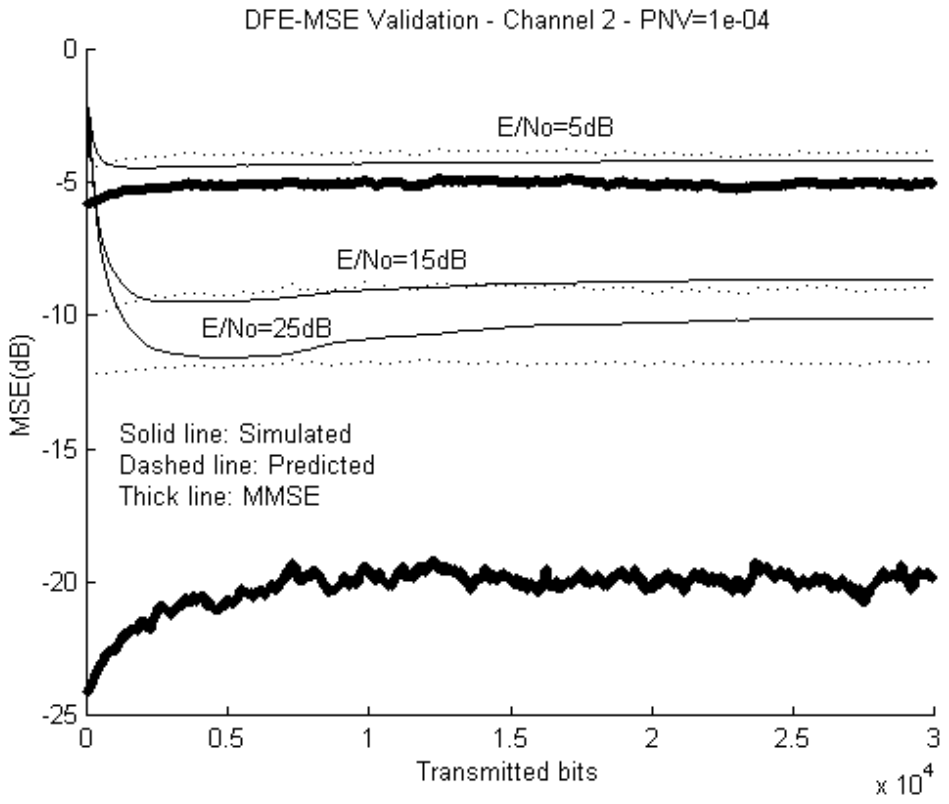

**Figure 6.9**: DFE LMS MSE Validation, Channel model 2, $\sigma_q^2=10^{-4}$.

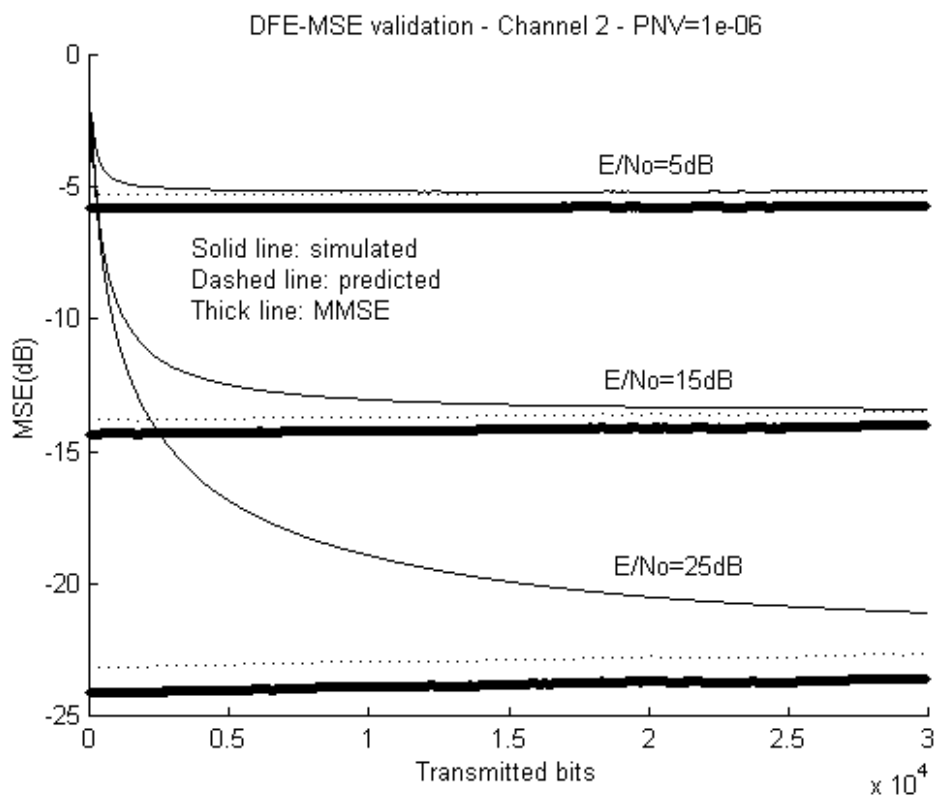

**Figure 6.10**: DFE LMS MSE Validation, Channel model 2, $\sigma_q^2=10^{-6}$.

In figures 6.11 and 6.12 the accuracy of equation (6.16) is verified when using different number of taps in the FFF and the FBF. One comment is needed on the results shown in figure 6.11. This figure compares the predicted with the simulated MSE for different values of $N_f$. As was explained at the beginning of this section, the length of the FFF needs to be at least as long as the CIR in order to guarantee that the FFF is able to capture the dominant path. For this channel model (N=11) and using a 6-tap FFF, this condition is not met and in some simulations where the main path in the channel moved beyond the $6^{th}$ tap in the CIR, a huge increase in the MSE level appeared.

In order to make a fair comparison, these simulations have been removed from the ensemble averaging process for any of the $N_f$ values. The results presented in figure 6.11 show that adding taps to the FFF tends to reduce the MSE but a point is reached where the MSE reduction becomes very small. Eventually, as in the LE case and for the same reasons, after a certain number of taps making the FFF any longer will start increasing the MSE. The important fact to appreciate in figure 6.11 is that equation (6.10) successfully predicts the impact the FFF length has on the overall MSE.

In figure 6.12 the effect of the FBF length is studied. Some known results regarding the DFE-MMSE are discussed in section 6.8; one of them is particularly relevant to the results shown in figure 6.12 and is therefore anticipated now: assuming a decision delay $D=N_f-1$ the combined channel-FFF impulse response will have N-1 postcursors. Therefore, any FBF longer than N-1 taps will not improve the MSE in any way. This can be clearly seen in figure 6.12. Knowing that the channel has 11 taps (N=11), we can predict that there will not be any improvement when using more than 10 taps in FBF.

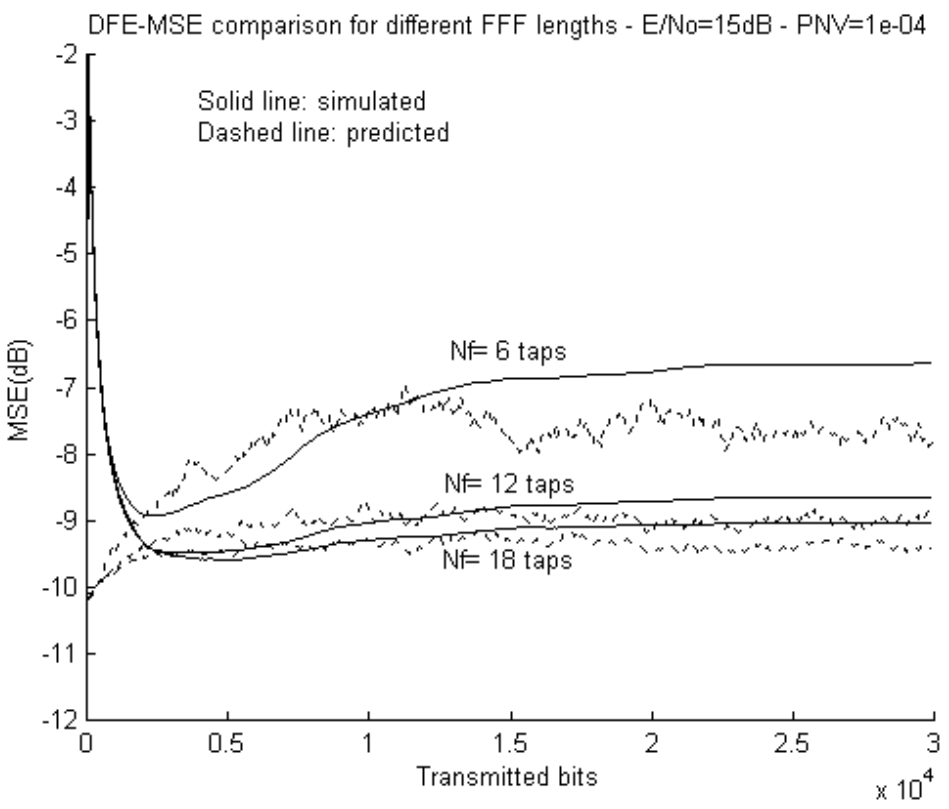

**Figure 6.11**: DFE LMS MSE Comparison for different number of FF taps, Channel model 2, E/No=15dB, $\sigma_q^2=10^{-4}$.

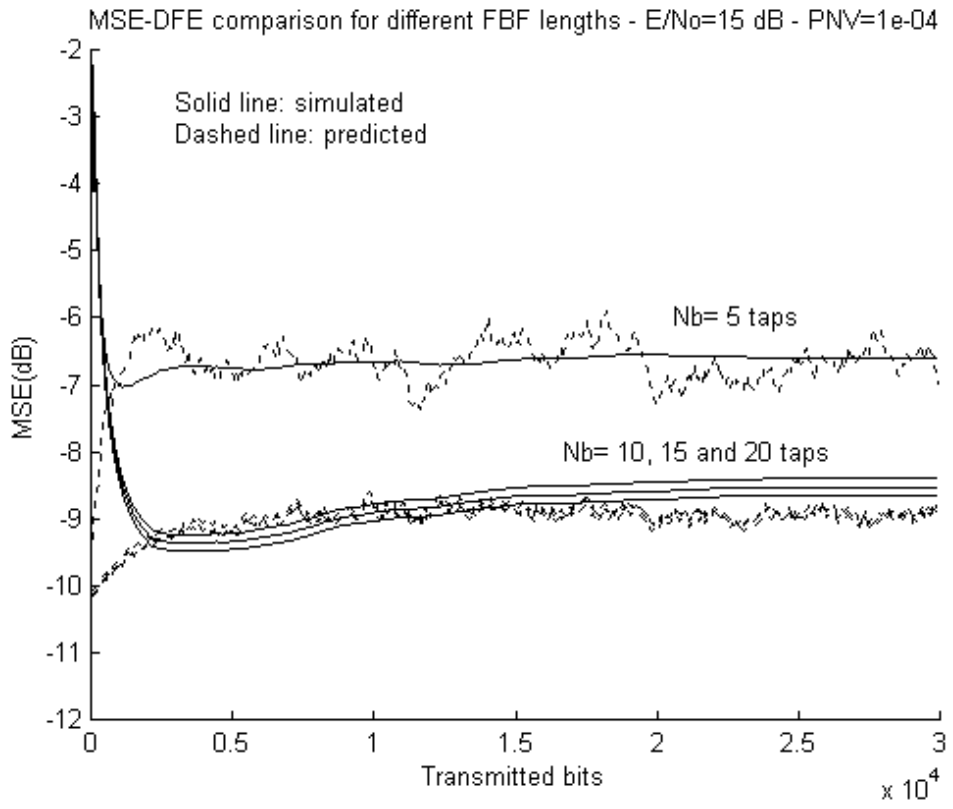

**Figure 6.12**: DFE LMS MSE Comparison for different number of FB taps, Channel model 2, E/No=15dB, $\sigma_q^2=10^{-4}$.

The curve for $N_b=5$ taps is clearly worse than for the other $N_b$ values as some of the postcursors are not cancelled. However the curves for 10, 15 and 20 feedback taps offer very similar performance as there are only 10 postcursors to cancel. In fact, although barely distinguishable in the graph, the MSE for $N_b=10$ is lower than for $N_b=15$ which in turn is lower than $N_b=20$. This indicates the correctness of the 3rd term in equation (6.16) and proves the hypothesis that using too many taps in the feedback filter not only wastes computations but also increases the MSE level.

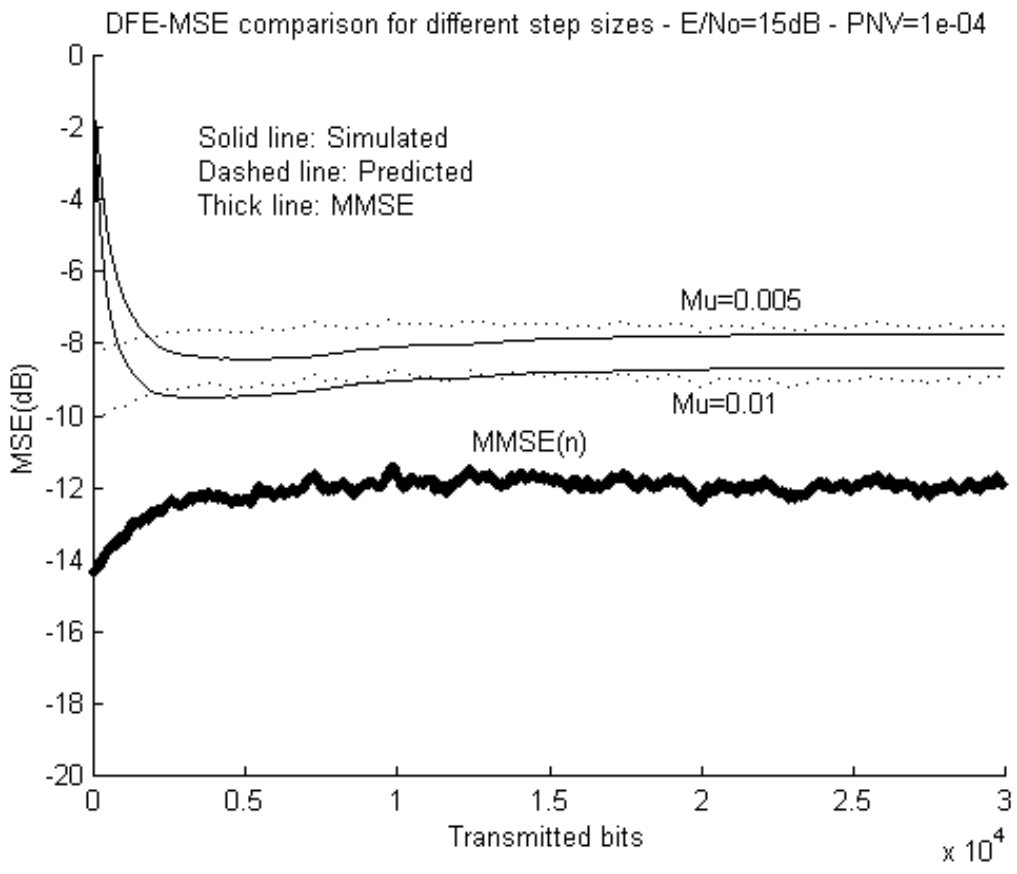

**Figure 6.13:** DFE LMS MSE Comparison for two different values of $\mu$, Channel model 2, E/No=15dB, $\sigma_q^2=10^{-4}$.

Finally, the effect of varying the step size has also been studied. Figure 6.13 shows the simulated and predicted MSE curves for two different values of $\mu$ when E/No=15dB and $\sigma_q^2=10^{-4}$. Again the prediction agrees very accurately with the experimental results. It is interesting to verify that reducing the step size produces an increase in the MSE: although the smaller step size reduces the misdjustment errors of the FFF and FBF, it increases the tracking misadjustment producing an overall larger MSE.

## 6.4.3 Channel model 3

As has already been said (chapter 3), Channel model 3 is a fairly pessimistic scenario due to the existence of a null in the passband. Figure 6.6 shows that the hypothesis about the optimum equaliser variations holds only in a rather loose way, therefore it is expected that the predictions made by equation (6.16) are not as exact as in the other channel models, especially at high E/No levels.

Figures 6.14 and 6.15 present the results for $\sigma_q^2=10^{-4}$ and $\sigma_q^2=10^{-6}$ respectively. For the fast varying channel ($\sigma_q^2=10^{-4}$) a general good agreement exists in the low to medium E/No range. For the high E/No case, the discrepancy becomes more noticeable. However, the predicted MSE is far more accurate than simply using the MMSE as an approximation of the real MSE performance. The same applies for the slowly varying channel ($\sigma_q^2=10^{-6}$). In fact, at a high E/No, the differences between the prediction and the measurements become quite large. Notice however that the E/No=25dB curve in figure 6.15 has not yet reached the steady state after 30,000 samples. If the simulation is allowed to run for a larger number of samples, the measured MSE will tend to get closer to the predicted value although a gap between the two will still exist. The reason for such a slow convergence is to be sought in the dependence of the LMS transient on the channel spectrum (see chapter 2). Both the in-band null and the large E/No contribute to the poor power spectral characteristics of this particular scenario. The source of the prediction inaccuracy is to be attributed to the incomplete satisfaction of the optimum equaliser variations hypothesis (see figure (6.6)).

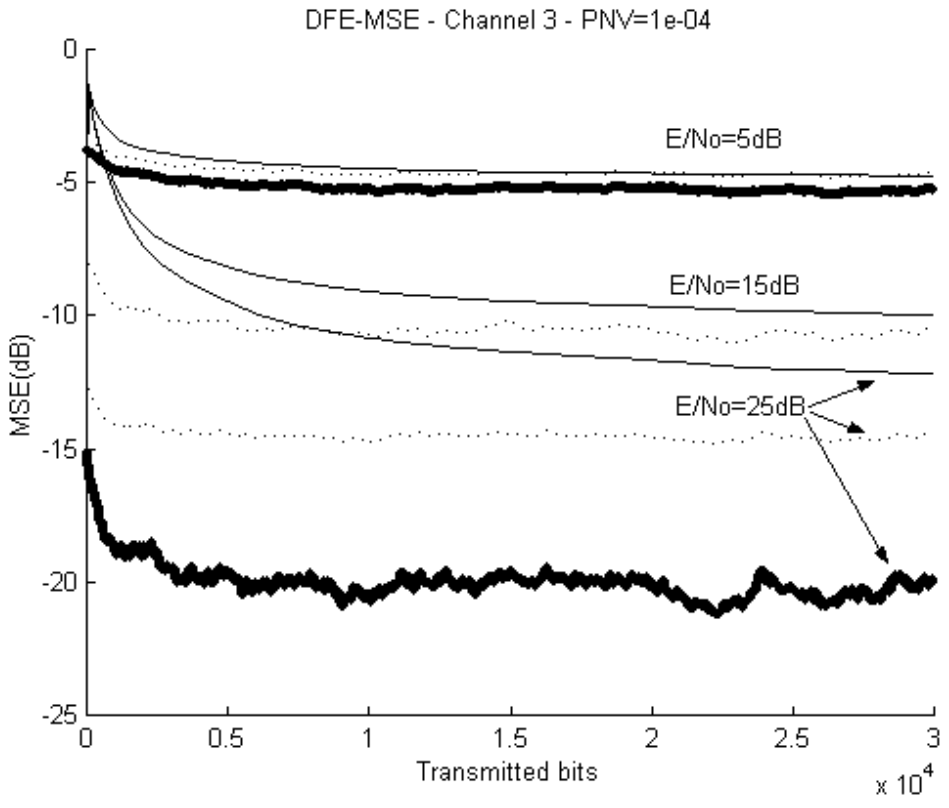

**Figure 6.14**: DFE LMS MSE Validation, Channel model 3, $\sigma_q^2=10^{-4}$.

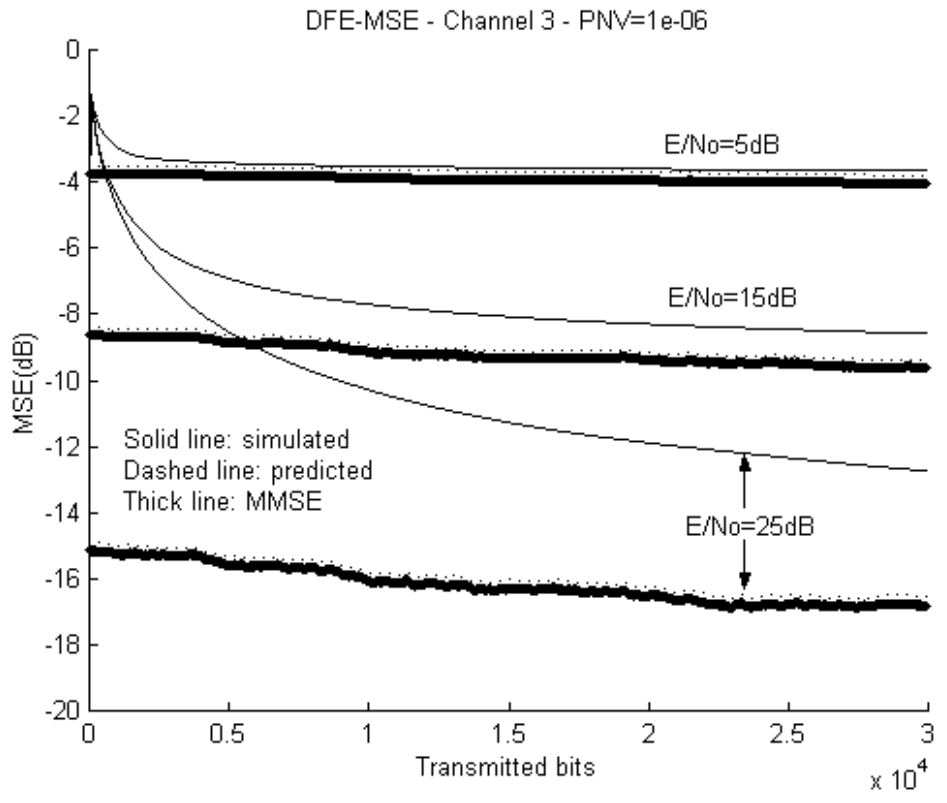

**Figure 6.15**: DFE LMS MSE Validation, Channel model 3, $\sigma_q^2=10^{-6}$.

# 6.5 Decision delay effect on the LMS-DFE

While studying the LMS-DFE, an interesting novel result has been found concerning the choice of decision delay. This topic has already been considered in section 3.5.4. There, it was mentioned that some specific techniques exist ([Voois96], [Al-Dhahair96]) in order to choose the decision delay that minimises the attainable MMSE. These methods, although optimum from an MSE point of view, do not take into account the characteristics of the algorithm used to perform the filter adaptation.

In the particular case of the LMS algorithm, as has been mentioned in section 2.3.1, the convergence of the algorithm is sensitive to the eigenvalue spread of the input autocorrelation matrix, $\chi(\mathbf{R})$. The larger $\chi(\mathbf{R})$ is, the longer it takes the algorithm to converge.

We have found that, in a typical channel, $\chi(\mathbf{R})$ strongly depends on the decision delay, D. By typical channel we understand one whose impulse response contains a clear peak of energy followed and preceded by other pulses of smaller amplitude (multipaths).

The relation between D and $\chi(\mathbf{R})$ arises from the way the decision delay affects **R**[19], whose structure is shown in equations (6.7) and (6.8). In these equations, it can be seen that **R** is composed of four clearly defined submatrices: **F**, **V**, $\mathbf{V}^T$ and **D**. The choice of decision delay affects only **V** (and logically $\mathbf{V}^T$) leaving the other submatrices unaffected. The elements of **V** are just coefficients of the channel impulse response and the decision delay determines which ones appear within **V** and where.

It can be shown ([Riera-Palou02]) that $\chi(\mathbf{R})$ is minimised by choosing D so that the largest channel coefficient (main peak) does not appear in **V**. In minimum phase channels this is accomplished by setting $D \geq N_f$-1, where Nf is the FFF length. In maximum or mixed-phase channels, the delay must be set to $D \geq \mathbf{\Delta}_c + N_f$-1 where $\mathbf{\Delta}_c$ is the delay introduced by the channel.

If the delay is set according to this rule, the sample being detected is always located on the last tap of the FFF, independently of the FFF length. This choice of delay guarantees a low $\chi(\mathbf{R})$ but it does not assure its MSE optimality.

As an example, suppose that a channel given by **c**=[0.2 –0.5 1.0 –0.2 0.15 –0.05] is equalised with a (4,5)-DFE. Figure 6.16 shows the eigenvalue spread achieved as a function of the delay. According to the rule stated in previous paragraphs, $D \geq 5$ ($N_f$ =4 and $\mathbf{\Delta}_c$=2) which can be seen in the figure that is the delay where $\chi(\mathbf{R})$ drops abruptly.

It is also important to look at how the decision delay affects the MMSE. This information is shown in figure 6.17 and reveals that the minimum MMSE is achieved by choosing D=4. Also fundamental is the fact that when $D > \mathbf{\Delta}_c + N_f - 1$ the MMSE attained becomes extremely large. This is not surprising as with this choice of delay, the main peak of energy falls out of the span of the FFF and is therefore not captured. Consequently, the rule to minimise the eigenvalue spread while at the same obtaining a sensible MMSE can be simplified to just set the delay to D=$\mathbf{\Delta}_c + N_f - 1$, in this example, D=5.

---

[19] For simplicity, in this section the channel is assumed to be static, hence, **R** will also be static.

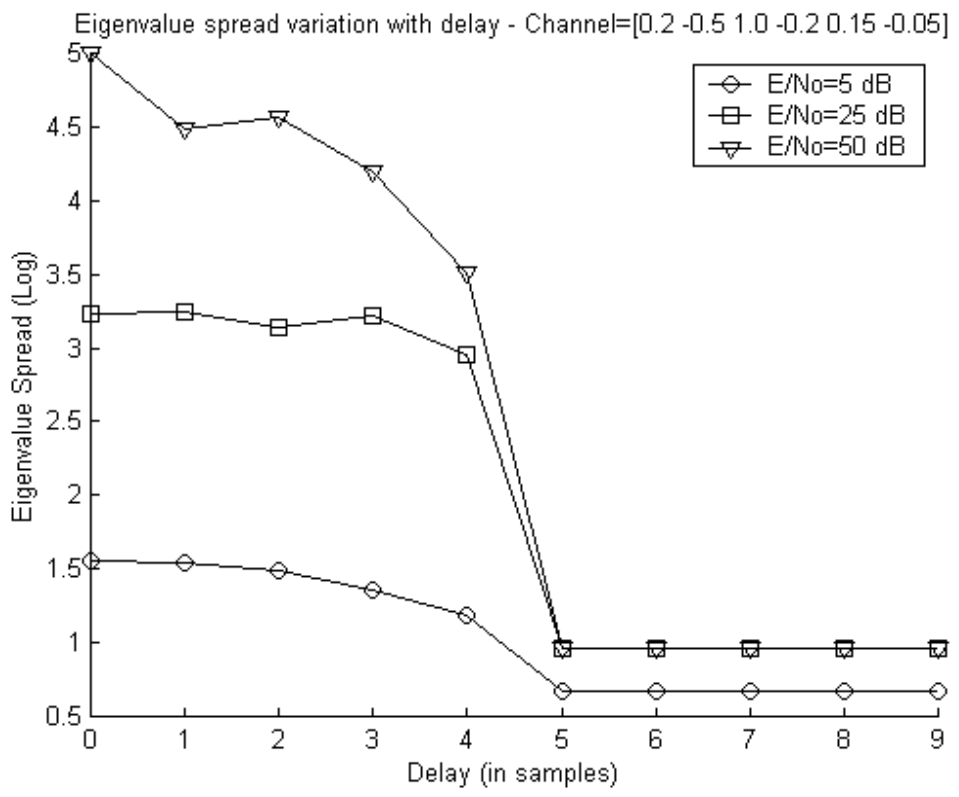

**Figure 6.16**: Eigenvalue spread for different choices of the decision delay.

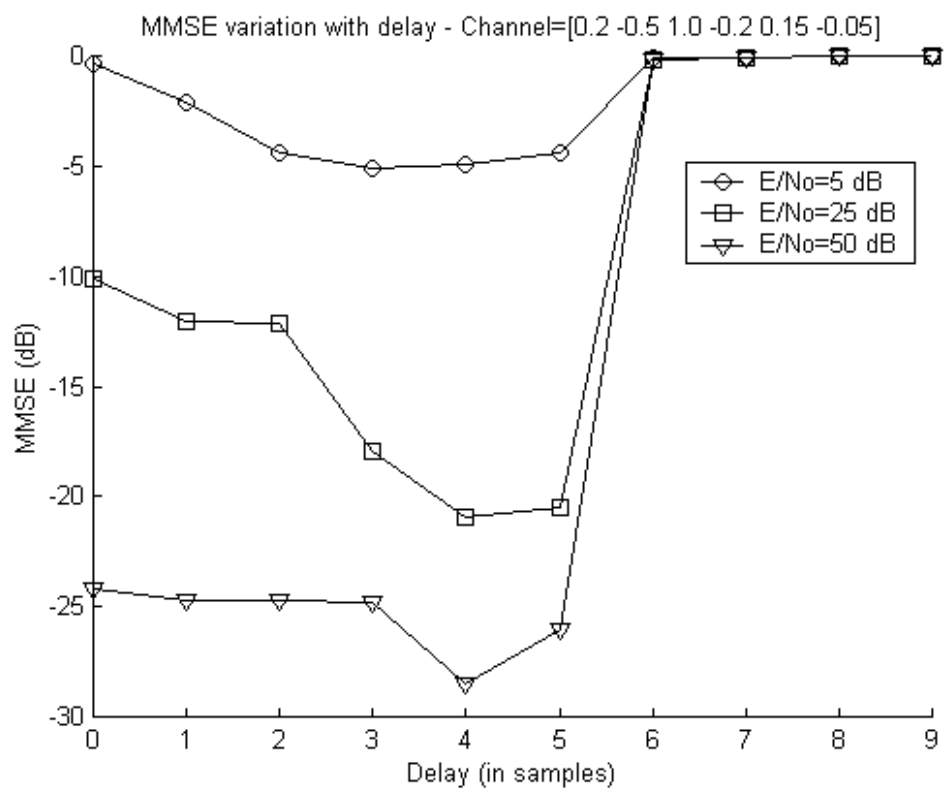

**Figure 6.17**: DFE-MMSE for different choices of the decision delay.

In order to verify theory with reality, simulations have been run using the LMS algorithm in the same scenario as figures 6.16 and 6.17. The delays chosen for the simulations are D=3, 4 and 5, which are the ones combining low MSE and $\chi(\mathbf{R})$.

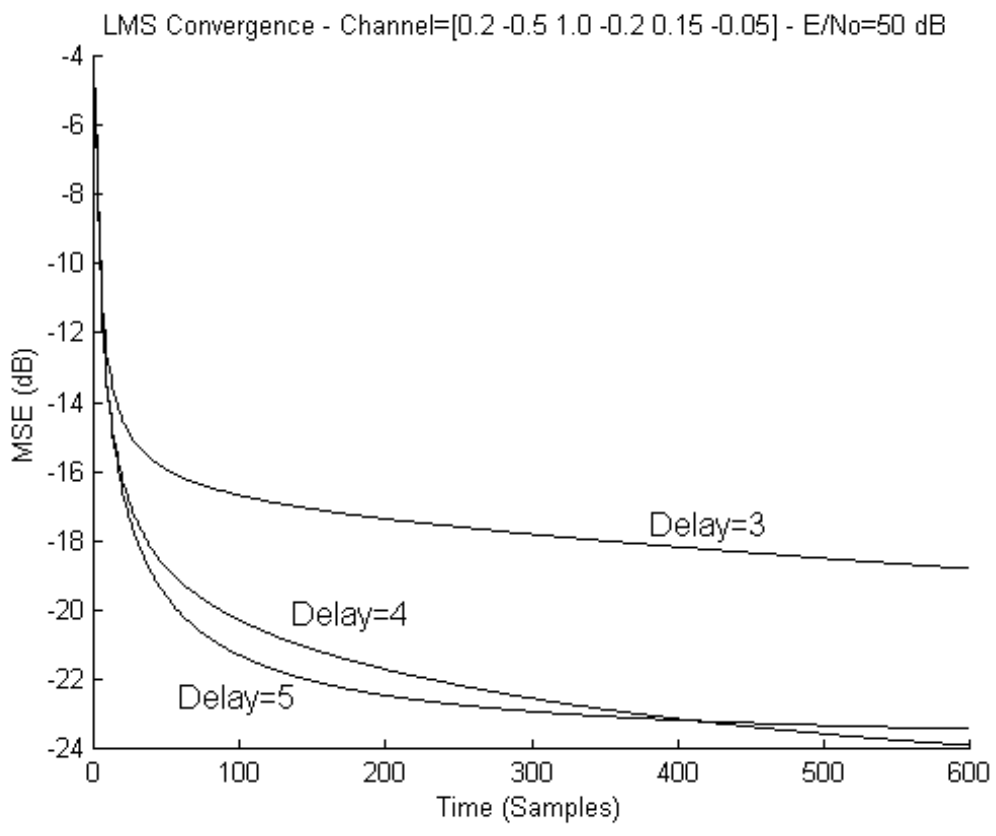

**Figure 6.18**: MSE curves for the LMS (3,4)-DFE for delays 3, 4 and 5.

It is seen in figure 6.18 that D=5 is the one with the fastest rate of convergence as could be already inferred from its low eigenvalue spread in figure 6.16. The curve for D=4 is the one achieving the lowest MSE as already predicted by the results in figure 6.17. Notice however, that this slight improvement in MSE with respect to D=4 comes at the prize of a significantly longer convergence time. Even more revealing is the curve for D=3. If the simulation is let run for long enough, this curve will end up converging very near the curve corresponding to D=5. Note in figure 6.17 that D=3 and D=5 have almost identical MMSE. However, and due to its large $\chi(\mathbf{R})$, the MSE convergence for D=3 is extremely slow. Obviously, this choice of delay should be avoided.

# 6.6 RLS-DFE analysis

The RLS equations for a DFE have exactly the same form as the corresponding LE equations given by (4.33)-(4.36). These equations are repeated here for convenience:

$$\mathbf{k}(n) = \frac{\lambda^{-1}\mathbf{P}(n-1)\mathbf{u}(n)}{1+\lambda^{-1}\mathbf{u}^{T}(n)\mathbf{P}(n-1)\mathbf{u}(n)} \tag{6.17}$$

$$\xi(n) = d(n) - \hat{\mathbf{w}}^{T}(n-1)\mathbf{u}(n) \tag{6.18}$$

$$\hat{\mathbf{w}}(n) = \hat{\mathbf{w}}(n-1) + \mathbf{k}(n)\xi(n) \tag{6.19}$$

$$\mathbf{P}(n) = \lambda^{-1}\mathbf{P}(n-1) - \lambda^{-1}\mathbf{k}(n)\mathbf{u}^{T}(n)\mathbf{P}(n-1) \tag{6.20}$$

Although identical in form, the individual variables bear important differences with respect to the RLS-LE counterparts. The $N_f$+$N_b$ input vector **u**(n) is defined as:

$$\mathbf{u}(n)=[\mathbf{r}(n)\ \mathbf{d}(n)] \tag{6.21}$$

where **r**(n) is made of $N_f$ consecutively received channel samples and d(n) are the $N_b$ most recently detected symbols. The equaliser settings vector, denoted by $\hat{\mathbf{w}}(n)$, is defined as:

$$\hat{\mathbf{w}}(n) = [\,\hat{\mathbf{w}}^{f}(n)\ \ \hat{\mathbf{w}}^{b}(n)\,] \tag{6.22}$$

where $\hat{\mathbf{w}}^{f}(n)$ and $\hat{\mathbf{w}}^{b}(n)$ represent the $N_f$ feedforward and $N_b$ feedback taps respectively. The ($N_f$+$N_b$) vector **k**(n) is the Kalman gain applied to each of the equaliser coefficients. The ($N_f$+$N_b$)x($N_f$+$N_b$) matrix **P**(n) is defined as the inverse of the input temporal correlation matrix:

$$\mathbf{P}(n) = \mathbf{\Phi}(n)^{-1} = \sum_{1}^{n}\lambda^{n-i}\mathbf{u}(i)\mathbf{u}^{T}(i) \tag{6.23}$$

In the previous definition, λ, is the forgetting factor used to improve the tracking of the algorithm.

As in the LMS case, the MSE derivation for a DFE using the RLS algorithm follows very closely its LE counterpart. In fact, the derivation presented in section 4.6, taking into account the definitions of (6.21) and (6.22), applies also for the DFE up to equation (4.44), restated here:

$$\mathbf{t}(n) = \left[\mathbf{I} - \mathbf{\Phi}^{-1}(n)\mathbf{u}(n)\mathbf{u}^T(n)\right]\mathbf{t}(n-1) + \mathbf{\Phi}^{-1}(n)\mathbf{u}(n)e_o(n) - \mathbf{p}(n) \qquad (6.24)$$

Recall from the LE derivation that $\mathbf{t}(n)$ is the error vector given by $\mathbf{t}(n) = \mathbf{w}_{opt}(n) - \hat{\mathbf{w}}(n)$ and $\mathbf{p}(n)$ is the process noise vector associated with the optimum equaliser variations.

Similarly to the previous section for the LMS, it is now important to examine the structure of the temporal correlation matrix $\mathbf{\Phi}(n)$. We start with an assumption already made in the RLS-LE analysis, that is:

$$\mathbf{\Phi}(n) \cong E[\mathbf{\Phi}(n)] = E\left[\sum_{1}^{n} \lambda^{n-i}\mathbf{u}(i)\mathbf{u}^T(i)\right] = \sum_{i=1}^{n} \lambda^{n-i} E\left[\mathbf{u}(i)\mathbf{u}^T(i)\right] = \sum_{i=1}^{n} \lambda^{n-i}\mathbf{R}(i) \qquad (6.25)$$

where $\mathbf{R}(n)$ denotes the ensemble correlation matrix. Therefore, from (6.25) it can be written: $\mathbf{\Phi}(n) \cong \sum_{i=1}^{n} \lambda^{n-i}\mathbf{R}(i)$. The structure of the ensemble input correlation matrix is given by equation (6.7), and, in a more detailed form, by equation (6.8). Notice however, that now we are interested in computing not only $\mathbf{R}(n)$ but also $\mathbf{\Phi}(n)$. Equation (6.25) can be expressed as:

$$\mathbf{\Phi}(n) = \lambda^n \mathbf{R}(0) + \lambda^{n-1}\mathbf{R}(1) + \cdots + \lambda \mathbf{R}(n-1) + \mathbf{R}(n) \qquad (6.26)$$

In order to proceed further it is necessary to establish the relation between **R**(n) and **R**(n+1). As shown in equation (6.7), **R**(n) is composed of four submatrices: **F**(n), **V**(n), $\mathbf{V}(n)^T$ and **D**. The submatrix $\mathbf{F}(n)=E[\mathbf{r}(n)\mathbf{r}(n)^T]$ corresponds to the correlation of the inputs to the FFF, so clearly the relation of **F**(n) and **F**(n+1) can be determined using equations (4.48), (4.49) and (4.50) corresponding to the LE. Therefore it can be written:

$$\mathbf{F}(n+1) = \mathbf{F}(n) + N\sigma_d^2\sigma_q^2\mathbf{I}_{Nf} \qquad (6.27)$$

where $\mathbf{I}_{Nf}$ is the $N_f x N_f$ identity matrix.

The entries of **V**(n) are just entries of the channel impulse response (see equation (6.8)), therefore, **V**(n) and **V**(n+1) are related directly by the channel process noise vector **q**(n) as follows:

$$\mathbf{V}(n+1) = \mathbf{V}(n) + \mathbf{\Delta}_V(n) \qquad (6.28)$$

where $\mathbf{\Delta_V}$(n) is given by:

$$\mathbf{\Delta_V}(n) = \begin{bmatrix} q_{Nf}(n) & q_{Nf+1}(n) & \cdots & q_{Nf+Nb}(n) \\ q_{Nf-1}(n) & q_{Nf}(n) & \cdots & q_{Nf+Nb-1}(n) \\ \vdots & \vdots & \ddots & \vdots \\ q_1(n) & q_2(n) & \cdots & q_{Nb}(n) \end{bmatrix} \tag{6.29}$$

As with the channel coefficients, any $q_i$ with $i \notin [0..N-1]$ is zero by definition, therefore many of the positions of $\mathbf{\Delta_V}$(n) are potentially zero. The last submatrix of **R**(n) to examine is **D**. This diagonal matrix, given its time independence, will remain constant.

Having examined the evolution of each of the submatrices of **R**(n), it is now possible to write **R**(n) in a recursive manner as follows:

$$\mathbf{R}(n+1) = \begin{bmatrix} \mathbf{F}(n+1) & \mathbf{V}(n+1) \\ \mathbf{V}(n+1)^T & \mathbf{D} \end{bmatrix} = \begin{bmatrix} \mathbf{F}(n) + N\sigma_d^2\sigma_q^2\mathbf{I}_{Nf} & \mathbf{V}(n) + \mathbf{\Delta_V}(n) \\ [\mathbf{V}(n) + \mathbf{\Delta_V}(n)]^T & \mathbf{D} \end{bmatrix} \tag{6.30}$$

Equation (6.30) can be written in a more compact way as:

$$\mathbf{R}(n+1) = \mathbf{R}(n) + N\sigma_d^2\sigma_q^2\mathbf{\Psi}(n) \tag{6.31}$$

with $\mathbf{\Psi}(n)$ being an ($N_b$x$N_b$) matrix given by:

$$\mathbf{\Psi}(n) = \begin{bmatrix} \mathbf{I}_{Nf} & \frac{1}{N\sigma_d^2\sigma_q^2}\mathbf{\Delta_V}(n) \\ \frac{1}{N\sigma_d^2\sigma_q^2}\mathbf{\Delta_V}(n)^T & \mathbf{0}_{Nb} \end{bmatrix}$$

where $\mathbf{0}_{Nb}$ is an $N_b$x$N_b$ zero matrix.

The recursion given by equation (6.31) allows us to expand the individual terms on the r.h.s. of equation (6.26) as follows:

$$\begin{aligned} &\mathbf{R}(0) \\ &\mathbf{R}(1) = \mathbf{R}(0) + N\sigma_d^2\sigma_q^2\mathbf{\Psi}(1) \\ &\mathbf{R}(2) = \mathbf{R}(0) + \mathbf{R}(1) + N\sigma_d^2\sigma_q^2[\mathbf{\Psi}(1) + \mathbf{\Psi}(2)] \\ &\vdots \\ &\mathbf{R}(n) = \mathbf{R}(0) + \mathbf{R}(1) + \cdots + \mathbf{R}(n-1) + N\sigma_d^2\sigma_q^2[\mathbf{\Psi}(1) + \mathbf{\Psi}(2) + \cdots + \mathbf{\Psi}(n-1)] \end{aligned}$$

When n is large, we can assume that:

$$\sum_{i=1}^{n}\mathbf{\Psi}(i) \cong n\begin{bmatrix} \mathbf{I}_{Nf} & \mathbf{0} \\ \mathbf{0} & \mathbf{0} \end{bmatrix} \overset{def}{=} n\mathbf{\Omega} \tag{6.32}$$

as the entries in the submatrices $\mathbf{\Delta_V}(n)$ and $\mathbf{\Delta_V}(n)^T$ have zero mean and will tend to cancel when n is large. Equation (6.31) can also be expressed "backwards" as:

$$\mathbf{R}(n) = \mathbf{R}(n+1) - N\sigma_d^2\sigma_q^2\mathbf{\Psi}(n) \tag{6.33}$$

Using equation (6.33) and the approximation of equation (6.32), $\mathbf{\Phi}(n)$ can be expressed as:

$$\mathbf{\Phi}(n) = \sum_{i=1}^{n}\lambda^{n-i}\mathbf{R}(i) = \lambda^{n-1}\left[\mathbf{R}(n) - N(n-1)\sigma_d^2\sigma_q^2\mathbf{\Omega}\right] + \lambda^{n-2}\left[\mathbf{R}(n) - N(n-2)\sigma_d^2\sigma_q^2\mathbf{\Omega}\right] + \cdots + \mathbf{R}(n) \tag{6.34}$$

Rearranging (6.34):

$$\mathbf{\Phi}(n) = \mathbf{R}(n)\left[\lambda^{n-1} + \lambda^{n-2} + \cdots + \lambda^0\right] + N\sigma_d^2\sigma_q^2\mathbf{\Omega}\left[\lambda^{n-1}(n-1) + \lambda^{n-2}(n-2) + \cdots + \lambda^0\right] \tag{6.35}$$

The summation terms within brackets in equation (6.35) can be recognised as the sums of two geometric series allowing the previous equation to be written in a compact way:

$$\mathbf{\Phi}(n) = \frac{1}{1-\lambda}\mathbf{R}(n) + \frac{N\sigma_d^2\sigma_q^2\lambda}{(1-\lambda)^2}\mathbf{\Omega} \tag{6.36}$$

Recall that in chapter 4, during the development of an steady state MSE expression for the RLS-LE, a similar expression to (6.36) was also derived (see equation (4.54)). There, it was justifiable to ignore the second term in the expression for $\mathbf{\Phi}(n)$ because, under normal conditions, it was very small when compared to the first term. The same reasoning is now applied to equation (6.36). Assuming the forgetting factor ($\lambda$) is chosen consistently with the magnitude of the channel variations ($\sigma_q^2$), the second term in (6.36) will be one order of magnitude smaller than the first term. In order to ease the analysis it is assumed that the temporal correlation matrix is given by:

$$\mathbf{\Phi}(n) = \frac{1}{1-\lambda}\mathbf{R}(n) \tag{6.37}$$

Given that our main interest is in the inverse of $\mathbf{\Phi}(n)$, we write:

$$\mathbf{\Phi}(n)^{-1} = (1-\lambda)\mathbf{R}(n)^{-1} \tag{6.38}$$

We are now in a position to continue the analysis from equation (6.24) making use of (6.38):

$$\mathbf{t}(n) = \left[\mathbf{I}_{Nf+Nb} - (1-\lambda)\mathbf{R}(n)^{-1}\mathbf{u}(n)\mathbf{u}^T(n)\right]\mathbf{t}(n-1) + (1-\lambda)\mathbf{R}(n)^{-1}\mathbf{u}(n)e_o(n) - \mathbf{p}(n) \tag{6.39}$$

As in the LMS derivation we assume that $\mathbf{u}(n)\mathbf{u}^T(n) \cong E[\mathbf{u}(n)\mathbf{u}^T(n)] = \mathbf{R}(n)$. This further simplifies the previous equation:

$$\mathbf{t}(n) = \lambda\mathbf{t}(n-1) + (1-\lambda)\mathbf{R}(n)^{-1}\mathbf{u}(n)e_o(n) - \mathbf{p}(n) \qquad (6.40)$$

The derivation follows now exactly the same steps as those given for the RLS-LE in equations (4.57) to (4.61) and a very similar expression is found for the RLS-DFE EMSE:

$$\mathrm{EMSE}(n) = \frac{(1-\lambda)\mathrm{MMSE}(n)}{2}\mathrm{Tr}[\mathbf{I}_{Nf+Nb}] + \frac{1}{2(1-\lambda)}\mathrm{Tr}[\mathbf{R}(n)\mathbf{P}] \qquad (6.41)$$

This equation, although identical in form to equation (4.61) for the LE, will be different when the traces have been computed due to the different structure of the correlation matrix **R**(n). **P** will be a ($N_f$+$N_b$) diagonal matrix with elements $\sigma_p^2$. Using the definition of **R**(n) from equation (6.8) and assuming $\sigma_d^2$=1, we can compute Tr[**R**(n)**P**]:

$$\mathrm{Tr}[\mathbf{R}(n)\mathbf{P}] = \mathrm{Tr}\begin{bmatrix} (1+\sigma_v^2)\sigma_p^2 & 0 & \cdots & \cdots & \cdots & \cdots & \cdots & 0 \\ 0 & (1+\sigma_v^2)\sigma_p^2 & 0 & \cdots & \cdots & \cdots & \cdots & 0 \\ \vdots & \ddots & \ddots & \ddots & & & & \vdots \\ \vdots & & 0 & (1+\sigma_v^2)\sigma_p^2 & 0 & & & \vdots \\ \vdots & & & 0 & \sigma_p^2 & 0 & & \vdots \\ \vdots & & & & 0 & \ddots & \ddots & \vdots \\ \vdots & & & & & \ddots & \ddots & 0 \\ 0 & \cdots & \cdots & \cdots & \cdots & \cdots & 0 & \sigma_p^2 \end{bmatrix} = N_f(1+\sigma_v^2)\sigma_p^2 + N_b\sigma_p^2$$

(6.42)

The trace of $\mathbf{I}_{Nf+Nb}$ is obviously $N_f$+$N_b$. Substituting both traces in equation (6.41), the EMSE is obtained:

$$\mathrm{EMSE}(n) = \frac{(1-\lambda)(N_f+N_b)}{2}\mathrm{MMSE}(n) + \frac{N_f(1+\sigma_v^2)\sigma_p^2 + N_b\sigma_p^2}{2(1-\lambda)} \qquad (6.43)$$

It is only left to express the second term on the r.h.s. of (6.43) as a function of the channel parameters (N and $\sigma_q^2$) by means of the assumption expressed in section 6.2 which is restated here:

$$\sigma_p^2 = \frac{N\sigma_q^2}{(N_f+N_b)} \qquad (6.44)$$

Finally, using (6.44) and adding the MMSE term we obtain the formula for the steady state MSE for a DFE using the RLS algorithm:

$$\mathrm{MSE}(n) = \mathrm{MMSE}(n) + \frac{(1-\lambda)(N_f+N_b)}{2}\mathrm{MMSE}(n) + \frac{N\,N_f\sigma_q^2(1+\sigma_v^2)}{2(1-\lambda)(N_f+N_b)} + \frac{N\,N_b\sigma_q^2}{2(1-\lambda)(N_f+N_b)}$$

(6.45)

Some comments are now in order regarding equation (6.45). First of all notice that there are four terms: the first one corresponds to the MMSE, the other three terms correspond to the EMSE due to the use of a forgetting factor less than one (2nd term), and the channel variations (3rd and 4th terms). If we set $\lambda$=1, which would correspond to a situation where the channel is static and therefore $\sigma_q^2$=0, the last three terms of (6.45) will cancel and the steady state MSE will be equal to the MMSE. This agrees with the RLS theory presented in chapter 2 which states that in an stationary environment, the least squares solution will converge to the optimum Wiener solution. It also reassuring to check that when the FBF is suppressed (i.e. $N_b$=0), the resulting expression is the same as equation (4.63) derived for the RLS-LE. As in the LMS, it should also be noticed that the FBF does not produce any noise enhancement as there are not any terms containing $\sigma_v^2$ proportional to $N_b$.

In the next section, simulation results are presented to give an idea of the accuracy of equation (6.45).

# 6.7 Validation of the RLS-DFE equation

The simulation results use the same channel models and simulation conditions as in previous sections. As in the DFE-LMS, the FFF length has been set to N+1 where N is the channel length. The number of feedback taps has been set to N-1. RLS-DFE performance in practical scenarios is deferred until chapter 7.

## 6.7.1 Channel model 1

MSE curves for Channel model 1 are shown in figures 6.19 and 6.20 for $\sigma_q^2=10^{-4}$ and $\sigma_q^2=10^{-6}$ respectively. In general, and for this particular channel, good agreement exists between the predicted and measured MSE. However, notice that in scenarios subject to fast variations (figure 6.19), the predicted MSE tend to be larger than the simulated MSE. This tendency is more accentuated as the E/No gets larger, but over a reasonable range of E/No, such as from 5 to 25 dB, the maximum divergence between prediction and simulation is only about 1 dB.

When the channel varies slowly (figure 6.20) the prediction equation becomes very accurate. In this figure, MMSE curves are not shown in the plot because they lie on top of the other two sets of figures (predicted and simulated MSE). Notice how in this particular channel model, the MMSE could serve as a good indication of what would be the true MSE performance of the equaliser. This is not always the case as is shown in the other channel models where equation (6.45) models far more accurately the actual performance of the equaliser than the MMSE.

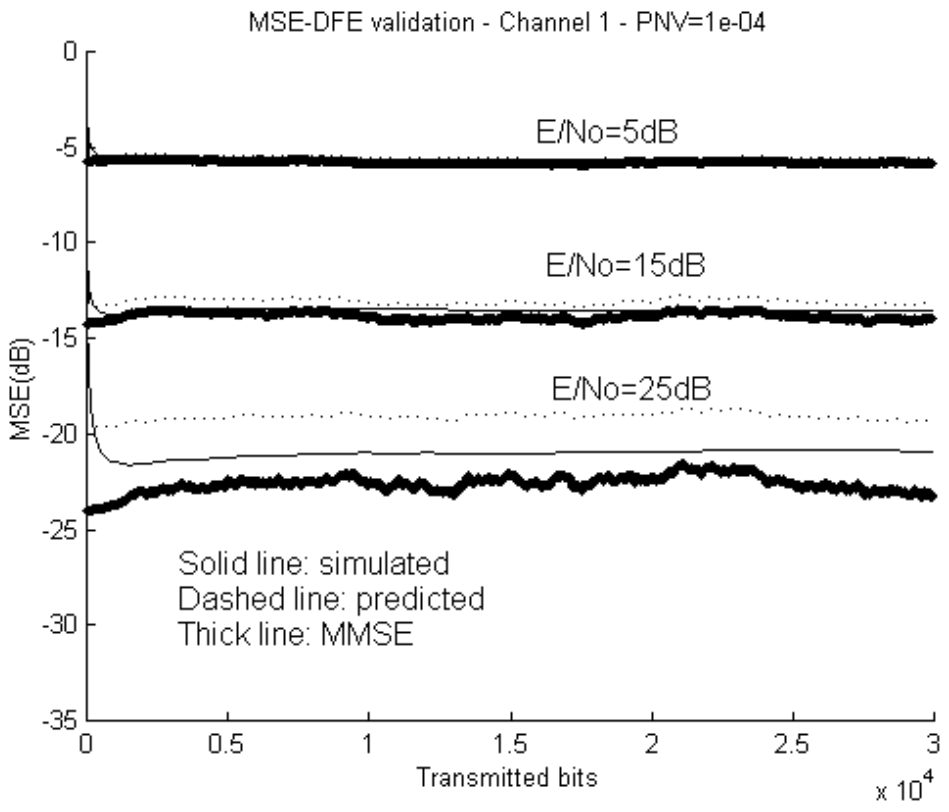

**Figure 6.19**: DFE RLS MSE Validation, Channel model 1, $\sigma_q^2=10^{-4}$.

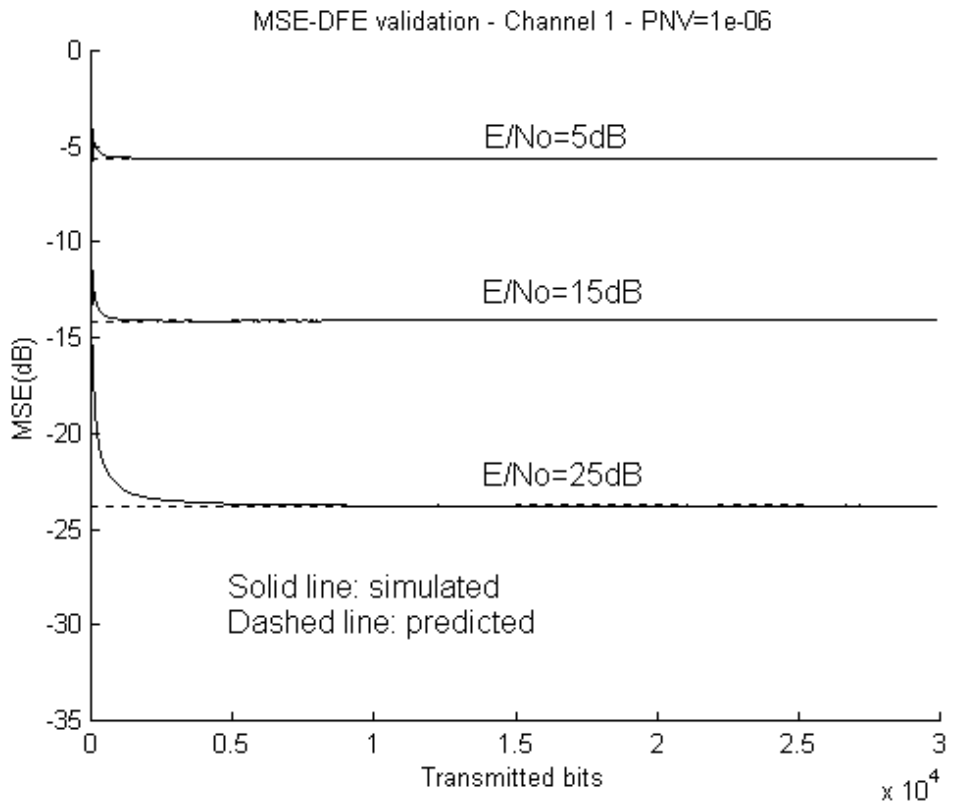

**Figure 6.20**: DFE RLS MSE Validation, Channel model 1, $\sigma_q^2=10^{-6}$.

## 6.7.2 Channel model 2

In figures 6.21 and 6.22 results are shown for a DFE when equalising Channel model 2. Again, some discrepancy (still within a dB) between the predicted and simulated MSE is found in a large E/No scenario with rapid channel variations. Notice however, that unlike the previous channel model, the predicted MSE is much closer to the real performance than the plain MMSE. This indicates that in this particular scenario the EMSE components predicted in equation (6.45) are significant, about 5-6 dB over the MMSE and therefore their effects should be taken into account.

When this channel varies slowly, as happened also in Channel model 1, the true MSE performance and also its predicted level, approach the MMSE.

In figure (6.23) the effect of varying the FFF length is shown. The channel variations were set at $\sigma_q^2=10^{-4}$ and the E/No level was set to 15 dB. Three equaliser lengths have been tested: 6, 12 and 18 taps. As has been explained for the LMS simulations (see section 6.4.2), the simulation runs where the dominant tap of the channel had moved beyond the 6$^{th}$ tap in the CIR have not been included when computing the averaged results shown in figure (6.23). The difference between the predicted and simulated MSE is about 1-1.5 dB. So, for example, it can be inferred from the predicted MSE curves that there is a significant improvement when the equaliser is expanded from 6 to 12 taps. On the other hand, it can also be deduced that extending the equaliser to 18 taps will only reduce the MSE level by a small amount. Both conjectures are seen to be true by examining the simulated results.

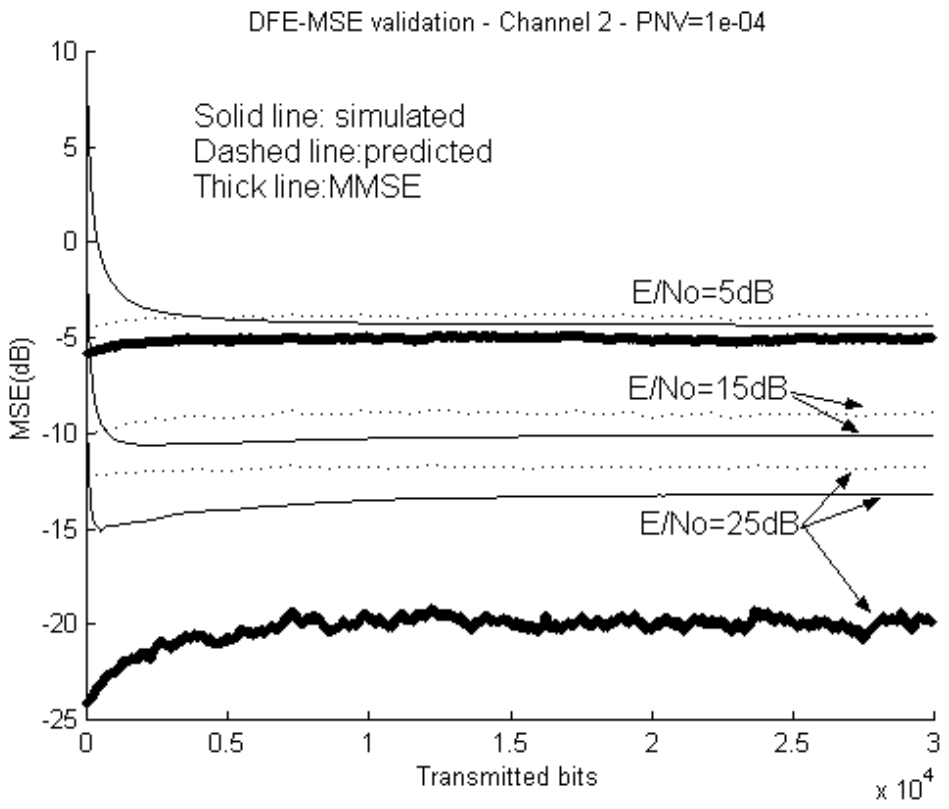

**Figure 6.21**: DFE RLS MSE Validation, Channel model 2, $\sigma_q^2=10^{-4}$.

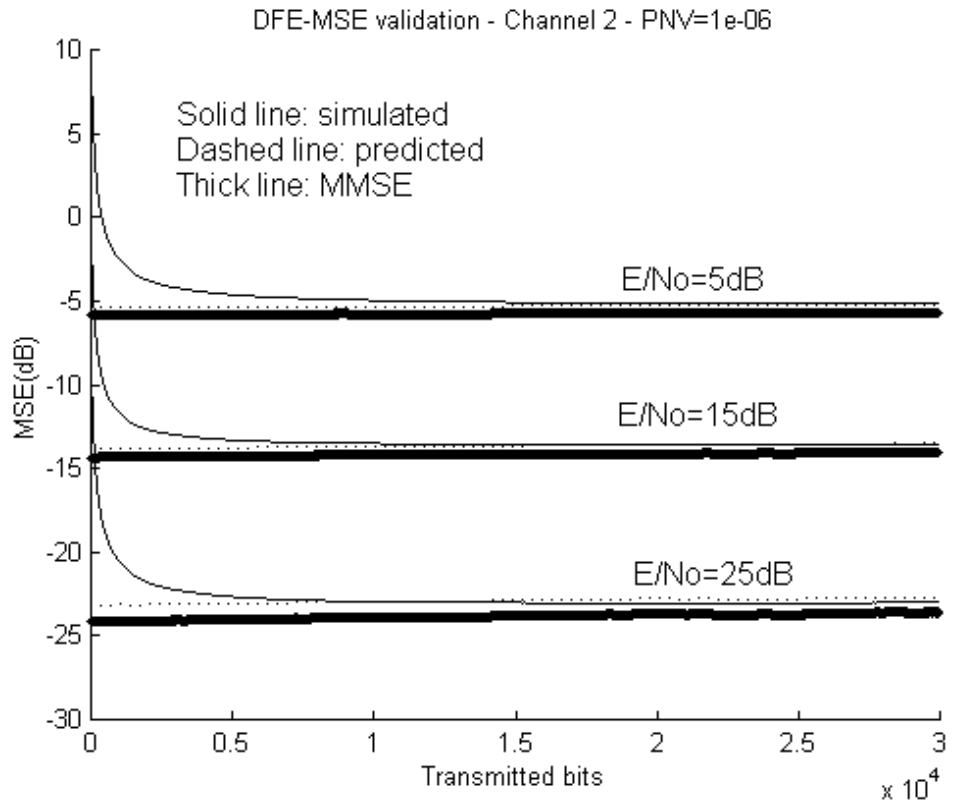

**Figure 6.22**: DFE RLS MSE Validation, Channel model 2, $\sigma_q^2=10^{-6}$.

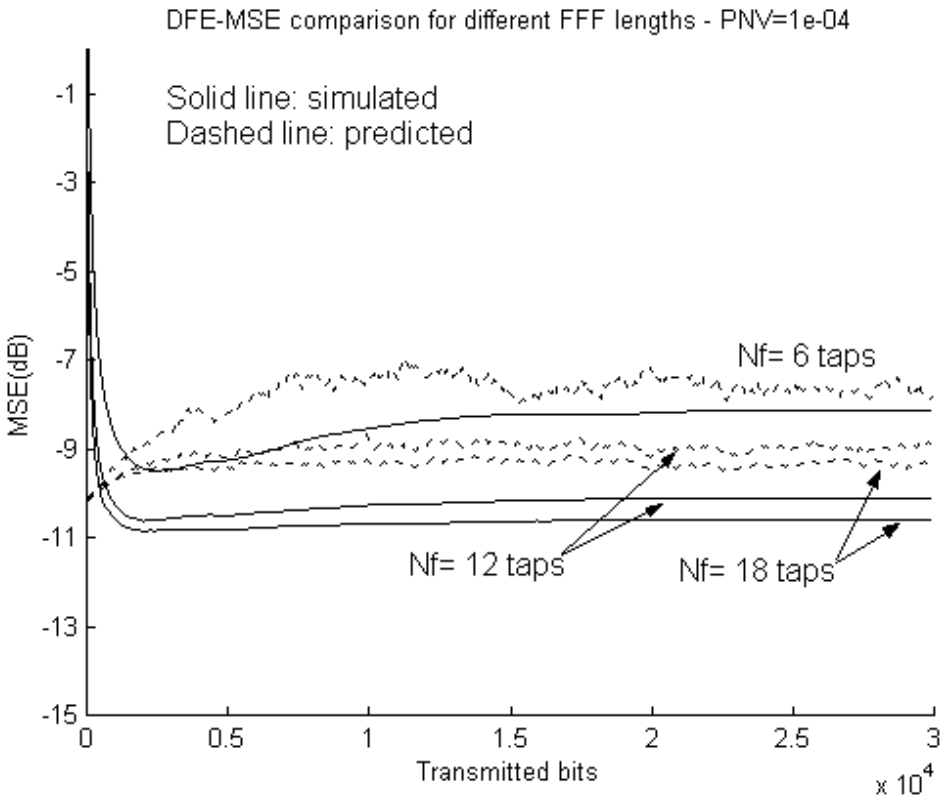

**Figure 6.23**: DFE RLS MSE Comparison for different number of FFF taps, Channel model 2, E/No=15dB, $\sigma_q^2=10^{-4}$.

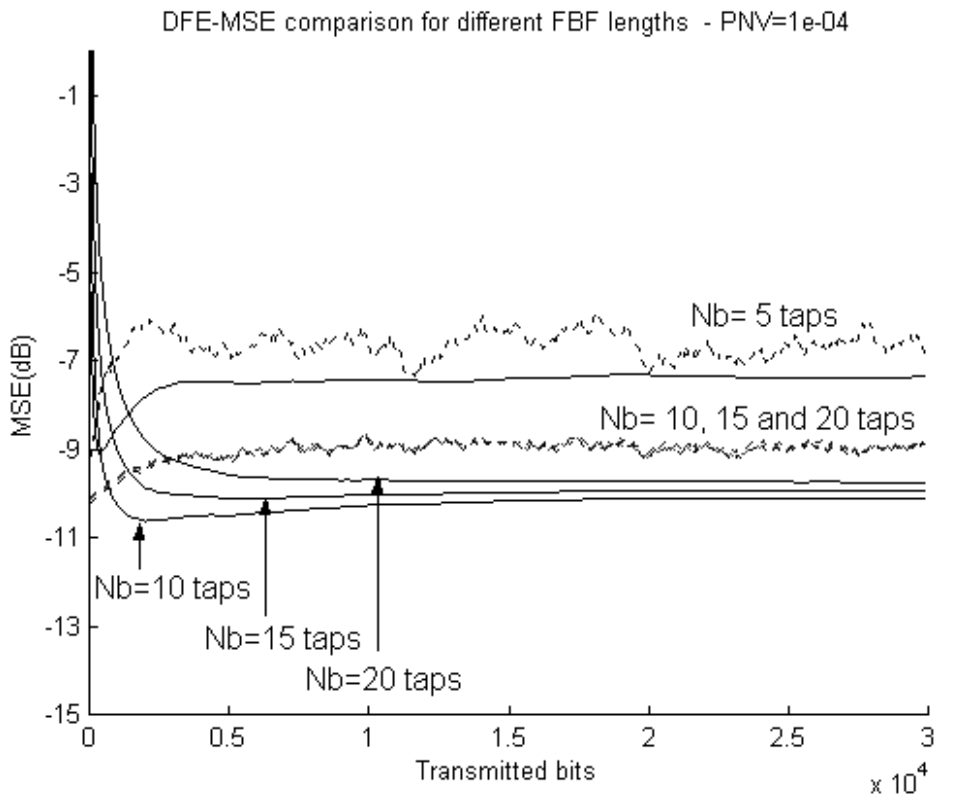

**Figure 6.24**: DFE RLS MSE Comparison for different number of FBF taps, Channel model 2, E/No=15dB, $\sigma_q^2=10^{-4}$.

In figure (6.24) the effect of the FBF length is studied under the same conditions as described for figure (6.23). As in the FBF length study for the same channel model when using the LMS algorithm, it can be seen that, when the number of taps in the FBF exceeds N-1 (N= channel length), there is not any reduction in the MSE level. Although there is a gap of about 1 dB between the predicted and simulated MSE, it is clear that equation (6.45) already confirms this phenomenon.

### 6.7.3 Channel model 3

Validation of equation (6.45) is concluded by presenting the results obtained for Channel model 3. The curves presented in figures 6.25 ($\sigma_q^2=10^{-4}$) and 6.26 ($\sigma_q^2=10^{-6}$) reflect again the good accuracy of the MSE prediction equation. Notice that when $\sigma_q^2=10^{-4}$ and as in the previous channel model, the predicted MSE (dashed curve) is much closer to the real MSE (solid curve) than just the MMSE alone (thick line). In low E/No and/or slowly varying channels (figure 6.26) that the MMSE alone could serve as a good estimate of the MSE. Overall, a good agreement is observed between predictions and simulations.

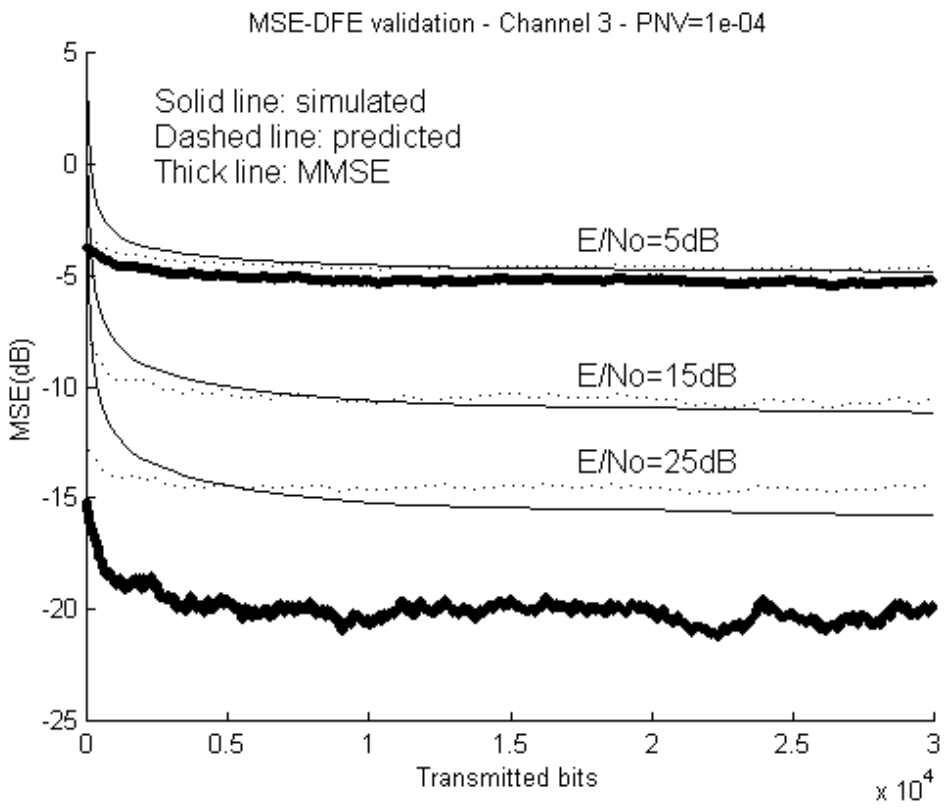

**Figure 6.25**: DFE RLS MSE Validation, Channel model 3, $\sigma_q^2=10^{-4}$.

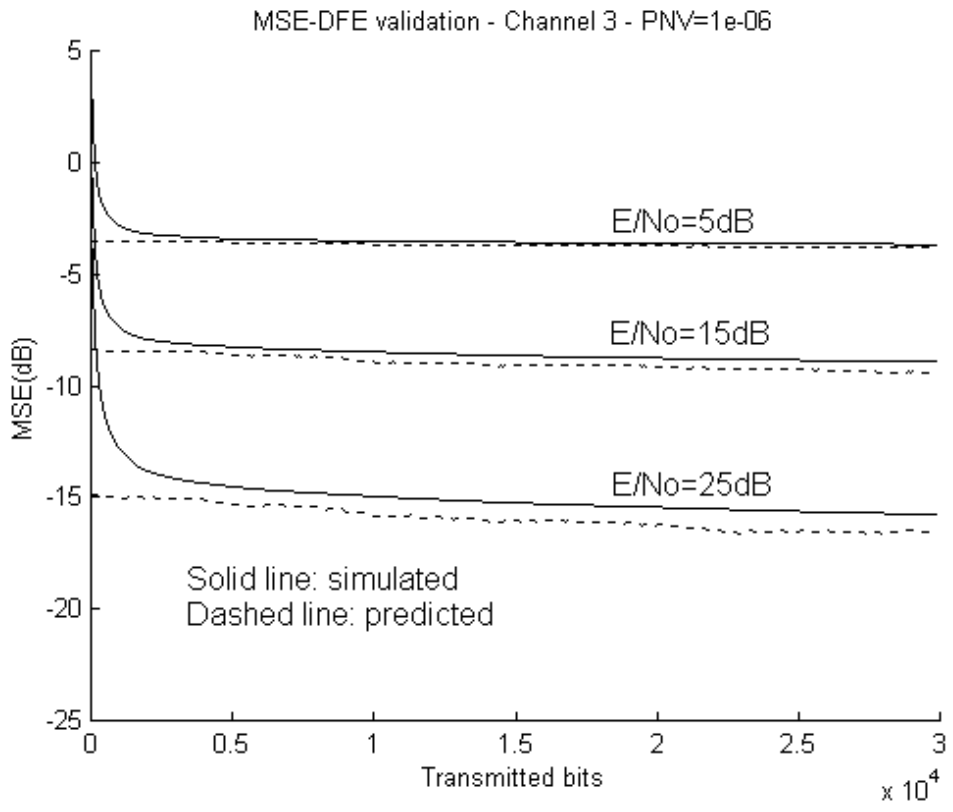

**Figure 6.26**: DFE RLS MSE Validation, Channel model 3, $\sigma_q^2=10^{-6}$.

## 6.8 Conclusions on the derived MSE equations

So far, this chapter has been devoted to finding and validating expressions for the steady-state MSE of a DFE when using the LMS (equation (6.16)) and RLS (equation (6.45)) algorithms to perform the coefficients update. Given their importance, these expressions are re-stated next:

$$\mathrm{MSE}(n) = \mathrm{MMSE}(n) + \frac{\mu N_f \left[1+\sigma_v^2\right]}{2}\mathrm{MMSE}(n) + \frac{\mu N_b}{2}\mathrm{MMSE}(n) + \frac{N\sigma_q^2}{2\mu}$$

LMS-MSE

$$\mathrm{MSE}(n) = \mathrm{MMSE}(n) + \frac{(1-\lambda)(N_f + N_b)}{2}\mathrm{MMSE}(n) + \frac{N\,N_f\sigma_q^2(1+\sigma_v^2)}{2(1-\lambda)(N_f + N_b)} + \frac{N\,N_b\sigma_q^2}{2(1-\lambda)(N_f + N_b)}$$

RLS-MSE

It might be argued that both expressions are useless in practical scenarios as some of the parameters in them will not be known a-priori. For example, it is impossible to know the instantaneous MMSE as this would require perfect knowledge of the channel impulse response at any instant. Additionally, the practical equivalent of $\sigma_q^2$, the Doppler spread, might also be difficult to estimate. Nonetheless, these MSE equations do reveal some important information regarding the parameters influencing the MSE. Even if the expressions are not very accurate under certain channel conditions, it is useful to know what parameters are available to the designer to influence the steady state MSE level. It can be seen that given a particular scenario ($\sigma_v^2$, N and $\sigma_q^2$), three parameters can be used to adjust the performance of the equaliser, namely: $N_f$, $N_b$ and the algorithm parameter ($\mu$ or $\lambda$). In mobile radio systems, where the environment is subject to a wide range of potential variations, these parameters will rarely be optimum all the time. Therefore, it would be desirable to adjust them accordingly in response to changes in the environment.

As has been said in section 4.10, adaptive algorithms with time varying parameters have already been studied in the past. On the other hand, and as in linear equalisation, dynamic variations of the FFF and FBF lengths have received no attention at all[20]. Our interest for the rest of this chapter is to investigate in detail how $N_f$ and $N_b$ influence the MSE performance in a DFE and also to study the potential reconfiguration of the filters' length to better withstand changes in the environment.

---

[20] To the best of author's knowledge.

## 6.9 DFE MMSE results

The steady state DFE-MSE expressions derived for the LMS and RLS in previous sections show a clear dependence on the MMSE level. In this subsection some known results regarding the DFE-MMSE are presented. These results, jointly with equations (6.16) and (6.45), provide some insight on how the filters' length can be dynamically adjusted in an efficient manner.

Given an N-tap static channel, **c**, and a ($N_f$,$N_b$)-DFE and assuming the detected symbols are correct, the optimum DFE settings can be computed by solving the Wiener-Hopf equation ([Smee97, [Proakis95]) applied to the DFE system shown in figure 6.2:

$$\mathbf{w}_{opt} = \mathbf{R}^{-1}\mathbf{p} \tag{6.46}$$

where $\mathbf{w}_{opt}$ is the vector containing the set optimum feedforward and feedback coefficients:

$$\mathbf{w}_{opt} = [\mathbf{w}^{f}_{opt}\ \mathbf{w}^{b}_{opt}] \tag{6.47}$$

**R** is the autocorrelation matrix of the input vector, **u**(n), which is formed by $N_f$ channel samples and $N_b$ detected symbols:

$$\mathbf{u}(n) = [\mathbf{r}(n)\ \mathbf{d}(n)] \tag{6.48}$$

Lastly**, p** denotes the cross-correlation between the current sample being detected, d(n), and the input vector **u**(n). The solution given by $\mathbf{w}_{opt}$ will produce the minimum mean squared error (MMSE) between the output of the equaliser and the detected symbol. Notice that in equation (6.46), a determined decision delay D, is implicitly assumed. In [Voois96] it is shown that under the assumption that the FBF is made sufficiently long, the optimum delay from an MSE point of view, approaches D=$N_f$ -1. Later on in this chapter the issue of delay choice is covered in greater detail.

The combined FFF (feedforward filter) and CIR (channel impulse response), **z**, is given by:

$$\mathbf{z} = \mathbf{c} * \mathbf{w}^{f}_{opt} \tag{6.49}$$

where * denotes the linear discrete convolution. Notice that **z** will have $N_f$+N-1 taps. Figure 6.27 shows a generic FFF-CIR combined response. In the figure it can be seen that the precursor ISI is very much reduced by the action of the FFF. The FBF will take care of suppressing the postcursor components.

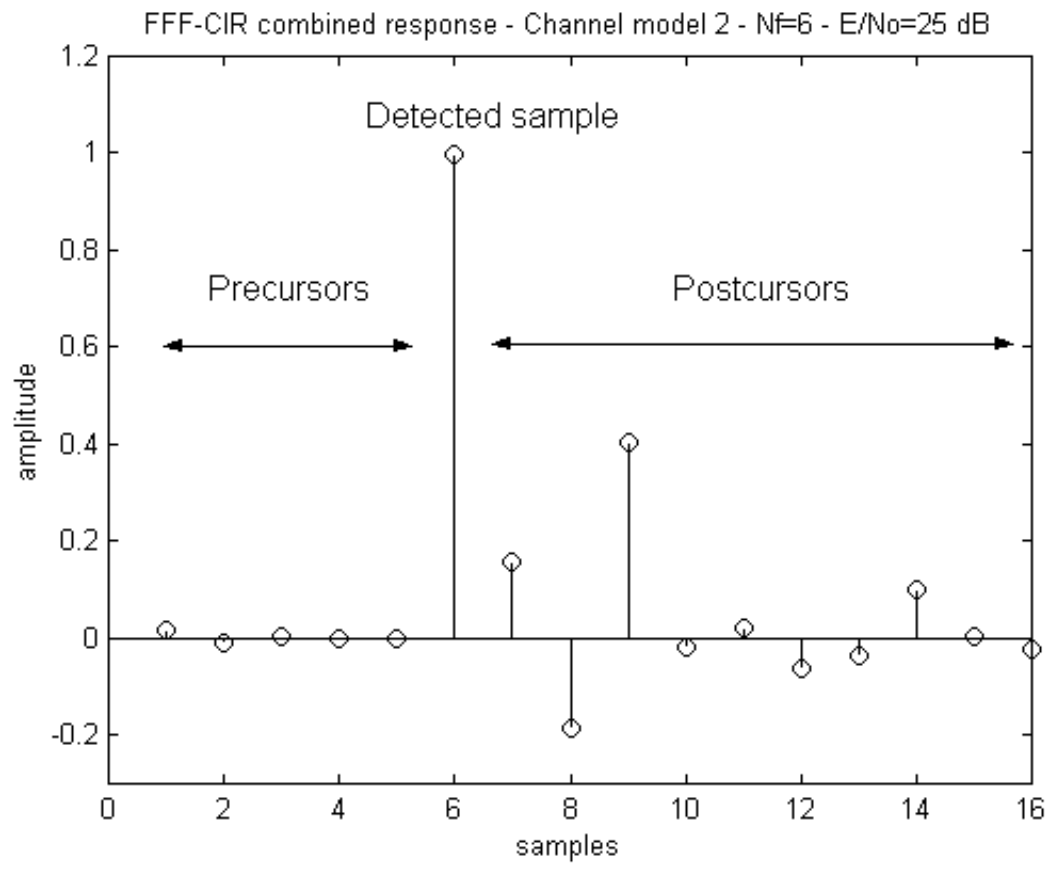

**Figure 6.27:** Typical FFF-CIR combined response for Nf=6 and N=11.

Assuming the source symbols have unit power, we can define the equaliser output SNR as:

$$SNR_{out} = \frac{z_D^2}{\sum_{i=0}^{Nf-1} z_i^2 \sigma_v^2 + \sum_{i=0, i \neq D}^{N+Nf-1} z_i^2} \qquad (6.50)$$

The first summation term in the denominator of (5) accounts for the filtered noise power ($\sigma_v^2$) while the second summation represents the residual ISI energy. The ISI can be further divided into precursor ISI energy and postcursor ISI energy as follows:

$$ISI_{pre} = \sum_{i=0}^{D-1} z_i^2 \qquad (6.51)$$

$$ISI_{post} = \sum_{i=D+1}^{N+Nf-1} z_i^2 \qquad (6.52)$$

The FFF settings are set so that a compromise is reached between the levels of $ISI_{pre}$ and filtered noise. $ISI_{post}$ is completely eliminated provided the FBF has enough taps. Given the finite length of the channel and FFF and having set the delay to $N_f$ -1, it is easy to see that there are N-1 postcursor samples (see figure 6.27). An FBF longer than N-1 taps will not decrease the MMSE level any further.

The length of the FFF depends very much on the position of the main peak of energy within the CIR. In practical applications, the strongest paths are usually concentrated at the beginning of the CIR and therefore a short FFF is enough to capture most of the energy ([Ariyavisitakul97]) and significantly reduce the amount of $ISI_{pre}$ without excessive noise enhancement.

The following set of graphs (6.28 to 6.31) shows the dependence of the MMSE level on the number of taps in the FFF and FBF. The results shown in figures 6.28 and 6.29 correspond to a DFE equalising Channel model 2 for E/No=5 dB and E/No=25 dB respectively. The delay in both cases was set to $N_f$ -1=5. In both cases, a FFF of 5-6 taps is enough to achieve the minimum attainable MSE level. Regarding the FBF length clear differences can be appreciated. While for E/No=5 dB, 3 taps is enough to approach the minimum possible MMSE, for E/No=10 dB the FBF needs to be expanded up to 10 taps. Notice that for this particular channel, and as has been explained in a paragraph above, the longest the FBF needs to be is 10 taps as this is the number of postcursors in the FFF-CIR combined response.

Attention is now turned to figures 6.30 and 6.31 where the same data is presented but now equalising Channel model 3. As before, a FFF 5-6 taps long is enough to cancel out most of the precursor ISI. As for the FBF, 4 taps are required to cancel out all the postcursors.

What can be inferred from these simple examples and has been covered in previous studies ([Ariyavisitakul97]) is that the FFF will typically be short, just long enough to capture the main peak of energy of the received signal whereas the FBF length depends very much on the particular CIR. If the CIR is very long, many taps will be needed to reduce the postcursor interference. On the other hand, if the CIR is short, only a few FBF taps are required.

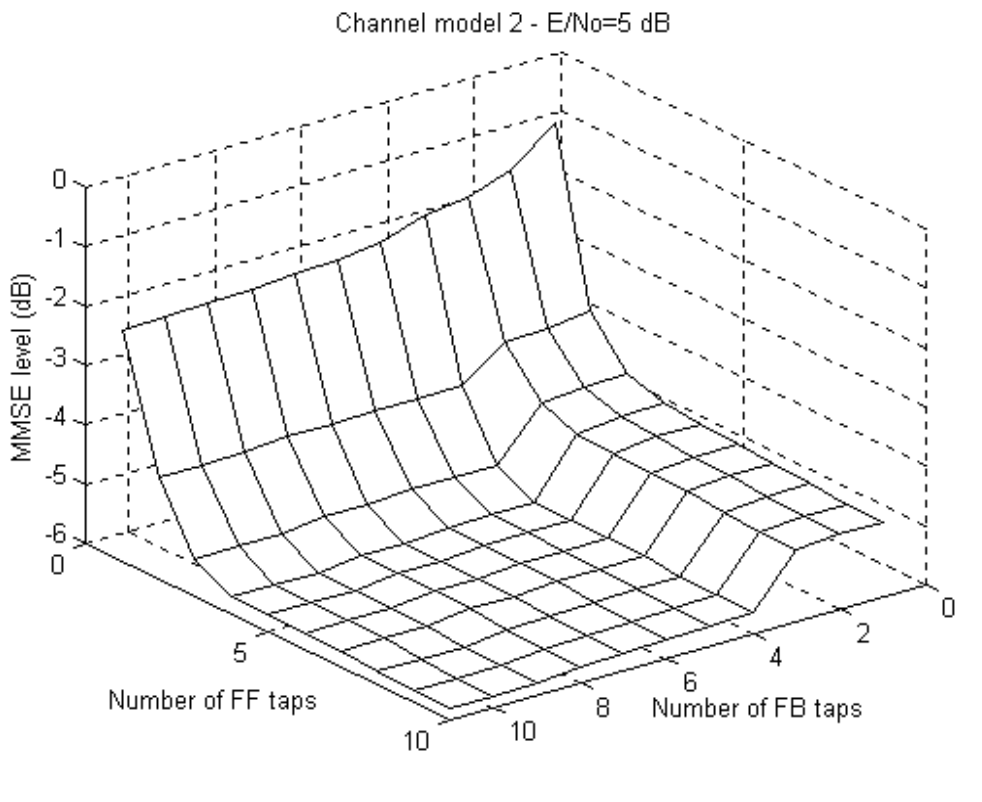

**Figure 6.28**: DFE-MMSE for Channel model 2(Static) for various $N_f$ and $N_b$. E/No=5dB.

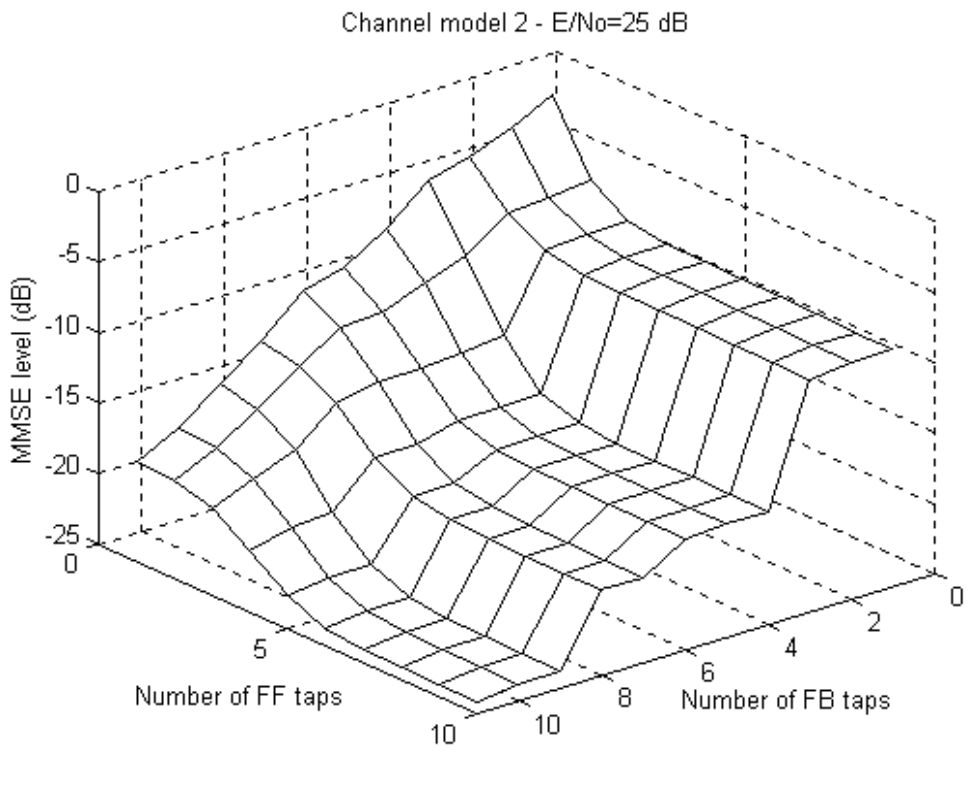

**Figure 6.29**: DFE-MMSE for Channel model 2 (Static) for various $N_f$ and $N_b$. E/No=25dB.

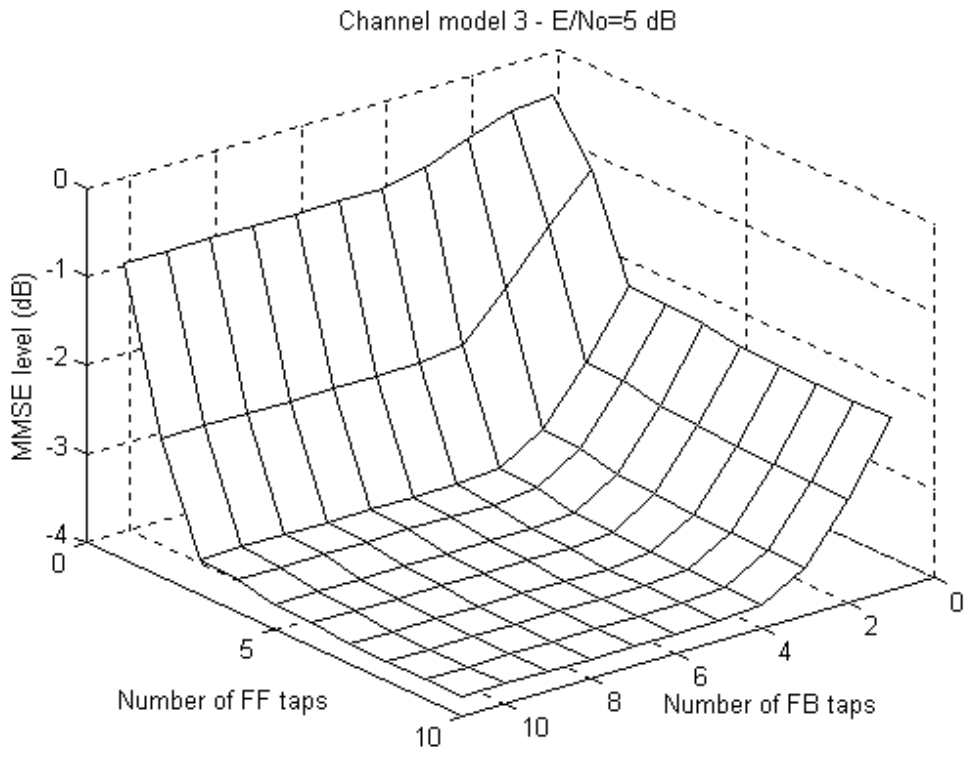

**Figure 6.30**: DFE-MMSE for Channel model 3 (Static) for various $N_f$ and $N_b$. E/No=5dB.

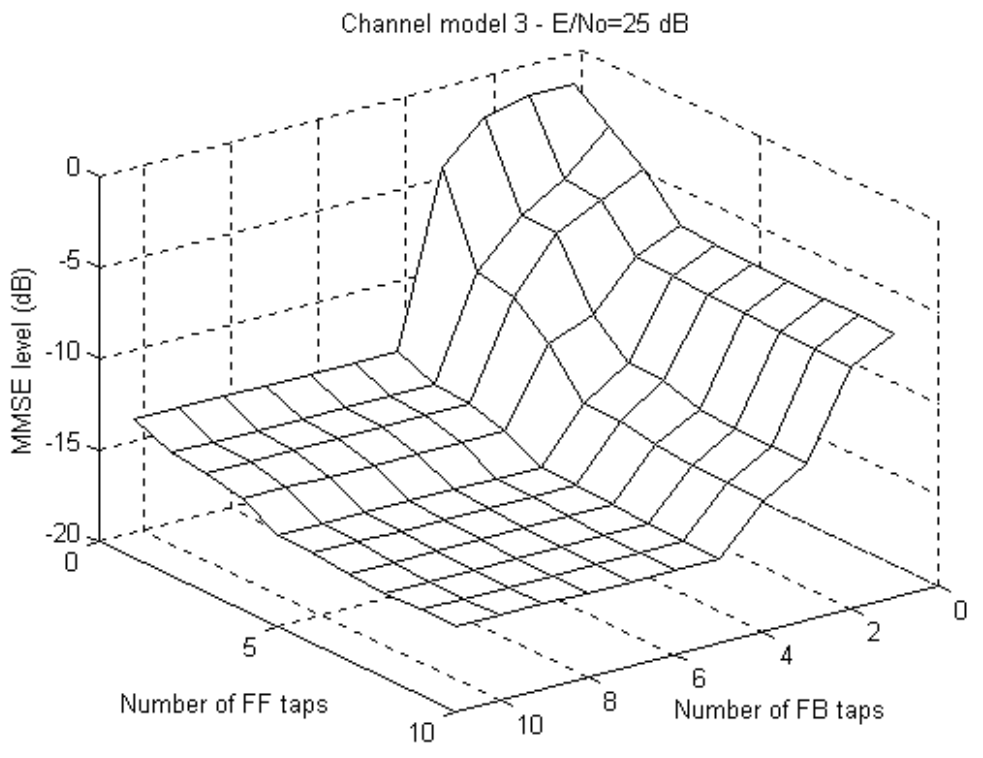

**Figure 6.31**: DFE-MMSE for Channel model 3 (Static) for various $N_f$ and $N_b$. E/No=25dB.

In order to get an idea of the variety of possible channels found in a mobile system, it is useful to look at the COST207 channel models (pp. 728 in [Steele99]) developed during the GSM study phase. The delay spread of these models varies between 0.5 $\mu$s in rural areas up to 20 $\mu$s in hilly terrain.

The number of taps required to cancel out all the postcursor interference will obviously depend on the signalling rate, however it should be clear that the spread of $ISI_{post}$ is very variable. This suggests that in a DFE, the FBF is more suitable for reconfiguration given its potentially wide range of taps whereas a short fixed-length FFF is able to cope with most of the practical scenarios.

# 6.10 Motivations for a DFE with variable length FBF

Given that the required number of feedback taps to cancel out all the $ISI_{post}$ is deemed to be N-1, and that N depends very much on the particular environment, we could design a "worst-scenario" DFE with enough taps in the FBF so as to cancel the longest possible channel. This approach presents three problems. The most obvious and important inconvenience is the waste of resources this method implies, as in many situations the equaliser will be oversized.

The second problem arises from the use of adaptive algorithms such as the LMS and RLS to perform the coefficient update. These algorithms do not produce the exact optimum solution but instead, as is seen in equation (6.16) and (6.45), they introduce an excess mean squared error (EMSE). The FBF taps beyond N-1, which in theory should be zero, will not be zero in practice due to the use of a step size greater than zero (LMS) or a forgetting factor smaller than one (RLS). These excess taps introduce a third source of ISI, which we call Excess ISI energy and is given by:

$$ISI_{excess} = \sum_{i=N}^{Nb} w_i^b(n)^2 \qquad (6.53)$$

Equation (6.53) indicates that excess taps (any tap beyond N-1) not only do not decrease the steady state MSE but in fact increase it. Typically $ISI_{excess}$ will be small but it can be significant in rapidly varying scenarios where the LMS algorithm will use a large step size and the RLS algorithm a small forgetting factor. Figures 6.32 and 6.33 show the steady state MSE achieved when using a DFE with LMS and RLS respectively to equalise Channel model 3 with $\sigma_q^2=10^{-4}$ when using different FBF lengths. The FFF length was fixed at six taps. Theoretical results (see previous section) predict that 4 taps are necessary to cancel all the $ISI_{post}$. In both figures can be seen that the minimum steady state MSE is achieved, as predicted, when the FBF has four taps. Making the FBF longer than 4 taps, does not decrease the MSE but on the contrary, the curves show slight increases in the MSE level for larger $N_b$ values as a result of the $ISI_{excess}$.

Lastly, there is another reason to keep the FBF as short as possible which cannot be appreciated when just looking at the MSE, that is, error propagation. Error propagation has been briefly mentioned in section 2.2.2. Many researchers have studied this phenomenon and have found bounds on the performance of a DFE when taking into account error propagation ([Duttweiler74], [Altekar93], [Smee98]). Some solutions have been proposed to mitigate its effects ([Fertner98]).

Erroneous symbols in the FBF reduce the noise margin for future decisions, increasing in this way the probability of error. The classic paper of [Duttweiler74] shows that in high E/No, the error probability is multiplied by at most a factor of $2^{Nb}$ relative to the error probability in the absence of decisions errors. In [Gitlin92] an intuitive explanation of this result is presented on the basis of the following three facts:

1. All errors are cleansed from the FBF when $N_b$ consecutive corrections are made.
2. The probability of making an error is no bigger than ½.
3. Let K denote the number of symbols needed to make $N_b$ correct decisions (K= error propagation length). Since a single error produces on average K/2 errors, the average error rates is $(K/2)P_0$ where $P_0$ is the probability of making an error when all the symbols in the FBF are correct.

It is clear that if the FBF is unnecessarily long (i.e. longer than N-1 taps), the potential error propagation will also be increased.

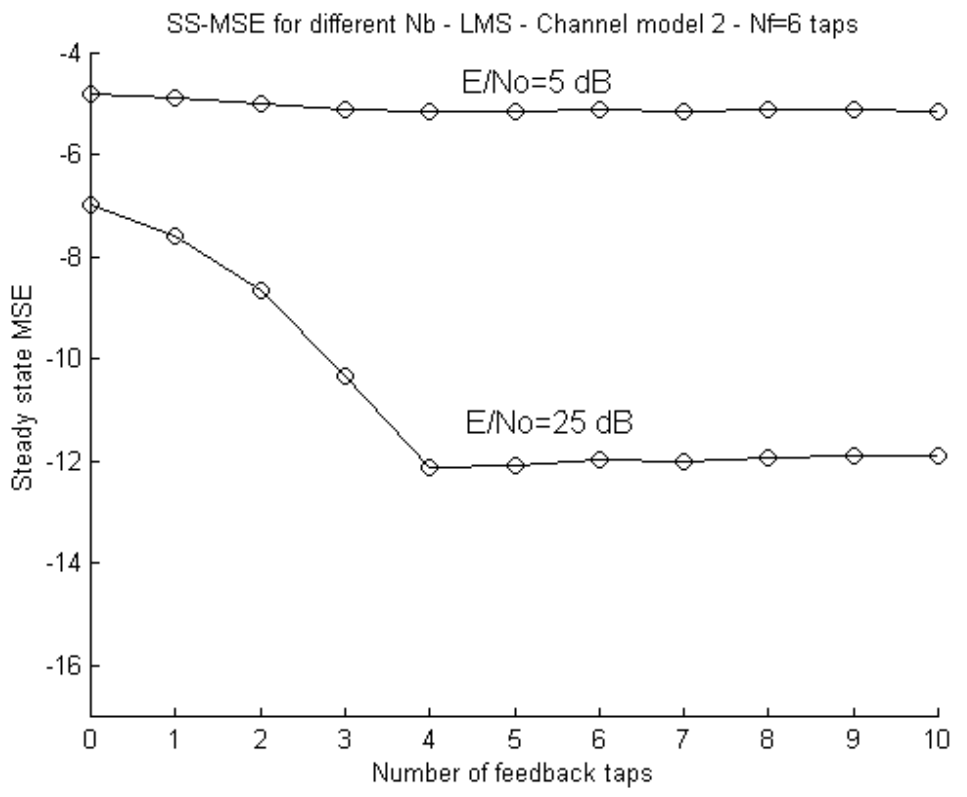

**Figure 6.32**:SS-MSE using LMS for Channel 3 ($\sigma_q^2=10^{-4}$) for various $N_b$. $N_f$=6. $\mu$=0.01.

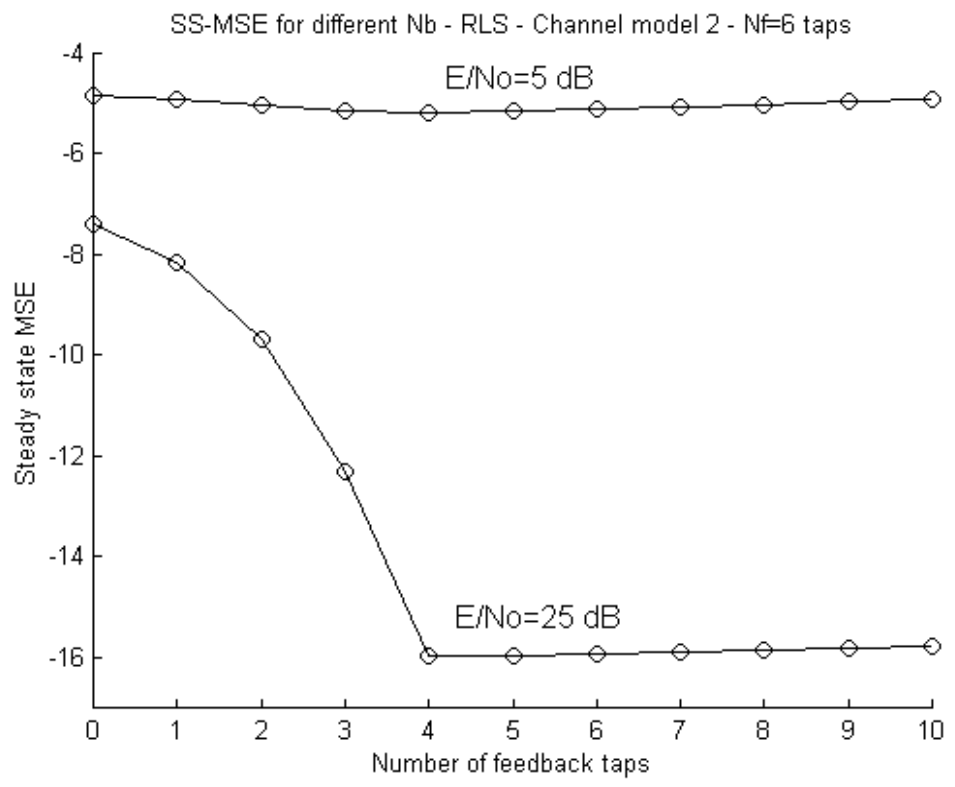

**Figure 6.33**: SS-MSE using RLS for Channel 3 ($\sigma_q^2=10^{-4}$) for various $N_b$. $N_f$=6. $\lambda$=0.99.

From previous paragraphs, it seems desirable to be able to adjust the length of the feedback section depending on the instantaneous channel characteristics. This would help to avoid the unnecessary update of taps (reduction in computational complexity), eliminate the $ISI_{excess}$ component and reduce the potential error propagation.

# 6.11 FBF length update algorithm

The algorithm controlling the length of the FBF filter, as in the case of the linear equaliser length update, should have two properties: first, it should only use information which is available at the receivers and secondly, it should be computationally simple. The solution we propose is to use the following test iteratively to control the length of the FBF:

$$\textbf{if} \quad E[w^b_{Nb}(n)^2] > \chi MSE(n) \Rightarrow N_b = N_b + 1 \qquad (6.54)$$

$$\textbf{if} \quad E[w^b_{Nb}(n)^2] < \chi MSE(n) \textbf{ and}$$
$$E[w^b_{Nb-1}(n)^2] < \chi MSE(n) \Rightarrow N_b = N_b - 1 \qquad (6.55)$$
$$\text{with} \quad \chi << 1, \quad \text{typically} \quad 0.1 > \chi > 0.001$$

The heuristic of the algorithm is very simple: if the power of the last active tap in the FBF is significant with respect to the MSE level, an extra tap is added. If the power of the last two taps in the FBF is significantly smaller than the MSE level, the last tap in the FBF is switched off.

The idea of the algorithm is to detect the length of the CIR-FFF combined response and adjust the FBF length appropriately. The parameter $\chi$ acts like a "tuner" to control whether a tap power is significant or not. If $\chi$ is very small, for example $\chi$ =0.001, the algorithm tends to enlarge the FBF so as to cancel all postcursors of the CIR-FFF response. When $\chi$ is large, for example $\chi$ =0.1, the algorithm considers the small components of the CIR-FFF response as noise and therefore the FBF is kept short. The algorithm is based on using the last two active taps of the FBF to estimate the length of the postcursor ISI. After convergence and recalling the definition in (6.49):

$$w^{b\,2}_i \cong z^2_{D+1+i} \quad i > 0 \qquad (6.56)$$

Suppose that $N_b$>N-1, then any $z_i$ with i>N-1 will be zero and the corresponding FBF tap $w^b_i$ should, in theory, be zero too. In practice due to the EMSE, these $w^b_i$ (i>N-1) will not be zero. The algorithm detects the point from which the $w^b_i$ are small compared to the steady state MSE level and it switches off the FBF taps from that point onwards as they are considered excess taps.

It is important to recognise that the algorithm always uses one more tap than those considered as "true cancelling" taps. This additional coefficient serves as a watchdog to detect when the CIR is getting longer. When this happens, this last tap will start to have some significant value with respect to the MSE and an extra tap is added.

The problem with the test/algorithm given by equations (6.54) and (6.55) is that the true ensemble values of the FBF taps will not be available in practice. However, assuming ergodicity of the tap values, the ensemble averages can be substituted by time averages. Given that the environment is likely to change, the time averages are performed over a time window of a certain number of samples. The practical version of the FBF length control algorithm is given by:

*Parameters*: Nb = Current FBF length

W = Averaging window length

*For each received symbol do:*

$$\text{TapPower}_{Nb}(n) = \text{TapPower}_{Nb}(n-1) + w^{b}_{Nb}(n)^2 \quad (6.57)$$

$$\text{TapPower}_{Nb-1}(n) = \text{TapPower}_{Nb-1}(n-1) + w^{b}_{Nb-1}(n)^2 \quad (6.58)$$

*When* n=kW (k *is an integer*):

$$\text{if} \quad \text{TapPower}_{Nb} > W\,\chi\,\text{MSE}(n) \Rightarrow N_b = N_b + 1 \quad (6.59)$$

$$\begin{aligned}&\text{if} \quad \text{TapPower}_{Nb} < W\,\chi\,\text{MSE}(n) \quad \textbf{and} \\ &\quad \text{TapPower}_{Nb-1} < W\,\chi\,\text{MSE}(n) \Rightarrow N_b = N_b - 1\end{aligned} \quad (6.60)$$

*Restart* $\text{TapPower}_{Nb}$, $\text{TapPower}_{Nb-1}$

**Algorithm 6.1**: FBF length update algorithm.

The algorithm given by (6.57)-(6.60) is the temporal and windowed version of the one given by equations (6.54) and (6.55). Notice that divisions are avoided in the algorithm by making the comparisons of the accumulated tap power with a scaled version of the MSE (scaled by W). The length of the window, W, will depend on the type of environment in which the receiver has to operate. In the context of mobile systems where variations may take place in a very rapid and sudden way, it is advisable to use short windows. In the simulations presented in the next section, W was set to 150 samples.

As a final point, notice that the computational complexity of the algorithm given by equations (6.57)-(6.60) is very low. Only (6.57) and (6.58) are computed for each symbol and each of them only performs one multiplication and one addition.

## 6.12 Simulations using a FBF with variable length

In this section, results obtained when using the FBF length update algorithm are presented. First, the algorithm is used to predict the adequate number of feedback taps to be used in a static unknown environment. In subsection 6.11.2 the behaviour of the algorithm is verified when there are sudden changes in the channel impulse response. Finally, in subsection 6.11.3 the response of the algorithm to sudden jumps on the signal level is investigated.

## 6.12.1 Static channels

The aim of this set of simulations was to check whether the algorithm was able to correctly estimate the right number of feedback taps to be used when confronted with an unknown static channel.

The simulations have been performed using the same channel models used previously. In theory, any adaptive algorithm can be used to perform the equaliser adaptation. Here, the LMS with a variable step size (VSLMS) was used. This algorithm has been shown to be numerically robust and offers a better performance than the conventional LMS. Of course, care must be taken when changing the FBF length to ensure that the stability conditions of the adaptive algorithm are satisfied. In the case of LMS type algorithms, this condition limits the maximum value of the step size (see chapters 2 and 3) which is a function of the equaliser length.

Stability when using a least squares algorithm (such as RLS) is covered in more detail in section 6.12. Two different E/No levels have been used, 5 and 25 dB. The parameter of the FBF length update algorithm, $\chi$, has been checked for a wide range of values and the choice offering the best performance is when $0.001<\chi<0.1$. The results shown here are for three different values of $\chi$ (0.1, 0.01 and 0.001). The FBF had an initial (minimum) length of two taps and could potentially expand up to 25 taps. The averaging window length of the algorithm, W, has been set to 150 samples, that is, every 150 samples the algorithm checks if the current FBF length is appropriate. The FFF was fixed to six taps in all simulations. The decision delay in all situations has been set to $D=N_f-1=5$. As in nearly all the simulations in this work, the curves shown are the average of 30 independent runs for each set of parameters.

Figures 6.34 and 6.35 show the results of the FBF length evolution when equalising Channel model 1. Recall that this channel had only two taps (see chapter 3). According to section 6.8, the FBF needs only one tap to cancel out all the postcursors of the CIR-FFF combined response. Notice in figures 6.34 and 6.35 that in both situations, E/No=5 dB and E/No=25 dB, the FBF attains a length of two taps (or very close). This extra tap results from the strategy of the FBF length update algorithm of using an extra tap as a detector to identify when the number of postcursors is increasing.

Note also that when $\chi$ has a small value such as $\chi$=0.001, the FBF length presents some oscillations. These fluctuations are due to the high sensitivity of the algorithm when $\chi$ is so small. Finally, notice also that in this case, $\chi$=0.001, for both E/No levels the FBF length is on average slightly greater than two taps (around 2.3 for E/No=5dB). This is again due to the high sensitivity of the algorithm, which sometimes considers the noise in the last tap as a true postcursor and therefore increases the length. This effect is more obvious for the low E/No level as the noise samples are much larger. Nonetheless this simulation shows that even in the most extreme situations (high sensitivity, large noise) the algorithm can still determine the right FBF length with very good accuracy.

In figures 6.36 and 6.37, the same results are presented when equalising Channel model 2. In this case, and given that the channel has 11 taps, the FBF should have length 10 to cancel all postcursor ISI. These simulations illustrate very clearly the effect of the parameter $\chi$. When $\chi$=0.001, the FBF is expanded so as to cancel all, or almost all, postcursor ISI. For E/No=25, the FBF length is set to 11 taps, that is 10 cancelling taps plus the additional detector coefficient. Comparing the lengths attained by the FBF for both E/No levels for any value of $\chi$, it can be observed that for the low E/No, the FBF is always kept shorter than for the larger E/No.

It is interesting to compare the curves for $\chi$=0.01 at both E/No levels with the graphs in figures 6.28 and 6.29. The FBF lengths achieved when operating the algorithm with $\chi$=0.01 are 4 and 10 taps for 5 and 25 dB respectively. Looking at figures 6.28 and 6.29 for $N_f$=6 taps, it can be seen that these $N_b$ values correspond to the lengths where any further FBF enlargement produces very marginal MSE reductions. On the other hand, when $\chi$=0.001 the FBF is always kept long, and when $\chi$=0.1 the FBF is always kept short.

In figures 6.38 and 6.39 the corresponding learning curves (MSE) for figures 6.36 and 6.37 are shown. It is clear from these pictures that from a performance point of view it is best to use always a small $\chi$ such as 0.001 as in this way the lowest and fastest-converging MSE is achieved. However, there might be situations, especially in a mobile handset, where power consumption needs also to be maximised. In this case, using a value such as $\chi$=0.01 might be more appropriate as, although it provides a slightly degraded MSE performance, it offers important savings in the number of computations due to the use of much shorter FBF filters.

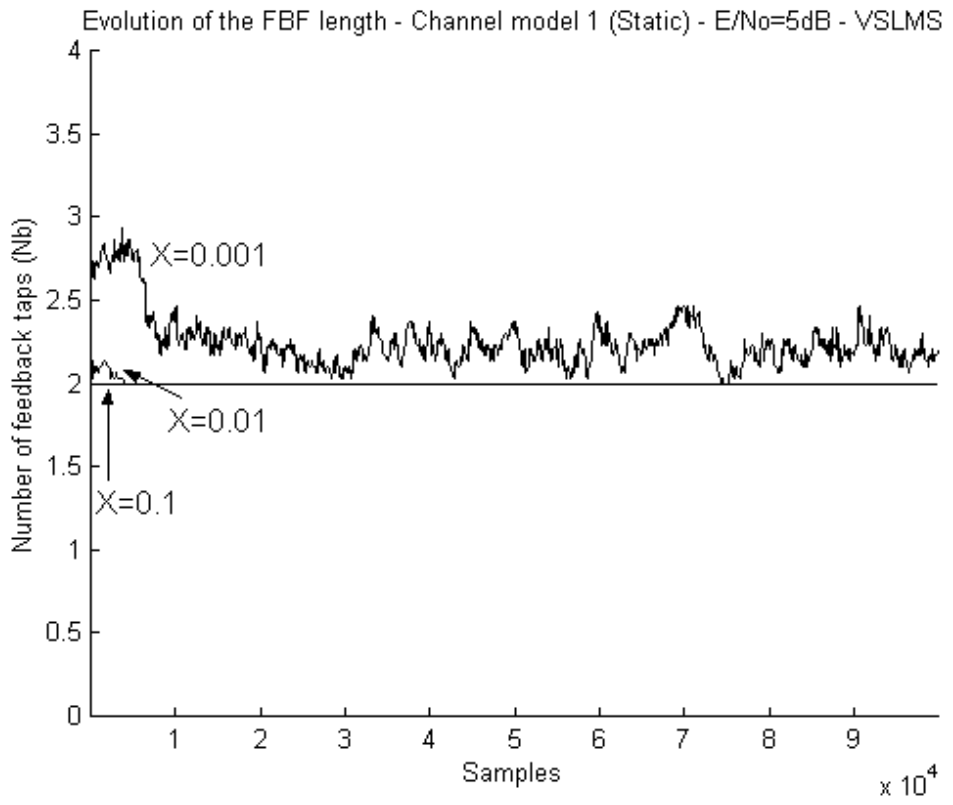

**Figure 6.34**: FBF length evolution using VSLMS for Channel model 1 (Static). E/No=5dB.

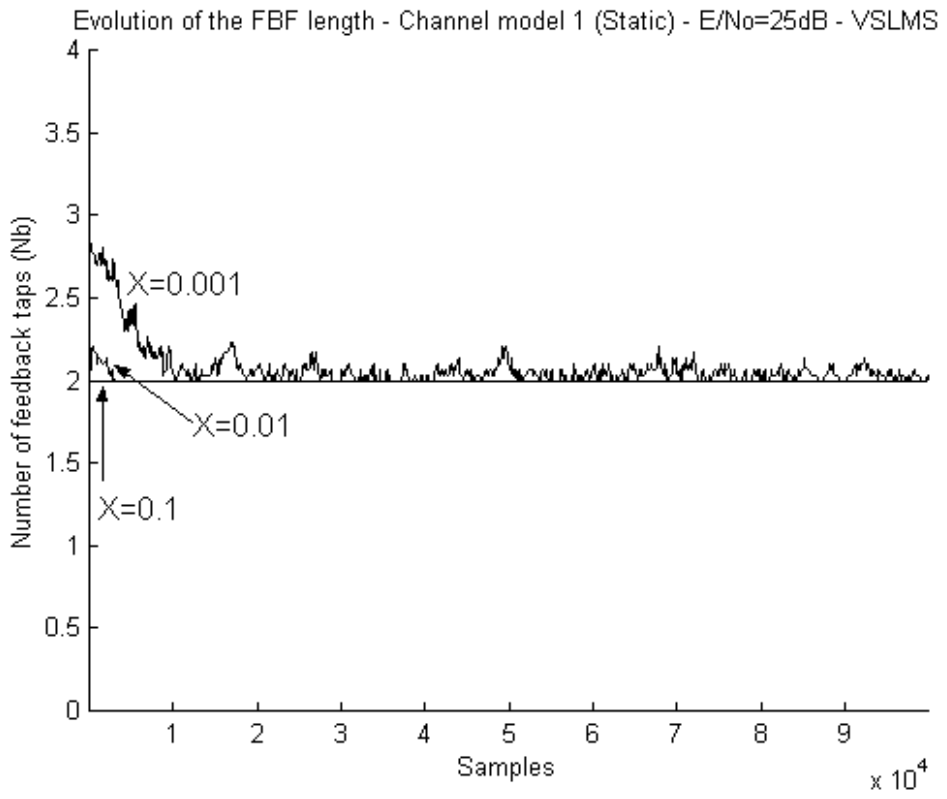

**Figure 6.35**: FBF length evolution using VSLMS for Channel model 1 (Static). E/No=25dB.

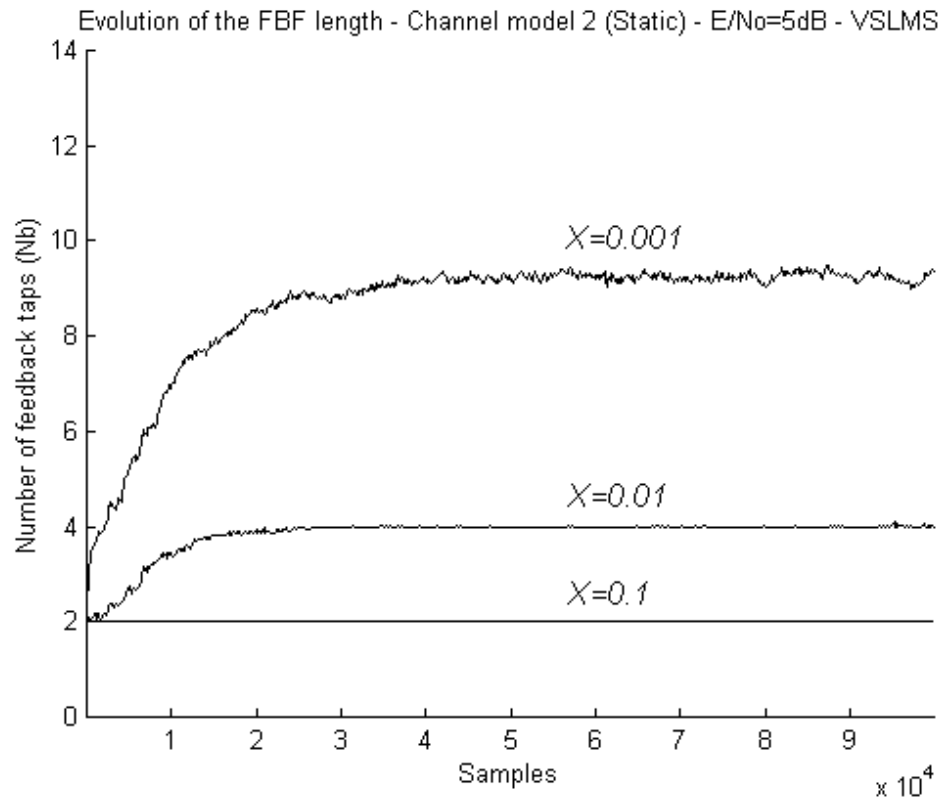

**Figure 6.36**: FBF length evolution using VSLMS for Channel 2 (Static). E/No=5dB.

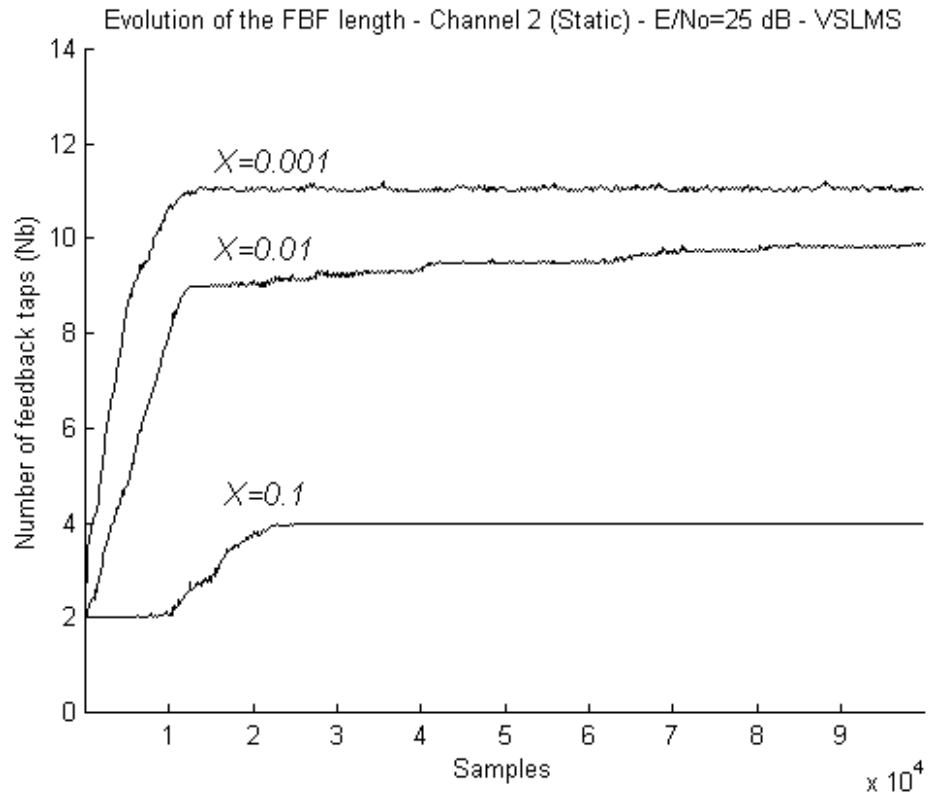

**Figure 6.37**: FBF length evolution using VSLMS for Channel 2 (Static). E/No=25dB.

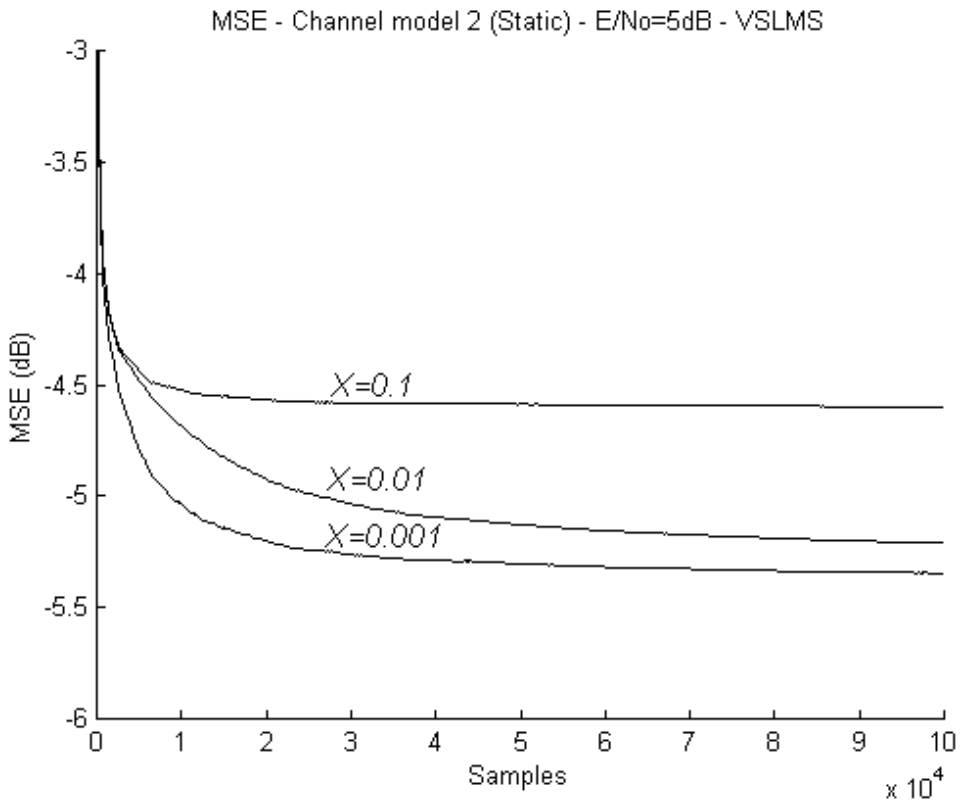

**Figure 6.38**: MSE evolution using VSLMS for Channel 2 (Static). E/No=5dB.

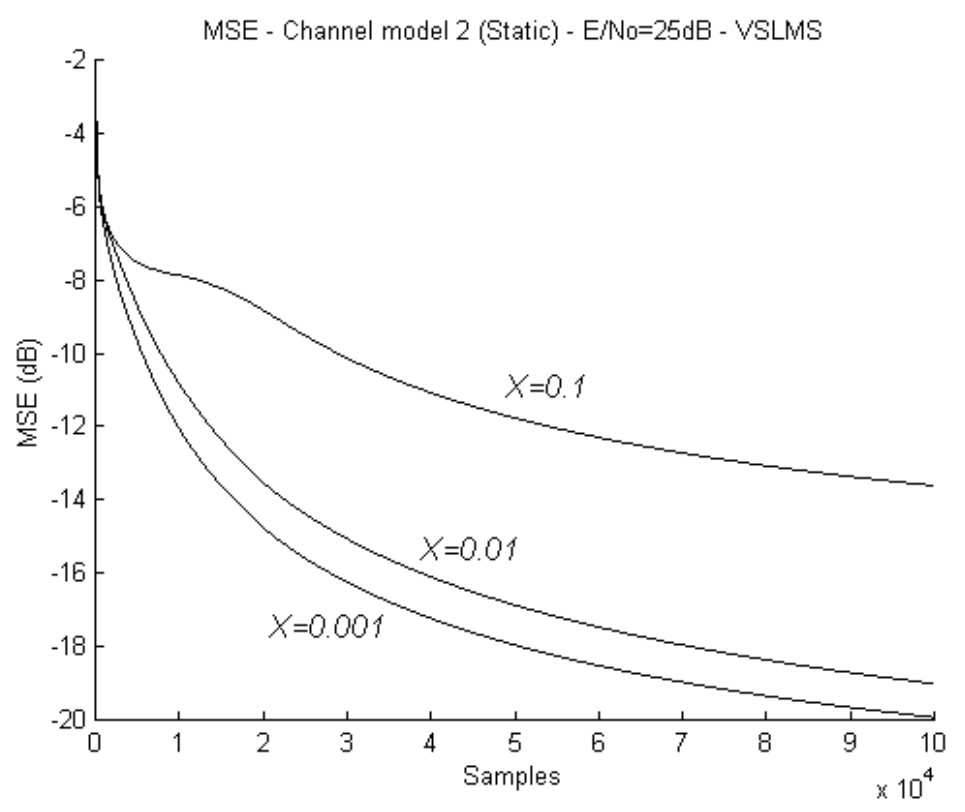

**Figure 6.39**: MSE evolution using VSLMS for Channel 2 (Static). E/No=25dB.

Finally, the length update algorithm has been checked with Channel model 3. For this channel the optimum FBF should have four taps, so in theory the algorithm should set the FBF length to five taps (four cancelling taps + one detector tap). Figure 6.40 shows the results for E/No=5dB. For medium and large $\chi$ values, the length is set to 3-4 taps as the last postcursors of the CIR-FFF combined response are regarded as noise terms.

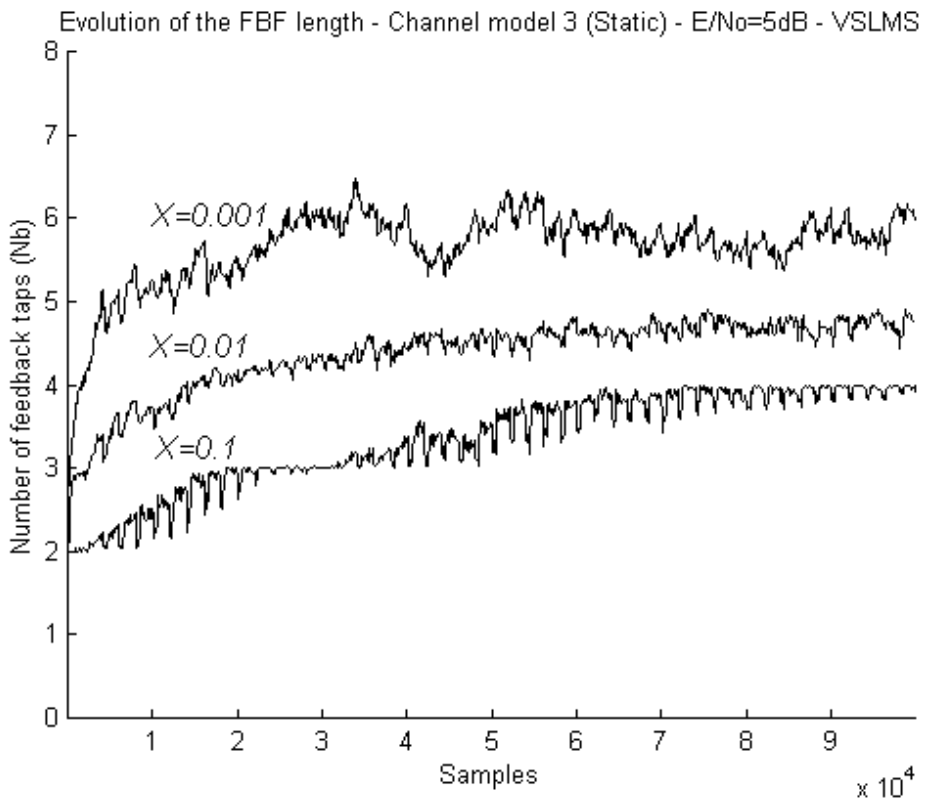

**Figure 6.40**: FBF length evolution using VSLMS for Channel 3 (Static). E/No=5dB.

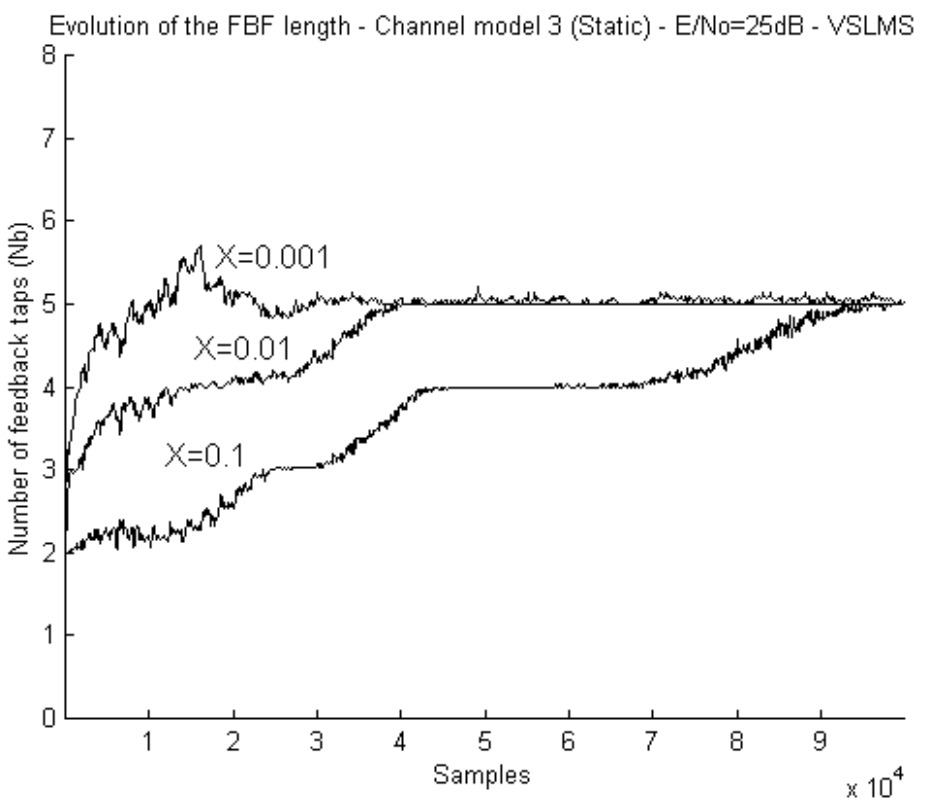

**Figure 6.41**: FBF length evolution using VSLMS for Channel 3 (Static). E/No=25dB.

When $\chi$ takes a small value, the algorithm sets the FBF length to an average of around 5.5 taps. This slight increase with respect to the optimum length is due to the high sensitivity of the algorithm, which is accentuated by the resulting large steady state MSE level when equalising this particular channel with a very low E/No. When E/No=25 dB (figure 6.41), any $\chi$ value makes the FBF enlarge up to five taps. Notice that the larger $\chi$ is, the longer it takes the FBF to achieve the definitive length. This illustrates the fact that the choice of $\chi$ influences both the length attained and the amount of time required to get to that length.

From the results presented in this subsection, it can be concluded that the FBF length update algorithm is able to predict, with very good accuracy, the number of taps to be used in an unknown environment.

## 6.12.2 Abrupt changes in the channel profile

In this section, the behaviour of the FBF length update algorithm is investigated when confronting sudden changes in the channel impulse response. It is with this type of scenario that the technique has proven to be most useful, as the FBF is able to adjust its length according to the current channel spread characteristics. The types of changes simulated here have the same form as those already shown in section 5.4.5 which follow the pattern of changes given by:

1) Channel model 2-Channel model 1-Channel model 2
2) Channel model 2-Channel model 3-Channel model 2

As in the previous section, the VSLMS algorithm is used to perform the coefficient adaptation although any of the other adaptive algorithms could have been used provided the respective stability conditions are met.

The results are compared with those obtained using a fixed 11-tap feedback filter in terms of performance and number of computations. The simulation length was set to 180,000 samples with the abrupt changes taken place at iterations 60,000 and 120,000. As in previous sections dealing with abrupt changes and given its wide variance, the MSE is windowed over a temporal length of 2,000 samples. In all situations the FFF filter has been set at 6 taps and remains fixed during all simulations.

Figures 6.42-6.49 summarises the results obtained for the succession of changes given by Channel 2-Channel 1-Channel 2. The evolution of the FBF length when E/No=5 dB is shown in figure 6.42. Notice that when operating in Channel model 2, the FBF length, depending on the setting of $\chi$, takes the same values as those shown in figure 6.34. When the channel model changes to Channel model 1, the algorithm is able to recognise the shorter CIR and reduces the FBF length to just two taps. Notice that this happens for any value of $\chi$.

In figure 6.43 the corresponding MSE curves are shown. During the initial interval in Channel model 1, the performance for the different values of $\chi$ can be appreciated. In the same graph the performance of an 11-tap fixed equaliser is also shown as a dashed line. This curve is difficult to see due to the fact that is nearly coincident with the variable length FBF with $\chi$=0.001 curve.

It is crucial to notice that the performance when the system confronts Channel model 1 is the same for all equalisers. In particular, the 11-tap fixed equaliser does not perform any better than any of the variable length ones, which during this interval are reduced to just two FBF taps. This is the main benefit of the FBF length update algorithm: the potential savings in computation when the channel spread is short.

Another point to highlight is the flexibility offered to the designer of various options trading performance with computational complexity. For example, it may seem reasonable in this case to use $\chi$=0.01 as, although there is some performance degradation with respect to the $\chi$=0.001 case, this greatly reduces the number of the FBF taps when operating in Channel model 2. Figures 6.44 and 6.45[21] show exactly the same results but now for E/No=25 dB. The same conclusions as for the 5 dB situation can be drawn.

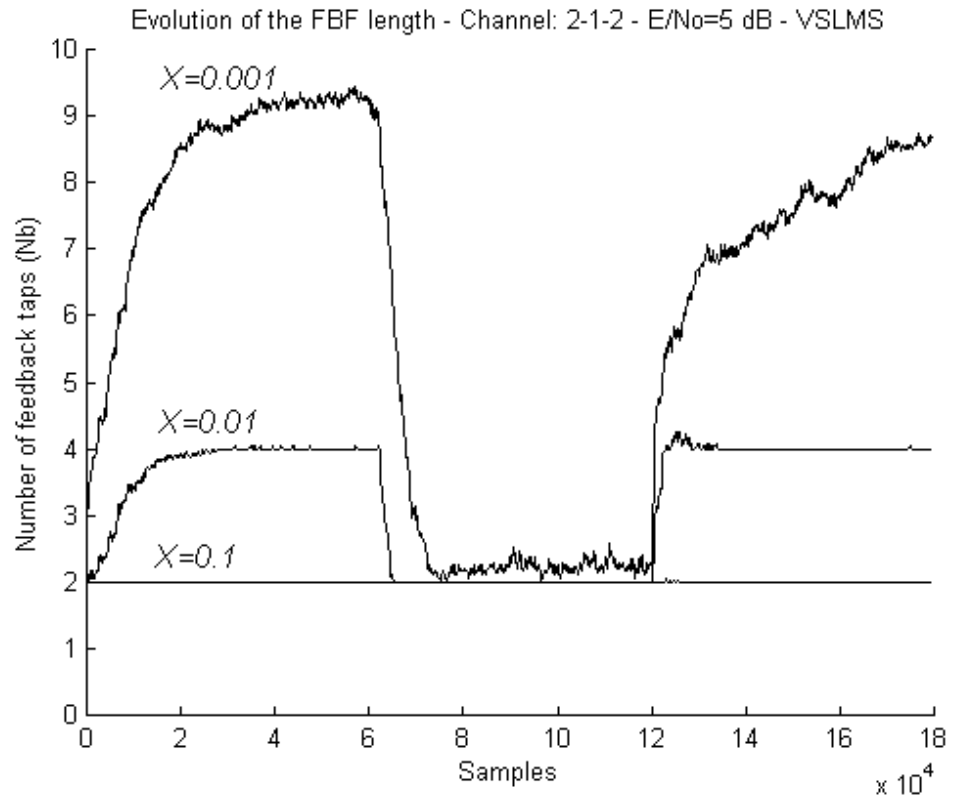

**Figure 6.42**: FBF length evolution using VSLMS for abrupt channel change: channel 2-channel 1-channel 2. E/No=5dB.

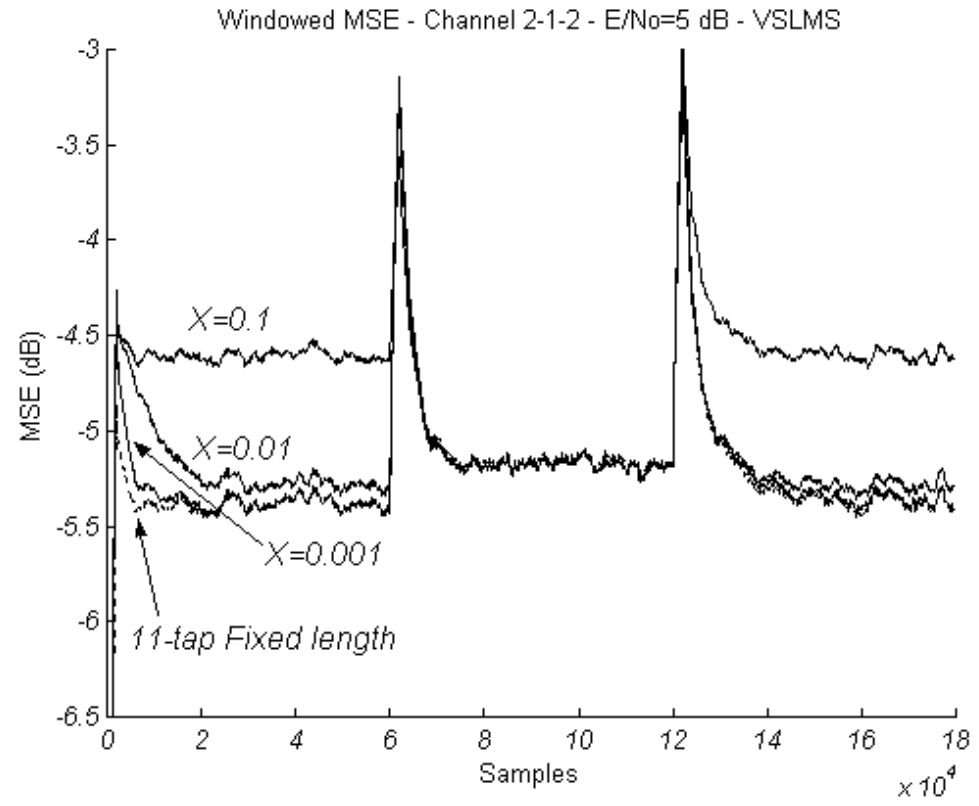

**Figure 6.43**: Windowed MSE using VSLMS for abrupt channel change: channel 2-channel 1-channel 2. E/No=5dB.

---

[21] In figure 6.42, curve for $\chi$=0.01 is not shown as it lied on top of the $\chi$=0.001 and fixed equaliser curves.

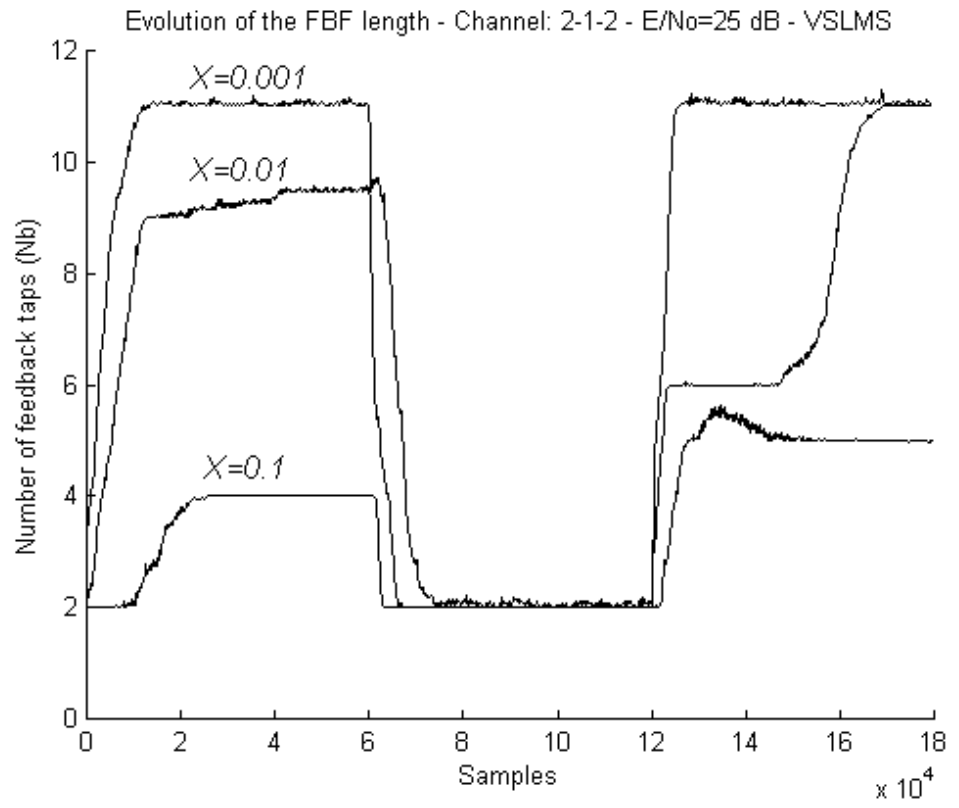

**Figure 6.44**: FBF length evolution using VSLMS for abrupt channel change: channel 2-channel 1-channel 2. E/No=25dB.

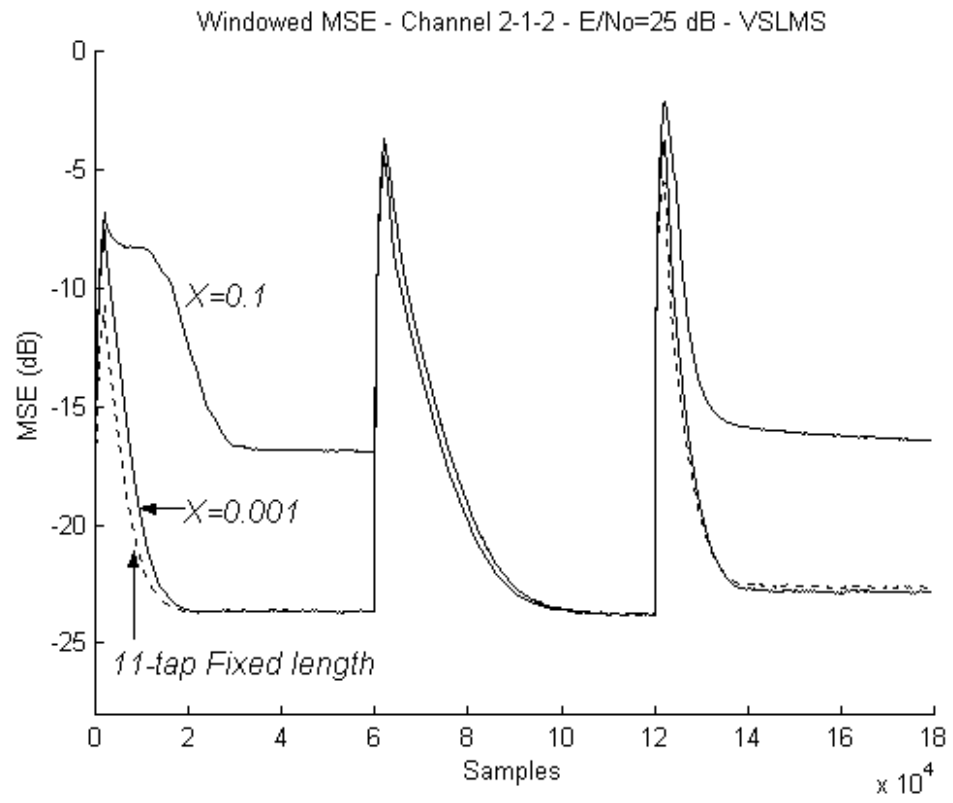

**Figure 6.45**: Windowed MSE using VSLMS for abrupt channel change: channel 2-channel 1-channel 2. E/No=25dB.

In general, it is impossible to quantify the computational savings obtained from the use of this algorithm as this depends very much on the particular channel in which the equaliser operates and also on the length of the fixed FBF used for the comparison.

Just to give an idea of the reductions achieved in a particular scenario, figures 6.46 and 6.47 show the number of products performed over the whole previous simulation (figures 6.42-6.45) for both E/No scenarios when using variable length FBF and an 11-tap fixed equaliser. In both E/No levels, all variable length FBF equalisers perform fewer products than the fixed length FBF equaliser by reducing the FBF length when operating in Channel 1. Even when $\chi$=0.001, the variable length FBF achieves a reduction of nearly 15% in the number of products required with respect to the fixed equaliser without incurring any performance penalty (see MSE curves in figures 6.43 and 6.45). These reductions are much more significant if a moderate MSE increase is allowed.

Using the FBF length update algorithm with $\chi$=0.01 reduces the number of products by around 35% with respect to the 11-tap fixed equaliser. Notice also that the computational savings are more evident when E/No is low, as the algorithm is able to recognise this condition and keep the FBF shorter. As can be seen in figures 6.48 and 6.49, similar reductions are also achieved in the number of additions performed over the whole simulation.

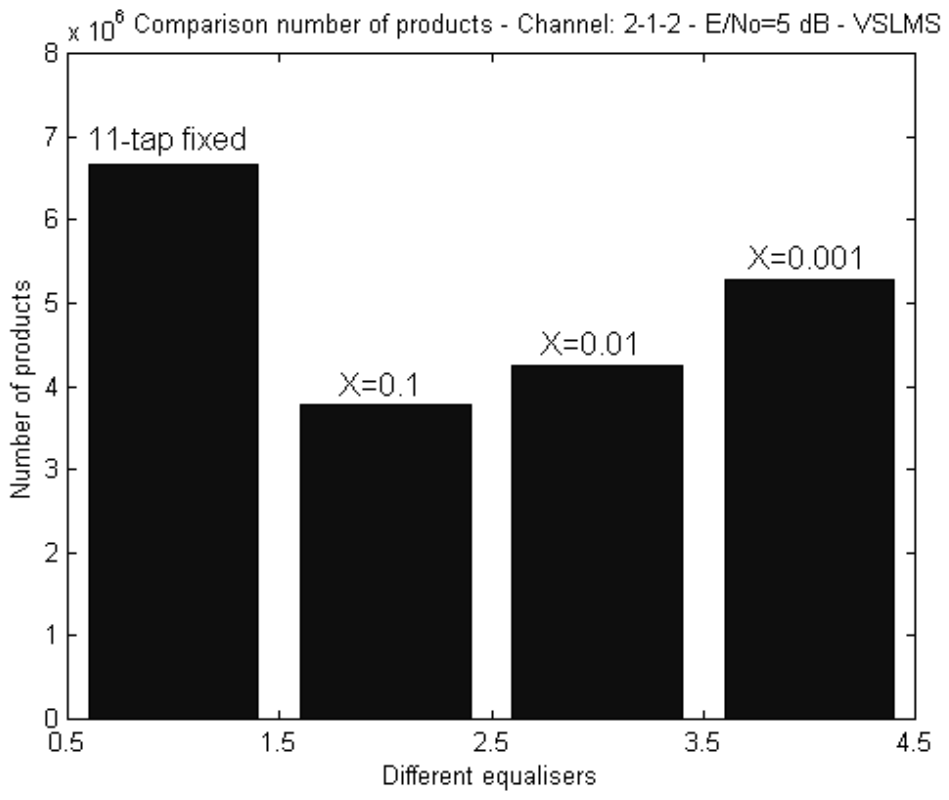

**Figure 6.46**: Comparison of the number of products for abrupt channel change: Channel 2 - Channel 1 – Channel 2. E/No=5dB.

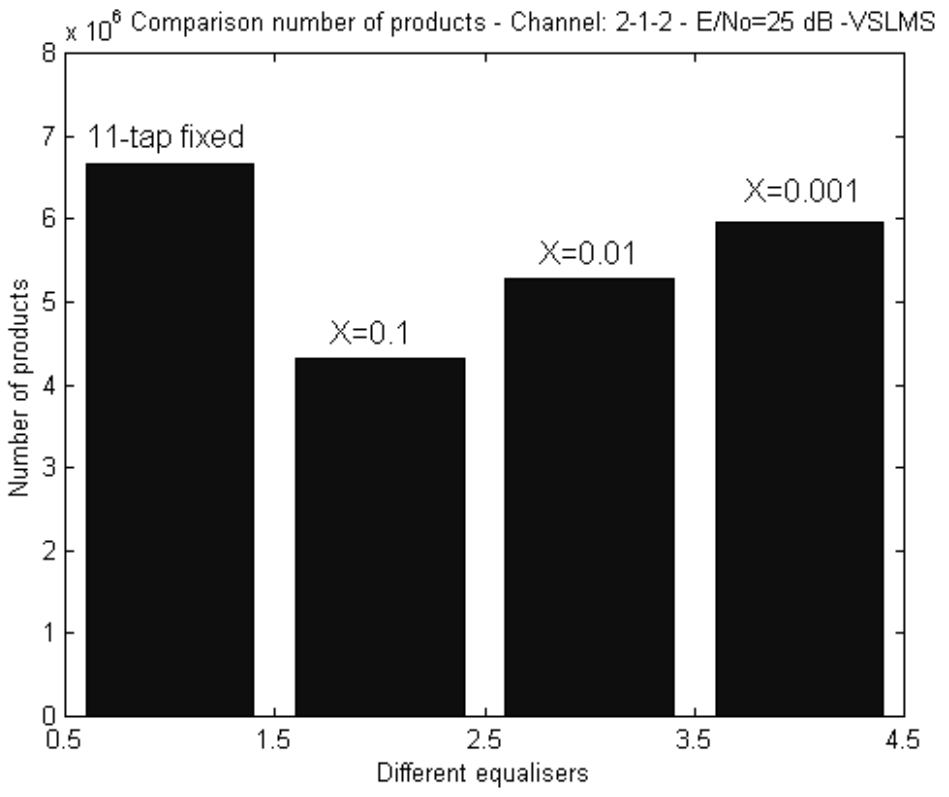

**Figure 6.47**: Comparison of the number of products for abrupt channel change: Channel 2 - Channel 1 – Channel 2. E/No=25dB.

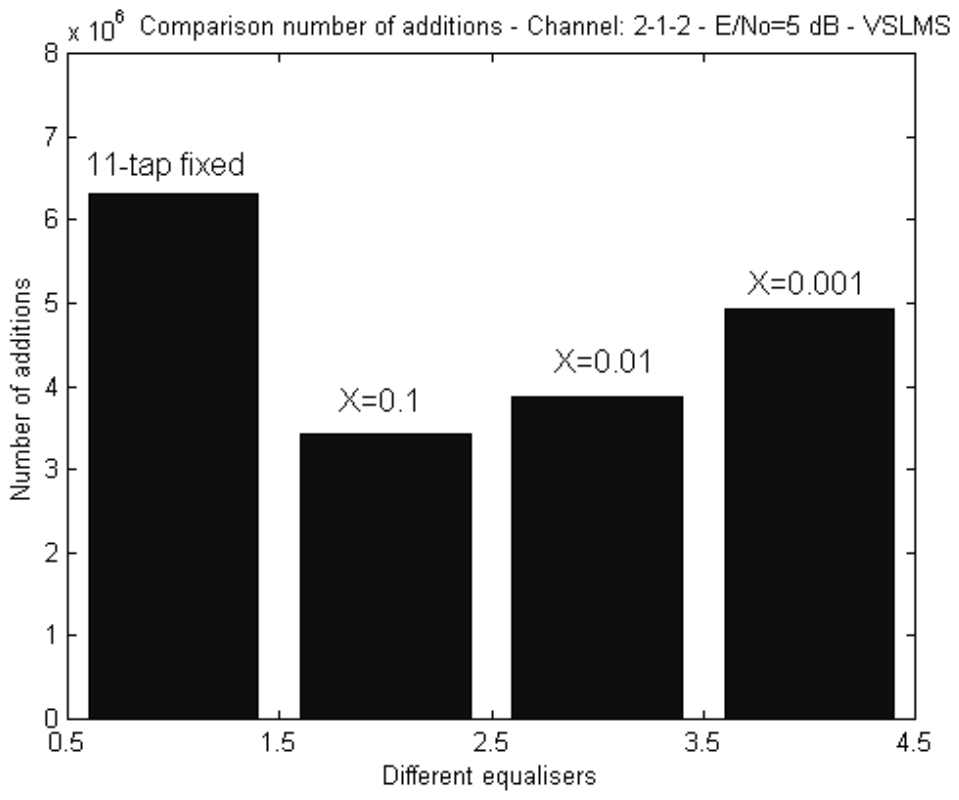

**Figure 6.48**: Comparison of the number of additions for abrupt channel change: Channel 2 - Channel 1 – Channel 2. E/No=5dB.

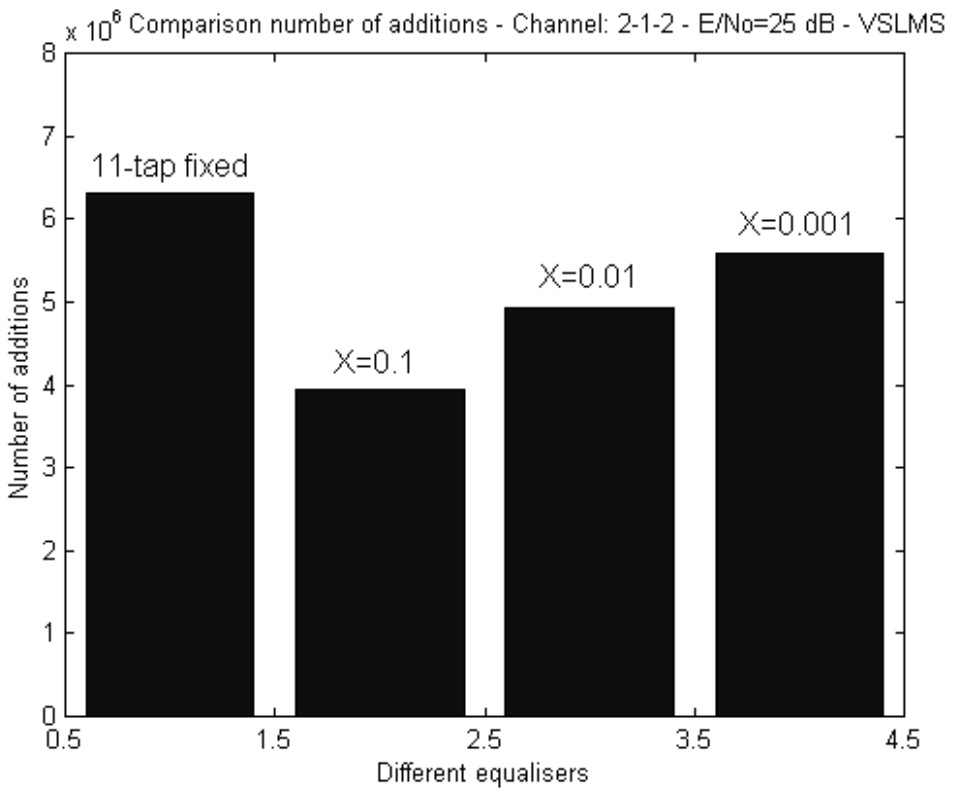

**Figure 6.49:** Comparison of the number of additions for abrupt channel change: Channel 2 - Channel 1 - Channel 2. E/No=25dB.

The FBF length update algorithm has also been tested with a different profile of abrupt changes, given by: Channel model 2-Channel model 3-Channel model 2. Simulations have been run for E/No=5 dB and E/No=25 dB. Results are summarised in figures 6.50 to 6.53. During the first 60,000 iterations the FBF length and MSE curves are exactly the same as in the previous profile as in both cases the systems starts operating in Channel model 1. Now the abrupt change causes the channel to be transformed to Channel model 3.

In figure 6.50, when E/No=5 dB, it can be seen that when $\chi$=0.001, the FBF length update algorithm recognises this change to a channel profile with a shorter delay spread and starts to decrease the FBF length. It should be noted however, that the FBF reduction takes place very slowly. This is because the steady state MSE when equalising Channel model 2 is fairly large, a condition accentuated by the low E/No in this particular case. This extremely large MSE (notice in the graph the MSE level is about –1.5 dB) slows down the length adjustment process.

The curve for $\chi$=0.1 in figure 6.50 shows another interesting property of the algorithm. After the transition to Channel model 3, the FBF grows. This may seem inconsistent with the fact that the channel delay spread is getting shorter. The explanation is that, although the new channel is shorter, its postcursor components are now stronger and so the FBF expands (compare power profiles for Channel models 2 and 3 in chapter 3).

Figures 6.52 and 6.53[22] present the same results for E/No=25 dB. It can clearly be seen that after the transition to Channel model 3, all variable length equalisers are adjusted to 5 taps (4 cancelling taps plus the detection tap). Now the FBF length adjustment is far quicker than before due to the large E/No level.

There is one final comment to be made on this set of simulations regarding the MSE performance of the 11-tap fixed equaliser. Looking at figures 6.51 and 6.53, it can be observed that during the interval of samples [60,000-120,000] corresponding to Channel model 3, the 11-tap fixed FBF achieves the same MSE as the better adjusted and much shorter variable length FBFs. In the previous chapter on the variable length LE, it was noticed that when the equaliser was overdimensioned, its performance became worse due to an increase in EMSE. This effect does not happen in the feedback filter as it operates with noise free samples greatly reducing the increase in EMSE even if the FBF is very long. Additionally the use of the VSLMS algorithm, which tends to reduce the step size in the steady state phase, helps in keeping the $ISI_{excess}$ (equation 6.53) to an insignificant level.

---

[22] As in the previous profile, in figure 3.50 the MSE curve for $\chi$=0.01 is not shown as it overlapped with some of the other curves.

From both profiles of abrupt channel changes it can be inferred that the algorithm given by equations (6.57)-(6.60) is able to determine which is the most appropriate number of taps to be used in the FBF according to the instantaneous channel profile.

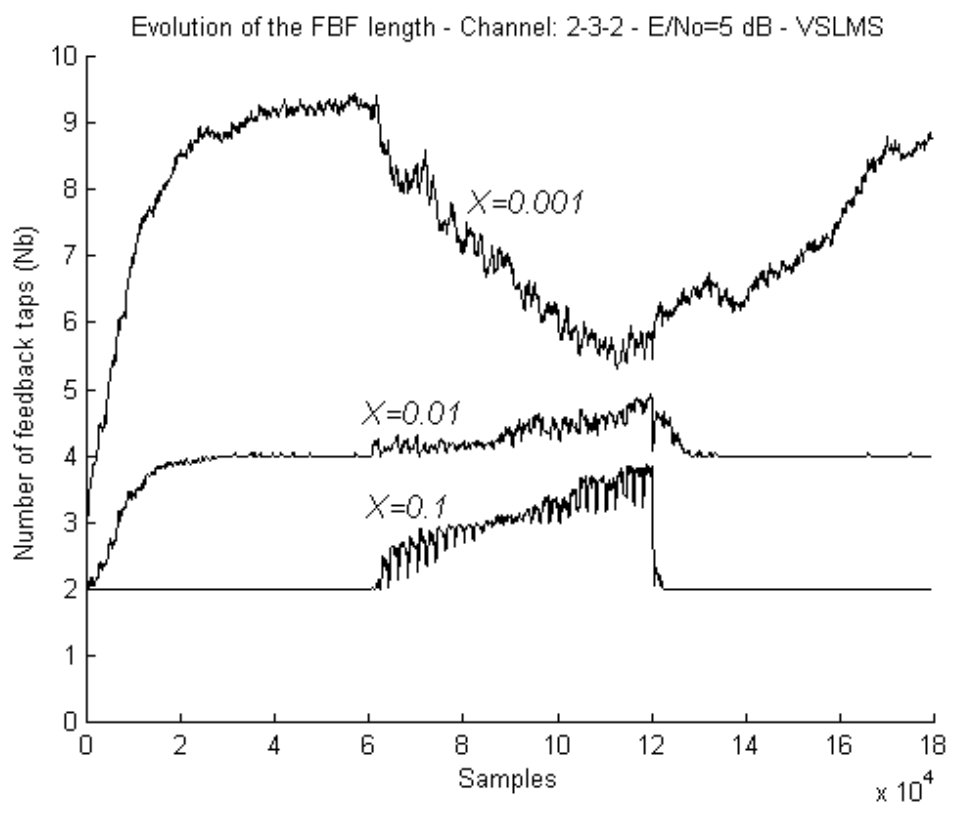

**Figure 6.50**: FBF length evolution using VSLMS for abrupt channel change: channel 2-channel 3-channel 2. E/No=5dB.

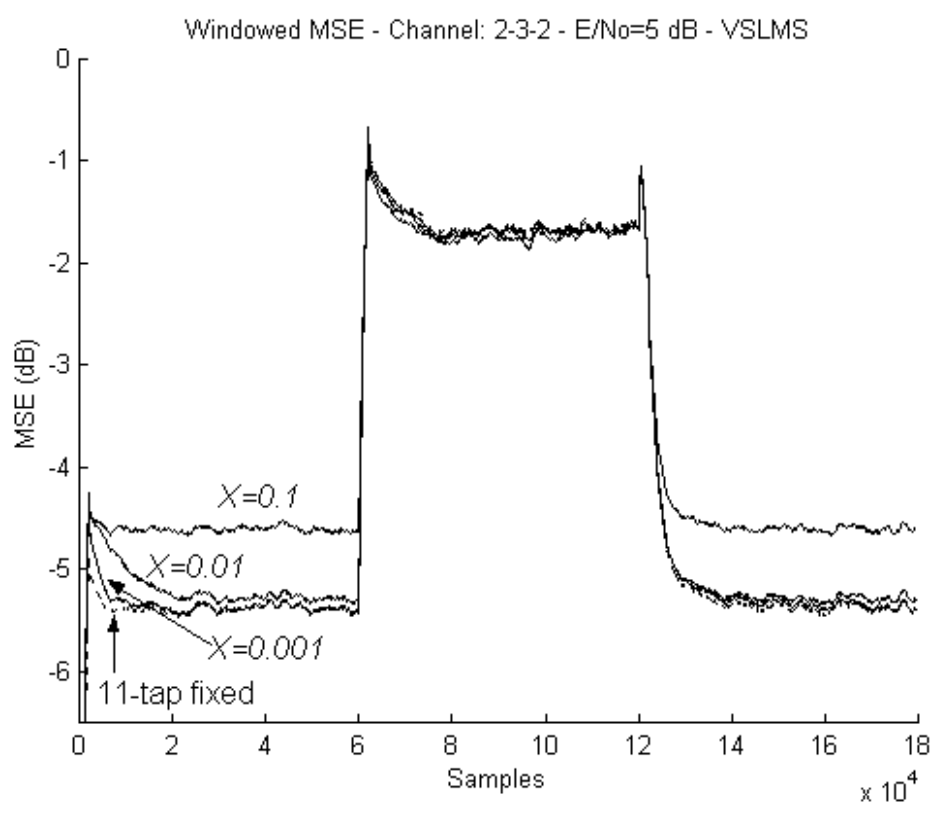

**Figure 6.51**: Windowed MSE using VSLMS for abrupt channel change: channel 2-channel 3-channel 2. E/No=5dB.

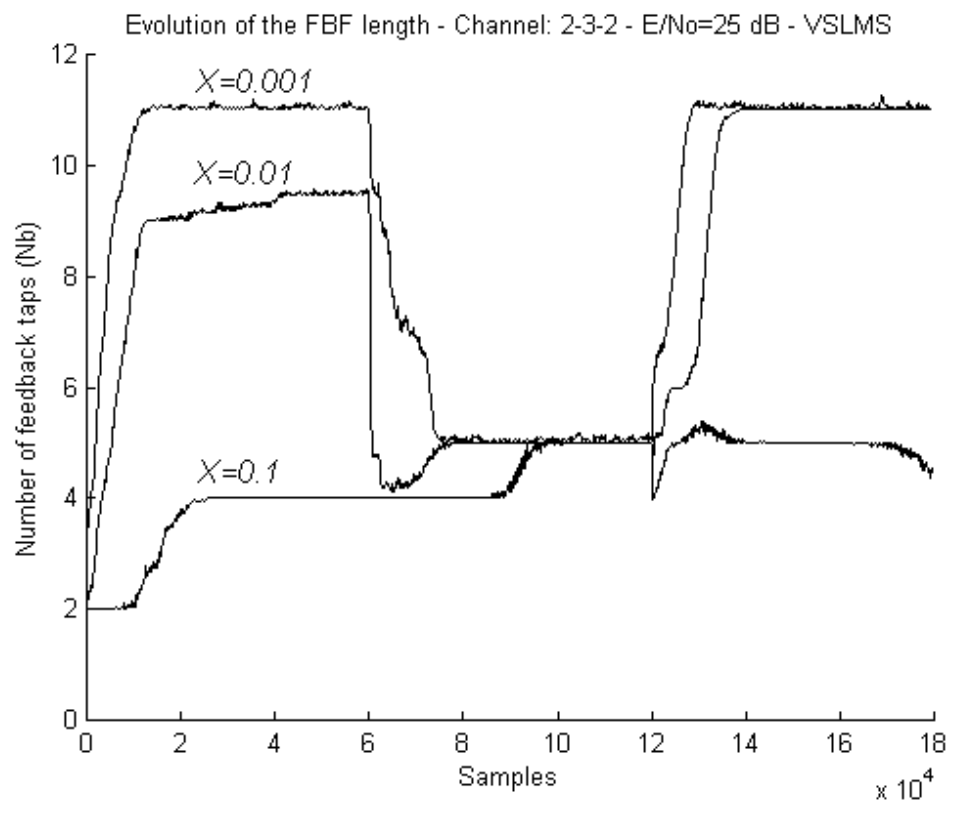

**Figure 6.52**: FBF length evolution using VSLMS for abrupt channel change: channel 2-channel 3-channel 2. E/No=25dB.

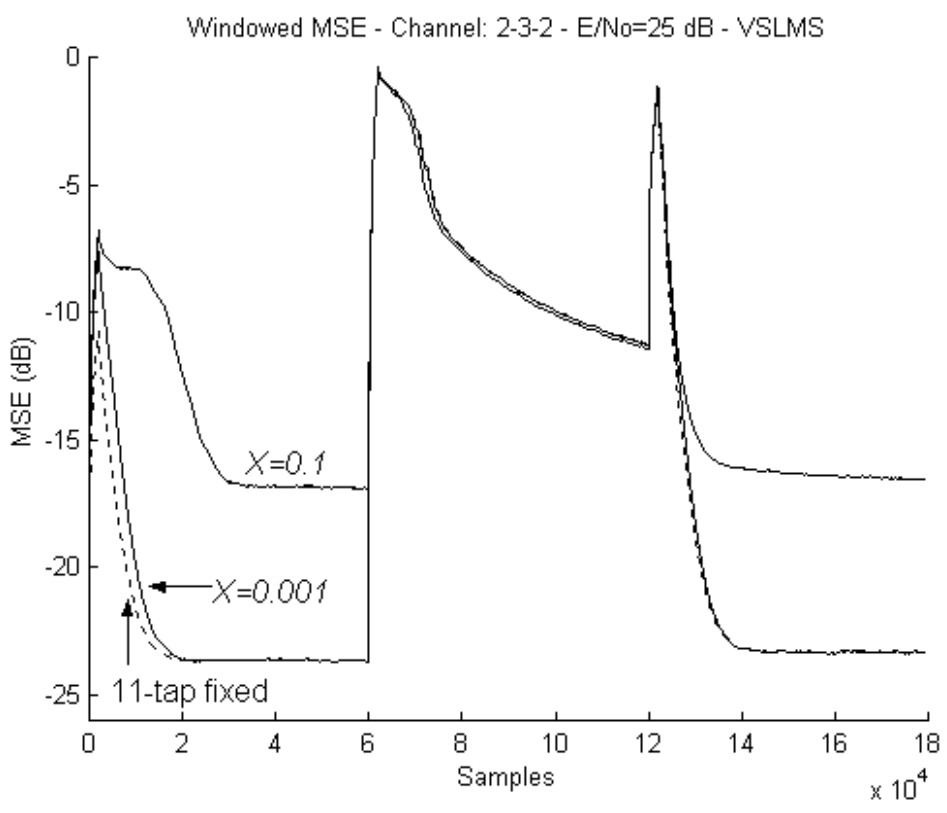

**Figure 6.53**: Windowed MSE using VSLMS for abrupt channel change: channel 2-channel 3-channel 2. E/No=25dB.

### 6.12.3 Variable E/No profile

Concluding the set of simulations for the FBF length update algorithm, the technique has been tested in a scenario with a variable E/No profile. The basic channel used is Channel model 3 with the dynamic E/No profile superimposed, as shown in figure 5.55 (Chapter 5). The FFF was again fixed to 6 taps with the decision delay set to 5 samples. In figure 6.54 the evolution of the FBF length can be seen for three different values of $\chi$. For $\chi$=0.1, the FBF is insensitive to any variation in the E/No level. Reducing the value of $\chi$ makes the FBF respond to the signal level variations. Observe how the smaller $\chi$ is, the faster the FBF responds to the changes. This was also seen in the abrupt channel changes. Notice however that the difference in number of FBF taps when E/No=25 dB and when E/No=5 dB is just 2-3 taps. This is a direct consequence of the fact that the FBF operates with noise free samples and therefore its functioning is quite independent of the E/No level. In figure 6.55 the corresponding (windowed) MSE curves are shown with the addition of the MSE curve obtained from an 11-tap FBF fixed equaliser.

It is clear that a variable length FBF with small $\chi$ achieves the same (or very close) performance as the 11-tap fixed equaliser. Again, it can be seen here that if a tiny increase in the MSE level and a slightly longer convergence time with respect to the optimum can be tolerated, a value like $\chi$ =0.01 can provide some computational savings.

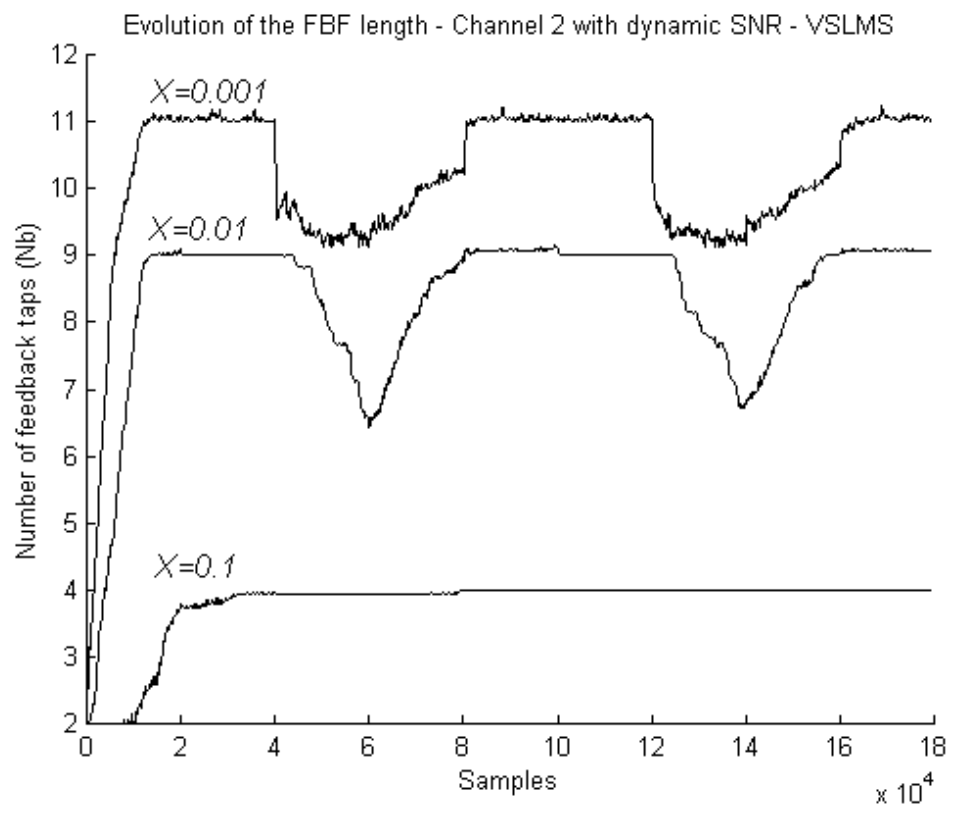

Figure 6.54: FBF length evolution using VSLMS for channel 2 with a dynamic E/No profile.

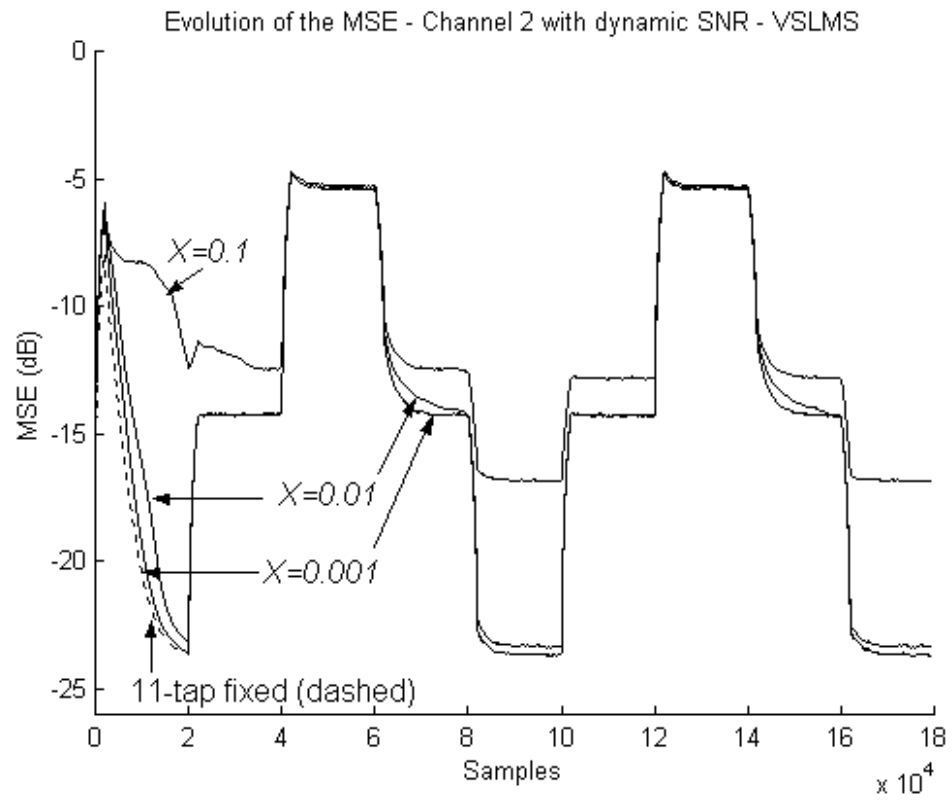

**Figure 6.55**: Windowed MSE using VSLMS for channel 2 with a dynamic E/No profile.

# 6.13 FBF length update algorithm: final remarks

In order to conclude the study of the DFE and more in particular of the FBF length update algorithm, some of its potential problems are considered. First, and as in the variable length LE, the stability of RLS algorithm when used on a DFE with a variable length FBF is discussed. Secondly, a scenario is presented which can cause trouble to the FBF length update algorithm, namely, sparse channel impulse responses. Some solutions are suggested to minimise this problem.

### 6.13.1 RLS-DFE stability with a variable length FBF

In section 5.3.2, the stability of the RLS algorithm when combined with a variable length LE has been considered. It was shown under what circumstances the algorithm can become unstable and what can be done to prevent it. It seems therefore mandatory to analyse the stability of the RLS algorithm, when driving the coefficients of a DFE whose FBF length can vary.

Recall from the LE case, that the source of instability was the appearance of inconsistencies in the recursive update of equations when the filter had been reduced. A priori, it may seem obvious that the same problem will occur in the DFE. However this is not the case. The justification for this is to be sought in the structure of the correlation matrix of the input vector for the DFE case and the way the FBF length update algorithm operates. This explanation is better illustrated by means of an example. Suppose a (6,11)-DFE operating on Channel model 2 (11-taps) with decision delay set to 5. The input correlation matrix will be given by[23]:

$$\mathbf{\Phi}(n)_{17x17} = \begin{bmatrix} \mathbf{\Gamma}(n)_{6x6} & \mathbf{\Lambda}(n)_{6x11} \\ \mathbf{\Lambda}^T(n)_{11x6} & \mathbf{I}_{11x11} \end{bmatrix}$$

Given that the FFF remains fixed, the dimensions of $\mathbf{\Gamma}$(n) will not change. However the FBF length changes will affect $\mathbf{\Lambda}$(n), $\mathbf{\Lambda}^T$(n) and **I**. The key point is to analyse the contents of $\mathbf{\Lambda}$(n) which in this case, (6,11)-DFE and Channel model 2, would be:

[23] Notice that the notation used for the submatrices of $\mathbf{\Phi}$(n) is the same as the one used for the submatrices of **R**(n) but in this case, the averages are temporal.

$$\mathbf{\Lambda}(n)=\begin{bmatrix} \sum_{i=0}^{n}\lambda^{n-i}c_6(i) & \sum_{i=0}^{n}\lambda^{n-i}c_7(i) & \cdots & \sum_{i=0}^{n}\lambda^{n-i}c_{10}(i) & 0 & 0 & \cdots & 0 \\ \sum_{i=0}^{n}\lambda^{n-i}c_5(i) & \sum_{i=0}^{n}\lambda^{n-i}c_6(i) & \cdots & \cdots & \sum_{i=0}^{n}\lambda^{n-i}c_{10}(i) & 0 & \cdots & 0 \\ \vdots & \vdots & & & & & & \vdots \\ \vdots & \vdots & & & & & & \vdots \\ \vdots & \vdots & & & & \ddots & & \vdots \\ \sum_{i=0}^{n}\lambda^{n-i}c_1(i) & \sum_{i=0}^{n}\lambda^{n-i}c_2(i) & \cdots & \cdots & \cdots & \cdots & \sum_{i=0}^{n}\lambda^{n-i}c_{10}(i) & 0 \end{bmatrix}$$

Notice the last column in $\mathbf{\Lambda}$(n) corresponds to the detection tap and will always be all zeros. Suppose that now the channel abruptly changes to Channel model 1 (2 taps). When this happens, and due to the effect of the forgetting factor, many of the entries of $\mathbf{\Lambda}$(n) will quickly tend to zero and eventually $\mathbf{\Lambda}$(n) will become:

$$\mathbf{\Lambda}(n)=\begin{bmatrix} 0 & 0 & 0 & 0 & 0 & 0 & 0 & 0 & 0 & 0 & 0 \\ 0 & 0 & 0 & 0 & 0 & 0 & 0 & 0 & 0 & 0 & 0 \\ 0 & 0 & 0 & 0 & 0 & 0 & 0 & 0 & 0 & 0 & 0 \\ 0 & 0 & 0 & 0 & 0 & 0 & 0 & 0 & 0 & 0 & 0 \\ 0 & 0 & 0 & 0 & 0 & 0 & 0 & 0 & 0 & 0 & 0 \\ \sum_{i=1}^{n}\lambda^{n-i}c_1(i) & 0 & 0 & 0 & 0 & 0 & 0 & 0 & 0 & 0 & 0 \end{bmatrix}$$

The fact that the length reduction of the channel causes some of the columns of $\mathbf{\Lambda}$(n) (from left to right) to become all zeroes (so do some of the rows of $\mathbf{\Lambda}^T$(n)) from bottom to top, allows the FBF to be reduced without re-starting the RLS algorithm.

Remember from the previous chapter that the source of the instability was in the update of the inverse of the correlation matrix, **P**(n). If the FBF reduced from to 11 taps to 2 taps, the correlation matrix, $\mathbf{\Phi}$(n), will now have dimension 8x8. However, when this happens and unlike in the LE situation, the inverse of the new correlation matrix, $\mathbf{P}_{8x8}$(n), can be extracted from the inverse of the higher order matrix $\mathbf{P}_{17x17}$(n) because, due to the new zero columns and rows:

$$\mathbf{P}_{8x8}(n)=\mathbf{\Phi}_{8x8}(n)^{-1}=\left[\mathbf{\Phi}_{17x17}(n)^{-1}\right]_{8x8} \tag{6.61}$$

This implies that when changing the FBF length, the only thing to be done is to re-size the vectors and matrices involved in the RLS algorithm. Notice that in this example and in order to simplify the presentation it has been assumed that the FBF shrank from 11 to 2 taps in one step. Using the proposed length update algorithm, this reduction will take place tap by tap, with a transient stage between reductions that would allow the correspondent column and row of $\mathbf{\Phi}(n)$ to converge to zero.

When the FBF is expanded, as the LE case, there are no stability problems and only re-sizing of the appropriate vectors and matrices is required.

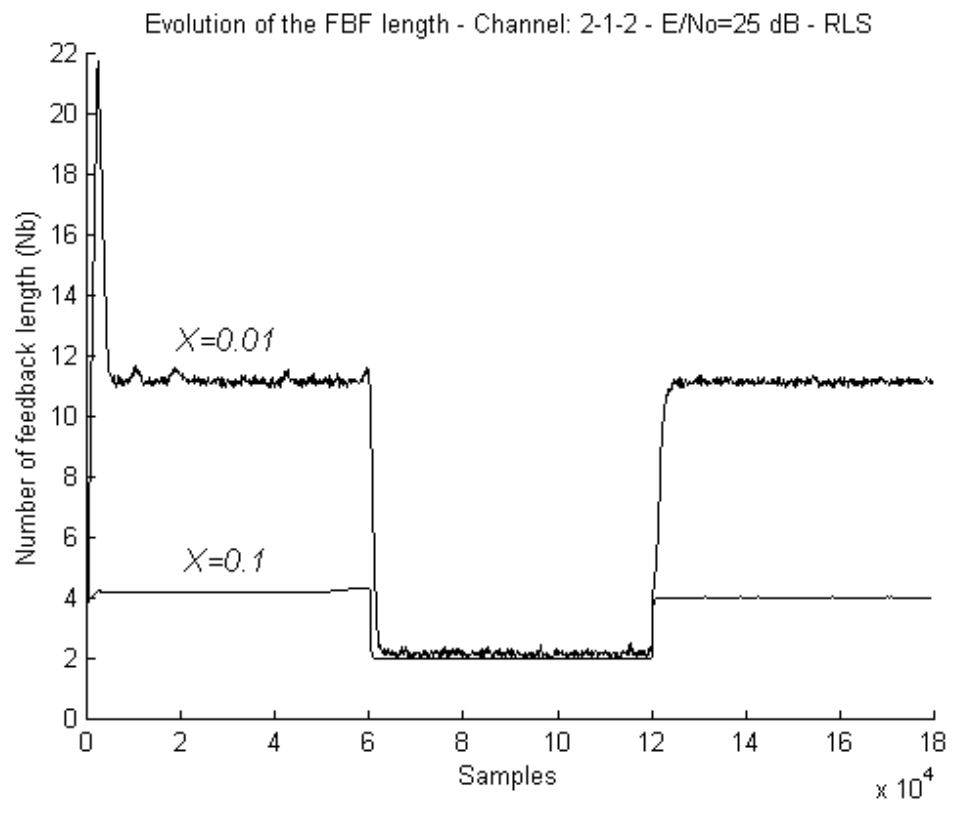

**Figure 6.56**: FBF length evolution using RLS for abrupt channel change: channel 2-channel 1-channel 2. E/No=25dB.

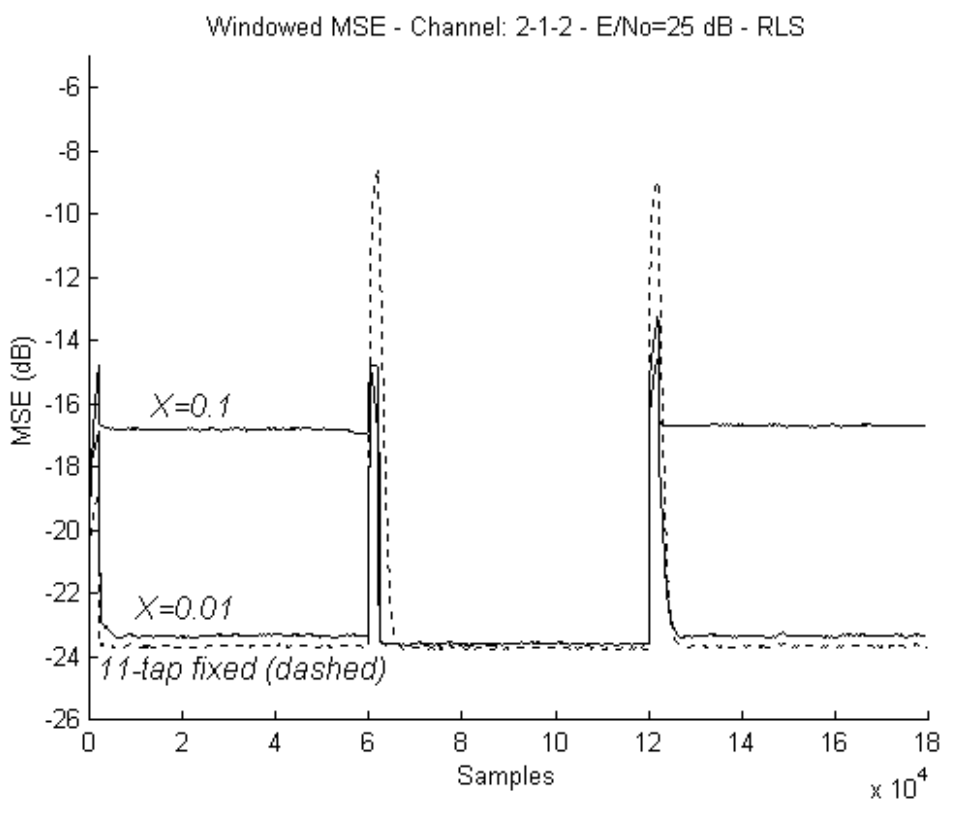

**Figure 6.57**: Windowed MSE using RLS for abrupt channel change: channel 2-channel 1-channel 2. E/No=25dB.

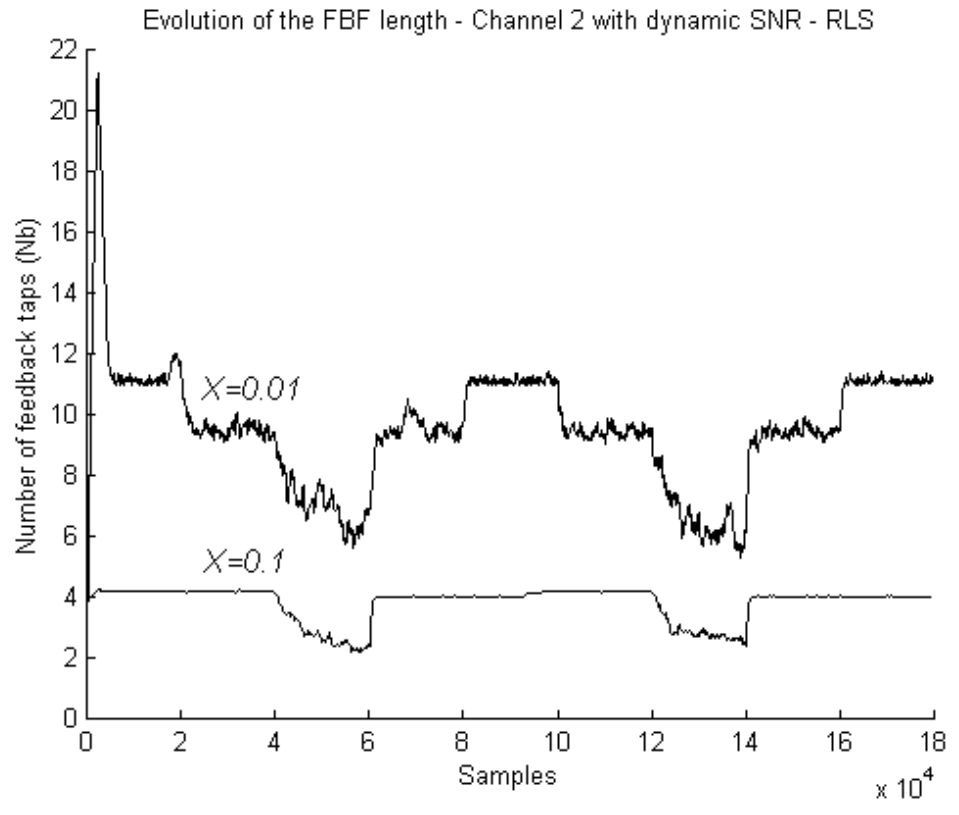

**Figure 6.58**: FBF length evolution using RLS for channel 2 with a dynamic E/No profile.

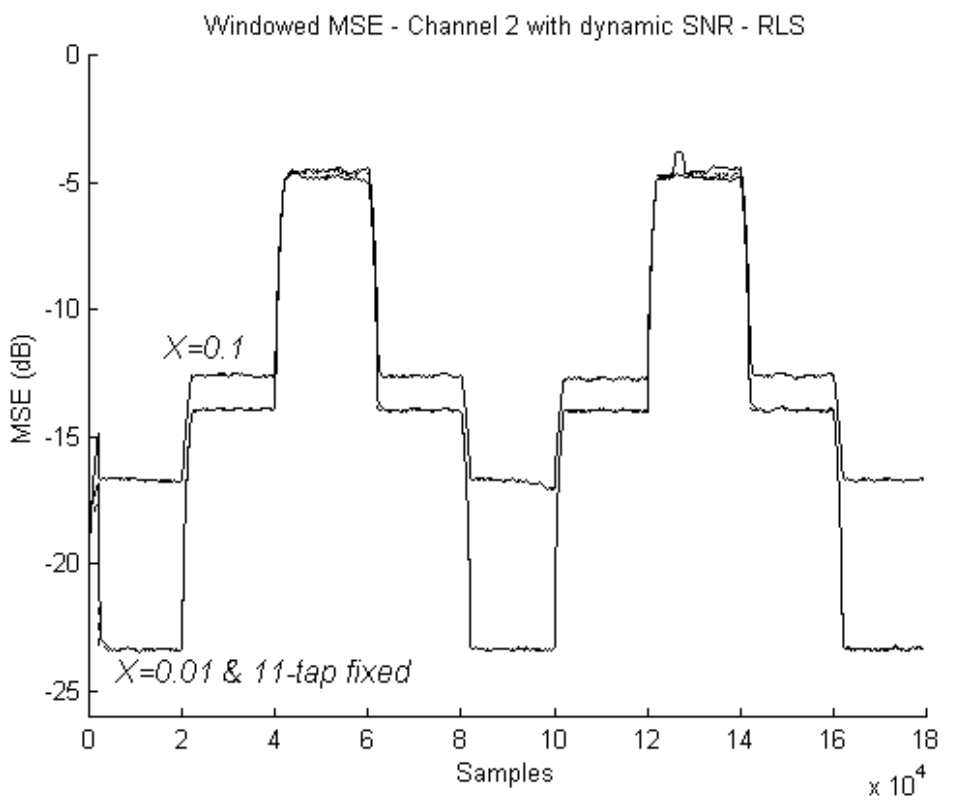

**Figure 6.59**: Windowed MSE using RLS for channel 2 with a dynamic E/No profile.

Figures 6.56-6.59 show the performance of the FBF length update algorithm when combined with the RLS algorithm to update the DFE coefficients. It has been noticed using the RLS algorithm, that the parameter $\chi$ should be set to larger values than in the case of the VSLMS as otherwise the FBF becomes too sensitive and tends to expand to the maximum number of taps. Note that now with $\chi$=0.01 the maximum performance is achieved (optimum expansion), whereas in the case of the VSLMS, $\chi$ needs to be set to 0.001 to expand the FBF so as to cancel all postcursors.

## 6.13.2 Sparse channel impulse responses

The algorithm has been shown to work on the basis of using a dummy tap (the last one) to detect when the channel grows. This mechanism works well when the channel expands in continuous way. To be more accurate, when the combined CIR-FFF evolves in a way such that it does not contain "holes".

If the channel is sparse, these holes are likely to appear and this is worth taking into account. Consider a situation where, due to changes in the CIR, the combined CIR-FFF response varies from figure 6.60 to figure 6.61:

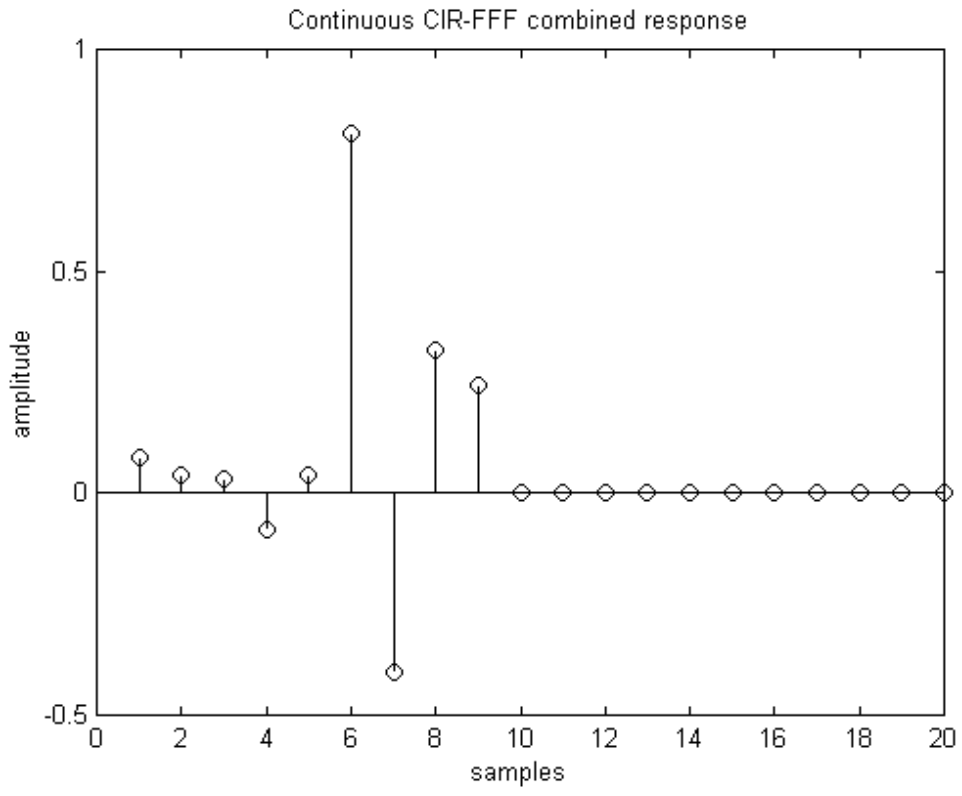

**Figure 6.60**: Continuous CIR-FFF combined response.

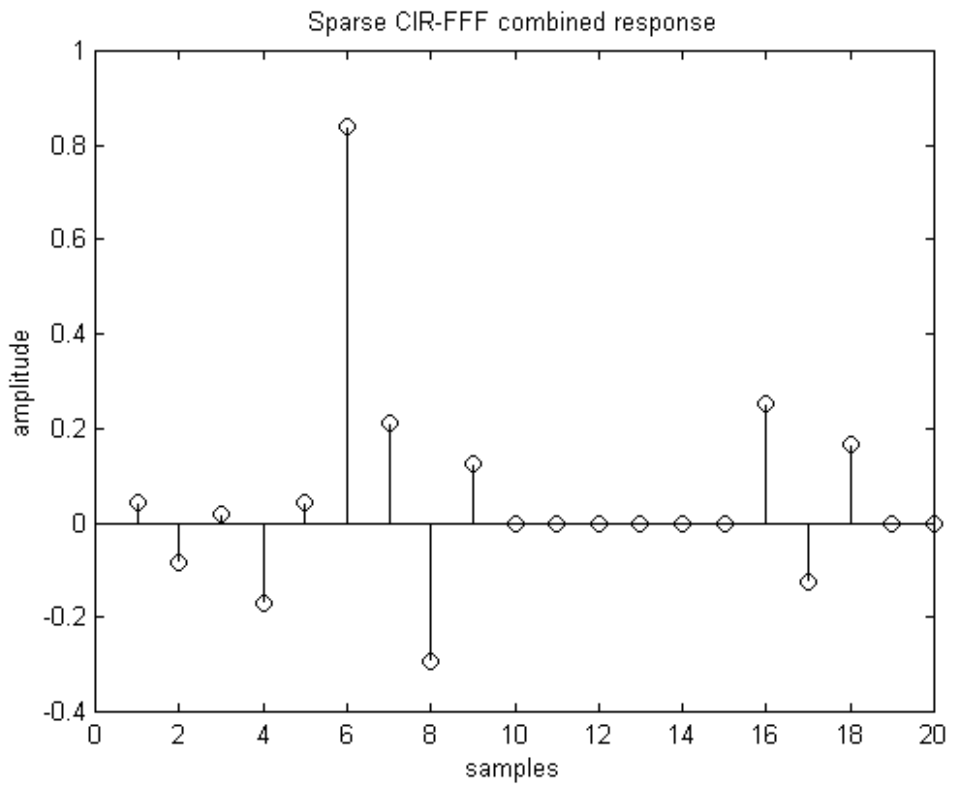

**Figure 6.61**: Sparse CIR-FFF combined response.

In these figures, the tap with highest magnitude corresponds to the detected sample, the previous samples are the precursor ISI while the posterior samples correspond to the postcursor ISI components. If the FBF length update algorithm is used when operating in the channel of figure 6.60, the FBF will be set to 4 taps (3 cancelling taps plus the detection tap). Suppose that the channel suddenly changes to the one in figure 6.61 where some sparse components appear towards the end of the CIR-FFF response. In this scenario, the FBF length update algorithm will not be able to expand the FBF to cancel the new postcursors. The reason is that the detection tap, which will be monitoring the sample number 10 in figure 6.61, will not register any increase in its power and will prevent the FBF from expanding. Notice that in order to get a sparse tap (one with zero magnitude) in the CIR-FFF combined response, there must be a hole of $N_f$ samples in the CIR response, assuming that the magnitude of the FFF coefficients are all greater than zero (albeit arbitrarily small).

Sparse channels do occur in practice. In the COST207 channel specifications there is a particular model corresponding to hilly terrain with a gap in the CIR of about 10 $\mu$s (pp. 628 in [Steele99]). One possible solution to allow the FBF length update algorithm to detect these sparse components would be to enlarge the FBF periodically up to its maximum length. From there, the algorithm would shrink the FBF down to the point where the last components, sparse or continuous, are located.

Exploitation of sparsity in adaptive filtering is a current topic of research, see for example [Martin01a], [Martin01b] and [Berberidis00].

# 7 RECONFIGURABLE STRUCTURES IN REALISTIC SCENARIOS

This chapter presents simulation results comparing the performance of the fixed length equalisers with the variable length structures introduced in chapters 5 and 6 when operating in realistic mobile scenarios. In section 7.1, the specific simulation environment is described, especially how the mobile channels have been generated. Sections 7.2 and 7.3 show results for the variable length LE (VL LE) and DFE with a variable length FBF (VL FBF DFE) respectively and how these compare, in terms of performance and computational complexity, with fixed length LE and DFE. In addition to the recurrent MSE graphs, BER curves are also presented for some of the simulations in this chapter to support the MSE results. The chapter finishes by summarising the conclusions to be extracted from these simulation results.

## 7.1 Realistic simulation environment

In chapters 5 and 6, novel techniques have been presented to adjust dynamically the number of taps in different types of equalisers. There, the simulation environment used was a rather simple one as the main objective was to highlight how the algorithms worked and the potential benefits of adjusting the equaliser length. It is however mandatory to verify that these algorithms and structures are also able to perform satisfactorily in a realistic mobile scenario.

As explained in chapter 2, equalisation is, or can be, applied in one form or another in nearly all cellular standards. Consequently, the simulation environment designed for this set of simulations is not specifically targeted to any particular standard although some of its parameters are taken from the UMTS (W-CDMA) specification ([3GPP00]). The basic blocks of the environment implemented are shown in figure 7.1.

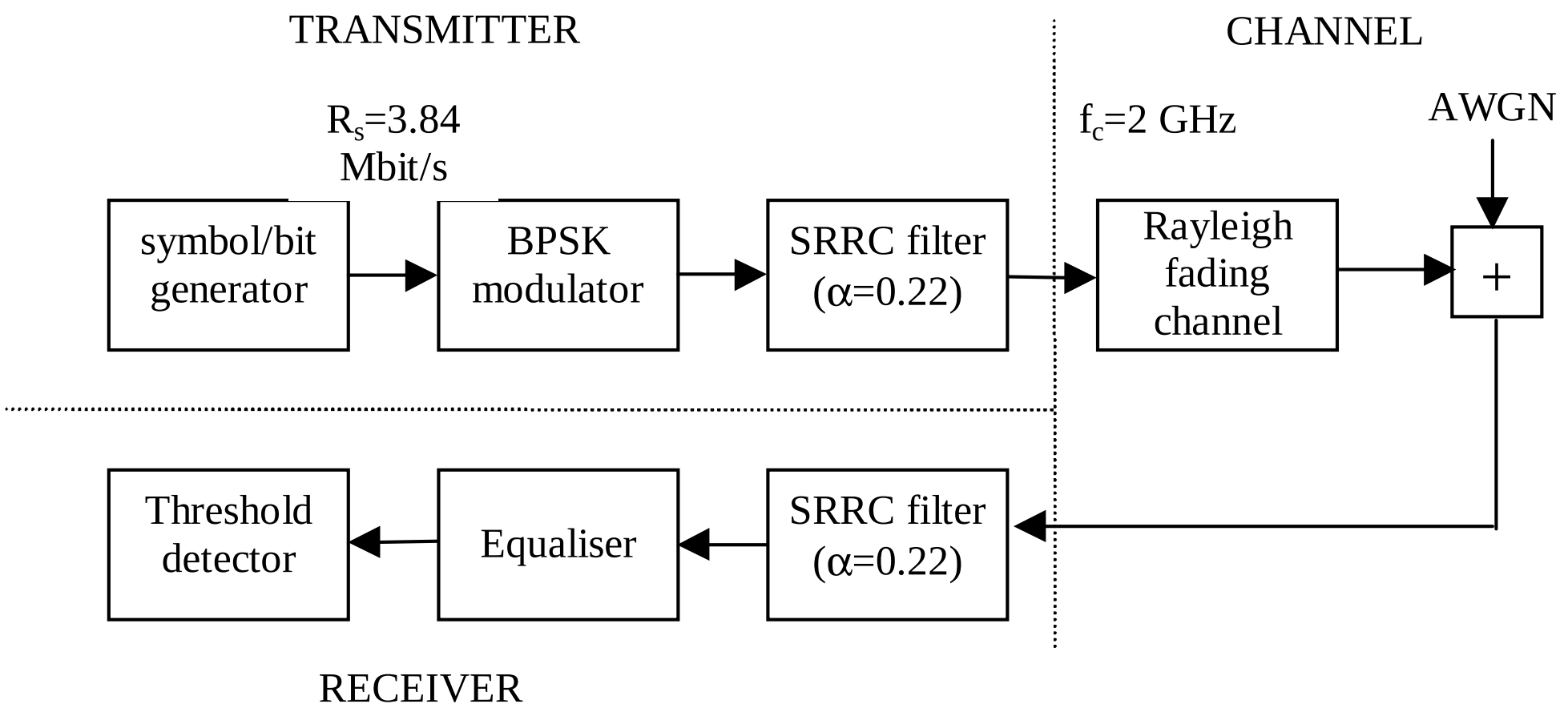

**Figure 7.1**: Realistic system model.

The symbol (bit) rate has been set to 3.84 Mbit/s, corresponding to the UMTS chip rate. Generated symbols are then BPSK modulated and subsequently band limited with a square root raised cosine filter (SRRC) with a roll-off factor of 0.22. The carrier frequency used for transmission is 2 GHz which is the frequency band selected for most 3G standards world wide. At the receiver, another SRRC filter is used to perform matched filtering on the incoming signal. The equaliser will reduce the effects of multipath propagation and will present "cleaner" samples to the threshold detector. In a practical system, error correction coding/decoding would also form part of the transceiver chain. However, in the context of this work, the focus lies only on the equalisation subsystem.

In order to get an idea of how the equaliser performs in real conditions, it is fundamental to use a channel that accurately models the physical channels encountered in practical mobile scenarios. For this purpose, a Rayleigh fading generator has been implemented ([Proakis95, pp 45-46, 767]).

In order to construct a Rayleigh process, two coloured Gaussian processes need to be generated. Here "coloured" means that the originally white Gaussian processes are filtered with a filter whose bandpass spectrum is given by the Doppler spectrum at a particular centre frequency and mobile speed. Two methods are available to generate such processes. The most obvious technique consists of filtering samples of white noise with a filter whose spectrum equals the Doppler spectrum ([Laurenson94]). This method presents the drawback that, given the sharp spectrum generated by the Doppler phenomenon, the filters are also required to have very sharp stopbands, which in turn implies using many coefficients.

The alternative method used in this work, known as the Sum of Sines or Jakes model ([Jakes74]), is based, as its names implies, on summing N sinusoids whose frequencies are chosen as samples of the Doppler spectrum. Suppose for example that a given Doppler spectrum extends from –50 Hz to 50 Hz around the carrier frequency and assume N=20 sinusoids. The frequencies of these sinusoids could be chosen to be –50 Hz, -45 Hz, -40 Hz, -35 Hz,…, 35 Hz, 40 Hz, 45 Hz and 50 Hz. In practice, N does not need to be this large and with just 8 sinusoids accurate models are obtained. The properties of the Sum of Sines model are thoroughly treated in [Patzold96].

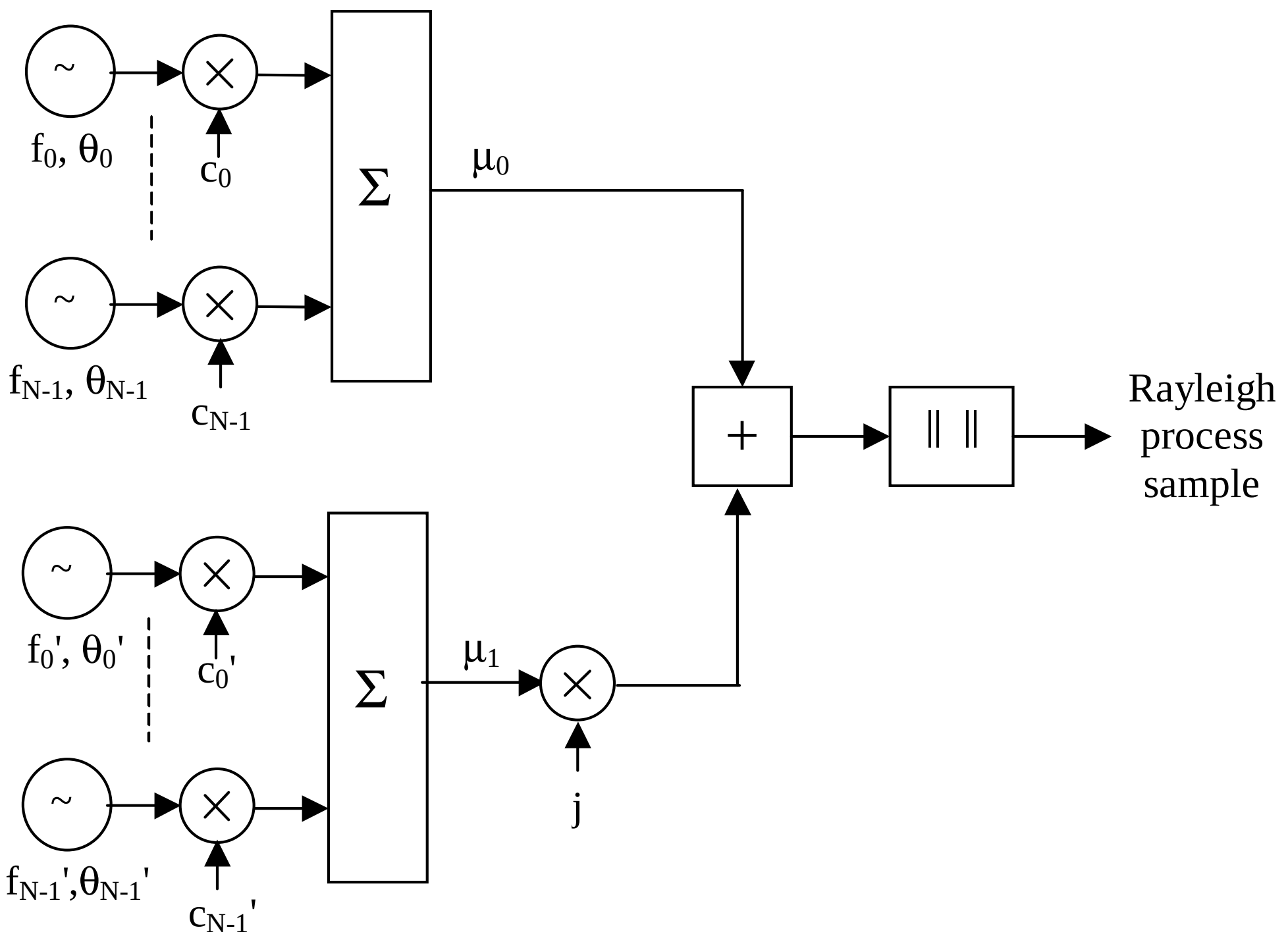

**Figure 7.2**: Generation of a Rayleigh process using the Sum of Sines model [Jakes74].

The figure above shows how the two independent coloured Gaussian processes, $\mu_0$ and $\mu_1$, are obtained using the Sum of Sines model. The outcomes from these processes are then summed in quadrature and the norm taken. The resulting output is a sample of a Rayleigh variable whose fading rate is given by the chosen Doppler spread.

The fading process described in previous paragraphs may apply to all or some of the paths in a multipath profile. In this chapter, extensive use has been made of the COST 207 Typical Urban profile depicted in figure 7.3 [Steele99]. The model specifies that, in this particular case, all six paths are subject to Rayleigh fading. It is therefore necessary to maintain an independent Rayleigh fading generator for each of the paths.

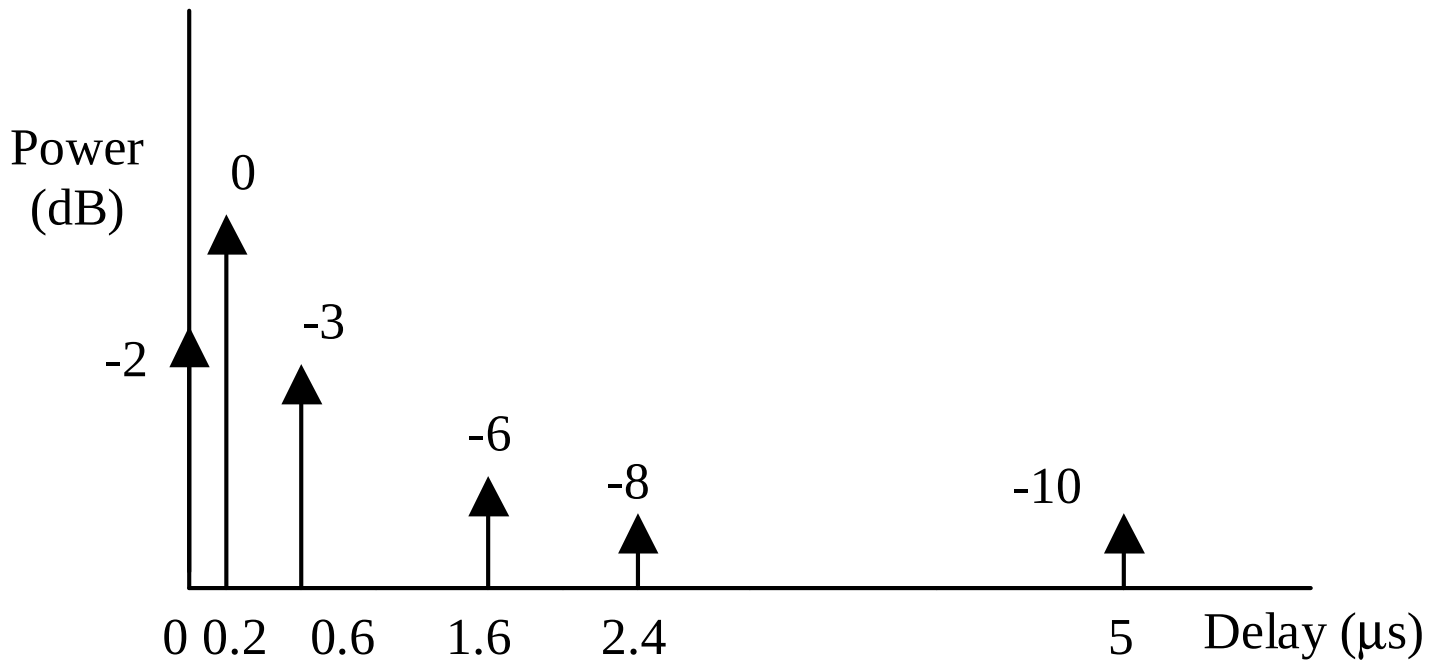

**Figure 7.3**: COST 207 Typical Urban channel power profile.

The sampling of a channel conforming to the power and delay distribution shown above at the specified bit rate (3.84 Mbit/s), results in a channel filter (FIR filter) with around 22-23 bit-spaced taps. In order to reduce the computational requirements of the simulations, the last path of the COST207-TU profile has been omitted. The resulting channel, referred as the reduced COST207-TU, has a maximum excess delay of only 2.4 $\mu$s and when sampled at bit rate, can be modelled with an FIR filter of just 11 taps.

Concluding the description of the realistic simulation environment, some comments about the amount of bandwidth devoted to training are appropriate. Systems like GSM or IS-54 devote some bits of a normal data frame to training purposes. In the case of GSM, training amounts to the 16.67 % of the total bandwidth. On the other hand, newer standards such as UMTS dedicate a specific physical channel (CPICh) for the transmission of a pilot signal known at the receiver. In this latter case, the assumption made so far of assuming a 100% training time matches with reality as the equaliser can use the signal from the pilot channel to update its values continually.

In the simulations presented over the next sections training time has been limited for most of the cases to 10% of the total transmission. In particular, 200 training symbols are transmitted every 2,000 symbols. This implies that in a block of 2,000 bits there are 1,800 bits corresponding to user data. These figures were all determined by simulation. The justification of just using a percentage of the available bandwidth rather that assuming the existence of a continuous training channel is to check how imperfect (i.e. wrong) decisions affect the performance of the length update algorithms.

# 7.2 Variable length LE in realistic scenarios

In this section, the behaviour of the variable LE in realistic conditions is presented. Results are compared in terms of performance and computational complexity with those obtained using fixed length LE. In all the simulations presented in this and forthcoming sections, the variable step size LMS (VSLMS) algorithm has been used to drive the filter coefficients. This algorithm has proven to be very robust in a wide variety of scenarios.

## 7.2.1 Effect of the equaliser length on a mobile channel

In chapter 5 the effect of the equaliser length has been studied for some static channels. In this section, a similar study has been conducted, using the reduced COST207-TU channel described before under different levels of Doppler spread.

The steady state MSE levels, measured after 100,000 iterations, are shown in figures 7.4 and 7.5 for two different E/No levels and different equaliser lengths. The decision delay was set to 8 samples, therefore the equaliser lengths were chosen to be longer than 8 taps as otherwise the LE would not be able to capture the main peak of energy.

There are a few points worth noting in these graphs. First, the steady state MSE for the different equaliser lengths are now much larger than when dealing with a static channel with similar characteristics. As an example, compare figures 7.4 and 7.5 with the curves in figure 5.3 corresponding to Channel model 2 being equalised using the LMS algorithm[24]. The MSE is larger due to two factors. First, channel variation introduces tracking misadjustment. Second, these variations may turn a fairly good channel into a very bad one with larger steady state MSE. Nonetheless, the important point to be noticed in these graphs is that again making the equaliser longer does not assure a reduction in the SS-MSE level. Under all E/No and Doppler conditions, a point is reached where additional taps do not decrease the MSE and, in fact, they may increase it.

[24] Remember that Channel model 2 was a "frozen" version of the COST207-TU profile used in this chapter.

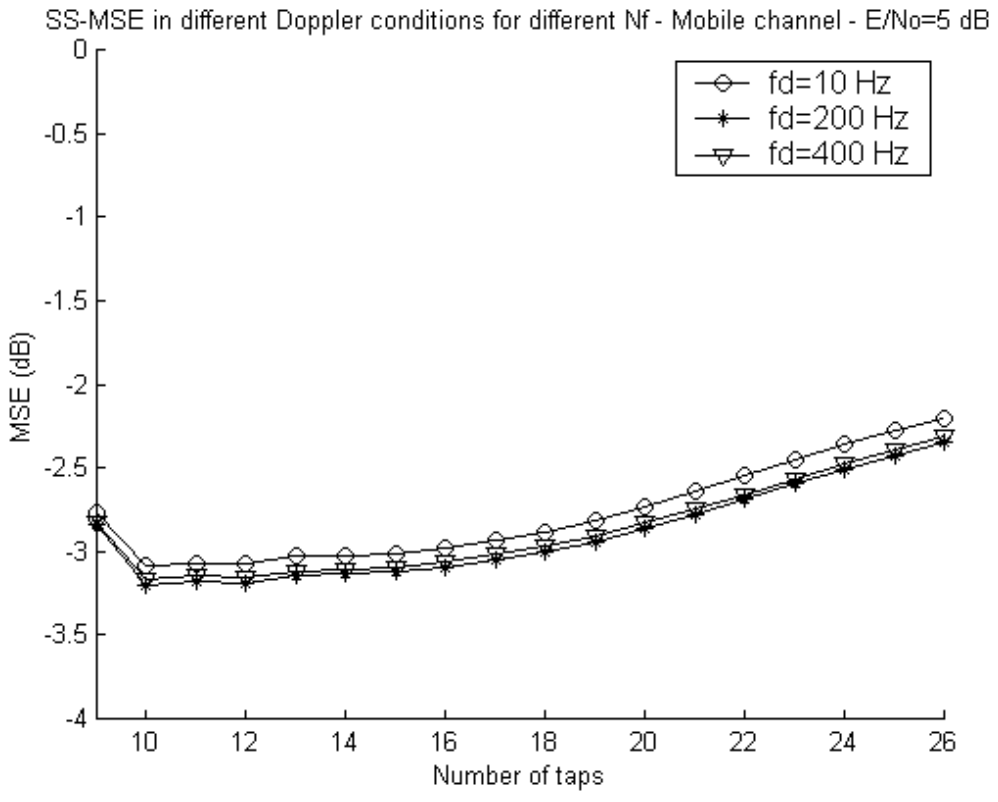

**Figure 7.4**: MSE for different length LE operating in a mobile channel with different Doppler levels. E/No=5 dB.

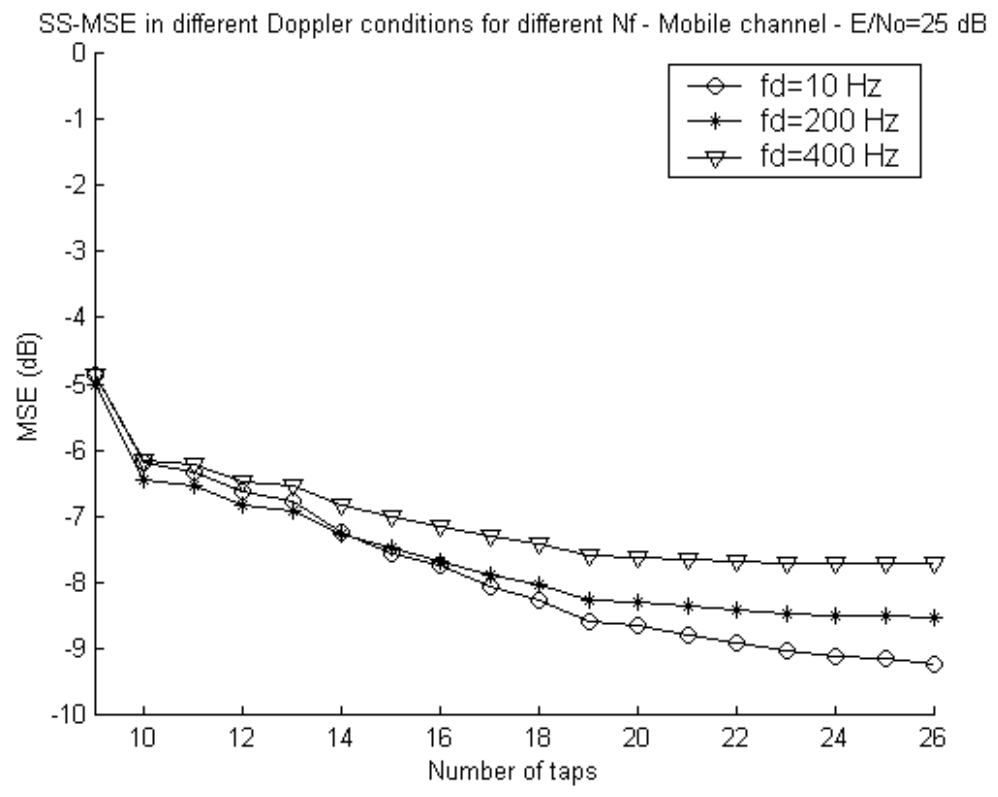

**Figure 7.5**: MSE for different length LE operating in a mobile channel with different Doppler levels. E/No=25 dB.

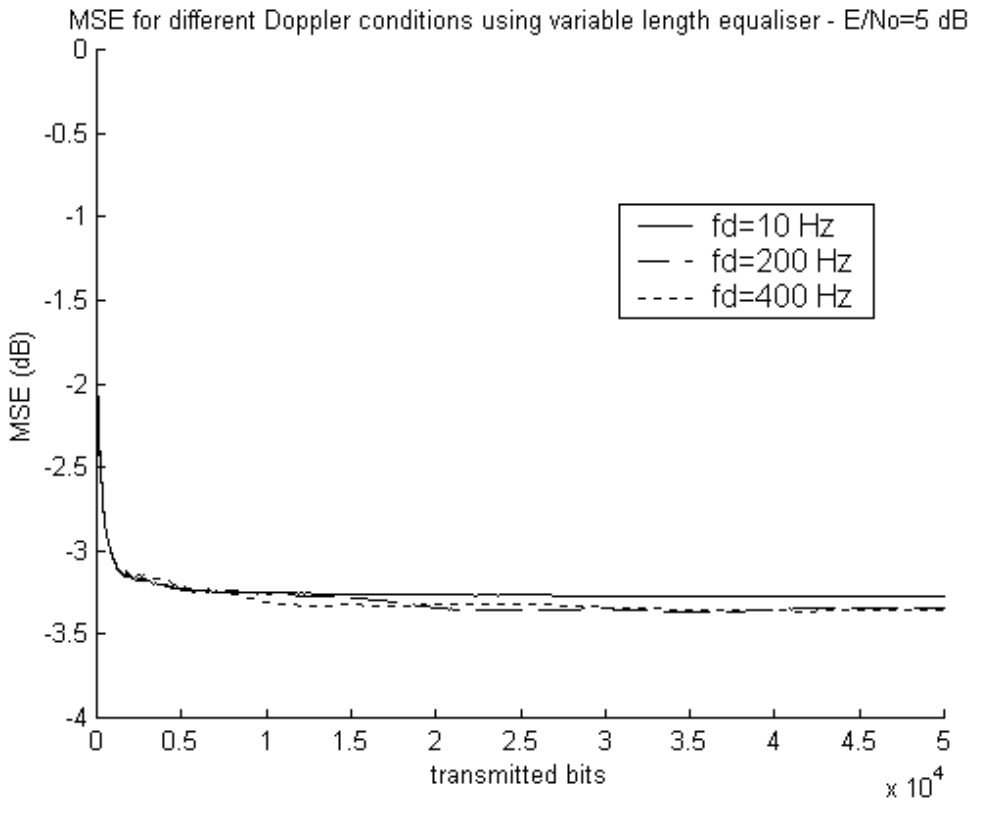

**Figure 7.6**: MSE for a VL LE operating in a mobile channel with different Doppler levels. E/No=5 dB.

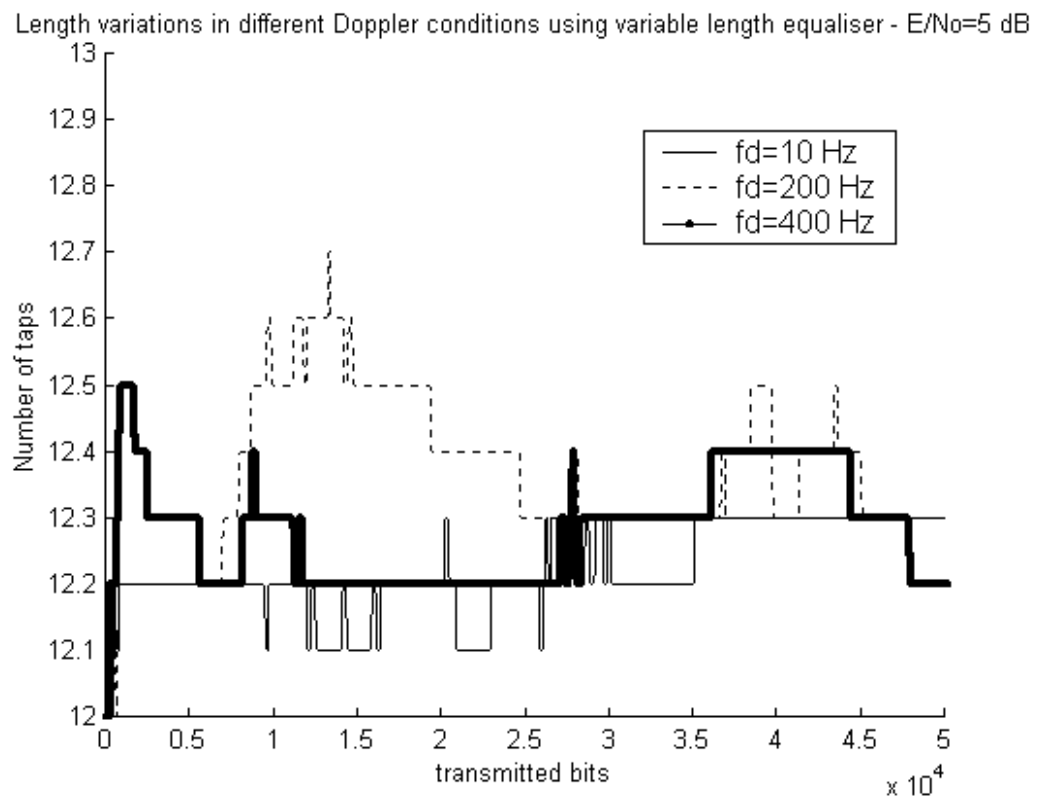

**Figure 7.7**: Length fluctuations of VL LE operating in a mobile channel with different Doppler levels. E/No=5 dB.

Figure 7.6 shows the learning (MSE) curve when E/No=5 dB for a segmented equaliser with 3 taps per segment, $\alpha_{up}=0.6$, $\alpha_{dw}=0.9$ and $\beta=0.999$. The initial number of taps was set to 9, since the decision delay was chosen to be 8. The MSE is seen to converge to around –3 dB which is the lowest MSE achieved by any of the fixed length equalisers. In figure 7.7 the length evolution of the equaliser is shown. The equaliser length can be seen to remain always between 12-13 taps, avoiding the MSE increase that would result if a longer equaliser was used.

In figures 7.8 and 7.9 the same results are presented but now for E/No=25 dB, using the same parameters as before for the length update algorithm. Notice in figure 7.9 that now the length update algorithm is able to recognise the better conditions of the link and expand the equaliser up to around 20 taps which is the optimum value for fd=200Hz and fd=400Hz. For fd=10 Hz, the optimum length would be around 22 taps but it can be seen in the graph that in this case the equaliser is still expanding. Making $\alpha_{up}$ larger would accelerate the equaliser expansion.

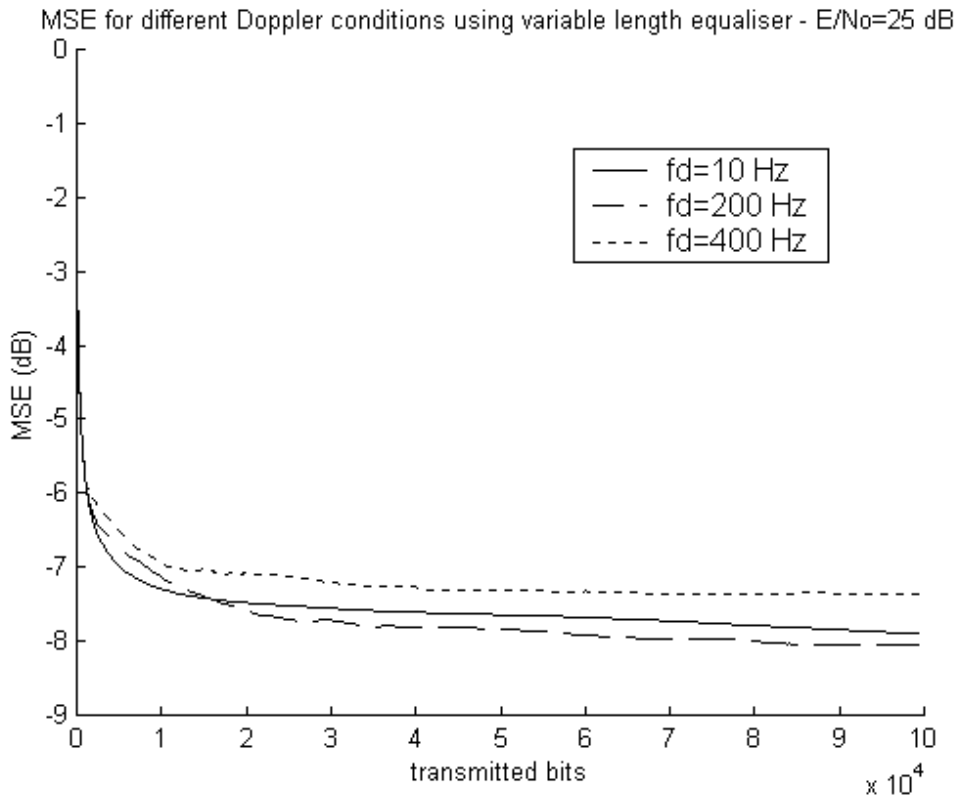

**Figure 7.8**: MSE for a VL LE operating in a mobile channel with different Doppler levels. E/No=25 dB.

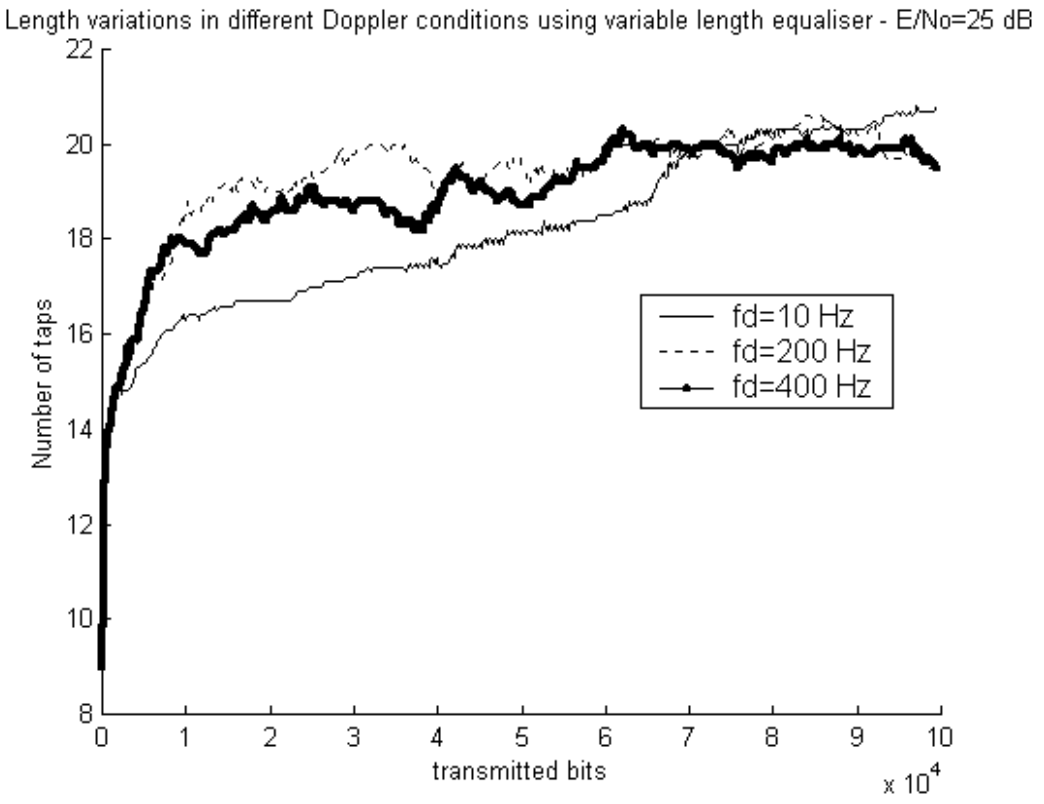

**Figure 7.9**: Length fluctuations of VL LE operating in a mobile channel with different Doppler levels. E/No=25 dB.

## 7.2.2 Dynamic E/No profile

As in section 5.4.4, the variable length LE has been tested in an environment where apart from the channel variations, the E/No level also varies as shown in figure 5.3. The channel impulse response is again based on the reduced COST207-TU and the Doppler spread is fixed to 100 Hz.

In figure 7.10 the windowed MSE (window length=2000 samples), when using a segmented equaliser with the same parameters as before (3 taps/segment, $\alpha_{up}=0.6$, $\alpha_{dw}=0.9$, $\beta=0.999$) is shown. Two curves are shown in the graph corresponding to two different degrees of training (10% and 25%). The 25% training curve has been included to show that an increase in the training does not improve performance significantly. Again the length update algorithm is able to appropriately select the most convenient length according to the E/No level and channel characteristics as can be seen in figure 7.11.

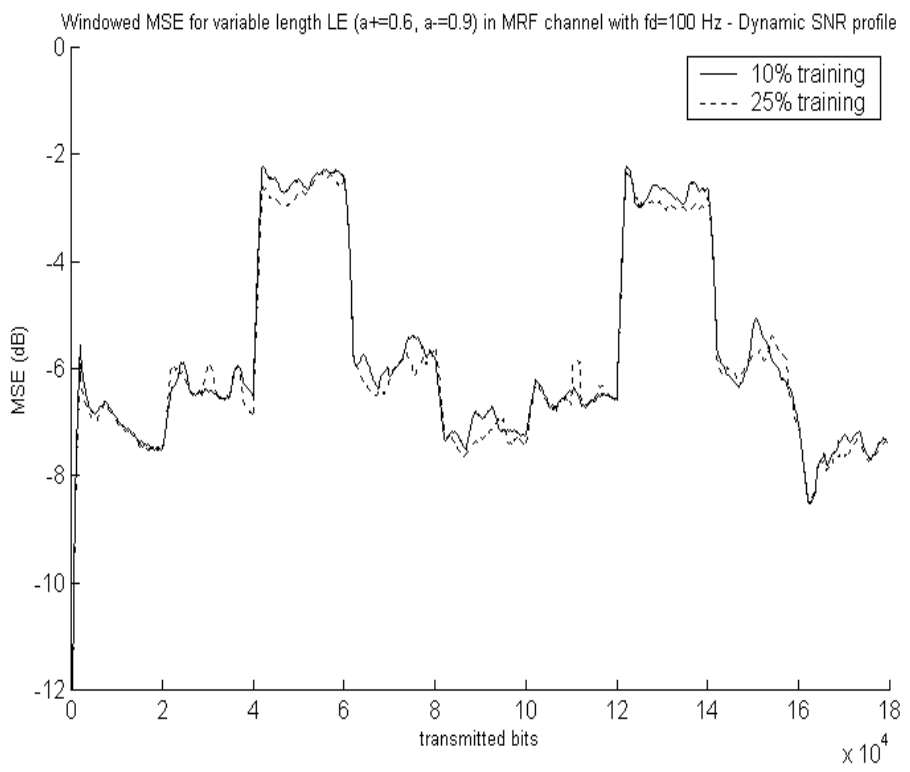

**Figure 7.10**: Windowed MSE for a VL LE ($\alpha_{up}=0.6$, $\alpha_{dw}=0.9$) operating in a mobile channel ($f_d$=100 Hz) and variable E/No profile.

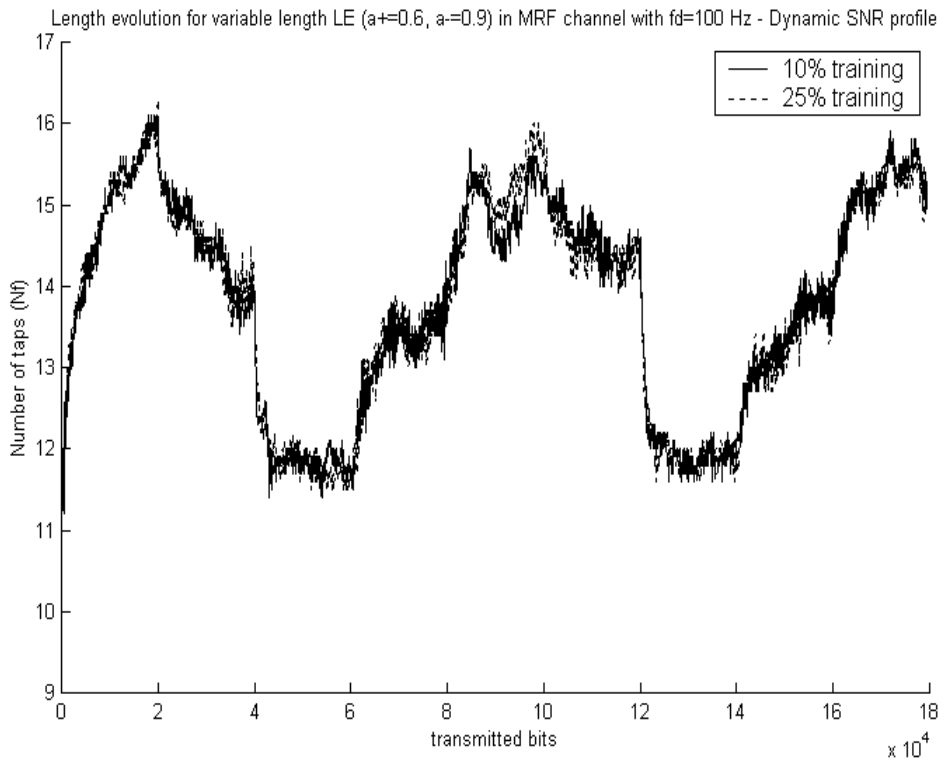

**Figure 7.11**: Equaliser length variations for a VL LE ($\alpha_{up}=0.6$, $\alpha_{dw}=0.9$) operating in a mobile channel ($f_d$=100 Hz) and variable E/No profile.

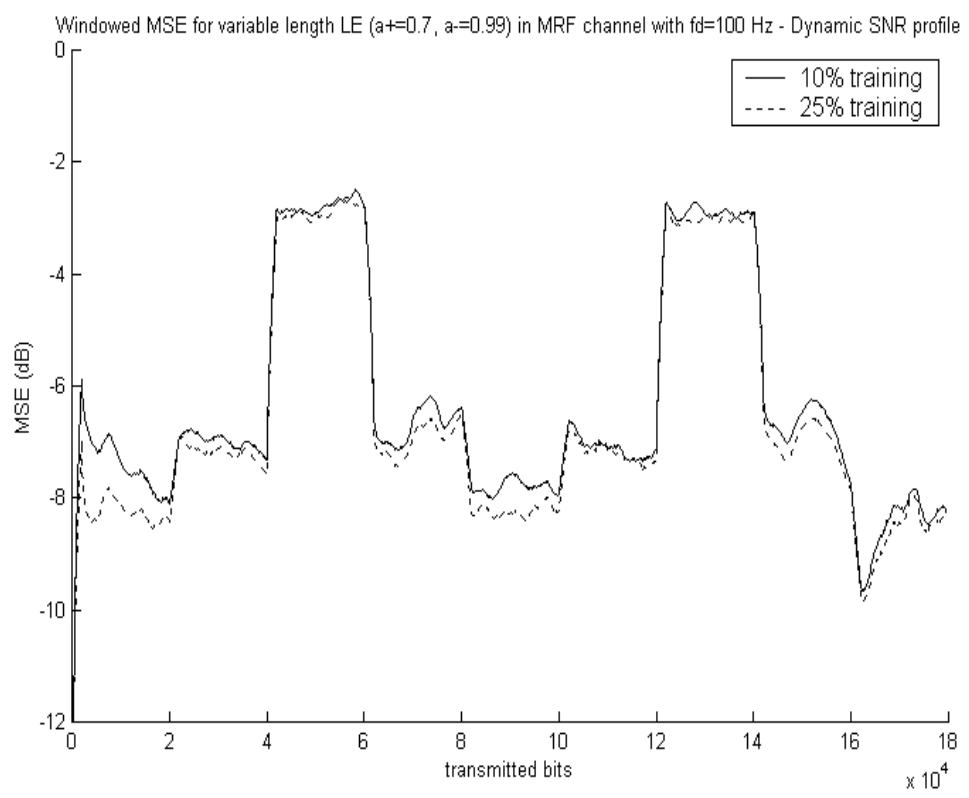

**Figure 7.12**: Windowed MSE for a VL LE ($\alpha_{up}=0.7$, $\alpha_{dw}=0.99$) operating in a mobile channel ($f_d$=100 Hz) and variable E/No profile.

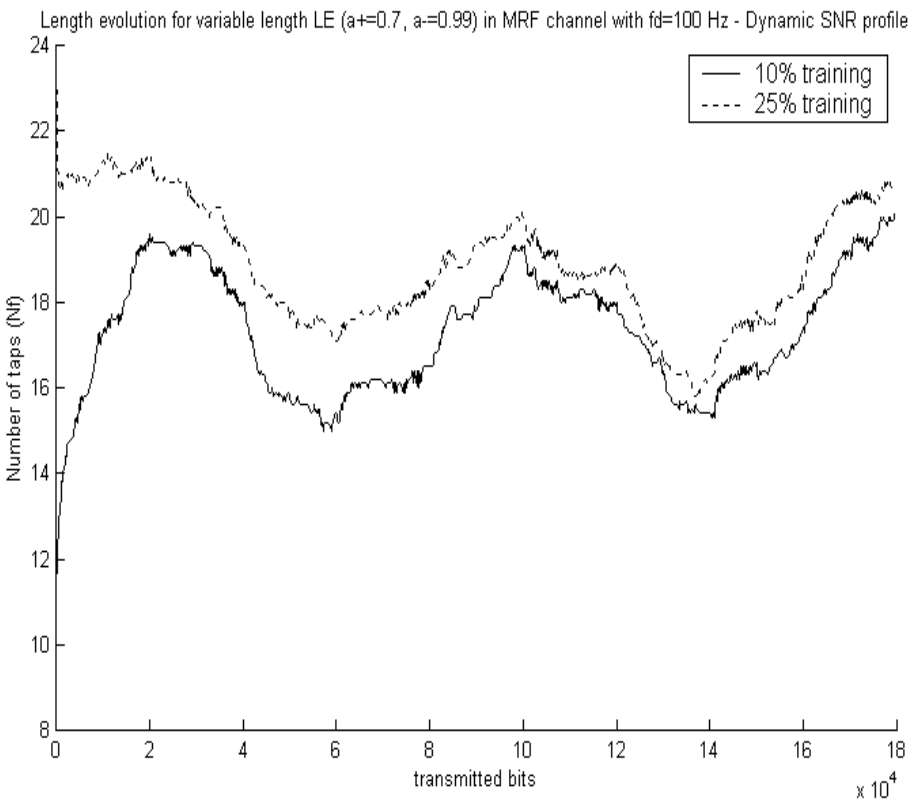

**Figure 7.13**: Equaliser length variations for a VL LE ($\alpha_{up}=0.7$, $\alpha_{dw}=0.99$) operating in a mobile channel ($f_d$=100 Hz) and variable E/No profile.

In figure 7.12 and 7.13 results are presented when $\alpha_{up}$ and $\alpha_{dw}$ are chosen to be 0.7 and 0.99 respectively. These particular values for $\alpha_{up}$ and $\alpha_{dw}$ tend to make the equaliser more prone to enlargement and less likely to contract. This mode of operation could be used when the maximum performance from the equaliser needs to be achieved and power consumption limitations are somewhat relaxed. Notice in figure 7.12 that the windowed MSE has slightly decreased with respect to the previous set of parameters at the cost of an overall increase in the equaliser length (figure 7.13).

In order to get an idea of how the variable length equalisers perform with respect to a classical equaliser some simulations need also to be run using a fixed length equaliser. The choice of the equaliser length is rather arbitrary and in this case, the common rule of thumb of making the equaliser 2N+1 taps long (N=channel length) has been used. Knowing that the channel is 11 taps long, the equaliser length was set to 23 taps.

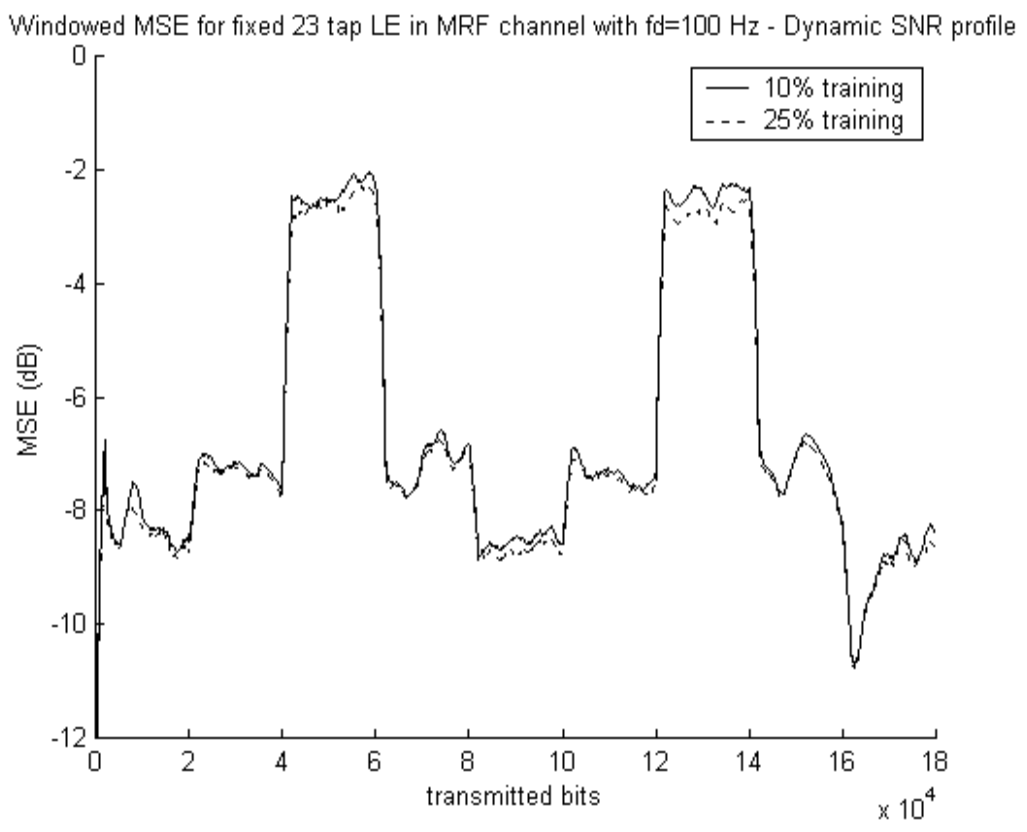

**Figure 7.14**: Windowed MSE for a fixed LE (23 taps) operating in a mobile channel ($f_d$=100 Hz) and variable E/No profile.

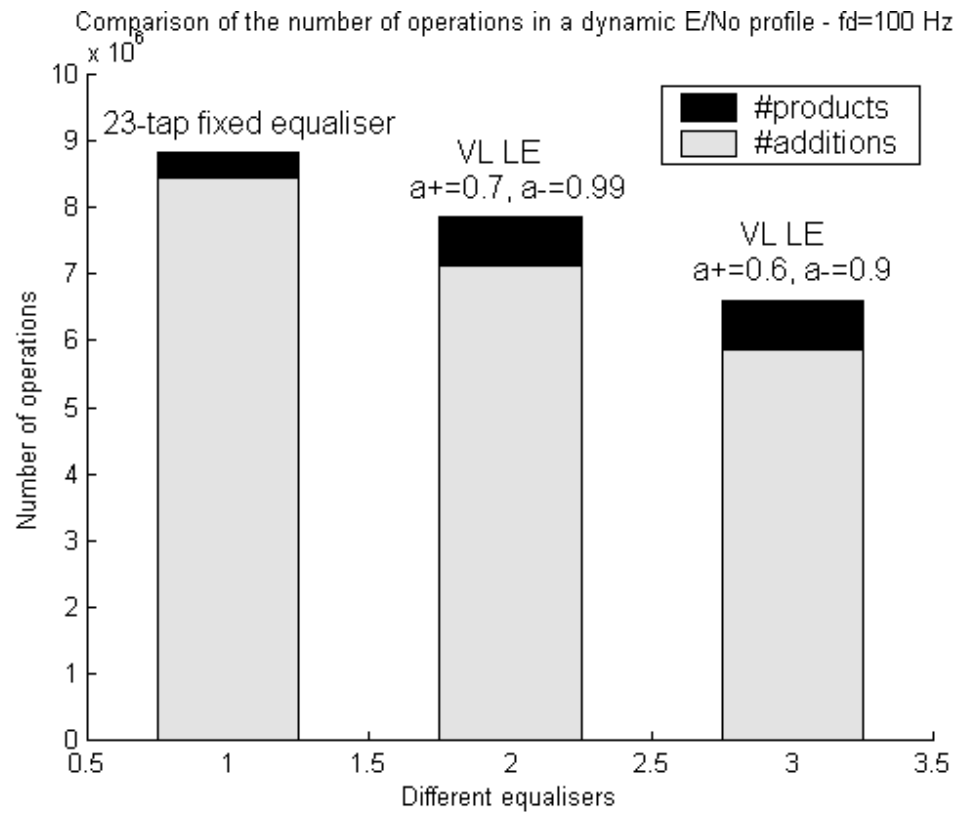

**Figure 7.15**: Comparison of the number of operation of the fixed and variable length linear equalisers.

The resulting MSE curves using the fixed equaliser are shown in figure 7.14. The performance is very similar to that shown in figure 7.12 and slightly better than the one in figure 7.10. In figure 7.15 comparisons of the number of operations performed during the whole simulation by the three equalisers are shown. The variable length LE with $\alpha_{up}$ =0.7, $\alpha_{dw}$ =0.99 performs around 15% fewer additions and products than a 23-tap fixed equaliser. The one with $\alpha_{up}$ =0.6, $\alpha_{dw}$ =0.9 offers a reduction of up to 30% in the number of operations with respect to the fixed equaliser, although in this case some performance degradation is observed. The savings of the variable length structure come from the fact that it is able to reduce its length when E/No is low.

It is worth stressing one point of this previous set of simulations: when choosing the length of the fixed equaliser, although using a rule of thumb, some knowledge has been assumed about the channel length. When using the variable length LE no such assumption needs to be made as the equaliser will adapt to an adequate length. Also, notice that the potential savings offered by the variable length LE depend strongly on the specific channel conditions. Nonetheless, the variable LE has been shown to work satisfactorily in many different scenarios.

### 7.2.3 Abrupt channel change

The next realistic scenario tested resembles that of section 5.4.5 in that there is also an abrupt change in the channel profile, but now the channel is also subject to Doppler spread. The simulation starts using the reduced COST207-TU channel model and after 100,000 iterations, it suddenly changes to a static single path channel. After another 100,000 iterations, the channel changes back to its original model.

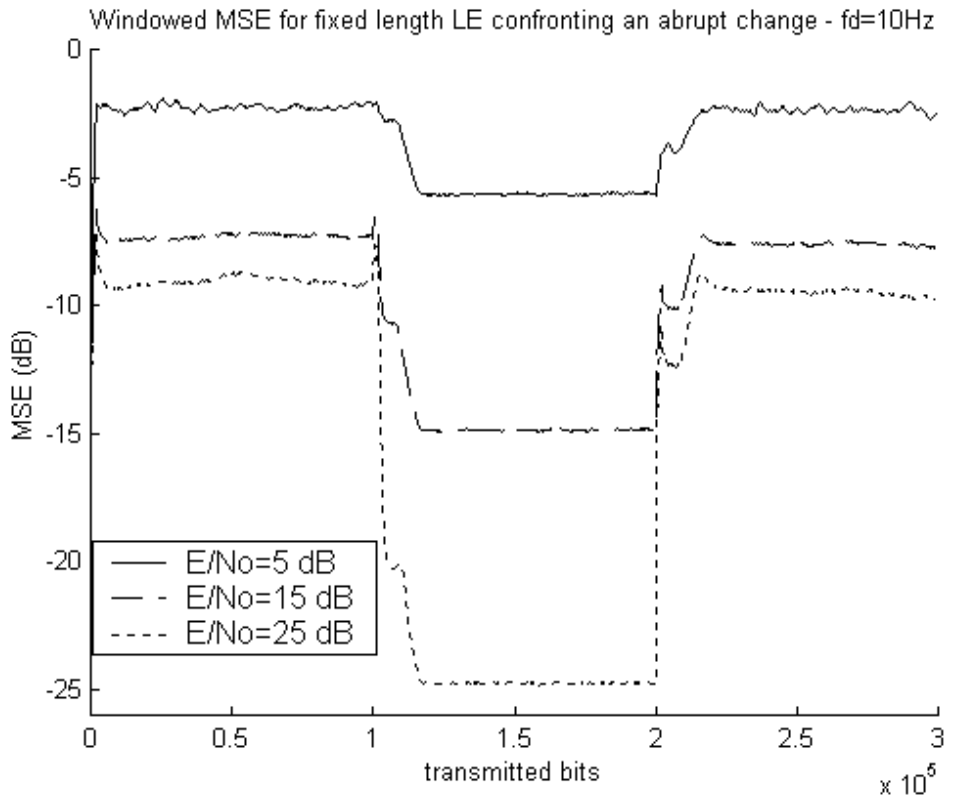

**Figure 7.16**: Windowed MSE for a fixed LE (23 taps) operating in a mobile channel ($f_d$=10 Hz) confronting an abrupt channel change.

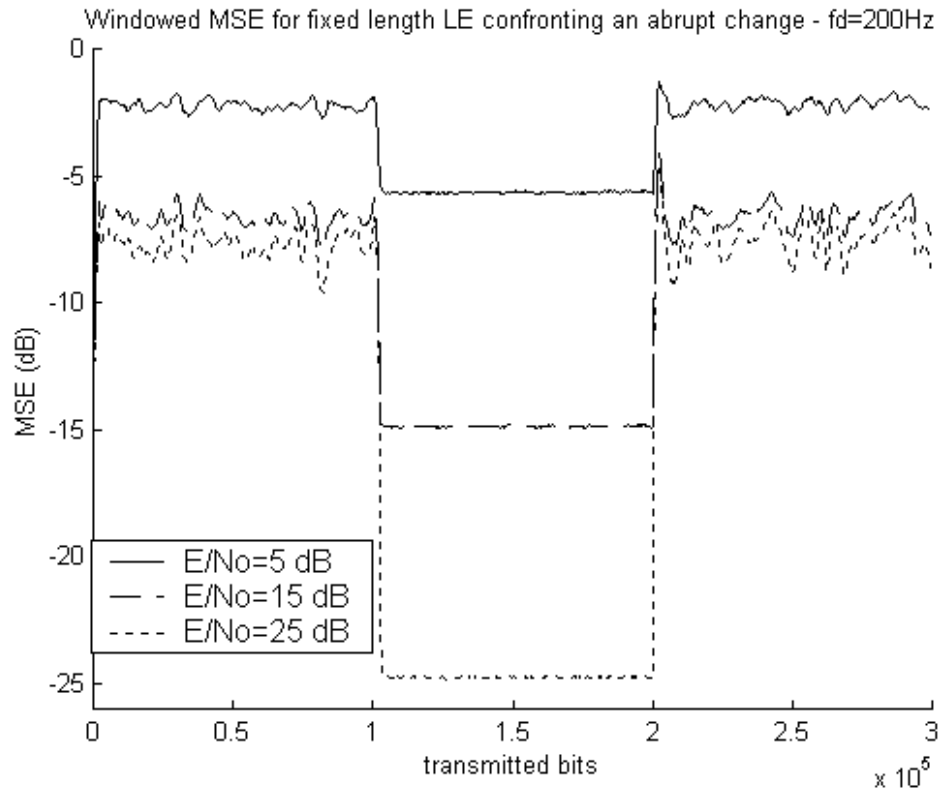

**Figure 7.17**: Windowed MSE for a fixed LE (23 taps) operating in a mobile channel ($f_d$=200 Hz) confronting an abrupt channel change.

Figures 7.16 and 7.17 show the windowed MSE curves for $f_d$=10 Hz and $f_d$=200 Hz respectively. These have been obtained when using a fixed equaliser whose length is again selected to be 23 taps. This type of abrupt channel change is common in mobile scenarios, as a consequence of a handover from one base station to another or a sudden change in the environment such as entering/exiting a building. Note the significant drop in the MSE after the first abrupt change as a result of the system operating in a far more favourable channel. It is also worth appreciating that the higher Doppler of the channel in figure 7.17 causes larger MSE fluctuations than in the results shown in figure 7.16.In figures 7.18 ($f_d$=10 Hz) and 7.19 ($f_d$=200 Hz) results are shown for the same simulation conditions but now using a variable length LE with $\alpha_{up}$=0.6 and $\alpha_{dw}$=0.9. Recall that this set of parameters will tend to keep the equaliser short.

The windowed MSE curves show the same pattern as in the case of the fixed length equaliser although in the higher E/No some increase in the MSE level can be observed, specially for E/No=25 dB.

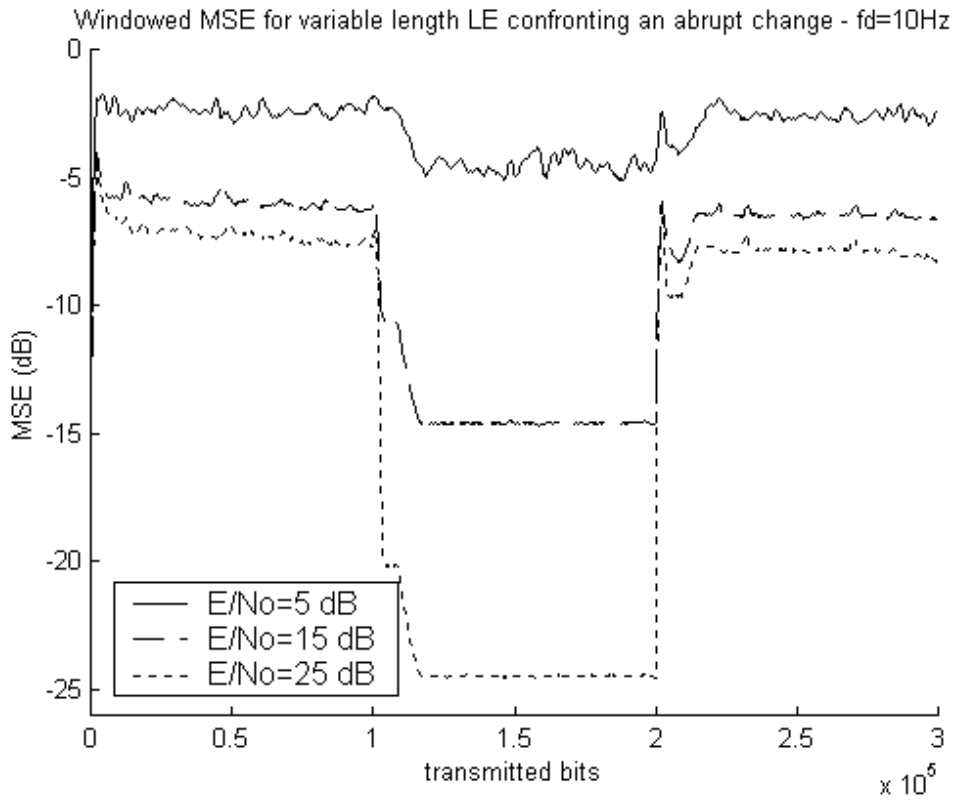

**Figure 7.18**: Windowed MSE for a VL LE ($\alpha_{up}$=0.6, $\alpha_{dw}$=0.9) operating in a mobile channel ($f_d$=10 Hz) confronting an abrupt channel change.

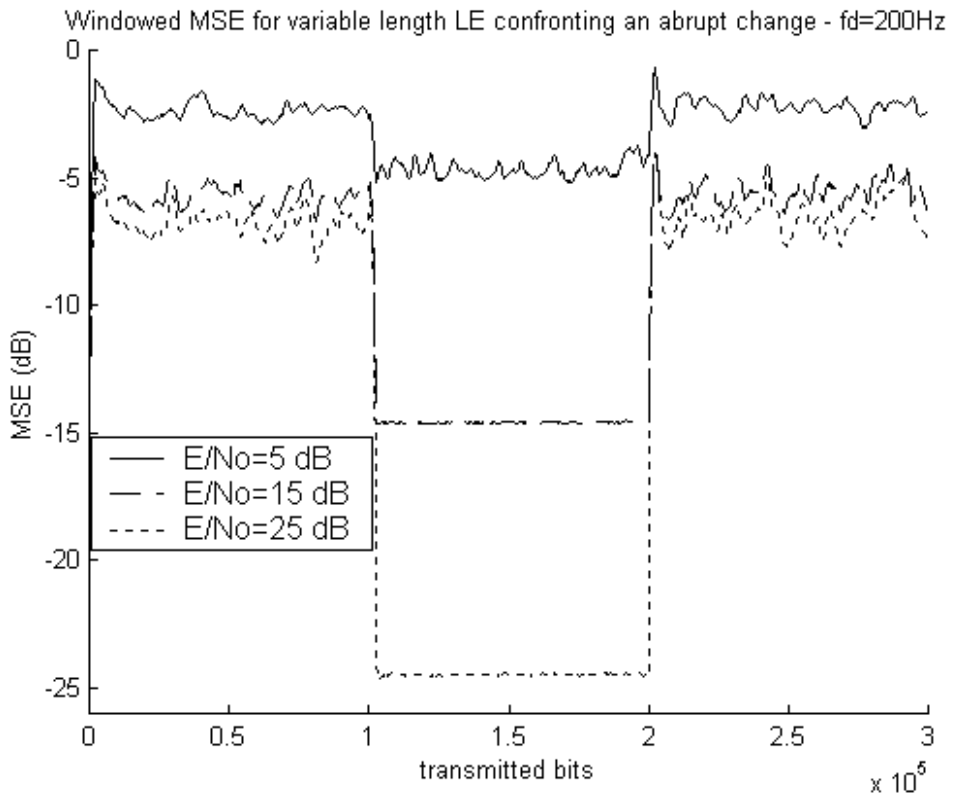

**Figure 7.19**: Windowed MSE for a VL LE ($\alpha_{up}$=0.6, $\alpha_{dw}$=0.9) operating in a mobile channel ($f_d$=200 Hz) confronting an abrupt channel change.

If the variable length equaliser parameters are changed to $\alpha_{up}$=0.7 and $\alpha_{dw}$=0.99, the MSE levels plotted in figures 7.20 and 7.21 are seen to approximate those obtained when using the 23-tap fixed equaliser. This is because these parameters tend to make the equaliser more sensitive to expansion. Still it can be observed that the variable length structure takes more time to converge to the lowest MSE level for each channel model when compared to the fixed equaliser. This is due to the fact that each additional segment is allowed to converge before adding another one, slightly increasing in this way the overall convergence time of the equaliser.

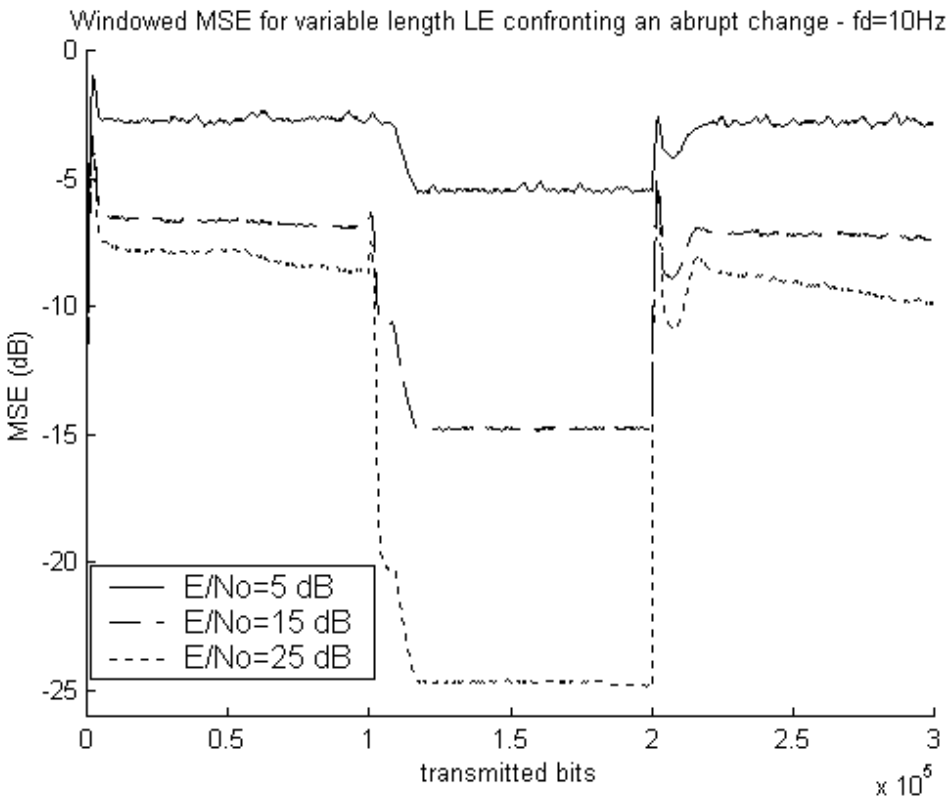

**Figure 7.20**: Windowed MSE for a VL LE ($\alpha_{up}$ =0.7, $\alpha_{dw}$ =0.99) operating in a mobile channel ($f_d$=10 Hz) confronting an abrupt channel change.

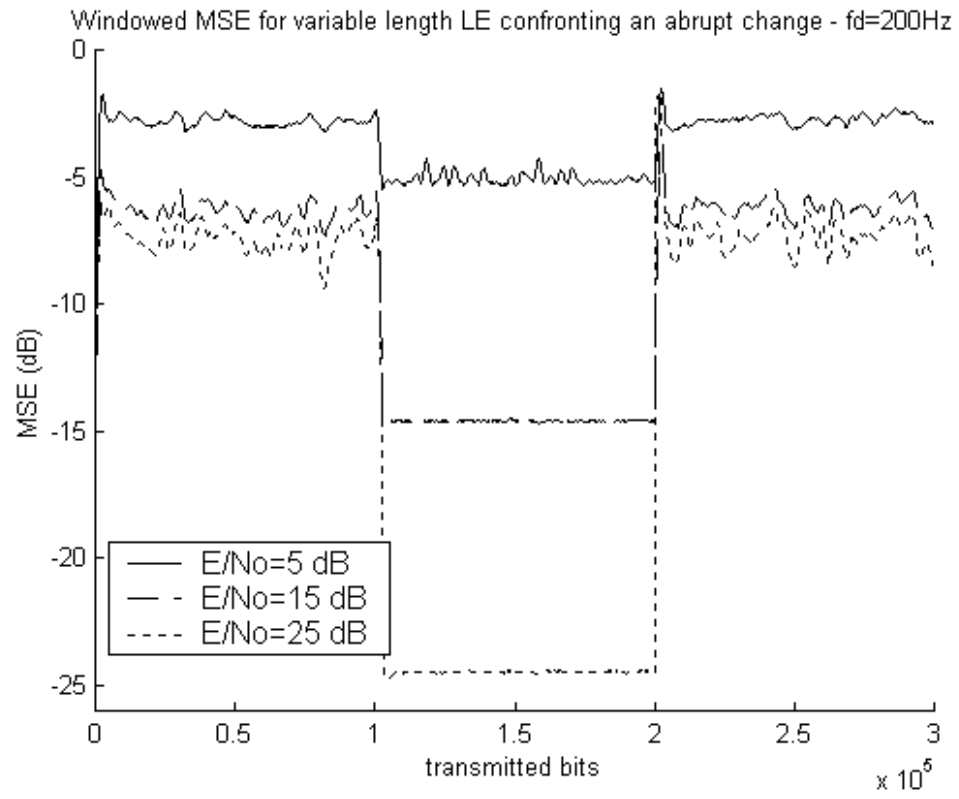

**Figure 7.21**: Windowed MSE for a VL LE ($\alpha_{up}$ =0.7, $\alpha_{dw}$ =0.99) operating in a mobile channel ($f_d$=200 Hz) confronting an abrupt channel change.

Convergence could be improved by choosing an even larger $\alpha_{up}$, such as 0.8 or 0.85. However, we have chosen to keep $\alpha_{up}$ at 0.7 because, as will be shown later, this slower convergence does not affect the equaliser performance in a significant way from the BER point of view.

It is interesting to check how the length update algorithm varies the number of taps of the equaliser depending on the channel in which the equaliser operates. The equaliser length variations for the different set of parameters and Doppler levels are shown in figures 7.22 to 7.25. In all four figures it can be clearly seen that the variable length LE recognises the channel changes and adjusts its length appropriately.

Focusing on figures 7.22 and 7.23 obtained when using $\alpha_{up}$ =0.6 and $\alpha_{dw}$ =0.9, and in particular in the case of $f_d$=10 Hz, notice that the equaliser length grows steadily. This effect is especially noticeable when E/No=25 dB.

Also important is to highlight that when E/No=5 dB, the algorithm correctly recognises that there is no improvement in expanding the equaliser and therefore is kept at 11-12 taps. Turning now attention to the case of $f_d$=200 Hz (figure 7.23), it is very clear that the equaliser changes length far more often than when fd=10 Hz. This is perfectly logical as the faster channel changes make the optimum number of taps more likely to vary. In the case of a large Doppler effect, the equaliser tends to be kept slightly shorter than when the Doppler spread is low. This could also be seen in figure 7.5, when fixed equalisers of different lengths were tested under various levels of Doppler spread.

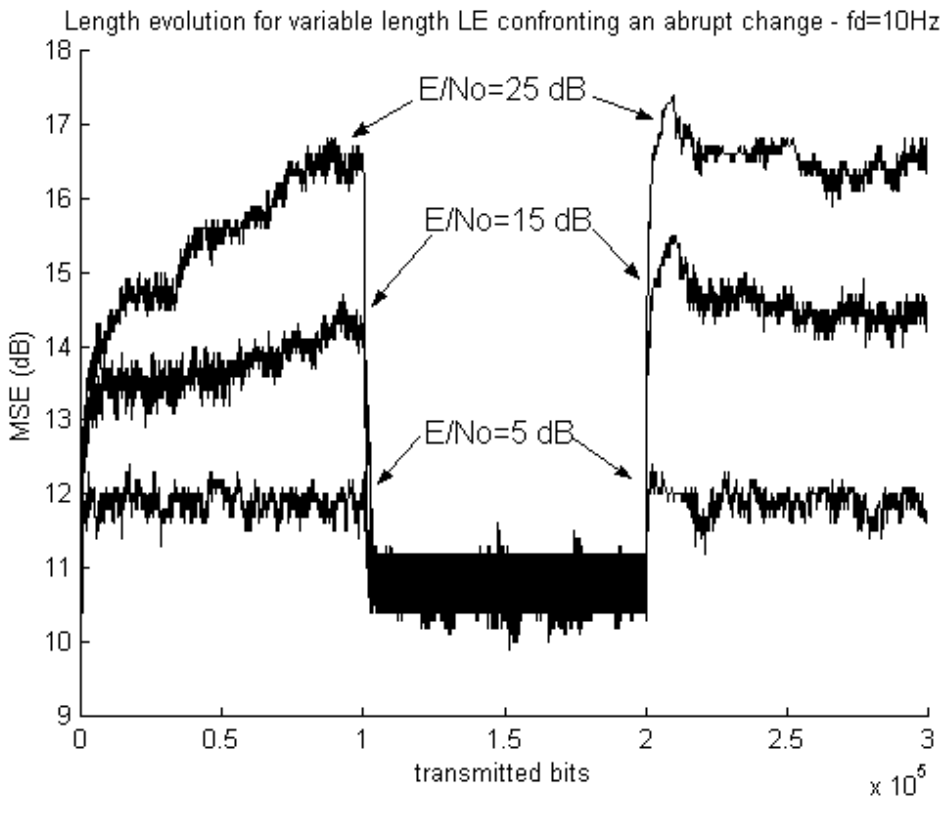

**Figure 7.22**: Length evolution for a VL LE ($\alpha_{up}$ =0.6, $\alpha_{dw}$ =0.9) operating in a mobile channel ($f_d$=10 Hz) confronting an abrupt channel change.

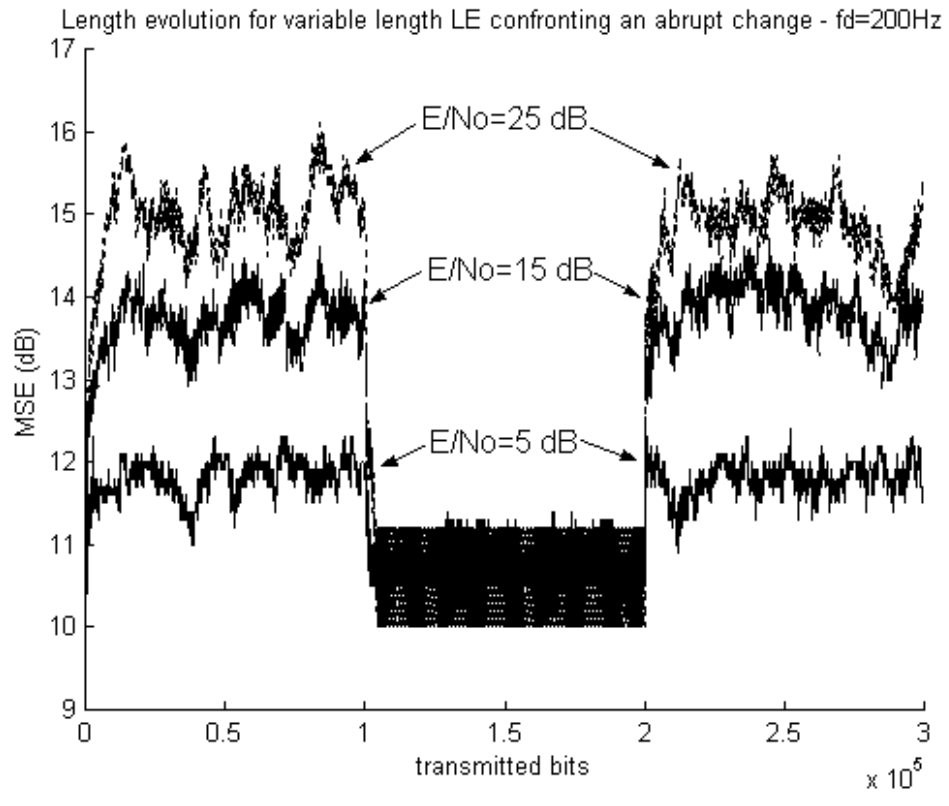

**Figure 7.23**: Length evolution for a VL LE ($\alpha_{up}$ =0.6, $\alpha_{dw}$ =0.9) operating in a mobile channel ($f_d$=200 Hz) confronting an abrupt channel change.

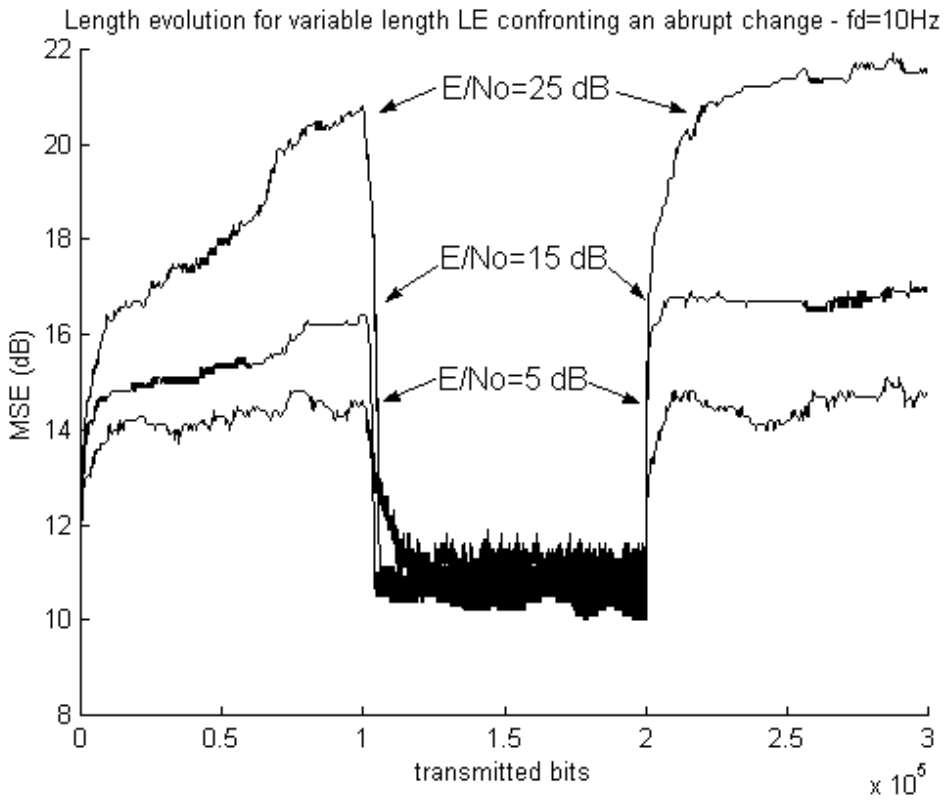

**Figure 7.24**: Length evolution for a VL LE ($\alpha_{up}$ =0.7, $\alpha_{dw}$ =0.99) operating in a mobile channel ($f_d$=10 Hz) confronting an abrupt channel change.

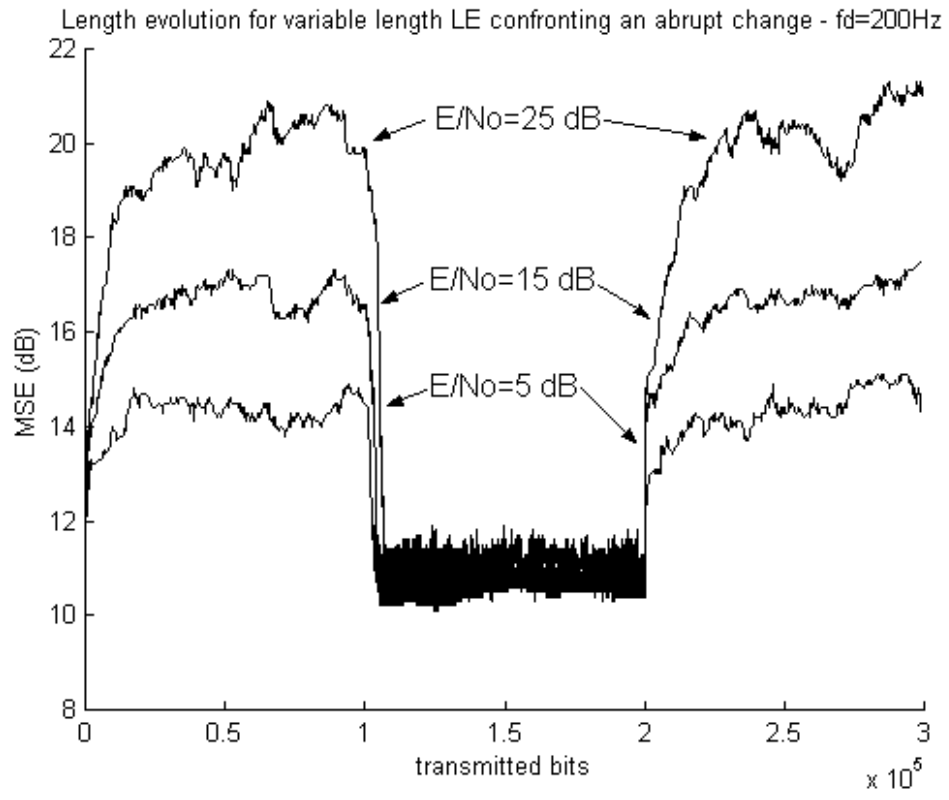

**Figure 7.25**: Length evolution for a VL LE ($\alpha_{up}$ =0.7, $\alpha_{dw}$ =0.99) operating in a mobile channel ($f_d$=200 Hz) confronting an abrupt channel change.

Figures 7.24 and 7.25 show the length evolution when $\alpha_{up}$ =0.7 and $\alpha_{dw}$ =0.99 are used to control the equaliser expansion. When compared with the previous set of parameters (figures 7.22 and 7.24) it can be appreciated that equaliser tends to get longer. Moreover, the equaliser expansion takes place more rapidly. Still it can be noticed that when E/No is large (25 dB), the expansion takes place very rapidly up to 16 taps and then slowly increases to 21 taps. This slowness in the expansion of the last taps is what motivates the slower convergence of the MSE level with respect to a fixed equaliser, as already noted in figure 7.20. However, it is important to point out that this expansion takes place slowly because these last taps have a marginal contribution to the overall MSE reduction.

Figures 7.26 and 7.27 present the measured BER at various E/No levels for the different equalisers used in both Doppler scenarios. When $f_d$=10 Hz (figure 7.26) several important issues must be pointed out. First, at low E/No levels, 0 to 10 dB, the three structures achieve the same or nearly the same BER. As the E/No gets larger, differences appear among the different equalisers. At E/No=25 dB, the VL LE(0.6, 0.9) BER is somewhat larger than for the 23-tap fixed equaliser and the VL LE(0.7, 0.99). These last two, although distinguishable, (practically) achieve the same BER for the same E/No levels. When the Doppler spread is set to 200 Hz, the same comments as for $f_d$=10 Hz apply, but now the differences among the three equalisers BER become even smaller.

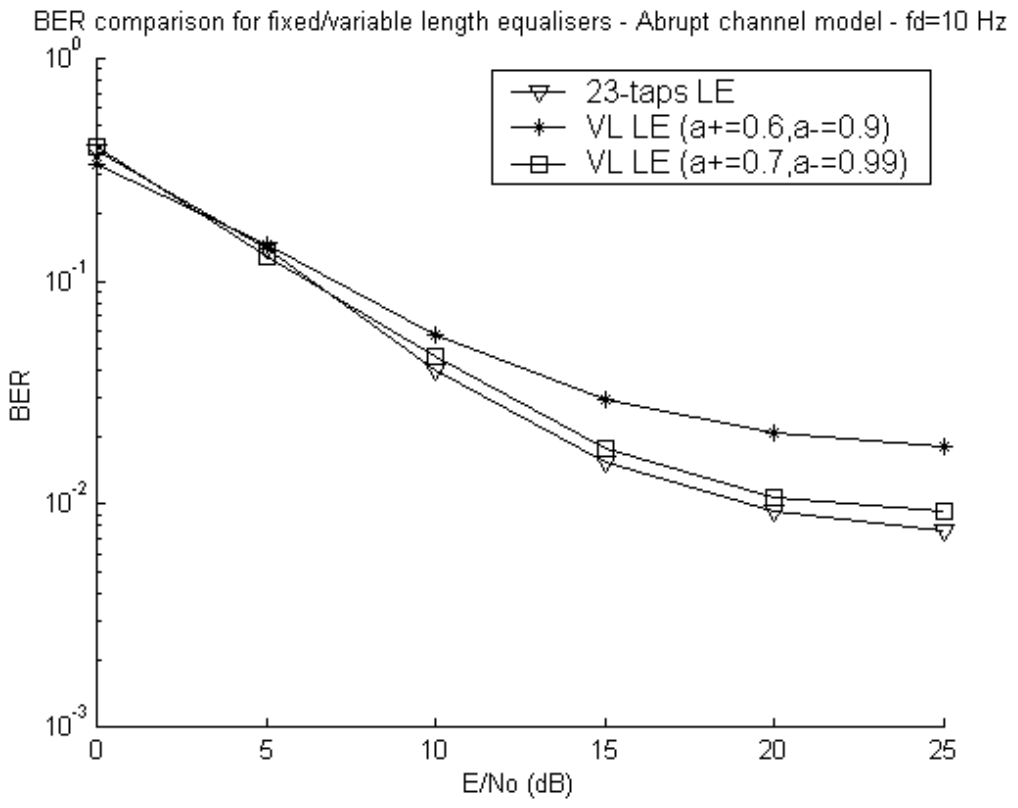

**Figure 7.26**: BER for the different LEs operating in a mobile channel ($f_d$=10 Hz) confronting an abrupt channel change.

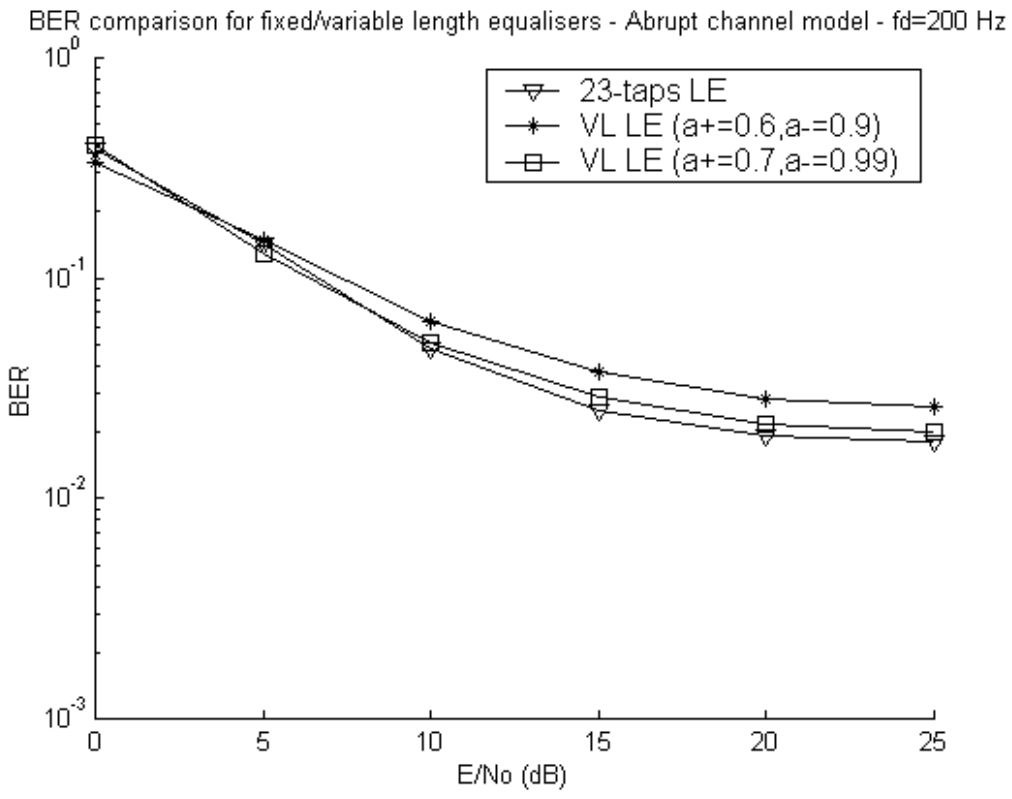

**Figure 7.27**: BER for the different LEs operating in a mobile channel($f_d$=200 Hz) confronting an abrupt channel change.

It is worth noting once more that, in a practical system, the equaliser would be followed by some form of channel decoding. Error correction schemes can easily provide a reduction of one order of magnitude to the BER curves shown in figures 7.26 and 7.27, making the differences among the different curves shown in 7.27 even less significant.

The BER results just shown call for a comparison of the number of computations for the different equalisers, in order to illustrate the significant advantage offered by the variable length structure.

Figures 7.28 to 7.31 compare the number of additions and products of the two variable length LEs used and the fixed 23-tap equaliser for two different E/No levels. When E/No=25 dB and $f_d$=10 Hz, figures 7.28 and 7.29 show that the VL LE(0.7,0.99) performs around 18% fewer additions and 12% fewer products than the fixed 23-tap equaliser, while offering a nearly identical BER. For $f_d$=200 Hz, the savings are slightly larger, as can be observed from figure 7.29.

When in a low E/No condition, such as E/No=5 dB, the savings of the variable length structure are even more significant than in a large E/No. Figures 7.30 and 7.31 show savings of around 35-40 % in the number of operations with respect to the 23-tap equaliser, independently of the Doppler level.

Once more, it is worth emphasising that the profiles used for these simulations are not tailored specially to favour the variable length structures. The nice property of the VL LE is its ability to adjust the number of taps to those that are significant to the equalisation process.

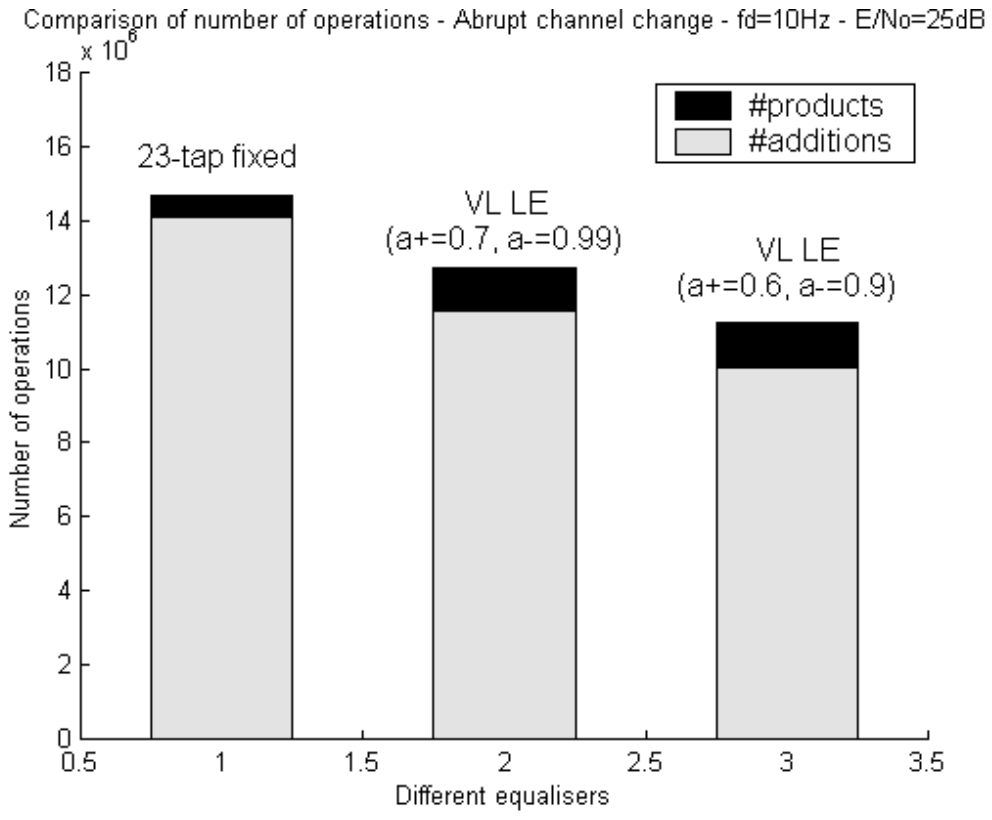

**Figure 7.28**: Comparison of the number of computations for different LEs operating in a mobile channel ($f_d$=10 Hz) confronting an abrupt channel change. E/No=25 dB.

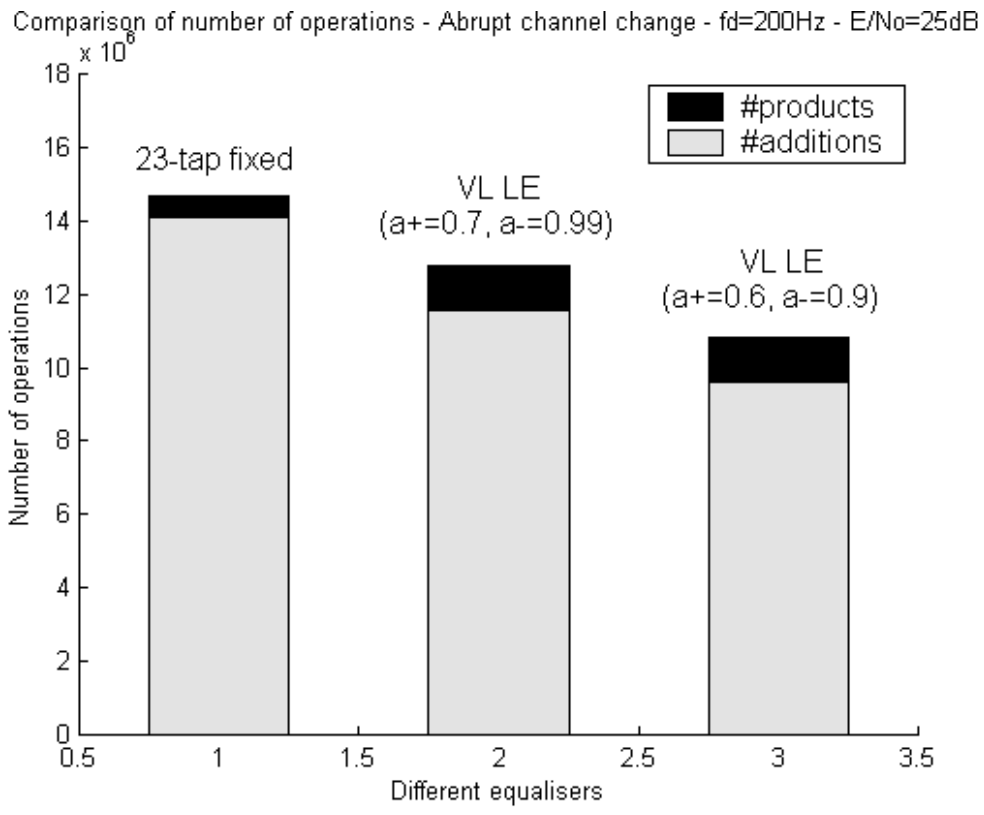

**Figure 7.29**: Comparison of the number of computations for different LEs operating in a mobile channel ($f_d$=200 Hz) confronting an abrupt channel change. E/No=25 dB.

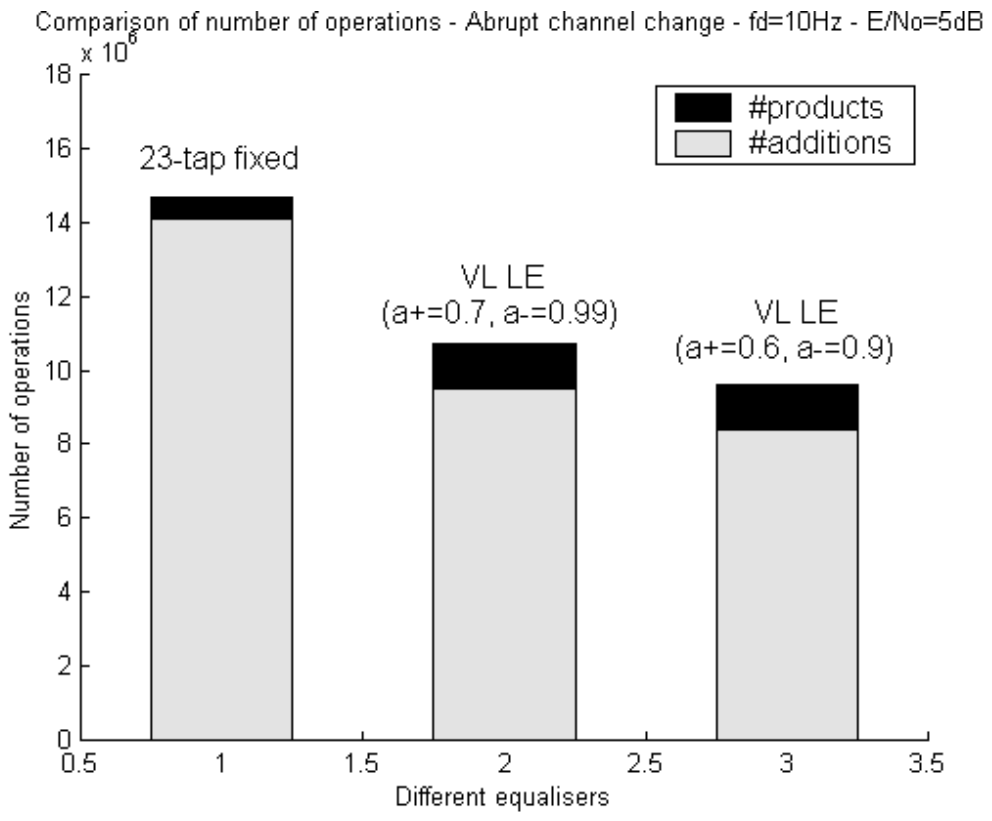

**Figure 7.30**: Comparison of the number of computations for different LEs operating in a mobile channel ($f_d$=10 Hz) confronting an abrupt channel change. E/No=5 dB.

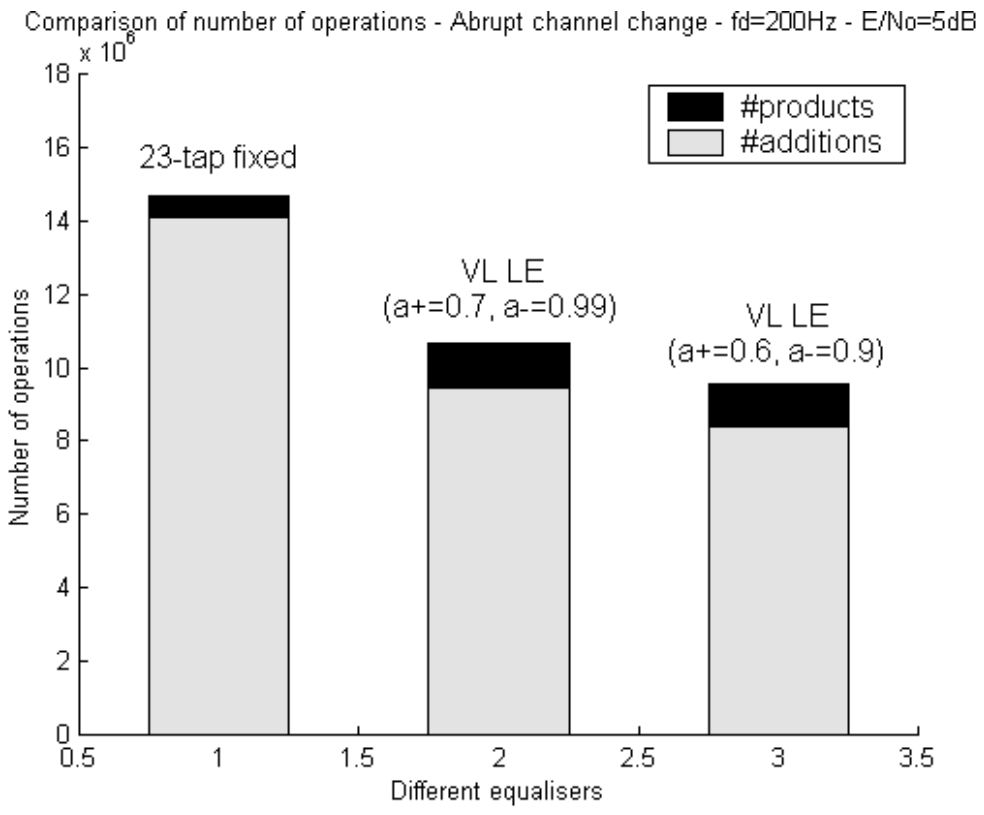

**Figure 7.31**: Comparison of the number of computations for different LEs operating in a mobile channel ($f_d$=200 Hz) confronting an abrupt channel change. E/No=5 dB.

# 7.3 Variable length FBF-DFE in realistic scenarios

This section uses simulation results to examine the performance of a DFE with a variable length FBF (referred to as VL FBF DFE) when operating in a realistic scenario. Its results are compared with those obtained using a DFE with a fixed length FBF.

## 7.3.1 Effect of the FBF length in a realistic scenario

In the last chapter, some simulation results were presented showing the effect the FBF length has on the MSE level when operating in a static channel. Now the study is repeated but with the additional feature that the channel is subject to Doppler spread. For the simulations in this and future sections, the FFF length has been set to 6 taps, as making the FFF longer will not reduce the MSE (see Chapter 6). The delay has been set to 5 samples ($N_f$-1). The training frequency had the same characteristics as in the simulations for the LE: 200 training symbols every 2000 symbols.

Figures 7.32 and 7.33 show the steady state MSE (after 100,000 iterations) at the output of the equaliser when compensating the reduced COST207-TU channel model. Notice that a priori, given that the channel has 11 taps and the particular choice of delay, there will be 10 postcursors to cancel in the combined channel-FFF impulse response. This means that making the FBF longer than 10 taps will not decrease the MSE at all.

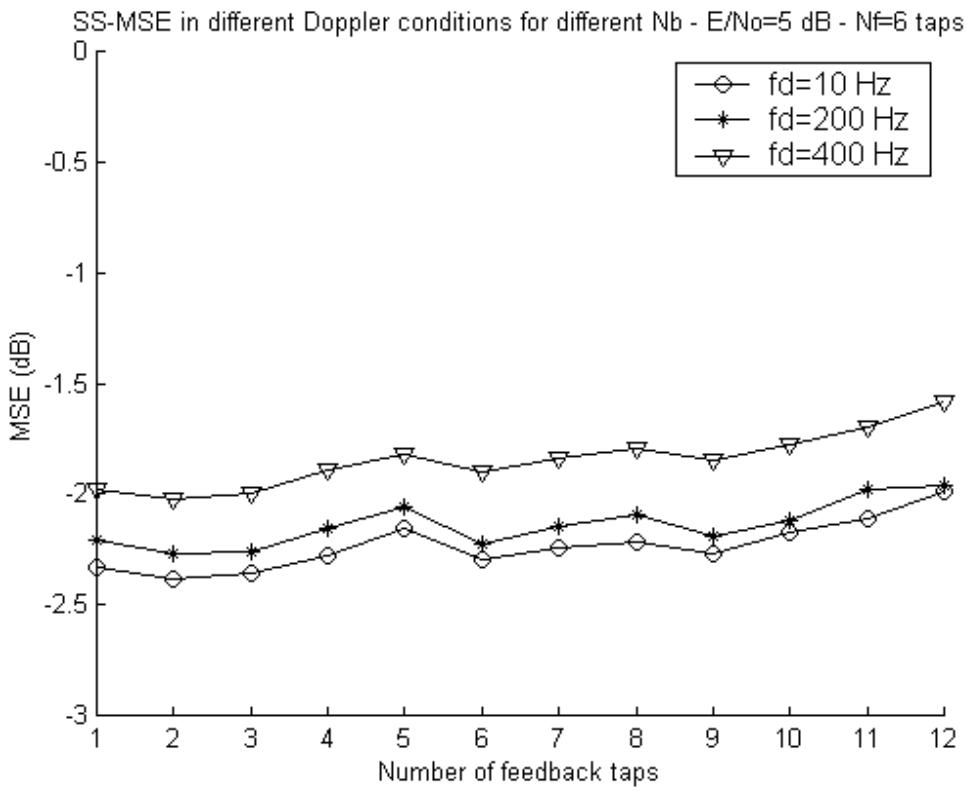

**Figure 7.32**: MSE for DFE using different lengths FBF operating in a mobile channel with different Doppler levels. $N_f$=6 taps. E/No=5 dB.

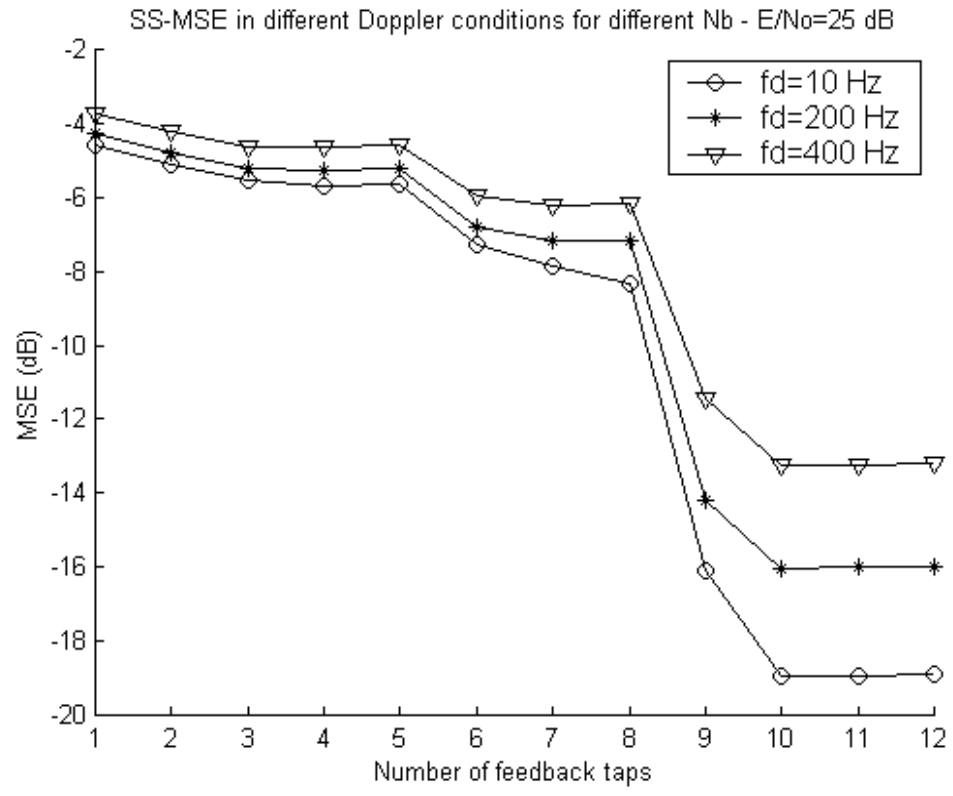

**Figure 7.33**: MSE for DFE using different lengths FBF operating in a mobile channel with different Doppler levels. $N_f$=6 taps. E/No=25 dB.

In figure 7.32 (E/No=5 dB) it can be seen that, indeed, there is no significant improvement in expanding the FBF beyond 2 taps. This is due to the fact that with such a low E/No, many of the fedback decisions are wrong. Feeding back incorrect decisions coupled with channel fluctuations make the convergence of the FBF rather difficult. Additionally, the longer the FBF, the longer the time wrong decisions will affect future outputs.

When E/No=25 dB, the situation is rather different and all FBF taps, up to 10, help to decrease the MSE level. Notice however that after 10 taps no reduction in MSE can be achieved by adding more taps.

Figures 7.34 to 7.37 show the results obtained when using a DFE with a variable length FBF. For the VL FBF DFE one parameter, $\chi$, need to be defined. This parameter controls the likeliness of the FBF to change its length. In this case, it was set to $\chi$=0.01. The initial length of the FBF was set to 2 taps. Figures 7.34 and 7.35 correspond to the MSE and length evolution curves respectively when E/No=5 dB. It can be seen that the FBF expands up to around 8 taps, whereas figure 7.32 predicted that taps beyond the first two would not reduce the MSE. The source of this discrepancy is due to the algorithm controlling the FBF length, which relies on imperfect decisions. However, the important point is that even in such a low E/No, the algorithm does not expand the number of taps beyond the channel length.

When E/No=25, the FBF expands up to 11 taps (figure 7.37) independently of the Doppler spread level, which is the result to be expected a-priori. In theory, only 10 feedback taps would be required to cancel all the postcursor interference. However, recall that the FBF length update algorithm is based on using an extra tap to detect changes in the channel impulse response.

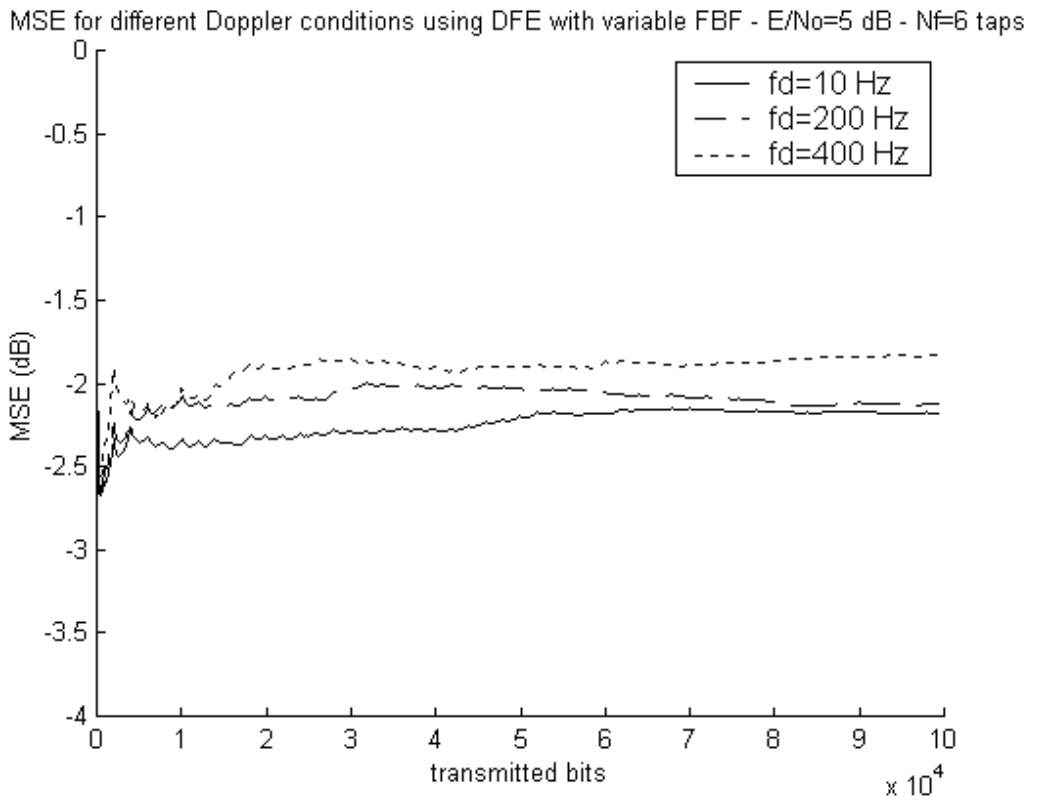

**Figure 7.34**: MSE for a VL FBF DFE operating in a mobile channel with different Doppler levels. E/No=5 dB.

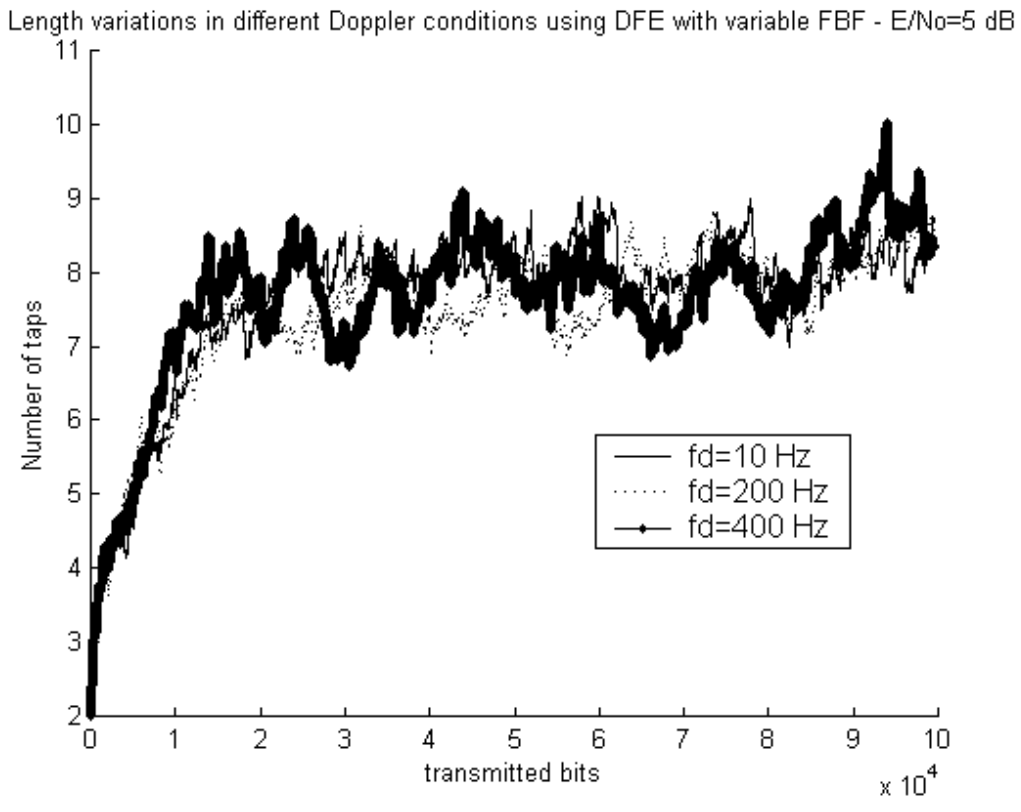

**Figure 7.35**: FBF length fluctuations when operating in a mobile channel with different Doppler levels. E/No=5 dB.

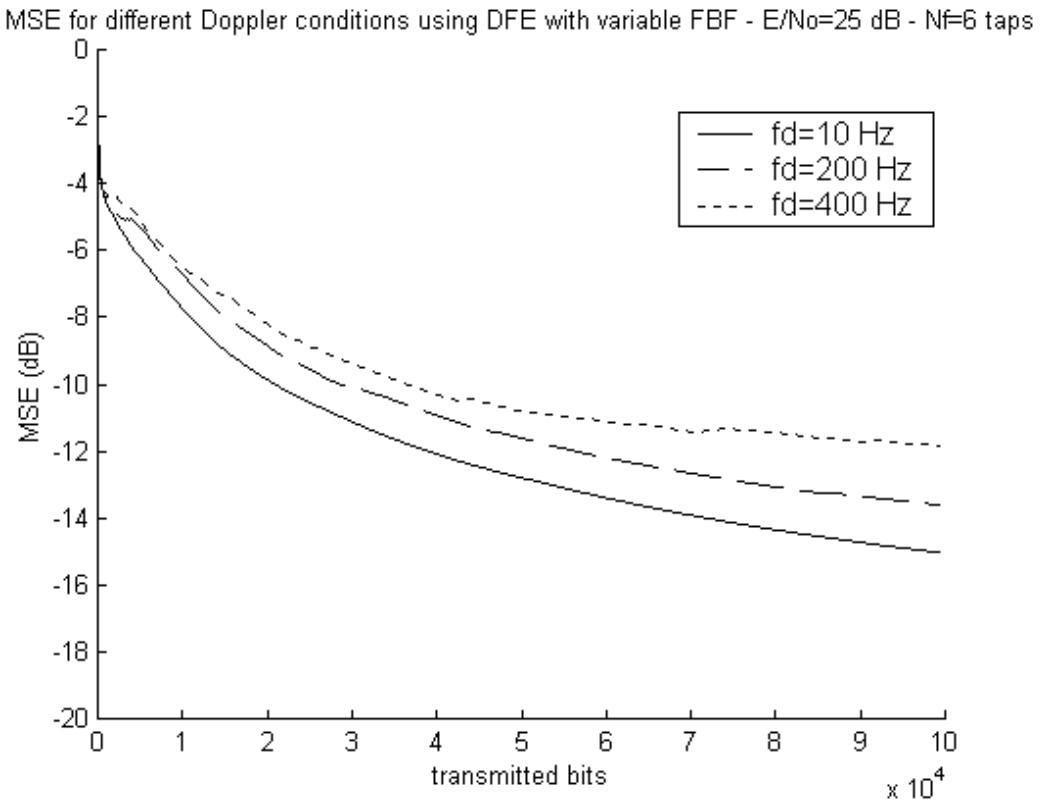

**Figure 7.36**: MSE for a VL FBF DFE operating in a mobile channel with different Doppler levels. E/No=25 dB.

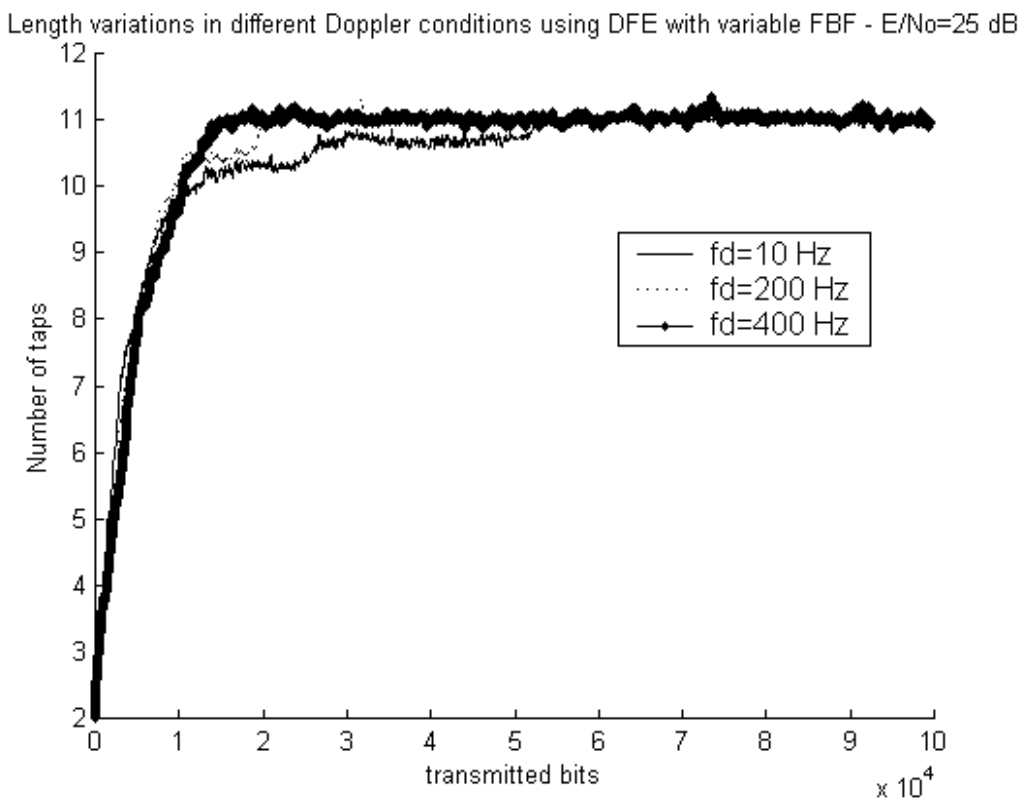

**Figure 7.37**: FBF length fluctuations when operating in a mobile channel with different Doppler levels. E/No=25 dB.

## 7.3.2 Abrupt channel change

To conclude this set of simulations, the behaviour VL FBF DFE in response to a realistic abrupt change in the channel profile has been checked. The sudden channel change is exactly the same as the one used in section 7.2.3, namely, starting with the reduced COST207-TU channel and changing, after 100,000 iteration, to a static single path model. After another 100,000 bits, the channel changes back to its initial profile. The scenario has been verified in two different Doppler conditions, $f_d$=10 Hz and $f_d$=200 Hz.

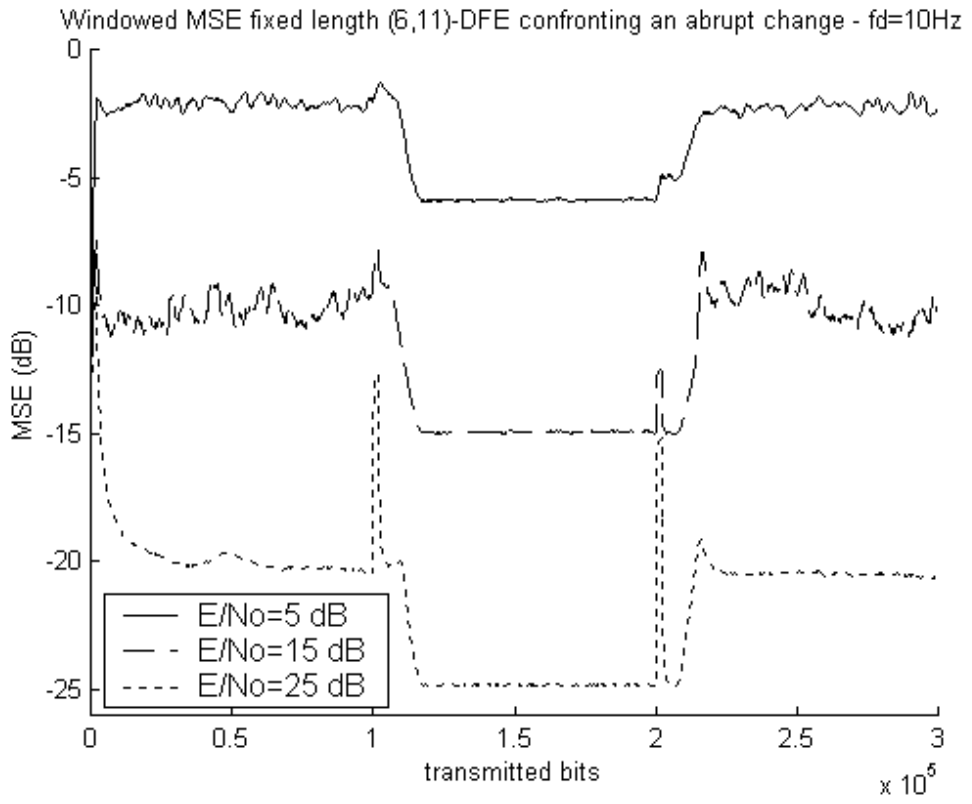

**Figure 7.38**: Windowed MSE for a fixed (6,11)-DFE operating in a mobile channel ($f_d$=10 Hz) confronting an abrupt channel change.

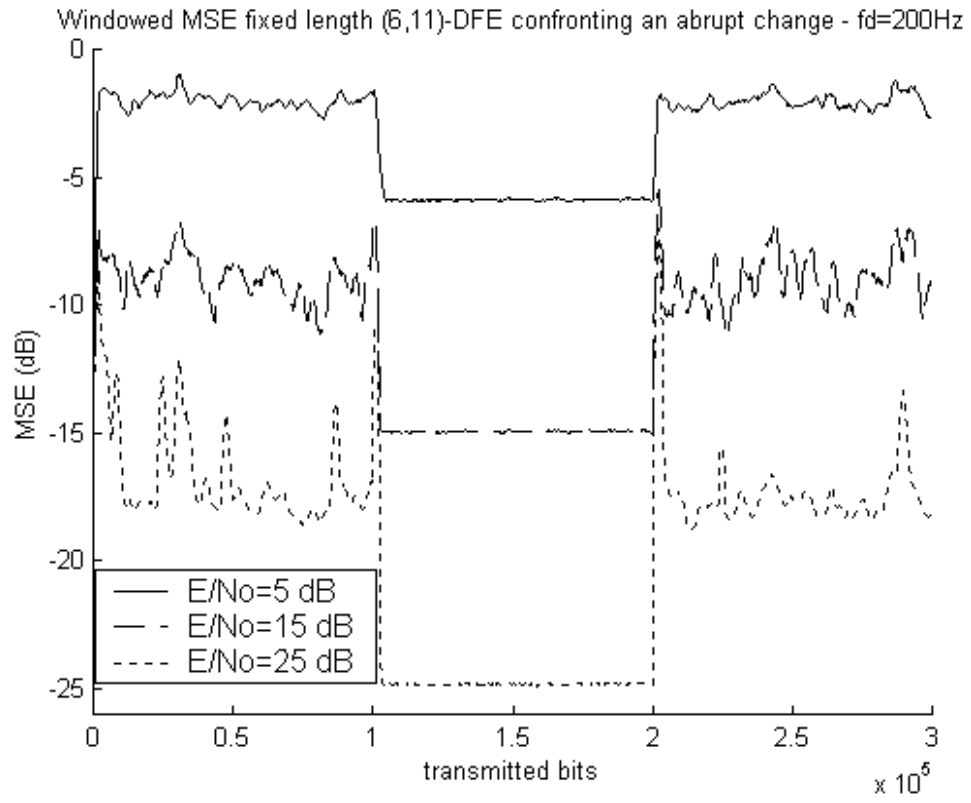

**Figure 7.39**: Windowed MSE for a fixed (6,11)-DFE operating in a mobile channel ($f_d$=200 Hz) confronting an abrupt channel change.

In order to be able to compare of the performance the VL FBF DFE with that of a conventional DFE, some simulations have been run using a fixed length (6,11)-DFE. The windowed MSE curves for $f_d$=10 Hz and $f_d$=200 Hz are shown in figures 7.38 and 7.39 respectively. Notice that the MSE oscillations are more significant in the second figure, due to the larger Doppler spread. The sudden peaks in the MSE curves, very apparent for E/No=25 dB, are due to abrupt changes of the channel.

Figures 7.40 to 7.43 present the results obtained when using the VL FBF DFE. Two values of $\chi$ have been used: $\chi$=0.1, which would correspond to an FBF only compensating the main postcursor components, and $\chi$=0.01, which would make the FBF more sensitive to smaller postcursors.

In figures 7.40 and 7.41 the curves for $\chi$=0.1 are shown. When compared with those of the fixed length (6,11)-DFE, a significant increase in the MSE can be appreciated, especially when E/No is large. This is a direct effect of not cancelling some of the postcursor components. When $\chi$=0.01 (figures 7.42 and 7.43) the MSE curves are seen to approximate those obtained with fixed length DFE. Still, and in the same manner as for the variable length LE, the MSE converges slower when using the variable length FBF. This means that the VL FBF DFE will tend to have a slightly larger number of errors during transient phases (i.e.: start of transmission and after abrupt changes).

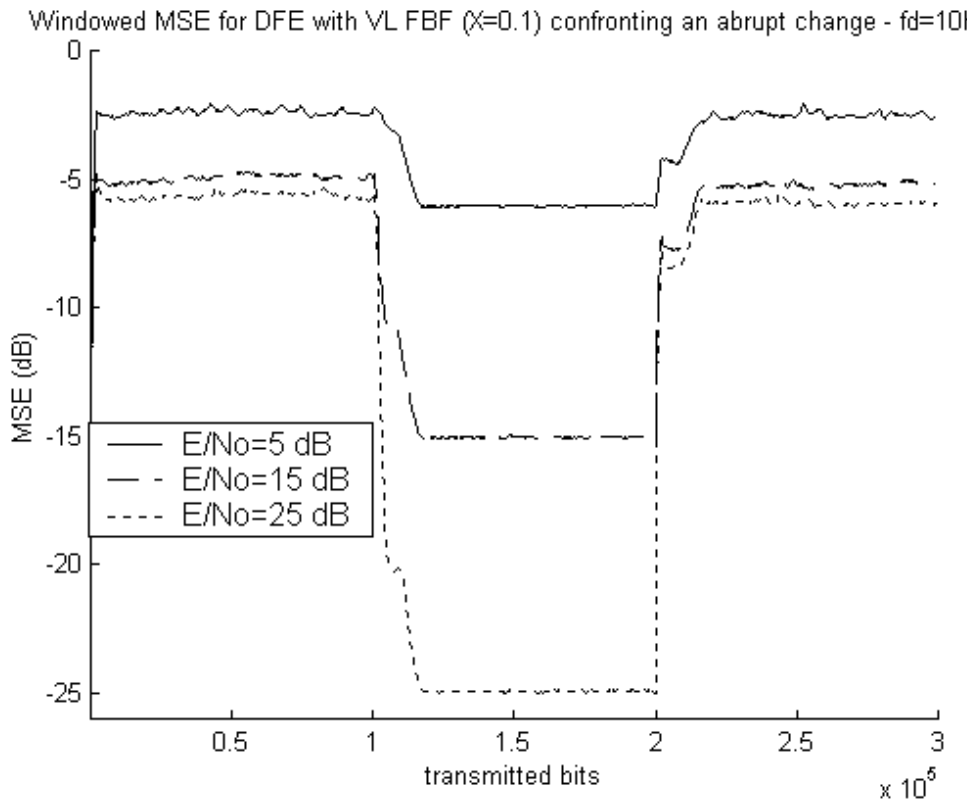

**Figure 7.40**: Windowed MSE for a VL FBF DFE ($\chi$=0.1) operating in a mobile channel ($f_d$=10 Hz) confronting an abrupt channel change.

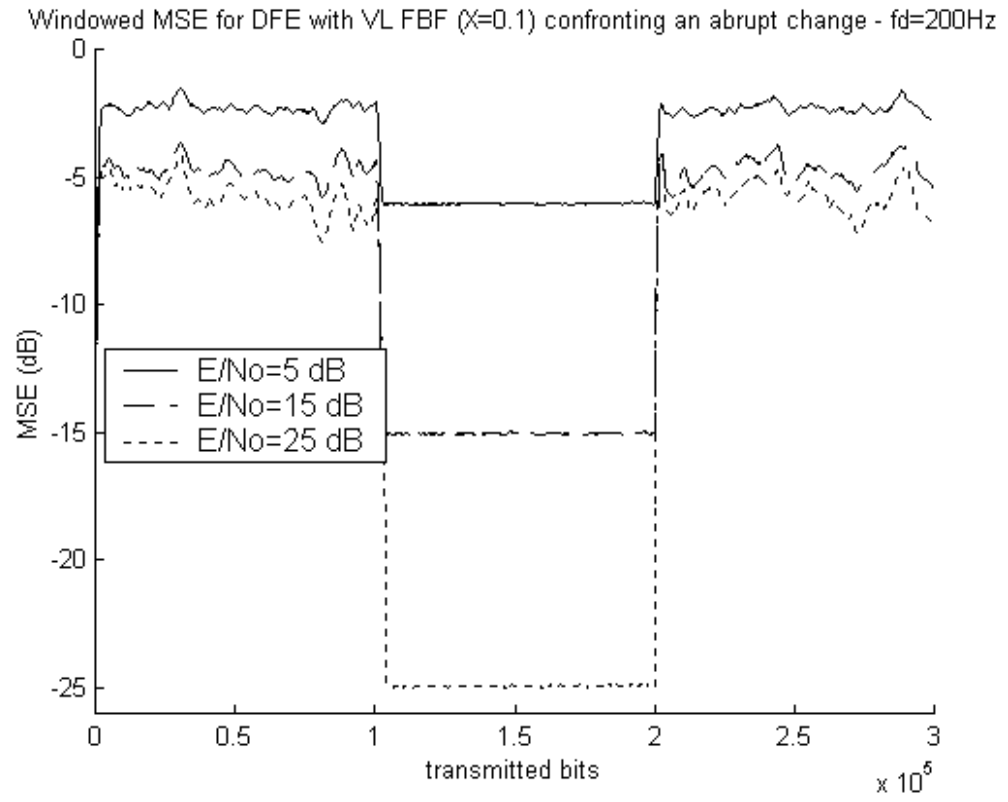

**Figure 7.41**: Windowed MSE for a VL FBF DFE ($\chi$=0.1) operating in a mobile channel ($f_d$=200 Hz) confronting an abrupt channel change.

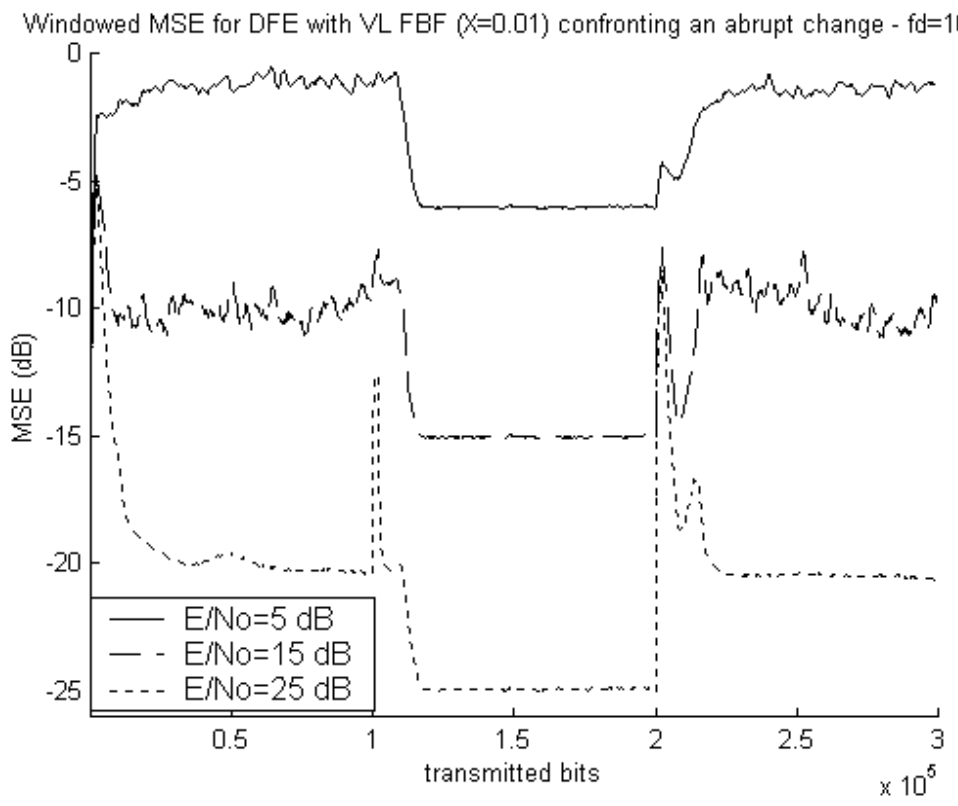

**Figure 7.42**: Windowed MSE for a VL FBF DFE ($\chi$=0.01) operating in a mobile channel ($f_d$=10 Hz) confronting an abrupt channel change.

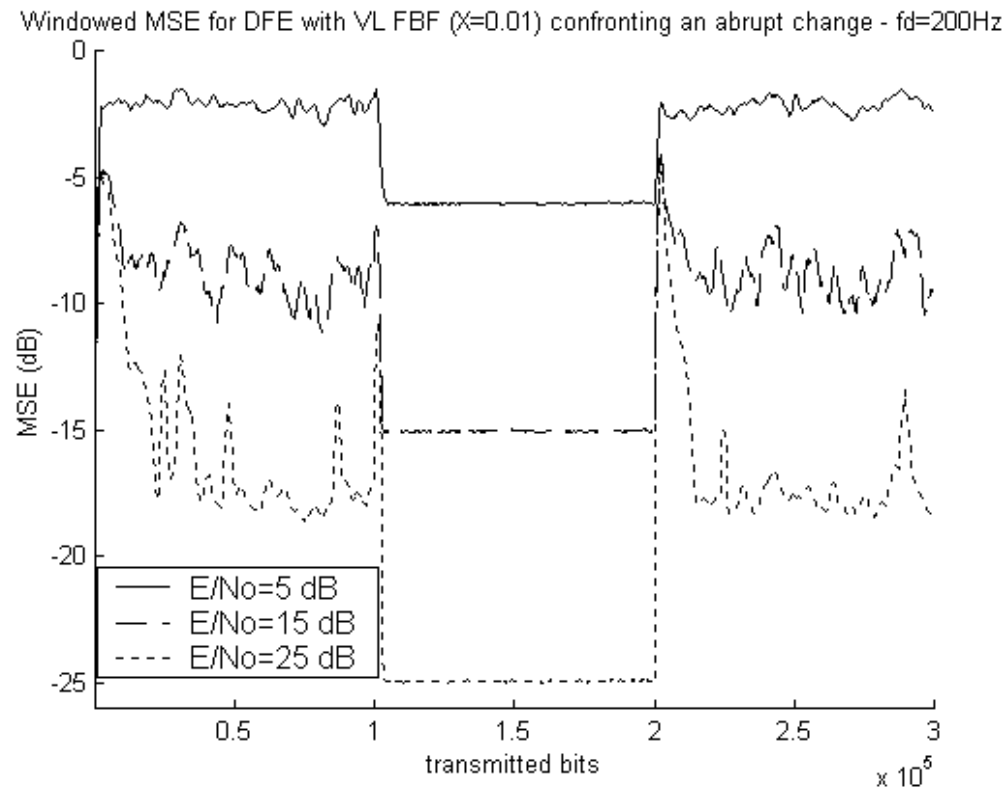

**Figure 7.43**: Windowed MSE for a VL FBF DFE ($\chi$=0.01) operating in a mobile channel ($f_d$=200 Hz) confronting an abrupt channel change.

In theory, and to a certain extent in practice, convergence time could be improved by further reducing $\chi$. However, in this realistic scenario, this has proved to be a difficult adjustment to do as using a very small $\chi$ tends to affect the performance in low E/No (the FBF expands to its maximum value). Therefore, a compromise value for $\chi$ must be determined, which in this case, we have found by simulation to be between 0.0075 and 0.01. An alternative solution to this convergence problem would be to initialise the FBF to its maximum length and let the algorithm controlling the FBF length reduce it appropriately.

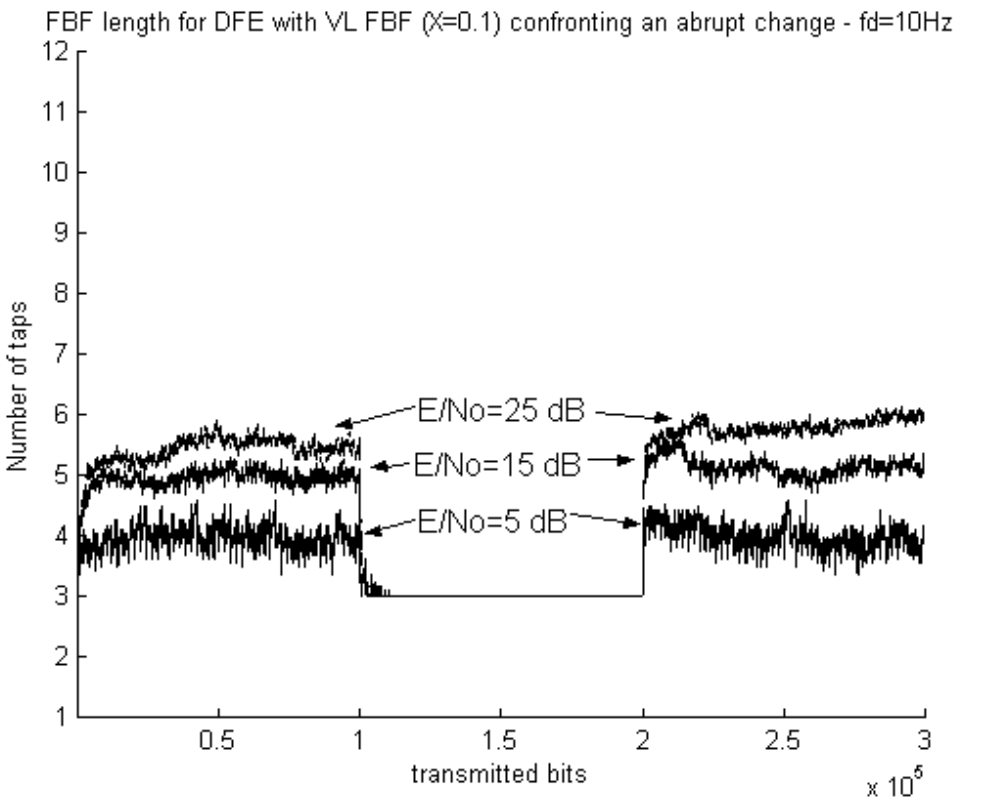

**Figure 7.44**: Length evolution for the FBF ($\chi$=0.1) operating in a mobile channel ($f_d$=10 Hz) confronting an abrupt channel change.

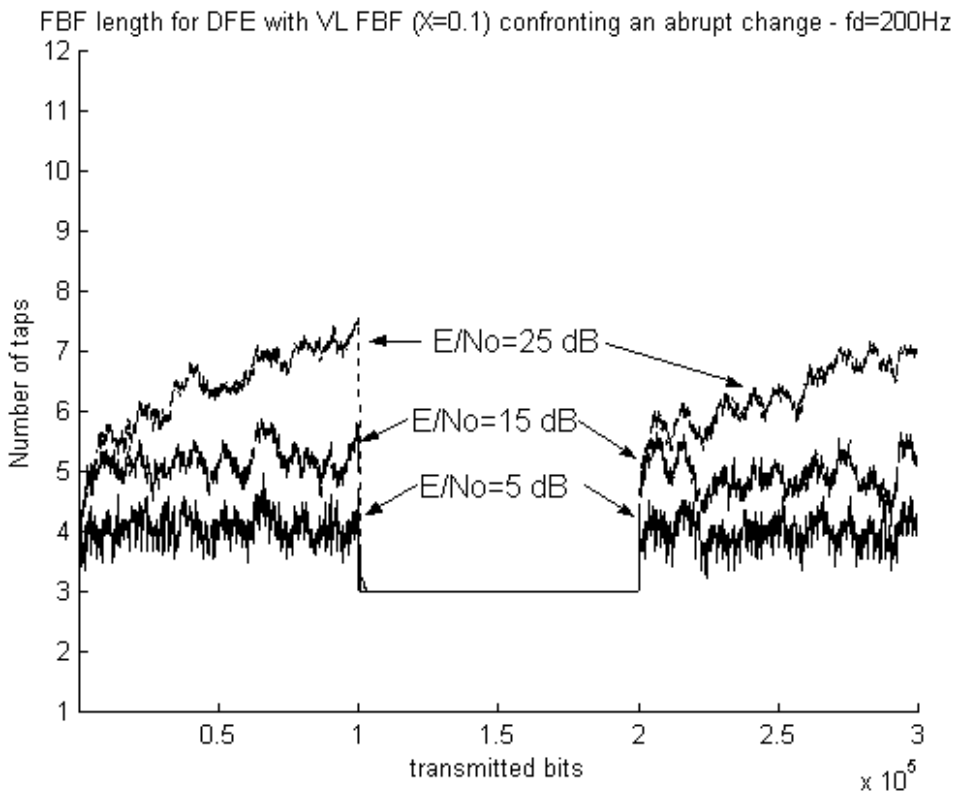

**Figure 7.45**: Length evolution for the FBF ($\chi$=0.1) operating in a mobile channel ($f_d$=200 Hz) confronting an abrupt channel change.

As with the variable length LE, it is important to observe how the FBF length varies. Figures 7.44 to 7.47 provide this information. Figures 7.44 and 7.45 display the results when using $\chi$=0.1 for the two simulated Doppler spreads. In both cases, the FBF is kept between 4 and 7 taps, depending on the E/No level, while operating in the COST207-TU model. When the channel changes to the single path static model the equaliser is able to recognise the transition and shrinks the FBF to just 3 taps, independently of the noise level.

When using $\chi$=0.01 (figures 7.46 and 7.47) the effect of the VL FBF DFE is much more noticeable. During the time the reduced COST207-TU model is active, the FBF achieves a length of 9-10 taps. When the channel changes to the static single path model, the FBF is reduced to 3 taps. In theory, when operating in the single path channel, the FBF should be kept with a length of 2 taps, but again decision errors tend to introduce some inaccuracy (albeit minor) in the length setting.

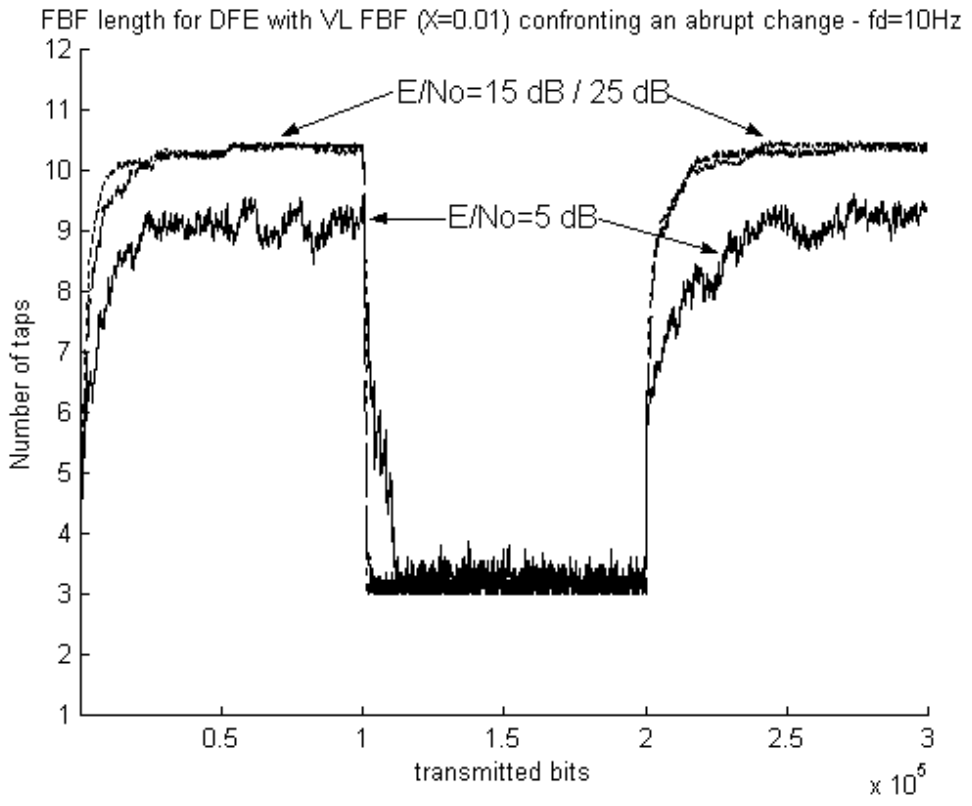

**Figure 7.46**: Length evolution for the FBF ($\chi$=0.01) operating in a mobile channel ($f_d$=10 Hz) confronting an abrupt channel change.

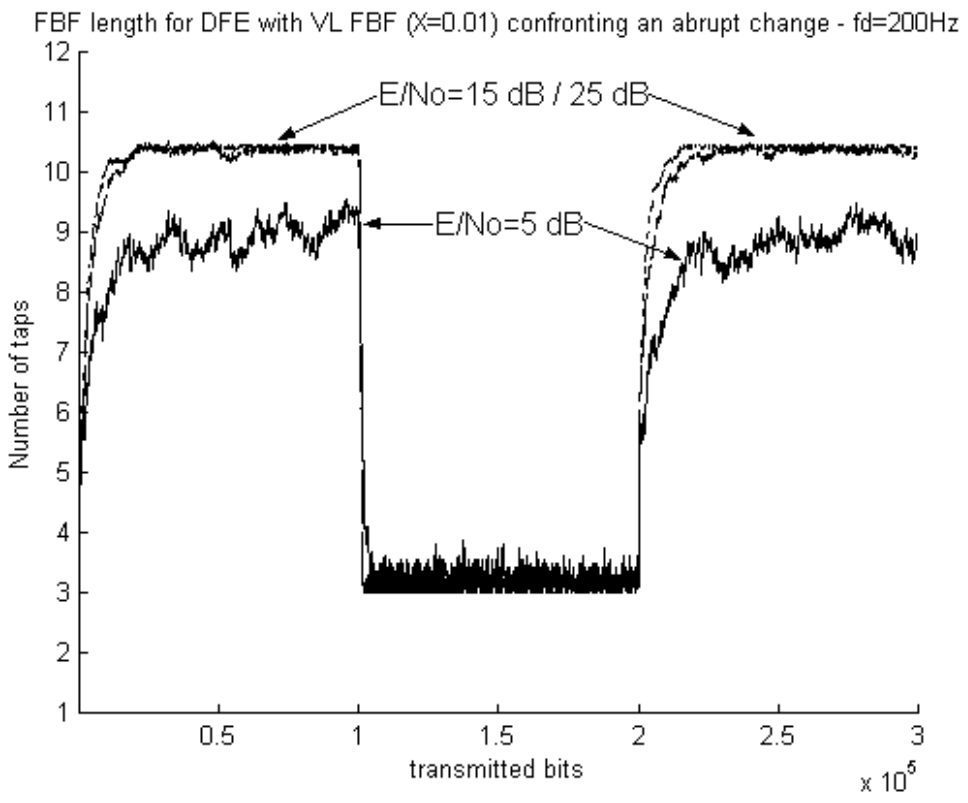

**Figure 7.47**: Length evolution for the FBF ($\chi$=0.01) operating in a mobile channel ($f_d$=200 Hz) confronting an abrupt channel change.

Figures 7.48 and 7.49 display the BER curves for the three different DFEs used in two different Doppler scenarios. Focusing on figure 7.48 ($f_d$=10 Hz), a very large difference can be observed between the VL FBF DFE ($\chi$=0.1) and the other two structures used as the E/No gets larger. This could have already been predicted from the corresponding MSE curves where the VL FBF DFE ($\chi$=0.1) converged to a significantly higher value. Comparing the BER of the fixed (6,11)-DFE with that of the VL FBF DFE ($\chi$=0.01), it can be seen that the fixed DFE achieves a lower probability of bit error for a large E/No. Nonetheless, the difference is not very much significant. This slight degradation of the VL FBF DFE ($\chi$=0.01), as has already been mentioned, is due to its slower convergence. If the number of bits of the simulation was increased[25], this difference would tend to vanish, achieving identical BER for both equalisers.

When fd=200 Hz (figure 7.49) the same comments apply, although now the differences between the different equalisers are much smaller. Comparing both figures, notice that the

[25] In a practical situation, the number of transmitted bits would be very much larger than the 300,000 bits used in these simulations. Obviously, to make simulation time reasonable, the simulation length had to be kept fairly short.

BER values at large E/No reflect the larger degradation provoked by the tracking misadjustment in the case of $f_d$=200 Hz. When E/No is low to moderate (5 to 10 dB), such a difference does not exist, as in this case, AWGN noise is the dominant source of error rather than tracking misadjustment.

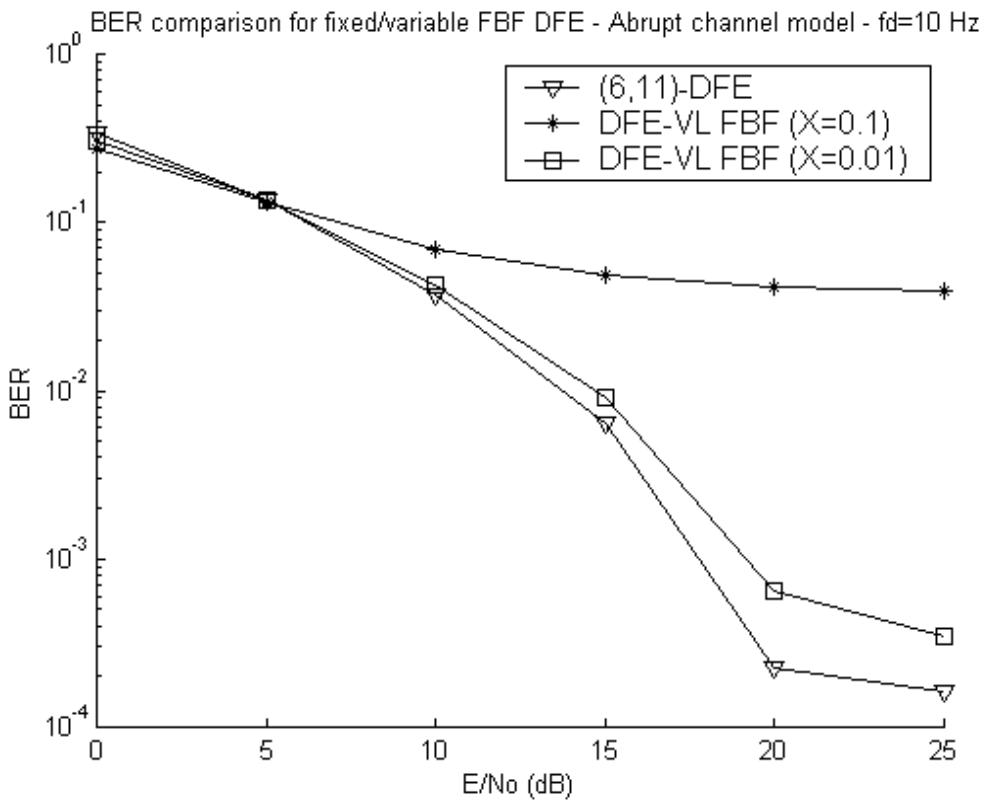

**Figure 7.48**: BER for the different DFE's operating in a mobile channel ($f_d$=10 Hz) confronting an abrupt channel change.

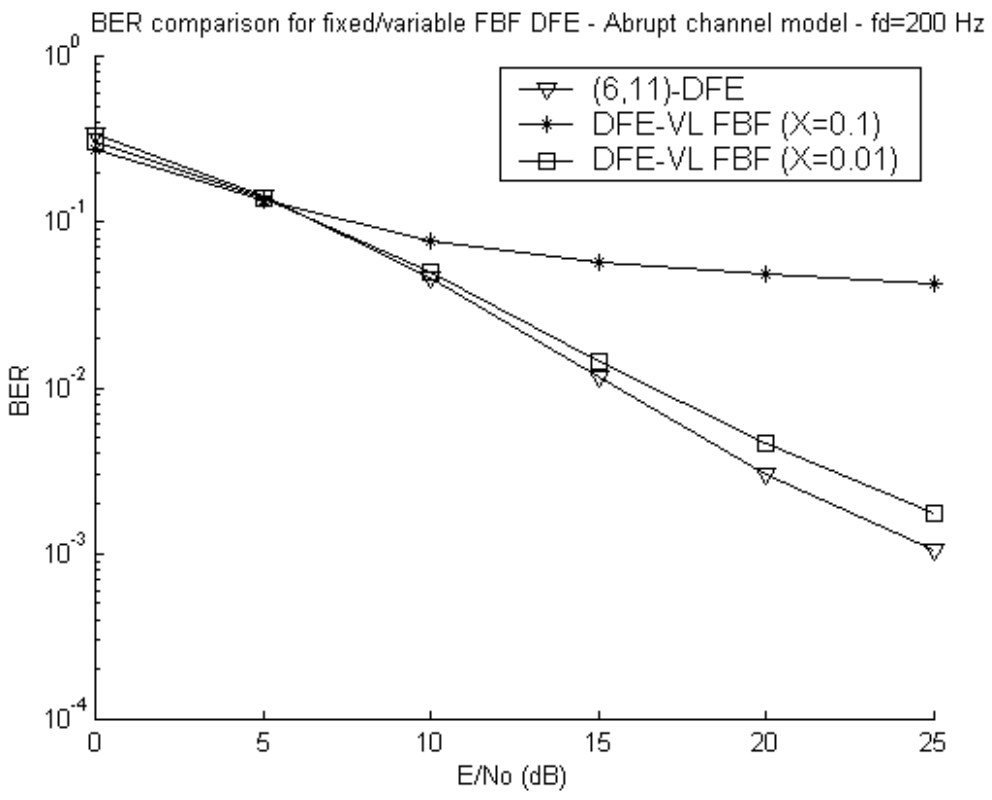

**Figure 7.49**: BER for the different DFE's operating in a mobile channel($f_d$=200 Hz) confronting an abrupt channel change.

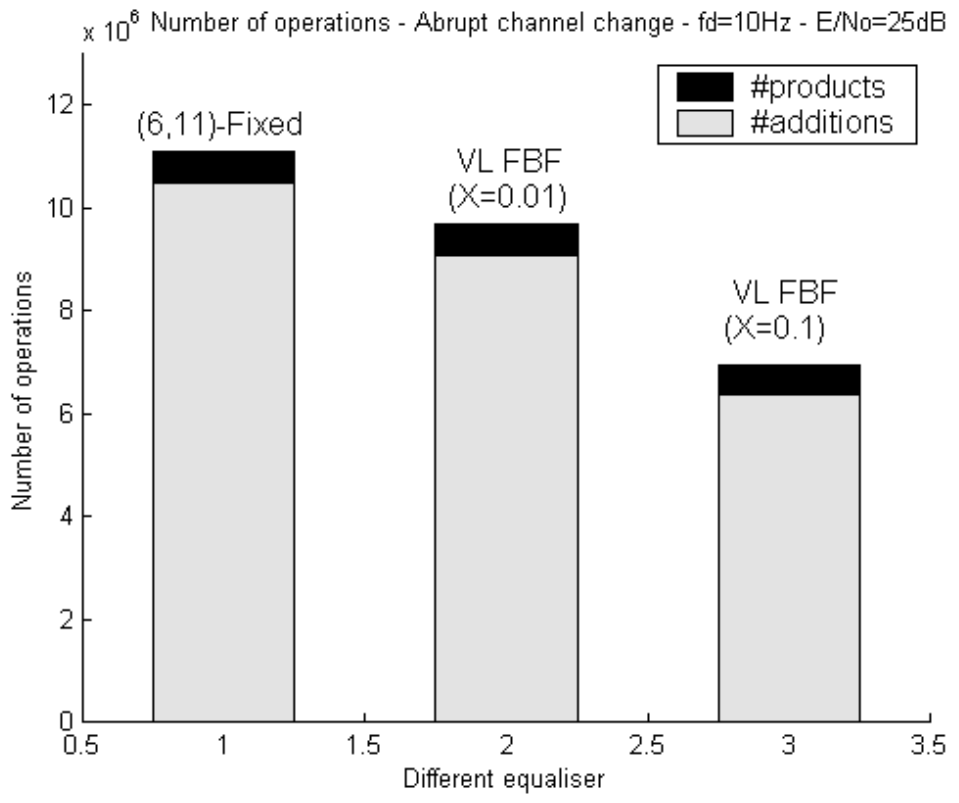

**Figure 7.50**: Comparison of the number of computations for different DFEs operating in a mobile channel ($f_d$=10 Hz) confronting an abrupt channel change. E/No=25 dB.

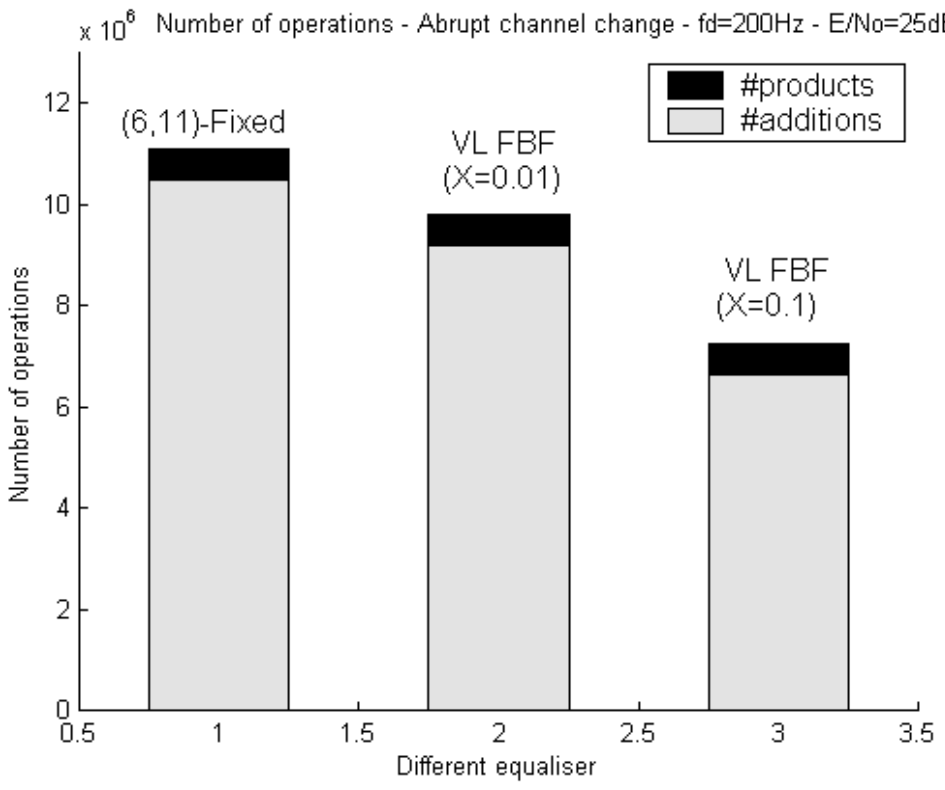

**Figure 7.51**: Comparison of the number of computations for different DFEs operating in a mobile channel ($f_d$=200 Hz) confronting an abrupt channel change. E/No=25 dB.

Complementing the BER results, comparisons of the number of operations for each of the equalisation structures are now presented. In figures 7.50 ($f_d$=10 Hz) and 7.51 ($f_d$=200 Hz) the number of computations for when E/No=25 dB are presented. In both graphs, the VL

FBF DFE ($\chi$=0.01) reduces the number of products and additions by 14 % approximately with respect to the fixed length DFE. The VL FBF DFE ($\chi$=0.1) offers a reduction of nearly 40%, although at the cost of a very significant increase in the BER.

When E/No=5 dB (figures 7.52 and 7.53), the savings of the variable FBF structures with respect to the fixed DFE are even more significant, with reductions in the number of operations of 24% and 45% for the VL FBF DFE ($\chi$=0.1) and VL FBF DFE ($\chi$=0.1) respectively. Moreover, these reductions do not imply any degradation of the BER.

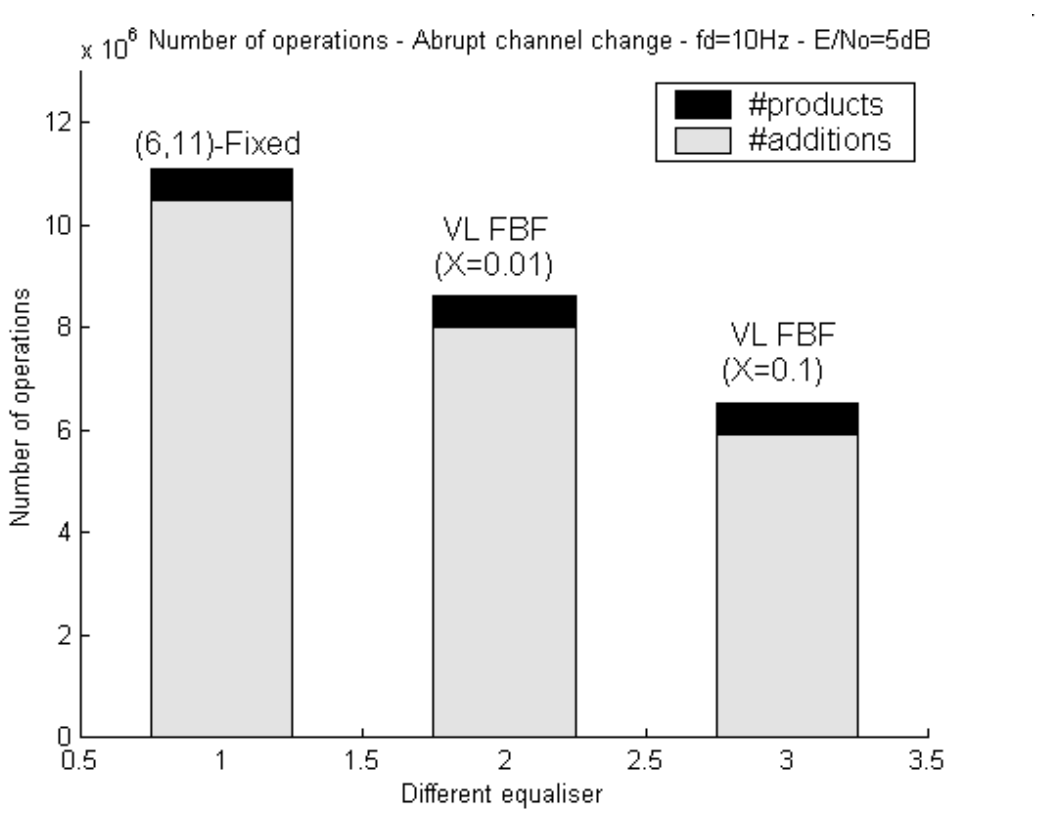

**Figure 7.52**: Comparison of the number of computations for different DFEs operating in a mobile channel ($f_d$=10 Hz) confronting an abrupt channel change. E/No=5 dB.

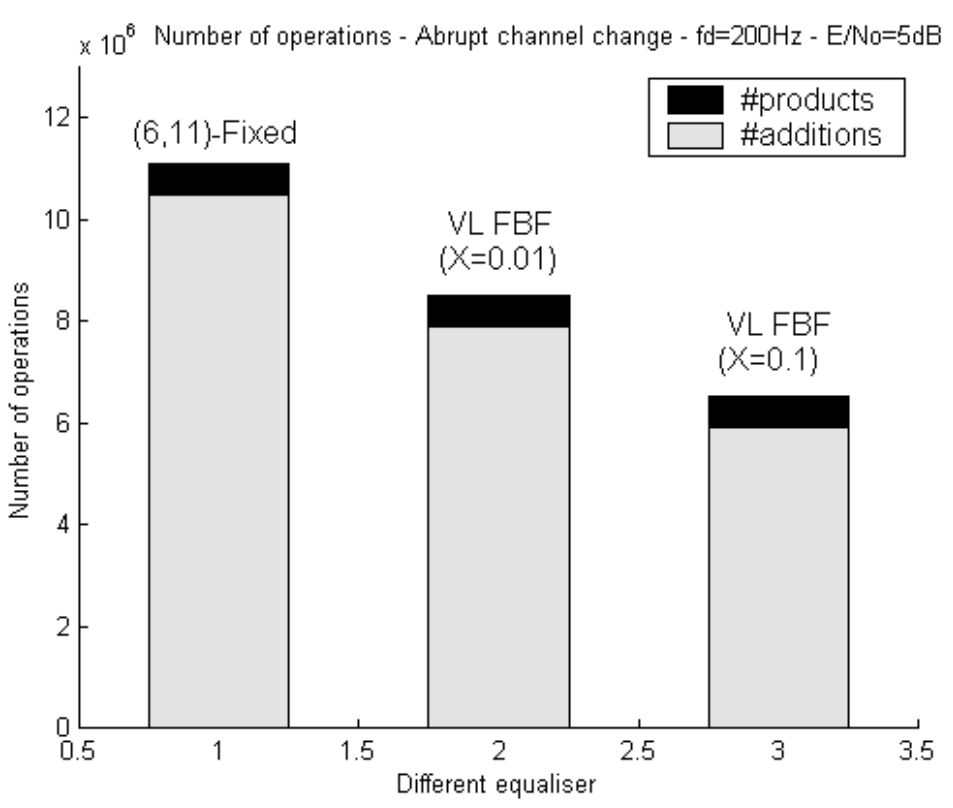

**Figure 7.53**: Comparison of the number of computations for different DFEs operating in a mobile channel ($f_d$=200 Hz) confronting an abrupt channel change. E/No=5 dB.

# 7.4 Conclusion drawn from the realistic scenario simulations

This chapter has shown results of the proposed variable length structures when operating in a realistic scenario. Overall, these results have confirmed that being able to vary the length of an equaliser (LE or DFE) depending on the instantaneous link conditions is a very attractive feature, as it may offer huge computational savings without degrading the BER. The length update algorithms have proved to be robust under practical conditions to factors such as Doppler shift and use of incorrect decisions.

# 8 IMPLEMENTATION OF RECONFIGURABLE EQUALISERS

Nowadays, most terminal handsets use fixed-point arithmetic to perform the different signal processing operations. The reason for this choice is that fixed-point devices (DSP, FPGA, ASIC), when compared with their floating-point counterparts, offer important reductions in power consumption, which in turn translates into extended battery times. It is therefore important when designing algorithms to be used in the handset side to verify that they work under fixed-point constraints.

The main objective of this chapter is to show that the length update algorithms for the LE and DFE presented in previous chapters also work when they are implemented on a fixed-point device. To verify this, the algorithms and simulation environment have been implemented on a Texas Instruments TMS320C5402 DSP. This particular DSP is widely used in many commercial handsets to perform the baseband processing. In a way this chapter complements the results presented in chapter 7. In there, we proved that the algorithms work under realistic physical conditions. Here we show that the algorithms perform well when implemented on a widely used commercial device.

In section 8.1 some background information is provided regarding fixed-point arithmetic and the problem it causes. Section 8.2 briefly comments on the different types of devices suitable to implement signal processing operations, namely ASIC, FPGA and DSP. Section 8.3 presents the development environment and processor used to implement the fixed-point version of the algorithms. It also shows and comments on the fixed-point assembly code generated to implement the length update algorithm for the LE and DFE. In section 8.4 simulation results are shown that prove the robustness of the algorithms to fixed-point inaccuracies. The chapter ends by summarising the main conclusions to be extracted from this implementation.

# 8.1 Fixed-point vs floating-point arithmetic[26]

The mathematical descriptions of signal processing functions such as the adaptive filtering and length update algorithms presented in previous chapters implicitly make use of infinite numerical precision. However any algorithm implemented on a digital processor can only utilise a finite number of bits to represent the different variables. This transformation from infinite to finite arithmetic, called quantisation, may influence significantly the performance of an algorithm. A clear example of this is the SFAEST algorithm (and other Fast Kalman algorithms) where finite precision effects cause severe instability problems (see chapter 2). Another example is the design of digital filters whose poles are very close to the unit circle. In this case, the quantisation error may move the poles outside the unit circle making the filter unstable.

There are two basic formats to represent numbers with a finite number of bits: floating-point and fixed-point. We will refer to the number of bits, typically 16 or 32, used to represent a number as a word. In floating-point representations, the binary point that separates the integer and fractional parts may vary its position within the word. In contrast, in fixed-point representation, the binary point is constantly fixed at the same position within the word.

Floating-point numbers have the form $n=M\ 2^E$ where M is called the mantissa and E is called the characteristic. The most widely used floating-point format is the ANSI/IEEE Standard 754-1985. In this standard, the 32-bit word is divided into 1 bit for sign, 23 bits for the mantissa and 8 bits for the exponent. The range of numbers that can be represented using this format is from $1.18 \times 10^{-38}$ to $3.4 \times 10^{38}$, either negative or positive. The main advantages of the floating-point representation are the very large dynamic range[27] the operands can have and also the possibility of performing operations very accurately without having to worry about overflow problems. On the other hand, the fact that the binary point might be in any position within the word complicates the design of the associated hardware to perform arithmetic operations ([Ackenhusen99]).

[26] The material on this section has been mainly compiled from [Mitra01], [Lyons97] and [Haykin96].

In fixed-point processors, knowledge of the location of the binary point is exploited when designing the hardware to implement the arithmetic operations. This results in a very efficient implementation with extremely low power consumption. However, this comes at the cost of a huge reduction in the dynamic range and the necessity to handle overflow situations. A very common 32-bit fixed-point format situates the binary point just after the most significant bit (MSB, the leftmost bit). The MSB is used to represent the sign of the quantity while the other bits represent a magnitude between 0 and 1, hence the number lies between –1 and 1. This implies that all the variables used in an algorithm must be first normalised to the interval [-1, 1]. Additionally, care must be taken to ensure that the result of any operation is still within this range, otherwise overflow may significantly distort the result of a signal processing operation. Overflow handling techniques are described in section 9.3 of [Lyons97]. The simplicity of the hardware implementation makes fixed-point devices the most suitable choice in power constrained environments.

The use of finite length arithmetic significantly influences the performance of adaptive algorithms such as LMS and RLS with the introduction of new error terms (chapter 17 in [Haykin96]). These factors can be seen as new forms of excess mean squared error (EMSE). Consequently, their influence must be taken into account when choosing algorithm parameters such as step size in the case of LMS and the forgetting factor for the RLS.

Additionally, undesired behaviour like gain stalling or explosive divergence may appear and therefore mechanisms to minimise their effects must be used. An example of one of these mechanisms is the leakage technique introduced in [Gitlin82] to make the LMS more robust when implemented on finite precision. Thus, it is very important to verify the performance of any signal processing algorithm when implemented on finite length arithmetic, especially if the target device uses fixed-point arithmetic. This will be the objective of sections 8.3 and 8.4, but first a brief survey of the available technologies to implement signal processing systems is given in the next section.

---

[27] The dynamic range of a numeric system is defined as the ratio between the largest and smallest magnitudes that can be represented.

## 8.2 Technologies for signal processing implementation

One of the most critical decisions when designing a signal processing system is deciding which semiconductor technology to use for its implementation. Nowadays the choice is typically one of the following:

- Digital signal processor (DSP)
- Application specific integrated circuit (ASIC)
- Field programmable gate array (FPGA)

The first two types of device have been the classical technologies used to implement signal processing functions whereas the FPGA is a relatively new type of device whose characteristics offer a compromise between the other two ([Tessier01]). Each of these technologies is now briefly discussed.

DSPs are microprocessors specifically targeted for signal processing operations. They usually have two separate buses and memory spaces, one for data and one for instructions (Harvard architecture) so that the fetching of operations and operands from memory can take place in parallel. Additionally, they include hardware structures to accelerate common signal processing operations such as multiplication/accumulation and circular convolutions. Other common features of DSPs are ([Ifeachor93], [Ackenhusen99]): pipelined architecture, data I/O facilities, on-chip memory and parallel units.

Their main advantage is flexibility as they can be programmed to do any task. The fact that the functionality of the DSP is mainly determined by the software running on top of it, means that modifying its function is just a matter of loading a new program on the DSP, easing in this way the maintainability of the system.

On the other hand, their major drawback is the lack of specialisation, which implies that they are not optimised for any particular DSP operation. This sets upper bounds to the performance of DSPs for certain operations.

The classic alternative to DSP implementation consists of using ASICs. As the name implies, an ASIC is a device designed to perform a very specific function. Its main advantage is the superior computational performance when compared to a DSP. Additionally, the fact that the ASIC contains just the elements needed may help in reducing the chip count of a system and, consequently, board area ([Ackenhusen99]). An important consequence of this reduction in the number of components is a reduction in the power consumption ([Tessier01]). The main drawback of ASIC technology is the long design cycle required to implement an algorithm on custom silicon ([Ackenhusen99]). This, coupled with its lack of flexibility, makes the time to market of ASIC based products considerably longer than in the case of software based systems such as DSPs.

Over recent years, considerable attention has been given to the signal processing capabilities of FPGAs[28] ([Tessier01]). An FPGA is a device containing a matrix of equal combinatorial logic blocks, with each block consisting of a small look-up table (ROM) and typically, a flip-flop ([Alfke98]). Metal lines of various lengths run horizontally and vertically between these logic blocks, selectively interconnecting them or connecting them to I/O blocks. A configuration sequence, stored in a special type of memory on the FPGA (configuration memory), directs the process of block interconnection. Typically, this sequence is uploaded to the FPGA from a programmable read only memory (PROM).

FPGAs offer levels of performance similar to those of ASICs as they are specifically configured to perform a specific task. On the other hand, the fact that their functionality can be altered, just by loading a new configuration sequence, confers them a high degree of flexibility.

Use of FPGAs for high performance signal processing applications has already been reported by many authors. In [Srikanteswara00], a single user (i.e. terminal) CDMA receiver was designed around the Xilinx XC4028EX. The tasks of the FPGA included equalisation and despreading using an adaptive filter whose coefficients were updated with the LMS algorithm. Thanks to the reconfiguration capability of the FPGA, parameters such as adaptive algorithm, step-size, filter length and spreading code could be arbitrarily selected. Efficient techniques to implement adaptive filters on FPGA can be found in [Allaire97].

---

[28] FPGA is a name coined by one particular manufacturer (Xilinx), nevertheless, it is often used to denote reconfigurable logic devices in general.

Figure 8.1 shows the trade-offs between performance and flexibility of the different technologies. It is unlikely, at least for the foreseeable future, that a single technology can be used to implement all the signal processing functions of a radio communications receiver. Instead, combinations of the three types of device are used ([Rabaey98]). In the context of CDMA receivers, the baseband processing at symbol level, such as source and channel coding or encryption, is typically performed on a DSP, whereas chip level processing such as de/spreading or multiplexing is implemented using either ASICs or FPGAs ([Tessier01]). Nonetheless, the trend is to move more and more of the processing onto the DSP and some implementations of chip processing functions on a DSP have already been reported ([Lange02]).

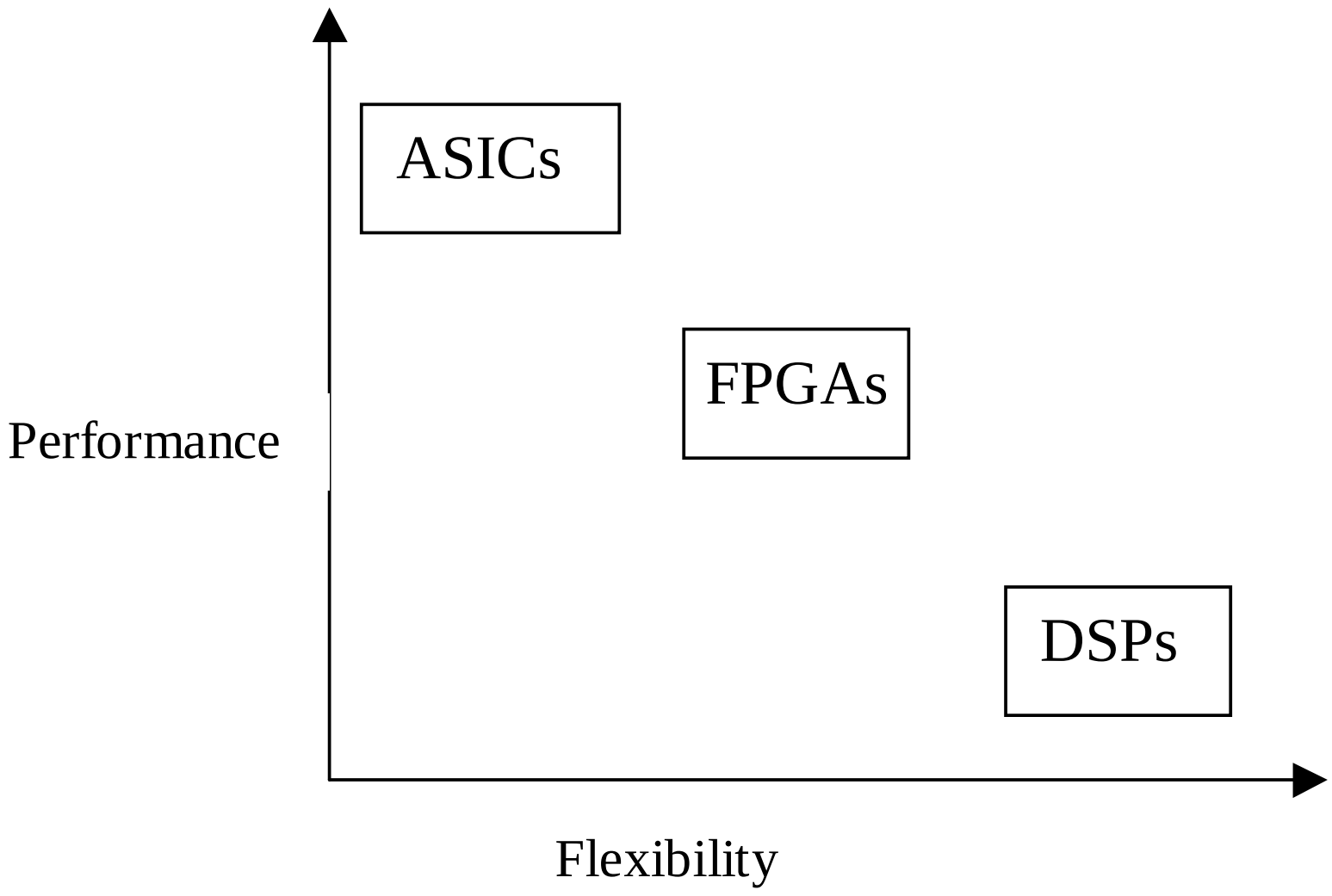

**Figure 8.1**: Trade-offs among different implementation technologies.

## 8.3 DSP environment and implementation

In this section details of the implementation of the length update algorithms for the LE and DFE on a fixed-point DSP are presented. First the development environment is briefly described. Then, fixed-point versions of the novel techniques introduced in previous chapters are presented and discussed.

### 8.3.1 The TMS320C5402 DSK

The TMS320C5402 development starter kit (DSK) is a package from Texas Instruments containing a DSP board based on the popular TMS320C5402 processor ([TI98]), with a development environment to implement and test real-time signal processing software on the board. The choice of using this particular DSP was mainly motivated by the fact that the hardware was already available in the Department and it served well for our purposes of checking the correct functioning of the algorithms under fixed-point arithmetic. This particular processor series (´C54XX) has been widely used in handset terminals because of its large processing power at low power consumption levels. The processor runs at 100 MHz with a core voltage of 1.8 volts and I/O voltage of 3.3 volts. The board also includes 128 kwords of external SRAM and 256 kwords of FLASH memory. Additionally it has different types of I/O connectors in order to interface with the outside world.

The software development environment, called Code Composer Studio (version 1.22), runs on a host computer connected to the board via the parallel connector. This environment contains a C/Assembly compiler, editor, linker and many other tools to debug and analyse the real-time execution of the system.

The generic procedure to implement a signal processing algorithm on the board is as follows: first the algorithm is coded in either C or Assembly. After compilation, the binary generated code is downloaded (via parallel port) to the memory on the DSP board, when execution can begin. During run time, several methods are available to monitor the execution of the program, for example, real-time graphics monitoring some particular parameter can be obtained. Once the program has finished, control returns to the development environment on the host computer.

### 8.3.2 Implementation of variable length equalisers with fixed-point arithmetic

In order to verify the performance of the length update algorithms for the LE and DFE on a fixed-point device, the simulation environment described in chapter 3 was implemented on the DSK in C language. The LMS algorithm has been used for all the simulations presented in this chapter. In theory, in order to avoid the undesired effects fixed-point arithmetic has on the LMS algorithm (see section 8.1), some changes to the algorithm would need to be included (such as the variation described in [Gitlin82]). However theory predicts that these effects, in the case of the LMS algorithm, only become significant after the algorithm has been operating for a long time. In this chapter, and in fact in the whole project, simulation lengths have been kept relatively short and therefore the plain version of the LMS can be used even when implemented in fixed-point arithmetic.

There is one important point to mention regarding the development of C programs for a fixed-point device. In theory, only fixed-point variables could be used. However in the environment used in this project, the use of floating-point variables was allowed provided a certain library was included. This library has the task of making the conversion floating-point to fixed-point arithmetic whenever one of these variables is used. This greatly eases the development of fixed-point applications, but at the cost of producing very inefficient code [Kuo01]. This inefficiency is caused by the numerous calls to functions in that library to make the conversions. Nonetheless, this method for fixed-point development is useful to verify the correct functioning of an algorithm in this type of arithmetic.

The resulting fixed-point assembly code of the length update algorithms for the LE and DFE is presented below. The algorithms were implemented in C using floating-point variables. Consequently this code cannot be regarded as very efficient but it serves the purpose of validating the algorithms under fixed-point arithmetic.

Some comments about this code are now in place. Each line in the assembly listings has one assembly instruction and a number between bars (|xxx|). This number corresponds to the C source line from which that assembly instruction has been generated. Obviously, one line of C source generates many more lines of assembly code. Also included in the listings below, are comments (in capital letters) relating groups of assembly instructions with the equations of algorithms 5.2 (variable length LE) and 6.1 (DFE with variable length FBF).

Finally notice in the assembly code the presence of many lines like this:

```
CALL      #F$$SUB              ; |396| ; call occurs [#F$$SUB]
```

These instructions are those calling routines to perform operations with floating-point quantities (in this particular case, a subtraction). In an efficient implementation, none of these calls would appear.

## FIXED-POINT ASSEMBLY CODE FOR THE VARIABLE LENGTH LE

INITIALISATIONS

```
;***************************************************************
;* FUNCTION DEF: _update_linear_equaliser_structure            *
;***************************************************************
_update_linear_equaliser_structure:
;* A     assigned to _equaliser_length
      .sym   _equaliser_length,0, 20, 17, 16
      .sym   _equalised_bits_segmented,8, 22, 9, 16
      .sym   _received_bit,10, 6, 9, 32
      .sym   _equaliser_length,2, 20, 1, 16
       PSHM      AR1
       FRAME     #-6
       NOP
      .line  4
       STL       A,*SP(2)             ; |391|
      .line  9
       DLD       *(_MSE_N),A          ; |396|
       DST       A,*SP(0)             ; |396|
       CALLD     #F$$MUL              ; |396|
       DLD       *(FL25),A            ; |396| ; call occurs [#F$$MUL]
       LD        A,B                  ; |396|
       MVDK      *SP(8),*(AR1)        ; |396|
       DLD       *SP(10),A            ; |396|
       DST       A,*SP(0)             ; |396|
       DLD       *AR1(2),A            ; |396|
       CALL      #F$$SUB              ; |396| ; call occurs [#F$$SUB]
       DST       A,*SP(4)             ; |396|
       DLD       *SP(10),A            ; |396|
       DST       A,*SP(0)             ; |396|
       MVDK      *SP(8),*(AR1)        ; |396|
       DLD       *AR1(2),A            ; |396|
       CALL      #F$$SUB              ; |396| ; call occurs [#F$$SUB]
       DST       A,*SP(0)             ; |396|
       DLD       *SP(4),A             ; |396|
       CALL      #F$$MUL              ; |396| ; call occurs [#F$$MUL]
       DST       A,*SP(0)             ; |396|
       CALLD     #F$$ADD              ; |396|
       NOP
       LD        B,A                  ; |396| ; call occurs [#F$$ADD]
       DST       A,*(_MSE_N)          ; |396|
      .line  10
       DLD       *(_MSE_N_1),A        ; |397|
       DST       A,*SP(0)             ; |397|
       CALLD     #F$$MUL              ; |397|
       DLD       *(FL25),A            ; |397| ; call occurs [#F$$MUL]
       LD        A,B                  ; |397|
       DLD       *SP(10),A            ; |397|
       MVDK      *SP(8),*(AR1)        ; |397|
       DST       A,*SP(0)             ; |397|
       DLD       *AR1,A               ; |397|
       CALL      #F$$SUB              ; |397| ; call occurs [#F$$SUB]
       DST       A,*SP(4)             ; |397|
       DLD       *SP(10),A            ; |397|
       DST       A,*SP(0)             ; |397|
       MVDK      *SP(8),*(AR1)        ; |397|
       DLD       *AR1,A               ; |397|
       CALL      #F$$SUB              ; |397| ; call occurs [#F$$SUB]
       DST       A,*SP(0)             ; |397|
       DLD       *SP(4),A             ; |397|
```

COMPUTATION $ASE_N$ (EQ. 5.7) — annotation beside line 396

COMPUTATION $ASE_{N-1}$ (EQ. 5.8) — annotation beside line 397

```
        CALL      #F$$MUL                  ; |397| ; call occurs [#F$$MUL]
        DST       A,*SP(0)                 ; |397|
        CALLD     #F$$ADD                  ; |397|
        NOP
        LD        B,A                      ; |397| ; call occurs [#F$$ADD]
        DST       A,*(_MSE_N_1)            ; |397|
       .line  14
        B         L59                      ; |401| ; branch occurs
L56:
       .line  17
        DLD       *(_MSE_N_1),A            ; |404| TEST FOR ENLARGING
        DST       A,*SP(0)                 ; |404|        (EQ. 5.9)
        CALLD     #F$$MUL                  ; |404|
        DLD       *(FL26),A                ; |404| ; call occurs [#F$$MUL]
        DLD       *(_MSE_N),B              ; |404|
        DST       B,*SP(0)                 ; |404|
        CALL      #F$$COMPARE              ; |404| ; call occurs [#F$$COMPARE]
        SSBX      SXM                      ;
        LD        *(AL),A                  ; |404|
        BC        L57,ALEQ                 ; |404| ; branch occurs
        MVDK      *SP(2),*(AR1)            ; |404|
        LD        #3,A                     ; |404|
        ADD       *AR1,A                   ; |404|
        LD        *(AL),A                  ; |404|
        SUB       #33,A,A                  ; |404|
        BC        L57,AGT                  ; |404| ; branch occurs
       .line  18
        LD        #3,A                     ; |405| EQUALISER ENLARGEMENT
        ADD       *AR1,A                   ; |405|        (EQ. 5.9)
        STL       A,*AR1                   ; |405|
       .line  19
        ST        #0,*(_state)             ; |406|
       .line  20
        LD        #0,A                     ; |407|
        DST       A,*(_transient_counter) ; |407|
       .line  21
        DLD       *(FL12),A                ; |408|
        DST       A,*(_MSE_N)              ; |408|
       .line  22
        DST       A,*(_MSE_N_1)            ; |409|
       .line  24
        B         L60                      ; |411| ; branch occurs
L57:
       .line  27
        DLD       *(_MSE_N_1),A            ; |414| TEST FOR REDUCTION
        DST       A,*SP(0)                 ; |414|        (EQ. 5.10)
        CALLD     #F$$MUL                  ; |414|
        DLD       *(FL27),A                ; |414| ; call occurs [#F$$MUL]
        DLD       *(_MSE_N),B              ; |414|
        DST       B,*SP(0)                 ; |414|
        CALL      #F$$COMPARE              ; |414| ; call occurs [#F$$COMPARE]
        SSBX      SXM                      ;
        LD        *(AL),A                  ; |414|
        BC        L60,AGEQ                 ; |414| ; branch occurs
        MVDK      *SP(2),*(AR1)            ; |414|
        LD        *AR1,A                   ; |414|
        ADD       #-3,A                    ; |414|
        LD        *(AL),A                  ; |414|
        SUB       #6,A,A                   ; |414|
        BC        L60,ALT                  ; |414| ; branch occurs
       .line  28
```

```
        LD        *AR1,A               ; |415|
        ADD       #-3,A                ; |415|
        STL       A,*AR1               ; |415|
       .line  29
        ST        #0,*(_state)         ; |416|
       .line  30
        LD        #0,A                 ; |417|
        DST       A,*(_transient_counter) ; |417|
       .line  31
        DLD       *(FL12),A            ; |418|
        DST       A,*(_MSE_N)          ; |418|
       .line  32
        DST       A,*(_MSE_N_1)        ; |419|
       .line  35
        B         L60                  ; |422| ; branch occurs
L61:
        FRAME     #6
        POPM      AR1
        RET ; return occurs
       .endfunc     434,000000400h,7
```

EQUALISER REDUCTION (EQ. 5.10)

END OF ROUTINE

**FIXED-POINT ASSEMBLY CODE FOR THE VL FBF DFE**

```
;***************************************************************
;* FUNCTION DEF: _update_dfe_structure                        *
;***************************************************************
_update_dfe_structure:
;* A     assigned to _FBequaliser
       .sym   _FBequaliser,0, 22, 17, 16
       .sym   _Nb,12, 20, 9, 16
       .sym   _desired,14, 6, 9, 32
       .sym   _equalised,16, 6, 9, 32
       .sym   _FBequaliser,2, 22, 1, 16
       .sym   _averageNb,4, 7, 1, 32
       .sym   _averageNb_1,6, 7, 1, 32
        PSHM      AR1
        FRAME     #-10
        NOP
       .line  2
        STL       A,*SP(2)             ; |440|
       .line  5
        DLD       *(_averaging_counter),A ; |443|
        ADD       #1,A                 ; |443|
        DST       A,*(_averaging_counter) ; |443|
       .line  7
        DLD       *SP(16),A            ; |445|
        DST       A,*SP(0)             ; |445|
        DLD       *SP(14),A            ; |445|
        CALL      #F$$SUB              ; |445| ; call occurs [#F$$SUB]
        DLD       *(FL18),B            ; |445|
        DST       B,*SP(0)             ; |445|
        CALL      #_pow                ; |445| ; call occurs [#_pow]
        DST       A,*SP(0)             ; |445|
        DLD       *(_windowMSE),A      ; |445|
        CALL      #F$$ADD              ; |445| ; call occurs [#F$$ADD] ;
        DST       A,*(_windowMSE)      ; |445|
       .line  8
        MVDK      *SP(12),*(AR1)       ; |446|
        LD        *AR1,B               ; |446|
```

INITIALISATIONS

COMPUTATION TAP POWER $N_B$ (EQ. 6.57)

```
 LD        *SP(2),A              ; |446|
 SUB       #1,B                  ; |446|
 ADD       B,#1,A                ; |446|
 STLM      A,AR1                 ; |446|
 NOP
 NOP
 DLD       *AR1,A                ; |446|
 MVDK      *SP(12),*(AR1)        ; |446|
 DST       A,*SP(0)              ; |446|
 LD        *AR1,B                ; |446|
 LD        *SP(2),A              ; |446|
 SUB       #1,B                  ; |446|
 ADD       B,#1,A                ; |446|
 STLM      A,AR1                 ; |446|
 NOP
 NOP
 DLD       *AR1,A                ; |446|
 CALL      #F$$MUL               ; |446| ; call occurs [#F$$MUL]
 DST       A,*SP(0)              ; |446|
 DLD       *(_tapNb),A           ; |446|
 CALL      #F$$ADD               ; |446| ; call occurs [#F$$ADD]
 DST       A,*(_tapNb)           ; |446|
.line  9
 MVDK      *SP(12),*(AR1)        ; |447|
 LD        *AR1,A                ; |447|
 SUB       #2,A,B                ; |447|
 LD        *SP(2),A              ; |447|
 ADD       B,#1,A                ; |447|
 STLM      A,AR1                 ; |447|
 NOP
 NOP
 DLD       *AR1,A                ; |447|
 MVDK      *SP(12),*(AR1)        ; |447|
 DST       A,*SP(0)              ; |447|
 LD        *AR1,A                ; |447|
 SUB       #2,A,B                ; |447|
 LD        *SP(2),A              ; |447|
 ADD       B,#1,A                ; |447|
 STLM      A,AR1                 ; |447|
 NOP
 NOP
 DLD       *AR1,A                ; |447|
 CALL      #F$$MUL               ; |447| ; call occurs [#F$$MUL]
 DST       A,*SP(0)              ; |447|
 DLD       *(_tapNb_1),A         ; |447|
 CALL      #F$$ADD               ; |447| ; call occurs [#F$$ADD]
 DST       A,*(_tapNb_1)         ; |447|
.line  11
 LD        #50,A                 ; |449|
 DST       A,*SP(0)              ; |449|
 DLD       *(_averaging_counter),A ; |449|
 CALL      #L$$MODS              ; |449| ; call occurs [#L$$MODS]
 SSBX      SXM                   ;
 SFTA      A,8                   ; |449|
 SFTA      A,-8                  ; |449|
 BC        L81,ANEQ              ; |449| ; branch occurs
 DLD       *(_averaging_counter),A ; |449|
 BC        L81,ALEQ              ; |449| ; branch occurs
.line  13
 DLD       *(_averaging_counter),A ; |451|
 CALL      #F$$LTOF              ; |451| ; call occurs [#F$$LTOF]
 DST       A,*SP(0)              ; |451|
```

COMPUTATION TAP POWER $N_{B-1}$ (EQ. 6.58)

PERIODICITY OF THE TEST

```
        DLD       *(_tapNb),A             ; |451|
        CALL      #F$$DIV                 ; |451| ; call occurs [#F$$DIV]
        DST       A,*SP(4)                ; |451|
       .line  14
        DLD       *(_averaging_counter),A ; |452|
        CALL      #F$$LTOF                ; |452| ; call occurs [#F$$LTOF]
        DST       A,*SP(0)                ; |452|
        DLD       *(_tapNb_1),A           ; |452|
        CALL      #F$$DIV                 ; |452| ; call occurs [#F$$DIV]
        DST       A,*SP(6)                ; |452|
       .line  15
        DLD       *(_averaging_counter),A ; |453|
        CALL      #F$$LTOF                ; |453| ; call occurs [#F$$LTOF]
        DST       A,*SP(0)                ; |453|
        DLD       *(_windowMSE),A         ; |453|
        CALL      #F$$DIV                 ; |453| ; call occurs [#F$$DIV]
        DST       A,*(_averageMSE)        ; |453|
       .line  17
        B         L78                     ; |455| ; branch occurs
L76:
       .line  19
        MVDK      *SP(12),*(AR1)          ; |457|   TEST FOR FBF ENLARGEMENT
        LD        #25,A                   ; |457|         (EQ.  6.59)
        SUB       *AR1,A                  ; |457|
        BC        L80,ALEQ                ; |457| ; branch occurs
       .line  20
        LD        *AR1,A                  ; |458|        FBF ENLARGEMENT
        ADD       #1,A                    ; |458|          (EQ.  6.59)
        STL       A,*AR1                  ; |458|
       .line  22
        B         L80                     ; |460| ; branch occurs
L77:
       .line  25
        MVDK      *SP(12),*(AR1)          ; |463| TEST FOR FBF REDUCTION
        LD        #2,A                    ; |463|          (EQ.  6.60)
        SUB       *AR1,A                  ; |463|
        BC        L80,AGEQ                ; |463| ; branch occurs
        DLD       *(_averageMSE),A        ; |463|
        DST       A,*SP(0)                ; |463|
        CALLD     #F$$MUL                 ; |463|
        DLD       *(FL25),A               ; |463| ; call occurs [#F$$MUL]
        DST       A,*SP(0)                ; |463|
        DLD       *SP(6),A                ; |463|
        CALL      #F$$COMPARE             ; |463| ; call occurs [#F$$COMPARE]
        SSBX      SXM                     ;
        LD        *(AL),A                 ; |463|
        BC        L80,AGEQ                ; |463| ; branch occurs
       .line  26
        DLD       *(FL12),A               ; |464|        FBF REDUCTION
        MVDK      *SP(12),*(AR1)          ; |464|          (EQ.  6.59)
        DST       A,*SP(8)                ; |464|
        LD        *AR1,B                  ; |464|
        SUB       #1,B                    ; |464|
        LD        *SP(2),A                ; |464|
        ADD       B,#1,A                  ; |464|
        STLM      A,AR1                   ; |464|
        NOP
        DLD       *SP(8),A                ; |464|
        DST       A,*AR1                  ; |464|
       .line  27
```

```
        MVDK      *SP(12),*(AR1)        ; |465|
        LD        *AR1,A                ; |465|
        SUB       #1,A                  ; |465|
        STL       A,*AR1                ; |465|
       .line  29
        B         L80                   ; |467| ; branch occurs
L78:
        DST       A,*SP(0)              ; |467|
        CALLD     #F$$MUL               ; |467|
        DLD       *(FL25),A             ; |467| ; call occurs [#F$$MUL]
        DST       A,*SP(0)              ; |467|
        DLD       *SP(4),A              ; |467|
        CALL      #F$$COMPARE           ; |467| ; call occurs [#F$$COMPARE]
        SSBX      SXM                   ;
        LD        *(AL),A               ; |467|
        BCD       L79,ALEQ              ; |467|
        NOP
        LD        #0,B                  ; |467| ; branch occurs
        LD        #1,B                  ; |467|
L79:
        LD        B,A                   ; |467|
        LD        *(AL),B               ; |467|
        BC        L77,BEQ               ; |467| ; branch occurs
        LD        *(AL),A               ; |467|
        SUB       #1,A,A                ; |467|
        BC        L76,AEQ               ; |467| ; branch occurs
L80:
       .line  35
        DLD       *(FL12),A             ; |473|      RESTART WINDOWING
        DST       A,*(_tapNb)           ; |473|
       .line  36
        DST       A,*(_tapNb_1)         ; |474|
       .line  37
        DST       A,*(_windowMSE)       ; |475|
       .line  38
        LD        #0,A                  ; |476|
        DST       A,*(_averaging_counter) ; |476|
L82:
        FRAME     #10                                END OF ROUTINE
        POPM      AR1
        RET ; return occurs
       .endfunc     482,000000400h,11
```

# 8.4 Simulation results for variable length equalisers using fixed-point arithmetic

In this section, simulation results obtained using the DSK are presented. In particular, the aim of the results shown next is to verify the correct functioning of the algorithms controlling the equaliser length when implemented on a fixed-point device. These results have been obtained when using the length update routines presented in the previous sections and using the LMS algorithm to drive the equaliser coefficients.

Figure 8.2 shows the measured length evolution when equalising Channel model 2 (static) using a LE with the length update algorithm operating in fixed-point arithmetic. The initial equaliser length was set to 6 taps with the decision delay selected to be 5 samples. The length update algorithm parameters were set to: $\alpha_{up}=0.6, \alpha_{dw}=0.99$ and $\beta=1.0$ (static channel). It can be observed that when E/No=5 dB, the filter grows up to 12 taps and when E/No=25 dB, it goes up to 27 taps.

These results should be compared with those presented in figure 5.3 where the MSE performance of equalisers with different fixed lengths was measured (using double precision arithmetic) for the same channel model. In there, it was observed that the optimum lengths for the LE would be 10-12 taps for 5 dB and 25-26 for 25 dB. Additionally, in section 5.4.1 it was shown that the length update algorithm (algorithm 5.2) was able to predict this optimum number of taps, but in that case double precision arithmetic was used.

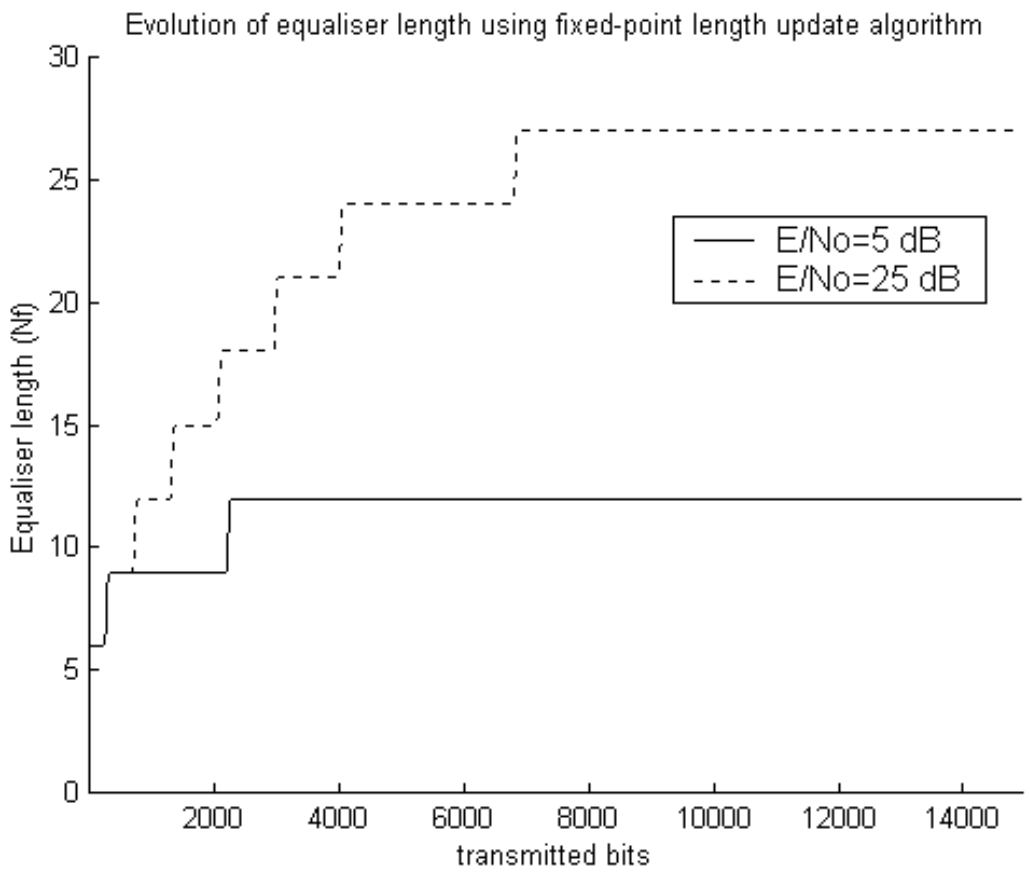

**Figure 8.2**: Measured length evolution for the variable length equaliser compensating Channel model 2.

Graph 8.2 is important because it proves that the algorithm controlling the equaliser length, even when using fixed-point arithmetic, accurately selects the optimum length the equaliser should have for a particular channel and E/No level.

The next scenario checked to verify the correct functioning of the fixed-point variable length LE is that with an abrupt and sudden change in the channel impulse response. Initially the channel corresponds to the profile given by Channel model 2. After 15,000 transmitted bits, the profile switches to a channel with just one single path and after another 15,000 bits the channel changes back to its original form (i.e. Channel model 2). The variable length equaliser parameters were set as: $\alpha_{up}$ =0.6, $\alpha_{dw}$ =0.99 and $\beta$=0.999. Figure 8.3 shows the MSE evolution for this simulation. One word of caution is needed for this graph. As in previous chapters when results were presented for similar situations, a windowed version of the MSE is plotted. Before, the window length was set to 2,000 samples. Now, due to a limitation in the declaration of arrays (memory limitations), the windowing used was only of 200 samples. This has the effect of distorting the resulting MSE curves by making them lower than their real values. Nonetheless figure 8.3 clearly shows the improvement experimented by the system when the profile turns to the single path channel.

More important is the information shown in figure 8.4. There, the length evolution of the equaliser is presented. It can be seen that when E/No=25 dB and the channel is given by Channel model 2 (iterations 1-15,000 and 15,000-45,000), the equaliser expands to 27 taps. When the channel changes to the single path profile, which in theory would require no equalisation at all, the length update algorithm shrinks the equaliser to 6-7 taps.

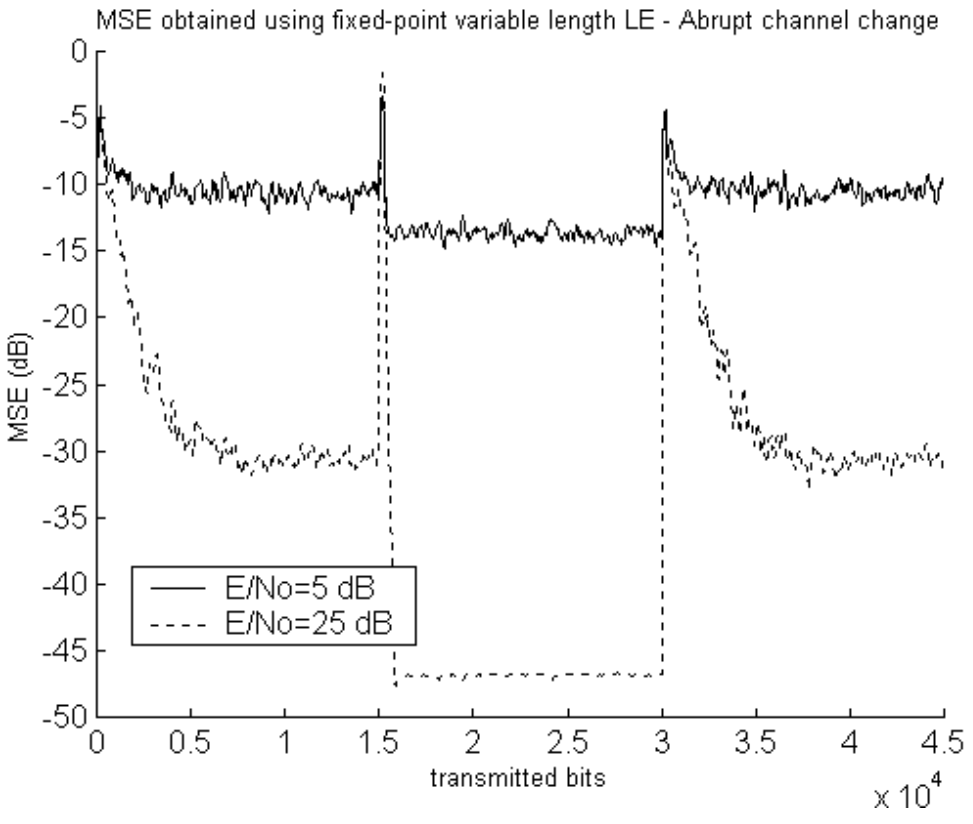

**Figure 8.3**: Windowed MSE for a VL LE confronting an abrupt channel change.

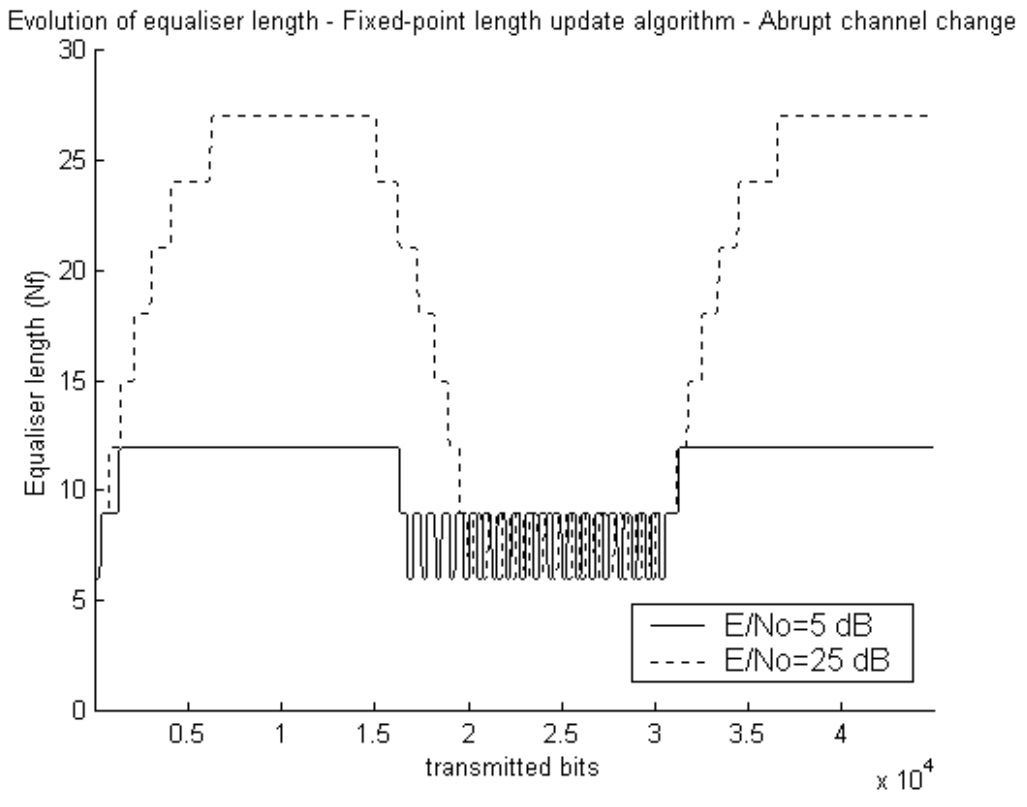

**Figure 8.4**: Length fluctuations of VL LE confronting an abrupt channel change.

Recall that the decision delay was set to 5 taps so the equaliser will never reduce below this number of taps. When E/No=5 dB, the length update algorithm detects the large noise level and only expands the equaliser up to 12 taps when operating in Channel model 2. The important point, once again, is that even when implemented on fixed-point arithmetic, the variable length LE is able to select dynamically the adequate number of taps to be used in the equaliser.

The next series of graphs show the results obtained when using the fixed-point versions of the algorithm controlling the FBF length in a DFE. The FBF had an initial and minimum length of 2 taps and the parameter $\chi$ was set to 0.01. The length of the FFF was fixed to 6 taps.

Figure 8.5 presents the evolution of the FBF length when equalising Channel model 2 for two different E/No levels. Recall that for a channel with N taps, N-1 taps are required in the FBF to cancel out all the postcursors from the channel-FFF combined response[29].

In these fixed-point simulations, the FBF is expanded to 11 taps for E/No=25 dB and 9 taps for E/No=5 dB, which in both cases is very close to the ideal FBF length, 10 taps. Comparing these results with those obtained using double-precision arithmetic (section 6.11.1), it is noticeable that the fixed-point results using $\chi$=0.01 are very similar to those obtained when using $\chi$=0.001 with double-precision arithmetic (figures 6.33 and 6.34).

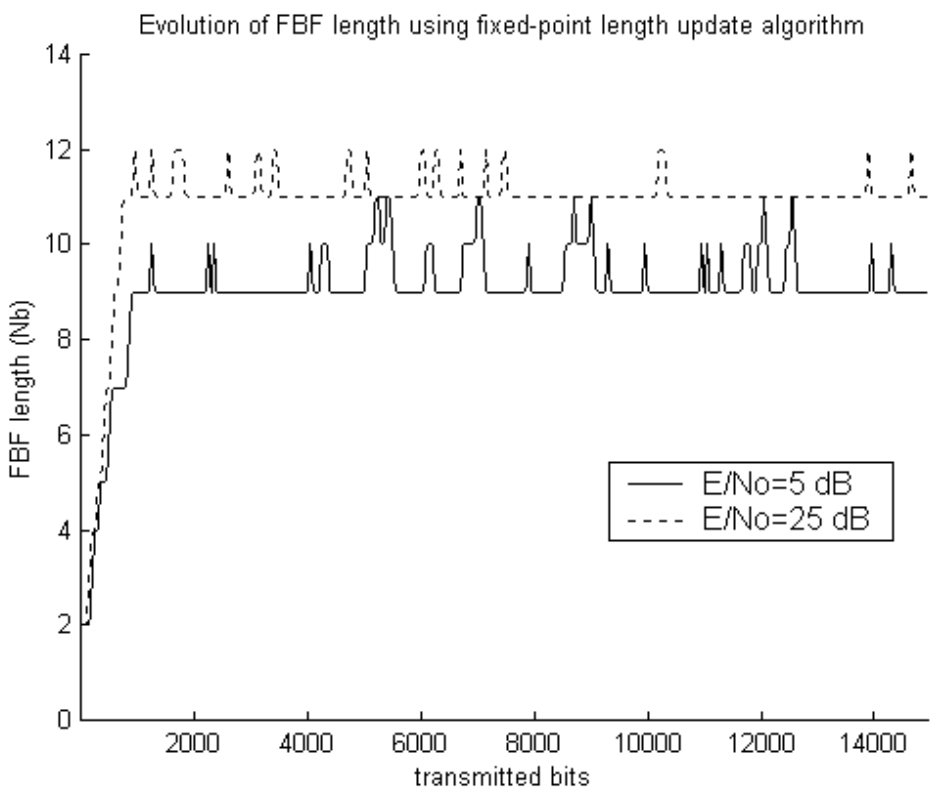

**Figure 8.5**: FBF length evolution for a VL FBF DFE compensating Channel model 2.

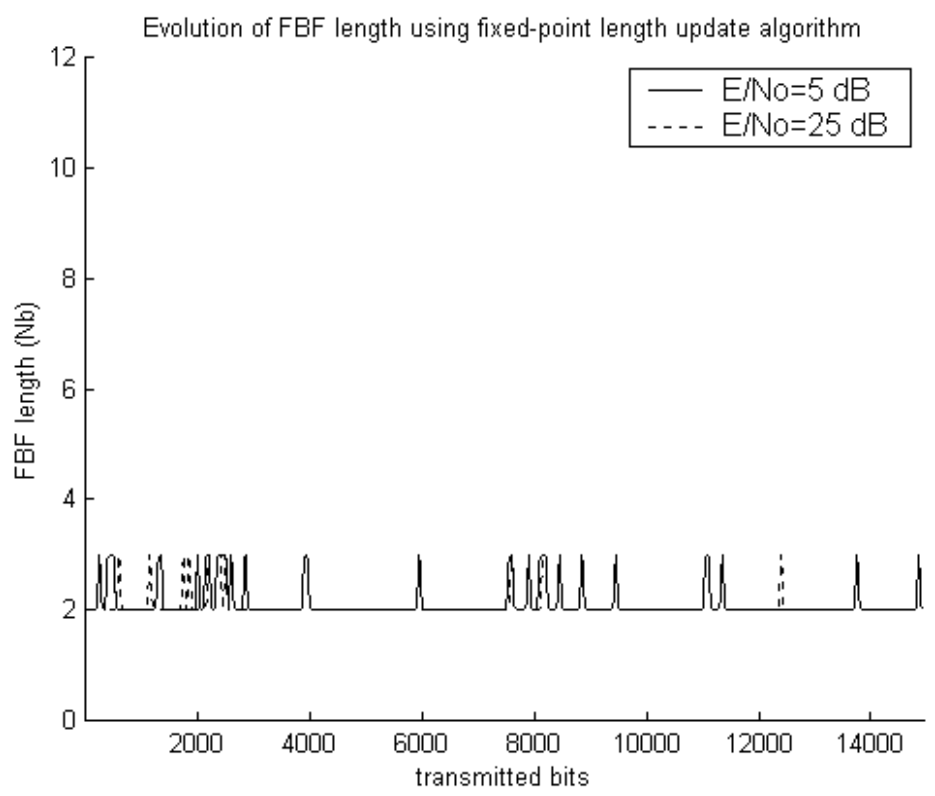

**Figure 8.6**: FBF length evolution for a VL FBF DFE compensating Channel model 1.

In figure 8.6 the length evolution of the FBF when compensating Channel model 1 is shown. For most of the time, the FBF has 2 taps. Recalling that this profile consisted of just 2 channel taps (i.e. only 1 postcursor), it can be concluded that the fixed-point FBF length update algorithm again selects the appropriate number of taps for the feedback section of the DFE.

---

[29] Assuming that the delay is set to Nf-1, where $N_f$ is the number of feedforward taps.

From figures 8.5 and 8.6, it is clear that the algorithm is able to “guess” how many taps must be used in each case without any a-priori knowledge of the environment.

Finally, the performance of the fixed-point VL FBF DFE has been verified in the context of an abrupt change in the channel profile. As in the VL LE case, initially the channel takes the form of Channel model 2 and after 15,000 transmitted bits changes to a single path channel. After another 15,000 iterations the channel changes back to the Channel model 2. The windowed MSE curve displayed in figure 8.7 reflects the changes in the channel impulse response. The evolution of the length of the FBF shown in figure 8.8 proves the effectiveness of the FBF length update algorithm to adjust the number of taps in the feedback section of a DFE according to the instantaneous characteristics of the environment.

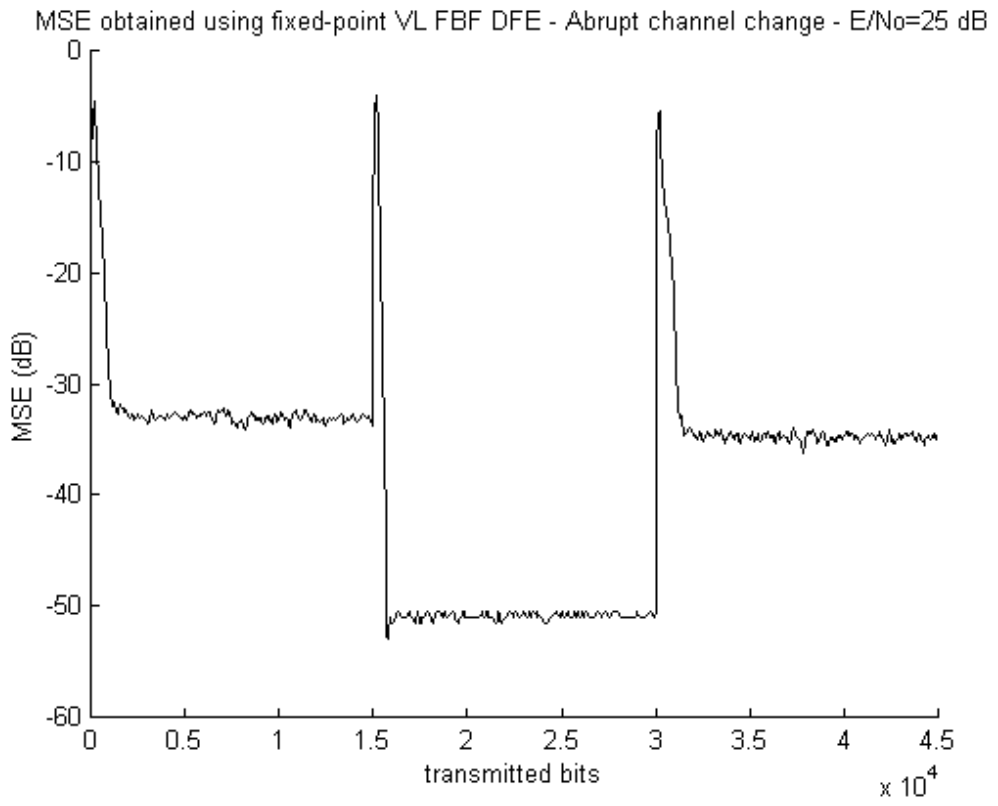

**Figure 8.7**: Windowed MSE for a VL FBF DFE confronting an abrupt channel change. E/No=25 dB.

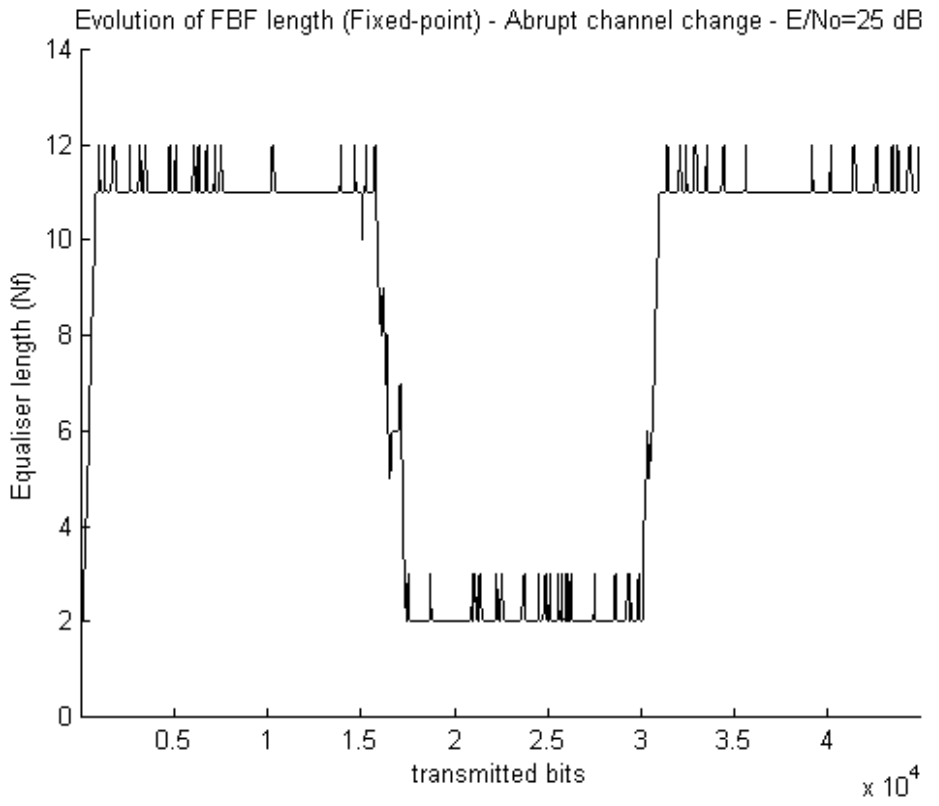

**Figure 8.8**: Length fluctuations of VL FBF DFE confronting an abrupt channel change. E/No=25 dB.

# 8.5 Implementation conclusions

This chapter has served various purposes. First, it has briefly considered some important issues, such as quantisation, that must be taken into account when implementing an algorithm on real hardware. Secondly a quick review of semiconductor technologies suitable to implement signal processing functions has been presented. Finally, fixed-point versions of the length update algorithms for the LE and DFE have been presented and tested in order to verify their correct functioning on a commercial DSP device. Simulation results have shown that variable length equalisers continue to perform very well when implemented with fixed-point arithmetic.

# 9 CONCLUSIONS AND FURTHER WORK

The final chapter in this thesis summarises the work done in this research project. Some suggestions for further research are also presented.

## 9.1 Conclusions

Our work has focused on possible forms of reconfiguration for adaptive equalisers, in particular linear equalisers (LEs) and decision feedback equalisers (DFEs), in the context of downlink mobile communications. These forms of equalisation have traditionally been used in TDMA mobile systems but recently their effectiveness in CDMA receivers has also been proved. Therefore, the results achieved are relevant to nearly any modern mobile standard.

LEs and DFEs are implemented using digital filters with a finite number of taps. The main question this work addresses is: "How long, in taps, should the equaliser in a mobile receiver be?". Even in the general context of digital communications, this apparently simple question has not found a definitive answer in the last 37 years since the adaptive equaliser was invented back in 1965. In the specific case of mobile handsets, and given that the equaliser length is directly related to power consumption, this issue becomes very relevant.

Our objective in this project has been the development of techniques to adjust, according to the channel conditions, the number of taps used in the equalising filters.

To that end, we have confirmed that the appropriate equaliser length, as was already known, depends strongly on the particular scenario in which the receiver is operating. Namely, it depends on the specific channel impulse response and noise/interference level. The adaptive algorithm (LMS and RLS have been considered) used to perform the equaliser coefficients adaptation has also been found to influence the length to be used.

In order to verify these dependencies, steady-state mean squared error (MSE) expressions for the LMS-LE and RLS-LE, typically used in channel estimation (system identification), have been re-derived for the case of linear equalisation (system inversion). These expressions, although only approximate due to the assumptions made, reveal some important information regarding the different parameters influencing the MSE. Broadly speaking, they are composed of an irreducible, channel-dependent component (MMSE) and an excess error component introduced by the adaptive algorithm (EMSE).

From the point of view of our work, it is important to observe how the equaliser length is related to each of these components. It is known from theory that adding taps to the equaliser will always provoke a reduction in MMSE level. However, it has been found by simulation that after a certain number of taps, the MMSE reduction is completely insignificant. On the other hand, the addition of taps will cause a reduction of the EMSE initially but a point is reached where the addition of more taps will cause the EMSE to start increasing. The tap whose addition does not reduce the MMSE any further and the tap whose addition causes the EMSE to start increasing are generally the same. If the equaliser is lengthened beyond this tap, the MSE increases and computations are wasted. One of our aims in this project has been to find a method to set the equaliser length at or near that critical number of taps.

A novel filtering structure that splits an N-tap FIR filter into K segments of P taps/segment (N=KP) has been introduced. This structure allows us to monitor how successive segments contribute to the reduction in MSE at the output of the equaliser. An algorithm has been devised that controls the number of segments to be used at each instant depending on the channel conditions.

This combination of segmented structure and supervisory algorithm is what we call a variable length linear equaliser (VL LE). The VL LE, in combination with LMS or RLS, has been tested in a wide range of different scenarios offering in all of them a very satisfactory performance. That is, the VL LE is able to estimate correctly the optimum number of segments (hence, taps) to be used in the equalisation according to the channel conditions.

As with the LE, a similar procedure has been followed with the DFE to assess its reconfiguration potential. First, the MSE expressions have been expanded to include the effects of the feedback filter, then the influence of the lengths of the FFF and FBF on the derived expressions have been analysed.

In the case of the DFE, some previous research had shown that the FFF, which is in charge of compensating the pre-cursor interference, can be kept fairly short. This is due to the fact that, normally, a mobile channel is composed of a few strong paths arriving first, being followed by a variable (potentially very large) number of weaker paths causing post-cursor interference. The FFF must take care of collecting most of the received energy in the strong paths while the FBF is in charge of cancelling the post-cursor interference. Given that the FFF is short, i.e. computationally inexpensive, our efforts have centred on the length adjustment of the FBF whose length can vary considerably depending on the specific environment in which is operating.

An algorithm has been presented to perform the FBF length adjustment. This algorithm works on a tap by tap basis and therefore does not require any special structure. The combination of the DFE with the algorithm controlling the length of the FBF is what we have called a variable length FBF DFE (VL FBF DFE).

The VL FBF DFE has also been tested on a wide variety of different situations and it has always adjusted the FBF length appropriately. Nevertheless, there is a situation where the proposed procedure might fail, the case of sparse channels. An additional strategy has been proposed in order to cope with this particular scenario.

As a side result, an interesting relation between decision delay and rate of convergence in the LMS-DFE has been found. A rule has been derived on how to chose the decision delay in order to minimise the convergence time of the LMS-DFE.

## 9.2 Further Work

There are quite a few ways to take the work presented in this thesis one step further. The four we believe are more interesting are explained next.

COMBINATION OF VARIABLE LENGTH LE/DFE WITH ITERATIVE DECODING

The first extension concerns the combination of the work presented in this thesis with modern channel decoding techniques such as turbo-decoding.

Turbo-codes can be decoded using iterative techniques, where at each iteration, a more reliable estimate is produced. In general, error control techniques perform well in uncorrelated channels, i.e. AWGN, whereas they are not adequate for linear distorting channels. Equalisers, on the other hand, work the other way round. They are effective in combating linear channel distortion but there is nothing they can do against AWGN. This complementary action of equalisation and decoding can be better exploited when both subsystems can be adjusted dynamically. This would be the case of a receiver combining the use of iterative turbo decoding (adjustable number of iterations) and the variable length equalisers presented (adjustable number of taps). In a system like this, a certain amount of computational power could be jointly assigned to the equalisation-decoding process. The specific amount of computation spent on each of the sub-processes would depend on the particular channel characteristics and would be able to change as channel conditions varies. This method ensures that each type of channel is tackled with the most effective countermeasure.

APPLICATION OF VARIABLE LENGTH EQUALISERS TO RECEIVERS WITH DIVERSITY

The second area for further work is the application of the principles of variable length equalisation to systems with spatial diversity. Advances in antenna technology start to make feasible the implementation of more than one antenna on a mobile handset.

In principle, and assuming that each diversity branch is connected to an independent equaliser, the techniques presented in this thesis could be applied to each individual branch. If the channel is subject to long fades, the different equalisers will have different lengths as all of them will see "different" channels (independently fading). Another interesting possibility to explore would be techniques that, given a fixed number of taps, distribute them optimally, and probably unevenly, among the different branches.

PERFORMANCE OF VARIABLE LENGTH STRUCTURES ON CDMA

A more straightforward and interesting task would be to test the variable length equaliser performance on a fully implemented CDMA environment. More specifically, it would be interesting to see how the variable length structure responds to variations in the number of users (interferers).

IMPLEMENTATION OF VARIABLE LENGTH EQUALISERS ON FPGA

The last area we propose for further work would cover the implementation aspects of variable length equalisers. In particular, the implementation of reconfigurable equalisers on reconfigurable logic devices such as FPGAs seems an attractive option worth considering in detail.

# References

[3GPP00] 3GPP R00 Technical Specifications. http://www.3gpp.org.

[Abdulrahman94] M. Abdulrahman, A. U. H. Sheikh and D. D. Falconer, “Decision Feedback Equalization for CDMA in Indoor Wireless Communications”, IEEE Journal on Selected Areas in Communications, Vol. 12, No. 4, 1994.

[Aboulnasr97] T. Aboulnasr and K. Mayyas, “A Robust Variable Step-Size LMS-Type Algorithm: Analysis and Simulations”, IEEE Transactions on Signal Processing, Vol. 45, No. 3, March 1997.

[Aboulnasr99] T. Aboulnasr and K. Mayyas, “Complexity Reduction of the NLMS Algorithm Via Selective Coefficient Update”, IEEE Transactions on Signal Processing, Vol. 47, No. 5, May 1999.

[Adachi98] F. Adachi, M. Sawahashi and H. Suda, “Wideband DS-CDMA for Next-Generation Mobile Communication Systems”, IEEE Communications Magazine, Vol. 36, No. 5, September 1998

[Ackenhusen99] J. G. Ackenhusen, “Real-Time Signal Processing”, Prentice Hall PTR, Upper Saddle River (US), 1999.

[Al-Dhahir95] N. Al-Dhahir and J. M. Cioffi, “MMSE Decision-Feedback Equalizers: Finite-Length Results”, IEEE Transactions on Information Theory, Vol. 41, No. 4, July 1995.

[Al-Dhahir96] N. Al-Dhahir and J. M. Cioffi, “Efficient Computation of the Delay-Optimized Finite-Length MMSE-DFE”, IEEE Transactions on Signal Processing, Vol. 44, No. 5, May 1996.

[Alberi98] M-L. Alberi, I. Fijalkow, J. D. Behm and T. J. Endres, “Fractionally-Spaced Equalization of Time-Varying Mobile Communications”, IEEE International Conference on Acoustics, Speech and Signal Processing, Seattle (US), May 1998 (ICASSP'98).

[Alfke98] P. Alfke, "Xilinx FPGAs: A Technical Overview for the First-Time User", Xilinx Application Note XAPP097, December 1998.

[Allaire97] B. Allaire and B. Fisher, "Block Adaptive Filter", Xilinx Application Note XAPP055, January 1997.

[Alouini99] M. S. Alouini, X. Tang and A. J. Goldsmith, "An Adaptive Modulation Scheme for Simultaneous Voice and Data Transmissions over Fading Channels", IEEE Journal on Selected Areas in Communications, Vol. 17, No. 5, May 1999.

[Altekar93] S. A. Altekar and N. C. Beaulieu, “Upper Bounds to the Error Probability of Decision Feedback Equalization”, IEEE Transactions on Information Theory, Vol. 39, No. 1, January 1993.

[Ariyavisitakul92] S. Ariyavisitakul, “A Decision Feedback Equalizer with Time-Reversal Structure”, IEEE Journal on Selected Areas in Communications, Vol. 10, No. 3, April 1992.

[Ariyavisitakul97a] S. Ariyavisitakul and L. J. Greenstein, “Reduced-Complexity Equalization Techniques for Broadband Wireless Channels”, IEEE Journal on Selected Areas in Communications, Vol. 15, No. 1, January 1997.

[Ariyavisitakul97b] S. Ariyavisitakul, N. R. Sollenberger and L. J. Greenstein, “Tap-Selectable Decision-Feedback Equalisation”, IEEE Transactions on Communications, Vol. 45, No. 12, December 1997.

[Ariyavisitakul99] S. Ariyavisitakul, J. H. Winters and I. Lee, “Optimum Space-Time Processors with Dispersive Interference: Unified Analysis and Required Filter Span”, IEEE Transactions on Communications, Vol. 47, No. 7, July 1999.

[Austin67] M. Austin, “Decision-Feedback Equalization for Digital Communication over Dispersive Channels”, M.I.T. Res. Lab Electron., Tech. Rep. 461, August 1967.

[Bahai99] A. Bahai and M. Rupp, “On the Learning Behavior of Decision Feedback Equalizers”, Asilomar Conference on Signals, Systems and Computers, Pacific Grove (US), November 1999.

[Baier94] A. Baier, U. Fiebig, W. Granzow, W. Koch, P. Teder and J. Thielecke, “Design Study for a CDMA-Based Third Generation Mobile Radio System”, IEEE Journal on Selected Areas in Communications, Vol. 12, No. 4, May 1994.

[Baines95] R. Baines, "The DSP Bottleneck", IEEE Communications Magazine, Vol. 33, No. 5, May 1995.

[Balaban91] P. Balaban and J. Salz, “Dual Diversity Combining and Equalization in Digital Cellular Mobile Radio”, IEEE Transactions on Vehicular Technology, Vol. 40, No. 2, May 1991.

[Balaban92a] P. Balaban and J. Salz, “Optimum Diversity Combining and Equalization in Digital Data Transmission with Applications to Cellular Mobile Radio – Part I: Theoretical Considerations”, IEEE Transactions on Communications, Vol. 40, No. 5, May 1992.

[Balaban92b] P. Balaban and J. Salz, “Optimum Diversity Combining and Equalization in Digital Data Transmission with Applications to Cellular Mobile Radio – Part II: Numerical Results”, IEEE Transactions on Communications, Vol. 40, No. 5, May 1992.

[Belfiore79] C. A. Belfiore and J. H. Park, “Decision Feedback Equalization”, Proceedings of the IEEE, Vol. 67, No. 8, August 1979.

[Bello63] P. A. Bello, “Characterization of Randomly Time-Variant Linear Channels”, IEEE Transactions on Communication Systems, Vol. CS-11, December 1963.

[Benedetto99] S. Benedetto and E. Biglieri, “Principles of Digital Transmission with Wireless Applications”, Kluwer Academic/Plenum Press, New York (The Netherlands), 1999.

[Berberidis00] K. Berberidis and A. A. Rontogiannis, "Efficient Decision Feedback Equaliser for Sparse Multipath Channels", IEEE International Conference on Acoustics, Speech and Signal Processing, (Istambul, Turkey), June 1999 (ICASSP'99).

[Berrou93] C. Berrou, A. Glavieux and P. Thitimajshima, “Near Shannon Limit Error-Correction Coding and Decoding: Turbo-Codes”, IEEE Proceedings of the Int. Conf. on Communications, Geneva (Switzerland), May 1993 (ICC’93).

[Boss98] D. Boss, K. Kammeyer and T. Peterman, “Is Blind Channel Estimation Feasible in Mobile Communication Systems? A Study Based on GSM”, IEEE Journal on Selected Areas on Communications, Vol. 16, No. 8, October 1998.

[Botto89] J-L. Botto and G. V. Moustakides, “Stabilizing the Fast Kalman Algorithms”, IEEE Transactions on Accoustics, Speech and Signal Processing, Vol. 37, No. 9, September 1989.

[Bottomley89] G. E. Bottomley and S. T. Alexander, “A Theoretical Basis for the Divergence of Conventional Recursive Least Squares Filters”, Proceedings of the International Conference on Acoustics, Speech and Signal Processing, Glasgow (UK), May 1989 (ICASSP'89).

[Bottomley91] G. E. Bottomley and S. T. Alexander, “A Novel Approach for Stabilizing Recursive Least Squares Filters”, IEEE Transactions on Signal Processing, Vol. 39, No. 8, August 1991.

[Braun91] W. R. Braun, “A Physical Mobile Radio Channel Model”, IEEE Transactions on Vehicular, Vol. 40, No. 2, May 1991.

[Butterweck95] H. J. Butterweck, “A Steady-State Analysis of the LMS Adaptive Algorithm without Use of the Independence Assumption”, IEEE International Conference on Acoustics, Speech and Signal Processing, Detroit (US), May 1995 (ICASSP'95).

[Carayannis83] G. Carayannis, D. G. Manolakis and N. Kalouptsidis, “A Fast Sequential Algorithm for Least-Squares Filtering and Prediction”, IEEE Transactions on Accoustics, Speech and Signal Processing, Vol. ASSP-31, No. 6, December 1983.

[Casas90] E. Casas and C. Leung, “A Simple Digital Fading Simulator for Mobile Radio”, IEEE Transactions on Vehicular Technology, Vol. 39, No. 3, August 1990.

[Casas97] R. A. Casas, F. Lopez de Victoria, I. Fijalkow, P. Schniter, T. J. Endres and C. R. Johnson, “On MMSE Fractionally-Spaced Equalizer Design”, International Conference on Digital Signal Processing, Santorini (Greece), July 1997.

[Chan94] N. L. B. Chan, “Multipath Propagation Effects on a CDMA Cellular System”, IEEE Transactions on Vehicular Technology, Vol. 43, No. 4, November 1994.

[Cioffi84] J. M. Cioffi, "Fast, Recursive-Least-Squares Transversals Filters for Adaptive Filtering", IEEE Transactions on Acoustics, Speech and Signal Processing, Vol. ASSP-32, 1984.

[Cioffi85] J. M. Cioffi, “When Do I Use an RLS Adaptive Filter?”, Asilomar Conference on Signals, Systems and Computers, Pacific Grove (US), November 1985.

[Cioffi95] J. M. Cioffi, G. P. Dudevoir, M. V. Eyuboglu, G. D. Forney, “MMSE Decision-Feedback Equalizers and Coding-Part I: Equalization Results”, IEEE Transactions on Communications, Vol. 43, No. 10, October 1995.

[Clark81] G. A. Clarck, S. K. Mitra and S. R. Parker, “Block Implementation of Adaptive Digital Filtes”, IEEE Transactions on Circuits and Systems, Vol. CAS-28, No. 6, June 1981.

[D’Aria91] G. D’Aria, R. Piermarini and V. Zingarelli, “Fast Adaptive Equalizers for Narro-Band TDMA Radio”, IEEE Transactions on Vehicular Technology, Vol. 40, No. 2, May 1991.

[Dahlman98] E. Dahlman, M. Nilsson and J. Skold, “UMTS/IMT-2000 Based on Wideband CDMA”, IEEE Personal Communications, Vol 47, No. 4, September 1998.

[Dembo88] A. Dembo, “Bounds on the Extreme Eigenvalues of Positive-Definite Toeplitz Matrices”, IEEE Transactions on Information Theory, Vol. 34, No. 2, March 1988.

[Douglas97] S. C. Douglas, “Adaptive Filters Employing Partial Updates”, IEEE Transaction on Circuits and Systems-II: Analog and Digital Signal Processing, Vol. 44, No. 3, March 1997.

[Duel-Hallen89] A. Duel-Hallen and C. Heegard, "Delayed Decision-Feedback Sequence Estimation", IEEE Transactions on Communications, Vol. 37, No. 5, May 1989.

[Duel-Hallen95] A. Duel-Hallen, J. Holtzman and Z. Zvonar, “Multiuser Detection for CDMA Systems”, IEEE Personal Communications, Vol. 2, No. 2, April 1995.

[Duel-Hallen00] A. Duel-Hallen, S. Hu and H. Hallen, "Long-Range Prediction of Fading Signals", IEEE Signal Processing Magazine, Vol. 17, No. 3, May 2000.

[Duttweiler74] D. L. Duttweiler, J. E. Mazo and D. G. Messerschmitt, “An Upper Bound on the Error Probability in Decision-Feedback Equalization”, IEEE Transactions on Information Theory, Vol. IT-20, No. 4, July 1974.

[Eweda94] E. Eweda, “Comparison of RLS, LMS, and Sign Algorithms for Tracking Randomly Time-Varying Channels”, IEEE Transactions on Signal Processing, Vol. 42, No. 11, November 1994.

[Eleftheriou86] E. Eleftheriou and D. D. Falconer, “Tracking Properties and Steady-State Performance of RLS Adaptive Filter Algorithms”, IEEE Transactions on Accoustics, Speech and Signal Processing, Vol. ASSP-34, No. 5, October 1986.

[Falconer78] D. D. Falconer and L. Ljung, “Application of Fast Kalman Estimation to Adaptive Equalization”, IEEE Transactions on Communications, Vol. COM-26, No. 10, October 1978.

[Farhang-Boroujeny98] B. Farhang-Boroujeny, “Adaptive Filters: Theory and Applications”, John Wiley & Sons, Singapore, 1998.

[Fertner98] A. Fertner, “Improvement of Bit-Error-Rate in Decision Feedback Equalizer by Preventing Decision-Error Propagation”, IEEE Transactions on Signal Processing, Vol. 46, No. 7, July 1998.

[Forney72] G. D. Forney, “Maximum Likelihood Sequence Estimation of Digital Sequences in the Presence of Intersymbol Interference”, IEEE Transactions on Information Theory, Vol. IT-18, No. 3, May 1972.

[Forney73] G. D. Forney, “The Viterbi Algorithm”, Proceedings of the IEEE, Vol. 61, No. 3, March 1973.

[Forney91] G. D. Forney and M. Vedat Eyuboglu, "Combined Equalization and Coding using Precoding", IEEE Communications Magazine, Vol. 29, No. 12, December 1991.

[Foschini82] G. J. Foschini and J. Salz, “Digital Communications Over Fading Radio Channels”, The Bell System Technical Journal, Vol. 62, No. 2, February 1983.

[Gersho81] A. Gersho and T. L. Lim, "Adaptive Cancellation of Interference for Data Transmission", Bell Systems Technical Journal, Vol. 60, November 1981.

[Gersho84] A. Gersho, "Adaptive Filtering with Binary Reinforcement", IEEE Transactions on Information Theory, Vol. IT-30, No. 2, March 1984.

[Gilhousen91] K. S. Gilhousen, I. M. Jacobs, R. Padovani, A. J. Viterbi, L. A. Weaver and C. E. Wheatley III, "On the Capacity of a Cellular CDMA System", IEEE Transactions on Vehicular Technology, Vol. 40, No. 2, May 1991.

[Gitlin77] R. D. Gitlin and F. R. Magee, "Self-Orthogonilizing Adaptive Equalization Algorithms", IEEE Transactions on Communications, Vol. COM-25, No. 7, July 1977.

[Gitlin79] R. D. Gitlin and S. B. Weinstein, "On the Required Tap-Weight Precision for Digitally Implemented, Adaptive, Mean-Squared Equalizers", The Bell System Technical Journal, Vol. 58, No. 2, February 1979.

[Gitlin81] R. D. Gitlin and S. B. Weinstein, "Fractionally-Spaced Equalization: An Improved Digital Transversal Equalizer", The Bell System Technical Journal, Vol. 60, No. 2, February 1981.

[Gitlin82] R. D. Gitlin, H. C. Meadors and S. B. Weinstein, "The Tap-Leakage Algorithm: An Algorithm for the Stable Operation of a Digitally Implemented, Fractionally Spaced Adaptive Equalizer", The Bell System Technical Journal, Vol. 61, No. 8, October 1982.

[Gitlin92] R. D. Gitlin, J. F. Hayes and S. B. Weinstein, "Data Communications Principles", Plenum Press, New York (US), 1992.

[Glentis99] G. O. Glentis, K. Berberidis and S. Theodoridis, "Efficient Least Squares Adaptive Algorithms for FIR Filtering", IEEE Signal Processing Magazine, Vol. 16, No. 4, July 1999.

[Glover98] I. A. Glover and P. M. Grant, "Digital Communications", Prentice Hall Europe, Hemel Hempstead (UK), 1998.

[Godard80] D. N. Godard, "Self-Recovering Equalization and Tracking in Two-Dimensional Data Communication Systems", IEEE Transactions on Communications, Vol. COM-28, No. 11, November 1980.

[Goodman97] D. J. Goodman, J. Borras, N. B. Mandayam and R. D. Yates, "INFOSTATIONS: A New System Model for Data and Messaging Services", Proceedings of the IEEE Vehicular Technology Conference, Phoenix (US), May 1997, (VTC'97-Spring).

[Grant99] P. M. Grant, S. Spangenberg, D. G. M. Cruickshank, S. McLaughlin and B. Mulgrew, "New Adaptive Multiuser Detection Technique for CDMA Mobile Receiver", Proceedings of the IEEE International Symposium on Personal, Indoor and Mobile Radio Communications", Osaka (Japan), September 1999, (PIMRC'99).

[Gray72] R. M. Gray, “On the Asymptotic Eigenvalue Distribution of Toeplitz Matrices”, IEEE Transactions on Information Theory, Vol. IT-18, No. 6, November 1972.

[Harashima72] H. Harashima and H. Miyakawa, "Matched-Transmission Technique for Channels with Intersymbol Interference", IEEE Transactions on Communications, Vol. COM-20, Vol. 8, August 1972.

[Haykin01] S. Haykin, “Communication Systems, 4th Edition”, Wiley, New York (US), 2001.

[Haykin95] S. Haykin, A. H. Sayed, J.Ziedler, P. Yee and P. Wei, "Tracking of Linear Time-Variant Systems", IEEE Conference on Military Communications, San Diego (US), November 1995 (MILCOM'95).

[Haykin96] S. Haykin, “Adaptive Filter Theory, 3rd Edition”, Prentice Hall, Upper Saddle Rive (US), 1996.

[Hanzo00] L. Hanzo, C.H. Wong and P. Cherriman, “Channel-Adaptive Wideband Wireless Video Telephony”, IEEE Signal Processing Magazine, Vol. 17, No. 4, July 2000.

[Holma01] H. Holma and A. Toskala, "WCDMA for UMTS, Revised Edition", Wiley, Chichester (UK), 2001.

[Hooli00] K. Hooli, M. Latva-aho, M. Juntti, “Performance Evaluation of Adaptive Chip-Level Channel Equalizers in WCDMA Downlink”, IEEE International Conference on Communications, Helsinki (Finland), June 2000 (ICC'00).

[Husson99] L. Husson and J. C. Dany, “A New Method for Reducing the Power Consumption of Portable Handsets in TDMA Mobile Systems: Conditional Equalization”, IEEE Transactions on Vehicular Technology, Vol. 48, No. 6, November 1996.

[Hubbing91] N. E. Hubbing and S. T. Alexander, “Statistical Analysis of Initialization Methods for RLS Adaptive Filters”, IEEE Transactions on Signal Processing, Vol. 39, No. 8, August 1991.

[Ifeachor93] E. C. Ifeachor and B. W. Jervis, “Digital Signal Processing: A Practical Approach”, Addison Wesley, Harlow (UK), 1993.

[Jakes74] W. C. Jakes, “Microwave Mobile Communications”, Wiley, (New York, US), 1974.

[Jeruchim92] M. C. Jeruchim, B. Balaban and J. S. Shanmugan, "Simulation of Communication Systems", Plenum Press, New York (US), 1992.

[Jung93] P. Jung, P. W. Baier and A. Steil, "Advantages of CDMA and Spread Spectrum Techniques over FDMA and TDMA in Cellular Mobile Radio Applications", IEEE Transactions on Vehicular Technology, Vol. 42, No. 3, August 1993.

[Kaaranen01] H. Kaaranen, A. Ahtiainen, L. Laitinen, S. Naghian and V. Niemi, "UMTS Networks: Architecture, Mobility and Services", Wiley, Chichester (UK), 2001.

[Kohno98] R. Kohno, “Spatial and Temporal Communication Theory Using Adaptive Antenna Array”, IEEE Personal Communications, Vol. 51, No. 1, February 1998.

[Klein94] A. Klein, G. k. Kaleh and P. W. Baier, “Equalizers for Multi-User Detection in Code Division Multiple Access Mobile Radio Systems”, IEEE Vehicular Technology Conference, Stockholm (Sweden), June 1994 (VTC-94 Spring).

[Koulakiotis00] D. Koulakiotis and A. H. Aghvami, “Data Detection Techniques for DS/CDMA Mobile Systems: A Review”, IEEE Personal Communications, Vol. 7, No. 3, June 2000.

[Kuo01] S. M. Kuo and B. H. Lee, "Real-Time Digital Signal Processing: Implementations, Applications, and Experiments with the Tms320C55X", Wiley, Chichester (UK), 2001.

[Kwong92] R. H. Kwong and E. W. Johnston, “A Variable Step Size LMS Algorithm”, IEEE Transactions on Signal Processing, Vol. 40, No. 7, July 1992.

[Labat98] J. Labat, O. Macchi and C. Laot, “Adaptive Decision Feedback Equalization: Can You Skip the Training Period”, IEEE Transactions on Communications, Vol. 46, No. 7, July 1998.

[Lange02] K. Lange, G. Blanke and R.Rifaat, “A Software Solution for Chip Rate Processing in CDMA Wireless Infrastructure”, IEEE Communications Magazine, Vol. 40, No. 2, February 2002.

[Laurenson94] D. J. Laurenson, D. G. M. Cruickshank and G. J. R. Povey, “A Computationally Efficient Multipath Channel Simulator for the COST207 Models”, IEE Colloquium on Computer Modelling of Communication Systems, 1994.

[Lee91] W. C. Y. Lee, "Overview of Cellular CDMA", IEEE Transactions on Vehicular Technology, Vol. 40, No. 2, May 1991.

[Lee98] I. Lee, “Optimization of Tap Spacings for the Tapped Delay Line Decision Feedback Equalizer”, IEEE International Conference on Communications, Atlanta (US), June 1998 (ICC'98).

[Lehne99] P. H. Lehne and M. Pettersen, “An Overview of Smart Antenna Technology for Mobile Communications Systems”, IEEE Communications Surveys, Vol. 2, No. 4, Fourth Quarter 1999.

[Lim98] T. J. Lim and S. Roy, “Adaptive Filters in Multiuser (MU) CDMA Detection”, Wireless Networks, Vol. 4, No. 4, 1998.

[Litwin99] L. R. Litwin, "Blind Channel Equalization", IEEE Potentials, Vol. 18, No. 4, October/November 1999.

[Lo91] N. W. K. Lo, D. D. Falconer and A. U. H. Sheikh, "Adaptive Equalization and Diversity Combining for Mobile Radio Using Interpolated Channel Estimates", IEEE Transactions on Vehicular Technologies, Vol. 40, No. 3, August 1991.

[Lo95] N. W. K. Lo, D. D. Falconer and A. U. H. Sheikh, "Adaptive Equalization for Co-Channel Interference in a Multipath Fading Environment", IEEE Transactions on Communications, Vol. 43, No. 2/3/4, February/March/April 1995.

[Lo99] B. C. W. Lo and K. B. Letaief, "Adaptive Equalization and Interference Cancellation for Wireless Communication Systems", IEEE Transactions on Communications, Vol. 47, No. 4, April 1999.

[Lucky65] R. W. Lucky, "Automatic Equalization for Digital Communication", The Bell System Technical Journal, Vol. 44, No. 4, April 1965.

[Lucky66] R. W. Lucky, "Techniques for Adaptive Equalization of Digital Communications Systems", The Bell System Technical Journal, February 1966.

[Lupas89] R. Lupas and S. Verdu, "Linear Multiuser Detectors for Synchronous Code-Division Multiple-Access Channels", IEEE Transactions on Information Theory, January 1989.

[Lyons97] R. G. Lyons, "Understanding Digital Signal Processing", Addison Wesley, Reading (US), 1997.

[Madhow94] U. Madhow and M. L. Honig, "MMSE Interference Suppression for Direct-Sequence Spread-Spectrum CDMA", IEEE Transactions on Communications, Vol. 42, No. 12, December 1994.

[Manolakis83] D. G. Manolakis, N. Kaloputsidis and G. Carayannis, "Fast Algorithms for Discrete-Time Wiener Filters with Optimum Lag", IEEE Transactions on Acoustics, Speech and Signal Processing, Vol. ASSP-31, No. 1, February 1983.

[Martin01a] R. K. Martin, W. A. Sethares, R. C. Williamson and C. R. Johnson, "Exploiting Sparsity in Adaptive Filters", The 2001 Conference on Information Sciences and Systems, Baltimore (US), 2001.

[Martin01b] R. K. Martin and C. R. Johnson, "NSLMS: a Proportional Weight Algorithm for Sparse Adaptive Filters", 35$^{th}$ Asilomar Conference on Signals, Systems and Computers, Pacific Grove (US), November 2001.

[Mathews93] V. J. Mathews and Z. Xie, "A Stochastic Gradient Adaptive Filter with Gradient Adaptive Step Size", IEEE Transactions on Signal Processing, Vol. 41, No. 6, June 1993.

[Mazo79] J. E. Mazo, "On the Independence Theory of Equalizer Convergence", Bell Systems Technical Journal, Vol. 58, No. 5, May-June 1979.

[Meyr95] H. Meyr and R. Subramanian, “Advanced Digital Receiver Principles and Technologies for PCS”, IEEE Communications Magazine, Vol. 33, No. 1, January 1995.

[Mitola95] J. Mitola, “The Software Radio Architecture”, IEEE Communications Magazine, Vol. 33, No. 5, May 1995.

[Mitra01] S. Mitra, "Digital Signal Processing: A Computer Based Approach", McGraw-Hill, New York (US), 2001.

[Morf76] M. Morf, L. Ljung and T. Kailath, "Fast Algorithms for Recursive Identification", Proceedings of the IEEE Conference on Decision and Control, Clearwater Beach (US), 1976.

[Moshavi96] S. Moshavi, “Multi-User Detection for DS-CDMA Communications”, IEEE Communications Magazine, Vol. 34, No. 10, October 1996.

[Mou87] Z. J. Mou and P. Duhamel, “Fast FIR Filtering: Algorithms and Implementations”, Signal Processing, Vol. 13, No. 4, 1987.

[Moustakides89] G. V. Moustakides, “Correcting the Instability due to Finite Precision of the Fast Kalman Identification Algorithms”, Signal Processing, Vol. 18, No. 1, September 1989.

[Moustakides91] G. V. Moustakides and S. Theodoridis, “Fast Newton Transversal Filters – A New Class of Adaptive Estimation Algorithms”, IEEE Transactions on Signal Processing, Vol. 39, No. 10, October 1991.

[Mueller75] K. H. Mueller and D. A. Spaulding, “Cyclic Equalization – A New Rapidly Converging Equalization Technique for Synchronous Data Communication”, The Bell System Technical Journal, Vol. 54, No. 2, February 1975.

[Mueller81] M. S. Mueller and J. Salz, “A Unified Theory of Data-Aided Equalization”, The Bell System Technical Journal, Vol. 60, No. 9, November 1981.

[Mulgrew98] B. Mulgrew, P.M. Grant and J. P. Thompson, “Digital Signal Processing, concepts and applications”- Palgrave MacMillan, London (UK), 1998.

[Niger91] P. Niger and P. Vandamme, “Performance of Equalization Techniques in a Radio Interference Environment”, IEEE Transactions on Communications, Vol. 29, No. 3, March 1991.

[Noble88] B. Noble and J.W. Daniel, “Applied Linear Algebra, 3rd Edition”, Prentice Hall, Englewood Cliffs (US), 1988.

[Noblet98] C.Noblet and A.H.Aghvami: "Accessing the over-the air software download for reconfigurable terminal", IEE Colloquim Digest, No. 242, 1998.

[Patzold96] M. Patzold, U. Killat and F. Laue, "A Deterministic Digital Simulation Model for Suzuki Processes with Application to a Shadowed Rayleigh Land Mobile Radio Channel", IEEE Transactions on Vehicular Technology, Vol. 45, No. 2, May 1996.

[Patzold98] M. Patzold, U. Killat, F. Laue and Y. Li, "On the Statistical Properties of Deterministic Simulation Models for Mobile Fading Channels", IEEE Transactions on Vehicular Technology, Vol. 47, No. 1, February 1998.

[Pickholtz91] R. L. Pickholtz, L.B. Milstein and D. L. Schilling, "Spread Spectrum for Mobile Communications", IEEE Transactions on Vehicular Technology, Vol. 40, No. 2, May 1991.

[Press93] W. H. Press, S. A. Teukolsky, W. T. Vetterling and B. P. Flannery, "Numerical Recipes in C, 2nd Ed.", Cambridge University Press, Cambridge (UK), 1993.

[Price58] R. Price and P. E. Green, "A Communication Technique for Multipath Channels", Proceedings of the IRE, Vol. 46, pp. 555-570, March 1958.

[Proakis91] J. G. Proakis, "Adaptive Equalization for TDMA Digital Mobile Radio", IEEE Transactions on Vehicular Technology, Vol. 40, No. 2, May 1991.

[Proakis95] J. G. Proakis, "Digital Communications, 3rd Edition", McGraw-Hill, New York (US), 1995.

[Qureshi73] S. U. H. Qureshi, "Adjustment of the Position of the Reference Tap of an Adaptive Equalizer", IEEE Transactions on Communications, Vol. COM-21, No. 9, September 1973.

[Qureshi77] S. U. H Qureshi, "Fast Start-Up Equalization with Periodic Training Sequences", IEEE Transactions on Information Theory, Vol. IT-23, No. 5, September 1977.

[Qureshi85] S. U. H. Qureshi, "Adaptive Equalization", Proceedings of the IEEE, Vol. 73, No. 9, September 1985.

[Rabaey98] J. M. Rabaey, R. Brodersenm W. Gaas and T. Nishitani, "VLSI Design and Implementation Fuels the Signal-Processing Revolution", IEEE Signal Processing Magazine, Vol. 15, No. 1, January 1998

[Raghavan93] S. A. Raghavanm J. K. Wolf, L. B. Milstein and L. C. Barbosa, "Nonuniformly Spaced Tapped-Delay-Line Equalizers", IEEE Transactions on Communications, Vol. 41, No. 9, September 1995.

[Rappaport96] T. S. Rapport, "Wireless Communications Principles and Practice", Prentice Hall, Upper Saddle River (US), 1996.

[Riera-Palou99] F. Riera-Palou, C. Chaikalis and J. M. Noras, "Reconfigurable Mobile Terminal Requirements for Third Generation Applications", IEE Colloquium on UMTS Terminals and Software Radio, Glasgow (UK), April 1999.

[Riera-Palou00a] F. Riera-Palou, J. M. Noras and D. G. M. Cruickshank, "Equalisation of HF Channels using the SFAEST Algorithm", Proceedings of the IEE International Conference on HF Radio Systems and Techniques, Guilford (UK), May 2000.

[Riera-Palou00b] F. Riera-Palou, J. M. Noras and D. G. M. Cruickshank, "Variable Length Equalisers for Broadband Mobile Systems", Proceedings of the IEEE Vehicular Technology Conference, Boston (US), September 2000 (VTC'00-Fall).

[Riera-Palou01a] F. Riera-Palou, J. M. Noras and D. G. M. Cruickshank, "Segmented Equalizers with Dynamic Length Selection", Proceedings of the Asilomar Conference on Signals, Systems and Computers, Pacific Groove (US), November 2001.

[Riera-Palou01b] F. Riera-Palou, J. M. Noras and D. G. M. Cruickshank, "Linear equalisers with dynamic and automatic length selection", Electronics Letters, Vol. 37, No. 25, December 6, 2001.

[Riera-Palou02] F. Riera-Palou, J. M. Noras and D. G. M. Cruickshank, "Analysis of the decision delay effect on the convergence of gradient recursive decision feedback equalizers", to appear in Proceedings of the IEEE International Conference on Acoustics, Speech and Signal Processing, Orlando (US), May 2002.

[Rupp99] M. Rupp and A. Bahai, "Equalisation Techniques for TDMA Systems", Lucent Internal Report (obtained directly from the author), 1999.

[Rupp97a] M. Rupp, "Adaptive DFE Algorithms for IS-136 Based TDMA Cellular Phones", IEEE International Conference on Acoustics, Speech and Signal Processing, Munich (Germany), April 1997 (ICASSP'97)

[Rupp97b] M. Rupp and A. Bahai, "Training and Tracking of Adaptive DFE Algorithms Under IS-136", IEEE Signal Processing Workshop on Signal Processing Advances in Wireless Communications, Paris (France), April 1997.

[Saltzberg68] B. R. Saltzberg, "Intersymbol Interference Error Bounds with Application to Ideal Bandlimited Signaling", IEEE Transactions on Information Theory, Vol. IT-14, No. 7, July 1968.

[Salz73] J. Salz, “Optimum Mean-Square Decision Feedback Equalization”, The Bell System Technical Journal, Vol. 32, No. 8, October 1973.

[Sari82] H. Sari, “Simplified Algorithms for Adaptive Channel Equalization”, Philips Journal of Research, 37, 56-77, 1982.

[Shannon48] C. E. Shannon, “A Mathematical Theory of Communication”, Bell System Technical Journal, Vol. 27, 1948.

[Shukla91] P. K. Shukla and L. F. Turner, “Channel-Estimation-Based Adaptive DFE for Fading Multipath Radio Channels”, IEE Proceedings-I, Vol. 138, No. 6, December 1991.

[Shynk89] J. Shynk, “Adaptive IIR Filtering”, IEEE Acoustics, Speech and Signal Processing Magazine, Vol. 6, No. 2, April 1989.

[Shynk92] J. Shynk, “Frequency-Domain and Multirate Adaptive Filtering”, IEEE Signal Processing Magazine, Vol. 9, No. 1, January 1992.

[Sklar97a] B. Sklar, “Rayleigh Fading Channels in Mobile Digital Communications Systems Part I: Characterization”, IEEE Communications Magazine, Vol. 35, No. 7, July 1997.

[Sklar97b] B. Sklar, “Rayleigh Fading Channels in Mobile Digital Communications Systems Part II: Mitigation”, IEEE Communications Magazine, Vol. 35, No. 7, July 1997.

[Sklar01] B. Sklar, “Digital Communications: Fundamentals and Applications, 2nd Edition”, Prentice Hall PTR, Upper Saddle River (NJ, US), 2001.

[Slock91] D. T. M. Slock and T. Kailath, “Numerically Stable Fast Transversal Filters for Recursive Least Squares Adaptive Filtering”, IEEE Transactions on Signal Processing, Vol. 39, No. 1, January 1991.

[Smee95] J. E. Smee and N. C. Beaulieu, “New Methods for Evaluating Equalizer Error Rate Performance”, IEEE Vehicular Technology Conference, Chicago (US), July 1995 (VTC'95).

[Smee97] J. E. Smee and N. C. Beaulieu, “On the Equivalence of the Simultaneous and Separate MMSE Optimizations of a DFE FFF and FBF”, IEEE Transactions on Communications, Vol. 45, No. 2, February 1997.

[Smee98] J. E. Smee and N. C. Beaulieu, “Error-Rate Evaluation of Linear Equalization and Decision Feedback Equalization with Error Propagation”, IEEE Transactions on Communications, Vol. 46, No. 5, May 1998.

[Smith86] J. Smith, "Modern Communication Circuits", McGraw-Hill, (New York, US), 1986.

[Srikanteswara00] S. Srikanteswara, J. H. Reed, P. Athanas and R. Boyle, "A Soft Radio Architecture for Reconfigurable Platforms", IEEE Communications Magazine, Vol. 38 No. 2, February 2000.

[Steele99] R. Steele and L. Hanzo (editors), "Mobile Radio Communications, 2nd Ed.", Wiley, New York (US), 1999.

[Stremler90] F. G. Stremler, "Introduction to Communication Systems, 3rd Edition", Addison Wesley, Wokingham (UK), 1990.

[Tessier01] R. Tessier and W. Burleson, "Reconfigurable Computing for Digital Signal Processing: A Survey", Journal of VLSI Signal Processing, Vol. 28, 7-27, Kluwer, 2001.

[TI98] Texas Instruments Manual – TMS320C54X DSP Reference Set, 1998.

[Tijdhof97] J. J. H. Tijdhof, J. van Bussel, C. J. E. van Heerde, C. H. Slump and M. J. Bentum, "On the Design and Realization of Adaptive Equalization for Mobile Communication", IEEE Signal Processing Workshop on Signal Processing Advances in Wireless Communications, Paris (France), April 1997.

[Tomlinson71] M. Tomlinson, "New Automatic Equaliser Employing Modulo Arithmetic", Electronic Letters, Vol. 7, No. 5-6, March 1971.

[Treichler96] J. R. Treichler, I. Fijalkow and C. R. Johnson, "Fractionally Spaced Equalizers: How Long Should They Really Be?", IEEE Signal Processing Magazine, May 1996.

[Turin80] G. L. Turin, "Introduction to Spread-Spectrum Antimultipath Techniques and Their Application to Urban Digital Radio", Proceedings of the IEEE, Vol. 68, No. 3, March 1980.

[Tuttlebee98] W.Tuttlebee, "Software radio – Impacts and Implications", IEEE Proceedings of the IEEE International Symposium on Spread Spectrum Techniques and Applications, Sun City (South Africa), July 1998 (ISSSTA'98).

[Ungerboeck72] G. Ungerboeck, "Theory on the Speed of Convergence in Adaptive Equalizers for Digital Communication", IBM Journal of Research and Development, November 1972.

[Ungerboeck76] G. Ungerboeck, "Fractional Tap-Spacing Equalizer and Consequences for Clock Recovery in Data Modems", IEEE Transactions on Communications, Vol. COM-24, No. 8, August 1976.

[Ungerboeck87] G. Ungerboeck, "Trellis-Coded Modulation with Redundant Signal Sets. Parts I and II", IEEE Communications Magazine, Vol. 25, No. 2, 1987.

[Voois96] P. A. Voois, I. Lee, J.M. Cioffi, "The effect of decision delay in finite-length decision feedback equalization", IEEE Transactions on Information Theory, Vol. 42, No. 2, March 1996.

[Vucetic91] B. Vucetic, "An Adaptive Coding Scheme for Time-Varying Channels", IEEE Transactions on Communications, Vol. 39, No. 5, May 1991.

[Webb91] W. T. Webb and R. Steele, "Equaliser Techniques for QAM Transmissions over Dispersive Mobile Radio Channels", IEE Proceeding-I, Vol. 138, No. 6, December 1991.

[Webb95] W. T. Webb and R. Steele, "Variable Rate QAM for Mobile Radio", IEEE Transactions on Communications, Vol. 43, No. 7, July 1995.

[Wepman95] J. A. Wepman, "A/D Converters and their applications in radio receivers", IEEE Communications Magazine, Vol. 33, No. 5, May 1995.

[Wesolowski92] K. Wesolowski, C. M. Zhao and W. Rupprecht, "Adaptive LMS Transversal Filters with Controlled Length", IEE Proceedings-F, Vol. 139, No. 3, June 1992.

[Wesolowski95] K. Wesolowski, "Adaptive Blind Equalizers with Automatically Controlled Paramaters", IEEE Transactions on Communications, Vol. 43, No. 2/3/4, February/March/April 1995.

[Widrow60] B. Widrow and M. E. Hoff, "Adaptive Switching Circuits", IRE WESCON Convention Record, Pt. 4, 1960.

[Widrow76] B. Widrow, J. M. McCool, M. G. Larimore, C. R. Johnson, "Stationary and Nonstationary Learning Characteristics of the LMS Adaptive Filter", Proceedings of the IEEE, Vol. 64, No. 8, August 1976.

[Wilson93] S. K. Wilson and J. M. Cioffi, "Equalization Techniques for Direct Sequence Code-Division Multiple Access Systems in Multipath Channels", IEEE International Symposium on Information Theory, San Antonio (US), January 1993 (ISIT'93).

[Winters98] J. H. Winters, "Smart Antenna for Wireless Systems", IEEE Personal Communications, Vol. 5, No. 1, February 1998.

[Woo98] H. C. Woo, "Improved Stochastic Gradient Adaptive Filter with Gradient Adaptive Step Size", Electronics Letters, Vol. 34, No. 13, 25th June 1998.

[Woodward98] G. Woodward and B. S. Vucetic, "Adaptive Detection for DS-CDMA", Proceedings of the IEEE, Vol. 86, No. 7, July 1998.

[Yates00] R. D. Yates and N. B. Mandayam, "Challenges in Low-Cost Wireless Data Transmission", IEEE Signal Processing Magazine, Vol. 17, No. 3, May 2000.